# THE BEGINNING
## AND
# THE END

## The Meaning of Life in a Cosmological Perspective

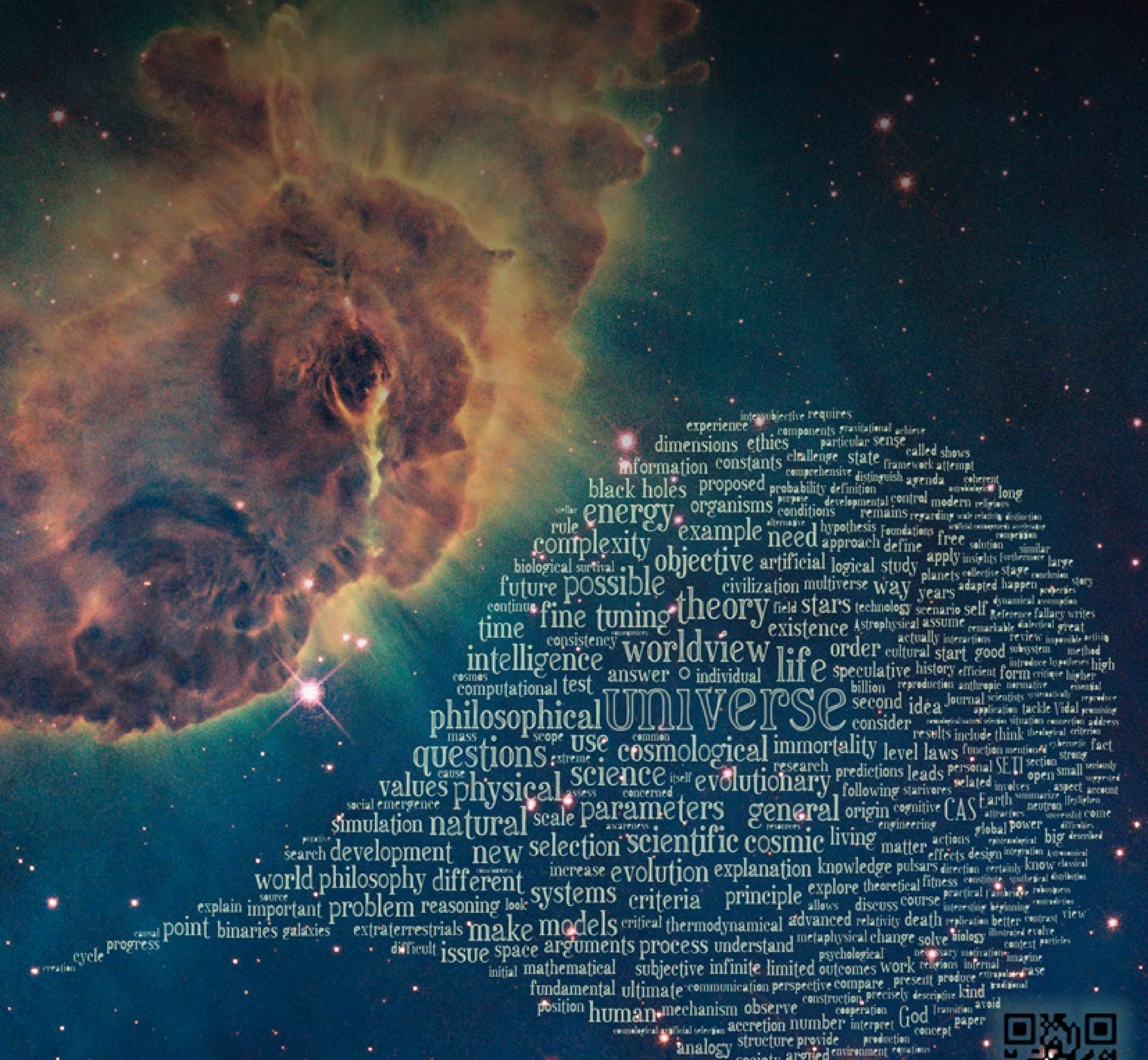

CLÉMENT VIDAL

# Dedication

*To every human, artificial or extraterrestrial intelligence in this universe.*

# Contents

# Preface - Psychiatry and Cosmological Speculation

After high school, when I told my aunt that I wanted to study philosophy at the university, she looked at me very empathically and said: “have you considered to consult a psychiatrist? they can be very helpful, you know.” I was shocked. What had the philosophical pursuit to understand humanity and the cosmos to do with psychic health? Maybe she confused philosophy and psychology. Or maybe she thought that studying philosophy leads nowhere socially or professionally and that I was simply experiencing a temporary existential crisis. Seeing a psychiatrist would put me back on the right social track.

But maybe she was right after all. Maybe asking fundamental and philosophical questions *is* an illness. In that case I am proud to be ill. Even more, my hope is that it is highly contagious, and that you, my reader, will want to pursue even further the intellectual journey I will now share with you. But first, a word of caution.

I would like to warn my readers that this work contains cosmological speculations[1]. The speculations I discuss are cosmological because they stretch over billions of years and billions of light years. How can we legitimate such speculations? Part I constitutes one third of this work and is dedicated to a study of the philosophical method at large. I argue that a major aim of philosophy is to construct *comprehensive* and *coherent* worldviews. To construct such worldviews demands to answer big questions like “where do we come from?”, “where are we going?” or “are we alone in the universe?” Motivated by our existential need to answer such big questions, we naturally tend to speculate. Isn't such an endeavor in striking contradiction with the rigor of the scientific enterprise? Is it just fantasy?

Certainly not! Speculating does not mean being unscientific. On the contrary, it means identifying and relying on the most fundamental scientific theories and principles, and then extrapolate them. In my speculations, mostly contained in Part III, I have done my best to focus on the most robust and general scientific theories such as principles of relativity theories, thermodynamics, systems theory, evolution, theoretical computer science or logic.

Of course, many speculations turn out to be wrong. As the multiple failures in the history of science show us, the risk for a speculative theory to become refuted is real. Indeed, cosmological speculations rely on and extrapolate from our current theories. Cosmological models in the next decades might refute speculations of our time, or lead to very different kinds of speculations. Speculating also means extrapolating the physical laws we can experiment with on Earth, to extreme regimes of large energies, high densities, as well as huge space and time scales. We need to be aware that such extrapolations are subject to strong uncertainties, for example because we do not have yet an established theory of quantum gravity.

Historian of cosmology Helge Kragh (2011, 2) wrote in the introduction of his book *Higher Speculations: Grand Theories and Failed Revolutions in Physics and Cosmology* :

> Speculations have always been an integrated part of the physical sciences, sometimes hidden under the more palatable term 'hypotheses'. Indeed, fundamental physics without some element of speculation is hardly conceivable.

---

1 To make this warning explicit, my PhD at the Vrije Universiteit Brussel was defended with the subtitle “Cosmological Speculation and the Meaning of Life”, instead of “The Meaning of Life in a Cosmological Perspective”.

> All scientists agree that speculations have a legitimate place in the construction of scientific theories, but there is no agreement as to *how* speculative a theory may be and still be counted as scientific.

To evaluate how speculative a theory may be, we need to be clear on *why* we speculate. What is our aim when we speculate? I distinguish three kinds of speculations to navigate into their variety (Vidal 2012a):

1. **Scientific**: a speculation is scientific if we have strong reasons to think that future observations or experimentations will corroborate or refute it.
2. **Philosophical**: a speculation is philosophical if it extrapolates from scientific knowledge and philosophical principles to answer some fundamental philosophical problems.
3. **Fictional**: a speculation is fictional if it extends beyond scientific and philosophical speculations.

Fictional speculations are found in counterfactual history or science fiction books. Their main goal is entertain a reader, but their value for history as a discipline or the scientific enterprise is limited.

Scientific or philosophical speculations stem from our urge to complete logically our knowledge in areas where it has not yet been probed. Importantly, scientific and philosophical speculations have a clear aim, namely to solve scientific or philosophical problems.

Of course, the status of a speculation can change through time, and even become normal science. For example, Giordano Bruno made the philosophical speculation that there were other solar systems in the universe. It was philosophical and not scientific because it was not clear at his time if we would ever be able to have telescopes and observational methods to discover exoplanets. With technological progress, the speculation becomes scientific, because observational techniques can empirically check such a claim. Finally, the status of speculation disappears altogether as we found the first exoplanets (Wolszczan and Frail 1992). Today, hunting and finding exoplanets is no speculation anymore, but part of normal astrobiological science. However, Bruno also believed in the common speculation that the universe is filled with a substance called “aether”, a theory which is now considered obsolete.

So any speculation has to be taken with a grain of salt. To this end, I tried to review many different speculations, and not to overstate the conclusions. I made clear the hypotheses on which some of my core reasonings hinge in Appendix II. This is presented as argumentative maps, and I hope they will facilitate rational and critical debate.

For readers familiar with my previous work, let me quickly outline how it connects with this thesis. Part I presents philosophical reflections on what philosophy is, and its method. It is a synthesis and expansion of several papers (Vidal 2007; 2008a; 2012b). Part II analyzes the origin of the universe, of which Chapter 5 is mainly based on (Vidal 2010a; 2012a); and section 6.3 The Cosmic Evolution Equation on (Vidal 2013). Part III is based on reflections about future cosmic evolution, Chapter 7 on (Vidal 2008b), Chapter 8 on (Vidal 2010a; 2012c; 2012a; 2012d) and Chapter 9 on (Vidal 2011). Good news, the rest is new.

Brussels, Vrije Universiteit Brussel, December 2012.

# Acknowledgements

The last few years, I have had the chance to interact with very inspiring, encouraging and unconventional thinkers. First and foremost, I show much appreciation for the atmosphere of free inquiry here at the Vrije Universiteit Brussel (VUB). It is a luxury for thinking deeply and I do not take for granted. I know that tremendous efforts of previous generations have been sustained to maintain the ideals of the enlightenment.

I thank the transdisciplinary research Center Leo Apostel (CLEA), the Evolution, Complexity and Cognition group (ECCO) and the Global Brain Institute (GBI) which provided me a unique environment to conduct free interdisciplinary thinking. In CLEA, I thank Diederick Aerts for his continuous support, encouragements, advices, and his almost endless passion for discussions. Jan Broekaert for being open minded to listen and try to understand my sometimes bold new theories and hypotheses. Alexander Riegler for discussions about constructivism, Karin Verelst for fascinating discussions about infinity and fractals, Karen De Looze for her insights into integral theory as well as the conceptions of death and immortality (that I discuss in section 10.4 Voyage to Five Immortalities, p293). I also thank Valérie Aucouturier, Sandro Sozzo, Gert Goeminne, Ellen van Keer, Nicole Note, Pieter Meurs, Wim Van Dale for various suggestions and discussions.

I thank the ECCO members Evo Busseniers, Petter Braathen, Øyvind Vada, Iavor Kostov, Nagarjuna G., Jan Bernheim, Marios Kyriazis, Tanguy Coenen, Marc Goldchstein, Tor Eigil Hodne, Nathalie Gontier and Shwetambara Sabharwal. I am grateful to Carlos Gershenson for discussions about immortality, Artificial Life, and the self-organizing philosophy; to Mark Martin for his passionate defense of rationality and evolutionary theory; to Klaas Chielsen for seeing the intersection of selection criteria in memetics and in philosophy, which was a start to develop metaphilosophical criteria for the comparison of worldviews (in section 2.2 Criteria for Worldview Comparison, p34); to Piet Holbrouck for introducing me to the basics of the theory of constraints, which lead me to enthusiastically draw the argumentative maps of the thesis (in Appendix II – Argumentative Maps, p322); to Marko Rodriguez for stimulating discussions and inspiring feedback; to Paul Iliano for discussions about worldviews and for helping me to find a personal vision, “providing people a meaning of life, in harmony with cosmic evolution”; to Jon Echanove for both warm and challenging discussions; to Viktoras Veitas for defending all positions on issues; to David R. Weinbaum (Weaver) for always offering enlightening new perspectives; to John Stewart for the many evolutionary inspirations, advices, insights and encouragements. I also thank Joseph Melotte, Jean-Philippe Cornelis for their curiosity and enthusiasm about the Evolution, Complexity and Cognition (ECCO) group.

I am glad to thank Georgi Georgiev and John Smart, co-directors with me of the Evo Devo Universe (www.evodevouniverse.com) research community. I am happy to thank the scientists who endorsed the project: Diederik Aerts, Robert Aunger, also for insightful discussions about big history; Yaneer Bar-Yam, Martin Beech, also for fascinating brainstorming about star lifting; Adrian Bejan, Howard Bloom, Tommaso Bolognesi, also for his interest in Artificial Cosmogenesis; Terry Bristol, José Roberto Campanha, Jean Chaline, David Christian, M. Ćirković, also for his support, encouragements and always very helpful critical feedback; James A.

Coffman, Peter A. Corning, Lee Cronin, Terrence W. Deacon, Jean-Paul Delahaye, also for his monthly intellectual delicacies published in *Pour la Science*, the French edition of *Scientific American* as well as for inspiring discussions on the limits of science and computation; Joël de Rosnay, also for his (1979) book *The Macroscope* which inspired me early in systems thinking; Steven J. Dick, also for encouragements and for producing and stimulating knowledge in so many aspects of cosmic evolution and cosmic culture, for his insights about the possibility of a postbiological universe, which was a decisive step in my reflections about the search for extraterrestrial life and intelligence (Chapter 9); Taner Edis, Börje Ekstig, George F. R. Ellis, also for a stimulating discussion in the philosophy of cosmology; Charles Fadel, Rainer Feistel, Carlos Gershenson, Richard Gordon, Jan Greben, Hari Mohan Gupta, William P. Hall, also for critical and constructive comments and insights; Stevan Harnad, Francis Heylighen, David Holcman, Hippokratis Kiaris, Osame Kinouchi, Otto Laske, John A. Leslie, also for critical and analytical feedback; Giuseppe Longo, Nicolás Lori, Jean-Pierre Luminet, Garimella Rama Murthy, Akop P. Nazaretyan, Laurent Nottale, Steve Omohundro, Alain Prochiantz, Carl Henning Reschke, Marko A. Rodriguez, Stanley N. Salthe, also or sharing his insights into developmental theories and thermodynamics; Tony Smith, Fred Spier for his inspiring big history work, Eric Steinhart for his philosophical sharpness, John Stewart, Robert E. Ulanowicz, Rüdiger Vaas for his critical interest in Cosmological Artificial Selection which stimulated much of the content of Chapter 8; Pushkar G. Vaidya, Jan Visser, Bruce H. Weber, Peter Winiwarter, also for his suggestion to apply my metaphilosophical framework to itself; Hector Zenil, also for mind-bending discussions about the computational perspective on the universe. I also thank the independent scholars who endorsed EDU, Mary Ann Allison, Robin Allott, Jon Beach, Rick Bradford, also for excellent feedback about fine-tuning; William R. Buckley, John O. Campbell, John Chelen, Michael Chorost, Bruce Damer, also for insights in large scale computer simulations, James N. Gardner, for having beautifully written the first book-length work on cosmological artificial selection (see Chapter 8), as well as his vision that Artificial Life would become more and more relevant for cosmology, a vision that I largely elaborated through the idea of Artificial Cosmogenesis in Chapters 6 and 7; William F. Harms, Hans Kuijper, John McCrone, Robert L. Oldershaw, Cecilia Abadie, Arnaud J. Blanchard for his precious help, Atanu Chatterjee and Stefan Pernar.

A special thank to all active participants to the mailing list EDU-Talk[2] who debate fundamental issues in an academic and interdisciplinary atmosphere. Researchers interested in the topics of this thesis (and others) are welcome to join the “Evo Devo Universe” research community at www.evodevouniverse.com.

I thank Alain Prochiantz for his early support in organizing the first international conference on the evolution and development of the universe (Vidal et al. 2009). I thank Laurent Nottale for his enthusiasm, time and patience in explaining some implications of scale relativity theory. I am happy to see this theory of fractal space-time more and more recognized as a promising basis for unifying quantum mechanics and general relativity. I am also glad he accepted to talk with Jean Chaline and Pierre Grou at a the “Workshop on Scale Relativity: Universe, Life, Societies”[3] I organized in Brussels. I also thank Jean Chaline for his stimulating books and explanations about biological evolution. I thank Bernard Goossens for sharing his

2 http://evodevouniverse.com/wiki/EDU-Talk_subscription_form
3 http://ecco.vub.ac.be/?q=node/114

visionary interest in scale relativity, without which this workshop could not have happened.

I thank Christophe Portier, Marc Megaides, Martin Monperrus, Marc Sylvestre and David Brin for insightful discussions.

I am pleased to thank David Allen for creating the action management method “Getting Things Done” without which I would never have had the courage to take up such a big project as co-organizing an international conference and editing the proceedings (Vidal et al. 2009). I also co-wrote an academic paper further explaining why the method works so well (Heylighen and Vidal 2008).

I'm grateful to Alain Jorissen for his patience in answering my sometimes naive, sometimes strange questions about our singular universe and its binary stars; to Jason Cawley for criticisms and discussions; to Tim Swanson for enthusiastic speculative discussions; to Edgar Gunzig for his early support; to Dominique Lambert for stimulating discussions; to Claudio Maccone for discussions on extreme possibilities of gravitational lensing and its implications for SETI; to Nicholas Rescher for his encouragements and interest in my philosophical work and to Eric J. Chaisson for challenging discussions.

I warmly thank the anonymous referees who accepted or rejected my papers, in either case providing very valuable feedback. I thank those who took the time to correct and improve my english, Gabrielle Bossy, Hans Smet, Otto Bohlmann, Charles Baden-Fuller, Luke Lloyd, Steve Sachoff and Elizabeth Moliere. I thank Tagxedo[4] for offering the tool to create the cover picture, credits goes to NASA, ESA, and the Hubble SM4 ERO Team for the background picture of Carina Nebula.

I warmly thank John M. Smart for being such an amazing colleague and friend, for being the co-founder with me of the Evo Devo Universe community, for being a rare polymath, never short of amazing and stimulating ideas. I owe him inspiration on many aspects of this thesis, such as his contagious passion for astrobiology or the general evolutionary developmental view on the universe. I am also glad we identified together the Barrow scale of civilizational development (section 9.2.2 Barrow Scale – the Inward Manipulation, p226).

My immense gratitude goes to my PhD supervisor Francis Heylighen for having been and still being an unfailing intellectual mentor. I am grateful that he trusted me to tackle big Cosmological questions despite the fact that I had no formal background in theoretical physics. After one master in Philosophy, another one in Logic and a third in Cognitive Sciences, I thought I started to have a solid intellectual background, but I was wrong. Learning in depth with Francis Heylighen evolution, cybernetics and complexity sciences led me to an intellectual rebirth. I felt almost angry to learn such powerful yet simple concepts so late in my intellectual life. In particular, I hope the basics of cybernetics and systems theory will soon be taught early in schools.

I warmly thank my parents for offering me a slice of space-time-energy-complexity we call life, through their biological and cultural legacy. A very special thanks to my father, for his immense support, patience and trust.

4 http://www.tagxedo.com/

# Abstract

Where does it all come from? Where are we going? Are we alone in the universe? What is good and what is evil? The scientific narrative of cosmic evolution demands that we tackle such big questions with a cosmological perspective. I tackle the first question in Chapters 4, 5 and 6; the second in Chapters 7 and 8; the third in Chapter 9 and the fourth in Chapter 10. However, where do we start to answer such questions wisely? Doing so requires a methodological discipline mixing philosophical and scientific approaches.

In Chapter 1, I elaborate the concept of *worldview*, which is defined by our answers to the big questions. I argue that we should aim at constructing *comprehensive* and *coherent* worldviews. In Chapter 2, I develop criteria and tests to assess the relative strengths and weaknesses of different worldviews. In Chapter 3, I apply those methodological insights to religious, scientific and philosophical worldviews.

In Chapter 4, I identify seven fundamental challenges to any ultimate explanation of the origin of the universe: epistemological, metaphysical, thermodynamical, causal, infinity, free parameters and fine-tuning. I then analyze the question of the origin of the universe upside down and ask: what are the origins of our cognitive need to find an explanation of this origin? I conclude that our explanations tend to fall in two cognitive attractors, the point and the cycle. In Chapter 5, I focus on the free parameters issue, namely that there are free parameters in the standard model of particle physics and in cosmological models, which in principle can be filled in with any number. I analyze the issue with in physical, mathematical, computational and biological frameworks.

Chapter 6 is an in depth analysis of the fine-tuning issue, the claim that those free parameters are further fine-tuned for the emergence of complexity. I debunk common and uncommon physical, probabilistic and logical fallacies associated with this issue. I distinguish it from the closely related issues of free parameters, parameter sensitivity, metaphysical issues, anthropic principles, observational selection effects, teleology and God's existence. I conclude that fine-tuning is a conjecture, and that to make progress we need to study how common our universe is compared to other possible universes. This study opens a research endeavor that I call *artificial cosmogenesis*. Inspired by Drake's equation in the Search for Extraterrestrial Intelligence, I extend this equation to the *Cosmic Evolution Equation*, in order to study the robustness of the emergence of complexity in our universe, and whether or to what extent it is fine-tuned. I then review eight classical explanations of fine-tuning (skepticism, necessity, fecundity, god-of-the-gaps, chance-of-the-gaps, weak-anthropic-principle-of-the-gaps, multiverse and design) and show their shortcomings.

In Chapter 7, I show the importance of artificial cosmogenesis from extrapolating the future of scientific simulations. I analyze two other evolutionary explanations of fine-tuning in Chapter 8. More precisely, I show the limitations of *Cosmological Natural Selection* to motivate the broader scenario of *Cosmological Artificial Selection*.

In Chapter 9, I set up a new research field to search for advanced extraterrestrials, *high energy astrobiology*. After developing criteria to distinguish natural from artificial systems, I show that the nature of some peculiar binary star systems needs to be reassessed because of thermodynamical, energetic and civilizational development arguments which converge towards them being advanced extraterrestrials. Since those putative beings actively feed on stars, I call them *starivores*. The question of their artificiality remains open, but I propose concrete research proposals and a prize to further continue and motivate the scientific assessment of this hypothesis.

In Chapter 10, I explore foundations to build a cosmological ethics. I build on insights from thermodynamics, evolution, and developmental theories. Finally, I examine the idea of immortality with a cosmological perspective and conclude that *the ultimate good is the infinite continuation of the evolutionary process*. Appendix I is a summary of my position, and Appendix II provides argumentative maps of the entire thesis.

# Introduction

*The great philosophers have always been able to clear away the complexities and see simple distinctions –simple once they are stated, vastly difficult before. If we are to follow them we too must be childishly simple in our questions –and maturely wise in our replies.*

(Adler and Doren 1972, 271)

Where does it all come from? It takes nothing less than a synthesis of modern science to answer this childish question. In a nutshell, modern science gives us the following story of our past. Everything started with a Big Bang, about 13.7 billion years ago. As the universe expanded and cooled down, atoms formed, stars structured in galaxies and clusters of galaxies. On a tiny solid planet, around an average star, some special conditions allowed the self-organization of molecules, and the first cell was born. Life started. Later, vegetation used the radiation of the Sun and contributed to the creation of an atmosphere. Gradually, more and more complex organisms emerged, competed and cooperated. Nowadays, human cities, societies and technologies are growing rapidly.

That's for our past. What about our future? Where are we going? Of course, we have –by definition– no data about the future. However, we do have physical scientific theories, which are temporally symmetrical and it is thus legitimate to apply them in the past as well as in the future.

Astrophysicists teach us that in about 5 billion years, our solar system will end, with our Sun turning into a red giant star, making Earth's surface much too hot for the continuation of life as we know it. The solution then appears to be easy: migration. However, even if life would colonize other solar systems, there will be a progressive end of all stars in galaxies. Once stars have converted the available supply of hydrogen into heavier elements, new star formation will come to an end. In fact, the problem is even worse. It is estimated that even very massive objects such as black holes will evaporate (see e.g. F. C. Adams and Laughlin 1997). The second law of thermodynamics, one of the most robust laws of physics, states that disorder or entropy of an isolated system can only increase. Eddington (1928) applied it to the universe as a whole and concluded that our universe is doomed to perish in a *heat death*. Modern cosmology confirms that in the long-term future we will need to deal with a cosmic doom scenario, heat death or other. Now, what do those insights about our past and future imply for the meaning of life, intelligence and humans in the universe? Most people I talk to, both colleagues and friends, think it is too early to think about such a long-term issue as cosmic doom. I strongly disagree. For, when are we to start worrying about the far-future, then? When will be a good time to take responsibility?

Humans are insignificant regarding the space they occupy in the universe. The Earth is ridiculously small compared to the universe. It is a tiny planet orbiting a common star, in a galaxy composed of billion of stars. And there are billion of galaxies. Humans are also insignificant in terms of universal time. This is illustrated by Carl Sagan's (1977) cosmic calendar, in which the ~15 billion year lifespan of the universe is compressed into one year. One second in that cosmic calendar corresponds to 475 years in our Western calendar. Then, the first humans would appear on December 31, the very last day of the cosmic calendar, very late, at ~10.30pm. Our spatial occupation and duration of existence are thus ridiculously small seen from a cosmological perspective. To sum up, the fact that in a single, tiny cosmic pocket, life and intelligence very recently appeared seems an accident without any significance, anyway doomed to be wiped out.

Is this murky story true? Is it correct in all its aspects? Is it possible that it misses some important aspects of cosmic evolution? My aim is to show you that this story is wrong. Not so much in its scientific content, but in its conclusions and limited perspective. In this thesis, I will tell you a very different story, where intelligence and complexity are the keys to unlock the universe's mysteries.

The questions of the beginning and the end of the universe are extremely difficult, because they require the utmost extrapolation of scientific models, whose results then become highly uncertain. Additionally, scientific models cannot directly answer metaphysical questions like “why is there something rather than nothing?”; “was there a beginning of the universe?”; “is our universe fine-tuned for life?” or “what is the meaning and future of life in a cosmological perspective?”

When dealing with such difficult questions, we have to acknowledge the limit of the scientific enterprise. For example, regarding the *ultimate origin*, there are observational limits to the early universe, i.e. data we will never get. Neither science nor philosophy will bring us certain answers and this is an invitation to humility, honesty, and carefulness when approaching those ultimate questions in cosmology (see e.g. Ellis 2007a, 1259; Vaas 2003, sec. section 7). This does not mean that science won't provide us an answer later. Scientific progress often surprised us in the past, and there is no reason it won't continue to do so.

Compared to the early universe, there are surprisingly few works on the *ultimate future* of the universe. However, mysteries about our far-future are as important and fascinating to explore as the ones of our past. Classical cosmic doom scenarios presuppose that intelligence is and will remain insignificant. But this is in contradiction with past cosmic evolution, which shows a complexity increase, from galaxies, solar systems, life, mind, intelligence, society, science and technology. What if intelligence could have a major impact on cosmic evolution? If we do not underestimate that complexity increase, the matter of cosmic doom becomes much more exciting because it remains unsettled.

The human world may be small in comparison to cosmic space and time, but what about the increasing complexity? Mind, intelligence, science, technology and culture *are* highly significant from a complexity point of view. There is even an acceleration in this rise of complexity (see e.g. Coren 1998; Livio 2000; Chaisson 2001). Could it be that mind, culture, science and technology are not accidents in the universe? Could they have a profound and important cosmological meaning?

Moreover, our day-to-day lives seem very much disconnected from astrophysical or cosmological events. Yet we are in the cosmos. And future generations ultimately depend on the future of our universe. How will this acceleration and increasing complexity evolve in the far future? Where will it end? Will the universe end in a cosmic doom? Or will it be our aspirations that will win out in the long term? Is it possible for life and intelligence to survive forever? Can searching –and maybe spotting– advanced extraterrestrials bring us a better understanding of the increasing complexity in our universe? What is goodness in a cosmological perspective? Could an evolutionary ethic inspired by cosmic evolution help us in guiding our actions?

How can we hope to answer such difficult questions? If they cannot be fully answered in a scientific framework, should we switch to a religious approach? Religions indeed offer creation myths and value systems, and thereby give practically and socially very valuable answers to those questions. However, they often lack a self-critical and scientific basis. Could a philosophical approach answer those questions, with open, non-dogmatic, rational and cautious answers? If so, how are we to answer them in a philosophical and not theological way? Is there a philosophical method to do so?

I always was fascinated by the beauty of the starry sky and images of nebula, galaxies and other astrophysical objects that modern telescopes bring us. But how do we fit into this scheme? In high school, my curiosity led me to buy a French translation of Paul Davies' (1991) *The Mind of God: The Scientific Basis for a Rational World*, which I found most fascinating even though I couldn't understand most of it. Yet, I loved Davies' profound and sincere quest for understanding, combined with up-to-date cosmological and scientific theories. I was especially fascinated by two ideas: the theoretical possibility of baby universes, and Davies' intuition that mind, intelligence and consciousness might not be mere cosmic accidents. I started to speculate that those very two ideas might well be connected. But this was just a vague intuition, and I do not trust or value intuitions as such. They are good starting points, but need further elaboration, clarification and argumentation. This thesis is my attempt to lift this intuition as precisely as possible to a carefully crafted argument and worldview.

Story 1: Minding rationality and the Mind of God.

I argue in this thesis that intelligence and complexity are essential components of our universe. My philosophical position is to remain rational, in agreement with science, yet attempting to go one step further than scientific inquiry, motivated by our childish curiosity and need to answer the big questions.

I will not attempt to write another comprehensive story of the universe. I assume that my reader is aware of and familiar with cosmic evolution (if not, see e.g. Sagan 1985; Jantsch 1980; Turchin 1977; Christian 2004; Laszlo 1987; Chaisson 2001; 2006; Dick 2009a; 2012).

What follows is instead an exploration and analysis of three *extreme* and ultimate points of this story: the beginning, the end, and the meaning of increasing complexity. If on large scales all our sciences are reduced to cosmology, and if on small scales all our sciences can be reduced to particle physics, then how do we link the two? The sciences of evolution and complexity have the potential to bridge the gap between these two reductions to cosmology and particle physics. They are anti-reductionist by nature and try to understand the emergence of new laws, of complexity transitions.

Inspired by Adler's and Van Doren's quotation above, my ideal is to answer childishly simple questions, in a maturely wise manner. These questions concern the ultimate origin, future and values from a cosmological perspective. To answer them in a wise manner, we will inquire into many intricate theories and discussions, but their point will always be to answer those simple questions. I will also balance this apparently overzealous ambition with an appeal to considered conclusions through an extensive study of the philosophical method in Part I.

The organization of this thesis is simple. *Part I deals with the philosophical method, Part II with the beginning of the universe and Part III with intelligence in the far-future universe*. By weaving insights in these three parts, we can find and refine a meaning of life in a cosmological perspective. Both the beginning and the end are extreme extrapolations, and it makes sense to treat them together, as we will face similar problems and solutions in exploring them.

**What is philosophy?** This work is of synthetical and speculative nature. It is an attempt to answer some of the deepest philosophical questions, by constructing a coherent and comprehensive worldview. In Part I, we inquire about the philosophical method, and show that there is an existential need to answer those big questions. Answering them might involve some speculations. Yet, speculating does not mean being unscientific. On the contrary, it means relying on and identifying the most fundamental scientific theories and principles, and then extrapolate them. In my speculations, I have done my best to focus on the most robust and general scientific theories such as principles of relativity theories, thermodynamics, systems theory, evolution, theoretical computer science or logic. The first step to discuss philosophical questions in a way that is compatible with scientific results is to reformulate those childish questions in a maturely wise scientific and philosophical manner.

**Where does it all come from?** In Part II, we examine in three steps the problem of the ultimate origin of the cosmos. Firstly, in Chapter 4 we deal with our *cognitive needs* for origins. We ask, what is a cognitively satisfying answer to the question of the origin of the universe? In Chapter 5 we then focus on the existence of free parameters in physical and cosmological models. The problem is that those parameters are not specified by our theories, yet all particle physics and cosmological models have such free parameters. We explore them with a variety of approaches: physical, mathematical, computational and biological. Those free parameters need to be filled-in in our models. But *filling-in* doesn't necessarily imply *fine-tuning*. We thus discuss in Chapter 6 whether or not those free parameters are fine-tuned. More precisely, the fine-tuning of the universe is a highly confusing and controversial issue at the intersection of physics, philosophy and theology. It is infused with physical and probabilistic fallacies; mixed up with other issues (e.g. free parameters, parameter sensitivity, metaphysical issues, observation selection effects, anthropic principles, teleology and God's existence); and it is most often ill defined. To clarify the debate, we first debunk common and uncommon fine-tuning fallacies. We then ask the question “fine-tuning for what?” to generate different definitions of fine-tuning. For this, we introduce a Drake-like “Cosmic Evolution Equation”, defining different *cosmic outcomes* we want to focus on. We then review classical and evolutionary explanations in light of our new framework. We conclude that to scientifically progress on the issue we need to explore the space of possible universes with the help of computer simulations. This includes not only simulating our universe but also other possible universes, with the nascent field of *Artificial Cosmogenesis*.

**Where are we going?** In Chapter 7 of Part III, we explore the future of scientific simulations, and further substantiate the need to pursue Artificial Cosmogenesis. We discuss in Chapter 8 cosmological selections and develop in detail

*Cosmological Artificial Selection* (CAS, Chapter 8), a wide ranging philosophical scenario covering the origin and future of the universe with a role for intelligence. Surprisingly, CAS leads to the idea that it is by better understanding our ultimate future that we will progress to understand our ultimate origin. The two may well be deeply intertwined.

There is a great uncertainty regarding two main trends in cosmic evolution. The one trend goes towards more disorder or entropy, the other towards more complexity. Which one will prevail in the long term? If the first prevails, it will be our end in the universe. But if the second trend prevails, there is hope to construct a meaning of life in harmony with the complexity increase in cosmic evolution.

**Are we alone in the universe?** Predicting the *long-term* future of humanity, I mean its fate in thousands, millions or billions of years, is a notoriously difficult attempt, often deemed impossible. But there is a workaround. The idea is to look at how *other* civilizations might have developed their complexity in the universe. In Chapter 9 we thus look for very advanced civilizations in the universe, and, to my great surprise and awe, my theoretical reasonings led me to the conclusion that we might already have spotted very advanced extraterrestrials!

**What is good and what is evil?** What are the ultimate values for intelligent life? By ultimate I mean, values both valid at all times, past and future, as well as valid in all places in the universe. In Chapter 10 we enquiry about values derived from a cosmological perspective. We develop a cosmological ethics and apply the framework to the idea of immortality, which is a constant longing in human cultures. We survey five kinds of immortalities and how they relate to the definition of the self. We argue that the ultimate good is the *infinite continuation of the evolutionary process*. We then discuss the possibility or impossibility of such a cosmological immortality.

To facilitate the navigation in this work, on the one hand I provide in Appendix I a straightforward summary of the worldview developed, in terms of positions, not arguments. In many ways, it spoils the content of this thesis, so I leave the responsibility to the reader at which point he wants to read it. On the other hand, Appendix II provides two argumentative maps. The first map describes the core problems we tackle; the second map summarizes our proposed solution. These two appendices will mostly benefit to professional academics familiar with the issues I tackle, but also to other readers when they need to take a bird's-eye view. In a less formal way, I also took the freedom to write ten short stories, which I hope will entertain some of my readers and show some psychological and social aspects of doing research. Academics who disdain such a practice can easily skip such stories already well contained in grey boxes. Importantly, they do not contribute to the general argumentation of the thesis.

Academic research is always work in progress. So, at the end of each Part or important Chapter, I point to the most important and challenging *open questions* I came up with. I hope researchers will pursue them –with or without my collaboration.

In this thesis, I put together pieces of a cosmic puzzle which form a worldview that I find magnificent. My objective now is to share it with you. I hope to establish a deep, enduring and yet evolving connection between intelligent life and the cosmos, which will provide people a meaning of life, in harmony with cosmic evolution.

# Part I - Overview of Worldviews

In december 2005, I moved to Brussels to start a PhD in philosophy with Francis Heylighen. A PhD is a long and difficult project. I had the ambition to inquiry about big philosophical questions like “where does it all come from?” or “what is the ultimate future of intelligence in the universe?” Before entering into these waters, I wanted to be clear on *how* to philosophize. Despite years of university studies in philosophy, I figured out I didn't know precisely what the philosophical method was. What a shame! The best answer I quickly found was that each philosopher has his own method (Passmore 1967). It made sense, but I was disappointed. Isn't it a pity if each philosopher needs to redefine the philosophical method even before starting to philosophize? Isn't it a waste of intellectual resources? Imagine if empirical scientists needed to reinvent the experimental methodology each time they started a new experiment. They wouldn't have any time to actually start their experiment! I was committed to find a better answer, and this Part I is my small contribution to this issue.

The dictionary tells us that a method is a set of procedures for accomplishing an aim. Consequently, to find the philosophical method(s), I needed to find what the aim of philosophy was. However, defining what the aim of philosophy is comes down to answering the most debated philosophical question: *What is philosophy?* I started to research this question and was astonished by its difficulty. I thought it would keep me busy for a few months, but it kept me busy until now. I find it the most difficult issue I've inquired so far; maybe as difficult as my attempt to prove or disprove the existence of advanced extraterrestrial intelligence! (See Chapter 9).

Story 2: An ignorant young philosopher.

Since the development of modern science, science has been taking over more and more issues from philosophy. For example, classical philosophical problems about the mind, time, space, or the cosmos are now investigated by scientific means. How can philosophers react to this? Either they can feel invaded by an intruder, and will take refuge in issues science will never touch; or they can be delighted. Indeed, scientific progress on philosophical issues means that we are getting new ideas, arguments and insights in our common quest for understanding the world.

Philosophy thus often needs to redefine its scope and also its relationship to science. Philosophy can also take the opportunity to embrace all this new knowledge. However, partly because of this takeover, today's philosophy collapsed in two main traditions, *analytic* and *continental*, with different drawbacks that we will quickly examine.

Thirty years ago, Paul Ricoeur (1979) directed a survey of the main trends of philosophy. He distinguished three main trends:

(1) Philosophy as a Weltanschauung (worldview)
(2) English and American analytic philosophy
(3) Subjectivity and beyond.

In trend (3), philosophy explores other forms of experience than objective knowledge (e.g. the young Hegel, Kierkegaard, young Marx, and certain developments of phenomenology.) This trend corresponds to *continental philosophy.* It is a stimulating intellectual approach, but it faces harsh critiques, notably its lack of methodology (see e.g. Shackel 2005).

On the other hand, even if analytical philosophy (2) brings precise methods of analysis and critic into philosophy, it still lacks a general guideline, a unifying agenda. And the use of logical methods is insufficient to constitute such an agenda. Analytic philosophy really needs something more than pure analysis; it also needs to be completed with a synthetic dimension. Synthetic worldview construction, as we shall see, can fill this gap. Our own philosophical position will thus tend towards the trend (1).

Still, distinguishing those three trends does not answer our question: *What is philosophy?* A fuzzy answer is that it is a quest to understand humankind and its world. However, for the most important questions, this enterprise overlaps with science and religion. Philosophy, science and religion share this quest of understanding, and they can build more or less strong relationships to pursue it (see e.g. R. J. Russell, Stoeger, and Coyne 1988). The result is that starting either from science, religion or philosophy, we end up with different worldviews.

We will argue that having a coherent and comprehensive worldview is the central aim of philosophy. But, what is more precisely a worldview? How can we compare very different worldviews? Specifically, what are the strengths and weaknesses of scientific, religious and philosophical worldviews?

To better grasp what philosophy is and to navigate its rich and complex landscape, I first introduce in Chapter 1 six philosophical *dimensions* along with the *worldview agenda.* This agenda invites us to tackle big questions, and our answers to them define what our worldview is. Furthermore, to meaningfully and critically tackle the big questions, we must be able to compare different worldviews. For this we need a set of *criteria* and a battery of *tests*. We introduce such criteria and tests in Chapter 2, which facilitate the difficult endeavor of worldview comparison. We conclude our analysis of worldviews and criteria by showing that science and religion both have *complementary* strengths and weaknesses (Chapter 3). By synthesizing them, we aim at coherent and comprehensive philosophical or theological worldviews.

This Part I provides an *ambitious* yet *considered* philosophical framework as an indispensable prelude to conduct our journey into the big cosmological issues: the beginning and the end, and the meaning of life.

# *CHAPTER 1 - The Six Dimensions of Philosophy*

**Abstract**: We introduce six dimensions of philosophy, whose first three deal with first-order knowledge about reality (*descriptive*, *normative* and *practical*); the next two deal with second-order knowledge about knowledge (*critical*, *dialectical*), while the sixth dimension (*synthetical*) attempts the integration of the other five dimensions. We describe and illustrate the dimensions with Leo Apostel's worldview program. Then we argue that we all need a worldview to interact with our world and to give a meaning to our lives. Such a worldview can be more or less explicit, and we argue that for rational discourse it is essential to make it as explicit as possible. With a cybernetic diagram, we then illustrate how the different worldview components dynamically interrelate.

While defining what a worldview is, it is useful to distinguish six dimensions in philosophy, as depicted in figure 1. We distinguish between first- and second-order knowledge (Adler 1993, 13–16). *First-order knowledge* is about "reality", and *second-order knowledge* is about knowledge itself. We add a third-order *synthetical* dimension (6), which is the integration of first- and second-order dimensions of philosophizing. The *descriptive* dimension (1) and the *normative* dimension (2) correspond to Adler's (1993) *metaphysical* and *moral* dimensions. The *critical* dimension (4) and the *dialectical* dimension (5) partially overlap with Adler's *objective* and *categorial* dimensions. Dimensions (4), (5) and (6) are also inspired by Broad (1947; 1958) who calls them *analysis*, *synopsis* and *synthesis*. Let us further dive into these dimensions.

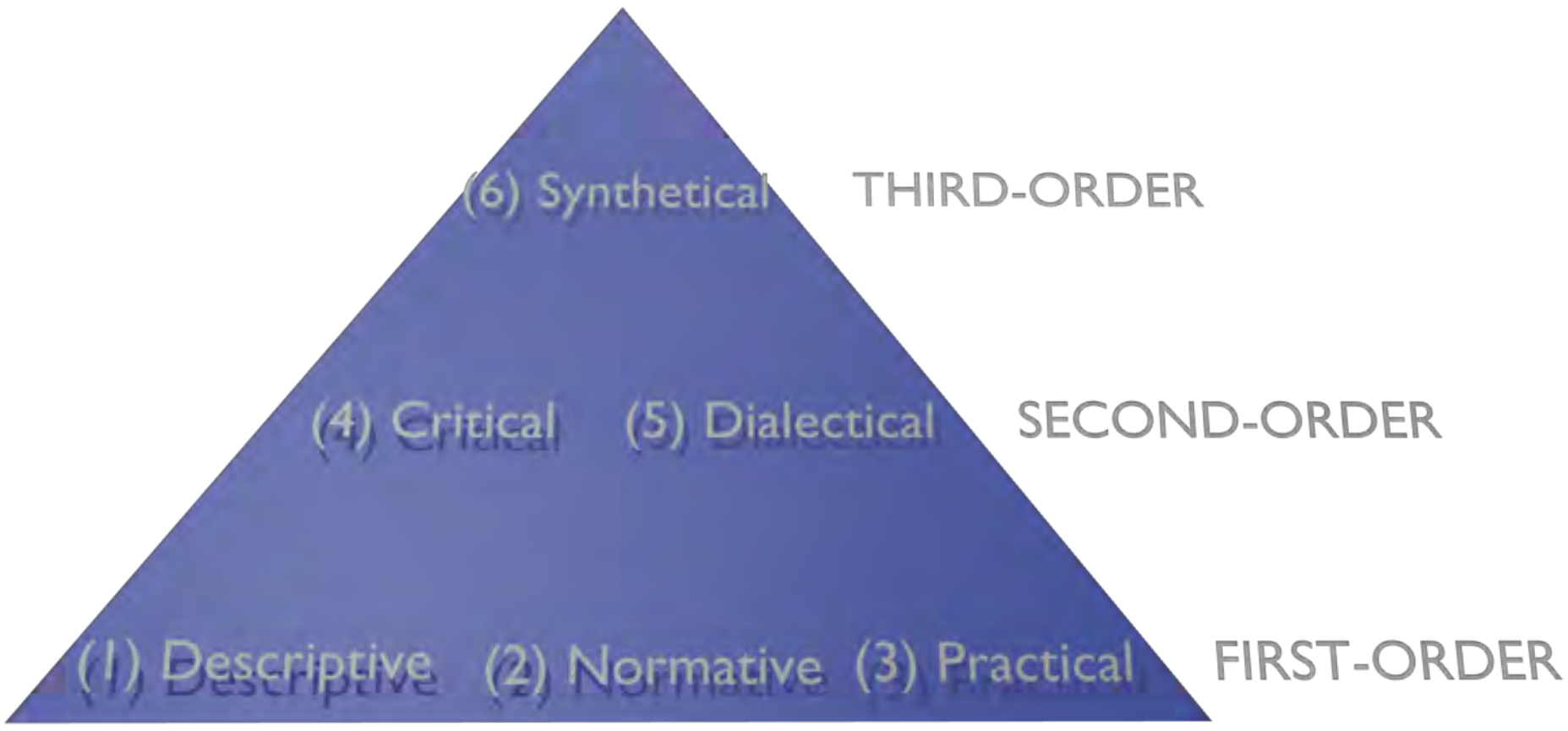

Figure 1: The Six Dimensions of Philosophy.

## 1.1 First-Order Questions

A philosophical agenda defines the range of problems and issues that are addressed by a philosophy. What are the most profound questions of existence? Those questions, *not their answers*, are surprisingly enduring throughout the history of philosophy (see e.g. Passmore 1961, 39; Rescher 2006, 91). The worldview approach developed by Leo Apostel elegantly makes the questions explicit (Apostel and Van der Veken 1991; trans. in Aerts et al. 1994); we can summarize them as:

(a) What is? *Ontology* (model of being);
(b) Where does it all come from? *Explanation* (model of the past);
(c) Where are we going? *Prediction* (model of the future);
(d) What is good and what is evil? *Axiology* (theory of values);
(e) How should we act? *Praxeology* (theory of actions).

These questions correspond to the "big", "eternal", or "age-old" philosophical questions. Each question corresponds to a first-order knowledge branch, in italics above. It is important to recognize that starting with this agenda is already a philosophical choice. We will discuss the agenda more in depth when describing the *scope in agenda* criterion (section 2.2.3, Scope p36). Although the six dimensions of philosophy are more general than this worldview agenda, I introduce it because it makes our philosophical framework more concrete.

Apostel's definition of a "worldview" is broader than just a representation of the world because it also includes theories of values and actions (questions (d)-(e)). An answer to a worldview question forms a *worldview components*. Articulated together, the components form a worldview that we define as a coherent collection of concepts "allowing us to construct a global image of the world, and in this way to understand as many elements of our experience as possible." (Aerts et al. 1994, 17).

Let us now clarify how this worldview agenda typifies three of the six philosophical dimensions. The *descriptive* dimension (1) attempts to describe the world as it is and thus corresponds to worldview questions (a), (b) and (c). The *normative* dimension (2) corresponds to worldview question (d), whereas the *practical* dimension (3) matches with worldview question (e), or praxeology.

The descriptive dimension (1) concerns "is-questions". Tackling this dimension is the task of an ontology, explanation and futurology. The first question (a) is the question of ontology, or a model of being. It can be typified with the question "*What is?*". The second question (b) explains the first component. Why is the world the way it is, and not different? What kind of global explanatory principles can we put forward? How did the Universe originate? *Where does it all come from?* The kind of explanation sought here is one in terms of antecedents. Answers to these questions explain how and why such or such phenomena arose. The third question (c) is complementary to the second one. Instead of focusing on the past, it focuses on the future. *Where are we going?* What will be the fate of life in the Universe? Answering these questions give us possible and probable futures. But there are many possible futures and their probability is most often hard if not impossible to assess. We need to cope with uncertainties, and this leaves us with choices to make. Which possible alternative should we promote and which one should we avoid? To answer this, we need values and thus the normative dimension (2).

Importantly, describing or modeling the world is an enterprise overlapping with science. The precise formulation of these first three worldview questions will thus vary from epoch to epoch. For example, current problems related to the ultimate constituents of matter (question (a)) highly depend on available scientific theories. It is thus mandatory to reformulate and define precisely those "big" questions in the context of a certain epoch. Such purely philosophical questions become *mixed questions* in the sense that they require scientific knowledge to formulate and to solve them (Adler 1993, 67; C. I. Lewis 1929, 4–8). Such mixed questions invite us to conduct “philosophy with” other disciplines, rather than the more common second-order “philosophies of” other disciplines (Hansson 2008). Anticipating what follows, considering mixed-questions is already part of the synthetical dimension (6) of philosophy.

The normative dimension (2) tackles “ought-questions”, typified with the fourth worldview question: *what is good and what is evil?*. How to live a good life? How to organize a good society? How do we evaluate global reality? What should we strive for? What is the meaning of life in a cosmological perspective? Axiology traditionally deals with those questions, including morality, ethics, and aesthetics. Also in the normative dimension, the questions are mixed. For example, the question of how to live a good life is mixed with the psychology of well-being; the question of how to organize a good society is mixed with political philosophy, sociology, etc. This worldview component gives us preferences, direction, purpose, a set of goals to guide our actions. Yet, it is not always clear how to connect values with actions.

The practical dimension (3) addresses “act-questions”. Given our model of the world and our axiology, *how should we act?* What are the general principles according to which we should organize our actions? We need such principles to implement plans of action according to our values, in order to solve practical problems. Importantly, such practical insights will remain implicit for most us, in the sense that we act without having a theory of how we act. Theorizing about action is the domain of praxeology, which is mixed with fields like operational research, problem-solving methods, management sciences, etc. Adler (1993) did not explicitly include this important dimension. It is however a notable kind of philosophizing, namely, philosophy as a way of life.

Most of today's philosophers would disagree that philosophy's task is still in dimensions (1) or (3). This is mainly because those questions which were once philosophy's territory gave birth to various modern sciences (James 1987, 993). For example, William James himself is one of the founding fathers of scientific psychology; Frege and Russell founded mathematical logic and Auguste Comte coined the the term “sociology”.

Those descriptive and practical philosophical dimensions were once at the core of the philosophical enterprise, and the fact that they are not anymore today is arguably only a historical accident (see Adler 1965, 1993). Let us now turn to second-order philosophizing.

## 1.2 Second-Order Questions

Apostel added two other questions:

(f) What is true and what is false? *Epistemology* (theory of knowledge);

(g) Where do we start to answer those questions?

They invite us to become aware of our current worldview, and to also ponder where our knowledge comes from. Yet these questions are of a different nature than the five others.

The five first worldview questions are first-order in the sense that they question directly our world and how to interact with it. By contrast, the sixth and seventh questions are about the origin of our answers to those first-order questions;

they are thus of a second-order nature. Let us now characterize the second-order dimensions more precisely.

The second-order critical dimension (4) is like an intellectual acid, which can attack anything. It has two traditions, *continental* and *analytical*. The two are critical approaches to philosophizing, yet in two very different ways. Continental philosophizing includes movements such as phenomenology, existentialism, critical theory, hermeneutics, structuralism, deconstruction, and postmodernism. First and foremost, it takes subjective and intersubjective perspectives as starting points. By contrast, analytical philosophy is mainly focused on objective aspects and emphasizes the use of precise definitions, sound arguments wrapped up in a rigorous logical analysis.

A third aspect of critical philosophizing is “philosophies of X”, where X can be almost any discipline. Those efforts which exploded in recent years are of a critical and second-order nature, contrasting with “philosophies with”, which are synthetical or first-order. Epistemology, typified with worldview question (f), is also a second-order inquiry, which is critical. Second-order philosophizing mobilizes a critical and reflexive attitude, typical to the philosopher. This Chapter 1 and Chapter 2 are of second-order nature, about the “philosophy of philosophy”.

Yet, even second-order questions are not disconnected from first-order ones. Answers to first-order questions, whether implicit or explicit, determine second-order analysis (Adler 1965, 45). For example, reflections in philosophy of mathematics, investigating what mathematical objects are, have implications in our epistemology (question (f)) and therefore on how to model and predict the world (questions (a)-(c)). Most lively debates are likely to be motivated by first-order questions. Platonists or constructivists disagree on the ontological nature of mathematical objects, and are thus ultimately busy with question (a). With this worldview agenda, we insist on reconnecting with first-order questions, whose corresponding dimensions are often neglected in contemporary philosophy (Adler 1965, 42–48).

The dialectical dimension (5) in second-order philosophizing describes different and sometimes contradictory positions on issues. Worldview question (g) requires that this dialectical dimension is properly answered. The concept of dialectic has a rich history in philosophy, but here, its etymological meaning will suffice: “the art of debate”. I do not use it in a Hegelian sense, nor in the derogatory sense of rhetoric or sophistry. The goal of dialectical philosophizing is to remain “point-of-viewless”. This philosophical activity consists in stating or reconstructing issues and a variety of positions towards them. Here, *dialectical* is opposed to *doctrinal*.

This can be illustrated by three great examples in the history of philosophy. In Antiquity, Aristotle in the first book of his *Metaphysics* describes in detail the positions of his opponents before developing his own. In the middle ages, Thomas Aquinas in his *Summa Theologica* (1265-1274) also represented other positions as objections. In modern times, with the two index volumes of “The Great Ideas: A Syntopicon of Great Books of the Western World” Adler and his team (1952, xxx) also had this ideal to remain position-neutral. They provided outlines and indexes of positions related to 102 great ideas in 443 books. Such a gargantuan work could be called a “Summa Dialectica” of the twentieth century (Adler 1952, xxxi).

As useful as it is, dialectical philosophizing alone still remains categorization, an exercise just slightly more difficult than philately. As Rescher (1985) argued, the temptation of *syncretism*, namely to accept all positions distinguished, is an insufficient philosophical accomplishment, since a mere conjunction of contradictory positions is of course self-contradictory. Syncretism stems from a confusion between first and second-order philosophizing.

A final dimension of philosophizing is needed to fully exploit this dialectical effort in a doctrinal way. As Broad (1947) noticed, philosophers doing such a dialectical investigation, what Broad calls synopsis, are most often motivated by synthesis. Assuredly, Aristotle, Thomas Aquinas or Adler are great synthetical philosophers.

The synthetical dimension (6) is the climax of philosophizing, but also its most arduous dimension. To be successfully conducted, it requires mastering and juggling with all other five dimensions. The great philosophers' feat is in providing a comprehensive and coherent synthesis of their time. It is so challenging that it is rarely attempted (Broad 1947). When we speak about "worldview synthesis", we refer to this dimension of philosophy.

This chapter and the next fall within the critical dimension (4), concerned with "the philosophy of philosophy". Nevertheless, my motivation in proposing evaluation standards and tests (in Chapter 2) is to help answer first-order questions and to encourage synthetical philosophizing. Faithful to the spirit of this synthetical dimension, there is a clear connection between my first and second order philosophizing. This is why at heart my analysis cannot be neutral, it can not be separated from my first-order philosophical position outlined in Appendix I, p315.

Even if synthesis remains an ideal, it is very important to note that each dimension of philosophizing can be pursued relatively independently. What is dangerous and ridiculous is when a philosopher claims that one of the dimensions is the only "real" or "true" way of philosophizing. For example, an historian of philosophy does very valuable work in dimension (5) when he clarifies, puts in perspective or corrects some misinterpretations of a great philosopher. The position of that philosopher is then faithfully represented. But this effort, however useful, remains at best one sixth of philosophizing. In section (2.3 Assessment Tests, p45) we will examine the interactions of the six dimensions, by proposing tests across each of them.

## 1.3 Necessity to Have a Worldview

In the section "The need for philosophy: humans as homo quaerens" Rescher (2001, 6–10) argued from an evolutionary point of view that humans' strength is in their capacity to acquire and use knowledge of the world:

> We are neither numerous and prolific (like the ant and the termite), nor tough and aggressive (like the shark). Weak and vulnerable creatures, we are constrained to make our evolutionary way in the world by the use of brainpower.

This leads to the practical need to acquire more knowledge, to be able to understand and thus predict features of our world. There is accordingly a need to have a worldview to *describe* the world and to act in it. What about the *normative* dimension of worldviews?

There are also psychological and sociological needs for a worldview giving values. Sociological research indicates that the feelings of insecurity and distrust are stronger among the people who least profess belief in a religious or philosophical worldview (e.g. Elchardus 1998). Psychologists researching life satisfaction have found that having such beliefs increases well-being, by providing a sense of life meaning, feelings of hope and trust, a long-term perspective on life's woes, and a sense of belonging to a larger whole (Myers 1993). If philosophy does not answer those questions, other realms of our culture will take advantage of the situation, and provide answers. These are principally religions, or, much more dangerously, cults, sects, extremist secular ideologies or fundamentalist interpretations of religion spreading irrational beliefs.

We all need a certain worldview, even if it is not made fully explicit, to interact with our world and give a meaning to our lives. There is a practical need to have at least an implicit, pre-ontological and for that reason "naive" answer to each of the worldview questions.

## 1.4 Implicit and Explicit Worldviews

Most people adopt and follow a worldview without much thinking. Their worldview remains implicit. They intuitively have a representation of the world (components (a)-(c)), know what is good and what is bad (component (d)) and have learned how to act in the world (component (e)). And this is enough to get by.

But some curious, reflexive, critical, thinking or philosophical minds wake up, and start to question their worldviews. They aspire to make it explicit. Articulating explicitly one's worldview is an extremely difficult task. It is so difficult that philosophical schools have tried to escape it, remaining in the comfortable armchair of second-order philosophizing. Two extreme positions are then possible; either to accept no philosophical doctrine at all (skepticism) or to accept them all (syncretism). Such positions are not tenable if we commit to answering first-order philosophical questions (Rescher 1985). At best, skepticism or syncretism can be useful philosophical critiques or dialectical descriptions.

Having a clear agenda is still not enough. What about the answers? Answering first-order philosophical questions explicitly is an enterprise which was traditionally philosophy's task. This took the form of comprehensive and coherent systematic philosophical treatises. Regrettably, this trend seems to have fallen out of fashion, since most of today's philosophy addresses second-order problems (see e.g. Adler 1965; Ricoeur 1979).

Before agreeing or disagreeing with someone, we need to explicitly understand our respective positions. Making explicit one's first-order *position* is extremely valuable to present one's philosophy immediately and truthfully. Unfortunately, this practice is not common amongst philosophers. But I choose and invite you to go against this trend. For intellectual transparency and honesty, I make explicit my current first-order position in the Appendix I, which is also a summary of my positions in this thesis. So, don't read it if you like suspense! Having a clear position on basic philosophical issues is the philosopher's identity card. Every thinker should have one, and be able to show it when entering the Agora of philosophical dispute.

In Appendix I, I have chosen only to state my *positions*, not to give *arguments*. Instead, I give main references to the works which most influenced me, where the curious reader will be able to find many detailed arguments. I also make explicit which criteria I value most to work out my position. Certainly this is not as satisfactory as a fully developed philosophical system (see e.g. the impressive work of Bunge 1974; or Rescher 1992). Yet, I am confident this effort will facilitate debate and critique of the positions and arguments presented here.

## 1.5 A Cybernetic Model of a Worldview

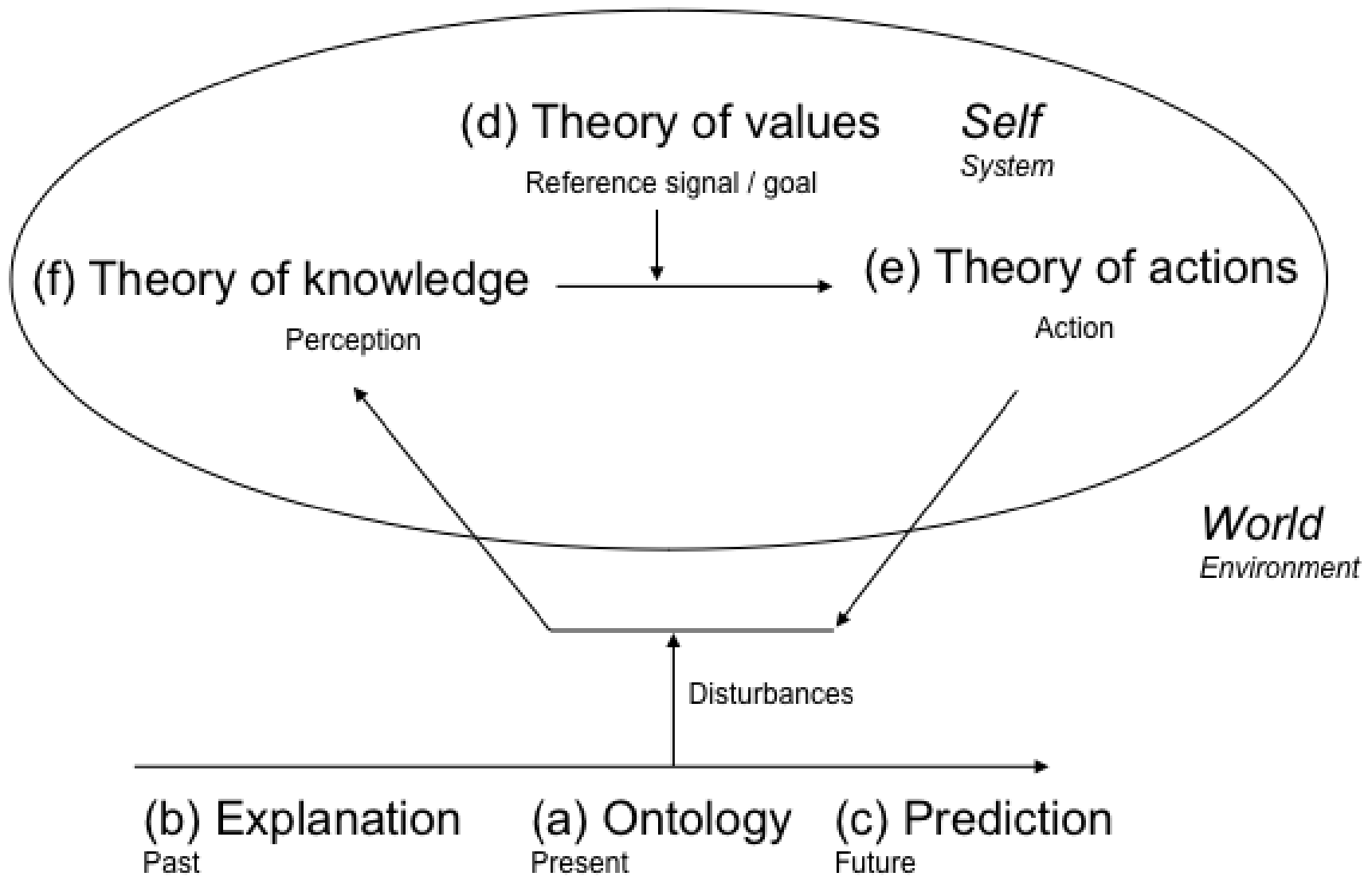

Figure 2: Worldview of an individual in a cybernetic system. Heylighen (2000a):

"The apparently disconnected components of a worldview can in fact be understood as part of an encompassing scheme describing the interaction between a system or self and the world or environment. In cybernetics an autonomous system or agent is conceptualized as a control system, which tries to achieve its goals or values by initiating the right actions that compensate for the disturbances produced by the environment. For that, it needs to perceive or get information about the effects of its actions and the effects of the events happening in the world. More specifically, it needs to understand how particular events (past) cause other events (future), that is to say it needs to have a model that allows it to explain and anticipate events. The first six components of a worldview cover all the fundamental aspects of this control scheme, as illustrated in the following figure. Worldview components (in [large font]) are written above the corresponding control scheme components." Reproduced with the permission of Francis Heylighen.

I reproduced in Fig.2 above the "Worldview of an individual in a cybernetic system" diagram of Heylighen (2000a). This cybernetic approach shows how the different worldview components dynamically interrelate.

When we look carefully at the diagram, we are struck by the centrality of the value component. The information we seek and the actions we do ultimately depend on our values. Murphy and Ellis (1996) also saw the importance of values and ethics when they argued that ethics is a science at the top of the hierarchy of social sciences. Likewise, our Chapter 10 which deals with cosmological ethics is a cornerstone of this work.

Let us also note that the seventh component does not appear here, since it is a second-order component. Also, the components need not to be explicit in each individual. I can act consistently according to some values, yet never think about a theory of values.

To illustrate this, we can take the far-fetched example of a bacterium's worldview. How can we interpret its worldview components? Its ontology is what it senses at present, its explanation is a kind of memory which can be its biochemical state, its prediction is a genetically-based feedback system, its axiology is mainly genetically determined: to find food, reproduce, move, eat and digest. Its perceptions are chemical gradients. Stuart Kauffman (2007, 909) also argued that “a bacterium swimming up a glucose gradient and performing work cycles is an agent, and glucose has value and meaning for the bacterium without assuming consciousness. Of course it is natural selection that has achieved this coupling. But teleological language has to start somewhere, and I am willing to place it at the start of life”.

This approach in terms of worldviews thus intricately links abstracts philosophical questions with an agent's personal experience. We do not just seek the most perfect model of the world; we also want it related to and embodied in individuals, thus providing rules to live and act meaningfully.

# *CHAPTER 2 - Criteria for Worldview Comparison*

**Abstract**: Philosophy lacks criteria to evaluate its philosophical theories. To fill this gap, this Chapter introduces nine criteria to compare worldviews, classified in three broad categories: objective criteria (*objective consistency*, *scientificity*, *scope*), subjective criteria (*subjective consistency*, *personal utility*, *emotionality*), and intersubjective criteria (*intersubjective consistency*, *collective utility*, *narrativity*). The Chapter first exposes the heuristic used in the quest for criteria. After describing each criterion individually, it shows what happens when each of them is violated. From the criteria it derives assessment tests to compare and improve different worldviews, which operate across the six dimensions of philosophy. These include the *is-ought*, *ought-act*, and *is-act* first-order tests; the *critical* and *dialectical* second-order tests; the *mixed-questions* and *first-second-order* third-order tests. The *we-I*, *we-it*, and *it-I* tests operate across objective, subjective and intersubjective worlds.

*If philosophical theories are all irrefutable, how can we ever distinguish between true and false philosophical theories?*

(Popper 1958)

Philosophers disagree. As the saying goes, philosophy is the field of unresolved controversies. There is no progress in philosophy comparable with progress in science. Philosophical disagreements are not replaced with agreements. Indeed, given the wide diversity of philosophical schools and traditions, it is difficult to point out even why or how two philosophers disagree.

Broadly speaking, philosophers have tried to understand the relation between humanity and the cosmos. But this enterprise is not philosophy's sole prerogative, it overlaps with science and religion. For this reason, the situation is even worse. Not only do philosophers disagree among themselves, but their answers to the biggest questions compete with answers provided by science and religion. The result is that humans use philosophical, scientific or religious insights –or a combination of them– to handle this quest for understanding, leading to very different kinds of worldviews. Instead of tackling the difficult task of synthesis, there is a trend of overspecialization which leads to a fragmentation of knowledge. The communication between worldviews becomes at best delicate and knotty and at worst, impossible. Can we compare and assess such different worldviews? How can we test their relative strengths and weaknesses? What criteria and tests can we use for arguing that such and such worldview is "better" than another? Furthermore, can we use those criteria and tests to construct synthetical worldviews?

The solution of synthesis can not come from science, which is empirical; nor from religion, which often relies on traditional dogmas. It must come from philosophy, whose very nature is reflexive. Indeed, this understanding of the relationships between different domains of knowledge is itself a (meta)philosophical dimension. There has been previous work in philosophy of science to assess the quality of scientific theories. Even this effort of finding clear criteria in science is not as easy as it seems (see e.g. T. S. Kuhn 1977; McMullin 2008). Surprisingly very few similar attempts have been made in philosophy. Indeed, finding criteria for a "good" philosophy or worldview seems even more difficult than in science. Why is this so? As Popper noticed in the quote above, criteria in science won't apply to philosophy because all philosophical theories are irrefutable. Even more than in science, there are in philosophy radically different aims, methods, schools and dimensions. This contributes to the richness of philosophy, but also to its confusion.

More specifically, philosophers disagree on the *agenda* and thus can not even agree on philosophy's task (Rescher 2001, chap. 3). In that sense, rather than saying that philosophers disagree, it is more accurate to say that philosophers *rarely disagree.* For they are beginning from different starting points and thus are simply talking past each other (Adler 1965, 165). This is a unique situation in the landscape of knowledge domains. To progress, one thus needs to propose a direction in the form of a philosophical agenda. Constructing a coherent and comprehensive worldview is such an agenda.

## 2.1 A Quest for Criteria

### 2.1.1 The Philosophy of "Meta-"

"Meta-" disciplines push reflection to another level. In mathematics for example, this gave rise to metamathematics and completely new kinds of insights. Indeed, proof theory which was initially called metamathematics, uses mathematical methods to study mathematical proofs. This leads to qualitatively new kinds of results, for example that a mathematical proposition is *not* provable in a particular axiomatic system. Such a proof is qualitatively distinct from the traditional mathematical activity consisting in proving statements. Another example can be found in historiography, which is the history of history. It asks "how is history written?" and leads to a new kind of reflection about history as a discipline.

In philosophy, this "philosophy of philosophy" endeavor is since a few decades explicitly studied (see e.g. Adler 1965; 1993; Rescher 1985; 2001; 2006; 2010 and the Journal *Metaphilosophy*). This questioning concerns the nature, scope, mission of the philosophical enterprise, and its relation to other knowledge domains. Our aim in this chapter is descriptive, to find and define criteria as much as possible independent of philosophical positions. This is why it is a work of metaphilosophical nature. This is of course an ideal, since no metaphilosophical approach is free of philosophical assumptions (c.f. Pepper 1945; Rescher 1985, chap. 8.1).

Our main philosophical assumption behind the criteria and tests we are about to propose is the endeavor of synthetical philosophizing (dimension 6). That is, to construct coherent and comprehensive worldviews, answering the philosophical agenda constituted by the five worldview questions. We call such a worldview which is both coherent and comprehensive, *synthetical*.

### 2.1.2 The Big Three

There are three perspectives we take into account to structure our criteria. We call them *objective*, *subjective* and *intersubjective*. In broad terms, they correspond to three aspects that many philosophers have distinguished. Let us take a bird's-eye view.

The term "worldview" itself comes in three different flavors and emphases:

(1) a world conception, systemic or objective;
(2) a life world, experienced or subjective;
(3) a world view, social or intersubjective.

In flavor (1) we find the rational scientific endeavor to construct a "world conception" *(Weltauffasung)*, as did the logical empiricists of the Vienna Circle (Carnap, Hahn, and Neurath 1929). Another comparable concept is the "world picture" (*Weltbild*) which insists on remaining consistent with scientific results. For example, Dilthey (1957, 25–27) speaks about an objective *Weltbild.* By contrast, a worldview (*Weltanschauung*) is based on this *Weltbild* to form values, ideals and norms for action, for individuals and society (i.e. subjective and intersubjective aspects). More on the definition and need of a worldview in this flavor can be found in (Aerts et al. 1994) and (Vidal 2008a).

Flavor (2) explores the "lifeworld" (*Lebenswelt)* with existential-phenomenological philosophies, which emphasize subjective experiences. The lifeworld stresses the personal aspect of a worldview. The inquiry is centered at the individual level, like in the existentialist philosophies of Kierkegaard, Heidegger, Jaspers, Sartre, or Merleau-Ponty. The drawback is that it does not emphasize higher levels of organizations (e.g. family, society, planet, universe). This is why it is crucial to go beyond the individual level, and answer those worldview questions with a wide *scope*, a criterion we will detail later.

In flavor (3) "world view" is used in a social and cultural sense, often in anthropology or social sciences (see e.g. Kearney 1975 for a review). The term then parallels "ideology", "symbolic order", "cultural code", etc. "Worldview" is also widely used in christian theology, generally between flavor (2) and (3). For a more thorough history and study of the concept, see (Naugle 2002; Koltko-Rivera 2004).

Speaking about "worldviews" can thus have at least these three possible nuances, depending on our emphasis in either objective, subjective or intersubjective aspects. This will become clearer and more detailed soon (in section 2.2 Criteria for Worldview Comparison, p34). My avowed bias goes towards flavor (1), but I will try to do justice to the two other flavors as well.

Turning to Kant's three critiques, we find them highly reflexive, epistemological and therefore second-order in approach. Yet, their themes concern three different philosophical realms. The *Critique of Pure Reason* concerns the possibility of objective judgments, the *Critique of Practical Reason* deals with intersubjective morality, and the *Critique of Judgment* is partly concerned with subjective aesthetic experiences.

In an attempt to go beyond monism or dualist philosophies, Karl Popper (1979) also proposed a three worlds pluralism. World 1 is "the world that consists of physical bodies"; world 2 is "the world of mental or psychological states or processes, or of subjective experiences" and world 3 is "the world of the products of the human mind". This world 3 is a wide category, including languages, myths, scientific theories and works of art such as songs, paintings and sculptures. He saw worlds 2 and 3 as successive evolutionary products of world 1. But he emphasized the difficulty of understanding interactions between the three worlds, because of the feedback processes going on between them (for a modern approach on the three worlds, see e.g.

Hall 2003). For a critical discussion and the limitations of this ontology from a sociological point of view, see (Habermas 1981, Vol. 1., 76–84).

Max Weber saw the birth of modernity with the distinction between several cultural spheres of value: science and technology (objective), law and morality (intersubjective), as well as art and criticism (subjective). As Habermas (1981, Vol.1, 340) describes, this leads to cognitive, normative and aesthetic validity claims.

In his influential theory of communicative action, Habermas (1981) took inspiration from Popper's three worlds and Weber's cultural spheres of values to define three validity claims. Actors evaluate their speech acts against three worlds (Habermas 1981, Vol. 1, 100):

> 1. The objective world (as the totality of all entities about which true statements are possible);
> 2. The social world (as the totality of all legitimately regulated interpersonal relations);
> 3. The subjective world (as the totality of the experiences of the speaker to which he has privileged access).

Those three worlds correspond to what we called objective, intersubjective and subjective worlds. Interestingly, this framework also inspired multimethodology research methods in information systems (Mingers 2001; 2003).

Ken Wilber (1995) made this tripartition popular, expressing it neatly with grammatical pronouns. The objective world corresponds to the “it”, the subjective to the “I” and the intersubjective to the “we”. He stressed the importance of taking perspectives from *inside* these worlds and not only describing them in an objective manner. It means, for example, that instead of striving to describe the subjective experience in a detached universal way, we can also experience it deeply from the inside. This makes a connection with meditative traditions which seek to explore the nature of inner experiences, as science tries to understand the nature of the outer world. Ken Wilber (1995, 211 and 538–539) also argued that integrating the “big three” is the central problem of postmodernity.

In cultural evolution studies, objective, subjective and intersubjective criteria to fit knowledge are also distinguished (Heylighen 1997a; Heylighen and Chielens 2008). Further developing the insights of Donald T. Campbell, these two papers distinguish three main classes of criteria to select “fit” knowledge. The selection concerns:

(1) Objective criteria – the object that knowledge refers to

(2) Subjective criteria – the subject who assimilates and remembers it

(3) Intersubjective criteria – the communication process used to transmit the knowledge between subjects.

### 2.1.3 Bootstrapping the Criteria

Are the criteria descriptive or prescriptive? The best way to answer this questions is to apply the “Meta-” philosophy to the criteria themselves. In other words, to bootstrap the criteria. This leads us to three principal applications of the criteria.

First, '*objectively*', the criteria can help the dialectical dimension of philosophy (5), by describing characteristics of different philosophical approaches and positions. This is partly the mission of the comparative history of philosophy, when it aims at what Rescher (1985) calls *descriptive metaphilosophizing*.

Second, '*subjectively'*, the criteria can be used to develop a clear substantive position. It is very insightful to recognize one's own cognitive values, and thereby *giving weights to the criteria*. In this section, I have tried to restrict my use of criteria in an 'objective' and dialectical manner. However, I do take a first-order position in Appendix I, (p 315), where I give weight to the criteria. The criteria can also be used as a self-critical checklist, to improve one's worldview, when one tries to maximize a number of criteria.

Importantly, when a philosopher says "philosophy should value only this criterion and not that one", he is just expressing his philosophical position. There is no absolute metaphilosophical position from which he could justify such a statement. *Prescriptive metaphilosophizing* is simply philosophizing (see Rescher 1985, chap 14).

Finally, the criteria can be used '*intersubjectively*', to compare worldviews, conduct debates and clarify disagreements. We emphasize this application in this Chapter and the next one. Importantly, even two thinkers adhering to the same descriptive metaphilosophical criteria list will certainly reach different conclusions. Indeed, they will most likely give different 'subjective' weight to different criteria.

To summarize, the criteria can be seen as tools for philosophers to describe the history of philosophy, to work out their own philosophical position, or to clarify disagreements.

### 2.1.4 Relativity, not Relativism

Of course it is possible to critique this tripartition. As Popper pointed out, there are many feedback loops between those three worlds. Yet, distinguishing the three worlds helps us to understand the different natures of various knowledge domains and traditions.

Taking into consideration the history of ideas as well as cognitive, social and communicative mechanisms, it is clear that knowledge and representations evolve. There is thus no "true" worldview, and it is fundamental to constantly criticize and improve our worldviews. There is therefore a fundamental *relativity* in our approach, in the sense that *we can only compare one worldview with one other.* To compare means exploring and assessing the strengths and weaknesses of different worldviews. There are thus no absolute criteria, nor any intrinsic "goodness" or "truthfulness" of a worldview, as there is no "sound" or "true" language. The French language may have some qualities to express emotions and convey poetry, but a formal mathematical language is indispensable to solve complicated financial problems. It would be vacuous to argue whether French or mathematics is the "better" language. A worldview pluralism is imperative.

This *relativity* does not imply *relativism* however. From a dialectical and second-order perspective, a philosopher can explore and understand a plurality of worldviews. But as she elaborates her first-order philosophical position, the same philosopher will still consider some worldviews to be better objectively, subjectively, or socially than others. The problem is to define what lies behind the word "better". When we use it, we implicitly use *cognitive values*. The role of our criteria is precisely to make them explicit.

For example, a scientist might argue that *objective* criteria are far more important than *subjective* and *intersubjective* ones, whereas a theologian would argue the opposite. This simple remark will lead us to suggest two directions in which the science-and-religion dialog can be enriched (see section 3.3.1 The Way to Philosophical Worldviews, p60).

How can we start to formulate criteria for “good” worldviews? A typical set of criteria would be to recommend good features for each of the worldview components. For example, we can ask: what is a true model of the world? what are the features of a good axiology or praxeology? These questions test worldview components and are certainly necessary to build a well-thought worldview (see section 2.3.1 Testing the Components, p46). However, this would not guarantee that the resulting worldview would make sense as a whole. For example, what if our representation of the world is in contradiction with one's values? We will discuss this *is-ought* assessment problem and other tests (in section 2.3 Assessment Tests, p45).

In formulating the criteria, we focused on *transversal* ones, as much as possible applicable to different worldview components. Which concrete criteria can we use to compare worldviews?

## 2.2 Criteria for Worldview Comparison

Nicholas Rescher (2001, 31) proposed an appealing list of evaluation standards for philosophical theories. Inspired by this list and the “big three” distinction, I propose in Table 1 below a list of criteria. I further explain and illustrate them in the following way. After a short description, for each of them, I attempt to answer: “what if this criterion is violated?”. I then point out abuses and limits of each criterion; and, where possible, suggest contrasting criteria. This balanced questioning will help us to better delineate both the importance and limitations of each criterion. When I refer to a criterion in the ensuing discussion, I italicize it.

| |
|---|
| Objective criteria<br>**Objective consistency** – The worldview exhibits internal and systemic consistency.<br>**Scientificity** - The worldview is compatible with science.<br>**Scope** - The worldview addresses a broad range of issues and levels, in breadth and in depth. |
| Subjective criteria<br>**Subjective consistency** - The worldview fits knowledge and experiences individuals already have.<br>**Personal utility** - The worldview promotes a personally rewarding life-outlook.<br>**Emotionality** - The worldview evokes emotions, so that it is more likely to be assimilated and applied. |
| Intersubjective criteria<br>**Intersubjective consistency** - The worldview reduces conflicts between individuals.<br>**Collective utility** - The worldview encourages a life-outlook and mobilizes individuals for what is socially beneficial.<br>**Narrativity** - The worldview presents its messages in the form of stories. |

Table 1: Criteria for worldview comparison.
One worldview is “better” than another, when, other things being equal, it better fulfills objective, subjective and intersubjective criteria.

### 2.2.1 Objective Consistency

*Objective consistency* requires us to hold a consistent worldview with the use of logic and rationality as a general way to understand, value and act in the world. This includes theorizing, a problem-solving attitude and use of arguments devoid of anomalies and contradictions. Applied to a worldview, this criterion makes us realize that answers to the different questions are interdependent, and must not contradict each other.

Argumentation theory helps in classifying and assessing arguments (see e.g. Weston 2000). Reading and producing complex arguments can greatly benefit from argumentation mapping techniques, which present an argumentation in a clear and accessible visual format, instead of a sometimes confusing lump of text (see e.g. Scheinkopf 1999; Twardy 2004; and Appendix II of this work).

If this *objective consistency* is violated, the result is an invalid or self-contradicting worldview, which is unacceptable. Adler (1965, 158-160) gave examples of self-contradictory theories in Lucretius, Descartes, Berkeley and Hume. In pure logic, the *ex falso quod libet* rule allows one to derive any proposition from a contradiction. But even that rule has two sides. First, it shows that the theory at hand is trivial, since it can derive anything and this is why logicians abhor contradictions. On the other hand, resolving a contradiction, precisely because it allows anything to happen next, can be seen as a great opportunity to question deeply rooted assumptions, and to try out radically new hypotheses or theories.

Yet, even if the worldview is perfectly self-consistent, one also needs to start with solid premises. The soundness of premises is as important as the validity of the reasoning itself. When consistency is taken too far, for example if we follow too closely the mathematical ideal, creative problem solving in broader contexts may be frozen by the requirement to comply with the formalism. To avoid this we need to maintain a wide *scope* (see 2.2.3 Scope, p36). Abusing *objective consistency*, we are naturally drawn into more formal thinking, and therefore into narrowing our creative potential. It seems that more supple tools for thinking are needed as this point (see e.g. the Dialectical Thought Form Manual by Laske 2008, 443–655).

### 2.2.2 Scientificity

Taking into account the advances of science is mandatory nowadays. A modern worldview is therefore expected to be compatible with all the natural sciences. The modeling of our world (questions (a) to (c)) is now mostly a scientific matter. A worldview with respect to the *scientificity* criterion constantly needs to be updated according to scientific progress. This criterion can also be seen as an *external* consistency criterion, whereas *objective consistency* was only an *internal* consistency criterion. By internal we mean a logical and systemic consistency, and by external we mean accuracy with regard to the external world.

Ignoring this scientific criterion leads to unscientific worldviews. This happens when we can study a subject scientifically but nonetheless ignore scientific methods and results. Importantly, Broad (1958, 103) distinguished between *nonscientific* and *unscientific*. Philosophy is certainly nonscientific, but this does not imply that it is unscientific. Indeed, philosophy, in contrast to science, is not an investigative enterprise; it does not question the world with observational or experimental methods. It is therefore nonscientific. However, it is possible and appropriate to conduct the philosophical endeavor in harmony with scientific progress, and thus avoid the unscientific pitfall.

What happens when *scientificity* is abused? Most likely, we fail to make the distinction between unscientific and nonscientific, dismissing both unscientific and nonscientific areas of knowledge. Such a worldview falls into *scientism*, as it displays an excessive trust in the power of scientific knowledge and techniques, applied to all areas of study.

Three general scientific approaches are key for building synthetical scientific worldviews (see section 3.2 Scientific Worldviews, p57). These are *systems theory* for an attempt toward a universal language for science; a general *problem-solving* perspective on scientific issues and *evolution*, broadly construed. To contrast and properly extend a scientific worldview, one needs to take into account the normative dimension of philosophizing in the agenda (e.g. worldview question (d)) as well as considering and integrating *subjective* and *intersubjective* criteria.

### 2.2.3 Scope

The *scope* criterion is particularly rich and vital. We can subdivide it into three: *scope in agenda; scope in level breadth* and *scope in level depth.*

***Scope in agenda.*** Other criteria being equal, one worldview is "better" than another when it has a larger *scope in its agenda*, tackling a wider array of issues. I have already mentioned that the philosophical agenda is a topic of critical importance and so of huge dispute. This dispute often remains implicit and therefore confusing. The worldview agenda covers the most important first-order questions. Here we have used five worldview questions as a prototypical first-order starting point, but more related philosophical questions might be added. To this end, it would be worth doing a history of philosophy based on a comparative analysis of philosophical agendas.

If the *scope in agenda* is violated, specific and narrow issues are considered, which leads to sectarianism and overspecialization (Bahm 1953, 423). What often happens in philosophy is that an intellectual conceptual world is built, criticized, refined, discussed again, and so on. With time, more and more complex distinctions emerge and the initial motivation for those distinctions is forgotten, as is the connection with first-order philosophical issues. This is precisely what has happened in modern philosophy in its insistence on second-order questions and knowledge. For example, American analytical philosophy after World War II tends to have a good internal consistency and a scientific aspect, but a very narrow scope in its agenda (Rescher 2001, 38). The *scope in agenda* is narrowed-down to second order problems (Adler 1965). It is remarkable that, according to Adler (1965), the commitment to first-order philosophizing is the only condition that is missing from analytical philosophy for it to become a respectable way of philosophizing.

Of course, the wider the agenda, the more difficult the synthetical integration. Such an integration has always been the achievement of a single philosopher. Those philosophical systems turn into unrevisable, untouchable personal constructions. At that point, there are no more common standards of truth that are applicable, and philosophizing becomes a personal enterprise, instead of a public one (Adler 1965, 55-56). Adler (1993, xx) describes this mode of validation as *poesis.* The mode of validation is *non-exclusionary*, where two philosophical systems are not more comparable than two poems. This contrasts with a logical and rational approach, which uses an *exclusionary* mode of validation, in which two contradictory propositions cannot be true at the same time. Ironically, this grand rational entreprise of philosophical systems then becomes like poesis, because philosophical systems are not comparable anymore. We thus need to build *revisable* philosophical systems, open to comparison and criticism. To progress in that direction, metaphilosophical agreement is required in the form of sharing a common agenda and list of criteria.

An analogy with thermodynamics is interesting here. We can distinguish between *closed* and *open philosophical systems*. If the philosophical system is *closed*, there is no communication with other philosophies, and philosophizing becomes dogmatic. Such system building (e.g. Descartes, Leibniz, Spinoza) is a dogmatic approach in the sense that it is like an artistic whole: one accepts it or not (Adler 1993, 244). Systems have a monolithic and deductive structure, rising from a firm foundation in unchangeable premises. Such closed systems, if we assume there is something like the second law of thermodynamics for ideas, are condemned to die out.

We thus need to build *open philosophical systems*, revisable, open to new ideas, comparison and criticism. Here, open means that new inputs are welcomed (e.g. data from science or from competing positions known from the practice of the dialectical dimension of philosophizing), as well as waste production (e.g. rejections of some old theses because they conflict with new scientific knowledge or thanks to the practice of the critical dimension of philosophizing). These inputs and waste production are necessary to maintain the system working. But there is a problem: what do we deem worth of including or rejecting? Having a list of criteria or explicit cognitive values is a key ingredient to take such a decision.

But even a wide *scope in agenda* is not enough. For example, Carnap (1928) initially had a very wide scope in his agenda. But it was reduced and translated in a very narrow way. It looked at every philosophical question solely from logical and empirical viewpoints. To avoid such reductionism, we also need to consider the *scope in levels*.

***Scope in level breadth.*** A worldview with a wide scope extends across many if not all domains of human experience. This synoptical dimension is fundamental to the philosophical enterprise (Broad 1947). Philosophical principles then apply to a wide variety of scales and aspects. Such philosophizing aims at unifying otherwise separate phenomena.

When the *scope in level breadth* is violated, philosophizing is restricted to one aspect. Such reductionism starts from a universal intuition such as “Everything is composed of atoms”; or “Everything can be analyzed logically”; or “All our thinking is embedded in language”, and so forth. When such and such insight is pushed in only one particular direction, thinking becomes reductionistic. The history of philosophy is full of such cases. For example, materialism assumes that everything is composed of elementary particles, which leads to difficulties. For example, how can we define what is beautiful or what is a morally good action if everything is determined by interactions between atoms? In the case of language, even if every expressible thought and idea must be expressed through language, does that mean that we can reduce every problem to a problem of language?

There is, however, an equally important danger in *abusing* a broad *scope in level breadth*. The worldview risks becoming too holistic, and might fall into vague New Age intuitions like “Everything is one field” or theories that are too abstract and useless. Accordingly, a delicate balance has to be found between *objective* *consistency* and the *scope in level breadth*. The broader our scope is, the harder it is to make it consistent. For example, Hume's work can be seen as mainly analytical, with *scientificity* and *objective consistency* as his main criteria, while Hegel's work is mainly synoptical, aiming at the widest possible scope (Broad 1947). Yet, Hegel's scope has a tendency to be too large. Some utility and pragmatic criteria can balance the holistic aspect (compare for example the *subjective* and *intersubjective* criteria).

Let me now mention one antidote to reductionism, Dooyeweerd's aspectual framework. In his unique philosophical approach, Herman Dooyeweerd (1953) introduces fifteen aspects through which we can make sense of the world. The aspects are quantitative, spatial, kinematic, physical, organic, psychic, analytical, formative, lingual, social, economic, aesthetic, juridical, ethical and “pistic”; pistic means a deep-seated faith, a kind of ultimate vision. This framework is very promising, and has already lead to applications in information science (see e.g. Winfield 2000; Basden 2007). If we systematically consider such different aspects, it is indeed difficult to fall into any kind of reductionism.

***Scope in levels' depth.*** A worldview with a wide scope extends across not only a wide diversity of levels, but also the extreme possibilities of each level. This is to be found in the idea that great philosophers go to the extremes by seeking the most universal issues, principles, theories and answers (Jaspers 1957, 1:intro).

If we maintain a worldview taking into account many different levels, it might still be reductionistic if all these levels are not pushed to their extremes. Let us take two examples violating the *scope in levels' depth*. If the *spatial scope* is violated in its depth, the worldview applies only to a very limited geographical area. How seriously can we take a philosophy that is based only on the life of a small village but claims to be universal? Similarly, when the *time scope* is violated in its depth, the worldview applies only to a very particular era. How seriously can we take a philosophy considering *only* what has happened in the past ten years of human history?

We need to consider the trade-off between depth-first or breadth-first in the *scope in levels*. Either we go in depth into a subject, with a particular methodology and aim, or we explore a wide variety of levels, aspects, and perspectives.

Even assuming we reach the broadest range of levels, and their deepest capacity, a fundamental issue remains, namely the *scalability* of the worldview, or its logical and scientific consistency *across* different levels. Scalability requires a dynamical hierarchical understanding of the world. We need to switch from *static* to *dynamical* hierarchical levels. Although Dooyeweerd proposes aspects that should be distinguished and taken into account, he doesn't convincingly explain their origin or their complex evolution and interrelations.

The dynamical and hierarchical understanding of different levels is key to understanding complex systems (see e.g. Salthe 1985). It is the ability both to analyze issues closely, and to have a broad perspective analyzing microscopic *and* macroscopic issues. In fact, even the micro- macro- terminology is misleading because we do not want to restrict the analysis to two levels only. We need to look at *n* relevant levels. If we seriously consider the relativity of scales, all scales might be equally important. Understanding the transitions between different levels of complexity arguably generates the hardest challenges in contemporary science. For example, how did space-time emerge in the Big Bang era? How did life, language, consciousness, society, and the rest emerge?

### 2.2.4 Subjective Consistency

*Subjective consistency* requires the worldview to fit the broader knowledge, or common experience individuals already have. It is an important theme in philosophy which is called with some variance “common sense” (Descartes), “immediate experience” (Whitehead), “macroscopic experience” (Dewey), “public experience” (Santayana), or “common experience” (Adler).

If an idea does not connect to existing knowledge, it simply cannot be learned. If *subjective consistency* is violated, knowledge becomes esoteric. Whatever its manifold benefits, if a simple and transmissible version of a worldview is not available, its qualities will not benefit large numbers of people.

There is a continuum between our day-to-day common experience and special experiences undertaken by empirical sciences. The *scope in level breadth* of common experience is much wider than the scope of the tightly controlled special experiences performed in science. So, even if one takes the position that common experience is not reliable for philosophical and scientific knowledge, we can't ignore it and need to interpret it somehow.

As with objective contradictions, subjective contradictions can generate a cognitive dissonance at the heart of a growth process. Radically new problems, ideas or theories hurt our basic expectations. For example, quantum mechanics is at odds with many of our intuitive ideas such as objectivity or causality, and many scientists work hard to interpret this theory consistently with our macro-world intuitions. If my worldview is not challenged by any experience, theory or person I encounter, I have no reason to change it. A contradiction with common-experience stimulates a quest for knowledge.

The *subjective consistency* criterion on its own has some limits. What might be obvious and consistent for a particular subject might not be so for another. This limits theorizing to particular events and subjects, not general theories and objects. Not surprisingly, it is in contrast with *objective criteria.*

### 2.2.5 Personal Utility

A worldview satisfying *personal utility* provides goals, values, or at least some preferences heuristic to choose between alternatives. It requires having a well functioning implicit or explicit theory of values (question (d)), which connects with ways to act (question (e)).

Life satisfaction research has shown that having clear goals or a personal vision is one of the key factors of happiness (e.g. Emmons 1986; Csikszentmihalyi 1990). In our time of information overload, we can easily get overwhelmed by a flow of possibilities. To navigate in this flow, many self-help books encourage one to make one's “vision” explicit (e.g. Nanus 1992; Covey 1999). In fact, even when a clear vision is found it requires a lot of courage and discipline to be faithful to it. Without a vision, one tends to be reactive instead of proactive. As Covey (1999, 72) puts it, reactive people “are driven by feelings, by circumstances, by conditions, by their environment. Proactive people are driven by values –carefully thought about, selected and internalized values”.

A vision in isolation remains idealistic and sterile if it does not help with day-to-day functioning. The challenge is thus to have practical means for coordinating one’s personal actions in harmony with one’s vision. Concretely, this can be achieved if the vision mobilizes one for action. In terms of real-world actions, what we need is a good method of managing action, answering question (e). Problem-solving methods, insights into operational research, management, logical-decision trees, criteria lists, and so forth, are all well-known tools to help us in complex decision making. The "getting things done" method proposed by David Allen (2001) is also a key building block for a praxeology. It is now widely used among knowledge workers, and its principles are supported by insights in cybernetics and cognitive sciences (Heylighen and Vidal 2008).

The doctrine of utilitarism centrally meets the criterion of *personal utility* –and its companion, *collective utility.* An action is right insofar as it promotes happiness. So *personal utility* could also be called "personal satisfaction" or simply "happiness". Yet, it is well known that you can be very happy but be stupid. In Bentham's (1789, chap iv) felicitous calculus, the *purity of pleasure* is essential to ensure further pleasures, and not to have a pleasure that could lead to a pain. This encourages us to build a good model of the world, using *objective criteria* to anticipate future pleasures or pains, within a wide time *scope*.

Abuse of this criterion leads to individualism. Everything becomes centered on the individual's gain in pleasures and decrease in pains. We want the pleasures to be not only personal, but also *scalable* to other individuals and larger systems. That is why we also need to take into account a larger *scope*, at least with an intersubjective or social perspective, as we will see later with regard to *collective utility.*

### 2.2.6 Emotionality

The rational attitude is unemotional (Bahm 1953, 14). It might then be surprising to include *emotionality* as a criterion for a good worldview. The trouble is that emotions often remain poorly recognized and discussed in many human interactions, even if their influence can be immense. Merely suppressing or leaving emotions unacknowledged is letting them intervene in more subversive and unconscious manners (e.g. Freud 1899). It would be foolish to dismiss their power over and impact on every aspect of our lives and worldviews. We definitely need a framework and tools to deal with them.

Emotional states of mind can be triggered by the environment or by interacting with others. It is therefore a criterion better categorized as both *subjective* and *intersubjective*. The interplay of emotions and higher cognitive functions, culture, education and personality is complex and intricate. It is the object of affective science to study motives, attitudes, moods, and emotions, and their effect in personal and collective behavior.

Emotions are basic cognitive mechanisms inherited through evolution. They can be viewed as basic survival "values" passed on *genetically* and *not culturally.* They have been successful over millions of years of evolution in achieving survival and reproduction (e.g. Darwin 1872; Ledoux 1996).

For example, they are indispensable to maintaining basic bodily functions (see e.g. Denton 2005). Such homeostatic emotions are feelings triggered by internal body states. Thirst, hunger, or feeling hot are all feelings engaging us to restore balance in bodily systems, by drinking, eating or moving into the shade. They can be distinguished from classical emotions triggered by external stimuli. Lust, anger and fear motivate us to copulate, fight or flee. More generally, beyond their original survival and reproductive merits, emotions are etymologically what moves us.

Emotions direct our attention and motivate our behavior, resulting in mobilizing us for action.

Finding a good emotional balance is fundamental for someone to be in good health and to be socially integrated. If not, he will experience emotional and behavioral disorders. Having too few emotions, like a psychopath; or having too many, as with some forms of neurosis, are both pathologies. Modern medicine and psychotherapies can help in dealing with such situations.

If *emotionality* is violated, it means that few emotions are involved (or only negative and low-energy emotions such as depression). Our worldview becomes bland and not engaging, whatever its other qualities may be. No motivation is found to accept or act according to a certain worldview rather than another. Not addressing emotions through psychological, social, educational and philosophical efforts leads to insufficient theories, missing a major aspect of our cognition.

But what if we abuse *emotionality* in expressing or communicating a worldview? To take an example in philosophy, Nietzsche's writing is at the border of philosophy and poetry. The result is a work of a great depth and beauty, which is very inspiring to many readers. His work is like a work of art, open to many different interpretations and ambiguities. This emotional or artistic approach to philosophy suffers from a downside –namely, that it does not use an argumentative approach (Rescher 2001, chap. 6.3). We thus need to balance *emotionality* with *objective consistency*, and also include a logical and argumentative attitude.

What are the limits of *emotionality*? The main problem is that emotions can take over our rational thinking mode (however, for a balanced discussion about cognitive and affective cognitive processes, see e.g. Damasio 2000; Davidson 2000). Reacting quickly, strongly and thoughtlessly was once an evolutionary advantage: you had better be scared and react if you see a hungry lion running after you. In our modern day-to-day life, this emotional reactivity is more and more a burden than an advantage. Uncontrolled strong negative emotions can have devastating effects in all kinds of social relationships, so that it seems worth learning how to tame them. On the other hand, positive emotions are crucial to creativity, which requires a relaxed, tolerant and open-minded attitude (see e.g. Fredrickson 2004).

Emotions thus need to be integrated with higher cognitive functions, towards an intentional emotional control (Stewart 2007). Emotional control is a theme that we find already in the Ancient Greece, for example in Epictetus' philosophy. The first step to emotional control is to be able to acknowledge, label and communicate them. Rosenberg's (2003) nonviolent communication method promotes this because it encourages listening and fully expressing universal human emotions, feelings, and needs. This communication method is also fruitful to start a constructive dialog between people holding very different worldviews (see section 3.3.4 Nonviolent Worldview Communication, p65).

Once we are able to recognize one's and other's positive and negative emotions, we can start to manage them. A simple idea for emotional control is to have functionally useful emotional reactions. For most purposes, it is useful to cultivate positive ones and tame negative ones (this is the theme of positive psychology, see e.g. Seligman 1998). Surely, it's easier said than done. Meditation teaches how to have emotions instead of emotions having you. Positive ones can be cultivated (e.g. love, joy, peace, compassion, etc.) but meditation can also help to deal with strong negative emotions (e.g. hatred, anger, fear, anxiety) by observing and releasing them. Such an activity can be seen as *intentional evolution*, where humans choose instead of being subject of their emotions.

In debates, emotional control leaves more space to an objective attitude by focusing on rational arguments instead of following instinctive impulses. The use of logical reasoning is what makes the rational inquiry a universal pursuit, beyond some subjective emotions which could hamper communication. Obviously, it doesn't imply that researchers shouldn't or do not experience emotions. The point is that pragmatically, positive emotions broaden the mindset, which will lead to more creativity than having negative emotions, which narrow the mindset (see again Fredrickson 2004).

From a collective perspective, a great issue in political theory involves the role of emotionality in human association. To promote or diffuse a worldview, which emotions should be initiated? There are four logically possible responses, either harnessing positive emotions, negative emotions, both or none.

First, we saw that relying only on rational arguments doesn't work for most people. This is a lesson of evolutionary psychology, which was since long before understood by rhetoric, the art to convince and persuade. Aristotle's *Rhetoric* famously divided it in three steps: *ethos*, *pathos* and *logos* (authority, motivation, and then only rational arguments). A politic using no emotions would most likely fail. It would not be engaging and it would not encourage mobilization. Mass media and politicians capitalize on this, applying many tricks to trigger and manipulate people's emotions. The abuse of rhetoric leads to verbosity and empty pomposity. This negative aspect is now a common use of the word “rhetoric”, but it is worth noting that rhetoric had in the past a much more positive connotation.

Second, using only positive emotions like love, generosity, compassion, etc. sounds like a good choice for a long-term engagement of humans. For example, an unprecedented global compassionate humanitarian response occurred after the 26 December 2004 Indian Ocean earthquake. But to generalize a positive attitude to any situation leads to a naive and gullible attitude. Emotions, even positive ones, need to remain functional.

Third, using negative emotions such as fear, hatred, etc. can also be effective. The fear of burning to hell helped to control the behavior of the faithful. However, such a strategy might not work on the long-term and is questionable ethically. Indeed, it is worth remembering that incitement to hatred is something which is legally punished in many countries and that history has and is still showing us the worrying results of politics based on fear.

The fourth option is to trigger both positive and negative emotions for different purposes. Even if it is probably the best option, it is important to note that the interpretation of emotions depends on the cognitive developmental stage individuals are (see e.g. Graves 1974).

Yet, harnessing emotions can very quickly be abused, because a strong emotional reaction will limit rational deliberation. History has taught us that emotional control of masses can lead to atrocities. Because emotions are hard-wired in our brains, they easily take over the more rational way of thinking, and can push us in unconsidered actions. Before moving masses we need to consider *utility* applied with a *wide scope* along with a “good” axiology and political project.

It is worth balancing the striving of political leaders to harness emotions with individual emotional control. If citizens are educated in emotional control, it becomes easier to engage –also– in rational argumentations, debates and actions. If their education led them to cultivate emotional awareness, expressing negative and positive emotions, they can more easily engage in informed long-term mobilization for a greater good.

### 2.2.7 Intersubjective Consistency

*Intersubjective consistency* calls for the reduction of conflict between individuals. In other words, it is an effort towards social peace. It requires human interactions flowing flawlessly. Moral philosophy, economics, ethics, politics and jurisprudence are mainly concerned with this criterion.

When *intersubjective consistency* is violated, conflicts between individuals and states occur and interactions are difficult. Yet, moments of high frictions can also be occasions for in-depth reforms and social learning. For example, too much friction in a society often leads to revolutions, along with a whole new way of functioning. A form of social learning occurred after World War II when states learned from their failures and set up new agreements and social bonds.

Abusing this criterion by avoiding conflict at all costs would only promote traditional ways of thinking and acting. This would hinder creative ways of reorganizing societies, because new ways of thinking necessarily imply friction with older ways. Fortunately, conflicts need not to be physical. As Karl Popper (1963, conclusion) eloquently wrote, "the role of thought is to carry out revolutions by means of critical debates rather than by means of violence and of warfare; (…) it is the great tradition of Western rationalism to fight our battles with words rather than with swords."

It is therefore important to balance *intersubjective consistency* with objective criteria, conducting conflicts in an Agora rather than on a battlefield. If communication is difficult, it may also be time to seriously take *emotionality* into account to open dialogue (see also section 3.3.4 Nonviolent Worldview Communication, p65). Additionally, it is fruitful to combine *intersubjective consistency* with a wide *scope*, considering a variety of levels from the individual to the ecological.

### 2.2.8 Collective Utility

*Collective utility* is a natural extension of *personal utility.* It encourages a life-outlook and mobilizes one for what is socially beneficial. We saw the importance of having a personal vision. Yet, if the vision is centered on the individual it runs the risk of being individualistic and opportunistic. Thus, the personal and collective visions should as much as possible be meaningfully integrated within a wide *scope*, leading to a personal life-outlook that is also beneficial for larger organizations. Those organizations range from the family, the social network, the country, to humanity as a whole, the planet as an ecosystem, or even the entire universe. It is of course a difficult challenge to integrate *personal* and *collective utility.*

The concept of *coordination* is central in this discussion. It can be defined as the organization of actions so as to maximize synergy and to minimize friction. To work properly, an organization needs individuals to coordinate their actions. Such coordination mechanisms can emerge more or less naturally, for example through cultural norms, linguistic conventions, traffic rules, and so on. For more elaborate purposes, however, the coordinating endeavor is much more difficult to achieve. How can we promote order and mobilize for collective actions, so that they are done smoothly and cooperatively? A promising compromise between individual freedom and collective interest is to design choice architectures that nudge people towards desired actions (Thaler and Sunstein 2009). In addition, information and communication technologies make this collective coordination endeavor technically workable on large scales (see e.g. Watkins and Rodriguez 2008). A famous illustration of collective coordination is the well-known Wikipedia encyclopedia, which coordinates millions of users to collaboratively write the largest encyclopedia ever.

The open-source software development community already functions with advanced collaborative coordination tools (Heylighen 2007). A central tool is the job-ticketing system, which stimulates the community to act. A user who finds a bug or a feature to implement leaves a message on a forum to which others have access. Other users can then work on this initial stimulus. Inspired by this success, and extending *personal utility*, personal action management systems could be extended to the collective, hinting at the possibility of a collaborative version of "Getting Things Done" (Heylighen and Vidal 2008). Collective problem solving through collaborative argumentation mapping methods also promises to significantly promote large scale rational decision making (Baldwin and Price 2008; Iandoli, Klein, and Zollo 2007).

If *collective utility* is violated, people strive to fulfill individualist values or basic needs. *Collective utility* can be interpreted as a consistency criterion, not on a theoretical level, to stay contradiction-free, but on a practical level, to achieve mutually beneficial actions. Idealized consistent systems are useless if they can not be applied in the real world. *Collective utility* is thus a strong pragmatic criterion to complement theoretical reasonings and constructions. Nonetheless, focusing on *personal utility* or *collective utility* leaves normative problems open. We need to find, define and refine what we deem is the most useful, personally and socially. An axiology is needed.

### 2.2.9 Narrativity

*Narrativity* calls for presenting the worldview's messages through stories. A story can be defined as a connected sequence of actions that follow from one to the next. Stories are everywhere, in every culture. Religious texts, newspapers, gossip, literature, movies and stage plays all use a narrative form to tell stories, real or imagined. Overall, stories constitute the vast majority of humanity's bookshelves.

*Narrativity* and *emotionality* go hand in hand, because both have a double subjective and intersubjective aspect. Subjectively, *narrativity* is essential because it makes the worldview emotional, motivating and easy to assimilate (see e.g. Oatley 1999; Heath 2007). Intersubjectively, it is also crucial for relaying messages effectively. Love stories are typically much easier to spread than mathematical theorems.

Stories are very efficient for passing messages on because our very thinking process works with stories. We are constantly constructing stories where we are the hero... or the victim. Even analytical philosophy applies *narrativity* when it presents moral dilemmas in the form of short stories. This is partly why it is so exciting to try to solve them.

When *narrativity* is violated, we are confronted with theoretical material. Theories are not only insipid emotionally, they are also hard to learn, hard to understand and hard to remember. Theories are disconnected from human concerns.

At the extreme of theorizing is mathematics, often painful to learn. A simple way to overcome difficulties in learning theoretical disciplines would be to include the history of disciplines in curricula more systematically. This would reestablish our natural tendency to be motivated and to learn better through stories. Philosophy and mathematics are not popular because *narrativity* is constantly violated. Notable popular exceptions such as *Sophie's World* (Gaarder 1994) and *Fermat's Last Theorem* (Singh 1998) both use a narrative form to make those disciplines widely accessible.

But leaving aside *narrativity* is also the price of good theorizing. Indeed, telling stories is antithetical to theorizing. The literary and the scientific are two different cultures, very hard to bridge (Snow 1959). Stories require a long message to convey a short idea, which may not be universally valid. Science and philosophy, because of their theoretical nature, seek generalities and *intentionally* avoid *narrativity.* Science aims at finding universal laws, supposed to be certain, independent of time, contexts or individual subjects; whereas stories narrate a sequence of actions at a particular time, in a particular context and with an uncertain outcome (Heylighen 2010a).

## 2.3 Assessment Tests

The simplest application of the criteria is to use them as a checklist to improve or to compare worldviews. Yet, such a worldview assessment is not to be understood as an issue-resolving algorithm. Rather, the criteria are cognitive values that influence the preferring of one worldview over another. As Kuhn (1977) and McMullin (2008) emphasied, the incentive behind such lists of criteria is to maximize each of them *simultaneously.* Rather than a mechanical process, constructing worldviews is an art that needs to balance contrasting and sometimes conflicting criteria.

Individually, criteria are also imprecise. Individuals may differ about their application in concrete cases (T. S. Kuhn 1977, 322). For example, philosophers of science have shown that even if scientists are in principle driven by objective criteria, subjective considerations (Is the theory simple to understand?) or intersubjective considerations (Do authorities believe in this theory?) play an equally important role in constructing and assessing scientific theories (e.g. T. S. Kuhn 1970; Feyerabend 1993).

Can we derive more criteria from the nine we proposed? Surely. For example, *sustainability* requires a wide time scale (*scope in level depth*) as well as *collective utility. Synthesis* requires a wide *scope* in agenda and levels, coupled with the striving of *objective consistency.* And so on, and so on.

Let us now turn from our criteria to evaluation tests and recover some of Adler's first (1965; 1993, 31). Adler argued that there are three families of tests: *empirical*, *pragmatic* and *logical*. *Empirical* tests, such as Popper's falsificationism, are clearly included in *scientificity.* Furthermore, if we follow Adler's distinction between special experience and common experience, *scientificity* would be useful to assess the special experience that science involves, whereas *subjective consistency* assesses the common experience that we undergo. *Pragmatic* tests are represented by *subjective* and *intersubjective utility* criteria. *Logical* tests are included in our *objective consistency* criterion.

Using each criterion individually is relatively easy, but the outcome of such a use is limited. How can we use *several* criteria at the same time? Combining more and more criteria, we face a combinatory explosion, especially as we enlarge our *scope*. Let us see why.

### 2.3.1 Testing the Components

A natural use of the criteria is to test the worldview components (ontology, explanation, prediction, axiology and praxeology). Let *X* be a worldview component and C*n,* criterion *n.* The general question is then:

"*What is a good X according to $C_1$, $C_2$ ,..., and $C_9$ ?*"

For example, "What is a good explanation according to *scientificity, scope, subjective consistency* and *collective utility*?" Arguably, the most important combination to keep in mind is the *scope in levels*. As I argued when describing the *scope* criterion, its use is fundamental to grasp complex realities and to avoid reductionism.

So, the criteria can be used analytically, to improve each worldview's components. But the comprehensiveness ideal urges us to think about several worldview questions and components *simultaneously.* So, if $X_n$ is component *n*, the problem becomes:

"*What are good $X_1$, $X_2$ ,..., and $X_5$ according to $C_1$, $C_2$ ,..., and $C_9$ ?"*

In other words, "What are a good ontology, explanation, prediction, axiology and praxeology according to every objective, subjective and intersubjective criteria?" Or more simply, "*What is a good worldview according to all criteria?"* The task is daunting. The purpose of the following assessment tests is to identify the most salient and useful tests among the most significant combinations. Some tests partially overlap, which is not ideal but can be a way to double-check one's worldview.

### 2.3.2 Testing the Dimensions

I will now describe first-order (*is-ought*, *ought-act* and *is-act*); second-order (*critical*, *dialectical*) and synthetical or third-order tests (*mixed questions*, *first-second-order*) operating across the six dimensions of philosophy. When there is an effort to make descriptive and normative dimensions consistent, we are in the domain of the *is-ought* test. Similarly, combining normative with practical dimensions is covered by the *ought-act* test. Finally, coupling descriptive and practical dimensions leads to the *is-act* test. Key questions summarizing the tests are given in table 2 below.

The *is-ought* problem (Hume 1739) reminds us that philosophy is a unique discipline concerned with questions both about what is the case and about what ought to be. In other words, it is busy with both descriptive and normative issues. Combining descriptive and normative theories leads to the is-ought problem, "the central problem in moral philosophy" (Hudson 1969, 11).

Let me illustrate the *is-ought* problem with the classical issue of determinism and freedom. If we assume at a descriptive level that everything is completely determined, can we then defend on a normative side the view there is such a thing as human freedom? This is a typical complication of the philosophical doctrine of determinism. Until this is-ought problem has received an adequate answer, the doctrine is not satisfying (see also e.g. Adler 1965, chap. 11; and 1993 for more details on the *is-ought* test).

Even if the worldview under consideration successfully passes the *is-ought* test, it tells us nothing about how to act in concrete situations. How well are the normative and practical dimensions holding together? How are moral principles and ethical theories applied in practice, individually and collectively?

The *ought-act* test concerns consistency between values (worldview question (d)) and actions (question (e)). Philosophy as a discipline is rarely considered to be occupied with this problem. In the *ought-act* test, efficiency in action is not primarily what matters. What matters is that individual or collective actions are in line with normative principles. How can we apply normative theories in specific cases and contexts? This is the central problem of *applied ethics*. For example, fields like medicine, business, engineering and scientific research are all confronted with difficult ethical choices to perform (see e.g. LaFollette 2007). To act meaningfully, a normative theory is largely insufficient. To act we also need practical realizable and concrete means consistent with normative rules. More realistically, to tackle complex moral decision making, applied ethics has developed sophisticated models such as case-based reasoning or Rawls's (1971) reflective equilibrium. In such an endeavor, the philosophical enterprise is mixed with the moral and political one, and needs insights from strategic action, management theories, and so forth, to conduct it.

Let us illustrate a failure of the *ought-act* test with Kant's (1785, 4:421) categorical imperative, “Act only on that maxim through which you can at the same time will that it should become a universal law.” Beautiful. But how are we supposed to apply it in practice? This normative imperative doesn't take into account any real-world complexity in decision making. It doesn't much help one to act. For example, it doesn't help a young doctor to decide if a fourteen-year-old pregnant girl should have an abortion or not. Another example is the value of achieving world peace. Almost everybody would see it as a valuable enterprise. But what are the most urgent and important actions to do now to accomplish it as soon as possible? To stop famine? To fight corruption? To provide all people in the world with energy by building more nuclear power plants? And so on. A lot of deliberation will be needed to reach agreement on these matters.

Often, philosophers are reluctant to embrace action, notably because they feel more comfortable with second-order philosophizing. A notable exception is Karl Marx who famously wrote to Feuerbach that “philosophers have hitherto only interpreted the world in various ways; the point is to change it” (Engels and Marx 1903). Mises (1949) also developed a theory of action, which is sometimes considered as the capitalist equivalent of Marx's *Capital*. The problem is that abuses of applying a philosophical theory in the moral or political sphere easily occur. Therefore, it is worth asking if the philosopher should not be more worried and present when this critical transition from value to action occurs. The *ought-act* test aims to cohere values with concrete actions and is as crucial to adress as the *is-ought* test.

Passing the *is-act* test successfully is essential for effective and efficient action. From a cybernetic viewpoint, this is obvious. The more accurate the model is, the more precise and effective will action and control be (Conant and Ashby 1970). This test entails an engineering attitude, and is of a technical utility. When the *is-act* test is neglected, action doesn't work. In fact, in science and engineering there is a constant feedback between modeling and experimenting (acting). An action that does not produce good results will not be selected. A good model of the world enables us to make predictions of our actions' outcomes.

However, considered alone, the *is-act* test short-circuits the normative dimension. The *only* implicit value here is efficiency. We might call such a shortcut the "normative fallacy," where the normative dimension is simply dropped. This is very important to acknowledge, because if we want to bring in our values, we need other dimensions of philosophizing. The most obvious solution is to combine this test with the *ought-act* or *is-ought* test.

Let us now turn to second-order tests. First, with the *critical* test. The key question is: "Did you critically analyze your worldview with objective, subjective and intersubjective criteria?" We can use both analytical and continental traditions to perform these tests. The analytical tradition will foremost use *objective consistency* and *scientificity* to perform its critique; while the continental tradition will focus on *intersubjective* (social) and *subjective* aspects of the worldview under consideration. Failing to pass this test, philosophizing is unreflective and possibly self-contradictory.

The *dialectical* test is summarized with the question: "Did you review all major positions on this issue?" All good academic research starts by reviewing as impartially as possible all positions and related issues on a certain topic or idea. The *Syntopicon* is a very advanced and useful example of such an effort. Once a wide review is made, it is possible to precisely define and join issues. When this dimension of philosophizing is skipped, we are likely to generate naïve theories, by reiterating mistakes made through the history of ideas. But this dialectical work ultimately helps philosophers occupied with first-order questions, to synthesize different conflicting positions, or to show why their position is "better" than others.

*Synthetical* tests are crucial for anyone concerned about synthesis, or third-order philosophizing, gluing together the previous five dimensions. In synthetical philosophy, we can distinguish at least two tests: the *mixed question* test and the *first-second-order* test.

The *mixed question* test asks: "Is your worldview consistent with and working with other branches of knowledge?" It demands for coherence between different disciplines, when each of them can make a contribution to the issue at hand. It requires an awareness of relationships between disciplines, their subject matter and their limits. For example, a *mixed question* test involving historical or scientific knowledge can discredit philosophical theories. Adler (1965, chap. 12) described the *mixed question* test in operation, by comparing our common experience of material objects with the scientific description of elementary particles. He concludes that one "measure of the soundness of a philosophical theory or doctrine is its ability to (…) reconcile what truth there is in a scientific theory with what truth there is in a common-sense opinion and in the philosophical elucidation of that opinion, when these several truths appear to come into conflict."

Today all first-order philosophical dimensions are mixed with other disciplines. To conduct such an interdisciplinary effort, we need to pass the *mixed question* test. We must know which disciplines we need to involve to solve which problem. The distinction between philosophical dimensions does not imply their separation (Bahm 1980, 4). The same can be said for the distinction between philosophy and science. We can distinguish them, but this does not imply that we should separate them. This point is essential to tackle complex problems, and this explains why the *scope* criteria are fundamental. Such synthetical philosophizing is more than the "philosophies of" typical of its *critical dimension* (4). As I noted earlier, this synthetical dimension connects to first-order dimensions, which are successfully conducted by doing "philosophy with", side by side with other knowledge branches (Hansson 2008). If the *mixed question* test is violated, it leads to monodisciplinarity, a naïve approach to complex problems or inconsistencies between disciplines.

The *first-second-order* test asks: "Is your second-order philosophizing ultimately working for first-order philosophizing or synthesis?" Critical philosophizing most often fails to connect with first-order issues, and thus leads to esoteric knowledge. For example, when studying epistemology, is our goal strongly committed to the effective production of knowledge, to explain, predict and control our world? Or are we engaged in a debate amongst second-order knowledge experts? It is easy to lose sight and sense of the traditional first-order philosophical enterprise. When this second-order philosophizing is unduly developed, several things happen. First, the *scope in agenda* is considerably narrowed down. Second, no connection with common sense is found –that is, it violates *subjective consistency.* Third, only one philosophical dimension out of the six is performed. A similar reasoning holds for the dialectical dimension. It needs at some points to reconnect with first-order issues to be of any use. In summary, the second-order critical and dialectical dimensions of philosophy work in the final instance at operating a synthesis between descriptive, normative, and practical philosophy.

Both continental and analytic philosophies fail this test. In continental philosophy, first-order philosophizing is ignored or drawn in an inaccessible conceptual vocabulary. Today's analytic and linguistic philosophies are focused on a technical second-order philosophizing and will most often fail to connect their analyses to first-order dimensions. In both cases, philosophy becomes an esoteric practice, reserved to a few intellectuals.

| **Test** | **Question** |
|---|---|
| *Is-ought* | Is your description of the world consistent with your values? |
| *Ought-act* | Do you connect your values with concrete decision making and action? |
| *Is-act* | Is your model for action efficient? |
| *Critical* | Do you critically analyze your worldview with objective, subjective and intersubjective criteria? |
| *Dialectical* | Do you join issues and review all major positions on ideas related to your worldview? |
| *Mixed question* | Is your worldview consistent with and working with other branches of knowledge? |
| *First-second-order* | Is your second-order philosophizing ultimately working for first-order philosophizing or synthesis? |

Table 2: Summary of worldview assessment tests across the philosophical dimensions.

### 2.3.3 Testing the Big Three

Let us now turn our attention again to the three worlds. How do they interact? How may we deal with this tension between the objective, subjective and intersubjective –not merely as independent sets of criteria but in their systemic interaction? I will offer some tests to tackle this issue of integrating the three worlds (summarized in table 3 below).

Let us take a bird's-eye view on our criteria. We humans are involved in three kinds of conflicts: against nature (objective), against ourselves (subjective) and against others (intersubjective). We want to minimize those conflicts, or at least we want tools to deal with them. More precisely, *objective criteria* require that the worldview not be in friction with the outside world; *subjective criteria* require that the worldview not be in friction with an individual's common knowledge and actions; and *intersubjective criteria* require that the worldview minimizes friction between individuals, and maximizes their synergistic interactions. In comparative philosophy, Huston Smith (1957, 8) recapitulated that, generally, the West has emphasized the natural problem (objective); India the psychological (subjective) and China the social (intersubjective). This indicates that comparative philosophy can be regarded as a pivotal starting point for satisfying criteria in the three worlds.

A worldview that fits well in the three worlds has more chances to be accepted, appealing, and useful. Ideally, it would give rise to the following benefits. A consistent conception of the world (objective benefit); a lifeworld providing a meaning of life, useful for living a good life (subjective benefit); and a world view whose foundations are fit for a well-organized society, where few conflicts arise (intersubjective benefit). Most importantly, those three worlds would be synthesized as much as possible in a coherent and comprehensive framework, thus forming a synthetical worldview.

If we sum up the use of the three-perspectives criteria, we come to the thesis of *minimizing friction*: *a good worldview has a minimum of friction within and between objective, subjective and intersubjective worlds*. Can we specify more precise and concrete tests towards this ideal? Just as I did for the philosophical dimensions, I propose here tests across the three worlds. This leads to three main tests: *we-I*, *I-it* and *we-it*.

In the *we-I* test, we ask: "Is your worldview compatible with or in friction with the interests of society?" This question raises a classical problem in political philosophy, namely, the conflicting interests of the individual and the collective. Ideal solutions of such conflicts do not reach compromise (a zero-sum game), but achieve cooperation and synergy (non-zero-sum game). To achieve cooperation, we often develop empathy with our neighbor or our society. But applying the *scope in level breadth*, we need not to stop at society. Larger systems can also be embraced, from the whole planet to the entire universe (see our exploration of a cosmological ethics in Chapter 10).

There are two common ways in which the *I-we* test is violated. First, if the subjective world dominates, it leads to individualism and its downsides. Second, if the intersubjective world and values dominate, there are risks of surrendering to a political system, such as communism. There is no obvious solution to this tension and a delicate balance between the two needs to be found.

We can not build empathy for systems we are not even aware of. With the *it-I* test, we can assess the extent of this awareness: "Is your worldview compatible with or in friction with the most up-to-date scientific models?" This test requires the integration of subjective and objective worlds.

When the *it-I* test fails, we live in an unscientific illusion, or with a very limited objective view. Such a narrow awareness might work in the short-term of a single life-span, but is likely to fail on larger time scales. Interestingly, it might be beneficial to work both on our inner subjective awareness –or *involution*– and on the outer objective *evolution* of systems (H. Smith 1976).

With the *we-it* test we ask: "Is the society we are developing compatible with or in friction with the objective world?" Here, we combine *objective criteria* to serve *collective utility.* But, as the *is-act* test showed us, we need to be sure that our values are not short-circuited in such an endeavor. If we emphasize *intersubjective criteria* too much, we might hamper the quality of our world models. On the other hand, relying exclusively on *objective criteria* to take decisions leads to a scientistic worldview, ignoring the will and values of individuals, societies and larger organizations.

| **Test** | **Question** |
|---|---|
| *We-I* | Is your worldview compatible with or in friction with the interests of society? |
| *It-I* | Is your worldview compatible with or in friction with the most up-to-date scientific models? |
| *We-it* | Is the society we are developing compatible with or in friction with the objective world? |

Table 3: Summary of worldview assessment tests on the big three

### 2.3.4 Summary

Without navigation tools, it is difficult to choose a philosophical direction or to compare the pros and cons of two opposite philosophical theories. To navigate the rich and complex philosophical landscape, I proposed four key metaphilosophical concepts: a set of philosophical *dimensions*, a philosophical *agenda*, a list of *criteria* and a battery of *tests*.

I first distinguished six dimensions of philosophy, composed of three first-order ones (*descriptive*, *normative*, *practical*); two second-order ones (*critical* and *dialectical*) and one third-order (*synthetical*). I then introduced a clear, explicit and enduring philosophical agenda constituted by seven big questions, which define what a worldview is. Next, we started our quest for criteria from previous work in philosophy, cognitive sciences and cultural evolution and ended it up with the finding of nine major criteria. We saw that the criteria can be used in three different manners. First, to describe the history of philosophy; second, to describe one's own position by giving more or less weights to the criteria; and finally to clarify disagreements between different worldviews. I then discussed the criteria one by one, pointing out both their strengths and weaknesses.

To facilitate the comparison of different worldviews, I identified from the criteria and philosophical dimensions various assessment tests: the *is-ought*, *ought-act*, *is-act* first-order tests; the *critical* and *dialectical* second-order tests; the *mixed-questions* and *first-second-order* third order tests; and the *we-I*, *we-it* and *it-I* tests.

Recognizing a set of philosophical dimensions, a common agenda, a shared criteria list and a battery of tests is essential to encourage communication and debate amongst philosophical schools and thus make philosophy a public enterprise.

Specifically, I see the six dimensions, the worldview agenda, the criteria and tests as a metaphilosophical apparatus to understand, improve, compare and constructively criticize different worldviews. Such tools are vital for the demanding endeavor of constructing together comprehensive and coherent worldviews, in the spirit of synthetical philosophizing.

Comparing in details two philosophical systems or worldviews is obviously outside the scope of this work. For example, comparing Whitehead and Spinoza's philosophies is a huge scholarly enterprise. Nevertheless, our criteria and tests should be useful to clarify concrete issues. This is why we now turn to applications, to further characterize religious, scientific and philosophical worldviews.

# *CHAPTER 3 - Religious, Scientific and Philosophical Worldviews*

**Abstract**: This Chapter uses the philosophical dimensions, criteria and tests to better appreciate the respective strengths and weaknesses of religious, scientific and philosophical worldviews. Religious worldviews are illustrated with the conflict between Intelligent Design and Flying Spaghetti Monsterism. We recognize psychological and societal strengths of religions, but also their limitations and failures. The strength of scientific worldviews is illustrated with systems theory, a problem-solving attitude and universal darwinism, while its weaknesses stem from its focus on objectivity only and thus its neglect of values and action, essential components for psychological and societal functioning. Then, I present philosophical worldviews as an attempt to build coherent and comprehensive worldviews, in the spirit of synthetical philosophizing. I discuss the pros and cons of promoting uniformity or diversity of worldviews. To understand what it means to answer worldview questions, I compare them to axioms, systems of equations and problems to solve. I argue that non-violent communication can be very useful when worldview conflicts become emotional. I end by introducing an extreme worldview agenda, which embraces a maximal space and time scope, thereby naturally introducing the cosmological perspective of Part II and Part III.

## 3.1 Religious Worldviews

The science-and-religion debate is arduous, complex and multidimensional. There are many pitfalls to avoid in this debate (see e.g. Van Bendegem 2007). I do not aim here to go in the depth of the debate, but to show how our dimensions, criteria and tests can help to enlighten it.

### 3.1.1 Intelligent Design Versus Flying Spaghetti Monsterism

Let us examine the conflict between Intelligent Design (ID) and Flying Spaghetti Monsterism (FSMism). In 2005, the Kansas State Board of Education required the teaching of ID as an alternative to biological evolution in public schools. Astonished by this decision, Bobby Henderson protested against it. He created a satirical deity, the Flying Spaghetti Monster (FSM), supposedly at the origin of our universe. In an open letter to the Kansas School Board, he then proposed that science classes should include: "One third time for Intelligent Design, one third time for Flying Spaghetti Monsterism, and one third time for logical conjecture based on overwhelming observable evidence" (B. Henderson 2005). The purpose of this action was to show that it does not make sense to teach ID in schools, at least no more sense than teaching FSMism.

With the help of our criteria, let us see whether we can confirm our intuitive idea that FSMism is still "less valid" than ID. Although the theories are not presented as worldviews, they underlie very different worldviews. Our criteria and tests can thus be applied if we look at FSMism and ID in such a broader context.

**Testing the Components.** Regarding *objective consistency*, ID and FSMism are comparable: they postulate a designer-of-the-gaps that can solve any contradiction. Concerning the *scientificity* criteria, ID and FSMism are equally bad: no scientific evidence supports either, which is the main point of this FSMism satire. Both are *unscientific* theories: biological evolution is most effectively researched with the available scientific evidence, theories, conjectures, methods, and the like. Still, ID arguments are more subtle than the FSMism ones (e.g. using the notion of “irreducible complexity”, instead of the FSM showed to be not very bright in his unintelligent design).

The *scope in agenda* criterion tests the breadth of worldview questions tackled. FSMism tells us about how the world came about (question (b)), and perhaps where we are going (question (c)). However, the ramifications of ID are much richer. ID originated from creationism and thus has clear links with the God of monotheists. Therefore, implicitly, supporters of ID have a religious agenda, which makes the theory appealing. However imperfect and self-contradictory religions sometimes are, they are full of recommendations and rituals concerning values (question (d)) and actions (question (e)). FSMism does not propose comparable values or ways of acting, gradually gathered by religious traditions over centuries. Both FSMism and ID are feeble in answering questions (a), (b) and (c), but FSMism has not much to say about questions (d) and (e). Therefore, ID has a greater *scope in agenda*.

In terms of *subjective consistency*, ID also scores much higher than FSMism. Indeed, in ID, the identity of the designer is not even systematically related to a God. The designer is thus a fuzzy concept open to many possible interpretations, and such vagueness can contribute creating a mysterious “guru effect” (Sperber 2010). In contrast, the Flying Spaghetti Monster is a very specific entity, with “his noodly appendage” and his meatballs, defying common sense. In this respect, ID is more subjectively consistent than FSMism. No enriching *personal utility* is offered in FSMism, except perhaps to fulfill a need for humor. It would rather be scary and disgusting to think that the world really originated from such a monster. This disgusting aspect of FSMism triggers a negative emotion, and makes it score low on the *emotionality* criterion. In contrast, ID points out that the scientific enterprise is limited. This is precisely the connotation behind terms such as “irreducible” which contributes to give a feeling of mystery and awe. Of course, this applies only to people not sensitive enough to *objective criteria*.

Regarding *intersubjective consistency*, it can be argued that FSMism is better than ID, because it has “never started a war and never killed others for their opposing beliefs” (B. Henderson 2006, 65). Concerning *collective utility*, both are quite good, although for different reasons. ID, because of its links with religions, which can potentially weave a social web. However, even if FSMism has some collective success because of its satirical, humorous and provocative aspects, its potential for *collective utility* is far behind ID. Both ID and FSMism use stories and thus apply *narrativity.* Yet ID can rely on hundreds of well known biblical stories to convey its messages, whereas FSMism has just a few freshly elaborated stories.

Summing up, both score equally low on *objective criteria* (except for the *scope*, which is larger in ID). Otherwise, ID generally scores better on *subjective* and *intersubjective criteria*. Therefore, from this analysis, we can conclude that ID underlies a “better” worldview than FSMism. I have few doubts that Pastafarians, the devoted members of the FSM Church, will work hard to improve this situation.

**Testing the Dimensions and the Big Three.** Both ID and FSMism score low on the *is-ought* test. FSMism because it is still in its infancy to provide values and moral principles to guide day-to-day life. ID because the oughts derived from creationism are often inconsistent and unscientific, violating two objective criteria, *objective consistency* and *scientificity.*

The *ought-act* test succeeds better in the case of ID, because social structures have already been in place since more than two thousands years with churches, priests, rituals, and the rest, supporting its values with practices. By contrast, FSMism is only some half-dozen years old and needs more time to seriously compete with ID. Because they are not really occupied with *objective criteria*, both ID and FSMism ignore the *is-act* test, which leads to inefficient acting.

ID and FSMism are likely to argue which of them is better when it comes to the *we-I* test. What is in the best interest of people and society? To believe in ID or in FSMism? Both fail the *it-I* and the *we-it* test, because they are not compatible with scientific theories, and will therefore maintain a naïve world conception.

All second-order and synthetical tests fail –or are rather not even applicable, since both ID and FSMism have no ambition to be that thoughtful.

### 3.1.2 Psychological and Societal Strengths of Religions

This brisk comparison between ID and FSMism shows that worldviews inspired by religions score high on many criteria *at the same time*, and this helps to explain their success. Why do so many people believe in religious worldviews? Let us now examine in more details the strengths of religions with our assessment tests.

**Testing the Components.** Religions are not afraid to frontally tackle taboo subject-matters, such as the origin of our world or what happens after death. Religions give meaning by providing answers to such big questions. Even if such answers are rarely up-to-date with recent scientific discoveries, at least, they do provide answers and do not evade, complicate or diffuse those questions as any well-trained second-order philosopher would do. And, as we will see in Part II, there are extremely difficult metaphysical issues regarding the origin, which can indeed be successfully tackled with an idea of God.

Religions excel both in *personal* and *collective utility* by offering ways to resolve internal conflicts or to improve social bounds. Indeed, even if a contradiction arises in the life of a believer, religious personnel will be available to support him. It has also a strong internal logic. As a last resort, there is always the authority of scripture or the will of a God-of-the-gaps. It will never end in uncertainty. The result is that religions convey a strong feeling of security, essential psychologically. Religions also bring a meaning of life, which is as important psychologically. They bring meaning to major events in human life, like birth with baptism, love relationship with marriage, death with funeral.

Religions also have time-proven and socially accepted value systems –except in some of its fundamentalist interpretations. Religious values ensure social cohesion. For example, the fear of burning to hell could entail believers to act according to religious values rather than others. This social role, although still important, is less and less central since the rise of secular governments.

The *emotionality* criterion is central in religions, with major concerns for subjective emotional experiences and the religious experience. Religions take the human dimension fully. *Narrativity* is also a key elements of all religions, since they all tell stories about heroes or martyrs to pass on messages.

**Testing the Dimensions and the Big Three.** Let us now discuss religions with our worldview assessment tests. The *is-ought* test rarely succeeds in popular religions, because they prefer tradition to novelty and adapt very slowly to new scientific discoveries. They maintain that they don't change essentially. They value tradition and authority (intersubjectivity) or personal experiences (subjectivity) more than they value scientific progress (objectivity).

The *ought-act* test succeeds quite well when applied to religions, which are quite focused on concrete actions and practices. At the individual level, there are rituals like prayer, meditation or confession that help one deal with difficult cognitive or emotional situations. At the social level, people go on pilgrimages or benefit from the wisdom of religious authorities to make sense of day-to-day life during Mass.

In a religious worldview, the most important of the big three tests will be the *we-I* test, because it is concerned with the cohesion of the individual with society and other human beings. It will more rarely be concerned with the *we-it* and *it-I* tests, which involve objective knowledge.

It is reductionist to speak about “the religious worldview”, as if it were unique. Of course, different world religions have different emphases, especially Western, Indian and Chinese religions (see e.g. H. Smith 1991). Within a same religion, there are also important inter-individual differences in belief systems. Psychologists have showed that there are as many ways to believe in God as there are psychological developmental stages. Indeed, Fowler (1981) has showed that there are “stages of faith” which correspond to Kohlberg's (1981) stages of moral development and other developmental theories in psychology.

To summarize, a religious worldview gives meaning, provides answers to fundamental questions, has a pragmatic value both in terms of psychological benefits and social cohesion. So, we have to be religiously attached to objective values not to acknowledge those benefits!

### 3.1.3 Failures of Tradition

Of course, it is also easy and healthy to criticize religions. The religious worldview has few rational and objective mechanisms to resolve issues or disagreements. This may explain why it can easily lead to the most primitive way to solve conflicts: war and physical violence. A religious worldview proposes few connections to new scientific developments, and is thus non-adaptive. It was not until 1992 that Pope Jean-Paul II issued a declaration acknowledging the errors committed by the Catholic Church tribunal at Galileo's time. In our era of accelerating scientific and technological change, we must recognize and acknowledge our mistakes much more quickly. In cases of doubt, a religious person also runs to risk to fall back into fundamentalism, i.e. the literal interpretation of century-old scriptures.

As we already saw, religions score very low on *objective criteria.* Dawkins's (2008) recent book *The God Delusion* is a detailed and systematic illustration on how religions fail satisfying *objective criteria*. Religions also fail second-order tests, and such a failure is grave. It means that the second-order philosophical mindset is ignored. There is no critical dimension, no dialectical dimension. Only some professional theologians will attempt to remedy those limitations, generally with great difficulties. We will discuss such a more philosophically minded attempt towards a *comprehensive theological worldview* in section (3.3 Philosophical Worldviews, p60).

A religious worldview is often weak when attempting to describe the world (worldview questions (a)-(c)). Furthermore, it will often use the gaps in scientific knowledge to substantiate its position. The God explanation of the fine-tuning issue is such an example and is widely used by theologians to suggest or prove the existence of a creator (see Chapter 6 for an in depth and critical discussion). Since religions are generally more concerned with questions about values (question (d)) and actions (question (e)) both from a subjective and intersubjective point of view, *objective criteria* are much less central than the *subjective* and *intersubjective* ones.

## 3.2 Scientific Worldviews

A scientific worldview is mainly concerned with modeling the world, that is answering questions (a), (b), and (c). Furthermore, two common requirements for a scientific worldview are to provide explanatory power based on and verified by observations and experiments. The requirement of an explanatory power includes for example the ability to make predictions, but also the ability to connect consistently new scientific theories to the rest of science, as the scientific *scope* is enlarged. The empirical requisite furthermore demands that predictions can be tested, or falsified (Popper 1962). Most scientists thus hold a critical realist worldview, and believe that "experimental and empirical activity can lead to truths about nature."

In the last centuries there has been an explosion of the scientific activity. The total number of papers in scientific journals increases exponentially. Along with this tendency of information overload, new scientific disciplines spread out, leading to more specialization. The scientific landscape thus becomes more and more fragmented. In this section I address the problem of bridging the different sciences, from a worldview construction perspective. What concepts should we emphasize to build a scientific worldview filling the gaps between different sciences?

Although such a question would deserve much analysis and development, we argue here that three very general scientific approaches are keys for this endeavor, namely, *systems theory* for an attempt towards a universal language for science; a general *problem-solving* perspective and *evolution* broadly construed. I end by pointing out limitations of scientific worldviews.

### 3.2.1 Systems Theory as a Universal Language for Science

Is it possible to find a universal language for science? Leibniz had a famous program towards a universal language for the sciences (*scientia universalis*), composed of a universal notation (*characteristica universalis*) and a deductive system (*calculus ratiocinator*). However, despite that such a logical approach has the benefits of clarity and precision, it is unable to model complex evolving dynamical systems. Our world is embedded in time, a dimension not modelled by classical logic. But we need to understand the evolution of our world in time. Dynamical mathematical models have been and are still widely used in science, but they often prove insufficient when dealing with complex systems.

General systems theory and cybernetics are a modern attempt to propose a universal dynamical language for science (see e.g. von Bertalanffy 1968; Boulding 1956). They provide general modelling tools (e.g. state-space approach) and concepts like system, control, feedback, black-box, etc. which can be applied equally well in physics, chemistry, biology, psychology, sociology... Those concepts have proven their importance and fruitfulness in engineering.

Traditionally, mathematical models based on physical laws are used to predict the behavior of a system from a set of parameters, boundary conditions and initial conditions. These models are in fact reductionist and developed with analytical methods where the problem is split into easier subproblems. However, when systems become more complex and the number of interactions increases, a simple analytic solution of the mathematical expressions is not possible anymore. Computer simulations can then be used to predict the behavior of complex systems. These simulations are based on a discretisation of space (finite elements methods) and/or time (simulation methods). It is then possible to run a simulation many times, varying parameters and extract general statistical trends. Computer simulations are nowadays indispensable for the design of modern systems and structures. In Chapter 7, we will see that they hold great promises for the future of science.

Systems theoretic models and computer simulations are very successful in engineering science. Nevertheless, they have limitations when dealing with non-linear and very complex systems. In case of chaotic systems for instance, the predictability of the behavior is in practice very limited. More generally, if mathematical models are not available, a *qualitative* approach is first suitable. A general problem-solving perspective allows to logically structure and clarify this qualitative approach.

### 3.2.2 Problem-Solving Approach

In a system-theoretic and cybernetic perspective, a problem can be defined as a gap which is experienced by an agent from his current situation to the situation in which he would like to be. A problem is solved by a sequence of actions that reduce the difference between the initial situation and the goal. Eliyahu Goldratt and Jeff Cox (1984) conceived the "Theory Of Constraints" (TOC) which provides organizations thinking tools to achieve their complex and fast changing goals. TOC practitioners map the logical structure of problems, which considerably helps to make clear where inconsistencies appear (see e.g. Scheinkopf 1999). Because it is a very general problem-solving approach, it can also be applied with great benefit to scientific and philosophical argumentation, as I did in Appendix II.

Karl Popper (1958, 268–269) understood the importance of a problem-solving attitude in the rational enquiry when he wrote:

> every *rational* theory, no matter whether scientific or philosophical, is rational in so far as it tries to *solve certain problems*. A theory is comprehensible and reasonable only in its relation to a given *problem-situation*, and it can be rationally discussed only by discussing this relation.

Most of our difficult problems involve many aspects of "reality", or *scope in level breadth*. For example, an ecological problem must often take into account knowledge about chemistry (e.g. pesticides), biology (e.g. genetics), climatology, without mentioning political, economical, ethical and philosophical aspects. An interdisciplinary approach is essential. As our world becomes more complex and interconnected, it becomes very limiting if not impossible to restrict studies to only one discipline or one aspect of reality.

When one endorses this problem-solving perspective, borders between different sciences fade out. If we have a complex problem to solve, we should use every possible resource at our disposal to tackle it. A major difficulty is then to be able to communicate with scientists of other disciplines. That is why the endeavor of a universal scientific language we outlined above is so important. Even a minimal knowledge in systems theory would already help scientists from very different backgrounds to communicate.

Bridges between exact and human sciences can be constructed from the endeavor to solve problems. In this view, the scientific activity can be drew as a map of challenges, or problems (theoretical or practical) being solved or being tackled, instead of a traditional disciplinary map.

### 3.2.3 Universal Darwinism

The general idea of evolution, which Darwin expressed through the concept of natural selection (variation and selection), has infiltrated almost every scientific field. This general application of evolutionary principles is known as *Universal Darwinism* (e.g. Dawkins 1983; J. Campbell 2011) or *Universal Selection Theory* (Cziko 1995). This spread of evolutionary theories can be illustrated with disciplines like "evolutionary psychology" where mental and psychological traits are explained through evolution (e.g. Wright 1994; Barkow, Cosmides, and Tooby 1992); the closely related "evolutionary ethics" which focuses on the apparition of moral traits; "evolutionary economics", which emphasizes complex interactions, competition, and resource constraints (e.g. Boulding 1991); "evolutionary epistemology" arguing that knowledge can be seen as a result of a natural selection process (e.g. D. T. Campbell 1974; Gontier 2010); "evolutionary computation" inspired by evolutionary processes to design new kinds of algorithms (e.g. Fogel 1995); "neural Darwinism" in neuroscience also has been proposed to explain the evolution of the brain (Edelman 1987) and even in cosmology a theory of "cosmological natural selection" has been hypothesized (Smolin 1992) that we will discuss in detail in Chapter 8.

Evolution has thus largely crossed the border of biological evolution, and can be seen as a general theory of change. For example, complexity theorist Eric Chaisson wrote a history of our cosmos, based on scientific findings, where evolution is its core engine. He defines it as "any process of formation, growth and change with time, including an accumulation of historical information; in its broadest sense, both developmental and generative change." (Chaisson 2001, 232).

In fact, we should not be surprised by this situation, since thinking in evolutionary terms simply means thinking with time, and more precisely about how any kind of structure and function can emerge from interactions occurring in time.

### 3.2.4 Limitations of Scientific Worldviews

What are the limitations of purely scientific worldviews? We saw that the mission of science is traditionally focused on modelling the world, i.e. on answering worldview questions (a), (b) and (c) in an objective manner. We saw that a religious worldview is weak in its answers to those three worldview questions because it is generally more focused on the two other questions regarding actions and values. On the other side, a scientific worldview is incomplete in the sense that it lacks integration of the model it constructs with the more philosophical problems involving the nature and meaning of *values*, *actions* and *knowledge* (respectively questions (d), (e) and (f)). Here too those questions are not exclusively philosophical. But disciplinary boundaries are less of an issue if we take a problem-solving perspective. Indeed, we saw that there exists a field of "evolutionary ethics" (thus addressing question (d)), and "evolutionary epistemology" addressing question (f) and there is a lot of management literature addressing the question of how to act (question (e)).

**Testing the Components.** The scientist has a mindset focused *solely* on *objective criteria* and scientific worldviews score the highest on *objective criteria*. If we look at *subjective* and *intersubjective criteria*, a scientist telling stories of his personal emotional experiences is, fortunately, not taken seriously. This is an important limitation of purely scientific worldviews.

**Testing the Dimensions and the Big Three.** Because science is not busy with ought questions, both the *is-ought* and the *ought-act* tests fail. Some normative principles need to be developed to complete a scientific worldview, if only to explicit its commitment to efficiency values only, that is to admit that it is only interested in the *is-act* test. An axiology, whether philosophical or theological, is an indispensable complement to a scientific worldview.

Out of the big three tests, the *I-it* and *we-it* tests are directly applicable in a scientific worldview. Since science and religion both focus on first-order questions, the *critical*, *dialectical* and *synthetical* tests will only be attempted by philosophically-minded scientists or professional theologians. Let us describe philosophical worldviews in more details.

## 3.3 Philosophical Worldviews

### 3.3.1 The Way to Philosophical Worldviews

Given our short analysis, a fruitful open discussion between scientific and religious worldviews should ideally lead to either:

(1) A religious worldview more objective, consistent with scientific findings.

(2) A scientific worldview completed with subjective and intersubjective perspectives, with a larger *scope in agenda* to include an axiology and a praxeology and second-order dimensions.

The direction (1) is taken by theologians working towards integrating science and religion to build a comprehensive worldview. They invite to higher levels of spiritual intelligence. Notable examples of such developments are the religious philosophies of Teilhard de Chardin (1959) or Whitehead (1930). A similar modern attempt in this direction was proposed by Michael Dowd in his book *Thank God for Evolution* (Dowd 2007) where he proposes an accessible integration of science and Christianity. His directive line is to reinterpret Christianity in the light of evolutionary theory. The result is very inspiring, because it provides a synthesis of *objective, subjective* and *intersubjective* criteria. The same interpretative effort to integrate modern evolutionary thinking would certainly greatly benefit other world religions. From a cosmological perspective, an effort towards constructing a comprehensive theological worldview can be found in Murphy and Ellis' (1996) *On the Moral Nature of the Universe: Theology, Cosmology, and Ethics.*

What about the other option (2)? It is interesting to note that scientific popularization is part of the solution. Indeed, at best, it will make science meet some of the *subjective* and *intersubjective* criteria. Typically, science popularizers trigger emotions by telling fascinating stories around scientists, their lives and theories. But questions of values and actions will remain unanswered. How can we build a naturalistic worldview on rational grounds? This is normally the task of non-theist philosophical systems. One way is to start from a scientific worldview and extend it philosophically to integrate more philosophical propositions involving the nature and meaning of values and actions (respectively worldview questions (d) and (e)). For example, Laszlo (1972a, chap. 13) develops a framework for normative ethics, which fits in with scientific knowledge. In a similar manner, the praxeological component could certainly be enhanced by integrating insights from problem-solving, management sciences, operational research, etc.

In my opinion, it is urgent that efforts are coordinated to build such philosophical worldviews, firmly based on *objective* criteria, and yet taking seriously into account *subjective* and *intersubjective* criteria. Such a philosophical approach would be based on a scientific worldview, but completed with an axiology and praxeology, that time successfully passing the *is-ought* and *ought-act* tests, also helped with second-order dimensions of philosophizing, in a spirit of synthetical philosophy.

Both directions (1) and (2) aim at constructing more comprehensive and coherent worldviews, which then become synthetical worldviews. More precisely, this leads to two kinds of worldviews, a "*comprehensive theological worldview*" and a "*comprehensive philosophical worldview*" (Carvalho IV 2006, 123). Surprisingly, these two endeavors have a similar aim, they just use different starting points, means and criteria.

In Parts II and III, I will take the direction of a comprehensive philosophical worldview and not of a comprehensive theological worldview, which would require another PhD in theology. In the kind of philosophizing which follows, I use in priority objective criteria (*objective consistency*, *scientificity* and *scope*) to begin the construction of a coherent and comprehensive worldview. More precisely, the cosmological perspective requires that the *scope in level depth* is maximally wide in time and space, embracing the whole universe.

Afterwards, when those objective criteria are maximally satisfied, I turn to *subjective* and *intersubjective* criteria to make the worldview successfully applicable in the conduct of a good life and in the organization of a good society. The pursuit of a good life and a good society can then be harmonized with cosmic evolution. However in this work, those further themes won't be as central as the objective endeavor.

### 3.3.2 One or Several Worldviews?

Should we struggle for one single worldview or for several ones? The key to answer this question is in our distinction between first-order philosophizing (*descriptive*, *normative* and *practical* dimensions) and the second-order *dialectical* dimension. We saw that it is necessary to have implicit or explicit answers to first-order worldview questions (sections 1.3 Necessity to Have a Worldview and 1.4 Implicit and Explicit Worldviews, p25). But it is not necessary to have sophisticated second- or third- order philosophical insights to live. Most bacteria don't and most people don't.

At first sight, one might be afraid of a single worldview. Why? We all know the dangers of powerful worldviews, underlying totalitarianism or fanaticism, such as the communist or the nazi ones. Of course, it is very important to analyze the complex reasons for the success of such worldviews at a particular time, but this is not the place to do that here.

Those worldviews bring simple, efficient and straightforward answers to first-order dimensions. However, they won't welcome harsh criticism (*critical* dimension 4) nor a dialectical comparison with alternatives viewpoints (*dialectical* dimension 5). Those critical and dialectical dimensions are precisely the philosopher's main playground. A major role of the philosopher in society is thus to bring perspective, to make sure that worldviews remain open to criticism and comparison; the philosopher does his best to secure, maintain and promote the fundamental values of critical thinking and open discussion.

What if we all had the same worldview? We could fear that it would imply that we would all think the same. This would only be true only if we consider first-order knowledge of a worldview alone. However, as we encourage second- and third- order knowledge, thinking always changes and improves. Additionally, values in a worldview are more like a guide, giving very general recommendations. There are always different roads for a same destination, thus leaving freedom for action.

Furthermore, for the time being, the danger is rather in worldview fragmentation than in uniqueness. Archie Bahm (1979, 101) expressed it well: "the problems facing us today are more those of achieving greater unity, through a new complex organic synthesis, than of achieving more diversity". This dilemma between uniformity and diversity is also well expressed in (Aerts et al. 1994, 24) "we have learned to appreciate variety and multiformity as values, and hence we do not want to strive for one unique worldview. But neither do we want to resign ourselves to the present situation of fragmentation." We can solve this dilemma by aiming for a unique worldview in first-order philosophizing, but continuously practice the *dialectical* and *critical* dimensions of philosophizing, to constantly improve and integrate different worldviews.

On the other hand, what reasons can we find to argue for a unique worldview? First of all, we could say that if reality is one, and a worldview is an objective description of reality, then there can be only one sound worldview. We can immediately object that a worldview as we have defined it also incorporates values, which are chosen, and thus not objective. Still, scientific progress and the *scientificity* criterion leave us less choices for components (a), (b) , (c). An other argument is that a homogeneous society has fewer conflicts (Durkheim 1893). Thus, sharing values and aims will reduce conflicts, and enable to conduct more elaborated collaborative projects. In spite of post-modern emphasis on cultural relativity, there are values common to all civilizations. As supported by empirical research about the factors determining what makes people happy (Heylighen and Bernheim 2000) murder, theft, rape, lying, etc... are negative values in all societies, whereas health, wealth, friendship, honesty, safety, freedom or equality are positive ones.

Generally, a homogeneous system is easier to control and has fewer conflicts, because the elements have the same goals. The word "control", when used in a system theoretic sense has no negative or totalitarian connotation. Thus, less diversity is easier to control. However, a consequence of Ashby's (1956) law of requisite variety is that more diversity allows more adaptability (see also Gershenson 2007). Therefore, it seems that a trade-off between diversity and uniformity has to be found.

### 3.3.3 Analogies for Philosophical Worldviews

How can we start building comprehensive philosophical worldviews? In this subsection I examine three analogies for tackling the worldview questions. I will analyze worldview questions as an axiomatic system, as a system of equations and as problems to solve. Importantly, analogies are only fruitful cognitive tools (see section 5.1 Analogical Reasoning in Science, p87 for a discussion of analogies), I do not want to import the full formalism of mathematics or problem solving to the endeavor of worldview construction.

**Worldview questions as an axiomatic system.** Here, the analogue of a worldview question is an axiom. A first consequence of this mathematical analogy is that every (hidden) assumptions has to be made clear and explicit. We can do mathematics without axiomatization, but it will be much less precise, consistent and systematic. The same holds for philosophizing. We can philosophize vaguely about anything, but philosophizing greatly benefits when problems are clearly stated. If worldview questions are axioms, then worldview answers or components are the analogue of a model of axioms, a structure satisfying a set of axioms. We use the term "model" in the model theoretic sense, i.e. not in the sense of a simplified representation. As it is often possible for a set of axioms to have different models, the same worldview questions lead to different answers translated in different worldviews.

But if we want to build synthetical worldviews, we need to pass the *first-second-order* test. How can we make up our minds in the large landscape of possible worldviews? We already argued that we should only keep worldviews answering the seven questions in a *coherent* and *comprehensive* manner. In our analogy, this corresponds to two fundamental properties of formal theories: *coherence* and *completeness*.

A philosophical worldview should be *complete* in the sense that it should answer all the seven worldview questions. This is the spirit behind synthetical philosophizing, or Rescher's (2001) comprehensiveness criteria, or the remark that philosophical systems should be evaluated "on their capacity for maximal integration of the [worldview] fragments." (Aerts, *et al.* 1994, 41). I mean that a "complete" worldview is suitable, in the sense of a worldview not excluding questions even if some answers are still problematic or *ad hoc*.

Immanuel Kant took into account this requirement. Indeed, in the first *Critique of Pure Reason*, Kant (1781) recognized that pure reason systematically fails to answer metaphysical questions and ends in antinomies impossible to solve rationally. He did not stop here, however. He still sought to answer those fundamental questions, but chose a more hypothetical approach, saying that freedom, immortality of soul and God's existence are postulates. This is fully developed in his second *Critique of Practical Reason* where he introduces the "regulative principle of the pure reason" as a way to quench humanity's thirst to answer metaphysical questions.

Let us now remember that a system is *coherent* if it is not possible to derive a contradiction from it. This of course echoes the *consistency* criterion. In the real world, worldview contradictions are more or less ubiquitous. We should however be aware of the danger in emphasizing coherence too much. The ideal is to build an abstract system of concepts, very coherent, but that would end up too far from reality. So, we should certainly add that coherence must not only be *internal* to the system, but also *external*, with "facts" or "reality", a requirement we have characterized with the *scientificity* criterion.

An important question emerges. Assuming that it is very difficult to build a worldview that is both coherent and complete, which of the following two possibilities should we prefer?

(i) an incomplete but coherent worldview

(ii) a complete but incoherent worldview

The scientific worldview typifies situation (i). The answers it gives to a model of the world (a), an explanation (b) and predictions (c) are very coherent and with some epistemological additions, it can handle the questions of the theory of knowledge (f). Note however that coherence *between* different sciences is still pretty hard to see achieved, and that is why we proposed as a remedy *systems theory*, *problem solving* and *universal darwinism* (see section 3.2 Scientific Worldviews, 57). But we saw that the scientific worldview is incomplete, in the sense that it does not answer problems of values (d) or actions (e). If we start with a very coherent scientific worldview we can then try to complete it with an axiology and praxeology. But how can concepts initially developed for components (a), (b) and (c) be extended or made compatible with attempts for answering the worldview questions (d) and (e)? This might well be very difficult to achieve.

On the contrary, religious worldviews tend to be complete but incoherent (ii). They are most often criticized for their inconsistencies. Indeed, if they remain traditional, they are very poor at components (a), (b), (c) and (f). However, they focus on the practical dimensions of life and thus do give values (d) and guidelines for action (e). Yet even such guidelines can be confusing. For example, if we follow a litteral reading of the Bible, we find both "An eye for an eye, and a tooth for a tooth" (Matthew, 5:38) and "If someone strikes you on the cheek, offer the other cheek as well" (Luke, 6:29). We have to acknowledge that many theologians do great efforts to achieve coherence, by working hard on interpreting the texts, and by including the results of modern science. In this sense, this approach is much more appealing than a purely scientific worldview, which cruelly leaves fundamental worldview questions unanswered. Carvalho (2006, 122) also argued that comprehensiveness "cannot be achieved by a strictly scientific worldview". One can work towards a complete and coherent worldview not only philosophically, but also through a comprehensive theological worldview.

To conclude, to focus on the big picture by valuing comprehensiveness makes much more sense than focusing first on coherence. If we have a comprehensive worldview, we can start solving its contradictions, thus going towards a comprehensive and coherent worldview. The other way around would be to use concepts from a coherent worldview and extend them to make it more comprehensive.

I insist again that this analysis is based on an analogy. Let us therefore point out some of its limitations. The analogue of axioms here are worldview questions, not propositions. The analogy thus does not imply a presupposition of foundationalism, in the sense that there would be propositions that we hold as dogmas. The foundational aspect means that worldview questions are fundamental, but there are no presuppositions regarding how to answer them.

We can also object that the analogy breaks down because of the well-known limitation theorems, which state that no formal system containing at least Peano's axioms of elementary arithmetic can be both coherent *and* complete. We do not exclude that a trade-off may be needed to balance coherence and completeness. But we are looking for heuristics, and this analogy gives us some clues about what an *ideal* worldview should come close to.

**Worldview questions as a system of equations.** Another interesting mathematical analogy is to compare worldview questions with a system of equations. Worldview questions are related, as are the equations in a system of equations. Hao Wang (1986, 210) explicated this analogy when he wrote that solving philosophical problems is

> comparable to solving an intricate set of simultaneous equations which may have no solution at all or only relative solutions in the sense that we have often to choose between giving more weight to satisfying (more adequately) one equation or another.

This means that we might have to give more weight to one component or another when answering the questions. Ideally, the philosopher should limit this bias, or at least be aware of it. He might preferentially answer certain worldview questions, or more generally use a specific set of cognitive values, which can be translated in how he weights the nine criteria we examined.

This analogy also implicitly assumes that there exists a common language to the different equations. Thus, for the worldview questions, this would imply having a same language for answering the different components. We saw that systems theory could fulfill this role for bridging sciences. Its concepts can also be used in philosophy (see e.g. Laszlo 1972a; Heylighen 2000b).

**Worldview questions as problems to solve.** This third analogy may be the most interesting and useful way to look at the worldview questions. Nicholas Rescher argued that the most valuable history of philosophy to write would be one explicating the dialectic of problems (or questions) and answers (Rescher 2001, chap. 2). If we assume that philosophy is problem-solving, then why not use its principles (see e.g. Newell and Simon 1972; Polya 2004) to formulate and tackle philosophical problems? The classical literature on general problem solving methods considers that a problem is solved by following this sequence of steps:

(1) Understand the problem

(2) Conceive a plan

(3) Execute the plan

(4) Examine the solution

In the case of building a worldview, the problem is a very difficult one, because it is in fact the set of problems given by the worldview questions. We saw with Popper the importance of this problem solving attitude (section 3.2.2 Problem-Solving Approach, p58). Not only do we need to examine the problem situation in its objective aspect, with a wide *scope*, aiming at *consistency* and *scientificity*; but we also must study problems from the *dialectical* dimension, to see how the theory solves the problem compared to alternatives. We already introduced nine criteria to achieve just that. However, the communication and comparison of competing solutions often becomes difficult because of very human reactions: emotions.

### 3.3.4 Nonviolent Worldview Communication

More concretely and practically, how can we express aspects of our worldview and listen openly to different ones, especially when we feel involved emotionally?

The NonViolent Communication (NVC) process (Rosenberg 2003) is very well suited for this. Let us see why. NVC is a foremost a communication approach which insists on acknowledging the universality of human emotions, thus fulfilling the *emotionality* criterion. The four steps to speak in NVC are to:

(i) Observe (without judging) what affects you
(ii) Express your feelings triggered by this observation
(iii) Express the needs, values, desires, etc. that create your feelings
(iv) Request your interlocutor concrete actions to enrich your life

It is remarkable that these steps correspond to worldview components, admittedly at a some level of abstraction. More precisely, what we observe (i) reflects how we model the world (questions (a), (b), (c)). Our feelings, needs, values, desires, (ii) and (iii), constitute our genetically and culturally inherited value system (question (d)). The actions we do and request (iv) reveal our implicit or explicit praxeology (question (e)). The result is that when we use NVC to speak, we present others clearly an aspect of our worldview.

Similarly, when using nonviolent communication to listen, we try to decipher and understand the other's worldview, without judging it. The four steps are then to be attentive to:

(i) What he observes
(ii) What he feels
(iii) What his needs and values are
(iv) What actions he requests to fulfill his needs

In NVC, there is also a general requirement to avoid easy judging dichotomies such as “true/false”, “right/wrong”, “good/bad” etc. This induces a major shift in communication. Indeed, this forces us to explicitly justify our knowledge and beliefs in terms of needs or values. This communication process is utterly simple and very easy to understand. However, the real challenge is to apply it in practice, but can bring remarkable results. For example, let us imagine what a neuroscientist could say to a religious person:

> You are just plain wrong, there is no evidence whatsoever that God created human consciousness. Open your eyes to evolutionary theory: it's clear that a mechanistic scheme of explanation is under way.

In NVC, this can be reformulated in the following manner:

> I see that we both wonder about the existence of consciousness. However, I feel puzzled because I'm not in agreement about the God explanation you propose. I need objective explanations to be convinced. I am therefore more inclined to think that consciousness gradually appeared through evolution. Could you be more explicit about why you need to introduce God in this context? Or is there a way to make your belief in God compatible with my need to fit the overwhelming evidence of evolutionary theory?

By using such language, one can truly say what is on our mind and heart, and also listen to other viewpoints to start a constructive dialog. Implicitly, we can start to ask questions such as: “what is the worldview of my interlocutor?”; “what are his observations, feelings, needs, and requests telling me about his worldview?”; “what criteria are most important to him?"; "on which worldview components do we disagree?"; "why and where do we disagree?", etc.

### 3.3.5 The Extreme Worldview Agenda

According to Karl Jaspers (1957), philosophers are concerned by problems of limits and look for extremes. It is also my concern and my directive line when pushing the worldview questions to their *extremes*. In other words, I try to answer the worldview questions within a maximally wide *scope* in time and in space. For example to the question "where does it all come from?", I will not be satisfied with an answer of the kind: "from my mother's belly". I mean, "where does our Universe comes from?" In the same way, I will seek in Chapter 10 ethical values with universal validity, not just within a small community or a restricted context. Similar observations can be made about the other worldview questions.

Those extreme questions are the most difficult scientifically, because they require the utmost extrapolation of models, to conditions we can not experiment with on Earth. Yet, from a philosophical point of view, speculations are very much needed to answer fully and meaningfully the worldview questions. Since this work is of philosophical nature, I am not restricting my quest to the scientific standards of staying within empirically testable theories.

I am fascinated by extreme sports like off-pist snowboarding or sea surfing. Ultimately, such sports harmonize two extreme phenomena, the maximal capacity of the human body with the vagaries of nature. For example, the extreme snowboarder will attempt to ride the steepest slope, even knowing a high avalanche risks; big waves surfers like Laird Hamilton seek the biggest waves, even knowing that a wipe out can mean minutes of apnea in highly turbulent waters.

I occasionally practice extreme sports (although the "extreme" here is very relative...) and I like to see my intellectual pursuit also as a quest to the extremes. The extreme consequences of logical reasoning and scientific theories, the extreme potential of the quest for knowledge. Both physical and intellectual extreme sports are exciting. But they can also be dangerous and require a constant *ambition* balanced with *caution*.

Story 3: Extreme Physical and Intellectual Sports

We saw in section (1.1 First-Order Questions, p21) that first-order worldview questions are *mixed-questions*, meaning that we need both philosophical *and* scientific expertise to tackle them. It means that scientific and first-order philosophical speculations very much depend on our current knowledge.

What follows in Part II and III is thus potentially vulnerable to new discoveries. Indeed, it may well be that in five years a revolutionary cosmological theory would dismiss much of what follows. In that case, the scaffolding presented in this Part I would prove very useful to understand why the worldview I present failed, to correct it, or to build a new one. This Part I was intended to help in adapting or updating our worldviews in a lucid way. The agenda, criteria and tests provide scaffolding for constructing and repairing philosophical worldviews.

In Part II and III, we reformulate, contextualize and focalize those philosophical worldview questions into their version compatible with modern science.

The reader might wonder why I do not address directly the other worldview questions about *ontology*, *epistemology* and *praxeology*. They are also important, which is why I still did answer them by stating my positions towards them (see Appendix I). But the questions of the beginning and the end requires to focus on the mixed questions to provide models of the ultimate past (*explanation*) and ultimate future (*prediction*), which can become meaningful with a theory of value (*axiology*).

## Open Questions

An architect does not leave his scaffolding after he has constructed a building. But here, the situation is different. Worldview construction resembles more a building site, in constant construction and reconstruction. Leaving some scaffolding is not the most aesthetic practice, but is very useful to ease and stimulate further improvements. The "Open Questions" sections at the end of each Part or important Chapter provide important questions to further research. Such open questions are even necessary if we take our analogy with thermodynamics and the idea of *open philosophical systems* seriously (see section 2.2.3 Scope, p36).

- **Philosophical agendas**. Studying philosophical agendas is key to understand different philosophical schools, trends and traditions. It would be very valuable to do an history of philosophical agendas, to better understand their evolution.
- **More criteria**. The criteria list we presented is a starting point, to be further refined and elaborated by other philosophers, possibly with different or more criteria. The criteria may also be refined or improved with studies in the history of philosophy. Conversely, the criteria can help to describe the complexity of the history of philosophy.
- **Worldviews and developmental psychology**. It is important to understand how the worldview of an individual changes through his life. We have not strongly connected developmental psychology with our worldview framework. It is something which remains to be done. Some useful starting points could be the works of Gebser (1986), Kohlberg (1981; 1984), Koltko-Rivera (2004), Laske (2008), etc. It would thus be possible to tackle questions such as: How do you evolve and develop your own worldview? What will trigger a change in your worldview? How much are you attached to your worldview? How to change from one worldview to another?
- **Interdisciplinarity and cognitive values**. We focused in Chapter 2 on cognitive values in philosophy. However, the encyclopedist may be interested to define systematically cognitive values, criteria and agendas in all disciplines of knowledge, like mathematics, engineering, empirical sciences, history, art, religion, etc. Such a study would provide a different outlook on different domains of knowledge, where distinctions between disciplines would become continuous instead of discrete.
- In Chapter 2 we attempted to fulfill one of Adler's (1965) condition which would improve philosophy as a discipline, by proposing nine criteria as standards of truth. But this is just one out of the six conditions Adler identified. Two others are especially important to pursue.
  - (1) **Philosophy as a public enterprise.** Having criteria was a necessary step to encourage philosophy as a public entreprise. However, it is quite a challenge for the *synthetical* dimension of philosophy, which requires systemic consistency. A public entreprise supposes that questions or problems can be attacked piece-meal, one by one, so that it is not necessary to answer all the questions involved in order to answer any one or some of them.

    However, it has at least been proven possible for its *dialectical* dimension, with the construction of the *Syntopicon* by Adler, Hutchins and their

editorial team. Thanks to modern collaborative web technologies, like editable “wiki” webpages, it is possible to ease and scale up such a collaborative effort. The project of a “wikidialectica” would be a great complementation of the encyclopedia Wikipedia. Indeed an encyclopedia traditionally presents facts, not arguments. It could be kick started based on the Great Books and the Syntopicon (1952) when its copyright will expire.

  - (2) **Philosophy as a first-order inquiry.** Philosophy should reconnect with first-order questions; i.e. about that which is and happens or about what humans should do and seek. Part II and III of this thesis are an example of such an attempt. Indeed, we will now tackle first-order questions “where does it all come from”, “where are we going?” and “what is good and what is evil?” as mixed questions, reformulated in our present scientific context.

- **The practical way to philosophical worldviews**. An open question towards the way to philosophical worldviews (3.3.1 ) is to further develop non-religious practices, rites or prayers. Philosophers and thinkers did make such various proposals (see e.g. Comte 1890; Haeckel 1902; Sageret 1922; Huxley 1957; Apostel 1998 up to today’s secular humanism).

# Part II - The Beginning of the Universe

Modern science can successfully connect physical and chemical evolution with biological and cultural evolution (e.g. Chaisson 2001; De Duve 1995). Thus, it seems reasonable to assume that science is an effective method to understand cosmic evolution. The problem of harmony in the universe has thus shifted to its beginning: how did it all start? why did the universe start with these initial conditions, parameters and laws, and not others? Was the initial universe fine-tuned for the emergence of life and intelligence?

The belief in God allowed western thinkers to understand why the "Laws of Nature" are as they are and not otherwise. Scientific activity ultimately consisted of discovering the "Laws of Nature" set up by God. However, now that many scientists no longer believe in God, there is a lack of explanation in the origin of the "Laws of Nature" (Davies 1998).

Nicholas Rescher (1985, 230) summarized alternatives to answer "why is nature's law system as it is?":

> 1. The question is illegitimate (rejectionism)
> 2. The question is legitimate, but inherently unsolvable (mystificationism)
> 3. The question is legitimate and solvable. But the resolution lies in the fact that there just is no explanation. The world's law structure is in the final analysis reasonless. The laws just are as they are; that's all there is to it. And this brute fact eliminates any need for explanation (arationalism).
> 4. The question is legitimate and solvable, and a satisfactory explanation indeed exists. But it resides in an explanatory principle that is itself outside the range of (normal) laws - as it must be to avoid vitiating circularity (transcendentalism)

*Rejectionism* (1) will not make science and rationality progress. Although I am aware that it is a common philosophical position, I am committed to answer childishly simple first-order questions. I do not want to try to dismiss too quickly those questions as meaningless. *Mystificationism* (2) does not make science and rationality progress either. Only if it could be *proven* that the question is indeed unsolvable would it be an impressive result, similar to negative results in mathematical logic, like the proof of the impossibility to construct the quadrature of the circle. *Arationalism* (3) is equivalent to saying that all explanations will fail. Without arguments to support this view, we can't take it seriously. *Transcendentalism* (4) invites an external and most likely supernatural explanation, which is not something we presuppose in this thesis. Rescher mentions a fifth option, the position of *rationalism*. It states that the question is legitimate and solvable, and the resolution lies in the fact that there is an explanation, yet to be defined and found.

Where does it all come from? Before attempting to answer this question, we ask in Chapter 4, where will a satisfying solution to "where does it all come from?" come from? The answer is ... from our brain! This is why I conduct a cognitive and philosophical study to understand our cognitive expectations to explain the origin of the universe. Of course, answers to the origins also very much depend on our available scientific theories. But exploring and better understanding how our cognition functions in this ultimate quest will help us to unveil our biases and preferences in selecting explanatory models. Specifically, I argue in Chapter 4 that there are two cognitive attractors that we use to explain the beginning of the universe, the *point* and the *cycle*. Building scientific models is a process which involves two equally important items, an external system to be understood, and an observer which constructs a model of that system. By better understanding the structure and functioning of the observer-model relationship, we have more chances to avoid biases and confusions between reality and our models.

In Chapter 5, I focus on a common feature of all cosmological models: they bring in free parameters, not specified by the model. Can we reduce their numbers? How can we capture them? Which strategy should we use? I will examine physical, mathematical, computational and biological approaches, bringing different perspectives on this fundamental problem. This multiple analysis will prevent us to fall into any reductionism. Moreover, an understanding of free parameters is a necessary step to make sense of the fine-tuning debate.

In Chapter 6, we will see that some free parameters also have puzzling properties. If we vary them even slightly, no complexity as we know it in our universe emerges. Our cosmological models display parameter sensitivity. This suggests that our universe is somehow very special. These arguments are known as *fine-tuning* arguments and are widely debated in science, philosophy and theology. Unfortunately, they are most often confused with other related issues. Many researchers, including leading scientists, commit and repeat fine-tuning fallacies. I clarify and untangle those issues, which are necessary steps for the new research discipline of *Artificial Cosmogenesis*, a scientifically promising and concrete way to study the emergence of complexity and the fine-tuning issue.

# *CHAPTER 4 - Origins of the Origin*

**Abstract**: This Chapter first distinguishes five challenges for ultimate explanations: *epistemological*, *metaphysical*, *thermodynamical, causal*, and the issue of *infinities*. In a Kantian manner, I then turn the question of the origin upside down and ask: what do we cognitively expect to be a satisfying answer to the ultimate origin of the universe? I argue that our explanations fall into two kinds of cognitive attractors: the point-explanation (e.g. God or big bang) and the cycle-explanation (e.g. cyclical cosmological models). Exploring and better understanding how our cognition functions in this ultimate quest will help us to unveil our biases and preferences in selecting explanatory models. I critically discuss objections against cycles, such as infinite regress or that an infinite universe would necessarily imply that we would have identical copies of ourselves somewhere or somewhen in this universe or another. I conclude that cyclical explanations are more promising than point explanations, but also that less trivial cognitive attractors are logically possible.

*When thus reflecting [of man as being the result of blind chance or necessity] I feel compelled to look to a First Cause having an intelligent mind in some degree analogous to that of man and I deserve to be called a Theist. But then arises the doubt, can the mind of man, which had, as I fully believe, been developed from a mind as low as that possessed by the lowest animal, be trusted when it draws such grand conclusions?*

Charles Darwin (1887a, 282)

All civilizations have developed myths explaining the origin of the world. They provide answers to the fundamental worldview question: "where does it all come from?" (b). In our modern societies dominated by science, myths competition is replaced by a competition of a wide variety of cosmological models. It is not our aim to review those modern cosmological models (see e.g. Heller 2009 for a recent and excellent overview).

Tackling the question of our ultimate origins, we encounter five puzzling challenges, *epistemological*, *metaphysical*, *thermodynamical, causal*, and the dealing with *infinity* in physics. They can be summarized with the following questions: what are the epistemological characteristics of an ultimate theory? why not nothing? where does the energy of the universe comes from? what was the causal origin of the universe? and Is the universe infinite?

Instead of tackling those questions frontally, our approach in this chapter is cognitive. As Darwin reminds us, our brain is a product of evolution. To what extent can we trust it to draw conclusions about the origin of the universe? What do we cognitively expect to be a satisfying answer to the ultimate origin?

I will here mainly focus on the causal problem. Was there a first cause? If so, is it a point-like explanation? If not, should we seek a cycle-like explanation? What are the limitations and biases of those two explanations?

I first outline the five major challenges ultimate explanations must face. I then show that there are two cognitive attractors on which ultimate explanations tend to fall, the *point* and the *cycle*. They are similar to the *fixed point* and the *limit-cycle* in dynamical system theory. They can be described as cognitive attractors in the sense that our ultimate explanations tend to fall into them.

## 4.1 Five Challenges for Ultimate Explanations

When no empirical data is available, we are left with theoretical reasoning. Furthermore, when it is doubtful to use physical theories, we are left only with logic and metaphysics. Before the rise of modern cosmological models a century ago, talking about the origin of the universe was chiefly a metaphysical effort. It still is, but it is less recognized as such. The reason is that we also need to include major results of modern cosmology as we dive into metaphysical waters.

I do not aim to solve the five challenges that I present below. They are supremely difficult and each of them would deserve a PhD on its own. However, we will see them reappear later in Chapter 8, with some possible resolutions. My aim here is simply to formulate the challenges clearly, and to distinguish them from one another. This is *in se* a valuable step, because they are often confused.

Furthermore, there are two additional challenges to ultimate explanations that we will examine in greater detail, namely the unsatisfactory fact that cosmological models have free parameters (Chapter 5), and the open question whether those free parameters are fine-tuned or not (Chapter 6).

### 4.1.1 Epistemological

When we adventure into the idea of an ultimate theory, some basic explanatory principles can be shown to be mutually contradictory. Nicholas Rescher (2000) wrote a remarkable article entitled *The Price of an Ultimate Theory,* in which he carefully analyzed the logical and epistemological foundations of an ultimate theory, and concluded that it leads to an impasse. Let us summarize this argument[5].

Rescher starts from Leibniz' *Principle of Sufficient Reason* (PSR) which states that every fact is capable to be explained. Formally, let the variables t, t', t'', etc. range over the set T of factual truth about the physical world. The abbreviation t'∑t means that “t' explains t”, and we have:

$$\text{(PSR)} \qquad \forall t\, \exists t' \quad t' \textstyle\sum t$$

Now, what are in particular the explanatory desiderata of an ultimate theory? Rescher identifies two such fundamental properties, *explanatory Comprehensiveness* (C) and *explanatory Finality* (F). The first states that whenever there is a fact, the ultimate theory affords its explanation. This can be formalized in the following manner:

$$\text{(C)} \qquad \exists t'\, \forall t \quad t' \textstyle\sum t$$

5 The reader who dislikes logical formula can easily ignore them, since they are also explained in the text. The courageous reader may like to note the notational conventions I use:
∃: existential quantifier. Read “there exists”
∀: universal quantifier. Read “for all”
&: logical symbol of conjunction (AND operator).
¬ : logical symbol of negation.
⊬: logical symbol of non deducibility. $\varphi \nvdash \psi$, reads “ψ is not deducible from φ”.

It is striking that PSR and C only differ by an inversion of quantifiers. This difference between PSR and C is structurally similar to the difference between the *potential* and the *actual infinite* (Vidal 2003). Indeed, the potential infinite can be expressed with the idea that for any given number $x$, it is possible to find a bigger one $y$. Formally:

$$\text{(Potential Infinite)} \qquad \forall x\, \exists y \quad x<y.$$

Whereas the actual infinite is a considerably stronger claim, positing the existence of an actual infinite number $y$, that Georg Cantor (1883) called ω. If $\omega=y$, we have:

$$\text{(Actual Infinite)} \qquad \exists y\, \forall x \quad x<y.$$

Of course, one cannot derive the actual infinite from the potential infinite, as we can not derive explanatory comprehensiveness from the principle of sufficient reason. Formally, it means that:

$$\text{Potential Infinite} \nvdash \text{Actual Infinite}$$
$$\text{PSR} \nvdash \text{C}$$

There is thus a huge gap between the principle of sufficient reason and the ultimate theory, gap similar to the one between the potential and the actual infinite.

The second fundamental property expected from an ultimate theory is its *explanatory finality*. Let us call the ultimate theory T*. The *explanatory finality* states that there is no further or deeper explanation than T* itself. Formally:

$$\text{(F)} \qquad \neg\, \exists t\, (t \sum T^* \ \&\ t \neq T^*)$$

Now, there is a very basic principle that any explanation must respect, the *explanatory noncircularity*. It states that no explanation can invoke the fact that is to be explained. Formally:

$$\text{(N)} \qquad \neg\, \exists t\, (t \sum t)$$

The problem is that explanatory comprehensiveness (C) entails $T^* \sum T^*$, which contradicts (N). There is even another way to come to this contradiction:

1. $\exists t\, (t \sum T^*)$ by PSR
2. $T^* \sum T^*$ from 1 by Finality (F).

This reasoning simply means that if we apply the PSR to the ultimate theory, the ultimate theory must explain itself, if it aims to be final (F). And a theory which explains itself is circular and thus in contradiction with explanatory noncircularity (N).

We can conclude that a circular explanation or an infinite regress can not be avoided in the context of an ultimate theory. Rescher concludes that ultimate theorists must jettison explanatory noncircularity (N). Nevertheless, infinite regresses can but *need not to be vicious* (Gratton 1994). So, it is also logically possible to replace an ultimate theory by a non vicious infinite regress. For example, Quentin Smith (1987) has shown that an actual infinite past is logically possible.

Rescher takes another road, and develops a solution where he considers not only *inferential* explanations, but adds a wider *systemic* explanatory mechanism. With this added explanatory scheme, he avoids a vicious circle and brings instead a virtuous circle of self-substantiation.

#### 4.1.2 Metaphysical

*[Metaphysics] is the oldest of the sciences, and would still survive, even if all the rest were swallowed up in the abyss of an all-destroying barbarism.*
(Kant 1781, B XIV)

"Why not nothing?" Those three words compose the most puzzling metaphysical issue. They question the brute fact of existence. This formulation is a shorter version of Leibniz's (1714, §7) "Why is there something rather than nothing?". The best treatment I know of this question was provided by Leo Apostel (1999) which he wrote before he passed away. The article is precisely entitled: *Why not Nothing?* I refer the reader to Apostel's paper for further reflections and to (R. L. Kuhn 2007) for a panorama of possible answers to this question. As Kant famously wrote above, metaphysical questions are unavoidable.

#### 4.1.3 Thermodynamical

This pure metaphysical question of existence has its energetic counterpart. In physics, a class of fundamental laws are conservation laws. In particular, the *first law of thermodynamics* states that *energy cannot be created or destroyed*, but only transferred from one system to another. How can our ultimate explanation of the origin be compatible with this first law?

Recently, Krauss (2012, 174) wrote that "The metaphysical "rule," which is held as an ironclad conviction by those with whom I have debated the issue of creation, namely that "*out of nothing nothing comes,*" has no foundation in science." Too bad this bold claim contradicts the first law of thermodynamics, indeed a foundational law of science.

Furthermore, how can an ultimate theory be compatible with the *second law of thermodynamics*, which states that the entropy or disorder in a closed system can only increase? Applied to universe as a whole, the second law famously led to the idea that our universe will ultimately end in a *heat death*. Is this application legitimate and the heat death conclusion inevitable?

Another key thermodynamical issue regards the isolation or openness of the universe. Is the universe isolated or open thermodynamically? The application of thermodynamics to the universe will be very different depending on how we answer this question.

### 4.1.4 Causal

What did cause the universe to be? Did the universe cause itself? Was there a first cause, a God or another first cause? If not, is there a cyclic process at play? Are we allowed to imagine an infinite causal chain? Wouldn't it be an infinite regression fallacy?

Is the concept of “cause” itself applicable? Doesn't it presupposes the existence of time or space-time? Our usual meaning of “cause” or “time” have good reasons to be challenged at densities and energies occurring at the big bang era, where the structure of space-time is altered.

Following our philosophical criteria of *scientificity*, we must avoid unscientific theorizing. That is, when a subject matter can be treated with scientific means, we use those means. What is the scientific way to tackle the causal structure of the universe? It is found in cosmological models built with general relativity. It is thus fundamental to take relativity theories seriously, since they can violate our naive intuitions about space-time and causality.

### 4.1.5 Infinity

*Cosmologists see there is room for a lot of infinities in the Universe.*
*Many are of the ‘potential’ variety—*
*the Universe might be infinite in size,*
*face an infinite future lifetime,*
*or contain an infinite number of atoms or stars.*

(Barrow 2007a, 28)

The metaphysical challenge, the first-law of thermodynamics, and the causal challenge all implicitly have to deal with the issue of infinity. Is the universe finite or infinite? What do we mean with this innocent question? Do we mean spatially infinite? Are we speaking about the global geometry of the universe? Do we ask whether the expansion of the universe will be finite or infinite? Are we speaking about the quantity of matter-energy in the universe? Do we want to conceptualize a possible infinite number of causes, before the big bang, and in the far future? If so, do we represent an infinite causal chain with a line or with a cycle?

Furthermore, in mathematics, infinities exist in several powers. Since physical models heavily use mathematical tools to model our world, this raises profound questions. How should we interpret the fact that mathematical tools using infinities work so well to model our world? Could we actually do the same with finite mathematics? What can we do if infinities appear in our equations? Should we consider that something has gone wrong? Or should we simply avoid infinities given their impractical nature? We need to make clear our affinity for infinity, and these are arduous questions in philosophy of mathematics and physics.

## 4.2 The Point as a Cognitive Attractor

*only an infinite sequence of finite causes*
*may replace the notion of God.*

(Heller 2009, 105)

### 4.2.1 Origin with Foundation

Michel Bitbol wrote a very valuable paper in (2004) about origin and creation. However, because it was published in French, I recapitulate its core message in what follows. Bitbol is specialist both in quantum mechanics and in the philosophy of Immanuel Kant. As a result of this rare double expertise, he is sometimes considered jokingly as a “Kantum mechanician”! In this paper, Bitbol distinguishes origin with or without foundation. The “first” origin needs a starting point, a cause which itself does not need another cause. This idea of a starting *point*, which I call a point cognitive attractor, or simply the *point*, seems a satisfactory way to approach the question of the origins. Such a point takes shape with two successful ideas for the origins: God and the Big Bang.

### 4.2.2 Points in Everyday Life

The origin with a foundation has a *causal* and *juridical* meaning. The causal nuance demands a first or ultimate cause. Indeed, if the first cause is itself caused by something else, then of course it is no more a first cause. This is why this first cause must be *causa sui*, cause in itself. This is mandatory to avoid an infinite regress. This illustrates again the tension between the desire of a final and comprehensive ultimate theory and the desire to avoid a circular explanation.

Importantly, this avoidance of infinite regressions was Kant's chief justification of his theses to solve the antinomies he describes in the *Critique of Pure Reason*.

In its *juridical* meaning, the origin is a *creation* point, a deliberate *act* implying a *responsibility*. In legal reasoning, one attributes the responsibility of an act to the nearest intentional agent which provokes a particular chain of causes and effects. This is essential to stop somewhere the causal tree, which otherwise could be extended up to the origin of the universe. For example, in the Concorde's crash of July 25$^{th}$ 2000, a tire exploded after running over a 40 centimeters metal strip that had fallen during the previous plane's takeoff. Who should be held responsible? The tire manufacturer? The pilot who didn't see the metal strip? The airport's runway maintenance service? The court ruled that John Taylor, the mechanic who had attached the metal strip which fell from the previous plane, was responsible. This act of Taylor who did not attach the metal strip tightly enough initiated a series of causes which led to the crash of the Concorde.

In such situations, Bitbol speaks of *heteronomy,* since the triggering cause stands out against the normal causal chain. As we will see with the cycle cognitive attractor, this is contrasted with *autonomy*, where all causes are internal. Heteronomy implies an asymmetry in the causal chain, where an intentional act triggers a discontinuity in the causal tree. Such a juridical reasoning stops the foundation series to a point. In the case of the Concorde's crash, the court decided it all started with John Taylor. What about the universe? How did it all started? Two possible foundational points are God and the Big Bang.

### 4.2.3 God's Point

Judaism introduced God as a creator making the universe in the past, as a definite event. Furthermore, for Judaism, Christianity and Islam, time is linear, not cyclic. God's creation is revealed in a sequence of Creation, fall, incarnation, redemption and judgment. The creation story has thus a beginning, a middle and an end (Davies 2002, 42).

God has the power to put infinity in quarantine. Infinity is concealed in a three letters word: God. Furthermore, we can reassure ourselves if we can not fully grasp God's infinity, because he is of a supernatural nature. With our finite and limited mind, we can not really make sense of God's infinity. Such a line of thinking indeed successfully avoids an infinite regress, since infinity is condensed in a single concept, rather than in an unknown and hard to grasp infinite causal chain.

The problem of the ultimate explanation has then shifted to a theological one. Indeed, the inquiring believer can still ask: where did God come from? What was God doing before he created the universe? etc. To make this logic watertight, theologians add that God is *causa sui*, or self-caused. Furthermore, the thermodynamical issue remains. How did the energy transfer from God to the universe occur? Is there a separation? Is God a being, or being itself? Is God the energy-matter of the universe itself, like in some pantheistic interpretations? Although theological reasonings could resolve those issues, they involve a supernatural explanation, which a non-theist philosopher by definition does not address. Is there an alternative?

### 4.2.4 Big Bang's Point

The Big bang is often conceived as a space-time point or *singularity.* Note however that this interpretation is debatable, since depending on the specific cosmological models, a singularity may or may not have occurred (Ellis 2007a, 1235). But could big bang models have the equivalent of God's *causa sui*? Quentin Smith (1988) argued in a paper entitled *The Uncaused Beginning of the Universe* that one can solve the causal challenge without referring to a supernatural being. Continuing heated debates between atheists and theists, Smith (1999) wrote another paper *The reason the Universe exists is that it caused itself to exist* which presents three ways for the universe to cause itself to begin to exist.

A more precise cosmological model was developed by Richard Gott and Li-Xin Li (1998), which uses closed time-like curves solutions of general relativity, making the question of an earliest point in time meaningless. They write: "asking what was the earliest point might be like asking what is the easternmost point on the Earth. You can keep going east around and around the Earth — there is no easternmost point."

So it may be possible to quarantine or avoid an infinite set of causes at the beginning of the universe, as an alternative to Heller's proposal in the above quote that "only an infinite sequence of finite causes may replace the notion of God."

## 4.3 The Cycle as a Cognitive Attractor

### 4.3.1 Origin without Foundation

Cycles have remarkable properties. They are without bounds, without author, are self-sustaining and autonomous. Let me illustrate the importance of cyclical thinking in different disciplines. The root of the “Meta-” philosophy we described earlier (see section 2.1.1 The Philosophy of “Meta-” , p30) is based on cyclical and self-sustaining principles. In mathematics, recursive proofs are fundamental; in linguistics, Saussure realized that the meaning of words is given by a network of mutually-defining meanings; in psychology, Piaget emphasized that objects and representational schemes are mutually defined. In biology, Maturana and Varela introduced the concept of *autopoiesis*, which etymologically means self (*auto-*) creation or production (*-poiesis*). In systems theory and engineering, both positive and negative feedback loops are crucial concepts to understand and steer complex processes. In stoic cosmogonies, the world goes through cycles of change, from chaos to order, until a catastrophe brings everything to chaos again. Such a cyclic cosmogony admits no absolute beginning, no permanent background, no end.

### 4.3.2 Cycles in Everyday Life

The cycle is also an attractor for evolutionary psychology reasons (Davies 2002, 41–42). In our past, survival depended on harmonizing our lives with natural cycles in nature, such as day and night cycles, menstrual cycles, astronomical cycles or seasonal cycles. It is not surprising that many creation myths are cyclical, such as in Buddhism or Hinduism. Another fundamental cycle is the *life cycle.* In higher organisms, it is a more complex version of the cycle, which involves reproduction with a blueprint and variations. However, the life cycle is certainly less acceptable cognitively, because we can not predict fully its outcome. There is no strict repetition. In Darwin's (1859) famous words, “endless forms most beautiful and most wonderful have been, and are being, evolved.”

Yet, the famous chicken and egg paradox remains. What appeared first, the chicken who produced the egg or the egg which produced the chicken? We might unwittingly introduce such paradoxes by the very formulation of our questions about the origin of the universe. In the case of the chicken-egg problem, it was solved by evolutionary theory, which unfolds the long history of life on Earth before an organism such as a chicken appeared. In a similar way that evolutionary theories bring a broader context to settle the chicken and egg paradox, could a broader theory explain the origin of the universe? We will develop later such a philosophical scenario of reproducing universes in Chapter 8.

### 4.3.3 Big Bang(s) Cycles

Cyclical universes with successive Big Bang expansion phases and Big Crunches contracting phases are not favored by current observations. Since 1998, observations support not only that the universe is in expansion, but also that this expansion accelerates (Riess et al. 1998). However, the fashion of ever expanding and closed universes models seems also to be cycling (see e.g. Dyson 2002, 149). So it is certainly wise *not* to dismiss cosmological models too quickly.

In the 1920s, relativistic cosmology showed that a static eternal universe was more and more difficult to maintain without *ad hoc* fixes. Friedman (1922; 1924) showed that cosmological solutions to Einstein's equations were unstable, leading to prefered solutions with expanding or contracting universes.

It is tempting to accept the idea of an oscillating universe, because it seems to solve both the causal and the thermodynamical challenge. On the one hand, we can assume that all the energy of the contracting universe is reused in the next expansion phase. On the other hand, the causal challenge is solved by the cycle itself. However, Tolman (1934) studied oscillating universes and showed that as the cycle repeats, universes grow bigger and bigger. If we apply this result to the past, it would make universes smaller and smaller in radius up to a tiny point, like in classical Big Bang models. The infinite regression in the past doesn't succeeds, because we start over with a point. Therefore, even if oscillating universes may solve the thermodynamical challenge, the causal challenge has merely been shifted to a point and remains unsettled.

However, cyclical universes also open the way to avoid Tolman's assumption that thermodynamics would hold across universal cycles. As Davies (1994, 146) pointed out, there may be a way out:

> The conclusion seems inescapable that any cyclic universe that allows physical structures and systems to propagate from one cycle to the next will not evade the degenerative influences of the second law of thermodynamics. There will still be a heat death. One way to sidestep this dismal conclusion is to suppose that the physical conditions at the bounce are so extreme that no information about earlier cycles can get through to the next. All preceding physical objects are destroyed, all influences annihilated. In effect, the universe is reborn entirely from scratch.

This way to solve the thermodynamical challenge was also envisioned by Misner, Thorne and Wheeler (1973, 1215), in their classical book *Gravitation:*

> Of all principles of physics, the laws of conservation of charge, lepton number, baryon number, mass, and angular momentum are among the most firmly established. Yet with gravitational collapse the content of these conservation laws also collapses. The established is disestablished. No determinant of motion does one see left that could continue unchanged in value from cycle to cycle of the universe.

### 4.3.4 Objections against Cycles

Let us examine in more details uncomfortable ideas behind cycles. Are they satisfactory explanation schemes? Are they logically fallacious? Do they imply an eternal return with identical copies of ourselves appearing somewhere or somewhen? For each issue, we present some possible replies and remedies.

The first problem is that answering the causal challenge with a cyclic explanation may not satisfy us fully. Indeed, with cycles there is no endpoint in our quest for knowledge, so it seems that we have abandoned this quest of an ultimate explanation. In fact, the word “ultimate” comes from the latin *ultimare*, which means “come to an end”. With a cycle, we are never going to come to an end. We are never going to make our explanation converge into a firm foundation. It seems that the explanation is less encompassing than a point. For, *where and how did it the cycle start?* But of course, this very question betrays an attraction for the point explanation! It implicitly assumes that only a “point” explanation can satisfy us. Yet, in a truly cyclical way of thinking, this question has no point!

The second commonly perceived fallacy regarding cycles or circles is that they are *always* vicious and thus must be avoided at all price. Yet, this is wrong. Accordingly, there are circularly vicious definitions or reasonings. But we must not systematically attribute them to cyclical or self-referential conceptualizations. As we saw, a wide variety of knowledge domains made impressive progresses by developing and using cyclical, self-referential or bootstrapping principles. The purported viciousness of circles hides instead a fear of infinite regresses. A major point is that *circular explanations and infinite regresses are not necessarily vicious* (Gratton 1994). One attributes viciousness to such reasonings, but this is based on the assumption that "there is some obligation to begin a beginningless process or to end some endless process" (Gratton 1994, 295). To sum up, instead of trying to avoid an infinite explanatory regress, we can choose to embrace it, without any contradiction. Such a regress can take two forms, the cycle or the series.

In the case of an infinite series, it is a very unsatisfactory explanation. The same problem is typically re-introduced in the solution, and the initial problem will recur infinitely and will never be solved. Such an infinite series may be even more unsatisfactory than an infinite cycle because it shifts the problem to a totally inaccessible realm, whereas a cycle seemed more accessible, with some kind of repetition. But again, as Gratton argued, there is no *objective* contradiction with infinite series, although it certainly contradicts our *subjective* cognitive inclinations to reach either a point or a cycle. The idea of an infinite series constitutes an accepted infinite series of causes, instead of one which is quarantined in a *causa sui* God or ultimate theory.

The third problem with cycles is that they would *ipso facto* imply an endless cycle of repetition, a veritable *eternal return*. The idea of an eternal return is ubiquitous in primitive world civilizations, religions and myths (Eliade 1959). In some interpretations, you and I would have identical copies (doppelgängers) of ourselves somewhere or somewhen in this universe or another. Barrow (2005, 28) calls it the *infinite replication paradox*. Interestingly, Paul Davies (2002, 44–45) reports that in his public talks, people find a cyclic universe palatable, but *not* an endless cycle of repetition. Indeed, we have no experience whatsoever of endless cycles of perfect repetition.

Yet, this idea of endless recurrence has reappeared in modern cosmological discussions (see e.g. Ellis and Brundrit 1979; Tipler 1980a; Jaume Garriga and Vilenkin 2001; Knobe, Olum, and Vilenkin 2006; Vilenkin 2006a). But the infinite replication is not at all a necessary implication of infinite or cyclical models. As we will now see, it is doubtful for logical, thermodynamical, and cosmological reasons.

Logically, the infinite replication paradox is groundless because it stems from a confusion between *infinity* and *exhaustion of possibilities*. If we grant that the universe is infinite –whatever vague idea hides behind such a claim– the infinite replication is at most a possibility, but not a necessity at all.

This can be explained by the following analogy (see Rucker 2004, 295). Let us consider the set of even numbers. It is infinite, but it does not contain all kinds of numbers since it contains no uneven numbers. *Infinity does not necessarily means exhaustion of all possibilities*. Another example is given by Heller (2009, 103). Let us consider the infinite set of real numbers. Despite the fact that the set is infinite, every number occurs just once in this set. And each number has its own individuality, in the sense that each has a different decimal expression (ex: 3,14159...)

In 1871, Louis Auguste Blanqui, in a brochure entitled *Eternity by the Stars: Astronomical Hypotheses*, drew consequences of Newtonian's mechanics and the infinite size of the universe. He concluded that "each man possesses within the expanse an endless number of doubles who live his life, in absolutely the same way as he lives it himself." (cited in Luminet 2007, 131)

The idea of eternal recurrence could also be supported with an important *recurrence theorem* proven by Poincaré (1890). It is technically formulated, but in plain english it states that given:

(1) a finite mechanical system of material points,
(2) which are subject to forces depending only on position in space
(3) and where coordinates and velocities do not increase to infinity,

then,

(4) the system will return to its initial state an infinite number of times.

Ernst Zermelo noticed that Poincaré's recurrence theorem is in contradiction with the second law of thermodynamics. Indeed, a cyclic universe and an irreversible process towards heat death are incompatible (Heller 2009, 26). This illustrates the difficulty to solve both the *causal* challenge and respecting the second law in the *thermodynamical* challenge.

Interestingly, Tipler (1980a) has proven the more complicated general relativistic version of Poincaré's recurrence theorem. That time, the conclusion is the opposite, namely that recurrence *cannot* happen in a closed universe. Accordingly, current observations favor an open universe, so Tipler's theorem does not refute eternal recurrence.

Cosmologically, the eternal recurrence is unlikely. As Luminet (2007, 132) argues, one can underline

> that the hypothesis of the duplication of all beings would perhaps be acceptable in an eternal and *stationary* Universe, where the average physical properties do not evolve over the course of time, but that the stationary theory has been rejected: the Universe evolves rapidly, as attested by the observation of the fossil radiation. The physical conditions of the big bang are radically different than those which will hold in a trillion years. The various configurations of physical systems are therefore not equally probable in time.

A thermodynamical interpretation of the infinite replication paradox states that strict cyclicity would imply that we are doomed to repeat the same events endlessly. Paul Davies (2002, 45) clarified and refuted this interpretation. He argues that it does not hold because it stems from a conflation between (i) the subjective psychological impression that time is flowing and (ii) objective time asymmetries in the world, of which the typical example is the entropy increase.

Finally, we should be very cautious when extrapolating specific, well-defined physics theorems to the whole universe. We must remain aware that such models only hold under many simplifying assumptions. There is no guarantee at all that recurrence theorems would hold if we would take into account more of the universe's complexity. For example, would the intrinsic indeterminacy of quantum mechanics prevent the eternal return?

## 4.4 Points, Cycles and Beyond

### 4.4.1 More Attractors

The point and cycle cognitive attractors are certainly limiting our imagination of what an ultimate explanation may be. In fact, deterministic models always reach an attractor, and thus get locked in at some point in the future. For example, the pendulum will naturally tend to a point of equilibrium. More precisely, attractors appear under the following conditions (Wuensche 1998):

> Given invariant network architecture and the absence of noise, a discrete dynamical network is deterministic, and follows a unique (though in general, unpredictable) trajectory from any initial state. When a state that occurred previously is re-visited, which must happen in a finite state-space, the dynamics becomes trapped in a perpetual cycle of repetitions defining the attractor (state cycle) and its period (minimum one, a stable point).

If those insights are valid for our cognitive processes, it means that all our deterministic models will lack creativity! A typical example of such creativity failures in our models are the predictions of the Club of Rome (1972). They predicted a social collapse, because of the exhaustion of finite resources (e.g. that oil would run out in 1990). In such ambitious world-modeling, it is easy to miss many parameters, events, dynamics, non-linear effects, new energy sources, which did occur and made such gloomy predictions inaccurate.

It seems that similar biases occur in discussions about the origin of the universe where, as we saw, two attractors for explanations emerge: the point and the cycle. We should keep in mind that the full picture might well be more complicated. Indeed, in dynamical system theory, the fixed point (0-dimensional) and the limit cycle (1-dimensional) are just the simplest attractors. We have no reason to exclude n-dimensional attractors or strange attractors (noninteger-dimensional) whose nature are fractal.

To be more specific, even if we define the beginning of the universe as occurring in a singularity, there is room to interpret it. Quentin Smith (1988, 45) reminds us that the singularity is not *in* a three-dimensional space, "it is in a space either of 0 dimensions (if it is just one point), 1 dimension (if it is a series of points constituting a line or line segment) or 2 dimensions (if it is a series of points comprising a surface-like space)". Yet, we have no theoretical reason to stop at the surface. We could imagine a limit torus attractor, and indeed the space-time singularity inside a rotating black hole leads to a ring singularity (Kerr 1963). We could also envisage any kind of complicated fractal topologies, beyond our day-to-day euclidian, newtonian or einsteinian prejudices.

### 4.4.2 Line and Combinations

A way to avoid the complications of describing the beginning or the end of the universe is to suppose that it essentially stays in the same state. The steady state cosmological theory is such a model, which was first developed by Sir James Jeans (1928), then in greater detail by Hermann Bondi, Thomas Gold, Fred Hoyle and Jayant Narlikar. It assumes the universe remains unchanged, it has no cycle, no beginning with a point, no end. We could associate it with a "line" cognitive attractor. It may have its roots in some subjective conceptions of time, or in our vision of the horizon in a natural landscape. It is definitely an elegant way to avoid the difficult waters of the beginning and the end.

However, Hoyle had introduced an *ad hoc* “creation tensor” to make the steady state theory consistent, thereby introducing a violation of the conservation of energy. The steady state theory has later been progressively dismissed and replaced by big bang models. The full story however is richer and more subtle (see e.g. Kragh 1996).

The difference between the line and infinite cycles is that in cycles there is some repetition, and perhaps singularity points, while this is not the case with the line. Although it is certainly a healthy scientific attitude not to accept the big bang theory as a creation myth, but as it is, a successful scientific model, which might be improved or refuted by other models in the future, observations do converge to the idea that there has been a Big Bang. So this “line” cognitive attractor seems to be currently ruled-out.

Of course, it is also possible to imagine much more complicated scenarios combining points, cycles and lines. However, what do we have to gain by combining and producing more complex explanations, if anyway we cannot test them? Of course, if we do have strong theoretical or empirical evidence that we can test scientifically, we are justified to advance more sophisticated explanations. Otherwise, a pragmatic principle of *rational economy* should be at play, for example:

> **rational economy**: Never employ extraordinary means to achieve purposes you can realize by ordinary ones (Rescher 2006, 8)

### 4.4.3 Cosmological Models

It is crucial to take seriously our best theories to answer our questions about the origins. Major physical theories like quantum mechanics or general relativity can have counterintuitive consequences, which nevertheless we must take into account. Such theories are more reliable than intuitions coming from our brains, a mere product of evolution. The brain is well adapted to recognize cycles in natural environments, or to recognize starting points in human actions, but not to guess what happened at the big bang era.

For example, the point attractor and the idea of an origin with a foundation is deeply problematic if we want to make it consistent with the first law of thermodynamics. Indeed, if no energy can be created or destroyed, how did the causing “point” transfer the energy to the universe?

We saw reasons why we have inclinations towards certain point-like or cycle-like explanations. Now, how can we choose pragmatically between the two? We agree with Bitbol's (2004) conclusion that autonomy or a cycle is more promising, as its underlying principle, co-creation or bootstrapping has proved very helpful in many sciences. Remarkably, it is this very same 'meta-' analysis technique which triggered our inquiry in this Chapter into our “metaorigins”, or the origin of our intellectual preferences to the origin of the universe. Less obvious for our hunter-gatherer brains, *n*-dimensional attractors should also be kept in mind.

# *CHAPTER 5 - Capturing Free Parameters*

**Abstract**: We address the free parameters issue in cosmology, namely that there are free parameters in the standard model of particle physics and in cosmological models, which in principle can be filled in with any number. We do not know why they should have the value they do. We analyze the issue with physical, mathematical, computational and biological frameworks. We review important distinctions such as dimensionless and dimensional physical constants, and the classification of constants proposed by Lévy-Leblond. Then we critically discuss Tegmark's proposal of the mathematical universe. The idea that our universe might be modelled as a computational entity is analyzed, and we discuss the distinction between physical laws and initial conditions using algorithmic information theory. Finally, we introduce a biological approach to the free parameters issue which will be later developed in Chapter 8. Importantly, analogies are both useful and fundamental cognitive tools, but can also be misused or misinterpreted, which is why this Chapter starts with a preliminary study of analogical reasoning in science.

Our particle physics and cosmological models have free parameters. In particular, the standard model of particle physics, despite its great successes, has many adjustable parameters. This is embarrassing, since we do not know why they have the values they do. In principle, they could take any value. Some parameters specify the masses of particles, others the relative strength of forces. Since we do not have theories to decide their value, they are determined by experiments and we then fill them in our models. Lee Smolin (2006, 13) described this situation as a great problem in theoretical physics:

> *Explain how the values of the free constants in the standard model of particle physics are chosen in nature.*

We will call such quantities "parameters" and not "constants". Indeed, if they are "free", they are not constants anymore, but parameters which can –at least theoretically– vary.

More precisely, there are two families of free parameters (see e.g. Demaret and Lambert 1994, chap. 5; Stenger 2011). Following Stenger (2011), we call parameters of the standard model *physics parameters*; and parameters of cosmological models *cosmic parameters*. Cosmic parameters include for example the expansion rate of the universe, the mass density of the universe, the ratio of the number of protons and electrons or the cosmological constant. Together, the standard model of particle physics and the standard cosmological model require 31 free parameters to be specified (Tegmark et al. 2006). It is a main challenge of modern physics to build stronger theories able to reduce this number. The *free parameters issue* can now be defined as:

> **Free parameters issue**: *There are free parameters in the standard model and in cosmological models, which in principle can be filled in with any number.*

The role of physical and cosmological models is to reduce this number of free parameters, possibly to zero. Claiming that there is a free parameters issue, is claiming that at least one parameter will never be determined by pure theory.

Claiming that the free parameters problem is solvable is claiming that a future physical theory will decide every parameter. All cosmological models have free parameters except Tegmark's (2007) *Mathematical Universe* that we will soon discuss (in section 5.3 The Mathematical Universe, p94).

In this Chapter, I first review important distinctions such as the *dimensionless* and *dimensional* physical constants, and the classification of constants proposed by Lévy-Leblond. Generalizing Lévy-Leblond's insights, I argue that as physics progresses, the number of free parameters decreases. I argue that *free parameters will progressively be reduced to initial conditions of a cosmological model*. I then discuss Max Tegmark's radical proposal of the mathematical universe, which indeed has zero free parameters. Then I examine our universe modelled as a computational entity and discuss the distinction between physical laws and initial conditions using algorithmic information theory. Finally, I mention the view of the biological universe, suggesting biological analogies as fresh perspectives to tackle the free parameters issue.

## 5.1 Analogical Reasoning in Science

In this Chapter, we will be using mathematical, computational and biological analogies to better grasp the nature of free parameters in physical and cosmological models. As a preliminary study to this survey, we analyze in this section analogies as cognitive tools. How can analogies be used for scientific purposes? Many great scientific discoveries have been triggered by analogies (see Holyoak and Thagard 1995, chap. 8 for plenty of examples). This constitutes an important motivation to understand in greater detail the functioning of analogical reasoning. Yet, since analogies are easily abused, they also need to be carefully used.

What is an analogy? It is a structural or functional similarity between two domains of knowledge. For example, a cloud and a sponge are analogous in the sense that they can both hold and give back water. More precisely, we can give the following definition: "an analogy is a mapping of knowledge from one domain (the base) into another (the target) such that a system of relations that holds among the base objects also holds among the target objects." (Gentner and Jeziorski 1993, 448–449). In this very simple example, the relations "holding and giving back water" which are verified in the base (the cloud) are also verified in the target (the sponge).

Analogical reasoning is recognized to be a basic cognitive mechanism allowing us to learn and solve problems (e.g. Minsky 1986; Hofstadter 1995; Holyoak and Thagard 1995). Leary (1990, 2) even argued that language and every kind of knowledge is rooted in metaphorical or analogical thought processes. Indeed, when we do not know a domain at all, we must use analogies as a cognitive tool to potentially gain some insights from what we already know. In this manner, we can draw a map from the known to the unknown.

Specifically, Holyoak and Thagard (1995, 185–189) argued that analogical reasoning is helpful in *discovering, developing, educating,* or *evaluating* scientific theories. Indeed, they enable us to propose new hypotheses, and thus *discover* new phenomena. These new hypotheses trigger us to *develop* new experiments and theories. Let us note however that there is nothing automatic or easy in this process. The relational system should first be examined in both domains, and then a more precise analogy or disanalogy can be found worthy of testing.

The *educating* part of analogies is useful for diffusing scientific knowledge, both with colleagues and pupils. Indeed, it is well known that good teachers use analogies to help others grasp a new idea, based on what they already know.

The *evaluating* part confronts us with one of the main dangers of analogies. One should emphasize that *an analogy is not a proof*. Analogies can thus not properly be used to prove statements, but their main utility is in giving *heuristics* for discovering and developing scientific theories. To illustrate this point, let us consider the teleological argument of God's existence popularized with William Paley's (1802) watchmaker analogy. It goes as follows:

(1) A watch is a fine-tuned object.
(2) A watch has been designed by a watchmaker.
(3) The Universe is fine-tuned.
(4) The Universe has been designed by God.

In the base domain (1-2), we have two objects, the watch and the watchmaker. They are linked by a "designed by" relationship. In the target domain (3-4), the Universe is like a watch, and God, like a watchmaker. That the relation (1)-(2) is a verifiable fact does not imply at all that the same relation "designed by" in (3)-(4) should be true. There is no causal relationship between the couple (1)-(2) and (3)-(4). This reasoning at most gives us an *heuristic* invitation to ponder whether the universe is fine-tuned. Although it is a logically flawed argument, one can appreciate its strong intuitive appeal.

There are in fact other pitfalls associated with analogical reasoning. To avoid them, Gentner and Jeziorski (1993, 450) proposed six principles of analogical reasoning:

> 1. **Structural consistency**. Objects are placed in one-to-one correspondence and parallel connectivity in predicates is maintained.
> 2. **Relational focus**. Relational systems are preserved and object descriptions disregarded.
> 3. **Systematicity**. Among various relational interpretations, the one with the greatest depth - that is, the greatest degree of common higher-order relational structure is preferred.
> 4. **No extraneous associations**. Only commonalities strengthen an analogy. Further relations and associations between the base and target – for example, thematic connections –do not contribute to the analogy.
> 5. **No mixed analogies**. The relational network to be mapped should be entirely contained within one base domain. When two bases are used, they should each convey a coherent system.
> 6. **Analogy is not causation**. That two phenomena are analogous does not imply that one causes the other.

Is it possible to further generalize the use of analogical reasoning into a science which would focus only on the structural or functional aspects of systems? Studying different models in different disciplines having structural or functional similarities leads to the development of very general interdisciplinary scientific frameworks, like network or systems theory. Indeed, Ludwig van Bertalanffy defined general systems theory as an interdisciplinary doctrine "elaborating principles and models that apply to systems in general, irrespective of their particular kind, elements, and 'forces' involved" (quoted in Laszlo 1972a, xvii). Analogies can be mathematically defined and specified to become different kinds of homomorphisms.

In a similar fashion, the study of networks is independent of the nodes and types of relations considered.

To conclude this section, we can use Hesse's (1966, 8) pragmatically valuable distinction between *positive*, *negative* and *neutral* analogies. The positive analogy addresses the question: *what is analogous?* and constitutes the set of relations which hold in the two domains. The negative analogy addresses the question: *what is disanalogous?* and constitutes the set of relations which do not hold in the two domains. Finally, neutral analogies trigger the question: *are the two domains analogous?* To answer this last question, one has to examine or test if such or such relation holds in the target domain.

In which way are modern cosmological models analogous? They at least have one gross characteristic in common, they have free parameters. That is, parameters not specified by the model. Given this analysis of analogical reasoning, we can now carefully explore the free parameters issue, aided by mathematical, computational and biological analogies. Then we will analyze in Chapter 6 whether those parameters are fine-tuned.

## 5.2 The Physical Universe

In physics, it is important to distinguish between *dimensional* and *dimensionless* physical constants (see e.g. Michael Duff's contribution in Duff, Okun, and Veneziano 2002). If a constant has a unit after its value, it is dimensional. Dimensional constants depend on our unit system choice and thus have a conventional aspect. The velocity of light *c*, the reduced Planck constant *ħ* or the gravitational constant *G* are all dimensional constants. Their respective dimensions are, for example, $m.s^{-1}$, eV.s and $m^3.kg^{-1}.s^{-2}$. Certainly, we can for example make the velocity of light equal to 1, and thus apparently dimensionless. However, this applies only to a particular unit system, and the constant will be dimensional again in another unit system.

By contrast, dimensionless constants are dimensionless in *any* unit system. They are ratios between two physical quantities, such as two forces or two masses. For example, the electron-proton mass ratio is $m_e/m_p$ = 1/1836.15... Since the two quantities are masses, we can get rid of the units (i.e. the dimension), and keep only a pure number. Other dimensionless constants are deduced by a similar *dimensional analysis*. If the analysis leads to a pure number without dimension, we have a "dimensionless" constant.

Along with this dimensional versus dimensionless distinction, Jean-Marc Lévy-Leblond (1979, 238) proposed another complementary classification of physical constants. Three types are distinguished, in order of increasing generality:

> A. **Properties of particular physical objects** considered as fundamental constituents of matter; for instance, the masses of "elementary particles", their magnetic moments, etc.
> B. **Characteristics of classes of physical phenomena**: Today, these are essentially the coupling constants of the various fundamental interactions (nuclear, strong and weak, electromagnetic and gravitational), which to our present knowledge, provide a neat partition of all physical phenomena into disjoint classes.
> C. **Universal constants**, that is constants entering universal physical laws, characterizing the most theoretical frameworks, applicable in principle to any physical phenomenon; Planck constant *ħ* is a typical example.

The classification is only intended to be a basis for discussing and analysing the historical evolution of different physical constants. For example, the constant *c*, the velocity of light, was first discovered as a type-A constant. It was a property of light, as a physical object. With the work of Kirchhoff, Weber, Kohlrausch and Maxwell, the constant gained type-B status when it was discovered that it also characterized electromagnetic phenomena. Finally, it gained type-C status when special and general relativity were discovered, synthesizing concepts such as spatio-temporal intervals, or mass and energy. For a detailed account of the status change of *c* see (Lévy-Leblond 1979, 252–258).

What happens next, when a constant has reached its type-C status? The fate of universal constants (type-C), explains Lévy-Leblond (1979, 246), is to "see their nature as concept synthesizers be progressively incorporated into the implicit common background of physical ideas, then to play a role of mere unit conversion factors and often to be finally forgotten altogether by a suitable redefinition of physical units." More precisely, this remark leads him to the distinction of three subclasses of type-C constants, according to their historical status:

> (i) the *modern* ones, whose conceptual role is still dominant (e.g. $\hbar$, *c*);
>
> (ii) the *classical* ones, whose conceptual role is implicit and which are considered as unit conversion factors (e.g. thermodynamical constants *k*, *J*);
>
> (iii) *archaic* ones, which are so well assimilated as to become invisible (e.g. the now obvious ideas that areas are square of lengths).

If all dimensional constants follow this path, then they all become "archaic", and thus integrated in the background of physical theories. The fate of dimensional constants seems then to fade away. Is it possible to seriously consider this remark, and try to relegate all dimensional constants to archaic ones? Michael Duff (Duff 2002; Duff, Okun, and Veneziano 2002) convincingly argued that the number of dimensional constants (type-C) can be reduced to ... zero! Thus, he considers constants like *c*, *G*, $\hbar$, which are often considered as "fundamental", as merely unit conversion factors. According to his terminology, only dimensionless constants should be seen as fundamental. Victor Stenger (2011) also describes dimensional constants as trivial and arbitrary parameters.

A dimensionless physics approach is also proposed in the framework of scale relativity (L. Nottale 2003, 16). Following the idea of relativity, one can articulate any physical expression in terms of ratios. Indeed, *in the last analysis a physical quantity is always expressed relative to another.* Of course, experimentalists still need to refer to metric systems, and often to many more dimensional physical constants than just the common *c*, *G* and $\hbar$. The point here is that it is possible to express the results in physical equations without reference to those dimensional constants (see also Lévy-Leblond 1979, 248–251).

What are the consequences of these insights for the free parameters problem? If the fate of dimensional constants is to disappear, then we obviously reduce the number of free parameters. Considering what would happen if a type-C dimensional constant would have a different value has to be considered very skeptically. Such a scenario has unfortunately been famously popularized by Gamow's (1939) book *Mr. Tompkins in Wonderland: or, Stories of c, G and h*. Mr. Tompkins is subject to a world where the arbitrary constants c, G and h vary, and the world changes accordingly. Again, as Duff argued, the problem is that dimensional constants are conventions, and changing them is changing a convention, not physics. It is thus only meaningful to express the possible changes in terms of dimensionless constants.

Parameters involved in cosmological models can be explained by new physical theories. Such is the case with the dimensionless cosmological constant, whose value has been predicted by scale relativity (L. Nottale 2010, 123–124).

In January 2006, Bernard Goossens, an engineer with passionate interest for physics, cosmology, evolution, complexity and philosophy contacted me. He attracted my attention to the possible intersection of the work of my PhD supervisor Francis Heylighen with the work of Laurent Nottale, a physicist specialist of fractal space-time.

I was at that point very skeptical of Nottale's work, because I was unable to assess the technicalities of the theory he developed, *scale relativity.* I had not the necessary physics background. Two years later, Bernard and I were having a drink at the Free University Brussels (VUB) and I told him that my skepticism stemmed from the fact that any good theory needs to make and validate concrete predictions. But when Bernard responded that scale relativity makes a lot of (verified!) predictions, this was a turning point.

I then looked more closely at Nottale's work, starting with a popular book, *La relativité dans tous ses états* (L. Nottale 1998). I was astonished by the quality of this little book which tells the history of relativity theories from Copernicus, Galileo to Einstein and Poincaré. The last chapters introduce scale relativity as a logical continuation of relativity theories, which that time aims to unify quantum physics with relativity theories. The principle of relativity has been successfully applied to positions, orientations and motions, and it is now extended to scales. It is one of the best popular science book I've ever read, and if you can't speak French, it's worth learning to read it!

Yet, I still couldn't understand the maths behind the theory. And I still can't. But my training as a philosopher of science made me see clearly that the theory had all major ingredients of sound theorizing. For once, philosophy would prove useful. Here is why.

Scale Relativity extends theories of relativity by including the scale in the definition of the coordinate system, then to account for scale transformations in a relativistic way. How is it possible? And why did Einstein not found this extension before? As often in the history of physics, part of the answer lies in the mathematical tools.

Einstein struggled years to develop the general relativity theory of gravitation because it involved *non-Euclidean geometries*. These geometries were counter-intuitive to manipulate and understand, and they were not used in physics before. Similarly, scale relativity uses a fundamental mathematical tool to deal with scales: *fractal geometries*. Including explicitly scale transformations in equations leads to an extension of general relativity including its previous results, with the construction of a fractal space-time theory. Indeed, relativity theory equations are limited to *differentiable* equations; scale relativity allows an extension to *nondifferentiable* equations, using fractal geometries. The constraint of differentiability is released and this leads to a more general theory which can deal both with the differentiable and the nondifferentiable case. As non-Euclidean geometries were new for Einstein, fractals are –relatively!– new to physicists because they were only studied in depth by Mandelbrot in the 1950's, although they were known by mathematicians much before; for example with Georg Cantor's triadic set.

This simple yet fundamental approach generates a proliferation of results, which are both theoretical and practical, with validated predictions. Let us mention a few of them. A new light on quantum mechanics can be thrown, since it is possible to derive *postulates* of quantum mechanics from scale relativity principles (Laurent Nottale and Célérier 2007). As a philosopher also trained in mathematical logic, I knew that if a theory can derive axioms or postulates, it is a certainly the hallmark that it works at a more fundamental level. Furthermore, scale relativity theory derives a macroscopic Schrödinger equation, which brings the statistical predictability characteristic of quantum mechanics into other scales in nature. For example, the relative positions of (exo)planets can be predicted in a statistical manner. The theory successfully predicts that they have more chances to be found at such or such distance from their star.

Moreover, it is possible to capture *physics* and *cosmic* free parameters thanks to scale relativity. Special scale relativity can indeed predict the value of the strong nuclear force. This was predicted with great precision, and has been confirmed by experimental measures (L. Nottale 2010). On cosmological scales, reasoning with universal scales allow to predict with great precision the value of a fundamental cosmic parameter, the cosmological constant (L. Nottale 1993). A quantitative prediction on which our finer observations keep on converging (L. Nottale 2010, sec. 3.1.2).

Models constructed with the general idea of relativity of scales bring new insights not only in physics, but also in earth sciences, history, geography and biology (L. Nottale, Chaline, and Grou 2000; 2002). To further explore those themes, I invited Laurent Nottale and his colleagues Jean Chaline (evolutionary biologist) and Pierre Grou (economist) to a workshop held at the Free University Brussels (VUB) on May 5-6 2009 (see a photo below).

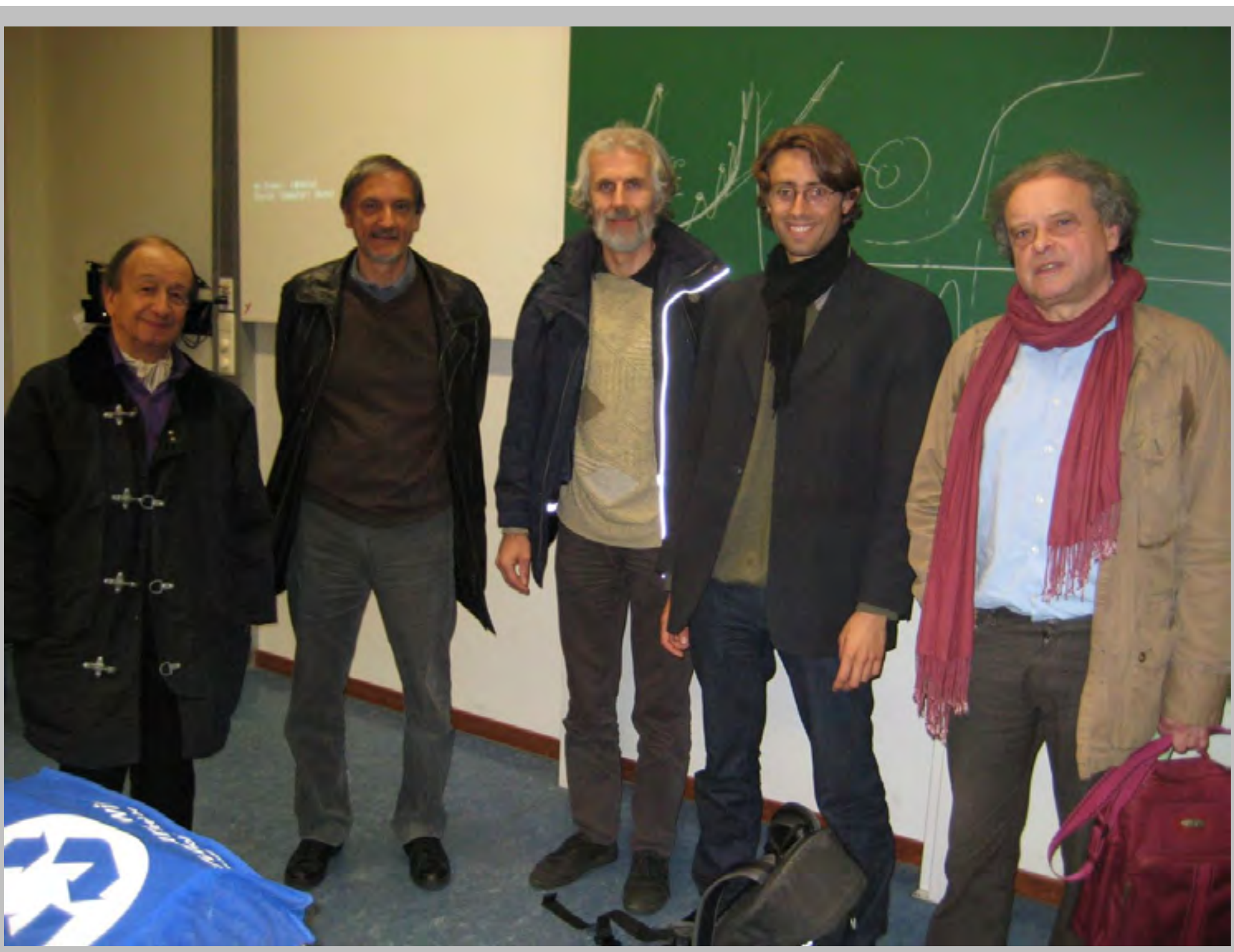
Figure 3: Workshop on Scale Relativity: Universe, Life, Societies, 5-6 May 2009. From left to right: Jean Chaline, Laurent Nottale, Francis Heylighen, Clément Vidal and Pierre Grou.

Scale Relativity is a fundamental approach which has consequences for nearly all sciences. It suggests that cosmology, fundamental particle physics, structure formation, biology, geology, economy and other fields might be approached with tools derived from the same few principles. Although, as Nottale explains in his 2010 paper, a lot of work still has to be done, the exposed vision is extraordinarily far reaching and inspiring. For these reasons, I was delighted to deliver Laurent Nottale the **Evo Devo Universe 2008 Best Paper Award** at the conference on the Evolution and Development of the Universe (Vidal 2010b).

Story 4: Capture of two free parameters in Nottale's notable theory of scale relativity.

Following Duff, Lévy-Leblond and Stenger we saw that type-C constants are bound to disappear. Another challenge I would like to propose is the following: could type-A and type-B constants emerge from initial conditions in a cosmological model? If we were able to explain all these constants in terms of a cosmological model, it would certainly be a great achievement. Smolin (1997, 316) also argued that fundamentally, progress in quantum mechanics must lead to a cosmological theory. Indeed, all particles ultimately originate from the big-bang, thus *a complete understanding of particle physics should include an explanation of their origin, and thus relate with a cosmological model.*

In a certain sense, progress in this direction has already happened, if we consider the discovery of big-bang nucleosynthesis. Properties of atomic elements could be thought as fundamental constituents of matter, and thus type-A constants, until we discovered they were actually formed at the big-bang era. If we extrapolate this trend, a future cosmological model may be able to derive many (or even all) type-A constants from initial conditions.

The same can be said about fundamental coupling constants (type-B). Special scale relativity can indeed predict the value of the strong nuclear force ($\alpha_s$) at the Z mass energy level. This was predicted with great precision, and has been confirmed by experimental measures (see L. Nottale 2010, 26–27). Thus, if physics continues its progress, it is reasonable to conceive that particle physics models would become integrated into cosmological models. The consequence for free parameters is that *physics parameters will progressively be reduced to cosmic parameters*.

We have outlined Duff's proposal that dimensional constants can be reduced to 0. We have suggested that fundamental coupling constants could be in future explained from more general principles, and that many apparent "fundamental constants" in the past can nowadays be explained by more general theories. Accordingly, a great number of free parameters have been and certainly will be explained by more advanced physical theories.

What about the role constants play in physics and mathematics? Consider the famous constant π. We can see a disanalogy, because mathematical constants are defined *a priori* by the axioms: they are *internal to the system* and are generally definable and computable numbers. For example, we have plenty of algorithms to calculate π. This is not the case with physical constants. Many of them remain *external to the system,* in the sense that they are not computable from inside the model. At some stage there has been a measurement process to get their values. Can we hope that science will allow us to understand or compute these constants from more fundamental principles? How far can we get in this direction? Let us now examine the mathematical universe.

## 5.3 The Mathematical Universe

Can we find a theory with *zero* free parameters? Is it just a dream of a theoretical physicist? Max Tegmark (1998; 2004; 2007) makes this dream true by arguing that the theory of everything (TOE) is –simply– the ultimate ensemble theory. The TOE in physics is the same as the TOE in mathematics. He argues that Newtonian gravity, general relativity or quantum field theory can all be seen as mathematical structures. More precisely, he assumes that there is a mathematically defined multiverse which actually exists, with all possible mathematical structures.

Tegmark further distinguishes four types of parallel universes, with greater and greater diversity (Tegmark 2004, 460). In level I, we are simply speaking about regions beyond our cosmic horizon. In level II, chaotic inflation (Linde 1990) is supposed to be correct, and so other post-inflation bubbles might exist. In level III, we assume Everett's (1973) many-worlds interpretation of quantum physics to be correct, where our universe keeps branching in parallel universes. In level IV, we assume a structural mathematical realism, where mathematical existence is the same as physical existence. More precisely, Tegmark (2007, 102) makes two very strong ontological and epistemological hypotheses:

> **External Reality Hypothesis (ERH)**: *There exists an external physical reality completely independent of us humans.*

> **Mathematical Universe Hypothesis (MUH)**: *Our external physical reality is a mathematical structure.*

There are many good reasons to be very skeptical about those two assumptions, and about multiverse levels II-IV. But for the sake of the argument, let us follow Tegmark's reasoning, and see its implications for initial conditions at Level IV. Tegmark (2007, 117–118) writes:

> There is nothing "initial" about specifying a mathematical structure. Whereas the traditional notion of initial conditions entails that our universe "started out" in some particular state, mathematical structures do not exist in an external space or time, are not created or destroyed, and in many cases also lack any internal structure resembling time. Instead, the MUH leaves no room for "initial conditions", eliminating them altogether. This is because the mathematical structure is by definition a complete description of the physical world. In contrast, a TOE saying that our universe just "started out" or "was created" in some unspecified state constitutes an incomplete description, thus violating both the MUH and the ERH.

This would indeed dissipate the problem of initial conditions. But how are we to solve concrete physics problems if physics *is* mathematics? Does it helps to know that there might be parallel universes with all possible mathematical structures?

Indeed, we may be speaking of a theory of *everything*, which unfortunately, has *nothing* to say. The literature aiming to solve challenges for ultimate explanations is often divided into two main classes of solutions: "God" or "Multiverse". Either it is God who created the Universe with all its parameters fit for life and intelligence; or there is a huge number of other universes with different parameters, so that it is very probable that there is one containing life and intelligence. The fact that it is the one we happen to inhabit is an observational selection effect which thus makes the free parameters and their possible fine-tuning less mysterious (e.g. B. Carr 2007; N. Bostrom 2002).

From a rational and scientific point of view, an appeal to God suffers from being a non-naturalistic explanation. Furthermore, God is often linked with the god-of-the-gaps assumption. If we can not understand a phenomenon, we use God to explain it, and we thus do not seek another explanation. This attitude can, by its very definition, *explain everything*. We can wonder if the hypothesis of a multiverse is any better. Indeed, the appeal to multiverse works everywhere and is not restricted by any limit; so it can also *explain everything*. Could it be that Tegmark is replacing the *god-of-the-gaps* with the *mathematics-of-the-gaps*?

As Barrow (2007a, vi) noticed, in contrast to the striving of theoretical physicists toward a TOE, the idea of multiple universes has on the contrary "undermined the naïve expectations of many, that a Theory of Everything would uniquely and completely specify all the defining quantities of the Universe that make it a possible home for life." By equating the TOE in physics with the TOE in mathematics, Tegmark has actually totally turned upside down the logic and hopes behind a TOE. Instead of having one unique theory, we have a space of possible theories as huge –and as hard to define– as the space of all mathematical structures! Such theorizing seems to me an interesting intellectual and metaphysical exercise, but it has lost touch with physics as a scientific discipline with its empirical desideratum.

Let us now outline a few more precise critiques, as Tegmark's extreme views constitute a unique opportunity to exercise a critical attitude on many issues. First, as I state in Appendix I, I hold a pragmatic, evolutionary constructivist epistemology. It took me years to see that constructivists are actually very realist –pun intended. This is a major disagreement with Tegmark's view, at the fundamental level of a worldview component. Indeed, if we study the history of science, and in particular the history of mathematics, the "unreasonable effectiveness of mathematics in the natural sciences" (Wigner 1960) which fascinates Tegmark becomes a truism. Of course mathematics is effective to describe nature, we construct it for this purpose.

The process of model construction is easier to see once we start studying other sciences. Psychologists, sociologists or businessmen are well aware that their models are always false, incomplete and make unrealistic assumptions (see e.g. the classic book by Sterman 2000). They will not hesitate to swap their model for a better one. But such models, despite all their limitations, are good only if they can make predictions or help to control the system at hand.

The situation with mathematics is a bit more complicated, because there are two selection mechanisms at play. As in empirical sciences, the first selection is for *external consistency.* We insist that our mathematics to describe the world accurately. For example, arithmetics must be respected. If you put 2+2 apples in a box, you expect to have four apples when you will open the box, not three. However, there is nothing really obvious in this. As Popper (1962, 211) analyzed, if you replace apples with rabbits or drops of water, you might end up having 2+2=5 or 2+2=1. In such cases, broader biological and physical models are needed to make reality consistent with arithmetic.

The second selection mechanism in mathematics is for *internal consistency.* A theory which derives a contradiction can derive any proposition (thanks to the "*ex falso quod libet*" logical rule). So a self-contradictory theory is trivial and useless. Over centuries, mathematics has made tremendous progress by using the internal consistency criterion *only.* This is remarkable, but this obscures the fact that mathematics is initially a construction of tools to deal with real world problems. The history of mathematics enlighten this situation. Substraction were first invented to deal with economical debts, divisions to deal with succession problems in families. This connection with real world problems holds not only for basic mathematics. For example, imaginary numbers were first introduced to solve problems in physics. Additionally, there is still a constant interaction between mathematics and physics, that we can see more clearly in the field of applied mathematics.

Importantly, Tegmark (2004, 465) takes Popper's (2002) criterion of falsifiability in a too liberal way when he writes:

> Containing unobservable entities does clearly *not* per se make theory nontestable. For instance, a theory stating that there are 666 parallel universes, all of which are devoid of oxygen makes the testable prediction that we should observe no oxygen here, and is therefore ruled out by observation.

To which Heller (2009, 89) makes the following caustic comments:

> We should realise, however, that not every statement (theory, model, hypothesis) the consequences of which may in principle be compared with observational or experimental results may be regarded as falsifiable, in the sense normally ascribed this concept in the philosophy of science. Taking Tegmark's style of comparison

> further, let’s imagine that someone before the age of space flight had claimed that the other side of the moon, which is not observable from Earth, is painted red and carries the inscription, “Coke is it!” in big white letters. It would definitely have been a falsified (and therefore falsifiable) prediction, but it could never be treated as a test of whether the hypothesis was scientific or not. It’s true that no theory or hypothesis which is not falsifiable even in principle may be regarded as scientific, but not all statements which are falsifiable (in the more colloquial sense of the word) may be regarded as scientific. The question of criteria distinguishing science from what is not science is a difficult methodological problem. Anyone who wants to write on this subject would do well to first look up the copious literature devoted to it.

There are many other problems related to Tegmark's proposal, such as the unclear definition of mathematical structures, problems related to infinities, to uncomputability of mathematical structures, to the inherent untestability of multiverse proposals, etc (see also Ćirković 2002 for critics based on physics). As Ellis (2007b, 401) writes, “claiming existence of something you cannot even properly characterize has dubious scientific merit.” We can also object that the first law of thermodynamics (see the thermodynamical challenge in section 4.1.3 Thermodynamical, p76) raises further problems, as realizing an infinite multiverse would require an infinite amount of energy.

However, if we saw that future physics may understand physics parameters in terms of a cosmological model, it is unlikely that this would also include the initial state of that model. Indeed, if we were to have a theory deciding all values of initial state in a cosmological model, it then leads to the idea of a “final theory” or a “theory of everything”. Besides the conceptual difficulties and objections we encountered in Tegmark's attempt to describe such a TOE, ironically, a TOE is an act of faith and is thus similar to the God explanation (e.g. Davies 2008, 170). Smolin (1997, 248) also wrote that the “belief in a final theory shares with a belief in a god the idea that the ultimate cause of things in this world is something that does not live in the world but has an existence that, somehow, transcends it.” Given our analysis of Chapter 4, maybe we should not be so surprised, since both God and a TOE are cognitive points attractors.

## 5.4 The Computational Universe

The idea that our universe is analogous to a computer is quite popular. We can see it as the modern version of a mechanistic worldview, looking at the universe as a machine. There are various ways to consider this analogy, with cellular automata, (see e.g. Zuse 1970; Wolfram 2002) with quantum computing (e.g. Lloyd 2005), etc. The analogy has been pushed so far that a modern version of idealism has even been considered, namely that our universe would actually be run by a computer, and we might be living in a computer simulation (e.g. N. Bostrom 2003; Martin 2006).

We saw that free parameters may ultimately be reduced to initial conditions of a cosmological model. Here, I first criticize an argument by Tegmark's claiming that a multiverse theory is more economical than a theory of one single universe. Then I analyze initial conditions from a computational perspective, and discuss the relation between physical laws and initial conditions. This is conducted within the framework of Algorithmic Information Theory (AIT, Chaitin 1974; 1987). I will conclude by pointing out some limitations of this computational analogy.

Tegmark has explored and replied to many critiques. A recurring objection Tegmark has to address is that multiverse theories are not simple and are ontologically wasteful, especially if they suppose the existence of an infinity of different worlds. Tegmark (2004, 489) uses algorithmic information theory to argue precisely the opposite, namely that a multiverse is actually extremely simple. Before seeing how he reached this conclusion, and how we can criticize it, let us summarize some basics of AIT.

AIT studies complexity measures on strings. The complexity measure –the Kolmogorov complexity[6]– of an object is the size of the shortest program able to specify that object. Below is a simple example originally presented in the Wikipedia encyclopaedia (2008) :

> consider the following two strings of length 64, each containing only lower-case letters, numbers, and spaces:
>
> abababababababababababababababababababababababababababababababab
> 4c1j5b2p0cv4w1 8rx2y39umgw5q85s7ur qbjfdppa0q7nieieqe9noc4cvafzf
>
> The first string admits a short English language description, namely "ab 32 times", which consists of 11 characters. The second one has no obvious simple description (using the same character set) other than writing down the string itself, which has 64 characters.

The first string has a low complexity, because the short program "write ab 32 times" can generate it, whereas the second one has a higher complexity because no short program can generate it.

Tegmark gives another exemple. If we pick up an arbitrary integer $n$, its algorithmic information content is of order $\log_2 n$, which is the number of bits needed to write it. But "the set of all integers 1, 2, 3, … can be generated by quite a trivial computer program, so the algorithmic complexity of the whole set is smaller than that of a generic member." Tegmarks concludes that *an entire ensemble is often simpler than one of its members*. Tegmark extrapolates this argument to physical theories:

> the set of all perfect fluid solutions to the Einstein field equations has a smaller algorithmic complexity than a generic particular solution, since the former is specified simply by giving a few equations and the latter requires the specification of vast amounts of initial data on some hypersurface. Loosely speaking, the apparent information content rises when we restrict our attention to one particular element in an ensemble, thus losing the symmetry and simplicity that was inherent in the totality of all elements taken together.

He also applies it to multiverse:

> In this sense, the higher-level multiverses have less algorithmic complexity. […] a multiverse theory is arguably more economical than one endowing only a single ensemble element with physical existence.

6 also known as algorithmic information, program-size complexity, Kolmogorov-complexity, descriptive complexity, stochastic complexity, or algorithmic entropy.

The argument is correct, but is highly biased. Indeed, the Kolmogorov complexity measure focuses on the shortest *length* of the algorithm, but does not take into account the *computating time*.

Fortunately, Charles Bennett (1988a; 1988b) defined another metric called *logical depth*. It is defined as the *computing time of the shortest program which can generate an object*. In the case of integers, it is clear that the time to compute a single number would be very short compared to the infinite time needed to compute all natural numbers. Similarly, no doubt that the time necessary to compute blindly the multiverse makes it an enterprise everything but pragmatic, realizable and economical.

What is the next best option after zero free parameter? It would be a theory with just one free parameter. Can we imagine such a theory? We could imagine the initial state of a universe contained in a single parameter indeed. But what about physical laws which take as input this initial state?

In the AIT framework, how can we interpret the difference between laws and initial conditions? Laws represent information which can be greatly shortened by algorithmic compression (like the "ab 32 times" string above); whereas initial conditions represent information which cannot be so compressed (like the second string above). If we import this analogy into physics, a physical law is to be likened to a simple program able to give a compressed description of some aspects of the world; whereas initial conditions are data that we do not know how to compress.

Can we interpret this distinction between physical laws and initial conditions in a cognitive manner? We either express our knowledge in terms of laws if we can compress information, and in terms of initial conditions if we cannot. In this view, scientific progress allows us to dissolve initial conditions into new theories, by using more general and efficient algorithmic compression rules.

In fact, *the distinction between laws and boundary conditions is fuzzy in cosmology* (Ellis 2007a, sec. 7.1; Heller 2009, 93). One can see boundary conditions as imposing constraints, not only on initial conditions (lower boundary of the domain), but also at the extremes of the domain. Both physical laws and boundary conditions play the same role of imposing constraints on the system at hand. Because we can not re-run the tape of the universe, it is difficult –if not impossible– to distinguish the two. In this view, some laws of physics might be interpreted as regularities of interactions progressively emerging out of a more chaotic state. The cooling down of the universe would progressively give rise to more stable dynamical systems, which then can be described by simple mathematical equations that we call physical laws.

A similar situation occurs in computer science. One can distinguish between a program, which is a set of instructions, and the data on which the program operates. The program is analogous to physical laws, and the data to initial conditions. This distinction in computer science can be blurred, when considering a self-modifying program, i.e. a program which modifies itself. Also, at a lower level, both the program and the data are processed in the form of bits, and here also the distinction is blurred.

In mathematics, Gödel's limitation theorems state that in any sufficiently rich logical system, there will remain undecidable propositions *in that system*. But using another stronger system, one can decide such previously "undecidable" propositions (even if new undecidable propositions will arise in the stronger system...). For example, the consistency of Peano's arithmetic cannot be shown to be consistent within arithmetic, but can be shown to be consistent *relative* to modern set theory (in the axiomatization of Zermel-Fraenkel and the axiom of choice).

There is a theorem similar to Gödel's incompleteness in AIT. Informally, it states that a computational system cannot compress structure in a system that is more algorithmically complex than this computational system. Let us assume again that physical laws represent compressible information, and initial conditions incompressible information. Are initial conditions in cosmological models algorithmically incompressible? There are two ways to answer this question.

First, we can interpret this incompressible data in an absolute way. This data is then “lawless, unstructured, patternless, not amenable to scientific study, incompressible” (Chaitin 2006, 64). Suggesting that those initial conditions are incompressible implicitly implies that we, poor humans, will never be able to understand them. This attitude freezes scientific endeavor and thus has to be rejected. Limitation theorems are only valid within formal systems, because one needs the system to be completely formalized and specific formal tools to be able to prove them. Therefore, we should be extremely careful when exporting limitation theorems into other less formalized domains. Moreover, the history of science has shown that it is hazardous to fix boundaries on human understanding. Let us take the example of infinity, which was for many centuries thought to be understandable only by a God who is infinite, and not by finite humans. A rigorous theory of infinite numbers which constitutes the foundations of modern mathematics has finally been proposed by the mathematician Georg Cantor. Therefore, boundaries are likely to be broken. We will see in Chapters 6 and 7 how the *multiverse hypothesis* or computer *universe simulations* bring us beyond the apparently incompressible initial conditions.

The second option is that incompressible information may reflect the limits of our theoretical models. If we are not able to account for the reasons of initial conditions, it is a hint that we need a broader theoretical framework to understand them. This situation can be illustrated by considering the problem of the origin of life. In this context, initial conditions for life to emerge are generally assumed without justification: chemical elements are assumed to be here, along with an Earth with water, neither too far nor too near from the Sun, etc. With these hypotheses (and others), we try to explain the origin of life. Now, what if we try to explain the origin of these initial suitable conditions for life? We then need a broader theory, which in this case is a theory of cosmic evolution. If we then aim to explain initial conditions in cosmology, we are back to the problem of free parameters.

Multiverse models such as Tegmark's are precisely attempting to introduce a broader theory to explain or dissipate initial conditions, by proposing the existence of various other possible universes. The problem is that the multiverse hypothesis is a *metaphysical assumption*. George Ellis (2007b, 400) expressed it well:

> There can be no direct evidence for the existence of other universes in a true multiverse, as there is no possibility of even an indirect causal connection. The universes are completely disjoint and nothing that happens in one can affect what happens in another. Since there can be no direct or indirect evidence for such systems, what weight does the claim for their existence carry? Experimental or observational testing requires some kind of causal connection between an object and an experimental apparatus, so that some characteristic of the object affects the output of the apparatus. But in a true multiverse, this is not possible. No scientific apparatus in one universe can be affected in any way by any object in another universe. The implication is that the supposed existence of

> true multiverses can only be a metaphysical assumption. It cannot be a part of science, because science involves experimental or observational tests to enable correction of wrong theories. However, no such tests are possible here because there is no relevant causal link.

To improve testability, Ellis further suggests examining a variation on the causally disconnected universes, considering multi-domain universes that are not causally disconnected (Level I parallel universes in Tegmark's terminology). Still, I would like to emphasize the *philosophical* importance of the multiverse hypothesis, because it is a logically consistent way to tackle the free parameters problem. How can we theorize more systematically about “other possible universes”? We will analyze this problem in Chapter 6.

In summary, if we assume that initial conditions are analogous to incompressible information, then there are two possible reactions. Either we claim that we reached the limit of scientific understanding; or we recognize that we need an extended framework. Multiverse and, as we shall see in Chapter 6, computer simulations of other possible universes are examples of such extended frameworks.

Let us now see some limits of this computational analogy. If we apply our insights about analogical reasoning, we can ask “what is disanalogous between the functioning of our universe and that of a computer?”. We can at least make the following restrictions. In a computational paradigm, space and time are assumed to be independent, and non-relativistic. Most of the well studied cellular automata even use only two spatial dimensions, which is of course a limitation for complexity to develop.

A fundamental difference between a physical and an informational-computational paradigm is that the former has at its core *conservation laws* such as the conservation of energy, where the total amount of energy remains unchanged in the various transformations of the system. By contrast, the bits manipulated by computers are largely not subjected to such conservation laws, even if running a computer has an energetic cost (Bremermann 1982). We neither create nor destroy energy, whereas we easily create and delete files in our computers.

Another limitation of this computational paradigm, which is similar to the Newtonian paradigm, is that when we have initial conditions and a set of rules or laws, then the evolution of the system is trivial and predictable: it is just an application of rules/laws to the initial conditions. We have understood nature, end of story.

The complexity of interactions such as synergies, feed-back loops, chaos, random errors, developmental processes, etc. is not in the focus of this approach. The biological analogy is more appropriate in exploring those complexities. Embryologists know that the formation of a fetus is a process of an incredible and fascinating complexity, leading from one single cell to the complexity of a billions-cells organism. The development of the individual is certainly not as easy to predict from the genome to the phenotype as was the case with the computational paradigm: we just needed the initial conditions and a set of rules to understand the dynamic. By contrast, in biology, phenomena of phenotypic plasticity have been identified, i.e. the acknowledgement that phenotypes are not uniquely determined by their genotype. This becomes particularly clear when considering genetically identical twins. They exhibit many identical features, but also a unique differentiation due to stochastic processes occurring during the development. As Martin Rees (1999, 21) noticed, cosmology deals with the inanimate world, which is in fact simpler than the realm of biology. A phenomenon is difficult to understand because it is complex, not because it has a huge extension.

## 5.5 The Biological Universe

In 2007, John M. Smart and I entered in contact discussing issues about universal change and broad cosmological and futuristic views. We noticed that scholars studying the cosmos where mainly into theoretical physics. It is of course an indispensable approach, but it does not strongly connects with life, intelligence and technology. Yet, we were also aware of dispersed insights in cosmology, theoretical and evolutionary developmental (evo-devo) biology and complexity sciences, which are providing ways to understand our universe within a broader framework.

We thought that these results and hypotheses deserved to be explored, criticized, and analyzed by an international interdisciplinary research community which we set up in 2008: '**Evo Devo Universe (EDU)**'. Such a framework promises to advance our understanding of both unpredictable "evolutionary" processes and predictable "developmental" processes at all scales, including the human scale. I welcome any researcher interested in these topics to join the research community!

I was surprised and delighted that cosmologist George Ellis (2007a, 1266) had very similar vision when he wrote:

> Thesis H4: The underlying physics paradigm of cosmology could be extended to include biological insights. The dominant paradigm in cosmology is that of theoretical physics. It may be that it will attain deeper explanatory power by embracing biological insights, and specifically that of Darwinian evolution.

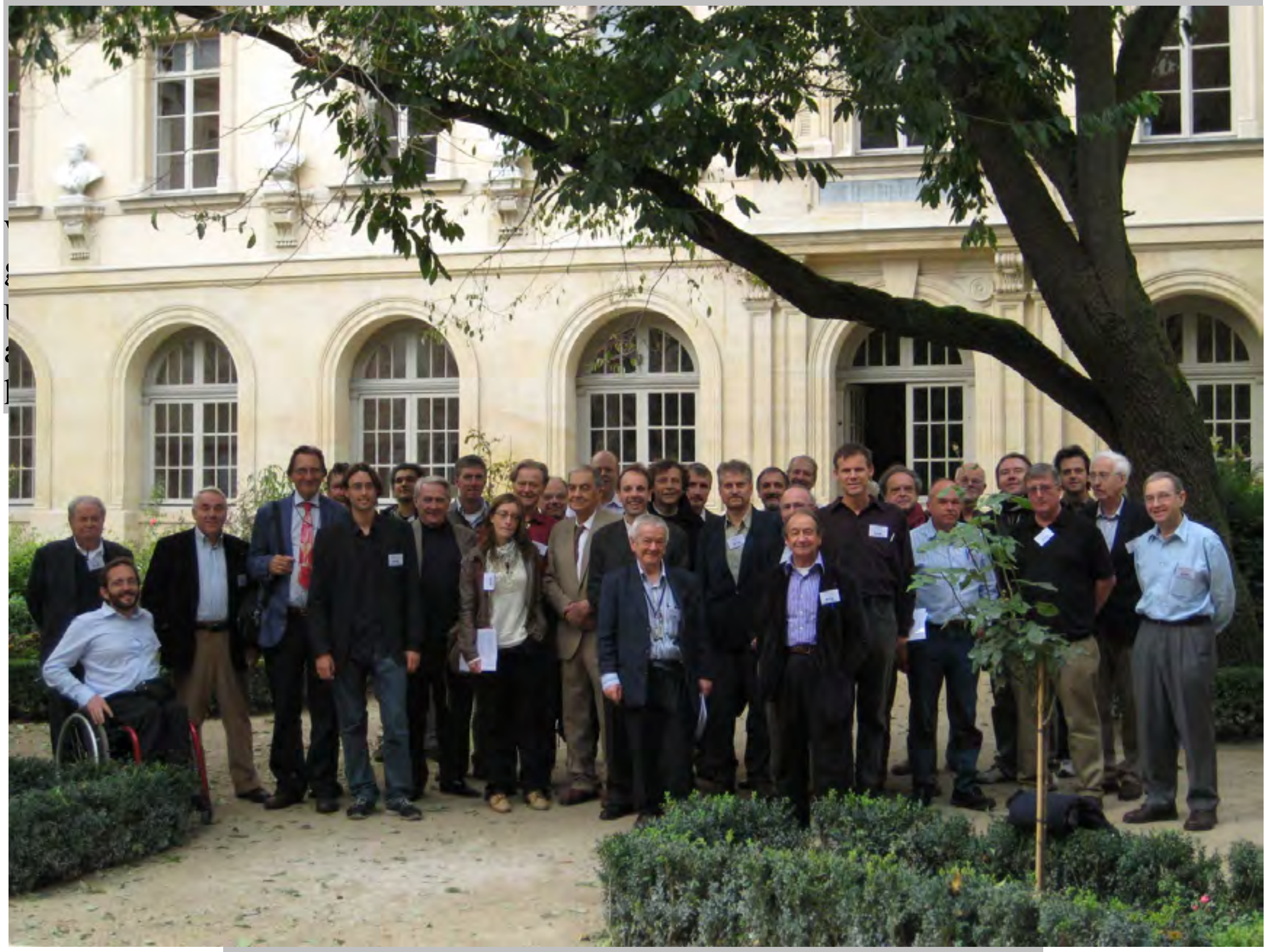

Figure 4: The First International Conference on the Evolution and Development of the Universe, 8-9 October 2008, Paris, France

John Smart and I first started to work actively on building the EDU website in 2008[7]. However, we understood that having a website and a virtual community was not enough. To effectively collaborate, human beings still need to meet in flesh. We thus focused our energy into setting up “The First International Conference on the Evolution and Development of the Universe”, which was held at the École Normale Supérieure, Paris, 8-9 October 2008[8] (see photo above). I edited the proceedings of the conference which were published as a special issue of the journal *Foundations of Science*, containing 12 peer-reviewed papers and 20 open commentaries and responses (Vidal et al. 2009).

Story 5: The Evo Devo Universe community.

The idea that our universe is similar to an organism has a rich and long story, which is still very inspiring. It can be traced back to the Ancient Greece (see Barrow and Tipler 1986 for historical aspects). One general aim of the “Evo Devo Universe” research community is to explore how traditional cosmology can be enriched by introducing a biological paradigm, as suggested by George Ellis (2007a, Thesis H4). More specifically, the field of evolutionary developmental (“evo-devo”) biology (e.g. Carroll 2005) provides a great source of inspiration, acknowledging both the contingency of evolutionary processes and the statistically predictable aspect of developmental processes.

A useful biological analogy to study the origin of the universe is to examine *other* origins, such as the origin of life. For example, biologists think it is unlikely that there has always been a single genetic code when life started. Rather, it is more likely that many genetic codes competed, maybe for millions of years, and the one we know outcompeted other codes. We have the impression to deal with a universal biological code, but it is probably just a selection effect, since other codes could have been as efficient. Could a similar mechanism have appeared at the origin of the universe? Could “universal” physical laws, initial conditions and free parameters have been selected by a yet unknown process?

In Chapter 8 I will describe two scenarios to analyze initial conditions and free cosmic parameters with a biological and evolutionary approach. The first model was proposed by Lee Smolin (1992) and is called “Cosmological Natural Selection”; the second was introduced by many authors, and I named it “Cosmological Artificial Selection”.

Even if Tegmark's model proves to be right in the future, our current physical and cosmological models need filling-in with free parameters. But *filling-in* doesn't imply *fine-tuning*. Although this distinction between free parameters and fine-tuning was not clear in (Vidal 2010a), it has few consequences for the general reasoning of that paper, because as we reduce free parameters we also reduce fine-tuning arguments associated with these parameters.

7 http://www.evodevouniverse.com

8 EDU 2008, http://evodevouniverse.com/wiki/Conference_2008

# *CHAPTER 6 - The Fine-Tuning Conjecture*

**Abstract**: The aim of this Chapter is to propose a scientific approach to find out whether our universe is fine-tuned or not. The first difficulty is to define fine-tuning, which requires three steps. First, to debunk common and uncommon fine-tuning fallacies which constantly sneak into the debate (section 6.1 ). Second, to distinguish fine-tuning from the closely related issues of *free parameters*, *metaphysical issues*, *anthropic principles*, *observational selection effects*, *teleology* and *God's existence* (section 6.2 ). Third, to provide a fallacy-proof definition of fine-tuning (section 6.3 ). We advocate that computer simulations are needed to address two key cosmological issues. First, the robustness of the emergence of complexity, which boils down to ask: "what would remain the same if the tape of the universe were replayed?" Second, the fine-tuning issue, which requires to answer the question: "are complex universes rare or common in the space of possible universes?" We first discuss definitions of possible universes and of possible cosmic outcomes – such as atoms, stars, life or intelligence. This leads us to introduce a generalized Drake-like equation, the *Cosmic Evolution Equation*. It is a modular and conceptual framework to define research agendas in computational cosmology. We outline some studies of alternative complex universes. Such studies are still in their infancy, and they can be fruitfully developed within a new research field supported by computer simulations, *artificial cosmogenesis*. Thanks to those new conceptual distinctions, I critically outline classical explanations of fine-tuning: *skepticism*, *necessity*, *fecundity*, *God-of-the-gaps*, *chance-of-the-gaps*, *WAP-of-the-gaps*, *multiverse* and *design* (section 6.4 ). Chapter 7 will further support the importance of artificial cosmogenesis by extrapolating the future of scientific simulations, while Chapter 8 will examine two additional evolutionary approaches to fine-tuning.

*We do not have any good way of estimating how improbable it is that the constants of nature should take values that are favorable for intelligent life.*

Weinberg (1993a, 221)

The fascination for harmony in Nature can be traced back to the dawn of civilizations. Such harmony has been most studied in natural theology, whose objective is to look for design in nature to infer the existence of a creator. For example, we can marvel at a proboscis of a butterfly, perfectly adapted or fine-tuned to eat pollen in flowers. The Earl of Bridgewater, a naturalist who truly loved the natural world –as his organization of diner parties for dogs can attest– commissioned eight famous *Bridgewater Treatises*. They were first published between 1833 and 1840 and intended to explore the goodness of God's creation in the natural world. For example, one treatise by Sir Charles Bell is dedicated to *The hand, its Mechanism and Vital Endowments as evincing Design.*

It is somehow surprising to see efforts to prove God's existence from the order in nature at this period, since Immanuel Kant's (1781) *Critique of Pure Reason* had famously refuted such proofs. He called such kind of proofs physical-theological, because they start from the physical order of the world, to infer the existence of a creator. It is however worth mentioning that Kant (1781, A627/B655) hold most respect for this proof, and made a subtle remark concerning its power:

> This proof can at most, therefore, demonstrate the existence of an *architect of the world*, whose efforts are limited by the capabilities of the material with which he works, but not of a *creator of the world*, to whom all things are subject.

This distinction between *architect* and *creator* is important and we will go back to it in Chapter 8. Today, thanks to modern evolutionary biology, we know such design arguments are wrong. When something seems fine-tuned, it surprises the curious scientist, and requires an explanation. The moral of the story is that when some fine-tuning is discovered, the scientific attitude is to invent mechanisms and to build theories explaining how it emerged.

The *scientific* elucidation of whether there is a cosmic design for life is more recent. It is generally traced back to Henderson's (1913) *The Fitness of the Environment* which points to the particular chemical properties of water for biology to work. Interestingly, he emphasizes the importance to study not only the fitness of *organisms*, like in Darwinian evolution, but also at the fitness of the *environment*. Indeed, both do co-evolve, and, as Francis Heylighen (private communication and 2007) has argued, the mechanism of *stigmergy* can be seen as the logical counterpart of natural selection (see table 4).

| | **Natural Selection** | **Stigmergy** |
|---|---|---|
| **Variation** | Organism | Environment |
| **Selection** | Environment | Organism |

Table 4: Natural selection and stigmergy as two fundamental evolutionary processes. In natural selection, the variation occurs at the level of the organism (e.g. genetic mutations), and its fitness is measured against a selection from the environment. By contrast, in stigmergy the environment varies, in the sense that each organism (e.g. ant, termite, etc.) will encounter different stimuli triggering them to perform different actions. In sum, the organism selects the action to perform, and the environment provides variation.

Let us start with a general definition of fine-tuning in cosmology proposed by Ellis (2007b, 388):

> if any of a number of parameters which characterize the observed universe – including both fundamental constants and initial conditions – were slightly different, no complexity of any sort would come into existence and hence no life would appear and no Darwinian evolution would take place.

However, the issue of fine-tuning in cosmology is so highly loaded, that it is worth first examining fine-tuning examples in non cosmological contexts. Let us switch on a good old radio tuner such as the one in Figure 5.

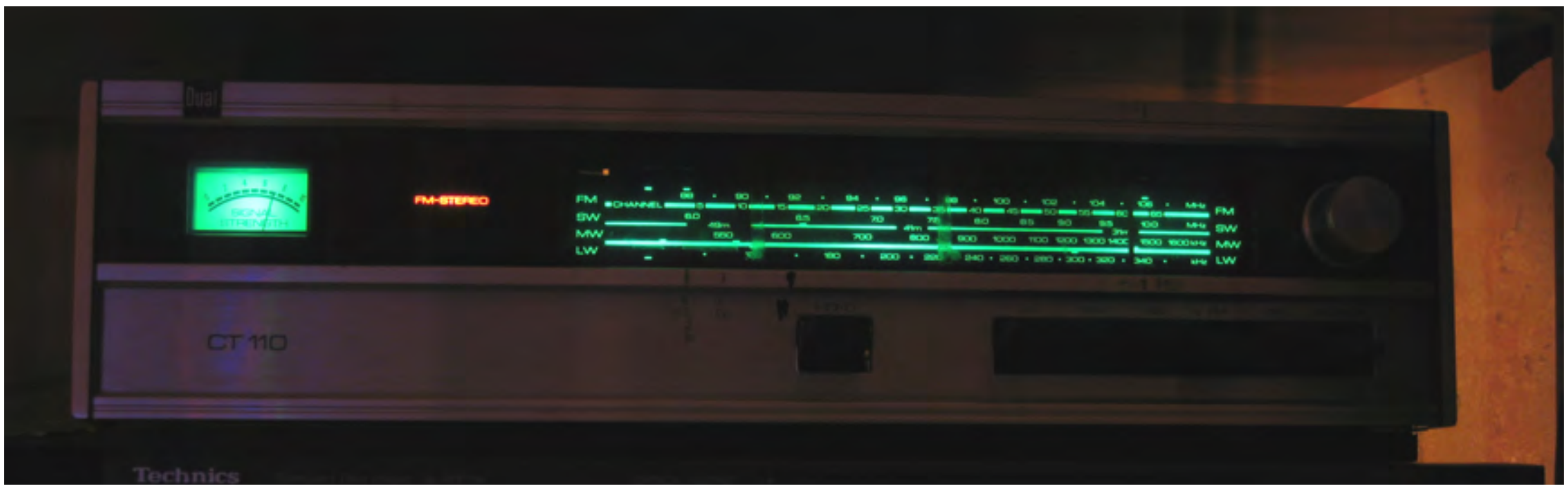

Figure 5: An analog radio tuner.
To find a radio station, you need to adjust one parameter, with the round button on the right.

To pick up a radio channel, you need to move the button in a delicate manner. There is just one parameter to vary between lower and higher radio frequencies (e.g. for FM Very High Frequencies, the signal varies from 30 MHz to 300 MHz). Let us suppose that your reception is of poor quality despite all your efforts to fine-tune the position of the button. What can you do? There are other parameters that you can try to vary in order to obtain the best possible reception. For example, you can move the antenna to improve the reception. Or maybe there are some interferences of your radio tuner with other electronic devices; so you might try to move your radio tuner in another room. Or you might also buy and install a roof antenna, which should bring you an even better reception. Such solutions to improve the reception were not obvious if you limited your attention to the only parameter of the radio tuner itself. This can illustrate the difficulty to capture all relevant parameters involved in a phenomenon which needs fine-tuning.

In engineering, one can fine-tune a system by trial-and-error or by using optimization methods. When one needs to adjust, 1, 2, 3, ..., *n* parameters the state space quickly becomes enormous to explore. Analytical solutions from mathematical equations become impractical, and this is where we need computer simulations (Sterman 2000, 37–39).

There are many natural ways which can lead to fine-tuned systems. In biology, organisms adapt and seek adaptation to their environment, for example as they look for a place which is neither too hot nor too cold, with food in the surroundings, etc. After some time, there is the phenomenon of *niche construction*. The logic is reversed since that time it is the organism which changes the environment in order to survive. For example, Inuits build igloos to be –relatively– at warmth.

*Is the universe fine-tuned?* The idea of fine-tuning is highly controversial. The large literature shows a wide diversity of diverging positions about fine-tuning. The issue stems from physics, but is often motivated by philosophical or theological agendas. We can find skeptics who insist that fine-tuning is impossible to define rigorously, physicists who either maintain that it is a central issue in theoretical physics, or that there is no need for fine-tuning, while natural theologians use fine-tuning arguments to prove the existence of God.

The difficulties are multiple because the precise formulation of the argument requires understanding of physics, cosmology, probability theory, dynamical systems theory, parameter sensitivity and philosophy.

Furthermore, fine-tuning is often mixed with related issues such as observation selection effects, teleology, anthropic principles and reasonings, metaphysical and theological issues. It is also a confusing domain because the question "*fine-tuning for what?*" is answered differently by different authors and approaches. The fine-tuning field is an intellectual minefield. I rarely encounter authors who do not commit a fine-tuning fallacy –I also was guilty of one in a previous paper. How can we define fine-tuning? How can we scientifically progress on this issue?

Our aim is to propose a scientific approach to find out whether our universe is fine-tuned or not. But the first difficulty is to define fine-tuning, which requires three steps. First, to debunk common and uncommon fine-tuning fallacies which constantly sneak into the debate (section 6.1 Fine-Tuning Fallacies, p107). The second step is to distinguish fine-tuning from closely related issues (section 6.2 Fine-Tuning and Other Issues, p115). Third, to provide a fallacy-proof definition of fine-tuning (section 6.3 The Cosmic Evolution Equation, p122). Inspired by Drake's equation in the Search for Extraterrestrial Intelligence (SETI), I present the *Cosmic Evolution Equation* as a skeleton to answer the question "fine-tuning for what?". This equation defines *cosmic outcomes* which can be used to generate different definitions of fine-tuning.

Thanks to those new conceptual distinctions, I critically review classical explanations of fine-tuning (section 6.4 Classical Fine-tuning Explanations, p144). Later in Chapter 7 I show how studying possible universes with computer simulations could guide the future of humanity. I will then examine two broad evolutionary approaches to fine-tuning: cosmological natural selection and cosmological artificial selection (Chapter 8).

## 6.1 Fine-Tuning Fallacies

On June 25th 2009, Victor Stenger kindly agreed to give a seminar about fine-tuning in our Evolution, Complexity and Cognition (ECCO) research group. Stenger's intellectual niche resides in skeptical inquiry, where he successfully debunks pseudoscience and religious arguments which use modern physics wrongly. This includes centrally fine-tuning arguments in cosmology.

Before meeting him, I had expected a close-minded atheist. I was dead wrong. Victor Stenger is not only a very nice person to talk with, truly open minded, but also an outstanding scientific popularizer and scientist. He is able to explain difficult and advanced scientific theories and concepts in a simple way, which I consider the hallmark of a bright mind.

Since I was both impressed and convinced by his debunking of fine-tuning arguments, I emailed him later to let him know that I was writing a paper about fine-tuning in which I wanted to include his important insights. He replied that he was just about to finish a book entitled *The Fallacy of Fine-Tuning*! We then pursued interesting discussions about a draft of this book and I tried in this chapter to take into account its lessons. I now consider Stenger's (2011) book an essential reading for anyone working on fine-tuning.

Story 6: Victor Stenger in Brussels

Victor Stenger is an atypical physicist. As an atheist, he has committed himself to seriously examine proofs of God's existence provided by natural theologians. Over the years, he has fine-tuned his arguments against fine-tuning, and concludes that fine-tuning arguments are unconclusive. Unfortunately, such fallacies are also sometimes committed by otherwise very respected and respectable scientists. Stenger's books are widely accessible, educational and with a scientific content using only mainstream science. In this section, I will include some of Stenger's (2011) most important arguments.

Let us first make the *logical structure of the fine-tuning argument* more precise (inspired by Colyvan, Garfield, and Priest 2005):

1. Models of our universe display parameter sensitivity for some cosmic outcome O, when varying one parameter at a time.
2. Our universe displays cosmic outcome O.
3. Outcome O is improbable.
4. Our universe is fine-tuned to produce outcome O.

Before getting our hands dirty in the fine-tuning machinery, let us make some general comments. Propositions (1) and (2) can be verified for many physics and cosmic parameters. By *cosmic outcome* I mean a milestone in cosmic evolution, such as the emergence of stable atoms, stars, galaxies, planets, life or consciousness. I will be more specific about cosmic outcomes (in section 6.3.2 Possible Cosmic Outcomes, p124). The inference from (1) and (2) to (3) is *not* conclusive. Indeed, we need many more assumptions to make probability claims. Proposition (3) is very questionable because it is hard if not impossible to define and quantify this purported improbability (see next subsection 6.1.1 Probabilistic Fallacies, p109). For example, proposition (1) explicitly acknowledges that we vary *one parameter at a time*. Is it serious scientifically? What happens if we vary more, ideally all free parameters of our models? Would outcome O still remain improbable?

If we grant proposition (3), the inference from (3) to (4) is still incorrect. We would have proven that our universe is *parameter sensitive*, which is not the same as proving that our universe is *fine-tuned*. As we will see, a proof of fine-tuning requires that this improbability is shown within a wide space of possible universes. Furthermore, it is easy to misinterpret improbabilities. For example, the odds of having a particular hand of cards in the game of bridge is roughly one to 600 billion. Can we conclude from this high improbability that our hand was not randomly dealt?

These four steps then can branch into two different arguments, either towards a *theological fine-tuning argument*:

5.1 The best explanation for this improbable fact is that the universe was created by some intelligence.
6.1 A universe-creating intelligence exists

Or towards a *multiverse fine-tuning argument*:

5.2 The best explanation for this improbable fact is that our universe is one in a much greater multiverse
6.2 A multiverse exists

Further inferences from (4) to (5.1 or 5.2) and from (5.1) to (6.1) or from (5.2) to (6.2) are speculations. We will see other explanatory mechanisms other than (5.1) and (5.2) (see section 6.4 Classical Fine-tuning Explanations, p144 and Chapter 8). Regarding propositions (5.1) or (5.2), we can ask: what makes creation by some intelligence or the existence of a multiverse "the best explanation"? According to which criteria? In propositions (6.1) or (6.2), the issue becomes philosophical: what are our criteria to state that something "exists"? Is it because something is the best known explanation that it necessarily exists?

### 6.1.1 Probabilistic Fallacies

Probabilistic fallacies in fine-tuning discussions lie in the lack of precise and sound mathematical definition of proposition (3). The difficulties are multiple. First because it is problematic to apply probabilities to a unique object, our universe; second, because we must decide the probability distribution of "interesting" universes; third because we must choose a variation range allowed for each parameter; fourth, because we must know how to deal with infinities; fifth because we must choose the resolution at which the parameter is allowed to vary and thereby addressing the coarse-tuning argument. Let us further examine these issues.

The first problem is that, as far as we know, our universe is unique. Obviously, speaking about the probability of a single object is problematic. *How can we define the probability of the universe?* As Ellis (2007a, 1218) wrote:

> **Thesis A4: The concept of probability is problematic in the context of existence of only one object.** *Problems arise in applying the idea of probability to cosmology as a whole — it is not clear that this makes much sense in the context of the existence of a single object which cannot be compared with any other existing object.*

If we want to assign probabilities to our universe, we thus need to assume a multiverse, *virtual or real*, so we can compare our universe to others. We thus implicitly assume in fine-tuning arguments that our universe's parameters could have been different. Otherwise, the probability of our universe would just be one, and there would be no need for fine-tuning. This is the *necessity* explanation we summarize in section 6.4.2 Necessity, p145.

Colyvan, Garfield and Priest (2005) did make an important skeptical critique on the use of probabilities in fine-tuning arguments. We now summarize some of their most salient insights. Let us assume our universe's parameters could have been different. Which modality do we mean when we ask "parameters *could* have been different?" There are two main options: *logical possibility* or *physical possibility.* In this section, we focus on the logical possibility. It is more general than the physical possibility we will discuss (in the next section 6.1.2 Physical Fallacies, p111) which is restricted by various physics consideration such as other parameters, physical laws or initial conditions.

If we want to assign probabilities, we need a set on which to assign them. This set can be either continuous or discrete. The *ideal analog radio tuner* gives the intuition of a continuous set. A *digital radio tuner* where you can increase frequencies by steps of 0.5Mhz illustrates the discrete case, as in Figure 6.

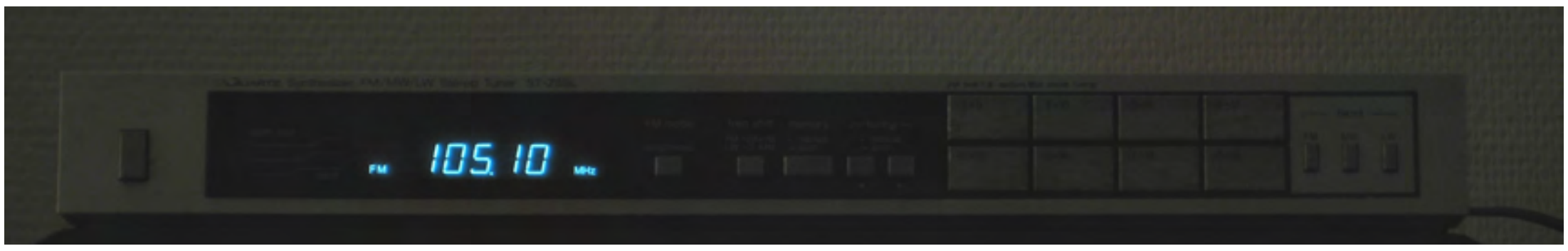

Figure 6: A digital radio tuner.

If the set is finite, then it is easy to use probabilities. With our digital radio tuner, there are on the FM band 540 possible frequencies you can try from 30 to 300Mhz. If you assume you live in a busy city where there is a radio station at every 0.5Mhz, then you have one chance in 540 to have your digital radio tuner tuned on your favorite station as you unpack it. Of course, in a more realist picture, there is no radio station at every half megahertz. How can we estimate in a rigorous way the probability to have a radio station at a particular frequency? We need to know the *probability distribution* of having a radio station at a certain frequency. How can we know this distribution *a priori*? There is no obvious way to answer or tackle this problem, neither for radio stations nor for universes.

An additional difficulty in the case of the universe is that there is no obvious well defined range to vary the parameter, as there is for the FM band, which is from 30 to 300Mhz. Without additional constraints, following Laplace's "principle of indifference", we must allow each physics and cosmic parameter to vary from $-\infty$ to $+\infty$.

Let us now imagine that you have bought an ideal analog radio tuner which is so sensitive that you can pick an infinite number of radio frequencies. There are two possibilities, either your radio tuner is "rationally sensitive", or "really sensitive". In the first case, the set is *dense* like an interval in the set of rational numbers; in the second case, the set is *continuous* like an interval in the set of real numbers. Both sets are infinite, but involve different mathematical treatments, the one with *discrete* probability tools, the other with *continuous* ones. In probability theory, one distinguishes on the one hand *discrete variables* which have either a finite or *countably* infinite number of possible values, from *continuous variables* which have an *uncountably* infinite number of possible values.

Here we face an ontological problem. Should we use discrete or continuous variables? Even with the analog radio tuner, we could argue that the number of possible frequencies is either huge but finite, countably infinite or uncountably infinite. Each option involves a different ontological position regarding the nature of the "real" world. A pragmatic approach is simply not to worry too much about this, and proceed with the mathematical tools which best help us to solve our problem.

Let us assume a continuous variable. The intuitive idea of fine-tuning, Colyvan, Garfield and Priest (2005) argue, is the following. Some physical constant $k$ must take a value in a very narrow range in order to a cosmic outcome to evolve. They write:

> Let us suppose that the constant in question has to lie between the values $v - \delta$ and $v + \varepsilon$. (Where $v$ is the actual value of $k$ and $\delta$ and $\varepsilon$ are small, positive real numbers.) The intuitive idea is that the interval $[v - \delta, v + \varepsilon]$ is very small compared to all the logically possible values $k$ might have taken (i.e., the whole real line), and since there is no explanation of why $k = v$ and not any other value, all possible values of $k$ should be considered equally likely. Thus, the probability of $k$ taking a value in $[v - \delta, v + \varepsilon]$ is also very small. That's the intuitive idea, but

> the problem is that it's not at all clear how this naïve intuition can be made rigorous.

In an attempt to clarify this intuition, they continue:

> One way to try to cash out this intuition is to compare the number of values that $k$ can take in $[v - \delta, v + \varepsilon]$ to the all possible values it could take in the real line. But in each case there are values so the relevant probability would appear to be one. That, however, is misguided. We are not interested in the number of values $k$ could take but rather the measure of the sets in question. Employing any standard measure (such as Lebesgue measure) will, as the fine tuning argument proponent suggests, indeed yield a very low probability for $k \in [v - \delta, v + \varepsilon]$. The problem for the proponent of the fine tuning argument, however is that the probability is too low! Such measures will yield a finite value for the interval $[v - \delta, v + \varepsilon]$ and an infinite value for the whole real line. The resulting probability for $k \in [v - \delta, v + \varepsilon]$ is zero (or infinitesimal if you prefer a non-standard analysis take on this). What is more, the probability that $k$ takes a value in any finite interval will be the same – even those we intuitively think of as being extremely large. So, for example, the probability of $k \in [v - 10^{10^{10}}, v + 10^{10^{10}}]$ is also zero (or infinitesimal).

In summary, the probability to have a certain value of a parameter in any finite interval is 0! Colyvan Garfield and Priest show that this constitutes a serious obstacle to rigorously argue that a parameter's particular value is improbable. The last remark also shows the importance to consider the *resolution* at which we consider the parameter's variation. This issue is often refereed to as the *coarse tuning argument*, because it is unclear what constitutes a "fine" or "coarse" interval. Consequently, the fine-tuning argument works as well –or as badly– as the coarse-tuning argument (see McGrew, McGrew, and Vestrup 2001; see also Manson 2000)!

In conclusion, the statement "outcome O in cosmic evolution is improbable" does not follow from existing fine-tuning arguments. Showing precisely and rigorously the improbability of a particular outcome is a difficult task.

### 6.1.2 Physical Fallacies

We just outlined fundamental probabilistic difficulties in formulating a rigorous quantitative fine-tuning argument. A natural way to overcome those issues is to take into account not only logical, mathematical and probabilistic considerations, but also physical and cosmological ones.

However there are even more fallacies lurking as we tackle the fine-tuning issue armed with physics and cosmology. I refer the reader to Stenger's (2011) book-length detailed debunking of fine-tuning claims such as the ratio of electrons to protons; the ratio of electromagnetic force to gravity; the expansion rate of the universe; the mass density of the universe; the cosmological constant; the Hoyle prediction or the relative masses of elementary particles. However, Stenger's book was recently criticized by an extensive review on the fine-tuning issue by Luke Barnes (2012). Indeed, scientists do not agree on specific instances of fine-tuning. And the debate has no reason to stop, as Stenger's (2012) reply to Barnes can attest.

Let us debunk some general sources of fine-tuning fallacies: the dimensional constants variation; the inconstant coupling constants; the free parameters capture; the variation of only one parameter and the one-factor-at-a-time paradox.

We already saw (in section 5.2 The Physical Universe, p89) that it only makes sense to vary dimensionless constants, and not dimensional ones like *c*, *G* or *h* (see also Stenger 2011, 59–62). Again, these are arbitrary constants useful to establish a system of units. The only meaningful parameters to vary are dimensionless ratios.

We also mentioned (section 5.2 The Physical Universe, p89) that coupling constants are defined at a certain level of energy, and change at different energy levels. For example, the weak nuclear force $\alpha_W$ drastically varies from ~ $3x10^{-12}$ at low energy (1 Gev) to ~ $3x10^{-2}$ at high energy (100 Gev) (Demaret and Lambert 1994, 4). Without this energetic consideration, it doesn't make sense to speak about a variation of coupling constants, which, contrary to their misleading name, are not always constant.

Another fine-tuning fallacy is to ignore the phenomenon of *free parameters capture.* It comes from overlooking readily available explanations in current physical and cosmological models. We already saw that the value of the (reduced) cosmological constant can be predicted in the framework of scale relativity (see story 4, p93). Since its value is constrained and predicted by theory, to speak of a free possible variation of it is precipitous. Victor Stenger refutes similarly other fine-tuning arguments, by taking seriously inflation theory. For example, the mass density of the universe is sometimes considered fine-tuned. If it were larger, the stars would burn too rapidly; and it if were smaller, too few heavy elements would form. But it is a prediction of inflationary theory that it should have its particular value.

Importantly, the point here is independent of whether scale relativity or inflation theory is true. It is just that if we want to speak meaningfully about fine-tuning of the cosmological constant or the mass density of the universe, or any other parameter, we can't simply ignore existing models predicting their values.

Fine-tuning arguments vary just one parameter, a fallacy which is nearly *always* committed. The underlying assumption is that parameters are independent. As Stenger (2011, 70) writes:

> This is both dubious and scientifically shoddy. As we will see in several specific cases, changing one or more other parameters can often compensate for the one that is changed. There usually is a significant region of parameter space around which the point representing a given universe can be moved and still have *some* form of life possible.

It is both the most common and the most serious fallacy. If the history of physics taught us something is that phenomena which where thought to be widely independent, turn out to have common underlying causes and principles. For example, our common sense fails to see a connection between the fall of an apple and the tides; magnetism and electricity; and even less between space, time and the speed of light. But all these phenomena have been unified thanks to physical theories.

Additionally, varying several parameters without care can lead to what is known as the *one-factor-at-a-time paradox* in sensitivity analysis. The problem with the one-factor-at-a-time (OAT) method is that it is non-explorative. Let us see why. At first sight, the method of OAT seems logical and rigorous, since it varies factors one-at-a-time while keeping the others constant. It seems consistent because the output from a change can be attributed unambiguously to the change of one factor. It also never detects non-influential factors as relevant. However, by construction, this method is non-explorative, with exploration decreasing rapidly with the number of factors. For a simple example, consider Figure 7, which shows clearly that OAT explores only 5 points forming a cross, out of 9 points in total.

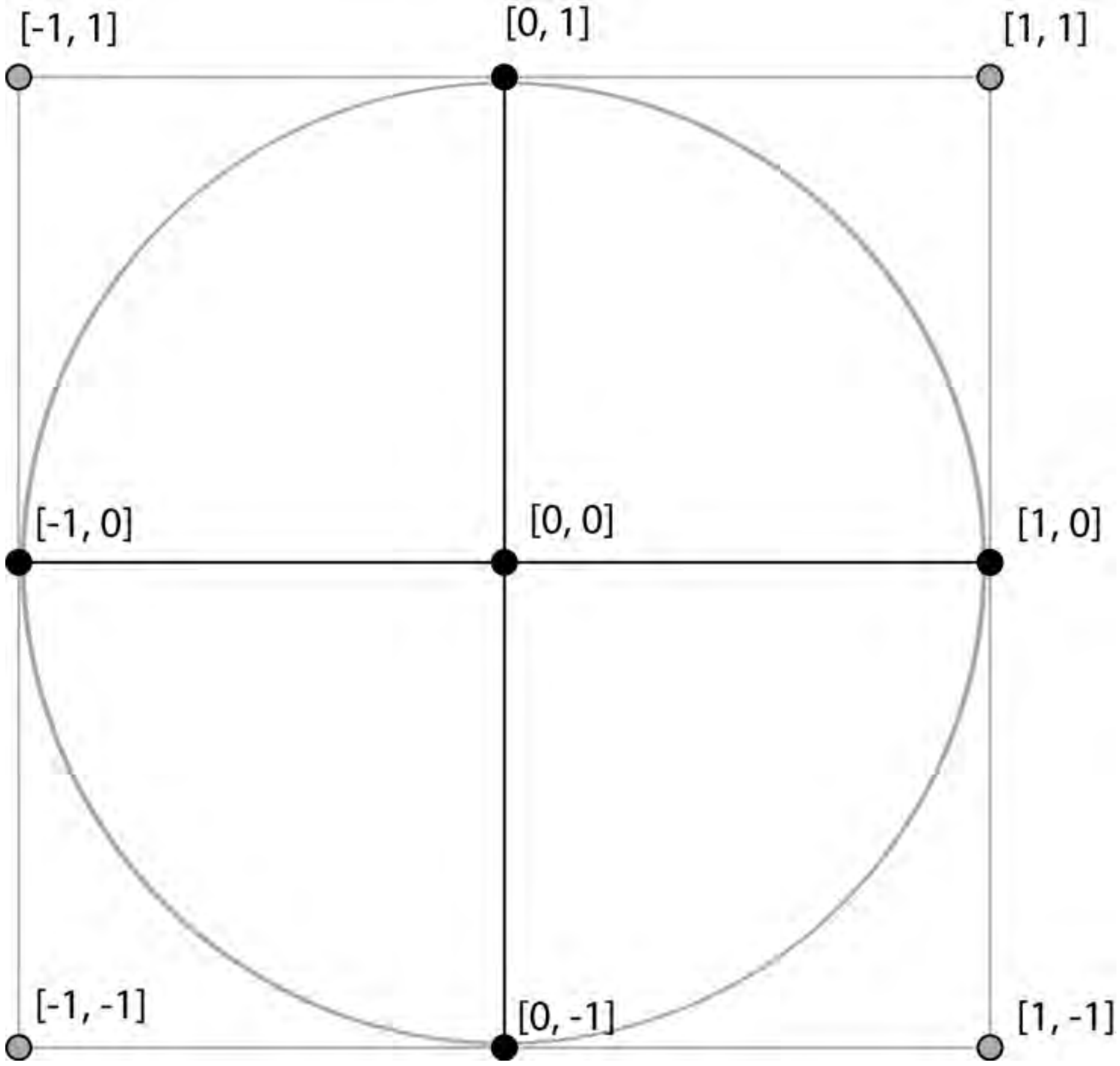

Figure 7: The one-factor-at-a-time method can only reach points on the cross. In this simple two-dimensional parameter space, each discrete factors can only take values 0, 1 or -1. OAT can reach [0, 0], [0, 1], [0, -1] (points on the vertical line); and [-1, 0], [1, 0] (points on the horizontal line). The points explored are thus on a cross. The points not explored are the corners [-1, 1], [-1, -1], [1, 1], [1, -1]. In a geometrical interpretation, note that the cross is by construction inscribed in the circle. But OAT actually restricts the exploration to points on the cross, not inside the circle because exploring points inside the circle would imply varying two parameters at the same time. Now, that cross itself is inscribed in the circle. In sum, OAT restricts the exploration to the cross, not the circle, but the cross is inscribed in the circle. And this circle is inscribed in the square (2-cube), which is why OAT can't reach the corners of the square.

Let us now generalize this example. In n-dimensions, the n-cross will necessarily be inscribed in the n-sphere. The problem is that this n-sphere represents a small percentage of the total parameter space defined by the n-cube. For example, in 2 dimensions, the ratio of the partially explored to the total area is $r \approx 0{,}78$. In the figure above, it means that the corners areas outside the circle are not explored. The problem gets quickly worse as we increase the number of dimensions. In 3 dimensions, $r \approx 0{,}52$ and in 12 dimensions, $r \approx 0{,}000326$ (see Andrea Saltelli and Annoni 2010 for those calculations, as well as critiques and alternatives to OAT).

Fine-tuning arguments typically vary one parameter at a time. So, they use the OAT method to explore the space of alternative universe by varying one by one some of the 31 fundamental physics and cosmic parameters, and actually explore only $r \approx$ $4,56 \times 10^{-15}$ of the parameter space. We conclude that such fine-tuning arguments have restricted their exploration to 0,00000000000000456 % of the relevant parameter space[9]!

### 6.1.3 Fine-Tuning Disproofs

More and more authors debunk logical, probabilistic and physical fallacies about fine-tuning. Scientific progresses provide the most elegant way to refute fine-tuning arguments. Indeed, we saw that physicists build new theories which capture free parameters (see Vidal 2010a and Chapter 5). This is the historical and logical scientific progress, which leads to a reduction of fine-tuning arguments.

One might object that some modern physics theories have *more* free parameters, not less. For example, the simplest supersymmetric extension of the standard model displays 125 free parameters (Smolin 2007, 330). How is this to be interpreted? Instead of the sudden emergence of an opposite historical trend where new physical theories should have more parameters, we can surmise that more work needs to be done at the level of the foundations of physics. Indeed, a theory with more free parameters is not a good one. It is well known that the more free parameters a theory has, the easier it is to make it fit any data.

Let us consider the cosmological constant as an example of fine-tuning disproof. For the sake of the argument, assume that we ignore Nottale's prediction of its value. Then, Don Page (2011) showed that the cosmological constant is not fine-tuned for life since “the fraction of baryons that become living organisms would be higher if the cosmological constant were lower.” This example shows that these kinds of arguments can also prove the opposite: that our universe could have been better fine-tuned for life! However, one could reply that fine-tuning would still remain. Such an argument would show that the parameter (here the cosmological constant) is not *optimized* for a particular outcome (here the production of baryons). Optimization can be distinguished from parameter sensitivity.

Once the basic fallacies are debunked, by varying more than one parameter, Stenger is able to show that our universe is not particularly fine-tuned. He actually avoids the OAT paradox by varying two parameters at a time, exploring a bit more of the parameter space. This is still largely insufficient if our standard is a systemic exploration of the parameter space, but enough to show that our universe is not uniquely fine-tuned.

To take another example, Stenger reminds us that weak nuclear reactions rates which give rise to the neutron-proton mass difference depend on a freeze-out temperature $T_F$. By varying both the neutron-proton mass difference and $T_F$ , a wide range of life-permitting parameter space is possible (Stenger 2011, 182).

By using computer simulations, it is possible to explore even more of the parameter space. Stenger (1995; 2000) has performed a remarkable simulation of possible universes, varying four parameters. His conclusion is that our universe is not fine-tuned and other universes are possible. We will discuss this attempt and see how

9 I used the formulae in (Andrea Saltelli and Annoni 2010, 1510) for this calculation. Note that this assumes that we can put upper and lower boundaries on each of the parameters, which is not at all warranted for physics and cosmic parameters. Note also that this is a very generous estimate, since the actual exploration of OAT will only be a tiny n-cross within the volume of the n-sphere, which itself represents only $4,56 \times 10^{-15}$ of the full parameter space defined by the n-cube.

to test rigorously fine-tuning with virtual universes soon (see section 6.3 The Cosmic Evolution Equation, p122).

## 6.2 Fine-Tuning and Other Issues

The fine-tuning issue is knotty and multidimensional. It is often mixed up with other closely related, yet different issues. The purpose of this section is to disentangle fine-tuning from other issues. I know few authors who have not confused at least once some of these issues... and I am no exception.

### 6.2.1 Fine-tuning and free parameters

In (Vidal 2010a; 2012a) I wrote about fine-tuning, while I was in fact tackling the free parameters issue. I confused the two and I apologize to my readers. However, this does not harm the argumentation of those papers, since reducing free parameters automatically reduces associated fine-tuning arguments (see Chapter 5). I already defined the free parameters issues as:

> **Free parameters issue:** *There are free parameters in the standard model and in cosmological models, which in principle can be filled in with any number.*

This issue is thus not concerned with the particular values the parameters take or could take, and the impact of their variation for the development of cosmic outcomes in the universe.

### 6.2.2 Fine-tuning and parameter sensitivity

Today, there is no rigorous evidence of fine-tuning. Nobody has run comprehensive simulations of possible universes to assess the likelihood of the emergence of life. *The majority of "fine-tuning arguments" are not about fine-tuning, but remain one-parameter-at-a-time sensitivity arguments.* All we can conclude with existing fine-tuning arguments is that our current models display parameter sensitivity when varying one parameter at a time. This is a considerably weaker claim than the full fine-tuning argument. Indeed, it is just *the first* step out of *six* in the fine-tuning argument we outlined above. The issue can be defined as follows:

> **Parameter sensitivity issue:** *Models of our universe display parameter sensitivity for some cosmic outcome O, when varying one parameter at a time.*

Even Stenger (2012) recognizes this fact when he writes: "I have never denied that life, as we know it on Earth, would not have evolved with slight changes in parameters." In other words, the parameter sensitivity issue asks: *why is the universe parameter sensitive?* Rick Bradford (2011) showed with toy models and a statistical entropy argument that *parameter sensitivity* is inevitable in any complex universe. This is an important result, and the general idea of the argument can be summarized as follows:

> The thesis has been presented that parameter sensitivity arises as a natural consequence of the mathematics of dynamical systems with complex outcomes. The argument is as follows: the emergence of complexity in a sub-system, ℵ, requires a sequence of entropy reductions of ℵ, which can be interpreted as a

> sequence of reducing phase space volumes. This leads, via a very general formulation of system evolution, to constraints on the set of universal constants. This is the origin of parameter sensitivity.

Bradford (2011, 1584–1585) had the remarkable insight to clearly distinguish parameter sensitivity from fine-tuning:

> We contend that fine tuning actually consists of two distinct phenomena.
>
> The first phenomenon is the parameter sensitivity of the universe. This is the (apparent) property of the universe that small changes in the parameters of physics produce catastrophic changes in the evolved universe. In particular the complexity of the evolved universe, and hence its ability to support life, would be undermined by small changes in the universal constants (in the pragmatic sense of 'small' changes discussed above). Thus, 'parameter sensitivity' is the claim that the target in parameter space which is compatible with a complex universe is small in some sense. The smallness of this target, if true, is a feature which requires explanation.
>
> The second, and quite distinct, phenomenon is that nature has somehow managed to hit this small target—which we will refer to as "fine tuning". The actual constants in our universe have to be fine tuned to coincide with the requirements for a complex outcome. In other words, given that only special values for the parameters will do (i.e., given parameter sensitivity), nature had to contrive to adopt these particular values (i.e., nature is fine tuned).

However, the first sentence of this citation confuses the two phenomena, since fine tuning is seen as *including* both parameter sensitivity *and* fine tuning. I make this distinction exclusive, so I see Bradford's argument showing "the inevitability of *parameter sensitivity* in a complex universe", and not "the inevitability of *fine tuning* in a complex universe", the latter being the title of the paper. As Bradford acknowledges himself, the "paper is concerned only with the first phenomenon: parameter sensitivity and how it arises".

### 6.2.3 Fine-tuning and metaphysical issues

Metaphysics sells, physics doesn't. This is why ambitious physicists, in their metaphysical quest pursued with scientific tools, will sometimes create confusion between what they actually do and the titles of their papers and books. For example, Vilenkin (1982) has a famous paper entitled *Creation of universes from nothing*; Krauss' (2012) latest book is entitled *A Universe from Nothing: Why There Is Something Rather than Nothing*. Another recent example is Hawking and Mlodinow (2012, 180) who wrote that because "there is a law like gravity, the universe can and will create itself from nothing in the manner described in Chapter 6. Spontaneous creation is the reason there is something rather than nothing, why the universe exists, why we exist." These titles and statements are misleading and philosophically naïve, because they confuse physics with metaphysics. It was to avoid such confusions that we made explicit the *metaphysical challenge* (see section 4.1.2 Metaphysical, p76):

**Metaphysical challenge**: *Why not nothing?*

Ted Harrison wrote in (1995) a speculative paper which is the basis of Chapter 8. He was then criticized harshly by John Byl (1996). Harrison (1998) replied by clearly distinguishing the metaphysical problem of *creation* from the problem of *fitness* of the universe for life, i.e. the fine tuning issue.

### 6.2.4 Fine-tuning and anthropic principles

*Anthropic principles serve only to obfuscate.*
(Swinburne 1990, 172)

To which I add:
*A serious discussion of the anthropic principle does not mention the anthropic principle.*

Statements about the "anthropic principle" raise passion in scientific circles. Some denounce its unscientific nature, others its tautological aspect, while yet others maintain that it is essential to scientific reasoning (see Dick 1996, 527–536 for an historical summary). A thing is certain, it is a confusing idea. When Brandon Carter (1974) coined the term, he intended to speak about *selection effects*. However, as Bostrom (2002, 6) reports,

> The term "anthropic" is a misnomer. Reasoning about observation selection effects has nothing in particular to do with homo sapiens, but rather with observers in general. Carter regrets not having chosen a better name, which would no doubt have prevented much of the confusion that has plagued the field. When John Barrow and Frank Tipler introduced anthropic reasoning to a wider audience in 1986 with the publication of *The Anthropic Cosmological Principle*, they compounded the terminological disorder by minting several new "anthropic principles", some of which have little if any connection to observation selection effects.
>
> A total of over thirty anthropic principles have been formulated and many of them have been defined several times over—in nonequivalent ways—by different authors, and sometimes even by the same authors on different occasions. Not surprisingly, the result has been some pretty wild confusion concerning what the whole thing is about.

Although Bostrom does not review the over thirty different anthropic principles, there are in practice two main uses of the anthropic principle today. Either the author wants to speak about *observational selection effects* or about *teleology*. There are good reasons to be careful and skeptical with both. The one refers to a basic scientific methodology, while the other may drift us into unscientific considerations. To summarize, when you see a mention of the anthropic principle, you should ask yourself, does the author means a selection effect, teleology or yet something different? Let us now discuss the two main meanings of selection effects and teleology, and thus follow my proposal not to use anymore the ambiguous "anthropic principle".

### 6.2.5 Fine-tuning and observational selection effects

To illustrate what an Observational Selection Effect (OSE) is, imagine you go fishing with a fishing net (see Leslie 1989 for extensive discussions of this example). You catch a fish of exactly 59.07 cm. Can you conclude anything about the fish population? If you want to make any serious theory about the fish population, you should at least be aware that holes in your fishing net have a certain size, and thus for example that you will never be able to catch fishes 2 cm long, because they are much smaller than the holes. Thus the very fishing net you are using induces an observational selection effect on the fishing outcome. Taking into account OSE is thus fundamental to draw meaningful conclusions. This holds for this example, but also in all scientific investigation.

Another example is the Malmquist bias in astrophysics (see e.g. Ellis 2007a, 1199). Let X be a population of luminous objects, with different luminosities. At a distance from X, we will only see the most luminous objects. Thus, the average luminosity will appear to *increase* with distance.

Now, we can reformulate Carter's famous “weak anthropic principle” and “strong anthropic principle” as two OSE:

**Observational Selection Effects:**

Weak Anthropic Principle (WAP) *: The existence of observers in the universe imposes temporal constraints on the positions of the observers in the universe.*

Strong Anthropic Principle (SAP): *The existence of observers in the universe imposes constraints on the set of cosmological properties and constants of the universe.*

It is important to note that Carter always intended his anthropic principles, both WAP and SAP as OSE, not as teleological principles. It is only later that Barrow and Tipler (1986) wrongly associated the idea of teleology with the SAP. This error is so widespread that it has made John Leslie (1998, 295) wonder if we should still try to correct it.

Philosopher John Leslie probably worked more than anyone else on the anthropic principle as an OSE. In his (1990) paper, *The Anthropic Principle Today*, he reviews thirteen common misunderstandings of the anthropic principle as a selection effect.

One of the key points he emphasizes is the distinction between *logical explanations* and *causal explanations*. OSE allow to make logical or probabilistic reasonings, but they do not aim at providing *causal* explanations. As Leslie (1989, 129) remarks, both WAP and SAP as OSE “are not in the least questionable, for of course the universe in which we observers exist now must be compatible with observership both *here and now* (Weak Principle) and *at some stage* (Strong Principle)”. This lack of *causal* explanation is a main reason why many scientists find those two anthropic principles useless or insufficient. This reaction stems indeed from the requirement that scientific theories must provide causal mechanisms. Thus OSE don't causally explain things. They can make things less mysterious (see Leslie 1989, 141). They can also allow some predictions, because they set *boundary conditions*. For example in the fishing story, you can predict that you won't catch fishes less than 2 cm long. And you won't be able to catch a whale either, because it would just tear your net apart.

It is not our aim to develop or discuss theories on how to model OSE (see N. Bostrom 2002 for an extensive treatment). A way to do it systematically is to consider science as the endeavor to construct models of the world, and then to *model the modeling process*. We can note that this is an other instance of “meta-” thought process, where the concept of modeling is applied to itself (as we had a metaphilosophical approach in Part I or a “metaorigin” examination in Chapter 4). Such a study is known as second-order cybernetics (see e.g. Heylighen and Joslyn 2001).

Let us now consider an important and less trivial discussion of OSE regarding the Eddington-Dirac large number hypothesis. Dicke proposed a *logical* explanation as an OSE which is now obsolete since Nottale advanced a more recent theoretical understanding from scale relativity.

Paul Dirac (1937) noticed some unexpected related ratios in cosmological and elementary particle scales. For example, the ratio between the scale of the observable universe and the microscopic scale are related to the ratio of the mass of the universe and the mass of an elementary particle.

However, the size and age of the universe depend on the Hubble constant, which led Dicke (1961) to state that it “is not permitted to take on an enormous range of values, but is somewhat limited by the biological requirements to be met during the epoch of man.” In a later reply, Dirac was not satisfied with this reasoning (see Dick 1996, 527–529 for a more detailed historical account).

Now, in scale-relativistic cosmology (L. Nottale 1993; 2003; 2011) the Eddington-Dirac large number hypothesis can be partially explained. Let us outline the general argument. The cosmological constant $\mathbf{\Lambda}$ has the dimension of a curvature. So it is the inverse of the square of some cosmic length: $\mathbf{\Lambda} = 1/\ \mathbb{L}^2$. $\mathbb{L}$ is identified as the maximum cosmic scale, in the same way that the planck scale $l_{\mathbb{P}}$ is the minimum scale. The ratio between the maximum and the minimum scale is then a new dimensionless number whose value is $\mathbb{K}= \mathbb{L}/l_{\mathbb{P}} \approx 5\text{x}10^{60}$. The next step is to replace the factor $c/H_0$, where $H_0$ is Hubble's constant, with the invariant cosmic scale $\mathbb{L}$. Eddington-Dirac's large number hypothesis can now be written in Planck units, without varying constants such as Hubble's, as:

$$\mathbb{K}= \mathbb{L}/l_{\mathbb{P}} = (r_0/l_{\mathbb{P}})^3 \equiv (m_{\mathbb{P}}/\ m_0)^3$$

where $r_0 = \hbar/m_0c$ is the Compton length associated with the mass scale $m_0$. What this relation says is simply that the scale of elementary particles $r_0/l_{\mathbb{P}}$ is at one third of the universal scale $\mathbb{L}/l_{\mathbb{P}}$ in the scale space. Note that the relation $(r_0/l_{\mathbb{P}})^3 \equiv (m_{\mathbb{P}}/\ m_0)^3$ can be verified from the very definition of $r_0 = \hbar/m_0c$ . Now, how can we *deduce* this conclusion? It is possible by reasoning about scales. The idea is to consider the vacuum as fractal, and the energy density not as a number, but as an explicit *function* of the scale. Without going into more details, this can be seen in Figure 8, where the $r^{-6}$ line (the scale variation of the gravitational self-energy density of quantum vacuum fluctuations) crosses the scale of elementary particles (*q*) at one third of scales in the universe. I encourage the curious reader to study Nottale's papers and books for more details and impressive results.

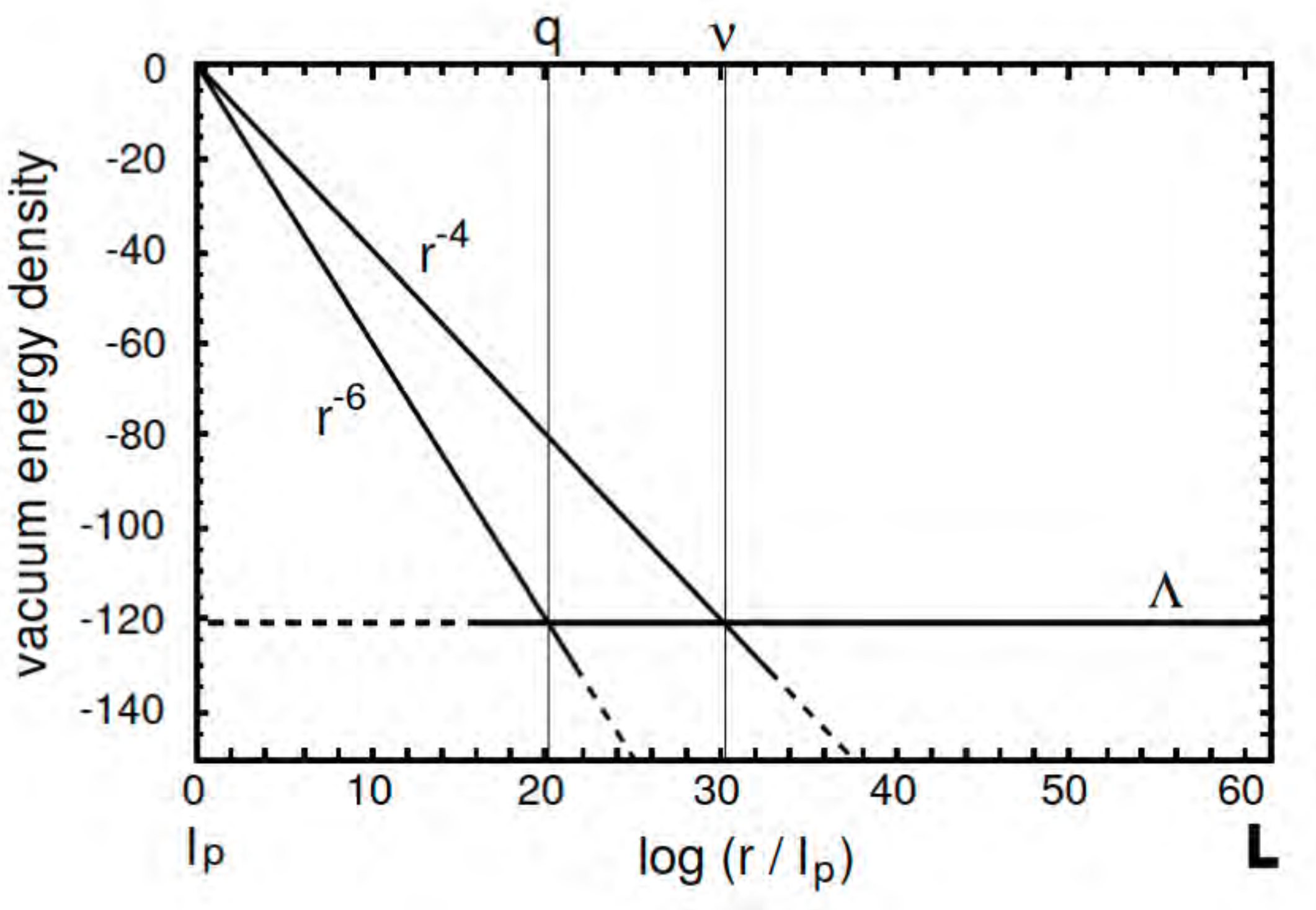

Figure 8: **Scale variation of vacuum energy density.**
From Nottale (2011, 545):"Variation of the vacuum energy density [in $(l_{\mathbb{P}}/r)^4$] and of the gravitational self-energy density of quantum vacuum fluctuations [in $(l_{\mathbb{P}}/r)^6$] in the framework of Galilean scale relativity. We have suggested in (L. Nottale 1993) that the $r^{-6}$ gravitational self-energy contribution crosses the geometric cosmological contribution **Λ** at a scale of 70 MeV (electron "classical" radius and quark confinement transition denoted by *q* in the figure). This scale lies at the third of the full interval going from the Planck scale $l_{\mathbb{P}}$ to the cosmic scale $\mathbb{L}$ (in scale space described by logarithms), which validates the Eddington-Dirac large number relation. (...)"

In conclusion, we can emphasize again that OSE are indeed no causal explanation. This was illustrated with simple examples, as well as with the Dirac-Eddington large number hypothesis. Dirac's intuition that Dicke's "explanation" in terms of OSE was insufficient is confirmed with a more recent partial deduction of the Eddington-Dirac large number relation. OSE should not obscure the fact that regular scientific work needs to be done.

The same limitation holds with the idea that an OSE (e.g. WAP), plus the hypothesis of the existence of a multiverse could "explain" the fine-tuning issue. The argument is to say that there exists a huge multiverse with various values of fundamental cosmic and physics parameters. It is thus not surprising that the only universe we can observe is fine-tuned. If it were not fine-tuned, observers like us wouldn't have emerged, and wouldn't be here to observe anything. The bottom line is that our universe might seem fine-tuned, but the multiverse needs not to be. We will criticize this dubious "explanation" later as "WAP-of-the-gaps" (see section 6.4.6 WAP-of-the-gaps, p147).

### 6.2.6 Fine-tuning and teleology

As with the expression "anthropic principle", scientists bristle at "teleology". And as we saw that selection effects behind anthropic principles are not cogent scientific explanations, let us see what problems are involved with teleology.

First, in theology, teleological considerations have been and are often used to prove the existence of God (see e.g. the physico theological proof of God's existence in Kant 1781). Such arguments attempt to show the existence of a plan, purpose, goal or design in the universe, and are known as *teleological arguments* or *arguments from* or *arguments to design* (Ratzsch 2010). Indeed, fine-tuning arguments are used to substantiate this teleological and theological view (see e.g. Collins 2009; and Stenger 2011 for debunking). So one can suspect attempts to prove the existence of teleology or purpose in nature to hide a theological agenda.

Second, teleology is associated with the Aristotelian categorization in four types of causes which were harshly criticized during the scientific revolution. Especially, final causes are bad explanations, as Francis Bacon (1620, book 2, II) denounces "the final cause rather corrupts than advances the sciences". The reason is that if we say that some phenomenon happens "be-cause-final" it tends toward a final state, it doesn't explain how this final state will be achieved, nor why the system acquired this goal-directedness in the first place.

However, teleology can be and should be addressed in a purely scientific manner. We can address the question to what extent systems are goal-directed without assuming a God or final causes driving the process. A major contribution by Rosenblueth, Wiener and Bigelow (1943) was to identify teleology with *purpose controlled by feedback*. Such an interpretation is fully compatible with a materialist and deterministic worldview. More generally, that paper laid the foundation for the science of *cybernetics*, which Wiener defined as the science of communication and control in the animal and the machine.

Obviously, there are fields of scientific study such as biology or sociology, where it is more than useful to use concepts related to teleology or intentionality (see e.g. Walsh 2008; Dennett 1988). In biology, to distance itself from the bad connotations of teleology, the term "*teleonomy*" is sometimes used instead.

Now, is there teleology at play in the universe? Has our universe a developmental dynamics? Would atoms, galaxies, stars, planetary systems, life, intelligence re-appear if we would re-run the tape of the universe (see Davies 1998, 317; Vidal 2008b)? How likely was it to have such or such outcome in cosmic evolution? We will see how we can address these difficult questions through a study of the *robustness* of the universe (see section 6.3.3 Robustness in Cosmic Evolution, p133).

### 6.2.7 Fine-tuning and God's existence

The study of teleology in the universe is often motivated by a religious agenda. Indeed, if we can prove that there is a purpose of the universe, it's only a small step to argue that this purpose was designed by God. This situation shouldn't obscure the fact that *the study of teleology is a legitimate scientific inquiry.* As we saw in the previous section, if teleology is defined as purpose controlled by feedback, then its study is simply a part of complexity science.

As Leslie (1989) reminds us, the argument that the universe is fine tuned is not based on the assumption that there is a fine-tuner. What it shows is that the emergence of life or complexity is parameter sensitive, one parameter at a time. Fine-tuning and teleology are thus quite distinct from the theological question of

> **God's existence**: *Does God exist?*

Ikeda and Jeffrey (2006) advanced a new critique against the fine-tuning argument used to prove God's existence. It is particularly interesting because it grants the most difficult premise, namely that our universe is indeed improbably fine-tuned (Step 3. "Outcome O is improbable." in our reconstruction of the fine-tuning argument above). The conclusion of Ikeda and Jeffrey is that if the universe is fine-tuned, then God's existence is *less likely*, not more likely! Why? They start with the following assumptions:

> a) Our universe exists and contains life.
>
> b) Our universe is "life friendly," that is, the conditions in our universe (such as physical laws, etc.) permit or are compatible with life existing naturalistically.
>
> c) Life cannot exist in a universe that is governed solely by naturalistic law unless that universe is "life-friendly."

Now the point is that "a sufficiently powerful supernatural principle or entity (deity) could sustain life in a universe with laws that are not "life-friendly," simply by virtue of that entity's will and power." Indeed, if God can intervene in the course of cosmic evolution then why bother fine-tuning? You fine-tune because you do not want to or you can not intervene. So, as Stenger (2011, 253) summarizes, the "more finely tuned the universe is, the more the hypothesis of a supernatural creation is undermined." I refer the reader to Ikeda and Jeffrey's paper for the stronger version of the argument which uses bayesian probabilities.

Of course, a professional theologian might laugh at this argument, because it interprets God in a simplistic way. It assumes the God of intelligent design, capable to intervene in the course of human and cosmic evolution. Theologians can have much more sophisticated conceptions of God (see e.g. Spinoza's or Whitehead's philosophical-theological views).

## 6.3 The Cosmic Evolution Equation

> ***Napoleon****: M. Laplace, they tell me you have written this large book on the system of the universe, and have never even mentioned its Creator.*
>
> ***Laplace****: I had no need of that hypothesis.*
>
> Cited in (Ball 1901)

Can we provide a general definition of fine-tuning avoiding all the logical, probabilistic and physical fallacies we spotted, as well as all the confusions with the seven other issues we outlined? This is my aim now.

Different discussions of fine-tuning focus on very different *cosmic outcomes*. We see fine-tuning discussions regarding the dimensionality of space (Rees 1999), the production of carbon atoms in stars (F. Hoyle et al. 1953), the existence of long-lived stars (Fred C Adams 2008); the number of black holes (Smolin 1992); biochemistry (Barrow et al. 2008); but also complexity of any sort (Ellis 2007b).

A key question to clarify the issue is thus to explicitly ask: *fine-tuning for what?* Which cosmic outcome are we interested in? To answer these questions, I now introduce the *Cosmic Evolution Equation.* It is a modular conceptual framework to discuss possible universes, possible cosmic outcomes as well as the extend of the robustness and fine-tuning of our universe.

To define it, I build on Drake's (1965) equation in SETI and on the thoughtful discussion of possible universes by Ellis, Kirchner and Stoeger (2004). The Drake equation estimates the number of communicative intelligent civilizations in our galaxy. By extension, one application of the generalized *Cosmic Evolution Equation* (CEE) is to estimate the likelihood of our particular universe in the space of possible universes. In other words, if Drake's equation allows to estimate the probability of life existing "somewhere in the galaxy"; one application of the CEE is to estimate the more general probability of life existing "*anywhere* in the space of possible universes".

The famous dialogue between Laplace and Napoleon above shows the strong scientific position of Laplace, who had no need of the hypothesis of God for his scientific work. Instead of God, do we need to assume an actual multiverse? No we don't. To study the fine-tuning issue, we need only *possible* or *virtual* universes, not actually realized ones. This interpretation still allows us to use the vast multiverse literature to define and explore possible universes, without making strong ontological claims regarding their actual existence.

### 6.3.1 Possible Universes

What are the possible universes? How can we describe the space of possible universes? These questions underly enormous logical, metaphysical, philosophical, and scientific problems. Although possible universes or possible worlds have been discussed centrally in the history of philosophy (see e.g. Leibniz 1710; D. K. Lewis 1986; see also Dick 1982 for a wider historical perspective), our aim here is to formulate the issue of possible universes so that it can progressively exit metaphysics and enter the realm of operational science.

We now follow Ellis', Kirchner's and Stoeger's (2004) definition of the class of all possible universes. Let $\mathcal{M}$ be a structural and dynamical space of all possible universes *m*. Each universe *m* is described by a set of states *s* in a state space $\mathcal{S}$. Each universe *m* is characterized by a set $\mathcal{P}$ of distinguishing parameters *p*, which are coordinates on $\mathcal{S}$. Such parameters will be logical, physical or dynamical. How will they dynamically evolve? The three authors elaborate:

> Each universe *m* will evolve from its initial state to some final state according to the dynamics operative, with some or all of its parameters varying as it does so. The course of this evolution of states will be represented by a path in the state space *S*, depending on the parametrisation of $\mathcal{S}$. Thus, each such path (in degenerate cases a point) is a representation of one of the universes *m* in $\mathcal{M}$. The coordinates in $\mathcal{S}$ will be directly related to the parameters specifying members of $\mathcal{M}$.

In such a procedure, we face a first major issue:

> **Possibility space issue:** *What delimits the set of possibilities? What is the meta-law or meta-cause which determines $\mathcal{M}$?*

As the three authors argue, we can't avoid the meta-law issue, because otherwise we have no basis to set up a consistent description of $\mathcal{M}$. We need to have a logic which describes $\mathcal{M}$. There are other difficult issues related to identifying which different representations represent the same universe models – *the equivalence problem* – and the problem of dealing with an *infinite space of possible universes*. I refer the reader to the three authors' paper for more in depth discussions of these issues.

More directly related to the fine-tuning issue is the remark of Jean-Philippe Uzan that "the larger the possibility space considered, the more fine-tuned the actual universe appears to be" (reported in Ellis, Kirchner, and Stoeger 2004, 923). Indeed, we can easily increase the unlikelihood of our universe simply by allowing the parameter space to grow. You could ask for example, did you explore if universes with 42 dimensions generate life? Do we really want to capture the radical idea of "all that can happen, happens"? There is much variation in the space of possibility we can delimit. Ellis (2007a, 1261) distinguishes four levels of variation, *weak*, *moderate*, *strong* and *extreme*:

> • "*Weak variation*: e.g. only the values of the constants of physics are allowed to vary? This is an interesting exercise but is certainly not an implementation of the idea 'all that can happen, happens'. It is an extremely constrained set of variations.
>
> • *Moderate variation*: different symmetry groups, or numbers of dimensions, etc. We might for example consider the possibility landscapes of string theory (Freivogel et al. 2006) as realistic indications of what may rule multiverses (Freivogel et al. 2006; Susskind 2005; 2007). But that is very far indeed from 'all that is possible', for that should certainly include spacetimes not ruled by string theory.
>
> • *Strong variation*: different numbers and kinds of forces, universes without quantum theory or in which relativity is untrue (e.g. there is an aether), some in which string theory is a good theory for quantum gravity and others where it is not, some with quite different bases for the laws of physics (e.g. no variational principles).
>
> • *Extreme variation*: universes where physics is not well described by mathematics; with different logic; universes ruled by local deities; allowing magic as in the Harry Potter series of books; with no laws of physics at all? Without even mathematics or logic?"

We indeed need to make a difficult choice between theoretical physics and magic... or anything in between.

### 6.3.2 Possible Cosmic Outcomes

Once we settle on a framework to define possible universes, a second major issue is to specify the parameters which differentiate possible universes:

> **Cosmic outcomes issue**: *What are the cosmic outcomes? What are the milestones of cosmic evolution? What parameters differentiate possible universes? How do we find those parameters?*

As the three authors mention, the values of the parameters may not be known initially. They may emerge out of *transitions* from one regime to another. For example, sociologists do not explore alternative sociological structures by varying the mass of elementary particles. They start from different, less fundamental parameters, such as the influence of population density, the climate or the media. *The challenge to understand complexity transitions in cosmic evolution is both of upmost importance and difficulty*. For example, how did atoms emerge out of the big bang era? How did planets form out of stars and stardust? How did life originate out of molecules? How did consciousness emerge from biological organisms? Etc.

The ideal of reducing such parameters is a major goal of science. The objective is to build a consistent theory and narrative of cosmic evolution, which explains a maximum of cosmic outcomes with a minimum of parameters. Scientific progress is achieved when new theories capture previously free and unexplained parameters. We saw the reduction of free parameters in physics and cosmology in Chapter 5, but we can now extend this attitude to attempt a reduction of other higher parameters (such as life) to fundamental physics and cosmic parameters. However, since we are still very far from such a feat, in our description of possible universes we must include explicitly those higher parameters. Typically, when researchers tackle the issue of the origin of life, they don't start from big bang nucleosynthesis, but they assume the existence of molecules.

Ellis, Kirchner and Stoeger categorize the parameters from the most basic ones to the most complex ones. They distinguish seven different categories of parameters $p_j$ , with $j = 1 - 2$ describing basic physics; $j = 3 - 5$ describing cosmology and a category of parameters $j = 6 - 7$ related to the emergence of life and higher complexity.

Each category $p_j$ is composed of different parameters $i$. For example, $p_1 (i)$ are basic physics parameters, such that the fine-structure constant; masses, charges and spins of particles, as well as other dimensionless parameters. I refer the reader to the detailed description of the parameters given by the three authors.

However, in each parameter category I would like to add explicitly some random, chance or noise parameters. For example, these could include for $j = 1 - 5$ quantum effects in the early universe; or nonlinear chaotic dynamics which might trigger catastrophic events, such as meteorites impacting planets for $j = 7$. This would certainly complicate the dynamics, but would also make it much more realistic. A dynamical argument can even be advanced that such random events might be essential to the open-ended growth of complexity. Indeed, this can be illustrated in engineering, with the heuristic of *simulated annealing*. One starts by adding important noise into the system and then gradually reduces it. The purpose of the noise is to shake the system to reach a maximally stable configuration and avoid being stuck in a local optimum.

Now, how do we decide which cosmic outcomes to keep, and which ones to leave out? At first, we can aim at including a maximum of parameters. Then, we would progressively reduce the number of parameters, as we get better and better insights on how they emerge from more fundamental principles and theories; i.e. from previous parameters. Robert Aunger (2007a, 1142–1144) did compile from many authors a list of more than 100 different cosmic outcomes. This is the most comprehensive review I am aware of, ranging from the big bang, the formation of atoms, stars, solar systems, life, DNA, multicellularity, sexual reproduction, fishes, to mammals, agriculture, modern science and space exploration. Table 5 summarizes this comprehensive review of cosmic outcomes, which Aunger calls "big history events". This list is certainly an excellent starting point for the endeavor of discussing the selection of cosmic outcomes.

| AUTHOR / EVENT | Sagan | Barrow / Silk | Chaisson | Christian | Modis | Coren | Spier | Maynard Smith/ Szathmary | Barbieri | Klein | Lipsey et al | Sanderson | Johnson/ Earle | Freeman |
|---|---|---|---|---|---|---|---|---|---|---|---|---|---|---|
| Big Bang! | X | X | X | X | X | X | X | | | | | | | |
| Planck era | | | X | X | | | | | | | | | | |
| Inflation | | | X | X | | | | | | | | | | |
| Gravity! | | | X | | | | X | | | | | | | |
| Nuclear forces! | | | X | X | | | X | | | | | | | |
| Electromagnetic forces! | | | X | X | | | X | | | | | | | |
| Hadrons | | | X | | | | | | | | | | | |
| Leptons | | | X | X | | | | | | | | | | |
| Nuclear particles | | | X | X | | | X | | | | | | | |
| Recombination | | | | X | | | | | | | | | | |
| Atoms* | | | X | X | | | X | | | | | | | |
| Stars! | | X | X | X | X | | X | | | | | | | |
| Black holes/ Quasars* | | X | X | X | | | | | | | | | | |
| Solar wind^ | | | X | X | | | | | | | | | | |
| Galaxies* | X | X | X | X | | | X | | | | | | | |
| Second-generation (Population II) stars | | X | | | | | | | | | | | | |
| Population I stars | | X | | | | | | | | | | | | |
| Solar system/ Planets* | X | X | X | X | X | | | | | | | | | |
| Cratering of planets | | X | | | | | | | | | | | | |
| Formation of Earth | X | | X | X | | | | | | | | | | |
| Oldest rocks form | X | X | | | | | | | | | | | | |
| First replicators^ | | | X | X | | | | X | X | | | | | |
| First life* | X | X | X | X | X | X | X | X | X | | | | | |
| Chromosomes | | | | | | | | X | | | | | | |
| RNA/DNA/ protein division^ | | | | X | | | | X | X | | | | | |
| Photosynthesis! | X | | X | | | | | | | | | | | |
| Eukaryotes (complex nucleus)* | | | X | X | | | | X | X | | | | | |
| Recombin-ation of DNA | | | | | | | | | X | | | | | |
| Sexual reproduction^ | X | | X | X | | | | X | | | | | | |
| Atmosphere oxygenated! | X | X | X | X | | | X | | | | | | | |
| Multi-cellular life* | | X | X | X | X | X | X | X | X | | | | | |
| Earliest fossils | X | X | X | | | | | | | | | | | |
| Intensive volcanism | X | | | | | | | | | | | | | |
| Worms | X | | | | | | | | | | | | | |
| Cambrian explosion (Vertebrates) | X | | | X | X | | X | | X | | | | | |
| Genetic *bauplane*^ | | | | | | | | | X | | | | | |
| Trilobites | X | | | | | | | | | | | | | |

| AUTHOR / EVENT | Sagan | Barrow / Silk | Chaisson | Christian | Modis | Coren | Spier | Maynard Smith/ Szathmary | Barbieri | Klein | Lipsey et al | Sanderson | Johnson/ Earle | Freeman |
|---|---|---|---|---|---|---|---|---|---|---|---|---|---|---|
| Brains^ | X | | X | X | | | X | | | | | | | |
| Fish | X | X | | X | | | | | | | | | | |
| Vascular plants | X | | | | | | | | | | | | | |
| Insects (Devonian) | X | | | X | | | | | | | | | | |
| Amphibians | X | | | | | | | | | | | | | |
| Reptiles | X | | | X | | | | | | | | | | |
| Dinosaurs (Permian) | X | | X | X | | | | | | | | | | |
| Paleozoic | X | | X | | | | | | | | | | | |
| Mammals | | | | X | X | X | | | | | | | | |
| Sociality* | | | | X | | | | X | | | | | | |
| Birds | X | | | X | | | | | | | | | | |
| Cretaceous period | X | | X | X | | | | | | | | | | |
| Primates | X | | X | X | X | | | | | | | | | |
| Orangutan | | | | | X | | | | | | | | | |
| Hominoids | X | | X | X | | | | | | X | | | | |
| Proconsul | X | | | | | | | | | X | | | | |
| Hominids | X | | | X | X | X | | | | X | | | | |
| Stone tools! | X | X | | X | X | | | | | X | | | | |
| Consciousness | | | X | | | | | | | | | | | |
| Fire! | X | X | X | X | | | | | | X | | | | |
| Most recent glacial period | X | | | X | | | | | | | | | | |
| Settlement of Australia | X | | | | | | | | | X | | | | |
| Family-level foraging! | | | X | X | X | | | | | X | | X | X | |
| Broad-spectrum food collection! | | | | X | | | | | | X | | | | |
| Neolithic | X | | | X | | | | | | | | X | | |
| Local group/ Bands* | | | | X | | | | | | | | | X | |
| Cultural learning^ | | | X | X | | | | | | X | | | | |
| Modern humans | | | | X | | X | X | | | X | | | | |
| Tool kits! | | | X | X | | | | | X | X | | | X | |
| Clan (tribe)/ Village* | | | | X | | | | | | | | | X | |
| Language^ | | | X | X | X | | | X | X | X | | | | |
| Wheel | | | | | | | | | | | X | | | |
| Pottery | | | | | | | | | | | X | | | |
| Sedentism | | | | X | | | | | | X | | | | |
| Horticulture! | | | | X | | | | | | | | X | | |
| Corporate group/ Big-man* | | | | X | | | | | | | | X | X | |
| Neanderthal Burial | | | | | | | | | X | X | | | | |
| Art (cave painting)^ | X | | X | X | X | | | | X | X | | | | |
| Agriculture! | | X | | X | X | | | | | | X | X | | |
| Domestication of animals! | | | | | | | | | | | X | | | |

| AUTHOR / EVENT | Sagan | Barrow / Silk | Chaisson | Christian | Modis | Coren | Spier | Maynard Smith/ Szathmary | Barbieri | Klein | Lipsey et al | Sanderson | Johnson/ Earle | Freeman |
|---|---|---|---|---|---|---|---|---|---|---|---|---|---|---|
| Pastoral society | | | | X | | | | | | | | X | | |
| Plow | | | | | | | | | | | X | | | |
| Chiefdoms* | | | | X | | | | | | | | X | X | |
| First cities* | X | | | X | | | | | | | | | | |
| First dynasties (Archaic states)* | X | X | | X | | | | | | | | | X | |
| Writing/ alphabet^ | X | | X | X | X | | | | | | X | | | |
| Hammurabic legal codes, taxation^ | X | | | X | | | | | | | | | | |
| Iron metallurgy!/ compass^ | X | | | | | | | | | | X | | | |
| Bronze metallurgy | X | | | | | | | | | | X | | | |
| Kingdoms (Asokan India, Ch'in Dynasty, Athens)* | X | | | | | | | | | | | | | |
| Christianity | X | | | | X | | | | | | | | | |
| Gunpowder | | | | | X | | | | | | | | | |
| Mayan and Byzantine civilizations | X | | | | | | | | | | | | | |
| Pulley | | | | | | | | | | | X | | | |
| World exploration/ Migrations | | X | | X | | | | | | | | | | |
| Three-masted ship | | | | | | | | | | | X | | | |
| Water mills! | | | | | | | | | | | X | | | X |
| Feudalism* | | | | X | | | | | | | | X | | |
| Market economy^ | | | | X | X | | | | | | | X | X | X |
| Rennaisance | X | | | | X | | | | | | | | | |
| Printing^ | | | | | | X | | | | | X | | | |
| Industrial Revolution (mechanization)! | | X | | X | X | | | | | | | X | | X |
| Steam* | | | | X | | | | | | | X | | | X |
| Modern science/ technology^ | X | | X | X | X | | | | | | | X | | |
| Democratic state | | | | | X | | | | | | | X | | |
| Factory systems^ | | | | | | | | | | | X | | | X |
| Railways^ | | | | X | | | | | | | X | | | X |
| Electrification! | | | | | | | | | | | X | | | X |
| International companies* | | | | X | | | | | | | | | | X |
| Telegraph/ Telephone^ | | | | X | | | | | | | | | | X |
| Motorization (internal combustion engine)! | | | | | | | | | | | X | | | X |
| Welfare state | | | | | | | | | | | | X | | |
| Mass education | | | | | | | | | | | | X | | |
| Multi-national agencies* | | | | X | | | | | | | | X | | X |

| AUTHOR / EVENT | Sagan | Barrow / Silk | Chaisson | Christian | Modis | Coren | Spier | Maynard Smith/ Szathmary | Barbieri | Klein | Lipsey et al | Sanderson | Johnson/ Earle | Freeman |
|---|---|---|---|---|---|---|---|---|---|---|---|---|---|---|
| Computing^ | | | | | | X | | | | | | | | X |
| Nuclear energy! | | | | X | X | | | | | | | | | |
| Globalization* | X | | | X | | | | | | | | X | | X |
| Internet* | | | | | X | | | | | | | | | X |
| Electronics^ | | | | X | | | | | | | | | | |
| Space exploration | X | | | X | | | | | | | | | | |

Table 5: List of Cosmic Outcomes.
The original title by Aunger is "Candidate Events in Big History".
! energy innovation
* organisational novelty
^ development in control
See References in (Aunger 2007a) for bibliographic details concerning the sources, and many more details regarding the selection of events.

However, we can already anticipate a fallacy lurking when considering a large list of cosmic outcomes. Similarly to Uzan's remark for the space of possible universes, we can note that the more cosmic outcomes we have, the more unlikely they will seem. The extreme case is to consider one single object as a cosmic outcome. For example, in intelligent design discussions, they consider a complex object such as a living organism or an airplane and try to assess the likelihood that it arose by chance. Of course this will be very unlikely! Additionally, as Dawkins (2008) argues, natural selection would still constitute a much better candidate explanation than design. A scientist will look for possible mechanisms and theories which can explain the emergence of complexity. The *a posteriori* probability of a single object isolated from its evolutionary or human context is of weak scientific interest.

To avoid such an error, we need to advance *theoretical reasons* to select certain cosmic outcomes and not others. This is rarely attempted. Most authors propose an arbitrary list without strong theoretical justification. Ellis, Kirchner and Stoeger did not justify their choice of distinguishing parameters; although it is clear that they included a lot of cosmological parameters necessary for their subsequent study of universes with different geometries.

A promising approach to select cosmic outcomes is to focus on thermodynamics. Indeed, all systems need to process energy, which is therefore a universal concept, applicable from the beginning of the universe to our energy hungry technological society. Robert Aunger (2007a; 2007b) built on a thermodynamical theory to select cosmic outcomes, *non-equilibrium steady-state transitions*. Each transition involves first an energy innovation, then a structural adjustment and finally a new control mechanism. He thus constructed a consistent selection of cosmic outcomes and evolutionary transitions (see Table 6).

| TRANSITION | ASPECT | NOVELTY | FUNCTION | DATE* | PLACE |
|---|---|---|---|---|---|
| Electron (Atomic) Transition | *energy* | electron capture by nuclei | neutralize atomic charge, separate matter and energy | 13.6997 billion | 'our' universe |
| | *organisation* | atoms (hydrogen, helium) | electrically neutral and hence complex, stable matter | 13.6997 billion | 'our' universe |
| | *control* | electro-magnetic forces | nucleus/electron structural mediation | 13.6997 billion | 'our' universe |
| Fusion (Stellar) Transition | *energy* | Proton-Proton reaction | ignition of proto-stars into stars | 13.5 billion | 'our' universe |
| | *organisation* | first generation stars | first large-scale structures | 13.5 billion | 'our' universe |
| | *control* | gravity vs gas pressure | debris removal and star shaping | 13.5 billion | 'our' universe |
| CNO (Planetary) Transition | *energy* | Carbon/Oxygen/ Nitrogen cycle | higher temperature fusion than Proton-Proton reaction (using heavier elements) | 13.3 billion (?) | 'our' universe |
| | *organisation* | solar systems with planets | first large-scale, hierarchically-structured clusters of matter | 13.3 billion (?) | 'our' universe |
| | *control* | gravitational torques vs solar wind | control of inter-planetary relations (migration) | 13.3 billion (?) | 'our' universe |
| Quasar (Galaxy) Transition | *energy* | quasars/black holes | high-intensity electromagnetic radiation | 13 billion | 'our' universe |
| | *organisation* | galaxies | super-scale, hierarchical structures (star clusters) | 12 billion (?) | 'our' universe |
| | *control* | gravity vs interstellar wind (produced by quasar) | force operational over long distances, galaxy shaping | 13 billion | 'our' universe |
| Metabolic (Cell) Transition | *energy* | 'metabolism' (e.g., photo-synthesis) | a chemical reaction transition that produces energy in a form harnessed by other processes | 3.75 billion | Planet Earth; deep sea vents? |
| | *organisation* | cell | protected micro-environment | 3.6 billion | ocean? |
| | *control* | genetic code (e.g., RNA, DNA) | self-catalyzing, durable inter-generational information storage system | 3.25 billion (?) | ocean? |
| Organelle (Complex Cell) Transition | *energy* | mitochondria (animals), chloroplasts (plants), lipids | use of free oxygen and photons as energy-rich source of 'food' | 2 billion (?) | ocean? |
| | *organisation* | eukaryote | nested protective envelopes (cell nucleus) | 1.75 billion | ocean? |
| | *control* | splicing codes (e.g. transfer RNA); genetic recombination [sex] | intracellular communication; division of labour and controlled trait recombination | 1.75 billion (?) | ocean? |
| Secondary Aerobic Reactions (Multi-Cell) Transition | *energy* | complex aerobic reaction cycles | (collection of) cells with improved long-term energy throughput and management | 700 million (?) | ocean? |
| | *organisation* | multi-cellular organism; sexual reproduction | greater variety of genotypes and phenotypes, including specialist tissues and organs | 650 million | ocean? |
| | *control* | pattern codes; neuronal networks (brains) | chemical messengers and regulatory signals for determining body plans; intra-generational memory system (learning) | 550 million | ocean? |
| Tool (Multi-Organism) Transition | *energy* | cooperative foraging; (learned) tool use and manufacture | extraction of inaccessible foods | 25 million | land |
| | *organisation* | parental unit cluster ('family') | loosely-linked group sharing kinship | 20 million (?) | land |
| | *control* | call system [e.g., birdsong] | coordination of groups through conventional signaling system | 15 million (?) | land |
| Fire (Band) Transition | *energy* | fire use [early *Homo*] | increased diet breadth and efficiency of consumption (cooked tubers) | 2.5 million | Africa |

| TRANSITION | ASPECT | NOVELTY | FUNCTION | DATE* | PLACE |
|---|---|---|---|---|---|
| | *organisation* | band | egalitarian collection of intermarrying parental clusters | 2 million (?) | Africa |
| | *control* | cultural traditions/ norms | use of arbitrary signs, symbols or behaviours to identify group membership | 1.5 million (?) | Africa |
| Multi-Tool (Tribal) Transition | *energy* | tool kits (wooden spears etc.) [*Homo heidel-bergensis*] | increased diet breadth (new prey species) | 500,000 | Africa |
| | *organisation* | tribe | large-scale affiliation, sharing common ancestry and culture with neither formalized nor permanent leadership | 400,000 (?) | Africa |
| | *control* | grammatical language ; abstractly decorated tools | sophisticated inter-personal information transmission system; aesthetics | 300,000 (?) | Africa |
| Compound Tool (Cultural) Transition | *energy* | horticulture; compound tools (e.g., bow-and-arrow), tools for making tools (e.g., burins) [*Homo sapiens*] | increased diet breadth (prey killed at distance); reduced variance in dietary intake | 50,000 | Africa |
| | *organisation* | 'Big Man' society | large-scale group with political leadership role based on personal ability (first division of labour) | 40,000 (?) | Africa, Europe |
| | *control* | iconic representation (cave art); common mythology (e.g., Venus figurines) | simple extrasomatic (environmental) memory system | 30,000 | Africa, Europe |
| Agricultural (Chiefdom) Transition | *energy* | cultural symbiosis (animal domestication/ plant cultivation); metallurgy; irrigation | increased regularity of dietary intake (domesticated species); stronger tools; increased ecological capacity | 10,000 | Middle East, Central America, Asia |
| | *organisation* | chiefdom/ city-state | institutionalized leadership with power to collect, store, and distribute surplus resources | 7,500 | Middle East, Central America, Asia |
| | *control* | symbolic representation (cuneiform writing, alphabet); legal system; mathematics; money | sophisticated extra-somatic memory; regulation of social relations on principles other than kinship; system for managing technical information; coordination of market exchange | 5,000 | Middle East, Central America, Asia |
| Machine *(Second Agricultural) Transition* | *energy* | watermill/ windmill; Medieval "agriculture transition" | muscles largely replaced as energy source; higher productivity levels | 1,000 | Europe/ China |
| | *organisation* | autocratic (feudal) state | centrally directed super-institution | 800 | Europe/ China |
| | *control* | "measuring instruments" (mechanical clock, astrolabe); printing press; science | "active" information processing by artifacts; widespread dissemination of information; organized knowledge acquisition | 500 | Europe/ China |
| Steam (Industrial) Transition | *energy* | steam | more efficient, portable power source | 250 | Europe, USA |
| | *organisation* | democratic nation state, corporation | increased institutional stability (through legitimation), financial risk reduction(limited liability) | 250 | Europe, USA |
| | *control* | canal, road and rail systems | greatly reduced transport costs | 200 | Europe, USA |
| Electricity (Cartel) Transition | *energy* | electricity | systematic urban infrastructure for power generation and distribution | 150 | Europe, USA |
| | *organisation* | international cartels; 'Taylorist' shop floors; | collaborative international production; scientific | 150 | Europe, USA |

| TRANSITION | ASPECT | NOVELTY | FUNCTION | DATE* | PLACE |
|---|---|---|---|---|---|
| | | industrial research laboratories | management of production | | |
| | *control* | telegraph/ telephone; bureaucracy; advertising | information transmission at a distance; rapid circulation of people; continuity management within state; mass manipulation of consumer motivation | 120 | Europe, USA |
| Engine (Multinational) Transition | *energy* | oil/internal combustion engine | efficient, portable power | 90 | USA, Europe |
| | *organisation* | multinational agency (e.g., UN); multi-national corporation (e.g., Standard Oil, Microsoft) | supra-national government; international capitalism | 70 | USA, Europe |
| | *control* | mass media (radio, TV); mass production; computer | fast, broad-scale information dissemination; standardized production; universal computation | 60 | USA, Europe |
| Nuclear (Globalization) Transition | *energy* | nuclear reactors | controlled atomic fission | 40 | USA, Europe |
| | *organisation* | global markets; World Wide Web | significant international capital flows and investment; globalized social and economic network | 30 | USA, Europe |
| | *control* | digital media | unified representation system for multi-modal data | 15 | USA, Europe |

Table 6: A theoretical selection of cosmic outcomes, by Robert Aunger (2007b).
Original title: “First known instances of non-equilibrium steady-state transitions”
* Dates in years before present; present taken to be year 2000. Uncertain dates include ‘(?)’.”

Which cosmic outcomes are contingent and evolutionary? Which ones are necessary and developmental? Are there attractors in the dynamic of cosmic evolutionary development? To answer these issues, we need to explore the *robustness* of the emergence of complexity. Stated otherwise, if we would re-run the tape of the universe, would galaxies, stars, biology and technology arise again and again? The straightforward way to answer those question, in parallel to a theoretical rational like Aunger's, is indeed to re-run the tape of the universe. Let us now examine how we can conceptualize and operationalize this idea.

### 6.3.3 Robustness in Cosmic Evolution

*what would remain the same if the tape of life were replayed?*
Stephen Jay Gould (1990)

*what would remain the same if the tape of the universe were replayed?*
Paraphrasing Gould's question to the universe (Vidal 2008b).

Answering this later question, Paul Davies (1998, 317) wrote that if “the universe were re-run a second time, there would be no solar system, no Earth and no people. But the emergence of life and consciousness somewhere and somewhen in the cosmos is, I believe, assured by the underlying laws of nature.” This claim, as Davies acknowledges, is only an informed intuition. How can we test this intuition or different ones scientifically? This is the issue of the *robustness of the emergence of complexity in cosmic evolution.*

A first analyze of the tape metaphor shows its limits. Indeed, if the tape and its player were perfect, we should get exactly the same results when re-running the tape. So, the thought experiment would be trivial. Yet if our universe self-constructs, one question is whether small fluctuations, chance events, noise or random perturbations would lead to slightly different outcomes, or very different ones. This makes the issue of robustness in cosmic evolution highly stimulating.

This issue is very hard to tackle because of a great weakness of cosmology as a science: it has only one object of study, our unique universe. More precisely, we can distinguish two fundamental limitations that Ellis (2007a, 1216) pointed out:

> **Thesis A1: The universe itself cannot be subjected to physical experimentation**. *We cannot re-run the universe with the same or altered conditions to see what would happen if they were different, so we cannot carry out scientific experiments on the universe itself.* Furthermore,
>
> **Thesis A2: The universe cannot be observationally compared with other universes.** *We cannot compare the universe with any similar object, nor can we test our hypotheses about it by observations determining statistical properties of a known class of physically existing universes.*

Our thesis is that it is possible to address those limitations and the issue of robustness by running computer simulations of our universe. It is important to note that if we replay the tape of *our* universe, we don't aim to actually explore the full space of possible universes. Here, we only aim to assess the robustness of the emergence of the different cosmic outcomes. We thus vary *only* nondeterministic dynamical parameters we discussed above (quantum mechanical effect, random perturbations, nonlinear chaotic dynamics, etc.). An open question is also how we vary the random parameters. How often? How strong is the variation? Various distributions can be tested, from gaussian distributions, where most random variations are of an average strength, few are weak or strong; to power-law distributions, where there are few very strong variations, some medium variations, and most of the time weak random variations.

Because of the inclusion of such unpredictable parameters, it makes sense to re-run the same universe simulation. By running a multitude of times the simulation, it is possible to make statistics on the emergence of complexity. An even more straightforward way to make such statistics would be to drastically intensify astrobiology –the search for extraterrestrials. If or when we will find extraterrestrials, we would be able to progressively study the “natural re-runs” of complexity. In particular, searching for extraterrestrials more complex than us would force us to break with the implicit anthropocentric assumption that life and humans on Earth are the highest development in cosmic evolution. Such search invites us to think about the existence of higher cosmic outcomes, and this opens the way to test our theories of the general evolution of cosmic complexity. We will discuss more in depth the search for extraterrestrials in Chapter 9.

An example of such ambitious simulations of *our* universe are the Millenium run simulations (Springel et al. 2005; Boylan-Kolchin et al. 2009; Guo et al. 2011). The authors studied the formation, evolution and clustering of galaxies and quasars within the standard (or concordance) model of cosmology. Although they did not run the same simulation in its full complexity many times, the volume space explored is large enough to extract meaningful statistical properties on the evolution of the distribution of matter.

Replaying the tape of our entire universe is still a much more ambitious project, which at present remains unrealistic. We should remain aware that our current models and their associated free parameters are most likely not the ultimate ones. Of course, new theories need to be developed to know what the key parameters of our universe are. In the meantime, a way to progress is to break down the issue into smaller solvable problems. For example, if we want to tackle the robustness up to the emergence of intelligent life, we can write a generalized Drake equation (Ellis, Kirchner, and Stoeger 2004, 925) that I call the *Cosmic Evolution Equation*:

$$\mathrm{N}_{life}(m^*) = N_g \cdot N_S \cdot f_S \cdot f_p \cdot n_e \cdot f_l \cdot f_i$$

where $\mathrm{N}_{life}(m^*)$ is the number of planets with intelligent life in our particular universe $m^*$; and

(1) $N_g$ is the number of galaxies in the model
(2) $N_S$ is the average number of stars per galaxy
(3) $f_S$ is the fraction of stars suitable for life
(4) $f_p$ is the fraction of such stars with planetary systems
(5) $n_e$ is the mean number of planets which are suitable habitats for life
(6) $f_l$ is the fraction of planets on which life originates
(7) $f_i$ is the fraction of life bearing planets with intelligent life.

There are many implicit assumptions in such a framework, for example that life-supporting stars will be Sun-like; or that life starts necessarily on planets and not on more exotic places. We also implicitly assume that the parameters are independent. To deal with dependent parameters, one would need to introduce a bayesian probability framework. Additionally, we may have clear definitions of what stars or galaxies are, but the issues of defining higher cosmic outcomes such as life or intelligence remain of huge scientific debate.

The factors $N_g$ and $N_S$ can nowadays be estimated, while the recent explosion of exoplanets discoveries is allowing us to estimate more and more precisely the factors $f_S \cdot f_p \cdot n_e$. However, huge uncertainties remain regarding the last two factors $f_l \cdot f_i$.

The main benefit of such a framework –whether we consider these seven factors to be most relevant or others– is that we can in a first approximation estimate the factors independently. Additionally, *the more we progress in our knowledge of the universe, the larger the distance between factors we can assess*. For example, assessing the number of planets with intelligent life knowing only the number of galaxies seems very hard. But shorter distances between factors are easier to assess. For example, Miller's (1953) famous experiment tells us that the probability to have amino acids out of a primordial soup and some energy source is high. Which is indeed an important insight to evaluate $n_e \cdot f_l$.

Let us now imagine that we run multiple times a model of our entire universe $m^*$. We would be able to interpret the results of the multiple runs of the simulations as a set of *virtual* universes. We would end up with a distribution function $f(m^*)$ combining the probability distributions obtained for each factor. However, we need to further specify a *possibility space*, which in this case is $\mathcal{M}^*$ resulting from the variation of random parameters only; and a measure π* on $\mathcal{M}^*$ . Such a virtual ensemble of simulated universes *V* would thus be defined as:

$$V = \{\mathcal{M}^*, \pi^*, f(m^*)\}$$

The number of planets with intelligent life would then be:

$$N_{life}(m^*) = \int f(m^*) \cdot N_g \cdot N_S \cdot f_S \cdot f_p \cdot n_e \cdot f_l \cdot f_i \cdot \pi^*$$

Note that the integral is necessary to normalize the result according to the measure π* and distribution function *f*(*m**). There are important and subtle issues to make this normalization sound and possible (see again Ellis, Kirchner, and Stoeger 2004).

Let us give some more concrete possible results such simulation studies would bring. We might conclude that our universe is robust in galaxy-formation, i.e. most simulation runs lead to galaxy formation. However, it might turn out that our universe is not robust for intelligent life, i.e. most simulations *do not* lead to the emergence of intelligent life.

We can now take a fresh eye on our question: are cosmic outcomes necessary or contingent? We can define a cosmic outcome as *necessary* if it appears again and again as we re-run the same universe, as *contingent* otherwise. For example, let us take the DNA code in biology: is it necessary that there is a unique DNA code for terrestrial or extraterrestrial biology? In biology, this general question is a dispute regarding on whether evolution is convergent (see e.g. Gould 1990; and Conway-Morris 2003 for contrary viewpoints and arguments). But here we generalize this issue to the whole of cosmic evolution. An other example in economy: is it a necessity in civilizational development that monetary systems converge to a common currency?

We can also compare the cosmic outcome selections. On the one hand we would have the ones resulting from "simulation experiments" (see e.g. Kleijnen et al. 2005 for a discussion of this idea of simulation experiments); and on the other hand the theoretical considerations (such as Aunger's). *Simulation experiments* in cosmology can play the role that *empirical experiments* play in other sciences. This approach can be called "cosmology in silico" or "computational cosmology". In fact, these endeavors are already developing quickly, as illustrated by the Virgo Consortium for Cosmological Supercomputer Simulations.

We have just begun to explore how robust the emergence of complexity in our universe is. If we want to understand it better, we need to perform computer simulations and use existing conceptual, mathematical and statistical tools to design the simulation experiments and to assess the results.

However interesting and important this enterprise is, it does not tackle the fine-tuning issue. Indeed, in studying the robustness of our universe, we try to understand the emergence of complexity in *our universe*, whereas to address fine-tuning we must study the place of our particular universe in *the space of possible universes*.

### 6.3.4 Artificial Cosmogenesis or the study of alternative Cosmic Evolutions

*Now, we create a considerable problem. For we are tempted to make statements of comparative reference regarding the properties of our observable Universe with respect to the alternative universes we can imagine possessing different values of their foundamental constants. But there is only one Universe; where do we find the other possible universes against which to compare our own in order to decide how fortunate it is that all these remarkable coincidences that are necessary for our own evolution actually exist?*

(Barrow and Tipler 1986, 6)

*you might end up having a future subject which is "comparative universality" – we have all these laws for the universe that cannot be eliminated as ours and you study them, you talk about them, you compare them, this could be a future subject. Students would be required to pass exams on their ten possible favorite universes ...*

Gregory Chaitin in (Chaitin et al. 2011, 339)

This first quote by Barrow and Tipler summarizes the core problem of fine-tuning. The second quote by Chaitin illustrates a core idea towards its resolution. With the robustness issue, we have focused on *our* universe. To assess to what extent our universe is fine-tuned, we must study the place of our universe in the *space of possible universes*. We call this space the *virtual multiverse*.

Let us first call a *fecund universe* a universe generating at least as much complexity as our own. *Are fecund universes rare or common in the multiverse?* This is the core issue of fine-tuning. Answering it demands to explore this virtual multiverse. Milan Ćirković (2009) and I both converged on this conclusion. Ćirković used the metaphor of sailing the archipelago of possible universes; I proposed to perform simulations of possible universes, an endeavor called *Artificial Cosmogenesis* (or ACosm, see Vidal 2008b; 2010a; 2012a). Such simulations would enable us not only to understand our own universe (with "real-world modelling", or processes-as-we-know-them) but also other *possible* universes (with "artificial-world modelling", or processes-as-they-could-be). We thus need to develop methods, concepts and simulation tools to explore the space of possible universes (the "cosmic landscape" as Leonard Susskin (2005) calls it in the framework of string theory). In (Vidal 2008b), I proposed to call this new field of research *Artificial Cosmogenesis* because it sets forth a "general cosmology", in analogy with Artificial Life which appeared with the help of computer simulations to enquiry about a "general biology". However, recent work on the EvoGrid[10] simulation project suggests that the growth of complexity is more likely to remain open-ended if stochastic, non-deterministic processing is used at the bottom, instead of deterministic rules, like in ALife. So I do not mean to strictly limit ACosm to deterministic rules.

10 http://www.evogrid.org

Now that we have a framework to define possible universes, we will need to generalize the "Cosmic Evolution Equation" we used to assess the robustness of our universe to explore not only our universe $m^*$, but also all universes $m$ element of the wider class of possible universes $\mathcal{M}$. This constitutes a rigorous approach to assess how fine-tuned our universe is. However, it is important to understand that the results of such studies would not *ipso facto* provide an *explanation* of fine-tuning. Only if it turns out that our kind of complex universe is common, then an explanation of fine-tuning would be a principle of *fecundity*: "there is no fine-tuning, because intelligent life of some form will emerge under extremely varied circumstances" (Tegmark et al. 2006, 4).

We saw that most fine-tuning arguments just change one parameter at a time and conclude that the resulting universe is not fit for developing complexity. We saw it leads to the "one-factor-at-a-time" paradox. What if we would change *several* parameters at the same time? Systematically exploring the multiple variation of parameters seems like a very cumbersome enterprise. As Gribbin and Rees wrote (1991, 269):

> If we modify the value of one of the fundamental constants, something invariably goes wrong, leading to a universe that is inhospitable to life as we know it. When we adjust a second constant in an attempt to fix the problem(s), the result, generally, is to create three new problems for every one that we "solve". The conditions in our universe really do seem to be uniquely suitable for life forms like ourselves, and perhaps even for any form of organic complexity.

Back in 1991, it indeed seemed very difficult to explore and find alternative universes. However, a way to overcome this problem is to use *computer simulations* to test systematical modifications of parameters' values. In varying just one parameter, parameter sensitivity arguments have only begun to explore possible universes, like a baby wetting his toes for the first time on the seashore. Surely, we had to start somewhere. But it is truly a tiny exploration. Furthermore, maybe there is a deep link between the different constants and physical laws, such that it makes no sense to change just one parameter at a time. Changing a parameter would automatically perturb other parameters (see Bradford 2011, 1581). Fortunately, recent research have gone much further than these one-parameter variations.

What happens when we vary multiple parameters? Let us first generalize the Cosmic Evolution Equation, which this time includes other possible cosmic *evolutions* –notice the plural! Let us imagine that we run multiple times simulations of different models of universes $m$. We interpret the results of the multiple runs of the simulations as a set of *virtual* universes. We end up with a distribution function $f(m)$ combining the probability distributions obtained for each factor of the CEE. Another way to choose distribution functions was developed by Schmidhuber (2000). He used theoretical computer science to study and choose distribution functions for possible universes.

The *possibility space* in this generalized case is the huge $\mathcal{M}$ resulting from the definition of possible universes; and we add a measure $\pi$ on $\mathcal{M}$. The resulting ensemble of simulated universes $E$ would thus be defined as:

$$E = \{\mathcal{M}, \pi, f(m)\}$$

The number of planets with intelligent life would then be:

$$N_{life}(m) = \int f(m) \cdot N_g \cdot N_S \cdot f_S \cdot f_p \cdot n_e \cdot f_l \cdot f_i \cdot \pi$$

We are now talking about cosmic outcomes in other universes. The topic becomes quite speculative, because it is not clear at all *which* cosmic outcomes are the most relevant to assess. The factors in the equation above might be totally irrelevant. What if other possible universes do not generate objects like galaxies, stars and planets, but completely different kinds of complex structures? Nothing *that we know* may evolve anymore... but other things might! We now see the fundamental importance to define cosmic outcomes and the emergence of complexity in a very general manner, so they can also apply to other possible universes. Bradford (2011) proposed such a framework when he analyzed sequences of entropy reduction. Aunger's (2007a) systems theoretical approach in terms of energy innovation, organization and control is also a higher-level approach. Valentin Turchin (1977) also proposed a cybernetic theory of complexity transitions with the central concept of *metasystem transition*. Mark Bedau (2009) did also articulate this issue in details in the context of artificial life. Theoretical computer science measures such as algorithmic complexity (see e.g. Li and Vitányi 1997) or logical depth (C. H. Bennett 1988b) are also precious tools to assess the complexity of systems in a universal manner. But these are just a few examples of frameworks to tackle the general, fascinating and fundamental problems of the evolution and measure of complexity.

We already saw that higher outcomes $f_l \cdot f_i$ are harder to assess. This is precisely where computer simulations can be very helpful. Typically, there are so many local interactions in the evolution of complex organisms that it is very hard to analyze them analytically with a deterministic or Newtonian approach. For example, there is not one single equation which allows to predict the development of an embryo.

Let us now outline some remarkable alternative complex universes that researchers recently studied. Gordon McCabe studied variations on the standard model of particles, by changing the geometrical structure of space-time. The result is not the end of any complexity, but just the beginning of a new set of elementary particles. McCabe (2007, 2:38) elaborates:

> Universes of a different dimension and/or geometrical signature, will possess a different local symmetry group, and will therefore possess different sets of possible elementary particles. Moreover, even universes of the same dimension and geometrical signature will not necessarily possess the same sets of possible particles. To reiterate, the dimension and geometrical signature merely determines the largest possible local symmetry group, and universes with different gauge fields, and different couplings between the gauge fields and matter fields, will possess different local symmetry groups, and, perforce, will possess different sets of possible particles.

It thus seems that we can vary basic physics parameters without compromising all kinds of cosmic evolutions. Who knows what kind of complexity can emerge from this new set of particles?

As an illustration of their framework to define the multiverse, Ellis, Kirchner and Stoeger (2004) did examine some parameter variations in Friedmann-Lemaître-Robertson-Walker (FLRW) models. They found life-allowing *regions* in a phase space described by the evolution of FLRW models. The fact that they found *regions* and not a *single point* in the phase space shows that there is room for some variation. So it seems that we can vary fundamental geometrical cosmological parameters without precluding the apparition of life.

Harnik, Kribs and Perez (2006) constructed a universe without electroweak interactions called the Weakless Universe. They show that by adjusting standard model and cosmological parameters, they are able to obtain

> a universe that is remarkably similar to our own. This "Weakless Universe" has big-bang nucleosynthesis, structure formation, star formation, stellar burning with a wide range of timescales, stellar nucleosynthesis up to iron and slightly beyond, and mechanisms to disperse heavy elements through type Ia supernovae and stellar mergers.

This is a truly remarkable result because the cosmic outcomes are numerous, relatively high and non trivial. Three factors in the CEE are addressed more or less directly: $N_g \cdot N_S \cdot f_S$. Maybe strong living creatures could live in the weakless universe? This remains to be investigated.

Anthony Aguire (2001) did study a class of cosmological models "in which some or all of the cosmological parameters differ by orders of magnitude from the values they assume in the standard hot big-bang cosmology, without precluding in any obvious way the existence of intelligent life." This study also shows that it is possible to vary parameters widely without obviously harming the emergence of complexity as we know it.

Robert Jaffe, Alejandro Jenkins and Itamar Kimchi (2008) pursued a detailed study of possible universes with modified quark masses. They define *congenial* worlds the ones in which the quark masses allow organic chemistry. Again, they found comfortable regions of congeniality.

Fred C. Adams (2008) has conducted a parametric survey of stellar stability. He found that a wide region of the parameter space provides stellar objects with nuclear fusion. He concludes that the "set of parameters necessary to support stars are not particularly rare."

An early attempt to explore alternative universes with simulations has been proposed by Victor Stenger (1995; 2000). He has performed a remarkable simulation of possible universes. He considers four fundamental constants, the strength of electromagnetism $\alpha$; the strong nuclear force $\alpha_s$, and the masses of the electron and the proton. He then analysed "100 universes in which the values of the four parameters were generated randomly from a range five orders of magnitude above to five orders of magnitude below their values in our universe, that is, over a total range of ten orders of magnitude" (Stenger 2000). The distribution of stellar lifetimes in those universes shows that most universes have stars that live long enough to allow stellar evolution and heavy elements nucleosynthesis. Stengers' initial motivation was to refute fine-tuning arguments, which is why he ironically baptised his simulation "MonkeyGod". The implicit idea is that even a stupid monkey playing with cosmic parameters can create as much complexity as God.

In conclusion, other possible universes are also fine-tuned for some sort of complexity! Those remarkable studies show consistently that *alternative complex universes are possible*. One might object that such explorations do not yet assess the higher complexity factors in the CEE. They do not answer the following key questions: would other interesting complex structures like planetary systems, life, intelligence or technology evolve in those other universes? However, these are only early attempts in conceptualizing and simulating other possible universes, and the enterprise is certainly worth pursuing. The fine-tuning issue could then be seriously tackled, because we would know more and more precisely the likelihood of having our universe as it is, by comparing it to other possible universes. Such pioneering studies are just a beginning, and future studies will certainly discover more and more complex alternative universes.

### 6.3.5 Summary

Let us now summarize the three main steps necessary to assess how fine-tuned our universe is.

(1) *Define* a space $\mathcal{M}$ of possible universes
(2) *Explore* this space
(3) *Assess* the place of our universe in $\mathcal{M}$

Let us review step (1). Our analysis on the historical trends of free parameters in Chapter 5 invites us to start by a *weak variation*, i.e. varying free parameters in physical and cosmological models. Why not vary the laws of physics themselves? It seems a very difficult enterprise, because we do not even know how to make them vary (see Vaas 1998)! It can also be dubious to do so, since we saw that the distinction between laws and initial or boundary conditions is fuzzy in cosmology (Ellis 2007a).

This suggestion to focus on weak variation makes most sense for the following reasons. First, it is concrete and operational, and has a clear meaning with well established physics. Second, we assume supernatural miracles happening in the middle of cosmic evolution to be –by definition– impossible. We assume there is a consistency and continuity in cosmic evolution. We hypothesize that higher level parameters are ultimately reducible to these physics and cosmic ones. The emergent higher levels occur naturalistically. Of course, this remains to be shown, and for practical purposes we might include as given such higher level parameters in our studies and simulations. New levels of emergence, new levels of complexity did historically emerge from lower levels, even if complicated top-down causation occurs (see e.g. Ellis 2008). Take for example an economic law like the law of supply and demand. It did not and could not exist before the apparition of organized human civilizations. It emerged out of such new organizations. It seems that what we call "natural laws" could be the result of more and more regular interactions. For example, as the universe cooled down, new organizations did emerge. Again, it is clear that a few billion years ago, there was no economic laws.

We also need to be more specific to apply probabilities to the ensemble of possible universes, and avoid probabilistic fallacies we described. For example, we must decide, arbitrarily or not, parameter's upper and lower bounds. This is necessary for all practical purposes, because we can't explore the parameter space of all parameters varying from $-\infty$ to $+\infty$. We thus need to define the maximum deviation allowed for each parameter.

We must beware of the one-factor-at-a-time paradox. We must define a probability measure on the parameter space. I refer the reader to (Koperski 2005) and (Ellis, Kirchner, and Stoeger 2004) for detailed arguments that measure-theoretical grounds can be specified to assess fine-tuning. It is also crucial to define *cosmic outcomes* to specify the object of fine-tuning we aim to address. Do we talk about fine-tuning for nucleosynthesis? atoms? Stars? Life? Intelligence? Or a more general complexity emergence?

Step (2) requires to explore this space. The simplest exploration is to re-run the tape of *our* universe. But this only tackles the issue of the *robustness* of the universe. If we want to address the fine-tuning issue we must also run and re-run tapes of *other possible universes*. This will bring us insights into how our and other universes are parameter sensitive, and how they generate complex outcomes. Although we always need good theoretical models to start with, it is necessary to use computer simulations to explore the huge parameter landscape we are talking about. That landscape is not just very big, but really huge. Because we do not want to and do not have the resources to explore the space blindly, it also makes most sense to use simulations to test particular hypotheses and theories. As we will see in Chapter 8, if we consider Lee Smolin's cosmological natural selection theory, and find alternative universes with more black holes (the cosmic outcome under consideration) by tweaking parameters, it is a way to falsify the theory.

The last step (3) is to compare the distribution functions of the cosmic outcomes of interests to the space $\mathcal{M}$ of possible universes. In other words, we assess the probability to find a universe with outcome O. Note that this is the crucial difference between tackling the robustness and the fine-tuning issue. In robustness analysis, we run multiple times the *same* universe simulation changing only the random dynamical parameters. We compare multiple runs of the same universe. In fine-tuning analysis, we run multiple *different* universe simulations, changing a wide number of parameters. We compare our universe to the set of possible universes. How typical or atypical is our universe in the space of possible universes? The results of such simulation experiments will enable us to answer this question. Ideally, we will be in a position to assess the likelihood or unlikelihood of complexity emergence in the space of possible universes. Even better than assessing specific cosmic outcomes, which might bias us to a universe-centric perspective, we can aim to assess the probability to find universes which display open-ended evolutionary mechanisms leading to ever increasingly complex cosmic outcomes.

To the traditionally trained cosmologist, this enterprise might seem totally unconventional. And it is. This is why I chose to give it a new name, *artificial cosmogenesis*. It might also seem out of reach given the computational resources needed. As we will see in Chapter 7, the sheer computational resources grow more than exponentially, this allows us in principle to increase accordingly the complexity and richness of our computer simulations.

However, computer simulations, even given huge computational resources, is not the whole of the story. It is only a necessary condition to solve these issues. We still need theories and fundamental understanding to set up simulations, know what to look for in the resulting data, and interpret the results.

Additionally, engineers and professional model makers have developed a wide variety of tools to test multiple variables, rarely used in cosmological contexts. Let us just mention of few of them. A starting point is to use the tools of global sensitivity analysis (see e.g. A. Saltelli, Ratto, and Andres 2008). These include advanced statistical approaches such as latin hypercube sampling, multivariate stratified sampling; Montecarlo simulations for finding dynamic confidence intervals. Systems dynamics and engineering have also many tools to offer such as phase portraits or probabilistic designs. The classic book by John D. Sterman (2000) remains a reference and quite comprehensive introductory book on complex systems modeling and simulations.

Let us now be scrupulous. What is a proof a fine-tuning? Let *n* be the number of free parameters. We have a logical and statistical version of what a proof of fine-tuning would be:

> **Logical proof of fine-tuning:** *If you vary one parameter, there exists no possible universe generating outcome O by adjusting the* (*n* - 1) *other parameters.*
> Which is equivalent to:
> *if you vary one parameter, there is no way whatsoever that all other possible universes can generate outcome O.*
>
> **Probabilistic proof of fine-tuning:** *If you vary one parameter, adjusting the* (*n* - 1) *other parameters won't make outcome O more likely.*
> Which is equivalent to:
> *if you vary one parameter, there is no way whatsoever that all other possible universes can generate outcome O with a higher probability.*

In sum, you need to have explored the relevant parameter space of possible universes to make serious claims about fine-tuning. Pretty hard to prove! This is even harder for outcomes as advanced as life or intelligence.

Our conclusion is *not* the one of Stenger (2011, 22) that “the universe looks just like it should if it were not fine-tuned for humanity”. However, note that this quote reflects Stenger's opinion. Stenger focuses on showing that theological arguments from design using fine-tuning are inconclusive. Criticizing and debunking arguments using fine-tuning claims is one thing, proving that our universe is not fine-tuned for intelligent life another.

Our conclusion is *not* either the opposite one of Barnes (2012, 561):

> We conclude that the universe is fine-tuned for the existence of life. Of all the ways that the laws of nature, constants of physics and initial conditions of the universe could have been, only a very small subset permits the existence of intelligent life.

Indeed, it is a much stronger claim compared to the parameter space he reviews, and the cosmic outcomes he considers. This is despite the fact that Barnes (2012, 531) has well understood that the scientific way to progress on fine-tuning is by exploring the space of alternative possible universes, i.e. to engage in artificial cosmogenesis:

> What is the evidence that FT is true? We would like to have meticulously examined every possible universe and determined whether any form of life evolves. Sadly, this is currently beyond our abilities.

So, Barnes' conclusion hinges on the assumption that we have explored enough of the space of possible universes to make meaningful extrapolations. He continues:

> Instead, we rely on simplified models and more general arguments to step out into possible-physics-space. If the set of life-permitting universes is small amongst the universes that we have been able to explore, then we can reasonably infer that it is unlikely that the trend will be miraculously reversed just beyond the horizon of our knowledge.

Given parameter sensitivity, fecund universes are likely to be rare, so this intuition may well be correct, but should certainly not considered as a proof, given the tiny exploration of the space that humanity did up to now.

Our conclusion is rather that *fine-tuning for life or intelligence remains a conjecture*. Like in mathematics, we have strong reasons to believe the conjecture is true, but a proof is out of reach and certainly requires a huge amount of work. As a matter of fact, the challenge of simulating possible universes and comparing them is overwhelming. This is why the concept of the cosmic outcome is so important to ease the process. Indeed, we can break down the problem and progress by tackling higher and higher outcomes, with more and more connection between outcomes. We don't need nor can assess all outcomes at once. As our understanding, modeling capacities and computational resources increase, we can be more ambitious in simulating more and more as well as higher and higher outcomes in cosmic evolution. Tomorrow's cosmology is not restricted to empirical observations or highly theoretical models. It is also the science of simulating and experimenting with alternative universes.

Surprisingly, as we will see in Chapter 8, this quest to understand our origins through the examination of parameter sensitivity and fine-tuning might give us clues, insights and tools to successfully deal with our far-future. We have now framed how to evaluate fine-tuning seriously, but still haven't seen possible explanations. We will now review classical explanations, and then in Chapter 8 two evolutionary explanations: Cosmological Natural Selection and Cosmological Artificial Selection.

## 6.4 Classical Fine-tuning Explanations

*God or Multiverse?*

*We don't need these hypotheses to* **study** *fine-tuning...*
*...but we may need them to* **explain** *fine-tuning.*

I went into much detail to define the fine-tuning issue precisely to allow its scientific study. I was addressing the question "Is there fine-tuning?". My conclusion is that at most it is a conjecture, a proof being largely out of reach. More specifically, we saw that parameter sensitivity is different from fine-tuning and is in fact not surprising. Rick Bradford has shown that parameter sensitivity can be expected in any complex universe. Therefore, it is not surprising that varying one parameter would not lead to an interesting complex universe. But the problem of why we happen to inhabit a parameter sensitive universe in the space of possible universes remains open. Suppose the fine-tuning conjecture is true. How can we explain fine-tuning? What are the possible explanations?

I now briefly review the main "classical" explanations: *skepticism*, *necessity*, *fecundity*, *god-of-the-gaps*, *chance-of-the-gaps*, *WAP-of-the-gaps*, *multiverse* and *design*. I will treat two additional evolutionary explanation in Chapter 8, *cosmological natural selection* and *cosmological artificial selection*. Once we analyze these possible solutions, we naturally tend to jump to the next arduous issues: the metaphysical ones (see e.g. Parfit 1998). Indeed, whatever X your answer to fine-tuning, you can ask: "Why is there X in the first place? Where did X come from?" To tackle those more speculative and metaphysical questions, we will in Chapter 8 not consider fine-tuning in isolation, but in connection with another great issue: our far-future.

#### 6.4.1 Skepticism

Skeptics challenge that fine-tuning would be an issue at all. As we saw, it stems from difficulties in defining probabilities rigorously (e.g. Colyvan, Garfield, and Priest 2005; McGrew, McGrew, and Vestrup 2001). However, we saw that with additional measure-theoretical considerations and hypotheses, authors are able to overcome this obstacle (see e.g. Koperski 2005; Ellis, Kirchner, and Stoeger 2004).

Furthermore, if something like ACosm is pursued, probabilities can be derived from datasets resulting from simulation runs. They can then be treated with statistical tools as any other data in science. There would be no room for such arguments anymore.

It is always possible to remain skeptic on any issue. Critical reasoning (philosophical dimension 4) is helpful for stimulating dialectical (philosophical dimension 5) discussions, but misses the connection with real world problems, and thus the first-order dimensions of our worldviews. The skeptical position is of second-order nature, and thus does not even consider the issue worth an explanation. In sum, it is not even a position. I included it here however for comprehensiveness.

#### 6.4.2 Necessity

The core of this position is that the key issue is free parameters, not fine-tuning. Since a lot of parameters have been captured in the past, there is no reason this trend would end. There is a mathematical or physical necessity in the fundamental parameters and laws that is still to be discovered. This question can be tackled by a scientific theory that could derive the values of these parameters. Therefore no parameters would be left to be fine-tuned, and the fine-tuning issue would be solved. It would lead to a "theory of everything" or ultimate explanation. That position is logically possible... but remains to be proven.

This reasoning is the implicit position of most physicists because it seems to be the most scientific approach. However, it is unlikely that *all* constants will be deduced from a theory. We need input from the physical reality at a certain point. The underlying assumption is that "nature is as it is because this is the only possible nature consistent with itself" (Chew 1968). On this camp, researchers might argue that it is certainly wiser to interpret the size of the possible multiverse as the size of our ignorance.

Furthermore, even assuming it is possible, we saw many problems of a zero parameters theory as we examined Tegmark's proposal of the mathematical universe (see section 5.3 The Mathematical Universe, 94). It leads to a strange situation where physics and mathematics merge.

Additionally, there is a tension between the logic of atomism and the desire of unification (Smolin 1997, 59). On the one hand, elementary particles have absolute, independent properties. On the other hand, unification requires that all particles and forces are manifestations of a single principle. Can we reconcile the two? Smolin argues that such a desire for unification should stop. We also saw that such a position is almost mystical with an underlying "dream of a final theory" (see section 5.3 The Mathematical Universe, 94).

On top of that, in the last decades, the most active attempts to find a unique theory were within the framework of string theory. But instead of producing a unique theory, it lead to a huge landscape of possible theories. Then we face the problem of choosing a theory! As Smolin (2006, 159) writes, if "an attempt to construct a unique theory of nature leads instead to $10^{500}$ theories, that approach has been reduced to absurdity."

But let's assume that those obstacles and objections can be overcome. What if we would end up with only one possible universe? Ellis (2007a, 1254) argues:

> Uniqueness of fundamental physics resolves the parameter freedom only at the expense of creating an even deeper mystery, with no way of resolution apparent. In effect, the nature of the unified fundamental force would be pre-ordained to allow, or even encourage, the existence of life; but there would be no apparent reason why this should be so.

This is correct, but that time, the mystery would be *metaphysical*: why something rather than nothing? Ellis is concerned about a genuine ultimate theory. However, we must acknowledge that such a hypothetical theory would solve the fine-tuning issue, albeit not the metaphysical one. Again, the metaphysical mystery of existence would remain whatever our explanation of fine-tuning (see also Vidal 2012a).

Actively pursuing the reduction of free parameters is the approach most faithful to physics. However, it is unlikely to succeed to its upmost extrapolation up to zero parameter.

In fact there are two opposite views on the matter: *necessity* and *fecundity*. Martin Gardner (1986) reminds us that the former is the view "that only one kind of universe is possible – the one we know. This was skillfully defended by the Harvard chemist Lawrence Henderson". The opposite fecundity view comes from Leibniz, who "argued exactly the opposite. He believed an infinity of universes are logically possible and that God selected the one he liked best." Let us consider the fecundity response more closely... and without God.

### 6.4.3 Fecundity

The principle of fecundity is that "intelligent life of some form will emerge under extremely varied circumstances" (Tegmark et al. 2006, 4). Let us imagine that we have enough data from simulations or unforeseen theoretical argument to precisely assess the fecundity explanation. Two results are logically possible. Either we find that fecund universes are common, or they are not. If they are common, no more fine-tuning explanation is needed, and it would indeed solve the fine-tuning issue. But the problem then shifts to a metaphysical one. Why would there be so many fecund universes in the first place? Is it due to a special universe generating mechanism? If so, doesn't it require further explanation?

On the contrary, if we find that fecund universes are rare, the fine-tuning issue is not solved, and this needs further explanation. The limit case is reached if it turns out that there is only one possible universe, ours. Then we are back to the *necessity* explanation. Our universe wouldn't be fine-tuned, because there was simply no other option. Note that the study of alternative complex universes (outlined in section 6.3.4 Artificial Cosmogenesis or the study of alternative Cosmic Evolutions, p137) already rules out this possibility for cosmic outcomes as high as star formation and heavy element dispersion. The question remains open only for higher outcomes such as life or intelligence.

As a warning, it is easy to manipulate the results of arguments regarding fecundity, simply by changing the definition of possible universes. If we *a priori* consider a small set $\mathcal{M}$ of possible universes, the fecundity principle may easily be satisfied. If on the other hand, we consider a very widely defined $\mathcal{M}$, then our kind of universe may be rarer in this space $\mathcal{M}$. To gain further insights, we will need to explore the space $\mathcal{M}$, both with theoretical considerations and computer simulations.

### 6.4.4 God-of-the-gaps

God as both the creator and the fine-tuner of the universe is a popular view. It is in a certain sense an elegant solution, because it solves both the existence problem (*metaphysical*) and the fine-tuning problem at the same time. One could even argue that God as the great architect has had a very positive influence on the development of science. In this view of God, popular in medieval times, God doesn't intervene. She just gives a blueprint, which is very different from a controlling God. Such a conception allows the development of science, because science consists in finding God's laws, which are nothing else than physical laws. Doing science was then equivalent to seek and worship God.

God as a key feature in a comprehensive theological worldview is a very consistent position. As we saw in details in Part I, a theological worldview scores high on many subjective and intersubjective criteria. In this philosophical –and not theological– work, I will not explore further this option because the problems of fine-tuning and creation shift to theology.

Obviously, as a rational explanation for fine-tuning, God suffers from the God-of-the-gaps critique. Whatever issue remains unexplained in our worldview, fine-tuning or other, God can fill the gap.

### 6.4.5 Chance-of-the-gaps

The laws of physics and the parameters of our universe just happened by chance. There is nothing more to explain here. Scientifically, this is an empty statement, which is not better than the God-of-the-gaps. As Ellis (2007a, 1258) wrote, "from this viewpoint there is really no difference between design and chance, for they have not been shown to lead to different physical predictions." It is logically similar to the God (or design) explanation, and so it is better qualified as a chance-of-the-gaps explanation.

### 6.4.6 WAP-of-the-gaps

According to this view, we first start by claiming a *universality principle* according to which "all that can happen, happens" (Rees 1999; 2001; Tegmark 2004).

Then we add a selection effect such as the weak anthropic principle (WAP), which bias us to think that our universe is special. But this is not the case. Our universe as a whole might seem fine-tuned, but the multiverse is not.

Such a reasoning can indeed make our existence less astonishing. But at what price! Obviously, such an option assumes the multiverse hypothesis, but does not support it (N. Bostrom 2002, 23). It does not *causally* explain existence of observers. As Swinburne (1990, 171) argues, the "laws of nature and boundary conditions cause our existence, we do not cause theirs." Again, selection effects don't causally explain things. The ambiguous expression "anthropic principle" is often used as a combination of both the weak anthropic principle as a selection effect and an actually existing multiverse. The two combined would indeed provide an explanation of fine-tuning.

In addition to the strong ontological claim, multiverse theories are notoriously hard if not impossible to test. As long as multiverse theories remain untestable, such theorizing can solve fine-tuning... or any other problem. There may be more subtle multiverse proposals to solve the fine-tuning issue, that we will now consider.

To sum up, from a scientific point of view, appealing to God, Chance or a universality principle with the WAP as a selection effect work everywhere and explain everything, anything ... and nothing. Those options are last-aid kits of intellectual survival.

### 6.4.7 Multiverse

*A multiverse must be a physically realized multiverse and not a hypothetical or conceptual one if it is to have genuine explanatory power.*

(Ellis 2007a, 1260)

Virtual universes generated by computer simulations we discussed do not constitute any *explanation* of fine-tuning. They provide a rigorous framework only for the *study* of fine-tuning. The bottom line is that we can study possible universes without making claims about their existence or nonexistence. It suffice to study a virtual multiverse, not an actual one. This is why so far we have not taken a position regarding the actual existence of a multiverse.

Now, how can we use the multiverse hypothesis to explain fine-tuning? As Ellis writes above, we must first assume an actually existing multiverse. Can we use the multiverse hypothesis in a more subtle way than the WAP-of-the-gaps? Yes we can! For example, the theory of eternal inflation (Linde 1990) allows to calculate the free parameters of other possible universes. The universe generation mechanism constrains the parameter variation in an actual multiverse space $\mathcal{M}$. Other multiverse options might also be considered (see e.g. Tegmark 2004; Smolin 2012).

The key and arduous issue is then to find smart ways to test such bold hypotheses. The testability is certainly very limited, maybe only available with multi-domain universes, and not to actually disconnected space-time regions (Ellis 2007b). Additionally, as with other options, the ultimate explanation is not addressed. Where does the multiverse comes from? How and why do the various universes exist and reproduce? From what? From which principles? This shifts the fine-tuning issue to the problem of the existence of a universe-generation mechanism.

### 6.4.8 Design

In this stance, our universe is the result of a purposeful design. Importantly, the agent responsible for design is not necessarily a supernatural theistic God (N. Bostrom 2002, 11–12). Of course, this is the most typical option, which leads to the God explanation. But logically, it needs not to be.

Can we imagine *naturalistic* and not supernatural design? Certainly. Systems exhibiting features of design such as teleology or parameter sensitivity can emerge both from *natural selection* processes, for example a living organism, (see Dawkins 1996) or from artificial processes such as engineering (e.g. a watch). A middle way between the two is found in *artificial selection*, where intentional breeding allows to select traits over others.

What if the same applies to the universe as a whole? In Chapter 8, we will see in more detail two such options, *Cosmological Natural Selection* and *Cosmological Artificial Selection*. As usual, we have to recognize lurking metaphysical issues. How could we prove naturalistic design? How scientific is this approach? Aren't supernatural and natural design explanations banned from science? Can we avoid the "design-of-the-gaps" trap?

## 6.5 Conclusion

We saw logical, probabilistic and physical fallacies around fine-tuning which are rarely avoided. The most striking logical fallacy is to vary just one parameter at a time, which drastically reduces the exploration of the parameter space. The use of probabilities is an essential step in formulating the fine-tuning argument. But it needs careful additional hypotheses to make it cogent (e.g. choices of probability theory framework, of a measure, of a distribution function). The physics underlying fine-tuning arguments is often wrong. Many fallacies occur and reoccur.

Many authors mix fine-tuning with other issues, such as free parameters, parameter sensitivity, metaphysical issues, anthropic principles, observation selection effects, teleology or God's existence.

Given the difficulty to assess fine-tuning *for life* seriously, it is too early to say whether there is fine-tuning of physics and cosmic parameters. However, we saw several examples of alternative possible universes which remarkably show complex cosmic outcomes and dynamics. The emergence of complexity is not our universe's sole prerogative. But regarding complex cosmic outcomes such as intelligent life, fine-tuning remains a conjecture.

Let us summarize three steps to clarify fine-tuning, once we have debunked fallacies and distinguished it from other issues. First, the scientific approach to tackle the source of fine-tuning is to capture free parameters. The role of science is to propose new theories which need less free parameters. This approach remains the most robust and promising. It should stay the absolute priority. Second, if free parameters remain, don't use explanations of the gaps! God-of-the-gaps, chance-of-the-gaps or WAP-of-the-gaps are last-aid intellectual survival kits. It is important not to confuse selection effects with causal explanations providing mechanisms to explain fine tuning. Third, we assess the typicality of our universe. To achieve this, it demands to explore the parameter space of possible universes. This includes not only simulating our universe (to assess its robustness) but also other possible universes (to assess its fine-tuning). The exploration of virtual universes will allow us to assess how likely the production of astrophysical bodies, biochemistry, life, intelligence or humanity are.

This exploration constitutes the *Artificial Cosmogenesis* field of research, which has already begun. But can we really hope to play or replay the tape of universes? How on Earth can we do that? To answer this, we now come to the third part of this thesis, a journey into our far future that we begin with an exploration of the future of scientific simulations.

# Part III - Our Future in the Universe

*the laws of the universe have engineered*
*their own comprehension*

(Davies 1999, 146)

We are now entering the third part of our journey. Our challenge is to answer the age-old question “Where are we going?” (Chapters 7, 8 and 9) and “What is good and what is evil?” (Chapter 10). As usual, we ask these questions in a cosmological context, with a maximal stretch in space and time scope. So we are concerned about where we are going in the extremely far future. The most extreme point is the “last” point, which leads us naturally to the field of *eschatology*. The word comes from the Greek *eskhatos* (last), and *logos* (doctrine or theory). The word “eschatology” introduces a bias similar to the word “ultimate”. We saw the later comes from *ultimare*, which means to “come to an end”, while the former is the doctrine of last things. Taken literally, those words *a priori* rule out cyclical views of the universe, where there is no past or future end point.

Our discussion of cognitive attractors to understand the origin (Chapter 4) also applies to the future. Let us give brief hints of why it is the case. What are our cognitive attractors for the future? What do we expect? In the optimistic case, civilizations long for a kind of immortality (see Chapter 10). It can take the form of a point (e.g. heaven) or a cycle (e.g. with reincarnation cycles). We find the idea of multiple reincarnation or resurrection not only in Eastern philosophies, but also in contemporary Christian theology (see e.g. Hick 1976; Steinhart 2008). From a physical perspective, as we will see below, the attractor point can be a cosmic doom scenario, where everything is stabilized to a uniform and lifeless state. Many cosmological models are cyclical, such as Tolman's (1934), the phoenix universe (Dicke and Peebles 1979), the famous chaotic inflation (Linde 1990), Smolin's (1992) Cosmological Natural Selection or Penrose's (2011) recent conformal cyclical cosmology. We also mentioned in Chapter 4 the model of continuous creation of Hoyle and Narlikar, which is more associated with a line rather than a point or a cycle. There are also cosmological models which include a role for intelligent life that we will examine in more details in Chapter 8. As with the origin, there are psychological difficulties to accept cycles in the future. But this is a problem only if one holds a point-like metaphysics which requires an ultimate beginning or end.

Importantly, different eschatologies focus on different “ends”. Do we mean the end of a human life? Of humanity? or of all things? Not surprisingly, we focus here on the end of all things, since we want to avoid anthropo- or species- centrisms. Inspired by Freitas (1979, chap. 22.4.4) I distinguish four kinds of eschatologies: *eternalistic*, *historical*, *naturalistic* and *physical*.

Eternalistic eschatologies see time as an endless cycle of eternal recurrence. We already discussed eternal return (section 4.3.4 Objections against Cycles, p81) and its many associated difficulties. For Stoics of the ancient Greek and for Indian thinkers, time moves in cycles. Buddhists and hindus believe in cycles of creations and destruction.

Historical eschatologies are grounded in linear time. Western traditions such as Christianity, Judaism or Islam believe in a beginning and an end of time. Even the title of this thesis “The Beginning and the End” indeed shows such a western bias. In Plato's *Republic*, death is accompanied with a judgment, where the immortal soul is given rewards and punishments before choosing the condition of its next existence. The nature of this new existence is a topic of theology.

Naturalistic eschatologies emphasize harmony with nature. Goodness is seen as unity with nature, while wrongness is seen as alienation from nature. Interestingly, the main concern is to be in harmony with nature here and now, and not the prospect of a far future state after death. This is illustrated in Taoism, where there is virtually no interest for the beginning or the end of the universe (Ward 2002, 235).

Eschatology has most often been discussed within religious doctrines. But this needs not to be. Milan Ćirković (2003) wrote a review of *scientific* approaches to this topic, *physical eschatology*, gathering more than 200 references. Still, if we consider from a symmetry argument that past and future studies should have equal importance in treatment, there are surprisingly few studies about the far future universe compared to studies of the early universe. Ćirković argued that physical eschatology is a part of science:

> Since the laws of physics do not distinguish between past and future (with minor and poorly understood exceptions in the field of particle physics), we do not have a *prima facie* reason for preferring “classical” cosmology to physical eschatology in the theoretical domain.

It is correct that most physical laws are time reversible... with the notable exception of thermodynamics. The reconciliation of classical, relativistic and quantum theories with thermodynamics is a major challenge of contemporary physics. It gives rise to thorny issues such as the arrow of time. Inspired by a science fiction novel by Gregory Benford (1978), Freitas proposed that an advanced civilization would focus on a “thermodynamic eschatology” striving to halt or reverse entropic processes in this universe. However, it is not necessarily the best strategy to fight frontally against such a widely confirmed physical law as the second law. It leads to the dream of a perpetual motion (Ord-Hume 1977). On the contrary, it was by accepting the laws of conservation of energy that engineers were able to design more and more efficient engines and machines.

Thanks to modern theoretical physics and astrophysics, many of the questions regarding the ultimate fate of the universe are thus now quantitatively addressed within the field of physical eschatology. What will happen to the Earth and the Sun in the far future? The story depicted by modern science is a gloomy one. In about 6 billion years, it will be the end of our solar system, with our Sun turning into a red giant star, making the surface of Earth much too hot for the continuation of life as we know it. The solution then appears to be easy: migration. However, even if life would colonize other solar systems, there will be a progressive end of all stars in galaxies. Once stars have converted the available supply of hydrogen into heavier elements, new star formation will come to an end in the galaxy. In fact, the problem is worse. It is estimated that even very massive objects such as black holes will evaporate in about $10^{98}$ years (F. C. Adams and Laughlin 1997).

Generally, the main lesson of physical eschatology is that in the long-term the universe will irreversibly go towards a state of maximum entropy, or *heat death*. This is a consequence of the second law of thermodynamics, one of the most general laws of physics. It was first applied to the universe as a whole by Hermann von Helmholtz in 1854. Since this heat death discovery, a widely spread pessimistic worldview sees the existence of humanity as purposeless and accidental in the universe (e.g. B. Russell 1923; S. Weinberg 1993b). The fatalism implied by this worldview may lead to a loss of the meaning of life.

Modern cosmology shows that there are some other models of the end of the universe (such as Big Bounce, Big Rip, Big Crunch..., see (Vaas 2006) for an up-to-date review). The point is that none of them allows the possibility of the indefinite continuation of life as we know it. If any of the cosmic doom scenario is correct, it means that the indefinite continuation of life is impossible in this universe. What is the point of living in a universe doomed to annihilation? Ultimately, why should we try to solve mundane challenges of our daily lives and societies, if we can not even imagine a promising future for life in the universe? If we recognize this fundamental issue, then we should certainly do something to avoid it, and thus try to change the future of the universe.

On the other hand, there is an apparent paradox between this increase of entropy and the accelerating *complexity increase* (e.g. Livio 2000; Chaisson 2001; Morowitz 2002; Kurzweil 2006). Chaisson (2001) showed with a thermodynamical analysis that this paradox can be solved. Indeed, it is the *expansion of the universe itself* which allows a decrease of entropy *locally*, while there remains an increase in entropy *globally*. Yet, which of the two trends will turn out to be dominant in the long term remains unsettled.

So, where are we going ultimately, towards entropy increase or complexity increase?

A few authors have proposed some speculative solutions, but we will see that they are insufficient because none of them presently allows the infinite continuation of intelligent life. I will instead argue that intelligent civilization could in the far future make a new universe (Chapter 8). Although it sounds like a surprising proposition resembling science fiction scenarios, I will consider it within a philosophical agenda, seriously and carefully addressing and replying to objections and comparing it with alternative options.

What are the limits of this complexity increase? The best way to test ideas regarding the future of complexity increase is to search for *advanced* extraterrestrial intelligence (ETI, see Chapter 9). For example, if we speculate that ETIs would play snooker with stars, we should see much more star collisions, and stars moving on unpredictable trajectories. This is not the case, so probably we don't miss the cosmic snooker. Philosophically, the search for extraterrestrials is also a topic of fundamental importance. Are we alone in the universe? If you want to know who you are as a human being, you must compare yourself to or interact with others. The same holds for humanity as a whole. If we want to understand our place in the universe, there is no other option than searching, and maybe finding and comparing ourselves to other life forms.

It is easy to predict that as humanity becomes more and more connected, more and more in harmony as a globalized entity, the question "who am I in the universe?" will only be more pressing. I chose to entitle this Part III "*Our* future in the universe" and not "*The* future of the universe", because we are immersed and implied in the universe. We are not merely spectators, we are actors in the great show of cosmic evolution. And to act in the universe, we need values and an ethics. This is why Chapter 10 is key to this Part III.

It should be noted that the proposition of involving intelligent life into the fate of the universe is at odds with traditional science. Indeed, the modern scientific worldview has often suggested that the emergence of intelligence was an accident in a universe that is completely indifferent to human concerns, goals, and values (e.g. S. Weinberg 1993b; Stenger 2007). I thus challenge this proposition, and another one that is commonly associated with it, which says that intelligent civilization can not have a significant influence on cosmic evolution. A recurring objection to the importance of this topic is that it is too far away to be worth of consideration. If this is your opinion, I invite you to read Story 7, “ Global warming and universal cooling”.

You are at a luxurious cocktail party, and you meet a rich CEO of a coal-fired power station. You attempt to start a conversation:

- You: What do you think about global warming?
- Him: It's not my job to think about it.
- You: There is a wide agreement that coal burning largely contributes to global warming.
- Him: So what?
- You: I am wondering how you can be morally comfortable about the major impact of your industry on the planet. Don't you care?
- Him: Not at all. Global warming is not my problem, it is for next generations. I focus on providing energy to people, and incidentally, making money.
- You: But... who is going to tackle the problem if each generation reasons like you?

What would be your opinion about this CEO? At the very least, you'll think that he's not highly morally developed because he doesn't care about future generations. I do have the same impression when friends or colleagues quickly dismiss cosmic doom scenarios as “not their problem” because too far in the future. The fact that so many people on Earth care about global warming is a truly extraordinary shift of mindset. It means that we have extended our worldviews to future generations and to planet Earth as a whole. As we extend further our worldview, why should we stop our sphere of compassion to the boundaries of our tiny planet? I predict that future generations will more and more care not only about global warming but also about the heat death of the universe, which is actually a universal cooling – or any cosmic doom scenario which could threaten the survival of life in the universe.

Story 7: Global warming and universal cooling.
This story is fictional.

Those issues are properly ethical. *What is good and what is evil?* I address this issue in Chapter 10. What lessons can we learn from our cosmological worldview? What does this cosmological perspective imply for our actions and values here and now? What is our purpose in the universe? What are the ultimate goals or results that intelligence is seeking in the universe? What is the meaning of life in this cosmological perspective? At first sight, evolutionary reasoning tells us that survival is the ultimate value. But survival of what? And for how long? Can we aim as high as immortality? If so, which kind of immortality can we long for?

Guessing the future is a notoriously perilous enterprise. Part III will thus be more speculative. Chapter 8 explores a philosophical extension of Lee Smolin's Cosmological Natural Selection (CNS), which in itself is already often considered as speculative. In Chapter 9, I will explore heuristics to search ETI, and even argue that we may well already have found ETIs much more advanced than us.

Speculating on those issues can easily lead us very far. For this reason, we need to have clear ideas on *why* we speculate. As I wrote in the preface, I distinguish three kinds of speculations to navigate into their variety (Vidal 2012a):

4. **Scientific**: a speculation is scientific if we have strong reasons to think that future observations or experimentations will corroborate or refute it.
5. **Philosophical**: a speculation is philosophical if it extrapolates from scientific knowledge and philosophical principles to answer some fundamental philosophical problems.
6. **Fictional**: a speculation is fictional if it extends beyond scientific and philosophical speculations.

Let us sum up how we will tackle those numerous issues. **Chapter 7** explores the future of scientific simulations, and its implications for our understanding of the universe. As Paul Davies wrote above, through cosmic evolution and the emergence of humans and science, "the laws of the universe have engineered their own comprehension". This self-awareness is dazzling. Where will this trend lead to the limit? We further develop and motivate the already mentioned extension of Artificial Life to *Artificial Cosmogenesis*.

**Chapter 8** presents cosmological selections. First, *Cosmological Natural Selection* (CNS), a remarkable theory applying ideas of evolutionary biology in cosmology. Second, *Cosmological Artificial Selection*, a philosophical scenario extending CNS with a philosophical agenda. I discuss the history of both and formulate critical objections. Through the CAS scenario, we will see surprising links between the study of the origin and our possible future of the universe!

**Chapter 9** addresses the big question: "Are we alone in the universe?" In Chapter 6, we unveiled the real difficulties behind the fine-tuning issue, namely, that we only know one universe, ours. To progress we concluded that we must study *other* possible universes. In a similar fashion, we know in details only one instance of the development of higher complexity: life on Earth. If we want to understand the general future of complexity in the universe, it would be invaluable to find *other* cosmic intelligences. The discipline of *astrobiology* promises to fill this gap, and we will put the foundations of the more specific sub-discipline of *high energy astrobiology*. I propose an observable hypothesis to test the existence of very advanced civilizations and suggest that we may already have observed them. Since they actively feed on stars, I call them *starivores*.

In **Chapter 10** we go back to Earth with cosmological wisdom. I explore foundations for ethics on a cosmological scale, a *cosmological ethics*. I build on thermodynamical, evolutionary and developmental values and argue that the ultimate good is the infinite continuation of the evolutionary process. I apply the framework to an ubiquitous longing of humanity: the will to immortality. I survey five kinds of immortalities and how they relate to the definition and development of the self.

To sum up, Chapter 7 emphasizes the importance of scientific simulations for understanding our universe; Chapter 8 presents cosmological selections and in particular CAS, a philosophical scenario linking the beginning and the end of the universe, providing a meaning of life for intelligence in the universe; Chapter 9 attempts to understand who we are in the universe, by searching and maybe finding extraterrestrials far more advanced than us; Chapter 10 outlines basic principles of a universal cosmological ethics, illustrated with five different kinds of immortalities.

# *CHAPTER 7 - The Future of Scientific Simulations*

**Abstract**: This Chapter explores the far future of scientific simulations. It is argued that the path towards a simulation of an entire universe is an expected outcome of our scientific simulation endeavors. I describe the exponential increase of computing resources in a cosmological context, using Chaisson's energy rate density complexity metric. Simulating the open-ended rise of levels of complexity in the physical, biological and cultural realms is the challenge of simulating an entire universe. However, such an effort will require to bridge the gaps in our knowledge of cosmic evolution, which is necessary to replay the tape of our and other possible universes. We elaborate the distinction between *real-world* and *artificial-world* modelling, the latter being at the heart of the artificial life and artificial cosmogenesis philosophy. We critically discuss the idea that we may be living in a computer simulation.

*I see no reason (in the really distant future) why all model-making,*
*and in this I include all "law-discovering",*
*should not be carried on, as a routine matter, inside computers.*

(Ashby 1981a, 353)

How important are scientific simulations if an intelligent civilization is to have influence on future cosmic evolution? It is increasingly clear that simulations and computing resources are becoming main tools of scientific activity. More concretely, at a smaller scale than the universe, we have already begun to produce, run and even play with artificial worlds, with the practice of computer simulations. In particular, efforts in the Artificial Life (ALife) research field have shown that it is possible to create digital worlds with their own rules, depicting agents evolving in a complex manner. We will see that such simulations promise to become more and more complex and elaborated in the future.

I argue that the path towards a simulation of an entire universe is an expected outcome of our scientific simulation endeavors. I will examine later in Chapter 8 how such a simulation could be realized (instantiated, made physical) and bypass cosmic doom scenarios, expected to happen at some future time.

## 7.1 Towards a Simulation of an Entire Universe

Simulating the general emergence of complexity is a long-term desiderata of our scientific simulation endeavors. The challenge is to simulate open-ended evolution not only in biology, but also to link it to physical evolution and to cultural evolution. To the limit, a simulation of an entire universe would allow us to probe what would happen if we would "replay the tape of the universe". In this Chapter, I will discuss in more depth the status and potential usefulness of a simulation of an entire universe, making a distinction between *real-world* and *artificial-world* modeling (see section 7.5 Real-World and Artificial-World Modelling, p160). I outline and criticize the “simulation hypothesis”, according to which our universe may be just a simulation (in section 7.6 The Simulation Hypothesis, p161). Let us first summarize the historical trend of exponential increase of computing resources.

## 7.2 Increase of Computing Resources

We may note two important transitions in the history of human culture. The first is the externalization of memory through the invention of writing. This allowed an accurate reproduction and a safeguard for knowledge. Indeed, knowledge could easily be lost and distorted in an oral tradition. The second is the externalization of computation through the invention of computing devices. The general purpose computer was inspired by the work of Church, Gödel, Kleene and Turing, and its formal specifications constitute the most general computing device (see Davis 2000 for a history of computation). The consequences of this last transition are arguably as significant –if not more– as the invention of writing. In particular, the changes induced by the introduction of computers in scientific inquiry are important, and remain underestimated and understudied (see however Floridi 2003 for a good starting point).

Computing resources have grown exponentially, at least for over a century. There is much literature about this subject (see e.g. Kurzweil 1999; 2006 and references therein). Moore's “law” famously states that the number of transistors doubles every 18 months on a single microprocessor. Exponential increase in processing speed and memory capacity are direct consequences of the law. What are the limits of computer simulations in the future? Although there is no Moore's law for the efficiency of our algorithms, the steady growth in raw computational power provides free “computational energy” to increase the complexity of our models and simulations. This should lead to longer term and more precise predictions. Apart from the computational limitation theorems (uncomputability, the computational version of Gödel's theorem proved by Turing), the only limit to this trend is the physical limit of matter or the universe itself (Bremermann 1982; Lloyd 2000; L. M. Krauss and Starkman 2004). As argued by Lloyd (2000; 2005) and Kurzweil (2006, 362) it should be noted that the ultimate computing device an intelligent civilization could use in the distant future is a maximally dense object, i.e. a black hole.

From a cosmic outlook, Moore's trend is in fact part of a much more general trend which started with the birth of galaxies. Cosmologist and complexity theorist Eric Chaisson proposed a quantitative metric to characterize the dynamic (not structural) complexity of physical, biological and cultural complex systems (Chaisson 2001; 2003). It is the *free energy rate density* (noted $\Phi_M$) which is the rate at which free energy transits in a complex system of a given mass (see Figure 9). Its dimension is energy per time per mass (erg $s^{-1}$ $g^{-1}$). Let us illustrate it with some examples (Chaisson 2003, 96). A star has a value ~1, planets ~$10^2$, plants ~$10^3$, humans ~$10^4$ and their brain ~$10^5$, current microprocessors ~$10^{10}$. According to this metric, complexity has risen at a rate faster than exponential in recent times. Along this complexity increase, there is a tendency to do ever more, requiring ever less energy, time and space; a phenomenon also called *ephemeralization* (Fuller 1969; Heylighen 2007), or “Space-Time Energy Matter” (STEM) compression (Smart 2009). This means that complex systems use increasingly less space and time, and become more dense in energy and matter flows.

In Tomas Ray's (1991) artificial life simulation *Tierra*, digital life competes for CPU time, which is analogous to energy in the organic world. The analogue of memory is the spatial resource. The agents thus compete for fundamental properties of computers (CPU time, memory) analogous to fundamental physical properties of our universe (energy, space). This design is certainly one of the key reasons for the impressive growth of complexity observed in this simulation.

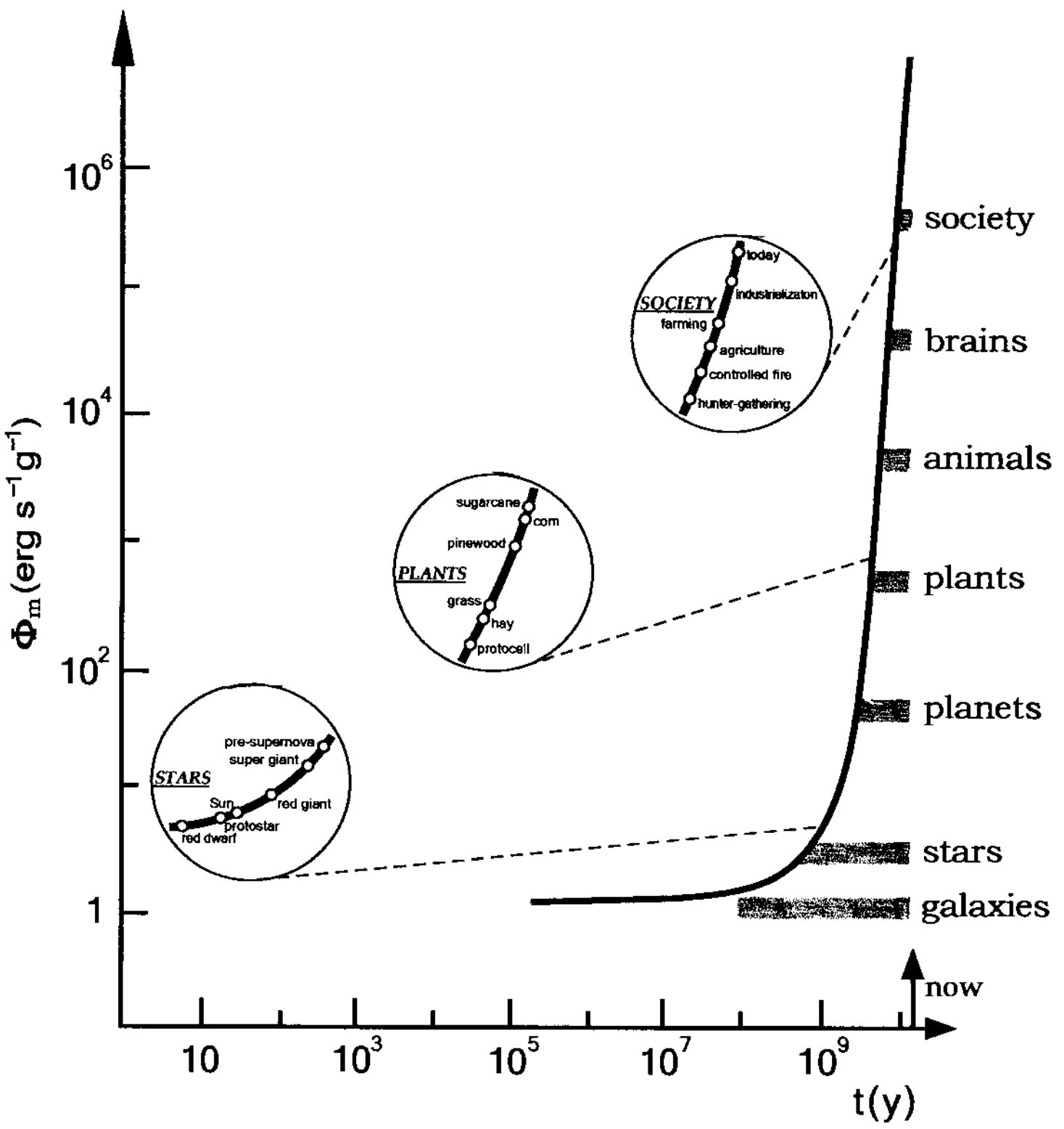

Figure 9 - The original caption is: "The rise of free energy rate density, $\Phi_M$, plotted as histograms starting at those times when various open structures emerged in Nature, has been rapid in the last few billion years, much as expected from both subjective intuition and objective thermodynamics. The solid curve approximates the increase in normalized energy flows best characterizing the order, form and structure for a range of systems throughout the history of the Universe. The circled insets show greater detail of further measurements or calculations of the free energy rate density for three representative systems - stars, plants and society - typifying physical, biological and cultural evolution, respectively. Many more measures are found in Chaisson (2001)." excerpted from (Chaisson 2003, 97). Note that microprocessors are outside the scale of this diagram since they appear at $10^{10}$ on the $\Phi_M$ axis.

## 7.3 Bridging Gaps in Cosmic Evolution

We saw that a metric can be found to compare complex systems traditionally considered as different in nature. This important insight is just a first step towards bridging physical, biological and cultural evolution, and ideally a maximum of cosmic outcomes.

Artificial Life (ALife) is a field of research examining systems related to life, its processes, and its evolution through simulations using either computer models (soft ALife), robotics (strong ALife), or biochemistry (wet ALife). A general challenge for ALife is to obtain an artificial system capable of generating open-ended evolution (M. A. Bedau et al. 2000). Some results have been obtained linking for example the evolution of language with quasi-biological traits (Steels and Belpaeme 2005). Working towards the design of a digital universe simulating the rise of levels of complexity in the physical, biological and cultural realms is the challenge of simulating an entire universe. We saw an important step in this direction (section 6.3.3 Robustness in Cosmic Evolution, p133), with the "Millenium Run" simulation. However, it stays on the physical level, since its starts from the beginning of the universe to generate the large scale structures of the universe.

However, we must acknowledge important difficulties of conceptual, methodological and cultural integration between the different disciplines involved. In such an endeavor, human-made social and academic boundaries between disciplines of knowledge must be overcome. In Part I (section 3.2 Scientific Worldviews, p57), I proposed to construct synthetical scientific worldviews with *systems theory*, *problem solving* and *evolutionary theory* as three generic interdisciplinary approaches. The ideal is to aim at a seamless link between simulations in physics, biology and social sciences (culture). If this would happen, we would have the basic tools to work towards a model and a simulation of the entire universe. In fact the search for such bridges is obviously necessary if we want to tackle such difficult issues as the origin of life, where we aim to explain the emergence of life out of physico-chemical processes.

## 7.4 Replaying the Tape of the Universe

Astronomy, astrophysics and cosmology are empirical, but *not* experimental sciences. It is possible to gather a lot of data about a wide variety of astrophysical systems, but unlike in experimental sciences, we can't design smart experiments to force nature's cosmic outcomes. However, computer simulations are progressively operating a revolution in science. They allow to conduct simulation experiments, even if such virtual experiments are imperfect compared to real experiments.

We saw (section 6.3 The Cosmic Evolution Equation, p122) that such an approach holds promises to tackle the question of the robustness of the emergence of complexity (by replaying the tape of *our* universe) and the fine-tuning issue (by playing and replaying tapes of *alternative* universes).

Let us take seriously the increase of computing resources. The simulation of an entire universe can be seen as perhaps the ultimate challenge of simulations in science. But what kind of simulation would it be? What could it be used for? To answer these questions we will now distinguish between two kinds of modelling.

## 7.5 Real-World and Artificial-World Modelling

A computer simulation can be defined as a model where some aspects of the world are chosen to be modelled and the rest ignored. When we run a simplified model on hardware more computationally efficient than the physical system being modelled, we can run the model faster than the phenomenon modelled, and thus predict our world. The paradigm of Artificial Life (ALife) strongly differs from such traditional modelling, by studying not only "life-as-we-know-it", but also "life-as-it-could-be" (Langton 1992, section 1). We propose to extend this modelling technique to any *process* and not just to life, leading to the more general distinction of *processes-as-we know-them* and *processes-as-they-could-be* (Red'ko 1999). I call the two kinds of modelling respectively *real-world* modelling and *artificial-world* modelling.

*Real-world* modelling is the endeavour to model *processes-as-we-know-them*. This includes traditional scientific modelling, such as models in physics, weather forecast models, but also applied evolutionary models, etc. The goal of such models is to better understand our world, and make predictions about it. For what would a *real-world* simulation of an entire universe be useful? We saw it would allow us to test the robustness of the emergence of complexity. What is more, we could think that it would provide us very good understanding of and predictive power over our world. However, this is not so simple. First, if the simulation is really of the entire universe, it should be "without anything left out", which is a strange situation. Indeed, it would imply that the model (simulation) is as complex as our universe. Such a simulation would thus not provide a way to systematically predict all aspects of our universe, because it would not be possible to run it faster than real physical processes. Another limiting argument is that more computational power does not necessarily mean better predictive abilities. This is pretty clear when considering chaotic systems such as the weather, which rapidly become unpredictable. A simulation still has to be simpler than reality if it is to be of any practical use. This means that in the context of "replaying the tape of our universe", we would still have to investigate a simplified simulation of our universe.

*Artificial-world* modelling is the endeavour to model *processes-as-they-could-be*. The formal fundamental rules of the system (of life in the case of ALife) are sought. The goal of ALife is not to model life exactly as we know it, but to decipher the most simple and general principles underlying life and to implement them in a simulation. With this approach, one can explore new, different life-like systems. Stephen Wolfram (2002) has a similar approach by exploring different rules and initial conditions on cellular automata, and observing the resulting behaviour of the system. It is legitimate to emphasize that this is a "new kind of science". Indeed, this is in sharp contrast with traditional science focusing on modelling or simulating reality. There is thus a creative aspect in the artificial-world modelling, which is why many artists have enthusiastically depicted imaginary ALife worlds. What would an artificial-world simulation of an entire universe be useful for? We would be able not only to "replay the tape of *our* universe", but also to play and replay the tape of *alternative* universes. We saw this endeavor constitutes a research program for tackling the fine-tuning issue in cosmology.

Should this artificial world modelling of an entire universe be interpreted as a *simulation* or as a *realization* (Pattee 1989)? Here we consider the first possibility, with the *simulation hypothesis*; the realization will be considered later in Chapter 8, with the philosophical scenario of Cosmological Artificial Selection.

## 7.6 The Simulation Hypothesis

Let us assume what we have argued in the previous section, i.e. that intelligent life will indeed be able at some point to simulate an entire universe. If such a simulation is purely digital, thus pursuing the research program of soft ALife, this leads to the *simulation hypothesis*, which has two main aspects. First, looking into the future, it means that we would effectively create a whole universe simulation, as has been imagined in science fiction stories and novels such as the ones of Isaac Asimov (1956) or Greg Egan (2002). Very well then! A second possibility is that we ourselves could be part of a simulation run by a superior intelligence (see e.g. N. Bostrom 2003; Barrow 2007b; Martin 2006). Although these scenarios are fascinating, they suffer from two fundamental problems. First, the "hardware problem": on what physical device would such a simulation run? Is there an infinity of simulation levels? Second, such an hypothesis is uninformative. Indeed, following Bateson's (1972) definition of information as "a difference which makes a difference", the simulation hypothesis makes no practical or theoretical difference. Unless we find a "bug" in reality, or a property that could only exist in a simulation and not in reality, this hypothesis seems useless, a mere fictional speculation. A more comprehensive criticism of these discussions can be found in (Polya 2004).

The ontological status of this simulation would be reflected by the states of the hardware running it, whatever the realistic nature of the simulation. From this point of view, we can argue that it remains a *simulation*, and not a *realization* (Harnad 1994). Is there another possibility for realizing a simulation of an entire universe? That is what we will explore in the next chapter.

## Open questions

- The idea of simulating whole universes raises the issue of computational requirements. What kind of computer is needed? What memory? What CPU speed? How many CPUs? Serial or parallel computing? What kind of software would be used? Where does the computer hardware exist? What power does it require?

These questions remain to further explore (see e.g. Bostrom (2003) for discussion and references to some of these issues).

Researchers spend a lot of time-energy writing research projects, which are, due to fierce competition often rejected. In 2010, I wrote a research project called "Big history and our future: extension, evaluation and significance of a universal complexity metric" (Vidal 2010c). The core idea is to update and further research Chaisson's (2001) energy rate density complexity metric. Unfortunately, despite good reviews and good academic support, it was not accepted. Disappointed, I decided to put the proposal on the website of the Evo Devo Universe community[11], in the hope that the research proposal would be taken up further. I was glad that it indeed happened and researchers later contacted John Smart and I to further work on it. Some of the key research questions are:

- Can we complete the curve to understand the past (early universe) and the future (acceleration of technology)?
    - What happens if we use this metric for the early universe? Indeed, Chaisson starts the measures with the birth of galaxies. But what about the energy rate density at the big bang era?
    - How well does the free energy rate density curve fits with Moore's law? If we extrapolate those two trends, do they have any functional relation?

- Has our universe a dynamics of reproduction? John Smart (2009) suggested so, but this needs further research. He hypothesized the general trend through time would be a "U-shaped" curve, with extremely high energy densities at the big-bang era, decreasing as the universe cools down. The energy rate density then starts to grow more than exponentially at the apparition of life. This energetic pattern bears a striking resemblance with the energy pattern of the life cycle of a living organism from its birth through its maturity stage.

On 3rd february 2010, I gave a seminar alone at my office about "Cosmic Embryogenesis" (Vidal 2010d)... Sad schizophrenia? No, cutting edge technology. Indeed, colleagues of mine were attending from several corners of the world. How was it possible? Richard Gordon did set up in the virtual environment "Second Life" an online "International Embryo Physics Course" to stimulate efforts to reverse

11 http://evodevouniverse.com/wiki/Research_on_free_energy_rate_density

engineering the process of embryological development. This is how the "Embryo Physics" seminar room looks like (figure 10):

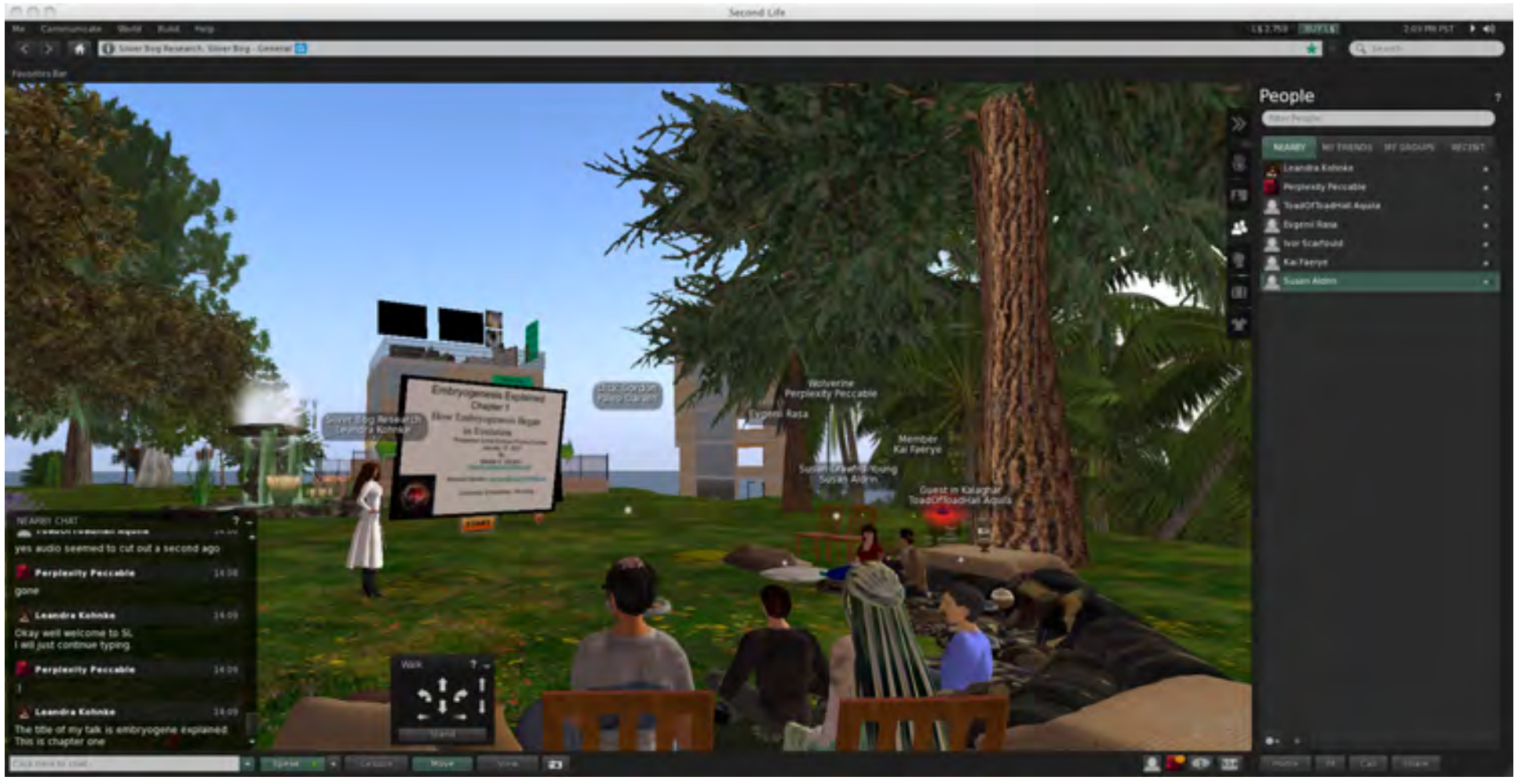

Figure 10 - The Embryo Physics Course virtual seminar room...
the seminar room of the future?

- to what extent can a biological view on the universe give us new insights?

In a landmark paper, Margaret Burbidge, Geoffrey Burbidge, William Fowler and Fred Hoyle (1957) showed the mechanisms responsible for the progressive synthesis of heavy chemical elements, or *stellar nucleosynthesis*. However, we all learn at school the stable table of periodic elements. It gives a wrong picture because those elements have an history and it took billions of years for them to stabilize. Yet, the formation of chemical elements is now largely stabilized. John Smart (2009) hypothesized that there might be a cosmic differentiation of chemical elements, which is now stabilized, as there is a cell differentiation in the development of multicellular organisms. Our blood cells don't suddenly turn into neuron cells, in the same way that hydrogen atoms don't suddenly turn into gold. We can also see the early universe nucleosynthesis as a progressive differentiation process, where protons and neutrons form atomic nuclei; then atom nuclei and electrons form atoms; and light elements nucleosynthesis produces hydrogen and helium.

# *CHAPTER 8 - Cosmological Selections*

**Abstract**: This Chapter first describes three fundamental evolutionary selection mechanisms which shape the biological landscape: natural, artificial and sexual selection. We then discuss two applications of selection mechanisms to cosmology. First the history, theory and limitations of the theory of Cosmological Natural Selection (CNS) proposed by Smolin. To remedy those limitations, we introduce Cosmological Artificial Selection (CAS) and review its history and theory. Since the theory is quite speculative, we propose six possible levels of universe making (from blind, accidental and artificial black hole production to cosmic breeder, cosmic engineer and God player). We defend it from a philosophical point of view, and address many objections which naturally arise. These include the design and creation terminology, the comparison of CAS with other fine-tuning explanations, the causal issue, the thermodynamical issue, epistemological issues, its feasibility, its underlying motivation and some implications for the idea of freedom. We then summarize four different roads leading to CAS and recapitulate the case for CAS.

*The problem of just how intelligence changes the universe*
*is a philosophical issue which (rather surprisingly)*
*few philosophers have addressed*

(Rescher 2009, 90)

*as the spirit wakens, it craves more and more to regard all existence not*
*merely with a creature's eyes, but in the universal view,*
*as though through the eyes of the creator.*

(Stapledon 1953, 150)

We saw classical explanations of fine-tuning (section 6.4 Classical Fine-tuning Explanations, p144). We also presented the general evolutionary thinking which has permeated most scientific disciplines, and is not restricted anymore to biology (section 3.2.3 Universal Darwinism, p59). Natural selection is the central mechanism to explain adaptation and fine-tuning in the biological domain. Could it be successfully applied to the universe as a whole?

How can we apply evolutionary thinking to the universe? Which kind of evolutionary selection can apply? How would such a selection function? Lee Smolin has attempted to apply the paradigm of adaptive evolution to the cosmos, with the theory of *Cosmological Natural Selection* (CNS). What are the strengths and weaknesses of CNS?

I first briefly discuss biological selection mechanisms and their possible counterparts as cosmological selections. Then I outline CNS as a promising evolutionary explanation of fine-tuning. However, I point out its weaknesses, especially its limited scope. I then take a wider philosophical scope to introduce

*Cosmological Artificial Selection* (CAS), an extension of CNS advancing an evolutionary explanation of fine-tuning which includes a role for intelligent life. CNS and CAS apply evolutionary logic to the universe as a whole. Given the speculative nature of those theories, I formulate objections and address limitations of both. Since evolutionary theories successfully explained adaptation and complexity in so many areas, they hold good promise to explain the complexity of our universe at large. I also present CAS as a promising scenario to give a long term meaning of life and intelligence in the universe.

## 8.1 Evolutionary Selections

### 8.1.1 Biological Selections

In biology, there are three selection mechanisms: *natural*, *artificial* and *sexual*. Let us have a quick overview.

*Natural selection* is the central mechanism of evolution. There are four conditions for natural selection to occur to any property of a species (Mark Ridley 2004, 74):

1. Reproduction. Entities must reproduce to form a new generation.
2. Heredity. The offspring must tend to resemble their parents: roughly speaking, "like must produce like."
3. Variation in individual characters among the members of the population. [...]
4. Variation in the *fitness* of organisms according to the state they have for a heritable character. In evolutionary theory, fitness is a technical term, meaning the average number of offspring left by an individual relative to the number of offspring left by an average member of the population. This condition therefore means that individuals in the population with some characters must be more likely to reproduce (i.e., have higher fitness) than others.

To know more about the formidable power of natural selection which explains complexity and adaptation of living organisms, I refer the reader to some excellent entry points (e.g. Darwin 1859; Dennett 1995; Dawkins 1996; Mark Ridley 2004).

Interestingly, Darwin introduced natural selection drawing an analogy with *artificial selection*. In artificial selection, breeders select organisms allowed to reproduce. In natural selection, it is the environment which plays this selective role. A striking example of artificial selection is found in the wide variety of dogs, as depicted in Figure 11. The artificial change from a wolf to a chiwawa took 5000 generations, while the natural change of Australopithecus to homo sapiens took 200,000 generations (Wright 1994, 26). Artificial selection thus goes much faster than natural selection. In the words of Bell (2008, 221) , the "exuberant diversity of dogs is a striking testimonial to the power of selection to direct adaptive change far beyond the limits of the original population within a few hundred generations." Understanding and mastering artificial selection is also a key to tackle global challenges that societies face. In effect, it is widely used in microbial genetic, molecular cloning, embryo screening and agriculture (Bell 2008).

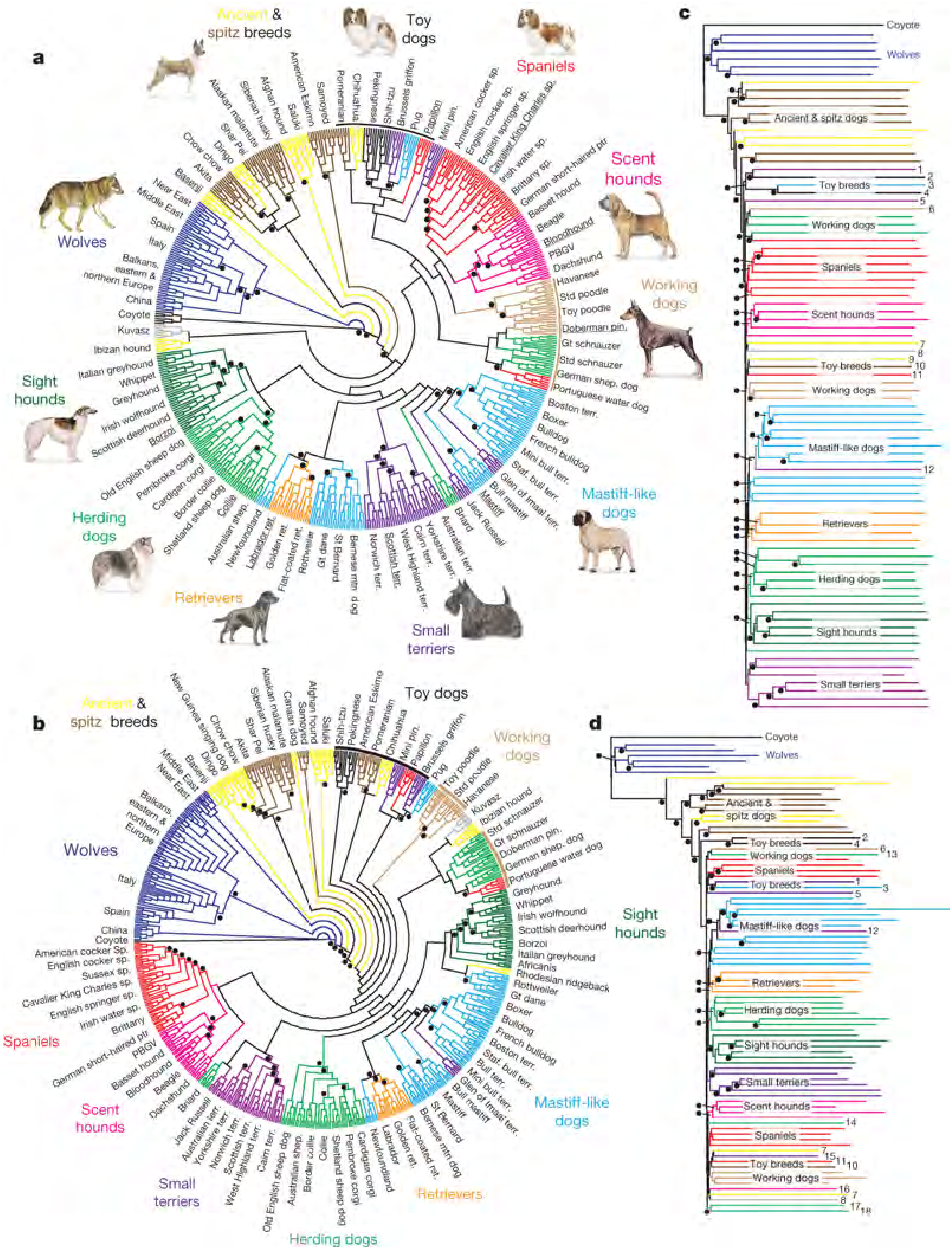

Figure 11 - Neighbour-joining trees of domestic dogs and grey wolves (see vonHoldt et al. 2010 for the legend and details )

The variety of dogs produced by humans illustrates the power of artificial selection.

Interestingly, *sexual selection* resembles artificial selection, but the selection doesn't occur through the breeder, but through *male combat* or *female choice* (Darwin 1876, 69). The male combat is the fight for the female, the matrix to reproduce genes, while female choice occurs when a female chooses a male with which to mate. Michael Ruse (2008, 19) compares natural and sexual selections with artificial selection: "natural selection is like choosing bigger and fleshier cattle, sexual selection through male combat is like two fighting cocks going at each other, and sexual selection through female choice is like choosing the dog that most closely fits the standards that one favors." Darwin (1876, 69) describes sexual selection in the following way:

> This form of selection depends, not on a struggle for existence in relation to other organic beings or to external conditions, but on a struggle between the individuals of one sex, generally the males, for the possession of the other sex. The result is not death to the unsuccessful competitor, but few or no offspring. Sexual selection is, therefore, less rigorous than natural selection. Generally, the most vigorous males, those which are best fitted for their places in nature, will leave most progeny. But in many cases, victory depends not so much on general vigour, as on having special weapons, confined to the male sex. A hornless stag or spurless cock would have a poor chance of leaving numerous offspring.

Of course, natural, artificial and sexual selection mechanisms can be combined and intertwined. Are there other selections we can think of? We should not confuse the three *mechanisms of selection* with *patterns of selection* which result from evolutionary selections dynamics. These latter patterns include balancing, extreme, directional, stabilizing or disruptive selections (see e.g. Mark Ridley 2004; Bell 2008).

In general terms, we can describe selection as blind-variation-and-selective-retention (D. T. Campbell 1974). Universal darwinist Gary Cziko (1995, 309–310) emphasized the power of multiple steps selection (constructive cumulative selection):

> This process of selecting and fine-tuning the occasional accidentally useful emergent system turns out to be so powerful that we should not be surprised that the adaptive processes of biological evolution, antibody production, […] learning, culture, and science all employ it, and that its power is now being explicitly exploited in the design of organisms, drugs and computer software by one of evolution's most complex and adaptive creations – the human species.

The importance and power of selection is also clear from a cybernetic principle. Indeed, the great cyberneticist Rosh Ashby went as far as to argue that intelligence is the power of appropriate selection. It is worth citing at length the closing words of his *Introduction to Cybernetics*, where Ashby discusses the amplification of intelligence and the importance of appropriate selection (Ashby 1956, 272):

> Now "problem solving" is largely, perhaps entirely, a matter of appropriate selection. Take, for instance, any popular book of problems and puzzles. Almost every one can be reduced to the form: out of a certain set, indicate one element. Thus of all possible numbers of apples that John might have in his sack we are asked to find a certain one; or of all possible pencil lines drawn through a given pattern of dots, a certain one is wanted; or of all possible distributions of letters into a given set of spaces, a certain one is wanted. It is, in fact, difficult to think of a problem, either playful or serious, that does not ultimately require an appropriate selection as necessary and sufficient for its solution.

It is also clear that many of the tests used for measuring "intelligence" are scored essentially according to the candidate's power of appropriate selection. Thus one test shows the child a common object and asks its name: out of all words the child must select the proper one. Another test asks the child how it would find a ball in a field: out of all the possible paths the child must select one of the suitable few. Thus it is not impossible that what is commonly referred to as "intellectual power" may be equivalent to "power of appropriate selection". Indeed, if a talking Black Box were to show high power of appropriate selection in such matters—so that, when given difficult problems it persistently gave correct answers—we could hardly deny that it was showing the *behavioral* equivalent of "high intelligence".

If this is so, and as we know that power of selection can be amplified, it seems to follow that intellectual power, like physical power, can be amplified. Let no one say that it cannot be done, for the gene-patterns do it every time they form a brain that grows up to be something better than the gene-pattern could have specified in detail. What is new is that we can now do it synthetically, consciously, deliberately.

### 8.1.2 Cosmological Selections

Can we get inspired by selection mechanisms in biology and find analogues at a cosmological scale? Pushing the analogy to cosmology, we logically find *Cosmological Natural Selection*, *Cosmological Artificial Selection* and *Cosmological Sexual Selection.* Such cosmological selections implicitly assume a multiverse on which selection operates. As we already mentioned, the idea of a multiverse is controversial scientifically –because it is hard if not impossible to test. For sure, those three extensions of selection mechanisms to the cosmos are increasingly speculative (as we defined three kinds of speculations at the beginning of Part III). Cosmological Natural Selection is a *scientific speculation* aimed to explain the fine-tuning issue. Cosmological Artificial Selection is a *philosophical speculation* aimed to explain the fine-tuning issue and the meaning of life and intelligence in the universe. Cosmological Sexual Selection is a *fictional speculation* which could take place if we make some further bold assumptions. It is fictional because it builds on the already speculative assumptions of CNS and CAS, and further assumes there are extraterrestrials competing for making baby universes with a dynamic similar to sexual selection in biology! I mention it however, for three reasons. First, it is a logical extrapolation of a fundamental selection mechanism in biology to cosmology, and it is the purpose of this Chapter to follow such a heuristic. Second, we will see in Chapter 9 some empirical indications that it might be happening as you read these lines. Third, even is it is plain wrong, since it is a fictional speculation, I hope it will inspire science fiction authors.

### 8.1.3 Biological and Cosmological Fitness

*Fitness* is a central concept in evolutionary biology. Generally, fitness characterizes the ability to *survive*, *grow* and *reproduce*. For example, to be fit, a wolf has to be adapted to survive its cold environment, find resources to be able to grow and ready to compete for successful reproduction.

Now, how can we interpret fitness in a cosmological context? Do we mean the survival of the universe as a whole, or of some of its constituents –like us humans?

Let us focus on the former. Modern physical eschatology shows that the death of the universe is guaranteed –not its survival. We indeed outlined various cosmic doom scenarios earlier. We know that the universe has been expanding exponentially fast at the big bang era and can roughly assimilate this as a growth process. The idea of universal reproduction seems at first sight more far-fetched. Yet, in the last three decades, theoretical physicists have advanced various mechanisms for natural production of offspring universes (see e.g. Sato et al. 1982; Linde 1984; Cornell 1989; Hawking 1993; B. Carr 2007). The interesting prospect with reproduction is the hope to tackle the thermodynamical challenge. We can hope that a new universe would reset its thermodynamical constraints (see e.g. Davies 1994, 146; and Penrose's 2011 Conformal Cyclical Cosmology). The biological analogy is a new-born baby who can start a fresh new life in contrast to the inevitable death of its parents.

Let us take the more precise definition of fitness Ridley gave above, "the average number of offspring left by an individual relative to the number of offspring left by an average member of the population." If we envisage a space of possible universes $\mathcal{M}$, and if we further assume some properties of the reproduction mechanism, we can indeed try to estimate the number of offspring universes in the framework of various multiverse theories, and thus the fitness of our universe. But let us now first introduce in more details CNS and CAS.

## 8.2 Cosmological Natural Selection

### 8.2.1 History

Smolin (2012) clearly articulated the root of the fine-tuning problem, which he calls more accurately the *landscape problem*. It is simply the fact that we can imagine a huge landscape of possible universes compatible with our physical and cosmological models. This begs the question: *why this universe?* Smolin points out that there are two ways to explain this:

> Either there are logical reasons it has to be that way, or there are historical causes, which acted over time to bring things to the present state. When logical implication is insufficient, the explanation must be found in causal processes acting over time.

Since we have no prospect to *logically* reduce the huge landscape of possible universes anytime soon, it is worth looking at the other way: *historical causes*. Smolin mentions that this point was clearly seen by Charles Sanders Peirce (1955, 318) in 1891:

> To suppose universal laws of nature capable of being apprehended by the mind and yet having no reason for their special forms, but standing inexplicable and irrational, is hardly a justifiable position. Uniformities are precisely the sort of facts that need to be accounted for. That a pitched coin should sometimes turn up heads and sometimes tails calls for no particular explanation; but if it shows heads every time, we wish to know how this result has been brought about. Law is par excellence the thing that wants a reason.
>
> Now the only possible way of accounting for the laws of nature and for uniformity in general is to suppose them results of evolution.

Broadly speaking, the hypothesis of historical causes in scientific domains has lead to major scientific advances. Let us see three example, in cosmology, biology and theoretical physics. First, when Georges Lemaître (1927) introduced the idea of an expanding universe, it meant that *the universe itself had an history*. Cosmology would then never be the same. Second, Lamarck and Darwin also introduced historical causes with evolutionary theories to explain the living world. This revolution of evolution is so important that it is still ongoing. Finally, Lee Smolin has theorized that the laws and parameters of the universe themselves may have had an history.

But this requires the idea that our universe is not unique –however paradoxical this sentence may seem at first sight. Our universe would be one in a vaster multiverse. We can trace the idea of the multiverse –or the broadly synonymous multiple universes, many worlds, pluriverse, megaverse, parallel worlds– back to Anaximander in Antiquity, Giordano Bruno and also Leibniz (see Kragh 2011, 256–259).

However, vaguely assuming a multiverse is not enough. We saw earlier that to define a multiverse we need to explain the logic of the universe generation mechanism. The most general way is to imagine a kind of random trial and error process. We may trace this idea back to Denis Diderot who wrote in 1749 a "Letter on the Blind, for the Use of Those who See" (Diderot 1875). Diderot imagines a debate between the blind Cambridge mathematician Nicholas Saunderson and a clergyman. The topic is the usual question whether order in nature implies the existence of God. After Diderot anticipates the darwinian idea of selection against the unfit in the animal world, he writes (Diderot 1937):

> But why should I not believe about worlds what I believe about animals? How many worlds, mutilated and imperfect, were perhaps dispersed, reformed and are perhaps dispersing again at every moment in distant space, which I cannot touch and you cannot see, but where motion continues, and will continue, to combine masses of matter until they shall have attained some arrangement in which they can persist. O philosophers, transport yourselves with me to the confines of the universe; move over that new ocean, and seek among its irregular movements some trace of the intelligent Being whose wisdom so astounds you here!

David Hume had a similar idea in 1779, when he wrote this passage in the famous *Dialogues Concerning Natural Religion* (Hume 2009):

> Many worlds might have been botched and bungled, throughout an eternity, ere this system was struck out; much labour lost, many fruitless trials made; and a slow, but continued improvement carried on during infinite ages in the art of world-making. In such subjects, who can determine, where the truth; nay, who can conjecture where the probability lies, amidst a great number of hypotheses which may be proposed, and a still greater which may be imagined?

Hume probably not meant this many worlds trial-and-error dynamics actually happened. Charles Pantin (1965, 94) envisioned –without development– that natural selection and a selection effect could explain the cosmic coincidences leading to organic life:

> If we could know that our own Universe was only one of an indefinite number with varying properties we could perhaps invoke a solution analogous to the principle of natural selection; that only in certain universes, which happen to include ours, are the conditions suitable for the existence of life, and unless that condition is fulfilled there will be no observers to note the fact.

The mechanisms of multiple universe production were later pioneered in modern cosmology. John A. Wheeler (1977, 4) proposed that the basic laws of the universe might fluctuate during a big bang or a big crunch (see also Misner, Thorne, and Wheeler 1973, chap. 44). In the 1980's theorists in quantum gravity began to theorize about multiple universes (see e.g. Hawking 1987; 1988; Coleman 1988). Hawking (1987) and Frolov (1989) even suggested that new universe production might happen in the singularity region of black holes.

Quentin Smith (1990) proposed a scenario to explain the very existence of basic physical laws. He developed a model where some black hole singularities could give birth to new universes. This would provide a statistical explanation of the existence of basic laws, since only a small fraction of black holes would be fit to produce universes. Two years later, cosmologist Lee Smolin started to develop this idea in much more details. Let us take a closer look.

### 8.2.2 Theory

If we take seriously universal darwinism, we can try to apply Darwinism to the universe itself. This is the visionary attempt of Lee Smolin's Cosmological Natural Selection (CNS) hypothesis to address the fine-tuning issue (Smolin 1992; 1997; 2007). Let us introduce this theory with an analogy (see table 7 below). The situation in contemporary cosmology is analogous to the one in biology before the theory of evolution, when one of the core questions was (1) *Why are the different species as they are?* It was assumed more or less implicitly that (2) *Species are timeless categories*. In present physics, the question behind the fine-tuning problem is (1') *Why are physics and cosmic parameters as they are?* Currently, it is usually assumed (probably from the remains of the Newtonian worldview) that (2') *Parameters are timeless*. It is by breaking assumption (2) that Darwin was able to theorize about the origin of species. Analogously, Smolin is trying to break assumption (2'), by theorizing about the origin of parameters.

| **Biology (yesterday)** | **Physics (nowadays)** |
|---|---|
| (1) Why are the different species as they are? | (1') Why are physics and cosmic parameters as they are? |
| (2) Species are timeless | (2') Parameters are timeless |

Table 7 Smolin's (1997, 260) analogy to present Cosmological Natural Selection. The situation of nowadays physics is analogous to the biologists' before Darwin.

According to this natural selection of universes theory, black holes give birth to new universes by producing the equivalent of a big-bang, which produces a baby universe with slightly different parameters. This introduces variation, while the differential success in self-reproduction of universes (via their black holes) provides the equivalent of natural selection. This leads to a Darwinian evolution of universes, whose parameters are fine-tuned for generating a maximum number of black holes, a prediction that can in principle be verified. Vaas (1998) did summarize the core argument of CNS:

> 1. If a new universe is born from the center of a black hole, i.e. if the bounce of a singularity leads to a new expanding region of spacetime,
>
> 2. and if the values of the fundamental free parameters of physical theories can change thereby in a small and random way, i.e. differ from those in the region in which the black hole formed (in particular, Smolin has in mind the dimensionless parameters of the standard model of particle physics),
>
> 3. then this results in different reproduction rates of the physically different universes.
>
> 4. Hence, our universe has been selected for a maximum number of black holes. It is a descendant of a series of universes, each of which had itself been selected for the same criterion.
>
> 5. Thus, the values of the parameters are the way they actually are, because this set of values leads to a (local) maximum of descendant universes.

### 8.2.3 Objections

However, the prediction that most changes in parameters would lead to less black holes has been challenged. Rothman and Ellis (1993) pointed out that changing cosmic or physics parameters may lead to either more black holes, or that the number of black holes can be insensitive to a parameter change. The problem is that in CNS, changing parameters should lead to a *decrease* of the number of black holes. This is necessary if we assume that our universe is fit in the sense that it is a typical member of the multiverse where cosmic natural selection takes place. Joe Silk (1997), Martin Rees (1997, 251) and Alexander Vilenkin (2006b) did also propose ways to tweak cosmic parameters to *increase* the number of black holes, thereby questioning the validity of CNS.

Yet, Smolin did reply to most of these issues in subsequent works (see especially the appendix in Smolin 1997; 2013). Smolin (2012) also noted that there is a trade-off to consider between primordial black holes and stellar black holes.

One might be surprised by the unconventionally speculative aspect of CNS. Although Smolin emphasizes the refutability of CNS and thus its scientific aspect in Smolin (2007), he himself is not proud that the theory talks about processes outside the universe (Smolin 1997, 114). This conjectural aspect of the theory puts it at the edge of science and philosophy (Vaas 1998). Let us stress once again that when we attempt to answer the question "why is the universe the way it is?", we must be ready to cross the border of the common experimental and observational science. Attempting to answer this issue leads to the construction of speculative theories. Thus, CNS should be compared to other multiverse speculations. It should also be put in perspective with respect to other attempts to explain the fine-tuning issue (see section 6.4 Classical Fine-tuning Explanations, p144; and story 8, p174). And in this respect, it is arguably the best variant amongst the speculative multiverse universe. It is parsimonious because it includes both a universe generation mechanism and a selection principle. This goes beyond a vague statement that "all other possible universe exist". Most importantly, it provides empirical tests such as the number of black holes observed or an upper-bound mass for neutron stars (Brown, Lee, and Rho 2008).

Nevertheless, the epistemological difficulties are real (see also Vaas 1998 for an in depth examination of CNS). Couldn't we find systematic *ad hoc* solutions to any objection against CNS? If the tests of CNS fail, couldn't we claim that we are simply not yet in the optimal universe generation, leading to the maximum number of black-holes? It seems that CNS can easily be saved with such *ad hoc* hypotheses.

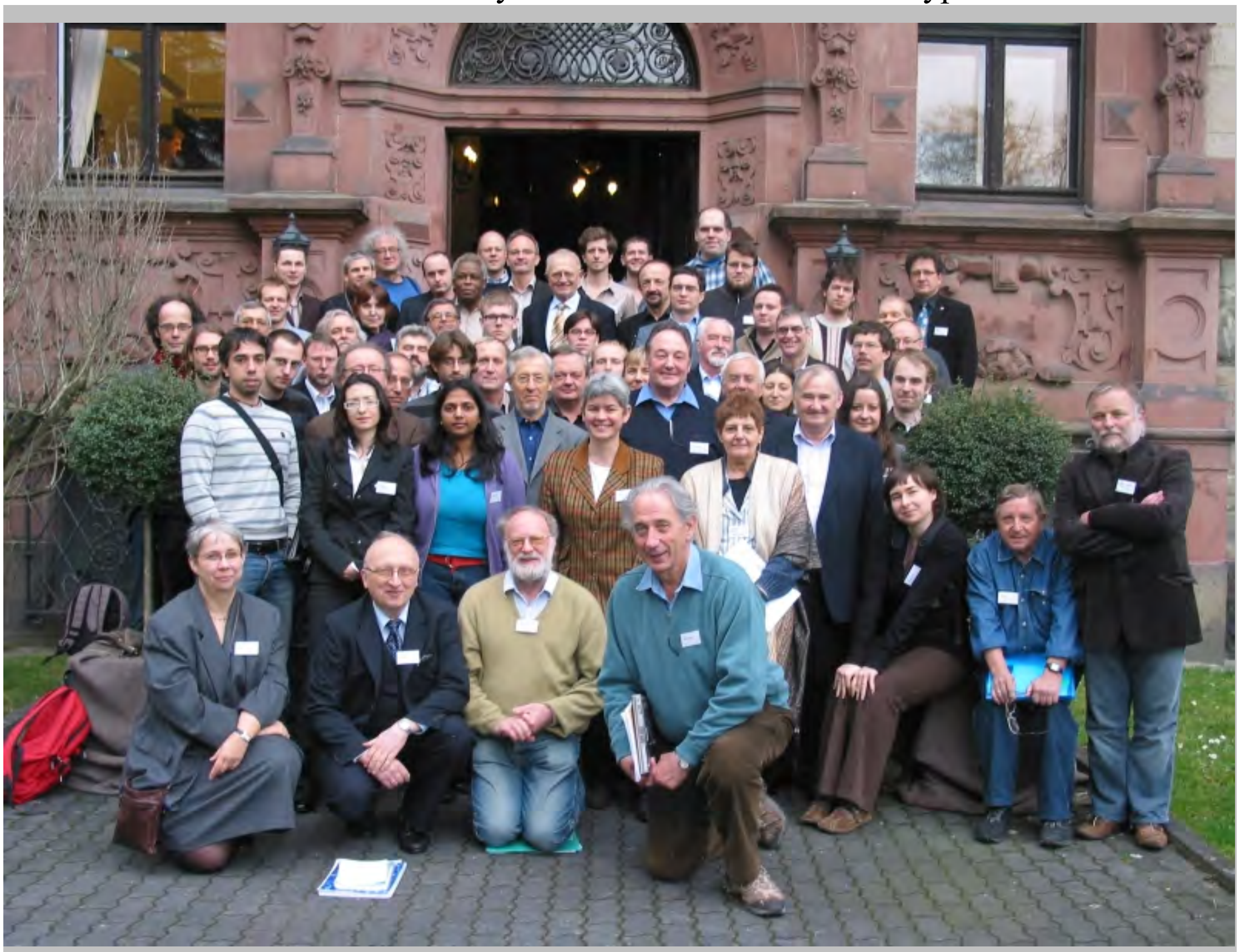

Figure 12 - WE-Heraeus-Seminar. Evolution and Physics - Concepts, Models and Applications. 21-23 January 2008 at the Physikzentrum Bad Honnef, Germany.

On 21-23 January 2008, I went to Bad Honneff in Germany to a workshop entitled “Evolution and Physics - Concepts, Models and Applications”. I presented a poster (Vidal 2008c) introducing CNS and its extension to CAS. Discussing with a scientist, he was uncomfortable with the speculative nature of those ideas. I replied that CNS and CAS should first be put in the context of the fine-tuning issue, so that we can compare them to alternative explanations. I then argued that CAS is “better” than the God or the multiverse hypothesis, because it is naturalistic and can be subjected to tests. Having made explicit my problem-solving attitude, and the weakness of alternative explanations, he seemed more open-minded. I finally won the best poster prize, even if I suspect the nice poster design helped more than the speculative content.

Story 8: CNS, CAS and alternative explanations.
Why speculate? It is important to be clear on which problem we tackle when we engage in speculations.

Let us now raise five more fundamental objections against CNS, besides the number of black holes which may not be optimal.

(1) CNS has no environment where selection operates
(2) CNS has no hereditary mechanism
(3) CNS unsuccessfully deals with parameter sensitivity
(4) CNS focuses on limited cosmic outcomes
(5) CNS does not address broader metaphysical issues

The first objection is that there is no environment where selection operates. This is a disanalogy with the darwinian mechanism of natural selection, where the environment plays the critical role of selection. Smolin defined natural selection as the selective growth of a population on a fitness landscape. The fitness is defined as the rate of reproduction. All this is correct, but, as Vaas (1998) argues, in biology the spread of populations is constrained by *external* factors which limit resources, such as space, food (energy) and mating opportunity (sexual selection). The problem is that there is no analog of these crucial elements in CNS. It seems that CNS could better be described as internal selection (or self-organization), rather than natural selection in a biological fashion. We will see in the next section (8.3 Cosmological Artificial Selection, p176) how CAS can remedy this issue.

The second objection is that CNS lacks a hereditary mechanism. Edward Harrison (1995) noted that CNS lacks a process that selects for reproduction only universes that are inhabitable by organic life, in order to solve fine-tuning for life. James N. Gardner (2003, 85) also pointed out the lack of hereditary mechanism in CNS. Why should baby universes resemble their parents?

The third issue regards parameter sensitivity. How reliable is universal reproduction? If there is a perfect reliability in the reproduction mechanism, there is no evolution. If there is slight variation on one single parameter, universe reproduction will most likely fail. This is indeed the point of existing fine-tuning arguments, which are in fact just *one-parameter sensitivity arguments*. Even Victor Stenger (2011), the most vigorous critique of fine-tuning, concedes this point of *parameter sensitivity*. This is why it is important to distinguish parameter sensitivity from fine-tuning. Indeed, if almost *any* slight variation in the physical constants renders a universe sterile to complex outcomes, then CNS is very likely wrong. Most small random variations will prevent the emergence of a complex universe. The rate of new universe production should be extremely high to randomly hit in the cosmic landscape congenial regions for complexity. It remains an open question whether the rate of black hole production and universe production would be high enough to meaningfully explore the landscape.

The fourth issue concerns the key question "fine-tuning for what?". Which cosmic outcome is CNS busy with? At first sight, only black holes matter, because they directly influence the amount of baby universes. But which black holes sizes are we considering? Primordial black holes? Stellar black holes? Supermassive black holes? In the case of stellar black holes, Rothman and Ellis (1993) made explicit the connection with star formation. Indeed, since stellar black holes are –usually– thought to be the remnant of a dead massive star, if we increase the number of massive stars, we would at the same time increase the number of black holes. So, it seems that CNS is mainly concerned with cosmic outcomes such as the formation of massive stars and the number of black holes. But what about higher cosmic outcomes such as life or intelligence? Do they conform or not with CNS? CNS is in fact indifferent to higher cosmic outcomes such as life or intelligence. Simply ignoring the growth of complexity in the universe up to life and intelligence may prove a too narrow scope when considering the problem of the origin and evolution of the cosmos.

The fifth objection is that CNS has a too narrow scope. Indeed, one can greatly vary the scope of cosmological issues, from observational cosmology up to speculative mathematical theories or broader cosmological scenarios and metaphysical issues (see Ellis 2007a, 1245–1247). In addition to the restriction to low cosmic outcomes, the scope of CNS is also too narrow from a philosophical point of view. What about "ultimate" explanations? Is there is a fine-tuning from generation to generation, in the sense that each universe becomes better at producing new offspring? If so, did universes progressively become more and more complex? Where does the "first" universe comes from? Was there a first universe at all? Of course, we saw that these are different issues from fine-tuning, and that all fine-tuning explanations we reviewed (see section 6.4 Classical Fine-tuning Explanations, p144) are incomplete in this regard.

There is no particular meaning or role for intelligent life in CNS. In this essay, since our primary objective is to understand life and intelligence in the universe in a broad cosmological and philosophical way, CNS is too restrictive. We will now see how Cosmological Artificial Selection, a variation on CNS with a wider scope, can remedy to these five objections.

## 8.3 Cosmological Artificial Selection

*It is wonderful what the principle of selection by man,*
*that is the picking out of individuals with any desired quality,*
*and breeding from them, and again picking out, can do.*
*Even breeders have been astounded at their own results.*

Darwin's 1858 letter to Asa Gray.
Reprinted in (Bajema 1983, 191–192)

### 8.3.1 History

Before remedying to objections to CNS (see next section 8.3.2 From Natural to Artificial Cosmological Selection, p179), let us dig the historical roots of Cosmological Artificial Selection. We saw (section 8.2.1 History, p170) that theorists have speculated that new universes could *naturally* emerge out of singularity regions inside black holes. Yet, other theorists made an even bolder step and explored the possibility of *artificially* making universes. Such an early study was conducted by Fahri and Guth (1987) who pointed out, as their paper's title indicates, "An obstacle to creating a universe in the laboratory". The obstacle being that the creation of such a universe in the laboratory would require an energy-density far too high with respect to what we can reach with our current technology. Others speculated that an advanced civilization could transfer information into their baby universe (see e.g. Ćirković and Bostrom 2000; J. Garriga et al. 2000). Ansoldi and Guendelman (2006) did review the literature about making child universes in the laboratory. Interestingly, they also advocated a more active approach to cosmology, not just with passive observations, nor with virtual simulations as in *Soft Artificial Cosmogenesis*, but also with attempts to build real child universes in the laboratory, or *Hard Artificial Cosmogenesis*.

However, we can find much deeper roots to the idea of a naturalistic intelligence advanced enough to make a universe (see also Dick 2008). This advanced intelligence is not a *supernatural God* but a *natural Demiurge*. In contrast with a God, a Demiurge is not omnipotent and, like an architect, has to work within the constraints of the material world. Such a *non ex nihilo* creation myth was elaborated in Plato's *Timaeus*. However, for Plato the demiurge made the cosmos out of an ideal bluebrint (see also Kragh 2007, 23). Plato's myth thus keeps a idealistic component.

More recently, Fred Hoyle (1983, 211–215) made an intriguing reasoning regarding the possible intelligent control from an advanced future intelligence. He noticed that the past-to-future sense of time is a special case of Maxwell's equations. The opposite time-sense from future-to-past is in principle not forbidden. He thus speculated that "biological systems are able in some way to utilize the opposite time-sense in which radiation propagates from future to past. Bizarre as this may appear, they must somehow be working *backwards* in time." This is an example of a superior yet natural intelligence at play in our universe.

However, even if we grant this provocative idea as plausible scientifically, it remains extravagant. Indeed, it is more cautious and natural to study the increase of complexity through the thermodynamics of open-systems and biological systems in particular, rather than assuming time travel of some kind of information from future to past.

In the context of the search for advanced extraterrestrial life, Allen Tough (1986, 497a) speculated that an advanced civilization would try to avoid a cosmic doom scenario, heat death or big crunch. For this grand purpose, "some way may be found to break out of this Universe into another one, either existing parallel to it or arising subsequent to it. That is, perhaps the best of our knowledge, consciousness, and genes can somehow be transferred to this other universe."

In 1994, Louis Crane published on the arXiv preprint repository a paper –later published in (2010)– with the intriguing title "Possible Implications of the Quantum Theory of Gravity: An Introduction to the Meduso-Anthropic Principle". Although I do not treasure the introduction of new anthropic principles, this one is worth closer examination. Indeed, Crane was the first to propose a variation on Smolin's Cosmological Natural Selection by introducing a contribution for intelligent life. He conjectured that "successful advanced industrial civilizations will eventually create black holes". He gave reasons why (for scientific purposes, energy source, waste disposal, starship propulsion) but we will get back to this idea more systematically later (see section 9.3 Black Holes as Attractors for Intelligence, p228). Such artificial black holes experiments would incidentally increase the overall number of black holes in the universe, and thus strengthen the basic mechanism of CNS. In the words of Crane, fine-tuning then reaches the cosmic outcome of "successful civilization", not only stars and natural black holes:

> If both Smolin's two conjectures and mine are true, then the fine tuning of physical constants would not stop with physics which produced stars and gas clouds which cool. Rather, the selection would continue until physics evolved which resulted in successful civilizations

One year later, eminent cosmologist Edward Harrison (1995) published in the *Quarterly Journal of the Royal Astronomical Society* a remarkable paper describing in much details the possible influence of intelligence in Smolin's CNS. Harrison was probably not aware of Crane's paper, since he doesn't cite it. Instead of the *meduso-anthropic principle*, Harrison named the scenario a *Natural Creation Theory*. Importantly, Harrison contrasts "natural" with "supernatural". So, natural is not opposed to artificial, it includes it. Instead of a Natural Creation Theory, it would thus be more accurate to speak of a "Natural and *Artificial* Creation Theory". This was also noticed by Barrow (1998, 175) who proposed instead the terminology of "forced breeding" or "artificial selection".

Harrison did articulate very clearly a very inspiring cosmogenic reproduction scenario. He did address in his paper the main issues this picture raises. We will come back to the details, but his core idea was simply to combine two lines of thought. First, the possibility of universe creation in the laboratory (Farhi and Guth 1987) and second, Smolin's cosmological natural selection hypothesis.

In January 1997, the Edge.org website published a short discussion between Smolin and Dawkins about CNS. Dawkins commented on the fact that CNS is only concerned with basic outcomes (Smolin et al. 1997):

> Smolinian selection may account for the fact that our universe has the necessary constants, dimensionality and laws to last for last for a long time (not fizzle out or crunch immediately its initiating bang), long enough to spawn daughter universes (and INCIDENTALLY long enough to breed life).

But what about higher outcomes? Dawkins continues:

> But Smolinian selection cannot account for the fact that our universe is specifically congenial to life, or to intelligent life, or to us. My negative conclusion would break down only if life itself is in the habit of engineering the spawning of daughter universes. As far as I am aware, this hasn't been suggested, but it is, I suppose, a theoretical possibility that daughter universes are generated as a consequence of the fooling around of highly evolved physicists.

As we saw, the idea was suggested and developed earlier, at least by Louis Crane and Ted Harrison. But it is remarkable that those three researchers all independently suggested to complete CNS with a role for high intelligence. However, Dawkins perhaps vaguely realizing that he is extrapolating evolutionary reasoning further than usual, ends his reply with: "But this may not be very coherent since I am suffering from flu".

In November 1998, Steven J. Dick (2000), reflected about Cosmotheology, or the impact of the new worldview of cosmic evolution. He also envisioned that a *natural God* could have had made our universe. He writes (Dick 2000, 204) that such "advanced intelligence could have fine tuned the physical constants". He did not link the idea with Smolin's CNS however.

James N. Gardner later developed the scenario much further, in a series of papers (J. N. Gardner 2000; 2001; 2005) and a beautifully written book, *Biocosm. The New Scientific Theory of Evolution: Intelligent Life Is the Architect of the Universe.* (J. N. Gardner 2003). He named the scenario the *Selfish Biocosm Hypothesis*, the cosmos itself selfishly aiming at its own replication, as genes want to replicate in Dawkins' famous metaphor of evolution.

In 2001, at the 12th Congress of the International Association "Cosmos & Philosophy", Baláz (2005) further elaborated the scenario that he named the *cosmological replication cycle* and made connections with the search for extraterrestrials.

Whereas most authors who developed this scenario focused on solving the fine-tuning issue, there is one exception. It is the work of futurist and developmental system theorist John M. Smart (2000; 2009; 2012). Extrapolating the trends of increasingly dense and energy intensive technologies, he came to the conclusion that ultimate technology should reach a black hole density. As we saw, black holes can then be hypothesized to generate new universes. It is then a small speculative step to assume that sufficiently advanced civilizations could actually foster the making of universes. He named the scenario *Developmental Singularity Hypothesis* and in the context of SETI, the *Transcention Hypothesis*. Importantly, the singularity Smart is talking about is not to be confused with the large literature about technological singularity. The latter technological singularity pertains to the exponential increase of computing resources and the predictable merging of humans, machines and the internet (see e.g. Heylighen 2005; Kurzweil 2006). Smart's singularity is a few million years ahead. His reasoning is very interesting, because it shows that we can arrive to CAS not only by thinking deeply about solving the fine-tuning issue (the beginning) but also by extrapolating the ultimate future development of civilizations (the end).

The topic has been further discussed from different perspectives in a variety of popular books (see e.g. Chown 2003; Kurzweil 2006; Martin 2006; J. N. Gardner 2007; Davies 2008; John Gribbin 2009) and papers (Barrow 2001; Dick 2008; Vidal 2008b; Stewart 2010; Vaas 2009; Vidal 2010a; Vaas 2012; Vidal 2012a).

To sum up, a variety of authors have discussed the same idea of intelligent life playing a role in a universal reproduction cycle. They used different terminologies to describe the scenario: meduso-anthropic principle, natural creation history, natural God, selfish biocosm hypothesis, cosmological replication cycle or developmental singularity hypothesis. I will use *Cosmological Artificial Selection* (CAS), because the scenario is a variation on Smolin's Cosmological Natural Selection (CNS) where *natural selection* is simply replaced by *artificial selection*. But how and why could we and should we shift from CNS to CAS?

### 8.3.2 From Natural to Artificial Cosmological Selection

Let us now see how CAS can remedy the objections we formulated against CNS, and thus understand the worth of extending CNS to CAS. We claim the following:

(1) CAS uses a *virtual multiverse* environment where selection operates
(2) CAS has an intelligence-driven hereditary mechanism
(3) CAS successfully deals with parameter sensitivity
(4) CAS reaches high cosmic outcomes
(5) CAS can be completed to address broader metaphysical issues

The first remedy to CNS is to re-introduce an environment, which is a vital component for selection to operate. The environment is here a set of virtual universes tested and fine-tuned by a naturally very advanced intelligence. This can in principle be conducted with artificial cosmogenesis, where virtual universes compete with each other. This also introduces competition (which you don't have in CNS) between artificially generated universe simulations. One can interpret this approach as a variation on the multiverse proposal. However, the selection of universes would take place on *virtual universes*, replacing Smolin's natural selection of *real universes* (Barrow 2001, 151). In CNS, we need many generations of universes in order to randomly generate an interesting fine-tuned universe. In contrast, these simulations would dramatically improve the process by artificially selecting (via simulations) which universe would exhibit the desired features for the next generation universe. This would facilitate the daunting task of making a new universe. In this case it is indeed appropriate to speak about a “Cosmological *Artificial* Selection” (CAS), instead of a “Cosmological *Natural* Selection”. This can be achieved as a far-off application of a mature Artificial Cosmogenesis research program. Accordingly, we are not only talking about *simulations* here, but also a much greater feat, which is the *realization* or the making of a new universe. Are these feats possible? We will discuss this objection soon (in section 8.3.9 Objection – Are Simulation and Realization Possible? , p190).

The second remedy is to introduce intelligent life as performing or helping the hereditary mechanism. The pioneer authors of CAS put forward the hypothesis that life and intelligence could perform this mechanism of heredity, thus playing an essential role in the Darwinian evolution of universes. To better grasp this extension of CNS, Baláz and Gardner proposed to consider von Neumann's (1951) four components of a self-reproducing automaton. I summarized this completion of CNS in table 8 below.

Let us describe these four components in more detail. Physics and cosmic parameters are analogous to DNA in biology, and to the *blueprint* of this self-reproducing automaton. The universe at large or the cell as a whole constitute the *factory*. When furnished with the description of another automaton in the form of a blueprint, the factory will construct this automaton. The *reproducer* reads the blueprint and produces a second blueprint. In the cell, these are reproduction mechanisms of the DNA. The *controller* will cause the reproducer to make a new blueprint, and cause the factory to produce the automaton described by the new blueprint. The controller separates the new construction from the factory, the reproducer and the controller. If this new construction is given a blueprint, it finally forms a new independent self-reproducing automaton.

| ***COMPONENTS*** | ***DESCRIPTION*** | ***BIOLOGY (cell)*** | ***COSMOLOGY (universe)*** |
|---|---|---|---|
| Blueprint | Gives instructions for the construction of the automaton | Information contained in the DNA | Physics and cosmic parameters |
| Factory | Carries out the construction | Cell | The universe at large |
| Reproducer | Reads the blueprint and produces a second blueprint | The reproduction of the DNA | CNS:? **Intelligence unravelling the universe's blueprint** |
| Controller | Ensures the factory follows the blueprint | The regulatory mechanisms of the mitosis | CNS:? **A cosmic process, aiming at universe reproduction** |

Table 8: Components of a von Neumann’s (1951) self-reproducing automaton.
The second column provides a general description of the automaton functions. The third and fourth columns propose examples respectively in biology –the cell– and in cosmology –the universe.

We now clearly see the limits of CNS, which is not specifying what the reproducer and controller are. Intelligence unravelling the universe's blueprint can precisely fulfill the reproducer's function. This reproducer component is indeed essential for providing a mechanism for heredity. Without heredity, there can be no Darwinian evolution. The controller in this context would be a more general process aiming at universe reproduction with the help of intelligence. In table 8, I completed in bold font these two missing components of CNS, thus including intelligence in this hypothesized cosmic reproduction process.

A consequence of this speculative theory is that intelligent life, unravelling the universe through scientific understanding, generates a “cosmic blueprint” (an expression used by Paul Davies 1989). The cosmic blueprint can be seen as the set of physics and cosmic parameters. In CAS, the fine-tuning of this cosmic blueprint would take place in “virtual universes”, that is in simulated universes (Vidal 2008b; 2013).

Third, CAS successfully deals with parameter sensitivity. The solution is simply that natural intelligences take good care of it. CNS is insensitive to parameter sensitivity. We saw that small variations of one or several parameters generally don't lead to fecund universes. Here is a way to distinguish CNS from CAS. We could predict that the number of off-spring universes produced in CNS would be insufficient to make it statistically probable that random variation would lead to an off-spring universe with complex outcomes (e.g. life, intelligence or technology). Of course, the number of offspring universes is hard to assess given our present knowledge of black holes and the speculative nature of universe generation inside black holes. For example, do we consider laboratory-sized black holes, stellar black holes and supermassive black holes? Could it be that rotating black holes generate several universes?

Fourth, CAS reaches high cosmic outcomes. Let us ask the key question: *fine-tuning for what?* In both CNS and CAS, the ultimate aim of the universe is to replicate. However, the cosmic outcomes which allow replication are different. In CNS, it is the sheer number of black holes which matters. In CAS, it is the number of intelligent civilizations actually making universes. Crane already noticed that fine-tuning would concern successfully advanced civilizations, and not just stars massive enough to collapse into black holes. More precisely, in CAS the universe is fine-tuned for recursive self-replication driven by intelligent life. The recursive aspect is fundamental. It involves the idea of fertility or immortality of the cosmic replication process (we will discuss this fundamental aspect at the end of this work, in section 10.4 Voyage to Five Immortalities, p293). Indeed, why bother making a new universe if it will be sterile and doomed to annihilation?

The fifth point regards the respective scopes of CNS and CAS. Since CNS is already a quite speculative theory difficult to test, why develop CAS, an even more speculative one? The answer is that the scope of CAS is different from CNS.

Rüdiger Vaas (2012) criticized CAS on the basis that CNS is "simpler" than CAS. I disagree. Since this claim is quite unexpected, I must first make an important epistemological remark about the concept of simplicity. It is well known that simplicity is very hard to define, and specialists consider it to be either subjective (Heylighen 1997a), or largely context dependent (T. S. Kuhn 1977; McMullin 2008). So we need to make explicit the cosmological context at play here, or the *scope* of the inquiry, as Ellis (2007a, 1245) already suggested. The scope we discuss here concerns four fundamental issues:

**(1) Why do the laws of physics, constants and boundary conditions have the form they do?**

As we saw in details in Chapter 6, this concerns the fine-tuning issue.

**(2) Why not nothing?**

This is certainly one of the deepest metaphysical question. The formulation here is a shorter version proposed by Apostel (1999) of Leibniz' "why is there something rather than nothing?". Again, we saw that it is an unavoidable metaphysical challenge (see section 4.1.2 Metaphysical, p76).

**(3) What is the meaning of the existence of intelligent life in the universe?**

This question asks about the meaning of intelligence in the universe. Here, meaning is interpreted as "purpose" or "significance". Are life and intelligence merely epiphenomena in cosmic evolution? Or could their presence have deeper, yet to be discovered implications? As Davies (1999, 246) formulates it, why did the laws of the universe engineer their own comprehension?

**(4) How can intelligent life survive indefinitely?**

The future of the universe is gloomy. Physical eschatology teaches us that none of the known scenarios seem to allow indefinite continuation of life and information processing in the very long term (Vaas 2006).

These four questions are more philosophical than scientific. Another way to put it is to see CNS as a *scientific speculation*, tackling question (1), whereas CAS is a *philosophical speculation*, tackling questions (1), (3) and (4). Question (2) has a special status, because it is metaphysical, and both CNS and CAS –and any "ultimate" explanation– has to deal with it.

To put it otherwise, it is worth mentioning that in this inquiry we reformulated and focused the philosophical worldview questions into mixed philosophical and cosmological questions. Arguing in favor of CAS is better seen as an exercise in synthetical philosophy, i.e. the construction of a worldview answering consistently and comprehensively the worldview questions. I translated the issue of the beginning of the universe into free parameters and fine-tuning (1). Regarding the "end of the universe", I focused on the future of scientific simulations, and the predictable end of the universe in a cosmic doom (4). The combinations of the answers should provide a synthetical worldview giving a meaning of life (3).

Looking at question (1) alone, CAS is indeed not a simple explanation at all, and CNS is much better. However, CAS is ultimately busy with those three (or four) questions together. Broadening the context is often necessary to solve complex and difficult problems. I insisted strenuously in my papers (Vidal 2008b; 2010a; 2012a) that CAS shall foremost be seen a speculative philosophical scenario, precisely because of its more philosophically ambitious and comprehensive scope than CNS. In one sentence, *CAS is a speculative philosophical scenario to understand the origin and future of the universe, including a role for intelligent life.*

Now, as Vaas notices, since CAS is a close relative of CNS, it might also have scientific aspects. However, they are quite difficult to define and assess. Still, indirectly, the line of thinking behind CAS and universe making will give us precious clues to look for advanced extraterrestrials (see Chapter 9). Interestingly, the Cosmic Evolution Equation can help to frame research agendas to test CAS. For example, we discussed the CEE in the context of studying how fine-tuned our universe is. But in the context of CAS, if the outcome is to artificially make a new universe, the CEE can be used in a different manner. We can reduce the parameter space of possible universes by focusing on *universes possible to make from our universe*. The space of possibilities may thus be reduced not only by the physical constraints of new universe formation (as in CNS or other multiverse models), but also by the limits of cosmic engineering capabilities of cosmic intelligence(s). What are these limits? What are the different levels of universe making we can foresee?

### 8.3.3 Six Levels of Universe Making

CAS raises the question: did our universe originate out of some kind of *natural* intelligence? If so, what is the level of influence of the previous intelligence?

Could there be a kind of upbringing of baby universes? This is highly doubtful. Indeed, transmission of information between universes is unlikely. In analogy with biological organisms, evolutionary theorist John E. Stewart suggested that a "parent universe" could transmit information to its offspring universe (Stewart 2010, 401). Although it is an exciting *scientific* and *philosophical* speculation, physical constraints are likely to rule out this possibility. Indeed, constraints related to the physics of hypothesized baby universes production are too strong to pass on messages. Let us assume that a whole new disconnected space-time structure is generated from one universe to another. Such a space-time has a different causal structure from the previous universe. Therefore, it is by definition impossible to make the two communicate, because the common causal relationship between the two vanishes after the reproducing event.

Yet, even if transmission is impossible, it is important to distinguish *six levels of universe making*, summarized in table 9. I draw inspiration from and extend the three levels of universe making that Gribbin (2009, 197) proposed.

| **Level** | **Description** |
|---|---|
| 1) Blind | The blind level of universe making is illustrated by CNS, where there is no role for intelligence. The random variation of parameters at each universe's bounce provides minimum ingredients to explore the parameter space. No intelligence is necessary in the process. |
| 2) Accidental | Louis Crane (2012), in response to my commentary (Vidal 2012d) on his 1994 paper was more explicit in his vision of artificial black hole production. He argued that black holes would be useful mainly for energy production. He wrote that "production of new universes is only a byproduct". At this level, universe making is a variation on Smolin's CNS: more black holes are produced *accidentally*, without explicit intention of universe making. As Crane writes, if artificial black holes "result in the creation of new universes, that explains the overall evolutionary success of fine tuned universes". |
| 3) Artificial black hole production | The next level is to manufacture *intentionally* –not accidentally– black holes and baby universes, without any attempt to influence the new universe. Of course, this raises the motivational question: why would you make a new universe? Furthermore, imagining that you can produce black holes and baby universes why not nudge their characteristics to a desired outcome? |
| 4) Cosmic breeder | Cosmic breeders have the ability to nudge the properties of the baby universes in a certain direction. This level is the most faithful to the metaphor of artificial selection. Indeed, breeders on Earth are totally incapable of designing from scratch a living organism. But they can selectively cross-fertilize and reproduce plants or animals to foster desired traits. In a similar way, cosmic breeder don't know how to design from scratch a universe. But they can play in a constrained way with physics and cosmic parameters. |

| Level | Description |
|---|---|
| 5) Cosmic engineer | Cosmic engineers have the ability to set precisely the physics and cosmic parameters in the baby universe, thereby designing the baby universe in detail. Parameters are then analogous to DNA in living organisms (see also J. N. Gardner 2003 who explores the analogy between DNA and parameters). Genetic engineers select and modify DNA to foster the making of new organisms in a certain direction. Cosmic engineers select and modify parameters to foster the making of baby universes. It is probably the level that Andrei Linde (1992, 440) had in mind when he wrote about artificial universe production in the laboratory and asked: "Does this mean that our universe was created by a physicist hacker?" |
| 6) God player | Could the level of cosmic engineers be outclassed? The "God players" have the ability to control every parameter of the universe produced. There are playing God because the whole space of possible universes is accessible to them. Not only they have the power to set values of free parameters, but any universe is possible (e.g. creating Harry Potter kinds of universes, see our discussion about possible universes, section 6.3.1 Possible Universes, p123). They can choose to create the best or the worst possible worlds. |

Table 9 - Six levels of universe making

In sum, we could speak of the "intelligent design" movement as an *intelligent reactive design*, where one passively wonder and admire order in nature. Nothing is done to further explain and explore the mechanisms at play in nature. Instead proponents of intelligent design try to show the insufficiency of current darwinian mechanisms and infer or suggest –hastily– the evidence of a divine creation. This is in stark contrast with those different levels of universe making, which we could call *intelligent proactive design.* This time, if CAS holds for the future, *we* will make new universes. The logic is turned upside down. Instead of looking for a creator, we speculate to what extent our future intelligence could approach the skills to breed or design baby universes. The extent of this design can vary a lot, from accidental production, to playing God. Obviously, the question remains open regarding which level may be correct. So, when we speak about *universe making* in what follows, all levels are possible (except the "Blind" one). Let us now turn to many objections against CAS, that we shall formulate and address.

### 8.3.4 Objection – Design and Creation

*Cosmological Artificial Selection:*
*Creation out of Something?*
Title of Vaas' (2012) paper.

CAS presents real epistemological, logical, and metaphysical difficulties. I did address them while replying to a critical paper that Rüdiger Vaas wrote in 2009, entitled "Life, the Universe, and almost Everything: Signs of Cosmic Design?" (Vaas 2009). A shorter version of these critiques was published under a no less provocative title: "Cosmological Artificial Selection: Creation out of Something?" (Vaas 2012) . What follows is based on my replies to Vaas' critiques (Vidal 2012a). I will show that the difficulties raised by Vaas can largely be weakened and put into perspective.

I start by inviting to carefulness when using terms such as "creation" and "design". Then I invoke a principle of rational economy to tackle fine-tuning, and to compare the competing explanations. I discuss the causal issue and propose a possible metaphysical framework to approach the question "how did the cosmic engineers emerge in the first place?". I also discuss the thermodynamical issue in the context of CAS. I then examine CAS epistemologically to determine wether it is part of science or philosophy. Then I discuss if universes can be simulated and realized. I discuss the motivations underlying universe making, although I will dig in more details into this key issue in Chapter 10. I briefly discuss the idea of freedom within CAS and argue that the scenario implies no fatalism. In the last two subsections, I first draw four converging roads to CAS and finally recapitulate the case for CAS. I hope this dialectic between objections and responses will help to clarify the scope and even beauty entailed by CAS.

First, I would like to forcefully stress that the whole CAS scenario is *naturalistic*, and, as Vaas notices, is fully compatible with ontological naturalism. This is why I would rather be careful with the term "*create*", because it generally supposes an origin out of nothing, whereas here it is indeed question of a "creation out of something". The roman philosopher Lucretius famously said that "nothing can be produced from nothing" (*ex nihilo nihil fit*) is a principle that not even the gods can violate. We already analyzed in details in Chapter 4 the difficulties associated with creation, more precisely with "point-attractor" explanations. For these reasons, instead of the verb "create", I will rather use "*produce*" or "*make*" (following e.g. Davies 2008; John Gribbin 2009).

Since Vaas speaks about "*design*" when discussing CAS, well trained scientists will surely brandish red flags. Design is associated with intelligent design or other bad explanations. Scientists indeed loath the term "design" in explanatory contexts. Why? Because it freezes scientific explanation. If we are confronted to something to explain, the "design" answer, like the god-of-the-gaps explanation, will at the same time always work, and explain everything (nay, rather nothing). We already discussed these limitations when discussing classical explanations of fine-tuning (see section 6.4 Classical Fine-tuning Explanations, p144). In a cosmological context, words like intentionality, purpose or design are immediately associated with supernatural causes. But this needs not to be!

Importantly, the intentional explanatory mechanisms involved in CAS do not interfere at all with normal scientific explanation. On the contrary, maybe surprisingly, CAS as a whole can be seen as an invitation to a fantastic scientific and technological challenge: making a universe. This is why in my first papers about CAS (Vidal 2008b; 2010a) I presented CAS as happening in the future. Ultimately, pondering the fundamental metaphysical uncertainties about the origin of it all, I think it is more fruitful to try to contribute to shape the future than to understand the past. However, the full CAS scenario is also about the origin of the universe and the meaning of intelligent life in it. So, let us have a closer look.

### 8.3.5 Objection – CAS versus other Fine-Tuning Explanations

Vaas describes four major responses to fine-tuning corresponding to the explanations we have given (in section 6.4 Classical Fine-tuning Explanations, p144): *fecundity*, *chance-of-the-gaps*, *necessity* and fine-tuning explained as a product of *selection*. This selection can be an observational selection or *WAP-of-the-gaps*; a divine selection (*God-of-the-gaps*); a natural selection (*Cosmological Natural Selection*) or an artificial selection (*Cosmological Artificial Selection*). From a logical point of view, he correctly points out that such explanations are not mutually exclusive. However, each of them was developed to be an independent and sufficient response. Since all of them lack definitive support, what benefit do we gain by combining them? Taking seriously those combinations as explanations resembles more *fictional* speculation than anything else. One might argue that such an attempt would please at the same time proponents of the different options, but even this is not certain. The principle of rational economy afore mentioned should be at play here:

> Never employ extraordinary means to achieve purposes you can realize by ordinary ones (Rescher 2006, 8)

In the complete CAS scenario, we assume an intentional cause, which is not present in other naturalistic scenarios. However, it is still logically possible to assume that CAS will happen in the future, but did not happen in the past. In that case, there would be no original intentional cause, no "natural intelligent design" involved to explain the origin of the universe. If we consider CAS as only valid in the future, it is perfectly possible to hold the following logically consistent positions:

(a) God as the first universe maker and CAS
(b) Fundamental theory and CAS
(c) Multiverse and CAS
(d) Skepticism and CAS

A religious person might go with (a), a scientist might like (b) or (c). The skeptic (d) might say that we should stop arguing about the origin of the universe, since anyway it is unlikely that we get unambiguous support for such or such option. Still, he could agree that CAS is an interesting prospect for the future of intelligence in the universe. Those four options would still allow intelligent life to take up the future of their universe.

However, as Vaas also remarks, adhering to one of these four options would violate the Copernican principle. Indeed our universe would be central in the supposed cosmological replication cycle. So, how can we avoid this bias? Following the Copernican principle and being faithful to the principle of rational economy against the combination of fine-tuning explanations, what could the scenario (e) "CAS and CAS" be?

Vaas points out that "CAS tries to explain something complex with something even more complex". This critique was also made in (Byl 1996; Barrow 1998, 132; Vidal 2012c). It is indeed correct, and Vaas explains:

> Furthermore one might wonder whether CAS has any convincing explanatory force at all. Because ultimately CAS tries to explain something complex (our universe) with something even more complex (cosmic engineers and engineering). But the usual explanatory scheme is just the converse: The *explanans* should be simpler than the *explanandum*.

This is correct, however the underlying fundamental problem is that the usual explanatory scheme does not hold when we bring a kind of "ultimate theory" at play (see our detailed analysis in section 4.1.1 Epistemological, p74). By ultimate theory, I do not necessarily mean a "theory of everything" like it is sometimes speculated in physics; but a general "all encompassing" scheme of explanation. Accordingly, the explanatory scheme of CAS is not usual, but comparing the scope of classical explanations and CAS, we can argue that the explanatory force of CAS is much wider (see also Vaas 2009 where Vaas acknowledges this broad view on CAS). We can now summarize three levels to interpret CAS, where each level includes the precedent:

**(i) CAS in the future**

This is the scenario I have described in my papers (Vidal 2008b; 2010a), which provides a response to cosmic doom, a promise to scientifically progress on the fine-tuning issue and a role for intelligent life in cosmic evolution. For what happened in the past, positions (a)-(d) are all logically possible options.

**(ii) CAS in the future and in the past**

This scenario chooses option (e) "CAS with CAS" to tackle the origin of the universe. This implies that our universe has been made and fine-tuned to some degree by an intelligent civilization.

**(iii) CAS in the future, past and a metaphysics**

However, position (ii) implies further metaphysical problems. A metaphysics for CAS is needed to avoid a *shift of the fine-tuning* issue, and to propose a framework to answer metaphysical questions like "who created the creators?" or "why not nothing?". I attempt in the following lines the sketch of one such possible framework.

Although it is at odds with our knowledge of cosmic evolution, to avoid a shift of the fine-tuning issue and a tower of turtles, one can suppose that the tuning of new universes is *not* enhanced as the universal reproduction cycle repeats. Indeed, if we assume a complexification between universes, we will automatically shift the fine-tuning problem. In addition, we must assume that there is no "first universe". This sounds strange for our inclinations towards point-like cognitive attractors. We are used to think in a linear way, with a beginning, a middle and an end. However, it is possible to postulate a cyclical metaphysics (the cycle-like cognitive attractor), where there is no beginning at all, only a cycle. To sum up, in this metaphysical picture (iii), CAS describes an infinite cycle of self-reproducing universes mediated by intelligent civilization.

Again, it is important to emphasize that circular explanations and infinite regresses are not necessarily vicious (Gratton 1994). One attributes viciousness to such reasoning, but this is based on the assumption that "there is some obligation to begin a beginningless process or to end some endless process" (Gratton 1994, 295). Again, instead of trying to avoid an infinite explanatory regress, we can choose to embrace it, without any contradiction.

### 8.3.6 Objection – The Causal Issue

"Who created the creators?" is typically a metaphysical issue steming from our metaphysical yearning to reach an ultimate point-like explanation. But whatever reply X will be to this question, we can always ask: "where does X which created the creator comes from?". In the last analysis, whatever we reply, at least the metaphysical question (2), "why not nothing?" will remain. These questions are metaphysical and should not be confused with the fine-tuning issue (E. R. Harrison 1998). The fine-tuning issue is concerned with question (1) above, and "who created the creator?" is of a metaphysical nature, like question (2).

If we take into account this distinction, then it follows that no response to fine-tuning escapes metaphysical issues. Indeed, if we could prove that there is indeed a fundamental theory, we could still wonder why it is here, and why it gave rise to our universe, rather than having just nothing. If we could prove that there is indeed a God, we run into the same problem of "who created the creator?" that time in a theological context. If we could prove that there is indeed a multiverse, we must answer: how did the multiverse start in the first place? Where do the generative mechanism, the space of possible universes and the variation of this space from one universe to another come from? In conclusion, to properly respond to "who created the creator?" in the framework of CAS, as with other options, we need to develop a metaphysics. Barrow (1998, 132) mocked Harrison's CAS scenario when he wrote:

> Unfortunately, this amusing idea cannot explain why the constants were such as to allow life to originate long before the ability to tune baby universes existed

However, it should now be clear that Barrow has introduced an implicit hypothesis. Namely, that *the ability to tune baby universes evolved progressively.* But this is not necessary for CAS. Indeed, it complicates the theory uselessly.

### 8.3.7 Objection – The Thermodynamical Issue

But where do we find the free energy to sustain an infinite cosmological replication cycle? I find this open question exquisitely difficult. There is indeed an essential tension between the first law of thermodynamics (no new energy can be created) and the second law (entropy can only increase). Taking the two together, it means that if the supply of energy is finite, energy is doomed to be dissipated.

On a closer analysis, one can criticize the applicability of thermodynamics on a cosmological scale. We could also try to relax either the assumption of the finiteness of energy available, or to remark that a rigorous formulation of the second law says that entropy can increase *or stay the same*. Regarding the first option, although not logically excluded, I find it hard to believe in an infinite supply of energy. Even with a finite supply of energy, the second option raises hope to stabilize the universe thermodynamically and stop generating entropy. But then it is not clear if something like life which needs an energy gradient could still subsist in an active state (see section 10.4.5 Cosmological Immortality, p304 for a discussion and a short critique of such a scenario using the concept of reversible computation).

But there is another way out. It is to assume that the densities involved in universe making are so high, that no thermodynamical information is retained in the process. Each new universe makes a fresh thermodynamic start. We already saw that Misner, Thorne, Wheeler and Davies pointed at such a possibility (see section 4.3.3 Big Bang(s) Cycles , p80).

### 8.3.8 Objection – Epistemological Issues: Science or Philosophy?

Is CAS science or philosophy? It is definitely philosophical by its wide scope, rather than scientific. It is a big picture philosophical scenario, which may in the future lead to more specific scientific predictions. In my opinion, we are still far from this. Yet, Gardner (2003, 135–136) proposed four "falsifiable predictions" of CAS (which he calls the selfish biocosm hypothesis). The first is the success of SETI; the second is convergent evolution toward sentience in nonprimate species; the third is the creation of a conscious artifact in Artificial Life evolution; and the fourth is the emergence of transhuman intelligence. Unfortunately, they are more general conjectures which, if verified, would tend to support the general CAS view. But they are not really specific and precise tests of CAS. Gardner (2005) also associated the Biocosm hypothesis with modern physical theories, such as M-theory or the ekpyrotic cyclic universe scenario (Steinhardt and Turok 2002). It is a laudable effort to try to fit speculative theories with scientific ones. However, scientific theories change and can be refuted, and it would be a shame to associate too strongly this inspiring view of the cosmos with specific scientific theories. For example, the epyrotic scenario requires a big crunch, which is not favored by our current knowledge of the accelerating expansion of the universe. String theory or M-theory is also under very serious criticisms (see e.g. Smolin 2006; Woit 2007).

Was our universe made by an intelligence? This question is in principle open to scientific investigation. More specifically, it is a field of research yet to be started that we call *Search for ExtraUniversal Intelligence* (SEUI). We will go back to this topic in Chapters 9.

Furthermore CAS is a philosophical theory because we are involved. We are actors, not spectators of cosmic evolution. Any theory conferring a cosmological role to intelligent life has a self-referential aspect. One could imagine proponents of CAS have a cosmo-political agenda thus making it a self-fulfilling prophecy! We humans are more than an independent subject looking to nature as an external object in an objective way. This limitation of objectivity is actually not new, since two revolutionary physical theories, quantum mechanics and general relativity had to include the observer to make any sense.

Furthermore, if we assume that we are alone in the universe (a big assumption, but still reasonable until a proof of the contrary), then the future of the cosmos ultimately depends on our choices and values. *What do we want to do with our intelligence in the cosmos?* This is a much broader question than a purely scientific one.

### 8.3.9 Objection – Are Simulation and Realization Possible?

Can universes be simulated and instantiated? Vaas asks whether CAS can be realized. The two underlying questions are:

(a) Can a universe be simulated?

(b) Can a universe be instantiated?

Those two questions underly major challenges, and efforts to answer them are still in their infancy. The domain of Artificial Cosmogenesis (ACosm) is meant to tackle those challenges explicitly. As in Artificial Life (ALife), ACosm can be divided in two efforts, *soft ACosm* and *hard ACosm*. It is clear however that the analogue of soft ALife (universe simulation) is only in its infancy, and the analogue of strong/wet ALife (universe realization) lies in the far future.

*Soft ACosm* consists in making computer simulations of other possible universes and is therefore busy with question (a). Indeed, as we saw, cosmologists have already started to simulate and model other possible universes (see section 6.3 The Cosmic Evolution Equation, p122). Performing simulations for this purpose does not require to simulate every detail. A simplified simulation to understand the general principles of cosmogenesis would suffice. It is useful here to remember the metaphor of artificial selection in biology, where the breeder does not need to understand all the biological processes involved. Knowing how to foster traits over others is enough.

*Hard ACosm* consists in actually making universes (question (b)). As Vaas mentions, there are already early attempts in this direction with preliminary propositions to make universes in the laboratory. Universe making is a challenge which is probably orders of magnitudes more difficult than soft ACosm, but not impossible. We should certainly not underestimate the continuing, accelerating and impressive progress of science, technology and engineering. Black holes are good candidate for the realization of universe making. First, from a physical point of view, they exhibit enormous densities (similar to the big bang singularity), which are susceptible to modify the structure of space-time and give birth to baby universes. Second, from a computational point of view, Lloyd (2000) argued that the ultimate computing device would be a black hole.

So one might speculate that, in the very far future, the hypothetical use of black hole computers will meet with universe making. Interestingly, Lloyd (2000, 1052) argued that at the density of a black-hole, the computation is entirely serial. Could the baby universe produced be computed as it expands? It would be an intriguing situation, where *computer hardware and software collapse*. The most advanced computation might ironically function like the first computing machines, where hardware and software were not distinguished. Indeed, the first computing machines were physical machines which could only run one software, which was embedded it its construction. Could a black hole ultimate computer run only one blueprint of a universe? But how exactly universe making may be achieved may be left to our descendants (E. R. Harrison 1995, 198). For other hints towards realizing universe making, see also (J. N. Gardner 2003, 167–173; Vidal 2008b; Smart 2009).

This idea of a giant computer raises a frightening question. Are we in a simulation? Indeed, Bostrom (2003) reasoned that given reasonable premises, we are likely to be living in a computer simulation.

I totally agree with Vaas that the question "how can something simulate something else which is comparably complex?" is highly problematic. This is why in soft ACosm we do not have to simulate every detail. As Vaas acknowledges, to run computer simulations, we also need some kind of *hardware*. As we saw (section 7.6 The Simulation Hypothesis, p161) the ontological status of a simulation would be reflected by the states of the hardware running it, whatever the realistic nature of the simulation. Because of this hardware requirement, running a simulation indefinitely in our universe seems very difficult, if not impossible. It means that even if we are in a simulation, we do need to worry about a cosmic doom. In the end, I consider the question of whether we are in a simulation or not as a *fictional speculation* which therefore does not deserve that much attention (see also Vidal 2012c).

### 8.3.10 Objection – What Motivation?

The question of motivation naturally arises when considering CAS. *Why would intelligent life want to make a new offspring universe?* Jim Gardner (2003, 224) argued that a high intelligence would produce a new universe because of altruistic motives. I do agree, but I think an accomplished cosmic wisdom would transcend the selfish v.s. altruistic alternative. It would identify with the whole of space-time-energy, as we sometimes wonder with awe that we are stardust. We will come back to such wisdom in Chapter 10.

Harrison (1995, 200) suggested three motives. First, to prove the theory is correct, and the technology adequate. This may be a good motive, but largely insufficient to engage in the great enterprise of universe making. Second, to make universes even more hospitable to intelligent life. This may indeed motivate an intelligent civilization. However, we should keep in mind that it leads to a shift of the fine-tuning issue to previous universes. And let us remember that explaining the fine-tuning issue is the main reason we engage is such speculations. The third motive is to inhabit the universe created. I think it is unlikely. First, we saw that the transmission of information would be difficult in the process of universe making. Again, as with Gardner's proposal of altruism, if we care enough for and identify with evolution at large, there is no need to inhabit the new universe. It would be a similar motivation as parents sometimes have towards their child: they want their offspring to be exactly like them. In short, to inhabit them. Of course, a more mature educational school would leave more freedom to the child. Since motivation has to do with our worldviews and values, we will further address these questions in Chapter 10 where we will deal with ethics on a cosmological scale.

Stewart (2010, 401) emphasized the importance to motivate intelligent life to take part into a supposed cosmic developmental process. However, even if we had the certainty to be in an developmental process, this would be just part of the motivation to produce a new universe. Two other drivers are likely to be central.

First, as described by Stewart (2010, 404), the most fundamental values an intelligent civilization evolves to are life-affirming and meaning-seeking. Those values are likely to be strongly connected to the idea of surviving indefinitely: *immortality* (Lifton and Olson 2004). A strong commitment to these values would reinforce the willingness of an intelligent civilization to participate actively to the evolutionary process. The will to immortality, which is nothing less than the drive to stay alive, is arguably a strong motive to make a new universe. Indeed, even if CAS is wrong, this search for infinite evolution will continue, whatever its form. Existing speculations to achieve infinite evolution include a simulated universe using reversible logical gates thus using no energy; hibernation near black holes (Dyson 1979); the speculative and controversial Final Anthropic Principle of Barrow and Tipler (1986); or the theological omega point theory of Tipler (1997). We will discuss these alternative options below (section 8.3.13 The Case for CAS, p197).

However, there is a second driver for intelligence to reproduce the universe. It is the growing awareness of a cosmic doom. This fate would arguably drive an intelligent civilization to act, and make a new universe. The core problem an intelligent civilization has to deal with in the very far future is the inevitable thermodynamical decay of stars, solar systems, galaxies and finally of the universe itself. In the realm of biology, the solution to aging is reproduction. Could it be that an analogous solution would take place at the scale of the universe? This is the proposal of CAS. Pursuing this cosmic replication process would in principle enable the avoidance of heat death in a particular universe (Vidal 2008b). Cosmic evolution would then continue its course indefinitely.

Is this issue far too far in the future to be a serious preoccupation? The situation here is analogous to global warming, except the problem concerns an even larger scale than the Earth. A few decades ago, few people were seriously concerned with global warming. Nowadays, people, various organizations and governments have started to seriously mobilize to tackle this complex issue. What produced this shift? Among other factors, the strong integration and globalization of societies contributes to this sensitivity about climate change. Indeed, it is only since recently that we have such numerous and precise channels of information from every corner of the world, providing us an unprecedented understanding and awareness of the planet as a whole. This leads us to a *global awareness* and compassion of what happens on a planetary scale.

Similarly, a *universal awareness* will eventually emerge from our increasing understanding of the cosmos. Only after such an awakening will an intelligent civilization start to actively tackle such a large-scale challenge. If this thesis contributes to this awakening, it will have achieved its purpose.

CAS implies a huge responsibility for intelligent life in the cosmos. Are we willing and ready to assume it? What if we fail? What if we self-destruct? Maybe we share this responsibility with our distant cosmic cousins, other extraterrestrial civilizations? It is worth pondering if more advanced civilizations would already have gone towards this way. In fact CAS will even give us fruitful heuristics to search for advanced extraterrestrials (see next Chapter 9).

### 8.3.11 Objection – No Freedom in CAS?

Is there freedom in CAS? Does CAS promote fatalism? This question would require a much longer treatment, since the idea of freedom has been debated across centuries (see e.g. Adler 1973). But let us mention some possible responses. CAS in its fullest form, i.e. (iii) CAS in the future, past and a metaphysics (see section 8.3.5 Objection – CAS versus other Fine-Tuning Explanations, p187) includes cycles. Old worldviews are cyclical and imply that nothing really changes. The same comes back again and again if we wait long enough for the cycle to complete. However, I rather see CAS as supporting a progressive evolutionary worldview than a fatalistic one. Indeed, the idea of progress in science and technology contrasts with the idea of destiny. Science and technology bring out radical novelties in our societies. Who did predict the Internet? Arguably not many thinkers.

The analogy of family is more appropriate here. Imagine you are 10 years old. Your parents ask you to sit down because they want to tell you something important. They tell you: “My dear child, when you will grow up, you will certainly find a lover and have children.” Somehow disappointed, your reply “That's all?”; and your parents: “Yes it is!”. This is indeed a likely outcome given the configuration of our society and our biology, but not even necessary. There is no fatality for you to have children. More importantly, your disappointment regarding the triviality of the statement means that it does not spoil the way. How, when or with whom this will be achieved and what the children will be like remains full of surprises –except for sociologists. Now, CAS is a comparable trivial statement at the level of the universe. It doesn't involve any fatality, just a general direction.

To conclude, in contrast to what Vaas wrote, I would like to stress that CAS is more than a "physical experiment", "a simulation" or an attempt to build a "rescue universe". The response of an intelligent civilization as they awaken to cosmic doom (heat death or another gloomy scenario) is likely to be a strong driver to make a new universe. Therefore, CAS is not about playing with virtual universes, nor making a physical experiment to see what it is like to produce a universe. The "rescue universe" idea remains interesting, although it would be more about rescuing cosmic evolution at large, rather than the memory of a particular intelligent civilization. But to care about cosmic evolution, we need a cosmological ethics (see Chapter 10).

### 8.3.12 Four Roads to Cosmological Artificial Selection

*Nature can never be completely described,*
*for such a description of Nature would have to duplicate Nature.*

Tao Teh King (Laozi 1958).

I chose to present CAS as a natural extension and remedy to CNS. Importantly, taking this road we focused on our *past*, since CNS was initially developed to provide a cogent answer to the free parameters and the fine-tuning issues. But reasoning from other starting points can also lead to CAS. Let us see three other roads. It is almost a truism that the major goal of science is to describe nature. Yet, there is a profound truth in the above opening lines of the *Tao Teh King*. Every description is an incomplete simplification of nature. So, to the limit, a full description of nature would indeed have to duplicate Nature. Instead of an impossibility, in CAS the tendency towards a duplication of Nature translates into a tendency towards universe making.

On 24th November 2007, at the Downing College in Cambridge was held a conference entitled "God or Multiverse". It was a laudable attempt to confront views from theologians and theoretical physicists. Bernard Carr organized the meeting and had recently edited an excellent book called *Universe or Multiverse?* (B. Carr 2007). At the end of the day, he invited the audience to propose short talks to complement the viewpoints presented during the day. I seized the opportunity and was excited to make a short talk about "God *and* Multiverse". I drafted some words to explain this idea. Indeed, CAS can be interpreted that way. From God, it keeps an ability to make universes –with, as we saw (section 8.3.3 Six Levels of Universe Making, p183) six different levels possible. From the Multiverse, it retains the possibility of other universes through the idea of the *virtual multiverse* –but not their actual existence. Instead of "God or Multiverse", we thus have "Natural God and Virtual Multiverse". Unfortunately my proposal was not picked up by the organizing committee. That day, the gap between God and Multiverse was only widened. Yet this "God or Multiverse" dichotomy is artificial. As Harrison (1995, 199) wrote, fortunately "for the inquiring mind, a natural creation theory offers a third choice".

Story 9: "God or Multiverse" or "God and Multiverse"?

As the story above relates, the second road to CAS is to choose the option of "Natural God *and* Virtual Multiverse". One might think that CAS also assumes a multiverse. This is not the case, since a real multiverse is not necessary, a virtual one is enough for the purpose of cosmic selection. Indeed, it might be that there is only one universe which recycles itself (like the phoenix universe), if a big crunch scenario is after all favored at the end of time. Another way is if intelligent life becomes powerful enough to reverse the expansion of the universe. But this is much much more speculative.

A third road to CAS is to follow the logic of evolutionary theory. The last chapter of Dawkins' (1995) *River of Eden*, is entitled "The Replication Bomb". Dawkins describes two kinds of bombs in our universe. Supernovae in astrophysics and replication in biology. He reports 10 replication thresholds which went on on Earth. In summary, these are:

> 1. Replicator threshold (self-copying system, e.g. DNA-RNA molecules)
> 2. Phenotype threshold (consequences of replicators that influence the replicators' success but are not themselves replicated).
> 3. Replicator team threshold (genes do not work in isolation, e.g. in bacterial cells)
> 4. Many-cells threshold (phenotypes and functions are on a much greater scale than the cell).
> 5. High-Speed-Information-Processing threshold (nervous system)
> 6. Consciousness threshold
> 7. Language threshold
> 8. Cooperative technology threshold
> 9. Radio threshold
> 10. Space travel threshold

Gardner (2003, 116) noticed that if CNS is validated, it would constitute an $11^{th}$ replication threshold, which he calls the "Cosmic Replication Threshold". However, since the replication threshold is a compositional (whole-part) hierarchy, the higher levels include the lower ones. So, it is more logical to assume that CAS would constitute this threshold, and not CNS. Indeed, CAS is in continuity with the evolutionary process of higher replication thresholds. John E. Stewart (2000) also came to the conclusion that evolution writ large goes towards cooperation on larger and larger scales. The last scale being the universe as a whole.

The fourth road to CAS is built by thinking about the future and meaning of science. As strange as it may seem, looking towards the far-future might provide new heuristics to approach those ultimate questions. As Gardner (2003, chap. 9) argued, CAS is indeed very much consistent with various futuristic visions of Kurzweil, Wheeler, Barrow-Tipler, Dyson and Dawkins.

I already argued that the future of scientific simulation goes towards a simulation of an entire universe (see Chapter 7). As an echo to Galileo, if the science of today reads the book of Nature, the science of tomorrow will write its next chapter. Futurist Michio Kaku (1997, 15) also shares this vision when he writes:

> For most of human history, we could only watch, like bystanders, the beautiful dance of Nature. But today, we are on the cusp of an epoch-making transition, from being *passive observers of Nature to being active choreographers of Nature*. It is this tenet that forms the central message of *Visions*. The era now unfolding

> makes this one of the most exciting times to be alive, allowing us to reap the fruits of the last 2,000 years of science. The Age of Discovery in science is coming to a close, opening up an Age of Mastery

This perspective offers a new way to think about the cosmos, which I call the *architect point of view*. It can be formulated as:

> **Architect Point of View**: *The more we are in a position to make a new universe, the more we will understand our own universe.*

If we extrapolate the steady growth of complexity and especially the amazing progress of science, then it may not be so difficult to imagine a stage at which intelligent life is able to conceive and make new universes. Accordingly, the question of motivation remains. We will see later that it can be grounded in a cosmological ethics and especially in the idea of cosmological immortality (see Chapter 10).

What is the meaning of science in CAS? Pondering the mysteries of the cosmos, we often marvel about the existence of life in the universe. Why are there so complex structures such as life or consciousness? But the most recent outcomes of complexity on Earth are not life nor consciousness. It is arguably the phenomenon of science, which is a natural continuation of the evolution of complex intelligent life (Turchin 1977; D. T. Campbell 1974).

Albert Einstein famously wrote that the "most incomprehensible thing about the universe is that it is comprehensible" (cited in Hoffmann 1972). What he meant is that matter self-organized up to a point where it can comprehend itself. Through humans and science, the universe is self-comprehending. This self-referential aspect is perplexing and awe-inspiring.

Scientists are by definition embedded in science. Somehow, this is quite ironic. Indeed, if we are very "scientific", meaning giving a maximum value to *objectivity* (i.e. observations independent of the particular properties of the observer) then we forget the very fact that there *are* observers! Which is itself a fundamental mystery. This is why I presented CAS as a philosophical theory and not as a scientific one. In CAS, we are involved... and other putative extraterrestrials too (see Chapter 9)! That is another reason why it is so important to try to search and find them. Our cosmic vision remains limited if we have only one example of the emergence of life. Obviously, it is difficult and scientifically dubious to assess any general trend out of one example.

What is the meaning of science in a cosmological perspective? Why did the cosmos generate structure not only capable of understand itself (self-consciousness); but also understanding and controlling its surroundings in a more and more developed and precise manner? What is the most complex model we can imagine, and which activity can provide it? We saw that it is a model of the whole universe, more and more assisted by computer simulations. As we extrapolate the phenomenon of science in the far future, we approach knowledge of the entire universe. CAS hypothesizes that this knowledge coupled with activities in soft and hard Artificial Cosmogenesis will provide all the ingredients for future universe making.

One can now throw a new light on the fact that cosmic evolution gave rise to scientific activity. In CAS, the increasing modelling abilities of intelligent beings is not an accident, but an indispensable feature of our universe, to ensure new offspring universes. In CAS, scientific activity does not seek an ultimate explanation but a pragmatic solution to a real problem: a lurking cosmic doom. The quest for an ultimate explanation is not anymore a quest for disinterested and absolute knowledge. Knowledge is useful and near complete knowledge about our universe should at some point be useful.

In the future, science is not anymore only a search for understanding the world; in the long run, it will tend to become a simulation or computation of the world. In the far future, such a simulation could be concretely implemented to make a new universe. CAS offers a fresh look to ponder the big questions Heinz Pagels (1986, 379) raised:

> Is it possible that life, or whatever it may become, can alter the program of the cosmic computer, changing the course of its destiny? It will take more than a metaphor to answer that important question; it will take a far deeper understanding of life and the cosmos than we currently possess. Yet the desire to know the answer to such questions about our destiny will never go away. And that desire is perhaps the profoundest program in our cosmic computer so far.

### 8.3.13 The Case for CAS

*When you have eliminated the impossible,*
*whatever remains, however improbable, must be the truth.*

"Sherlock Holmes' rule", by Arthur Conan Doyle (1890).

We saw in Part I that a worldview's merit can only be compared *relatively* to others. We now aim to show that CAS is, relative to other theories, the best solution to make sense of the beginning, the end and the meaning of life in the universe.

Inquiring into the mysteries of our universe is like a difficult crime investigation. You have few clues, but you need to build a consistent story to make sense of seemingly disparate pieces of a puzzle. Sherlock Holmes is of course expert in investigations and his rule above stimulates us to compare CAS with other theories.

Since we argued that CAS is foremost a philosophical theory and not a scientific one, we can use the criteria and conceptual tools we developed in Part I. Let us make explicit the metaphilosophical criteria we use. First, we consider a wide *scope in agenda*, looking for a framework to understand in a consistent manner the *origin*, the *future* and the *meaning of life.* In other words, from a philosophical point of view, we want to show that CAS is at present the "best" attempt to answer *at the same time*: "where does it all come from?"; "where are we going?" and "what is the meaning of life in a cosmological perspective?". Since our work is cosmological, our *scope in time and space* is also maximally large. We also are very much committed to the scientific method and the values of objectivity, and thus we do not assume any kind of supernatural forces.

Of course, the value judgment behind the word "best" must be read in the context of worldview construction, and is criteria-dependent. The word "best" is thus relative to a choice of criteria we discussed in Chapter 2 and a weighted choice of criteria I made explicit in Appendix I.

A classical example of conflicting criteria is found in the "God or Multiverse" debate. For example, Martin Rees (see e.g. his 1997 book) argues that a multiverse is more rational, while Leslie (1989) argues that postulating a God is more economical. Who is right? What should we value more? Rationality or economy? The core of the disagreement lies in our cognitive values. Are we ready to assume the actual existence of a huge –possibly infinite– number of universes? This is a natural explanation, but rather extravagant. Or do we prefer to assume one single all-mighty God at the origin of the universe? This seems indeed more economical... but at the price of a supernatural explanation. A price that most scientists are not willing to pay. My position was to send away both explanations when I called them "God-of-the-gaps" and "WAP-of-the-gaps" (see section 6.4 Classical Fine-tuning Explanations, p144). Let us take a bird-eye view on fine-tuning explanations and their broader implications for the future, the meaning of life and the associated metaphysics (see table 10 below).

| Question / Explanation | Origin | Future | Meaning | Metaphysics |
|---|---|---|---|---|
| *Skepticism* | - | - | - | - |
| *Necessity* | Unexplained | - | - | - |
| *Fecundity* | Explains fine-tuning, not the causal issue | - | - | - |
| *Chance-of-the-gaps* | No causal explanation | - | - | - |
| *WAP-of-the-gaps* | No causal explanation | - | - | - |
| *CNS* | Natural selection | Baby universe generation | - | Cyclical or branching ? (unspecified) |
| *Hibernation (Dyson)* | - | Infinite subjective time | - | - |
| *Reversible computation* | - | Infinite future information processing | - | - |
| *God-of-the-gaps* | Point explanation | Spiritual immortality | Religious values | Theologica-lly addressed |
| *Omega Point Theory (Tipler)* | God | Computational resurrection | Religiously inspired | Cyclical |
| *CAS* | Artificial selection | Universe making and cosmological immortality | Fundamen-tal role of intelligence in a cosmic replication cycle | Cyclical |

Table 10: Candidate explanations for the origin and future of the universe. The last two columns show the associated meaning of life and metaphysics

What is striking in the table above is that few explanations have a scope large enough to touch upon all the questions. In fact, there are only three, God, Tipler's Omega point theory and CAS.

The line for “skepticism” is empty and this reminds us that it is a second-order knowledge attitude and not an attempt to answer our big questions. I do very much value skepticism as a second-order knowledge attitude, but it is good to remember that it is silent if we want to take a first-order position on the big questions.

The *necessity, fecundity*, *chance-of-the-gaps, WAP-of-the-gaps* explanations are concerned only with the origin and do not aim to cope with a greater scope. We saw that the possible meaning of life and intelligence in the universe is not within the scope of Cosmological Natural Selection.

There are two promising options for life to continue in the far future. The first option is *hibernation* and was articulated in the landmark article about the future of civilizations written in 1979 by Freeman Dyson (1979). Dyson shows that even assuming a finite supply of energy, it would be possible for a civilization to live forever. The scenario takes profit of time dilation effects due to relativistic effects. An advanced civilization would hibernate near black holes, and use the energy of black holes. Despite the finiteness of the energy source, by hibernating longer and longer, and thus using less and less energy, a civilization would be able, to the limit, to live for as long as it wants in its subjective time. Of course, “life” is defined here not in terms of DNA and biochemistry, but in a more general information processing capability.

However, this scenario does not work if the universe continues its accelerated expansion (Dyson 2004, xv). A very stimulating debate arose between Dyson and Krauss and Starkman (2000). In their 2000 article “Life, The Universe, and Nothing: Life and Death in an Ever-Expanding Universe”, Krauss and Starkman criticized Dyson's proposal and showed that eternal life in our universe is impossible. Yet, Dyson showed that the core argument can be maintained if we replace digital computers by analog ones (see Dyson 2007).

The second option of *reversible computation* is highly promising for a civilization to endure forever. Rolf Landauer (1961) proved the theoretical possibility of logic gates that consume no energy. Given a computer built out of such gates, a possible solution to the problem of an ever-expanding and slowly dying universe would be to simulate a new universe on a collection of matter that would forever float into emptier and emptier space. Krauss and Starkman, although recognizing the theoretical possibility of this scenario, criticized it. They argue that “no finite system can perform an infinite number of computations with finite energy”, if it is to host evolved information processing. Why not? The main reason is that reversible computation is not possible for the operation of erasure. Deleting information has a thermodynamical cost, and the authors argue, consciousness or sophisticated information processing will certainly need to erase information (see also Zenil 2012). Note that these two options of hibernation and reversible computation remain neutral regarding issues of the origin, the meaning of life or a metaphysics.

The explanation of a God is satisfying from a worldview construction perspective. But it is unsatisfying if we aim at a naturalistic *philosophical* worldview and not a *theological* one. Of course, there is still the issue of God as filling our knowledge gaps, but this can be tempered with appropriate theological interpretations.

As a side remark, the God explanation induces a metaphysical asymmetry. Indeed, it is a point-like attractor for the origin, but supposes an immortality in the future, which is more like an infinite line of continuation. Why not an immortality in the past? And why not a point in the future?

Our reasonings about the origin of the origins in Chapter 4 largely holds for extrapolations into the future. It is thus not surprising to find the idea of a point-attractor in the future. Such a future ultimate point, the *omega-point*, was famously and beautifully articulated by the paleontologist and jesuit Pierre Teilhard de Chardin (1959) in his posthumous book *The Phenomenon of Man*. The vision of evolution Teilhard proposes is remarkably inspiring. Teilhard is arguably one of the very rare thinkers who foresaw the emergence of the Internet, when he introduced the *noosphere*, an evolutionary stage of development dominated by information processes on a global scale. This evolutionary vision was further elaborated into the concept of the *global brain* (see Heylighen 2005 for an historical review).

In a very controversial book, Frank Tipler (1997) revived Teilhard's vision of the omega point and articulated a theological vision using modern cosmology and computer science. His book is entitled *The Physics of Immortality: Modern Cosmology, God and the Resurrection of the Dead* and clearly has a theological agenda. This was further confirmed by Tipler's (2007) following book entitled *The Physics of Christianity.* Tipler's book on the physics of immortality has quite some value from a synthetical worldview perspective. It is highly ambitious and speculative, in order to solve not philosophical questions, but theological ones. The problem is that Tipler presents his work as a piece of science, whereas it is not. His use and abuse of *reductio ad absurdum* and general scientific logic is either wrong or very doubtful. In his book review, Ellis (1994) called Tipler's book a “masterpiece of pseudoscience”.

What worries me even more is a total and constant confusion of genres. One need only to read the opening words of the preface, where Tipler asserts that “theology is a branch of physics, that physicists can infer by calculation the existence of God and the likelihood of the resurrection of the dead to eternal life in exactly the same way as physicists calculate the properties of the electron.” That is for the confusion between theology and science. A few pages later he goes on to affirm that “reductionism is true”. Hard to swallow for any philosopher who worked more than half an hour in epistemology.

I see Tipler's essay as *remarkable modern theology*, *naïve philosophizing* and *unsound science*. Why is it remarkable theologically? Because Tipler tries to rescue traditional doctrines of Christianity in a modern cosmological and computational framework. And if you have faith in the arguments, it might even be inspiring. Of course, I am sure many theologians would still disagree with Tipler's approach.

Despite these critiques, there are some very interesting and provocative ideas in the book, provided that we exorcize Tipler's work from its theological inclinations. David Deutsch (1997, chap. 14) did reconstruct the core of Tipler's argument. Deutsch summarizes that the “omega-point theory is that of a class of cosmological models in which, though the universe is finite in both space and time, the memory capacity, the number of possible computational steps and the effective energy supply are all unlimited.” This assumes a closed universe which will re-collapse into a big crunch. Unfortunately for the omega-point theory, the empirical discovery in 1998 that our universe is in an accelerating expansion (Riess et al. 1998) refutes the core assumption of the theory.

We saw in the introduction of Part III that with the implicit assumption that intelligent civilization cannot have any significant influence on cosmic evolution, modern cosmology invites us to prepare for a cosmic doom. I share Darwin's (1887b, 70) reaction to the heat death of the universe when he wrote:

> Believing as I do that man in the distant future will be a far more perfect creature than he now is, it is an intolerable thought that he and all other sentient beings are doomed to complete annihilation after such long-continued slow progress.

What can we do? What if we challenge the assumption of intelligence having no influence on the future? What if life and *intelligence could have influence on the future of the cosmos*?

In the last Chapter of *The Anthropic Cosmological Principle* (Barrow and Tipler 1986) the two authors speculate about the far future of intelligence. They call the Final Anthropic Principle (FAP) the proposition that "intelligent information-processing must come into existence in the Universe, and, once it comes into existence, it will never die out" (Barrow and Tipler 1986, 23). The FAP is in fact an embryonic form of the omega point theory that Tipler later developed. Not surprisingly, it had already been subject to tough criticism, famously by Martin Gardner (1986), who called it the Completely Ridiculous Anthropic Principle (CRAP). I was very disappointed by Tipler's (1986) reply to Gardner which does not address most of the excellent objections raised by Gardner, and even more disappointed by Gardner's reply to Tipler: "I'm speechless." There is still so much to debate!

The reversible computation scenario is certainly to keep in mind for the extremely far future, but as long as billions of stars are shining in billions of galaxies, there is really no reason for an intelligent civilization to go on such a drastic energy diet.

In contrast to using less and less energy to endure forever, which is biologically a strategy of delaying senescence, there is another more radical solution. It is to replicate and start anew. Reproduction is a highly successful strategy that evolution uses to maintain and adapt living systems to their environment. Ashby's (1981b, 80) analysis showed that "reproduction is not something that belongs to living organisms by some miraculous linkage, but is simply a specialized means of adaptation of a specialized class of disturbances". On a cosmological scale, the proposal of CAS is indeed to aim for a cosmic replication. Assuming that the thermodynamical constraints are reset in the new universe made, there is hope for life *in another universe* to endure forever. Philosophically, CAS thus promises to fulfill subjective and intersubjective criteria, by giving a general direction and meaning for intelligence in the universe.

In sum, holding the scope in agenda, the criterion of naturalism, and the wide scope in time and space, there is no comparable theory to CAS. The closest attempt is arguably Tipler's omega point theory, albeit it is more focused on the future than on the past, and is largely theologically inspired.

I fail to see evidence for a God, a fundamental theory or a proof of a multiverse actually realized. Yet, I see overwhelming evidence of our exponential use of computing resources such as memory storage, computational power and bandwidth. Those advances have a tremendous impact on our lives and societies, and this is only the beginning. In particular, computers are more and more ubiquitous in scientific activities, for example in mathematics (to assist in proofs), in studying complex systems (by simulating them), in biology (e.g. with biotechnologies, and their databases of genomes or protein networks), in cosmology (with many projects of large-scale simulations of the universe) and of course with ALife and its legitimate successor, ACosm. If we choose and manage to successfully conduct soft and hard ACosm, (i) CAS in the future would be realized. It would then give us strong indications and inspirations to think that broader interpretations of CAS, (ii) or (iii) are accurate.

Can the general perspective of CAS help to search for extraterrestrials? Could it be that extraterrestrial intelligence is on its way to black hole manipulation and universe making? How can we further explore this speculative possibility? Can we find more general reasons towards this path, and confront the reasoning with observations? Let us now scrutinize these issues by searching for advanced extraterrestrials. I mean it, really advanced.

# *CHAPTER 9 - High Energy Astrobiology*

**Abstract**: This Chapter proposes a new concrete hypothesis to search and assess the existence of advanced extraterrestrial life. We first point out two methodological fallacies that we call *naturality-of-the-gaps* and *artificiality-of-the-gaps* and propose a more balanced *astrobiological stance*, which does not prejudices the naturality or artificiality of suspicious phenomena which we observe. We point out many limiting and implicit assumptions in SETI, in order to propose a "Zen SETI", thus opening the search space. In particular, we outline the case for postbiological evolution, or the probable transition from a biological paradigm to a nonbiological paradigm. We then discuss criteria to distinguish natural from artificial phenomena. We start with global criteria (*strangeness heuristic*, *non-exclusiveness heuristic*, *equilibrium heuristic* and *inverse distance-development principle*); then thermodynamical criteria (*thermodynamic disequilibrium* and *energy flow control*); and finally present living systems criteria (Miller's nineteen critical functional subsystems). Then we introduce a two-dimensional metric for civilizational development, using the Kardashev scale of energy consumption increase and the Barrow scale of inward manipulation. To support Barrow's scale limit, we argue with energetic, societal, scientific, computational, and philosophical arguments that black holes are attractors for intelligence. Taken together, these two civilizational development trends lead to an argument that some existing binary stars may actually be advanced extraterrestrial beings. Since those putative beings actively feed on stars, I call them *starivores*. I elaborate another independent thermodynamical argument for their existence, with a metabolic interpretation of some binary stars in accretion. We further substantiate the hypothesis with a tentative living systems interpretation. Ten critical living subsystems are suggested to apply to interacting binaries composed of a primary white dwarf, neutron star or black hole. We critically discuss the hypothesis by formulating and replying to ten objections. The question of artificiality remains open, but I propose concrete research proposals and a prize to further continue and motivate the scientific assessment of this hypothesis.

*Rather than propose a new theory or unearth a new fact,*
*often the most important contribution a scientist can make is*
*to discover a new way of seeing old theories or facts.*

Richard Dawkins (2006, xvi)

What are the general outcomes of the increase of complexity in the universe? We have only one such example: life on Earth. Unfortunately, we cannot carry out a scientific investigation with only one object of study. So, in order to know more about the ability of the cosmos to generate many times and in different circumstances complexity and intelligence, it is crucial to know whether or not we are alone. It is also very important for evolutionary and theoretical biology, to assess how convergent evolution is or the extent to which the origin of life was a unique cosmic event.

Steven J. Dick (2000, 196) summarized stages in cosmological worldview development from *Geocentrism*, *Heliocentrism*, *Galactocentrism* to *Biocentrism*[12]. We add two worldview stages, *Intellicentrism* and *Universecentrism*. Three "centrisms" have been refuted scientifically: Geocentrism, Heliocentrism and Galactocentrism. It is not my aim to develop the debate between Geocentrism and Heliocentrism featuring Ptolemy and Copernicus, nor the discovery by Shapley that our solar system is not at the center of our galaxy, nor that our galaxy is just one amongst many others. I invite the reader to consult the relevant literature in the history and philosophy of science. These were moments of great scientific advances, but it is good to remember that such changes in worldview did not happen overnight. Indeed, even if Copernicus' astronomical techniques were quickly appreciated by his peers, it took decades before his Heliocentric worldview would be taken seriously (T. S. Kuhn 1957). The reasons of such a resistance are not only scientific but also psychological, philosophical and religious. It is hard to shake a world where the Earth had always been considered the center of the universe.

The same holds for astrobiology. It will not be easy to quickly reach a consensus on what constitutes proof of extraterrestrial life. What is likely to happen is that we will be able to model some intriguing phenomena *both* from an astrophysical perspective as well as from an astrobiological perspective. As with Copernicus' refutation of Geocentrism, refuting Biocentrism or Intellicentrism will have a huge philosophical and religious impact. In the case of Intellicentrism, we would not be the only –let alone the most– intelligent species in the universe.

Certainly, there is still work to pursue the broad Copernican revolution because Biocentrism, Intellicentrism and Universecentrism still hold today. Indeed, we lack any definitive proof that life exists elsewhere in the universe, that *intelligent* life exists elsewhere, or that other universes exist. Let us say a few words about those three major challenges.

Biocentrism continues to hold in the sense that, even if most scientists subjectively *believe* the existence of extraterrestrial life is highly probable, we still haven't *proved* it yet. In recent years discoveries of exoplanets have grown exponentially, so there is a lot of hope that one of these rocky Earth-like planets will harbor life.

Intellicentrism sill holds, since, obviously, we haven't discovered intelligent life either. However, it is important to distinguish the two, not only because the impact of refuting one of them would be very different, but also because the search methods are different. In the search for Earth-like planets we look for signatures of a biosphere. But the main method in the Search for Extraterrestrial Intelligence (SETI) is to search for traces of an intelligent communicative signal in the electromagnetic spectrum. If Biocentrism is refuted, that is, if we find a primitive extraterrestrial lifeform, we will still hold to Intellicentrism and be proud to be the only and most *intelligent* species in the universe. The psychological and philosophical consequences of refuting Intellicentrism are much more radical and disrupting than those of refuting Biocentrism. Finding an extraterrestrial bacterium is indeed very different from finding advanced civilizations two billion years older than us.

12 Dick calls Biocentrism the "extraterrestrial/biophysical" worldview. It simply means that "life on Earth is unique in the universe", not the view or belief that the rights and needs of humans are not more important than those of other living things.

Universecentrism holds today in the sense that we lack proof that our universe is one of many. We have seen examples of multiverse theories (see e.g. section 5.3 The Mathematical Universe, p94), and it is a fact that many modern cosmologists are very open-minded about the existence of multiple universes. But, as with Biocentrism, being subjectively open-minded about the multiverse is not the same as having objective proof. We lack empirical evidence of eternal inflation or Cosmological Natural Selection. In the multiverse theory of conformal cyclical cosmology, Penrose (2011) proposed to look for subtle irregularities in the cosmic microwave background to find traces of a previous universal cycle. This is indeed one of the rare proposals for testing a multiverse theory, but the attempt remains preliminary. If we follow the theory of Cosmological Artificial Selection (see Chapter 8), proof of artificial universe making would also refute Universecentrism. In fact such a "Search for ExtraUniversal Intelligence" (SEUI) has already been hinted at (see e.g. Pagels 1989, 155–156; J. N. Gardner 2003; Dick 2008). Another way to refute Universecentrism is if access to other universes becomes possible.

In any case, refuting Biocentrism, Intellicentrism and Universecentrism requires more and more speculation. So it makes them increasingly difficult to refute.

But let us focus on our universe. We can summarize ten main possible detection scenarios with a level-attitude matrix (see Figure 13 below).

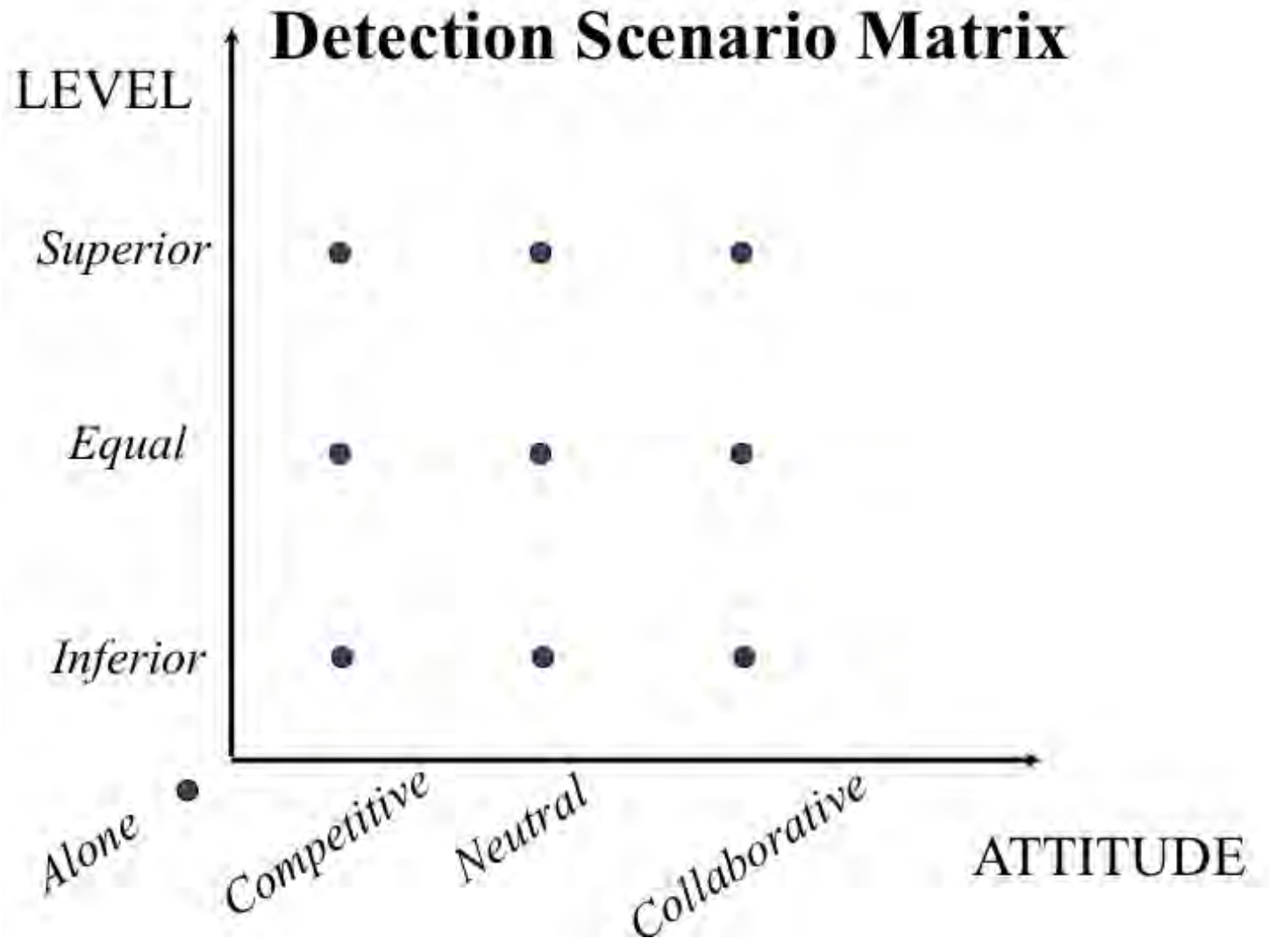

Figure 13: Ten possible detection scenarios of extraterrestrials

First, the point "alone" should not be forgotten because we could truly be alone in the universe. It is important to acknowledge that the dimension of "level" implicitly assumes that all life in the universe follows a similar developmental path. But if life starts, adapts and specializes to very different environments, it might be very hard or impossible to compare two biospheres. For example, even on Earth it is not easy to argue whether a shark or a bird is "superior". They are both very well adapted to water or to air. But let's assume we can find some kind of universal metric to assess the level of life or civilizational development. Extraterrestrials may be found to be *inferior*, *equal* or *superior* to us. Their attitude towards us can be *competitive*, *neutral* or *collaborative*.

What is the most interesting prospect for humanity? To find extraterrestrial bacterial life, or to discover a civilization with immense technology, know-how, wisdom and science? In what follows, I focus on *superior* extraterrestrial intelligence (ETI), and do not speculate much about their attitude towards us. Focusing on this search strategy is especially interesting because, if successful, it would kill a bird (Biocentrism) and a shark (Intellicentrism) with one stone.

How will the accelerating change we experience end up in thousands, millions or billions of years? An answer to this fascinating question could come if we find advanced ETI. Indeed, our cosmic cousins might give us a glimpse of our possible future on astronomical timescales (see e.g. Kardashev 1978). Such a search also makes sense scientifically, since, if ETIs exist, they are arguably much more advanced than us.

Why is ETI likely to be more advanced? To answer this question, we need to estimate the *maximum age of extraterrestrial intelligence*. I refer here the review of Steven Dick (2009b, 467–468). The starting point of the reasoning is our knowledge of cosmic evolution (see e.g. Delsemme 1998; Chaisson 2001). Dick elaborates:

> Recent results from the Wilkinson Microwave Anisotropy Probe (WMAP) place the age of the universe at 13.7 billion years, with one percent uncertainty, and confirm the first stars forming at about 200 million years after the Big Bang (C. L. Bennett et al. 2003; Seife 2003). Although these first stars were very massive—from 300 to 1,000 solar masses—and therefore short-lived, it is fair to assume that the oldest Sun-like stars formed within about one billion years, or about 12.5 billion years ago. By that time enough heavy element generation and interstellar seeding had taken place for the first rocky planets to form (Delsemme 1998, 71; Larson and Bromm 2001). Then, if Earth history is any guide, it may have taken another five billion years for intelligence to evolve. So, some six billion years after the Big Bang, one could have seen the emergence of the first intelligence. Accepting the WMAP age of the universe as 13.7 billion years, the first intelligence could have evolved seven and a half billion years ago. By the same reasoning, intelligence could have evolved in our galaxy four billion to five billion years ago, since the oldest stars in our galaxy formed about 10 billion to 11 billion years ago (Rees 1997).

You've read correctly, ETIs could maximally be 7.5 billion years our senior! More fine-grained estimates by Lineweaver and collaborators (Lineweaver 2001; Lineweaver, Fenner, and Gibson 2004) show that Earth-like planets on other stars are *on average* 1.8±0.9 billion years older than the Earth. Indeed, they showed that 75% of stars suitable for life are older than the Sun. Furthermore, Bradbury, Ćirković and Dvorsky (2011, 161) have a good point when they add that since “the set of intelligent societies is likely to be dominated by a small number of oldest and most advanced members [...], we are likely to encounter a civilization actually more ancient than 1.8 Gyr (and probably significantly more).”

What do these important insights imply? We need *not* be overcautious in our astrobiological speculations. Quite the contrary, we must push them to their *extreme* limits if we want to glimpse what such advanced civilizations could look like. Naturally, such an ambitious search should be balanced with considered conclusions. Furthermore, given our total ignorance of such civilizations, it remains wise to encourage and maintain a wide variety of search strategies. A commitment to observation, to the scientific method and to the most general scientific theories remains our best touchstone.

All right, but if these reasonings hold, shouldn't the universe be teeming with advanced ETI? How come we only observe a Great Silence (Brin 1983)? Could it be that we are hearing without listening? I mean, may ETI already be in the data? This was the opinion of Shvartsman (cited in Heidmann and Klein 1991, 393):

> I am convinced that among the several ten thousand radio sources in the catalogues of radio astronomy as well as among the several ten million optical sources on star maps there are plenty of artificial origin. These sources are recorded even today but they are misinterpreted since the recognition of ETI is not only a scientific, but also a global cultural problem.

Paul Davies shared this point of view when he wrote (Davies 2010, 124) that the " universe is a rich and complex arena in which signs of alien intelligence might be buried amid a welter of data from natural processes, and unearthed only after some ingenious sifting." I will attempt such sifting in this Chapter. Of course the idea that ETI is already in our data might seem premature. We have not explored all possible ways to see the universe. For example, we have not yet explored the spectra of neutrino radiation or gravitational waves. We might detect intelligent activity only in such spectra.

But humanity is not ignorant about the cosmos either. We have explored and are exploring the universe well beyond the visible. In fact we investigate the whole electromagnetic spectrum: radiowaves, microwaves, infrared light, visible light, ultraviolet, X-rays and gamma-rays. These new ways to observe the cosmos led to amazing progresses in astrophysics. Even if the cosmos is huge and very varied, we do see and study many kinds of stars, white dwarfs, neutron stars, pulsars, black holes, interstellar clouds, clusters of stars, planets, galaxies or clusters of galaxies in all wavelengths. If extraterrestrials are not particularly small or discrete, they may be observable or we *may have already observed them in the form of natural processes.*

Let us critically examine orthodox SETI. There is a double limitation in it, which stems from the famous Drake (1965) equation:

$$\mathrm{N} = R^* \cdot f_p \cdot n_e \cdot f_l \cdot f_i \cdot f_c \cdot L$$

where $N$ is the number of technological civilizations in the galaxy; and

(1) $R^*$ is The rate of formation of stars suitable for the development of intelligent life
(2) $f_p$ is the fraction of such stars with planetary systems
(3) $n_e$ is the mean number of planets which are suitable habitats for life
(4) $f_l$ is the fraction of planets on which life originates
(5) $f_i$ is the fraction of life-bearing planets with intelligent life
(6) $f_c$ is the fraction of planets with intelligent life that develop technological civilizations
(7) $L$ is the lifetime of a technological civilization

This equation has inspired much of our understanding of cosmic evolution and helped us to frame agendas for SETI. By extending and generalizing it to the "Cosmic Evolution Equation", it also was very helpful to frame new research agendas in cosmology (see section 6.3 The Cosmic Evolution Equation, p122). This equation is a tool to assess "the number of *communicative* civilizations which might exist in our *galaxy*" (my emphasis). However inspiring and helpful it has been, it has also introduced two fundamental biases in SETI.

First, it focuses on *communication*. This is the orthodox way of searching for messages coming from an ETI. This program has failed so far. One may advance many good reasons for this failure, but the bottom line is that we do not need to assume communication to conduct astrobiology. The equation introduces a second bias by focusing on our galaxy *only*. By endorsing the Drake equation's agenda uncritically, we study one object, our galaxy, out of the 170 billion ($1.7 \times 10^{11}$) galaxies estimated to shine in the observable universe (see e.g. Gott et al. 2005). For more critique on this limiting focus on our own galaxy, see also (Ćirković and Bradbury 2006; Vidal 2011; Bradbury, Ćirković, and Dvorsky 2011). The good news is that if we extend Drake's equation to the whole universe, then our detection chances increase. More precisely, looking at further and further away galaxies is like traveling through time, so it constitutes an opportunity to test wide-ranging scenarios for civilizational development at different periods (Kardashev 1997).

Those two biases have shifted SETI's fundamental question from (1) "Are we alone?" to (2) "Who wants to chat in the galaxy?". Of course, it would be much more enriching and fun to communicate or to have direct contact with ETIs. Accordingly, starting SETI in our own galaxy is also the first logical and practical step to take. Yet, if we really wish to find out whether we are alone or not (1), it demands to extend our search strategies.

Who are you? It is nearly impossible to have a meaningful answer without knowing and comparing yourself with others human beings. Similarly, when we ask "who are we?", where "we" refers to the human species, or better, as a cosmic complexity pocket on Earth, we will not find a meaningful answer without comparing ourselves to putative extraterrestrials. SETI then becomes a search for meaning. It is easy to predict that it will become a more and more important field of research, more and more taught in schools and universities. Indeed, as we are becoming a single unified globalized living entity, arguably a global brain (see e.g. Heylighen 2002; 2007), we will be more and more curious about life elsewhere. Visionary politicians and philanthropists have been and will continue to give support to astrobiology and SETI.

Besides improved self understanding at the level of humanity, there are many other benefits of astrobiology, both with a positive outcome –we find extraterrestrials– or with a negative outcome –we don't find extraterrestrials– (see e.g. A. A. Harrison 1997, 21–26; Tough 1986; 1998).

Equally informative would be to know that we are truly alone. There are some dubious such arguments, for example Tipler (1980b) argues that we are alone in the universe. Sagan's dictum provides a sufficient answer: "absence of evidence is not evidence of absence". Another much debated probabilistic argument from Brandon Carter (1983) suggests we are most likely alone (see Ćirković, Vukotić, and Dragićević 2009 for a review, a critical discussion and a possible resolution). The bottom line is that proving that there are no extraterrestrials also requires to explore the whole cosmos to confirm that there is no sign of life. This is an interesting point because it means that SETI pessimists should be as motivated as SETI optimists to explore the cosmos. They just have different hopes or expectations regarding the outcome of the search. As Arthur C. Clarke put it, “sometimes I think we’re alone in the universe, and sometimes I think we’re not. In either case the idea is quite staggering” (cited in Davies 2010, x).

Searching for ETI opens the field of the possible. The universe is so vast and so varied that it makes it delicate to choose where to start from or what to look for. I'll now tell you how I concluded that some peculiar interacting systems are excellent “candidates” of very advanced ETI, the *starivores*. Before that, I will tackle the difficult issue of finding criteria for artificiality (section 9.1 Criteria for Artificiality, p209), introduce a two-dimensional metric for civilizational development, using the Kardashev scale of energy use and the Barrow scale of inward manipulation (section 9.2 Two Scales for Civilizational Development, p224). Then we will make an excursion into the speculative topic of black hole technology and advance energetic, societal, scientific, computational, and philosophical arguments that black holes are attractors for intelligence (section 9.3 Black Holes as Attractors for Intelligence, p228). I will then apply the criteria for artificiality to starivores and propose a very straightforward thermodynamical argument for their existence. I will then address many objections against the arguments. If this new search strategy is further developed and confirmed it will be the refutation of two “centrisms” at once: Biocentrism and Intellicentrism! To know whether it is correct or not will require a lot of work and debate, within the *High Energy Astrobiology* research field. In the last section, I propose general predictions as well as concrete and refutable research proposals for high energy astrobiologists. I will be glad to offer a prize regarding the confirmation or refutation of these high energy astrobiology research proposals.

## 9.1 Criteria for Artificiality

### 9.1.1 What is Your Methodological Fallacy?

*I fully support Shklovsky's dictum*
*that every object must be assumed natural until proven otherwise.*
Freeman Dyson in (Dyson et al. 1973, 189)

*Extraterrestrial intelligence is the explanation of last resort,*
*when all else fails.*
Carl Sagan in (Dyson et al. 1973, 228)

To which I add the following statement:

> I fully disagree with Shklovsky, Dyson and Sagan and hold that every strange phenomena must be assumed natural or artificial unless proven otherwise.

During the first international SETI conference at Byurakan in Armenia, Shklovsky (1971) advocated a carefully reasoned search for extraterrestrials, but he may well have handcuffed the endeavor (see Rubtsov 1991). He advocated what I call the naturality-of-the-gaps principle:

> **Naturality-of-the-gaps**: *unless proven otherwise, assume phenomena to be of natural origin.*

If by "natural" we mean "respecting physical laws", everything we observe, living or non-living, will be natural. So the principle is trivially true. By definition, all systems in nature follow physical laws. In particular, every ETI must respect the laws of physics. The principle is in fact utterly useless for SETI, since living systems, like any other object in the universe, are subject to physical laws like gravitation and thermodynamics. Derogating to this principle is equivalent to searching for magic.

Of course what Shklovsky and Sagan had in mind was something else. They warned *against* the fallacy of seeing extraterrestrials when facing an unknown phenomenon. One should not quickly qualify a phenomenon as artificial without having first exhausted all the possible natural explanations. But are we really ever going to exhaust all the possible natural explanations? As Almar (in Heidmann and Klein 1991, 393) wrote, while "the principle is scientifically logical, it did not turn out to be constructive because one can never determine the moment when all natural explanations have been exhausted."

Even more difficult is the question "*on which conditions will we give up modelling the phenomenon as natural, and conclude that it must be artificial?*" If we consider the analogy with the situation between Geocentrism and Heliocentrism, the principle would be equivalent to say "unless proven otherwise, assume the Earth is the center of the universe". Today, who would dare to say this is a scientifically constructive principle? For the fruitfulness of scientific development, history of science has learnt us that we better be open-minded in allowing a wide variety of models to explain a same phenomenon, to allow them to co-exist and to rely on objective criteria to decide which model or theory to endorse. As Rubtsov (1991, 307) analyzed in the context of searching extraterrestrial astroengineering structures:

> In reality a "normal" astronomical investigation will never need an "artificial"(A-) approach to its object of study. Any refuted hypothesis will be replaced only with a new "natural"(N-) one. On the contrary, searches for astroengineering structures require the equal status of A- and N- explanations from the very beginning of the investigation. When studying an object or phenomenon selected by some preliminary criteria, one should bear in mind both of these hypotheses.

He adds that Aritificial (A-) and Natural (N-) research programs should

> develop, interact and enrich each other, seeking, on the one hand, for the most complete representation of the object of phenomenon in its description and, on the other hand, for the best possible conformity between the description and a theoretical explanation of the phenomenon. During this process, one of the two explanations will be gradually superseded by the other, and a correct explanation will result.

If by "natural" we mean a purely physical process, not living (biological) and not intelligent, then we need to define what we mean by "artificial" or non-natural. This implies to address the thorny questions of defining what life and intelligence are. In sum, the principle of naturality-of-the-gaps becomes useful only if we already have criteria to discriminate between natural and artificial. But proponents of naturality-of-the-gaps generally do not give such criteria.

The naturality-of-the-gaps has its exact logical counterpart, the artificiality-of-the-gaps principle:

> **Artificiality-of-the-gaps**: *unless proven otherwise, assume phenomena to be of artificial origin.*

If we hold that principle, we generate false positives; i.e. detection of extraterrestrials where there is only natural phenomena. It is hold by UFOlogists who quickly jump from some strange light or object in the sky to the conclusion that it was an alien. They make the mistake not to review systematically other explanations (unusual meteorological phenomenon, human artifact, secret military weapon or hoax). The reasoning is “we don't understand how it works, therefore it is ETI”.

The principle of artificiality-of-the-gaps underlies an unscientific attitude because every terrestrial or astrophysical phenomena hard to model would be driven by extraterrestrial intelligence, and this leads to an “extraterrestrials-of-the-gaps” explanations, which, like the god-of-the-gaps explanation, explains nothing and anything at the same time.

Importantly it would be normal not to fully understand a system more advanced than us by millions or billions of years. But it remains a fallacy to say that because we do not understand something, it is the manifestation of advanced intelligence. Difficulty of modelling is a *necessary condition* for advanced ETI, not a *sufficient* one. Advanced extraterrestrials will not be obvious to model with known astrophysics. As a counter example, if we would say that our Sun is an intelligent big yellow man who sneezes every 11 years, it would not be very credible, because its activity is better explained by traditional stellar physics.

To sum up, it is equally fallacious to hold *uncritically* and *by default* any of naturality-of-the-gaps or artificiality-of-the-gaps. The establishment of the scientific search for extraterrestrials focused on avoiding artificiality-of-the-gaps. This was of course in order to distinguish itself from the unscientific ufology field. Needless to say, Shklovsky, Dyson and Sagan had this threat in mind. Holding such a strong condition may even have been a necessary step to establish SETI as a legitimate scientific field of study, in contrast to unscientific ufology. But this was at the steep price of introducing the opposite bias of naturality-of-the-gaps.

Let us see another logical way to clarify this point. Harrison (1997, 44–48) elucidated the logical outcomes of our search for extraterrestrials. For this purpose he used two dimensions, first, the *interceptions* of signals and second, our *interpretation* of these interceptions. There are then four possible outcomes, summarized in table 11 below.

| **Interpretation** / **Interception** | **Positive** | **Negative** |
|---|---|---|
| Extraterrestrial | True positive | False negative |
| Other | False positive | True negative |

Table 11 - Possible interpretations of extraterrestrial and non-extraterrestrial signal interceptions

A *true positive* is when we detect extraterrestrials and we are right about it. A *false negative* is when we detect extraterrestrials, but our interpretation is that it is not an extraterrestrial. A *false positive* is when we think we have detected extraterrestrials, whereas the signal is either natural or man-made. A *true negative* is simply a normal interception of signals accurately known to be irrelevant to SETI.

Now, it is easy to see that artificiality-of-the-gaps leads to *false positives*. An over enthusiasm for the possible existence of extraterrestrials will make us see extraterrestrials everywhere. On the other side, the naturality-of-the-gaps brings us to *false negatives*. An over skepticism for the existence of extraterrestrials will make us blind to extraterrestrial manifestations in front of our telescopes.

It is worth mentioning a kind of false negative which arose from an overconfidence in natural explanations. In 1989, a team of astronomers discovered rapid pulses coming from a pulsar remain of supernova 1987A (Kristian et al. 1989). Another set of rapid pulses were later discovered. The astronomer teams concluded that it was again the pulsar, whereas it was only a human interference (Anderson 1990).

So, how do we avoid *both fallacies* of false positives and false negatives? How can we avoid contact with both artificiality-of-the-gaps and naturality-of-the-gaps? We can now formulate the more balanced astrobiological stance:

> **Astrobiological stance:** *unless proven otherwise, assume phenomena to be of either natural or artificial origin.*

Following this stance, astrobiologists explore as much the hypothesis that some astrophysical phenomena may be natural or artificial. However, searching for extraterrestrials need not to be the exclusivity of SETI researchers or astrobiologists. Any astrophysicist can take the astrobiological stance by asking: is the phenomenon I am studying easier and more cogently explained assuming naturality or artificiality?

That's enough for logic. What about practice? How can we recognize ETI? For a meaningful discussion regarding whether known phenomena are natural or artificial, we need criteria to correctly distinguish them. Where are we to search these criteria? From our best, most universal, most context-independent theories. These are foremost *physical laws* and *systems theory*. However, since we are interested in extraterrestrial life and intelligence, we can restrict the scope of systems theory to a subset of it, *living systems theory*.

At a closer look, the distinction between natural and artificial might well be ... artificial (see e.g. Davies 2010)! Could it be that the difference between natural and artificial phenomena is more of a continuous nature? Or should we rather consider criteria distinguishing simple versus complex processes? If so, what kind of more continuous criteria can we define? A thing is certain, we need to avoid to be too Earth-centric, and start the search with a strict minimum number of assumptions. So let us waive a maximum of our prejudices and enter the temple of Zen SETI.

### 9.1.2 Zen SETI

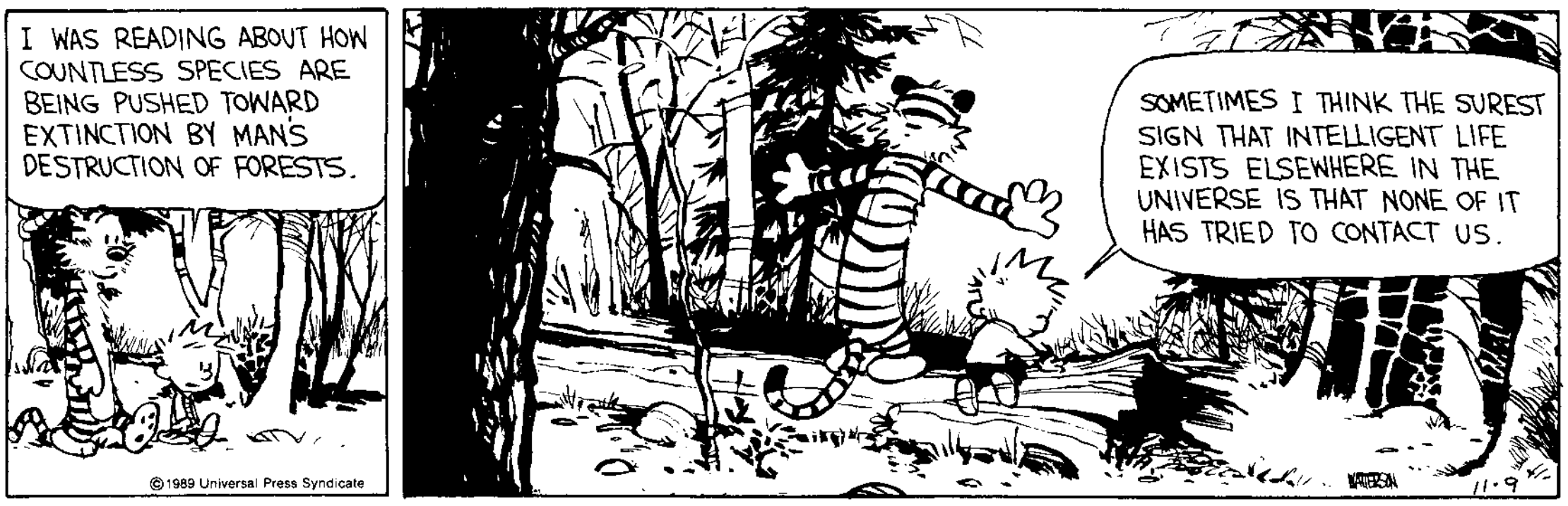

Figure 14 Calvin ponders a new proof of intelligent extraterrestrial life.
Calvin and Hobbes © 1989 Watterson. Dist. By Universal Uclick. Reprinted with permission.

There are two moments when we start to think about what Ćirković (2012) calls the *astrobiological landscape*. The first is to use divergent thinking and brainstorm about all the possible ways and places ETs could thrive. The list may become very long. This is an essential step to avoid a premature restriction of the search space. As we extend the possible, the exercise can become speculative and even fun. Such entertainment is actually the job of hard science fiction authors when they develop weird yet scientifically plausible scenarios. But it is not science. So, how do we make the second step and shift from science fiction to science? How would you put your speculations into the scientific track? If you were given 5 million euros to lead a search for extraterrestrials, where would you start? This practical constraint forces us to restrict the search space, or at least prioritize what we want to look at first. This is a critical step, prone to many biases. In his (1989) book *The Inner Limits of Outer Space*, psychologist John C. Baird criticized assumptions behind the SETI enterprise being “as much a function of the principles of human psychology as they are of engineering and physics.”

To make SETI scientific, the key is to connect speculations with what we can observe. Dyson (1966) thus advocated a focus on the most conspicuous manifestations of intelligence and technology, so that we have something big to observe. If we assume that ETIs are operating at a quantum scale using very few energy, they would leave virtually no trace, and we would not be able to search nor find them, even if they were thriving here on Earth. Such an idea and many others are quite credible, but if there is no hope to test them, they should remain in the province of science fiction. But if we say that an ETI uses the energy of stars in such or such particular way, and stars should be modified in that way because of this interaction, there is more hope for testability.

In his seminal paper, Dyson (1966, 643) assumed that ETIs would use technology we can understand. He qualified his own assumption as “totally unrealistic”. I do agree. If we look only at technology we understand, we restrict our search to civilizations roughly at our developmental level, not really higher. The search for ETI *more advanced than us* is bond to fail. If we release this rule, it will be hard –if not impossible– to argue that a phenomenon we do not understand is artificial, since its technology will, by definition, be alien to us. That is why we absolutely need criteria which minimally depend on our known technology.

Paul Davies (2010) advocated a renewal of search strategies and he calls for a "New SETI". Bradbury, Ćirković and Dvorsky (2011) did also call for a New SETI which they call *Dysonian SETI*. They summarize the salient differences with the table 12 reproduced below:

| | Orthodox SETI | Dysonian SETI |
|---|---|---|
| Main object of search | Intentional messages | Artifacts, traces, and signatures |
| Working ATC model | Biological, post-industrial analog existence | Postbiological, digital existence |
| Temporal window of opportunity | Narrow | Wide |
| Quantitative theoretical potential | Limited | Unexplored (large?) |
| Prejudicates ETI behavior? | Yes | No |
| Two-way communication? | Yes (?) | No |
| Interstellar travel | Irrelevant | Relevant |
| Operational risks? | Yes | No |
| Main working frequencies | Radio (cm) | Infrared |
| Natural mode of search | Active | Parasitic |
| Data resolution | High | Low (?) |
| Practical extragalactic SETI? | No | Yes |

Table 12 -A Comparison Between the Orthodox and the Dysonian Approach to SETI (by Bradbury, Ćirković, and Dvorsky 2011). Note that ATC means Advanced Technological Civilization.

It is important to emphasize that orthodox SETI and Dysonian SETI are not opposed. Accordingly, Dysonian SETI criticizes orthodox SETI, but does not dismiss it. It mainly extends the number of search targets. In what follows, I take such a *Dysonian approach* to SETI, emphasizing the search for extraterrestrial technological manifestations and artifacts (see e.g. Dyson 1960; 1966; Ćirković 2006). This approach is also in line with the framework of the *postbiological universe* introduced by Steven J. Dick, which includes insights from astrobiology, computer science and futures studies (Dick 2003; Ćirković and Bradbury 2006). This framework invites examination of new kinds of objects. For example, Seth Shostak (2010, 1029) recently proposed to widen the search to bok globules (cold molecular clouds), hot stars, neutron stars and black holes. Importantly, this means that the search recently widened not only to planets but to other objects in the universe. We thus get rid of the assumptions that ETI should necessarily thrive on a planet.

We take Calvin's remark seriously in the comic strip above, and will try to look for ETI also if they don't want to communicate with us: generally, they can be willing to communicate or not. Extending the search to non-communicative ETI also allows extragalactic SETI. As Sagan (1973b) wrote about advanced ETIs:

> There is a serious question about whether such societies are concerned with communicating with us, any more than we are concerned with communicating

> with our protozoan or bacterial forebears. We may study microorganisms, but we do not usually communicate with them. I therefore raise the possibility that a horizon in communications interest exists in the evolution of technological societies, and that a civilization very much more advanced than we will be engaged in a busy communications traffic with its peers; but not with us, and not via technologies accessible to us. We may be like the inhabitants of the valleys of New Guinea who may communicate by runner or drum, but who are ignorant of the vast international radio and cable traffic passing over, around and through them.

Carl Sagan did also debunk common chauvinisms in SETI. He identified oxygen, carbon, sun like-star (G-star), planetary and temperature chauvinisms (Sagan 1973a, chap. 6).

To sum up, we are now more Zen in SETI, because we have abandoned unnecessary assumptions: ETIs using oxygen, carbon, living on a planet around a sun like-star; using technology we know, striving on temperatures we know are suitable for life on Earth, ETIs communicating with us and searching ETIs in our galaxy only. We mentioned the radical proposal that advanced life might be in a *postbiological* form. This may seem odd, since it requires to abandon the idea that ETI must function on the biological substrate we know. I will now explore this important suggestion in more details, and then propose criteria to distinguish natural versus artificial, grounded in *general criteria*, *thermodynamics* and *living systems theory.*

### 9.1.3 The Case for Postbiology

*Surely the essence of humanity is what we do and think, not the chemical make-up of our bodies.*
(Davies 2010, 201)

Already in the 1980's, Feinberg and Shapiro (1980) stigmatized the carbon-and-water proponents as “carbaquists”. Those fail to imagine that basic building blocks of life could be very different. If not the chemical building blocks, what is the essence of life, then? It is the activity of a biosphere, which is itself (Feinberg and Shapiro 1980, 147):

> a highly ordered system of matter and energy characterized by complex cycles that maintain or gradually increase the order of the system through the exchange of energy with the environment.

It is very important to notice the high generality of such a definition. There is no mention of carbon, water or DNA. What remains are energetic exchanges leading to an increase of order. Free from limiting assumptions of carbaquists, the two authors conceptualize possible living beings living in lava flows, in Earth's magma or on the surface of neutron stars. The idea of life on neutron stars was explored in science fiction (see e.g. Forward 1980) but also by scientist Frank Drake (1973).

We can think much more systematically about life-as-we-don't-know it. This was done by Robert A. Freitas Jr. who wrote *Xenology* (1979), the most comprehensive and systematic study of extraterrestrial life, intelligence and civilization I am aware of. This long volume covers a much broader scope than the classical –and also very good– book by Shklovskii and Sagan (1966). I consider it a rare scientific masterpiece. Most of the book was written in 1979, although Freitas is constantly updating it, and made it freely available on the web (www.xenology.info).

Freitas (1981) takes a similar starting point as Feinberg and Shapiro and writes that life "requires metabolism, a systematic manipulation of matter-energy and information. But manipulation can only be accomplished by the application of force." Right, but which force? Freitas systematically analyses possible metabolisms of living systems based on the four fundamental physical forces: *nuclear force* (chromodynamics), *electromagnetism*, *weak force*, and *gravitation*. Those forces are weaker and weaker, and as Freitas argues, we "can imagine four broad classes of metabolic entities – chromodynamic or nuclear lifeforms, electromagnetic lifeforms (e.g., all Earth life, including humans), weak lifeforms and gravitational lifeforms. Each is most likely to evolve in those environments where the forces upon which they most depend predominate over all others."

Let us see how *chromodynamical lifeforms* would thrive:

> Chromodynamic creatures may evolve in an environment where nuclear forces are predominant. While the chromodynamic force is the strongest in nature, it is effective only over ranges of about $10^{-15}$ meter, so very special conditions might be required for such life to exist.

Where in the universe could such conditions be met? Freitas pursues:

> These conditions possibly could be found inside a neutron star. Neutron stars are heavy, rapidly spinning objects 10-20 kilometers in diameter with approximately the mass of a star. They have densities like nuclear matter, tremendous magnetic fields, surface gravities in excess of 100 billion Earth-gees, and are thought to be the energy source for pulsars. Neutron stars have atmospheres half a centimeter deep and mountains at most one centimeter high. Under the three-kilometer crust of crystalline iron nuclei a sea of neutrons circulates at a temperature of hundreds of millions of degrees. In this sea float a variety of nuclear particles including protons and atomic nuclei. Scientists believe that there may be neutron-rich "supernuclei" or "macronuclei" dissolved in the neutron sea. These macronuclei might contain thousands of nucleons (as compared to only a couple of hundred in normal matter) which could combine to form still larger supernuclei analogous to the macromolecules which make up earthly life. The neutron sea may be the equivalent of water in the primordial oceans of Earth, with macronuclei serving as the equivalents of amino acids, carbohydrates, and nucleotides in the prebiotic origin of life. It is possible to conceive of life evolving in neutron stars much as it did on our own planet nearly five billion years ago, but substituting atomic nuclei, supernuclei and neutrons for atoms, molecules and water.

To study *eletromagnetic lifeforms*, it suffices to open a terrestrial biological textbook. But not only. A computer science textbook is as relevant. Indeed, as Freitas writes, machine lifeforms also fall into this category, since "the advancing intelligence and versatility of electronic computers suggests that some sort of solid state "machine life" may be plausible. Such entities would survive by manipulating electron flows and fields in order to process matter-energy and patterns of information."

However, *weak lifeforms* hold weak promises. Freitas elaborates:

> Weak force lifeforms would be creatures unlike anything we can readily imagine. Weak forces are believed to operate only at subnuclear ranges, less than $10^{-17}$ meter. They are so weak that unlike other forces, they don't seem to play a role in actually holding anything together. They appear in certain kinds of nuclear collisions or decay processes which, for whatever reason, cannot be mediated by the strong, electromagnetic or gravitational interactions. These processes, such as radioactive beta decay and the decay of the free neutron, all involve neutrinos. A weak lifeform might be a living alchemist. By carefully controlling weak interactions within its environment, such a creature could cause its surroundings to change from a state of relatively high “weak potential” to a condition of low “weak potential” and absorb the difference into itself. A state of high “weak potential” might be characterized by extreme instability against beta decay – perhaps these beings are comprised of atoms laden with an excess of neutrons and become radioactive only when they die.

Gravitational lifeforms are very fascinating and promising if we look for the most advanced possible extraterrestrials. Indeed, if we follow the general theory of evolution, proportionately “to the decrease of bonding energies we find an increase in level of organization.” (Laszlo 1987, 27). Freitas elaborates:

gravitational lifeforms, should they exist, survive by making use of the most abundant form of energy in the universe. Gravity is also the most efficient – this is why a hydroelectric power station which converts the energy of falling water into electricity (essentially a controlled gravitational contraction of the Earth) can have an efficiency close to 100%. In theory, gravity beings could be the most efficient creatures in the universe. Their energy might be derived by arranging encounters of collisions between black holes, galaxies or other celestial objects, or by carefully regulating the contraction of various objects such as stars or planets.

We should also note that these four life forms are not necessarily *post*biological, but could simply be *non*biological. I mean, if extraterrestrial life *starts* based on another physical force than electromagnetism, instead of *developing* into another physical substrate and organizing around another physical force.

The theoretical study of such a wide scope of lifeforms is still in its infancy. But the field of Artificial Life of course has been busy with defining and producing life-forms based on general principles. A promising direction is to develop the study of possible lifeforms within the framework of *Chemical Organization Theory* (see e.g. Dittrich, Ziegler, and Banzhaf 2001). It started with chemistry, but it could be extended to nuclear reactions. The framework has indeed been generalized to describe any kind of dynamical systems as a network of reactants and reactions, where closed and self-maintaining networks are called organizations.

What does postbiology look like? Let us take our closest energy hungry and high information and computation processing relative, the computer. The material support on which computers operate has already changed five times since their invention! Kurzweil (2006, chap. 3) reminds us that computers went through several revolutions, from electromechanical calculators, relay-based computing, vacuum tubes, discrete transistors to today's integrated circuits. Of course, it won't stop here and Kurzweil predicts the sixth computational paradigm to be three-dimensional molecular computing. It is also natural to see that at some future point, we will use quantum computation, manipulating the smallest information markers physically possible.

Each computational paradigm uses quite a different material support or scale at which computations operate. What is the lesson for SETI? Imagine a computer engineer of the 1940's teleported in our modern world. There is only 70-odd years between his world and yours, but you challenge him: where is our technology? Wouldn't it be laughable to see him searching state-of-the-art computers in light bulbs? Searching *advanced* ETI only with biology as we know it is as naïve, restrictive and laughable.

The moral of the story is that in SETI, *matter doesn't matter* (much). What is important is the ability to manipulate matter-energy and information, not the material substrate itself. The case for postbiology is strong, and I invite skeptical readers to many more arguments in the literature (e.g. start with Dick 2003; or 2009b and references therein). Abandoning the hypothesis of ET using a biological substrate such as carbon, water, DNA molecules or proteins makes us focus on the *functional* aspects of living systems. This focus on function is the basis and conceptual strength of *systems theory*, aimed to be independent of a particular material substrate. This makes system theory the interdisciplinary research field *par excellence* and also an indispensable tool in astrobiology and SETI.

Let us now turn to a search for criteria to distinguish natural versus artificial astrophysical phenomena. I start with global and thermodynamical criteria and continue with living systems criteria.

### 9.1.4 Global Criteria

It is always extremely hard to do science with a unique object. We saw this clearly in the case of cosmology. I proposed to study general aspects of our universe by making statistics on possible universes resulting from computer simulations (see section 6.3 The Cosmic Evolution Equation, p122). This allows to scientifically study how robust the emergence of complexity is and how fine-tuned our universe is.

Focusing on one isolated object or phenomenon, it will also be very difficult to decide if it is natural or artificial. This invites us to take a more global approach in astrobiology, to look at several similar objects available to observation, and if needed, do statistics on them. We have the chance to live in a universe full of billion of stars and other structures. It's thus possible to gather a lot of data and do statistics. Let us see three general *heuristics* we can use.

> **Strangeness heuristic:** *Advanced extraterrestrials manifestations will not be easy to model.*

An ETI two billion years more advanced than us will not manifest a trivial behavior. As we mentioned, difficulty of modelling is a *necessary* condition for advanced ETI, but of course not a *sufficient* one (see Rubtsov 1991, 307). We should not commit the artificiality-of-the-gaps fallacy. So those strange phenomena should attract our attention and be carefully analyzed with an astrobiological stance (as defined above).

> **The non-exclusiveness heuristic**: *"diversity will tend to prevail unless there exists a mechanism to enforce conformity"* (Brin 1983, 287).

Indeed, we should not expect life or intelligence to look exactly the same from one side of the galaxy to the other, or from one galaxy to the other. Of course, we can imagine mechanisms to enforce some conformity, such as arguments from convergent evolution, or a more speculative "galactic club" (Bracewell 1974) which would regulate the activities of intelligent civilizations at a galactic scale.

Another heuristic Brin proposes is the following:

> **The equilibrium heuristic**: *"It is generally considered sound scientific practice to assume a state of quasi-equilibrium when beginning to explore a previously undeveloped field of knowledge, since most natural phenomena with long time-scales can be modelled as perturbations of an equilibrium state."* (Brin 1983, 287)

Brin then gives an example of a violation of this heuristic, criticizing an argument explaining that the absence of ETIs is explained by the fact that they have 'not arrived yet'. It would imply a situation of profound disequilibrium.

> **The inverse distance-development principle**: *"the more distant, the less developed we expect"* (Kardashev 1997)

This statement looks almost trivial, but for that very reason it constitutes a genuine and very important principle. It is well known that, because it takes time for light to travel unto us, the more distant astrophysical objects are, the younger they appear. Unfortunately this principle should also be balanced with the maximal age of extraterrestrial life, which we saw varies between 2 and 7.5 billion years older than us.

Futurists and science fiction authors might dislike the finiteness of the speed of light, but it constitutes a very informative constraint to model cosmic evolution in general, and the putative development of extraterrestrials in particular. As we already mentioned, this is a way to test wide-ranging scenarios for civilizational development at different periods. If we have a general theory of extraterrestrial life, we can test its various development stages by looking at farther and thus younger stars and galaxies. A very general search strategy derives from this principle:

1. Take a list of cosmic objects
2. Compare their number, distribution, properties, in *further and further away* galaxies.
3. Do patterns emerge? Compare the predictions of natural and artificial evolutionary models.
4. Beware of selection effects; further away objects will be more difficult to observe.

### 9.1.5 Thermodynamical Criteria

*In speculating about alien super-science, then,*
*the second law of thermodynamics should be the last one to go.*

(Davies 2010, 150)

We can read Davies' recommendation in another equivalent way. The second law of thermodynamics –and thermodynamics in general– should be the *first* tools to go with in SETI speculations. In fact, cosmologist Eric Chaisson (see e.g. Chaisson 2001; 2003; 2011a; 2011b) has championed the fruitfulness of an energetic view to describe the unfolding of 13.7 billion years of cosmic evolution. We already mentioned his free energy rate density metric, a quantitative complexity measure based on the energy flowing through a system of a given mass (see section 7.2 Increase of Computing Resources, p158), which allows to describe physical, biological and technological systems. Given such a billion-years applicability, we can reasonably hope that it would also apply to advanced extraterrestrials. Indeed the tool is allowed in the Zen SETI temple, since it uses only the very general concepts of energy, time and mass.

We saw that a universal feature of living beings is their having a metabolism, which implies a thermodynamic disequilibrium. Carr and Rees (1979) also maintained that thermodynamical disequilibrium is a strong necessary condition for life and intelligence elsewhere in the universe:

> life requires elements heavier than hydrogen and helium, water, galaxies, and special types of stars and planets. It is conceivable that some form of intelligence could exist without all of these features –thermodynamic disequilibrium is perhaps the only prerequisite that we can demand with real conviction.

The simplest thermodynamical criterion is thus:

> **thermodynamic disequilibrium**, a necessary condition to recognize life and intelligence

Let us be more specific. We can distinguish three kinds of more and more complex thermodynamical structures. First are *equilibrium structures* which are the subject-matter of classical thermodynamics, when it is applied to liquids or crystals. Then come the *dissipative structures* where the structure is in a nonequilibrium state, which generates self-organization (Nicolis and Prigogine 1977). A good example is the Belousov-Zhabotinsky (Belousov 1958; Zhabotinsky 1964) reaction. It is a fascinating chemical reaction, where the concentration oscillates periodically, leading to the formation of non-trivial patterns. However, as Nicolis and Prigogine (1977, 340) write,

> Eventually the oscillations die out, as the system remains closed to mass transfer and the raw materials necessary for the reaction are exhausted. Thus, although the initial mixture may be removed very far from thermodynamic equilibrium, it finally tends to the state of equilibrium where oscillatory behavior is ruled out.

Could a system sustain a non trivial behavior and stay in nonequilibrium? This leads us to the third kind of thermodynamical structures, *living structures*.

From the point of view of classical thermodynamics, life is a miracle. It is able to sustain a very far from equilibrium state, despite the second law of thermodynamics, which states that all systems tend to equilibrium. This seemed very paradoxical. The key to unlock the mystery of living systems was to consider them in a larger thermodynamical context. They should be modelled as *open systems,* meaning that a *flux* of energy goes through them, and not as *closed systems*. The second law only applies to closed systems, not to open systems. All in all, the second law is not violated because living systems increase *local* order at the expense of a more *global* disorder generated in the environment.

Additionally, *energy flow regulation or control is a necessary condition for the growth, maintenance, evolution and reproduction of complex systems* (see e.g. Aunger 2007b; Chaisson 2011a). For example, a stone processes virtually no flow of matter-energy, and most people will agree that it is dead. On the opposite side, we have a wild fire, which grows, uses a lot of energy, but is totally uncontrolled. Whatever shaman's view on the matter, scientists generally don't consider fire as alive. Living systems are in between these two extreme examples. They are able to regulate their energy flow. To take humans as an example, if we eat too few or too much, we die. We thus regulate the amount of food that we eat to stay alive.

> energy flow control: living systems control their energy flow to grow, maintain themselves, evolve and reproduce.

A living organism can be described broadly by three components: *a source of energy*, an organized entity, and a *sink* to waste (to export entropy). The living system increases its internal organization –or negentropy– thanks to this energy flow. As Freitas (1979, sec. 6.2.3) puts it, life "drives its environment to physical or chemical disequilibrium, establishing an entropy gradient between itself and its surroundings". He adds that all living systems "possess this feature, and it is contended that any system engaging in such negentropic operations must be considered "living" to a certain extent". This leads to the criterion of metabolism:

> **metabolism**: living systems maintains its *organization* by using a source of *energy* and producing entropy.

*The most straightforward astrobiological search strategy is thus to look for this kind of non-equilibrium systems in the universe*. We will soon apply these criteria to high energy astrophysics, and see that it leads to very promising and intriguing results (if you can't wait, see section 9.4 Signs of Starivores?, p232).

But of course, the thermodynamical criterion alone is insufficient. As Sagan (1975, 145) put it, thermodynamic disequilibrium "is a necessary but of course not a sufficient condition for the recognition of extraterrestrial intelligence". So, what else do we need?

#### 9.1.6 Living Systems Criteria

The game of SETI is to get rid of a maximum of assumptions about what we know regarding *terrestrial* life, to extract only life's essential characteristics. The hope is that our resulting concepts of life and intelligence will be so general that they will also apply to extraterrestrial life. James Grier Miller (1978) wrote a very impressive 1100 pages book entitled *Living Systems*. Miller shows that this general theory successfully applies to many different kinds of living systems at different levels, from cells, organs, organism, groups, organizations, societies to the supranational system. This *opus magnum* constitutes a very useful guide to think in general terms about extraterrestrial life (see A. A. Harrison 1997 for an extensive application).

Miller distinguishes 19 critical subsystems that all living systems have, which can be divided in three broad categories. First, subsystems which process both *matter-energy and information*; second, subsystems which process *matter-energy* and third, subsystems which process *information*. See Table 13 for details about these subsystems.

| MATTER + ENERGY + INFORMATION | |
|---|---|
| *1. Reproducer* | the subsystem which is capable of giving rise to other systems similar to the one it is in. |
| *2. Boundary* | the subsystem at the perimeter of a system that holds together the components which make up the system, protects them from environmental stresses, and excludes or permits entry to various sorts of matter-energy and information. |
| MATTER + ENERGY | |
| *3. Ingestor* | the subsystem which brings matter-energy across the system boundary from the environment. |
| *4. Distributor* | the subsystem which carries inputs from outside the system or outputs from its subsystems around the system to each component. |
| *5. Converter* | the subsystem which changes certain inputs to the system into forms more useful for the special processes of that particular system. |
| *6. Producer* | the subsystem which forms stable associations that endure for significant periods among matter-energy inputs to the system or outputs from its converter, the materials synthesized being for growth, damage repair, or replacement of components of the system, or for providing energy for moving or constituting the system's outputs of products or information markers to its suprasystem. |
| *7. Matter-energy storage* | the subsystem which retains in the system, for different periods of time, deposits of various sorts of matter-energy. |
| *8. Extruder* | the subsystem which transmits matter-energy out of the system in the forms of products or wastes. |
| *9. Motor* | the subsystem which moves the system or parts of it in relation to part or all of its environment or moves components of its environment in relation to each other. |
| *10. Supporter* | the subsystem which maintains the proper spatial relationships among components of the system, so that they can interact without weighting each other down or crowding each other. |
| INFORMATION | |
| *11. Input transducer* | the sensory subsystem which brings markers bearing information into the system, changing them to other matter-energy forms suitable for transmission within it. |
| *12. Internal transducer* | the sensory subsystem which receives, from subsystems or components within the system, markers bearing information about significant alterations in those subsystems or components, changing them to other matter-energy forms of a sort which can be transmitted within it. |
| *13. Channel* | the subsystem composed of a single route in physical space, or |
| *and net* | multiple interconnected routes, by which markers bearing information are transmitted to all parts of the system. |
| *14. Decoder* | the subsystem which alters the code of information input to it through the input transducer or internal transducer into a “private” code that can be used internally by the system. |
| *15. Associator* | the subsystem which carries out the first stage of the learning process, forming enduring associations among items of information in the system. |
| *16. Memory* | the subsystem which carries out the second stage of the learning process, storing various sorts of information in the system for different periods of time. |
| *17. Decider* | the executive subsystem which receives information inputs from all other subsystems and transmits to them information outputs that control the entire system. |
| *18. Encoder* | the subsystem which alters the code of information input to it from other information processing subsystems, from a “private” code used internally by the system into a “public” code which can be interpreted by other systems in its environment. |
| *19. Output transducer* | the subsystem which puts out markers bearing information from the system, changing markers within the system into other matter-energy forms which can be transmitted over channels in the system's environment. |

Table 13 - 19 critical subsystems of all living systems, according to Miller (1978, 3)

The biological world is robust and full of exceptions. This is why living systems will have *most* of the subsystems, but not necessarily all of them. Indeed, let us imagine a living system whose *reproducer* subsystem is absent. Should it be considered as dead? It would be harsh for the mule and a woman in post-menopause. Still, Miller maintains that if a system lacks a critical subsystem, it will be eliminated by natural selection. This is of course most obvious for the *reproducer*.

Traditional astrophysics is concerned more about matter-energy than of information. Orthodox SETI is looking for an *output transducer* (19). However, the other information components will be hard or impossible to guess unless we make contact with an alien and we indulge into its dissection. However, pursuing a Dysonian SETI, we can still look for the 10 first subsystems, without excluding informational subsystems, if ever they become available.

We can now raise a fundamental question for SETI. What are the criteria for artificiality (or complex living system)? Would we consider having a proof of ETI *if and only if* we recognize a system displaying these 19 subsystems? Ideally, yes. In practice, there might be shortcuts. Indeed, the orthodox SETI is very smart in this regard, because by looking only at one subsystem, the output transducer, there is the possibility –and hope– that we will be informed about the 18 others. Again, it requires many assumptions (ETIs want to communicate, ETIs communicate in a manner we can detect, we will be able to decode their message, ETIs send information about their 18 other subsystems, etc.).

Is orthodox SETI the only way to bring a sufficient condition for proving the existence of ETI? It may be. Indeed, even a vehement SETI skeptic would be complied to accept the evidence of a message from the stars saying “Hello Earth, here is the recipe for cold fusion: …. ”. However, there is hope for success in Dysonian SETI if we cleverly combine global, thermodynamical and living system criteria.

To sum up, these sets of criteria are just a starting point, and more advanced concepts in non-equilibrium thermodynamics, living systems theory, because of their general concepts and applicability, will certainly provide key conceptual frameworks. The *energy rate density* metric is certainly very informative because it suggests that the distinction natural versus artificial may be of a continuous nature. Artificial intelligence and artificial life in particular are also important fields to further explore because they face many similar problems as SETI (Dick 2003). For example, how do you decide that a software simulation is “alive” or “intelligent”?

Now that we have some general criteria, we need “candidate” advanced extraterrestrial civilizations to try them out. A typical strategy to find advanced ETI is to extrapolate general trends of our own development. Although it is admittedly Earth-centric, we have to start somewhere, and we have just one option: life on Earth. So, let us try.

## 9.2 Two Scales for Civilizational Development

*Our destiny is density.*

(Smart 2012)

We can distinguish two very general scales for civilizational development (Table 14). Kardashev's scale measures the energy consumption of a civilization. It has been refined since its original publication, but its original version will suffice for our purpose. Barrow's scale measures a civilization's ability to manipulate small-scale entities. It has been largely ignored up to now.

| Kardashev Scale | | | Barrow Scale | | |
|---|---|---|---|---|---|
| KI | – energy consumption at | ~ 4 x $10^{19}$ erg $s^{-1}$ | BI | – manipulates objects of its own scale | ~ 1 m |
| KII | – energy consumption at | ~ 4 x $10^{33}$ erg $s^{-1}$ | BII | – manipulates genes | ~ $10^{-7}$ m |
| KIII | – energy consumption at | ~ 4 x $10^{44}$ erg $s^{-1}$ | BIII | – manipulates molecules | ~ $10^{-9}$ m |
| | | | BIV | – manipulates individual atoms | ~ $10^{-11}$ m |
| | | | BV | – manipulates atomic nuclei | ~ $10^{-15}$ m |
| | | | BVI | – manipulates elementary particles | ~ $10^{-18}$ m |
| | | | BΩ | – manipulates space-time's structure | ~ $10^{-35}$ m |

Table 14: Energetic and inward civilizational development.
Kardashev's (1964) types refer to energy consumption; Barrow's (1998, 133) types refer to a civilization's ability to manipulate smaller and smaller entities. In section (9.4.1 Two Scales Argument, p232), I combine those two scales.

### 9.2.1 Kardashev Scale – the Energetic Increase

Our civilization uses more and more energy. Energy is all-purpose, so we don't even need to understand the how or the why of this energy use to see that this trend is robust. Extrapolating this exponential increase of energy consumption, Kardashev (1964) showed that this would lead our civilization to type KII in year ~5164 and to type KIII in ~7764. Although Kardashev's original scale is an energetic one, it has often been interpreted as, and extrapolated to a spatial one. This is probably because the order of magnitude of the energy processed is as follows. Type KI harnesses the energy of a Earth-like planet; type KII harnesses the energy of a star and type KIII the energy of a galaxy (see e.g. Baugher 1985, 116 for KIII speculations). We are currently a ~KI civilization. Let us examine, as a typical example, our possible transition from type KI to type KII. Where are we in this scale? Probably in an unstable transition phase between KI and KII, as clearly illustrated in Figure 15.

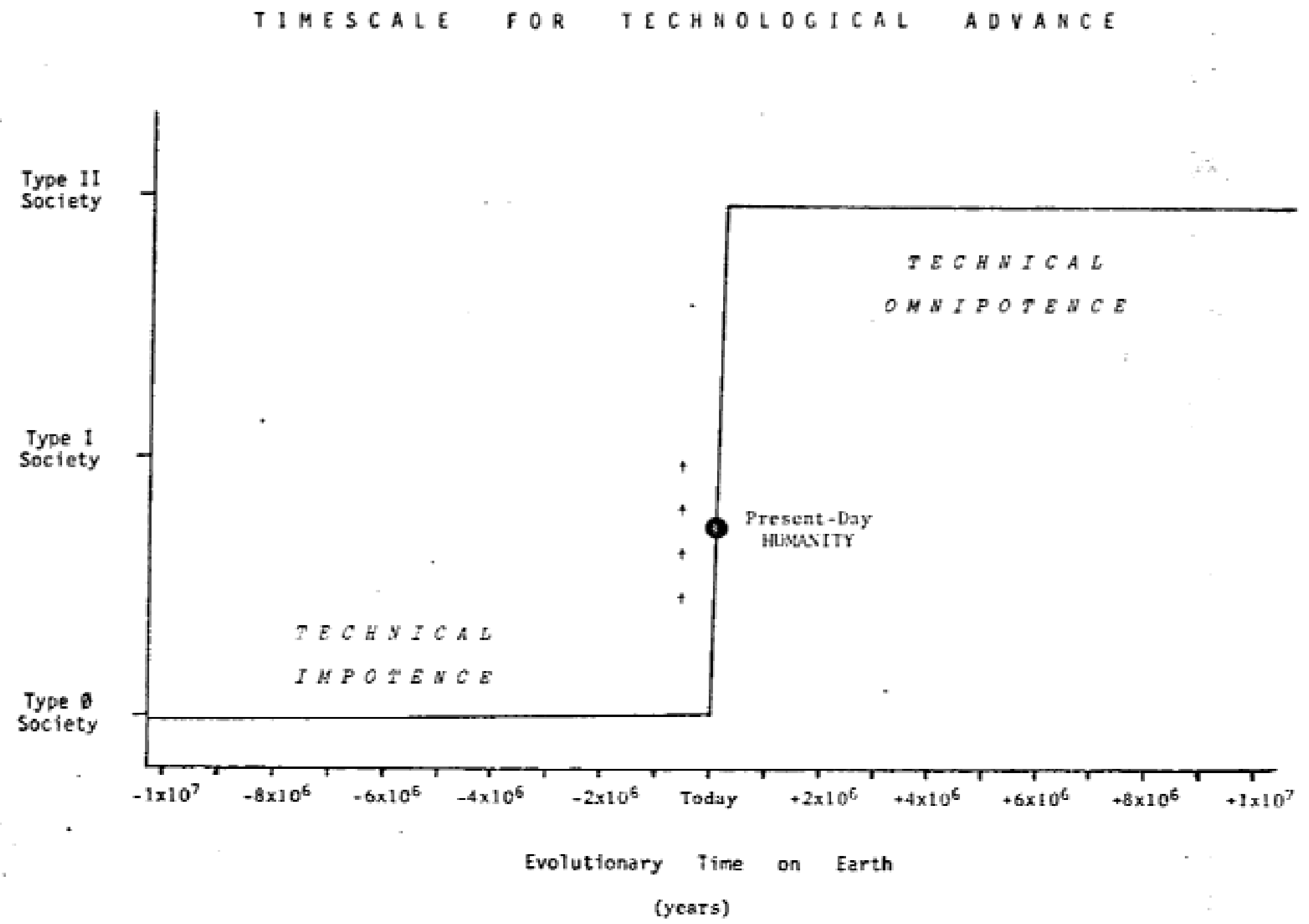

Figure 15 Freitas (1979, chap. 25.2.1) comments his figure: "the vast majority of sentient societies may lie on either side of the step (assuming humanity is a typical case) (Gunkel 1975). Most cultures may be regarded as "impotent" or "omnipotent" insofar as technical abilities are concerned. Only a tiny fraction of all evolving technological societies will be in the transition phase occupied by present-day humanity. Or, to put it in another more striking way, in any contemporary first contact situation humans are vastly more likely to encounter gods or animals, almost never peers. Indeed, it may be viewed as unethical for any omnipotent civilization to contact a society which is technologically impotent or in transition."

What motivations could we have to evolve to KII and harness the energy of the Sun? There are essentially two reasons. First, simply to meet our growing energy consumption needs; second, to avoid the predictable death of our Sun, associated with the destruction of life on Earth.

Let us first consider how to meet a civilization's growing energy needs. Einstein famously formulated the matter-energy equivalence formula $E=mc^2$. If we consider our solar system, where can we find most of its mass-energy? It is above all in the Sun, since 99.8% of our solar system's mass is in the Sun. That is, 99.8% of the energy in our solar system is to be found in the Sun. For any long-term use, the Sun is thus *the* obvious resource to harness energy from. Exploiting the energy of a star is an explorative engineering field known as *star lifting*, also called stellar mining or stellar engineering (see e.g. Reeves 1985; Criswell 1985; Beech 2008).

The second incentive to engineer our Sun is to avoid its red giant phase which will begin in ~5 billion years. This enterprise is vital if we are concerned about saving life on Earth. Various processes have been proposed for this purpose, resulting in an elimination of this red giant phase. The topic is treated extensively by Martin Beech (2008). From a SETI perspective, this leads to concrete and observable predictions. Beech (2008, 190–191) indeed proposes 12 possible signs of stellar rejuvenation in progress.

### 9.2.2 Barrow Scale – the Inward Manipulation

John Barrow (1998) classified technological civilizations by their ability to control smaller and smaller entities, as depicted in Table 14. This trend leads to major societal revolutions. Biotechnologies, nanotechnologies and information technologies are progressing at an accelerating pace and all stem from our abilities to control and manipulate small scale entities. This pivotal and omnipresent trend toward small spatial scales is largely overlooked in SETI, resulting in the Barrow scale being somewhat unknown. Barrow estimates that we are currently a type ~BIV civilization which has just entered nanotechnology. We could estimate that chromodynamical lifeforms Freitas described are type ~BV, weak lifeforms ~BVI and gravitational beings ~BΩ.

In contrast to the Barrow scale, many large-scale engineering or space travel projects have been proposed (see e.g. Badescu, Cathcart, and Schuiling 2006). However, there are strong obstacles against such grand projects. As Donald Tarter (1996, 292b) puts it:

> The unanswered question with these visionary proposals is economic, not technological. Our technology is fully capable of realising such projects, but our chequebooks forbid it. We are held on Earth not by the laws of physics but by laws of economics.

Taking economical factors into accounts, thoughtful speculations about space travel consider instead *small* scale self-replicating machines (see e.g. Bracewell 1960; 1962; Freitas Jr 1980).

Another argument for the importance of the Barrow scale is that, from the relative human point of view, there is more to explore in small scales than in large scales. As counter-intuitive as it is, space exploration offers more prospect in small scales than in large scales! This is illustrated in Figure 16.

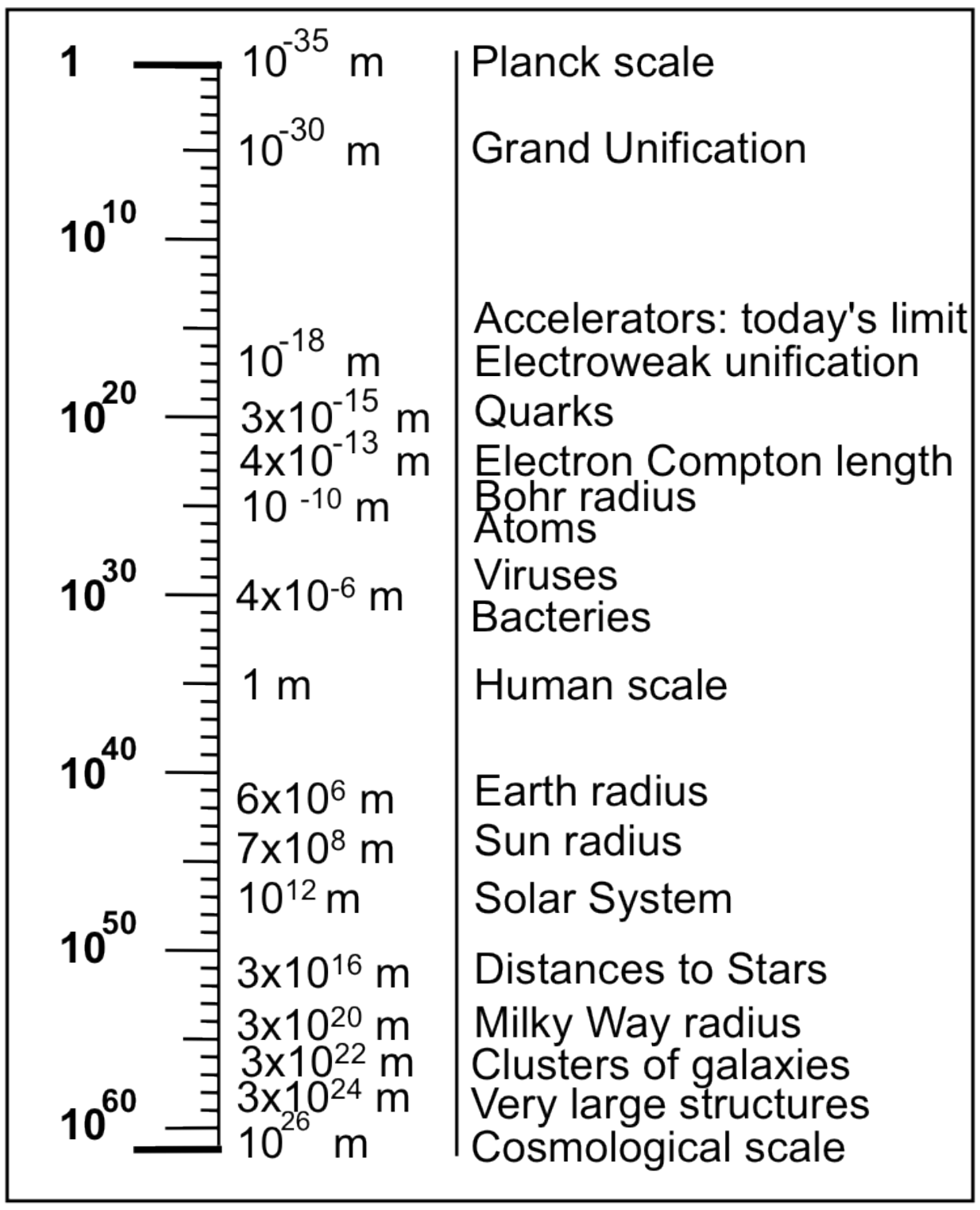

Figure 16 - Scales in the Universe.
That humans are not in the center of the universe is also true in terms of scales. This implies that there is more to explore in small scales than in large scales. Richard Feynman (1960) popularized this insight when he said "there is plenty of room at the bottom". Figure adapted from (Auffray and Nottale 2008, 86).

In contrast with large cosmological scales, manufacturing, testing, exploring and exploiting small scale technologies is easier, cheaper and more controllable. It is also more efficient energetically. Accelerating progresses on small scale engineering has no reason to slow down until we reach the Planck scale. Futurist and systems theorist John Smart (2009) characterized this trend as Space-Time-Energy-Matter (STEM) *efficiency* and *density*, or "STEM Compression". It can also be summarized by the motto "doing more with less".

I am more impressed when I look at a SDXC memory card the size of a post stamp, containing 256 GB of data, rather than contemplating the Great Egyptian Pyramids. And I can be sure that my children will find my astonishment for such a small amount of memory stored on such a big structure laughable. The feats of today's architects and engineers are to organize small scales, not large ones.

The two-dimensional metric complies with Zen SETI standards, because both the sheer energy use and the engineering scale are independent of goals or a particular technology. Putative ETIs can use energy for whatever they like; they can construct, organize and engineer at small scales whatever they deem useful. In SETI discussions, Kardashev's scale is widely accepted and used precisely because energy is technology neutral. The same holds for the scale-density, it is technology neutral. We can simply assume that an intelligent civilization will develop to type KII, KIII, and/or up to BΩ, whatever their purpose.

Still, it is intriguing and stimulating to speculate about what would further motivate an advanced civilization to climb those two developmental trends. I must warn the reader that the next section contains scientific and philosophical speculations, so the reader averse to such speculations, if any is still reading, might thus jump directly to section (9.4.1 Two Scales Argument, p232) for an application of the two-dimensional metric.

## 9.3 Black Holes as Attractors for Intelligence

*I love this topic, but it stretches my brain beyond its capabilities*

Internet user Aaron commenting on
Ray Villard's (2011) news report
about my (2011) paper.

*Les trous noirs, c'est troublant*[13]

### 9.3.1 Unknown Black Hole Technology

If we extrapolate the Barrow scale to its limits, we come to a civilization able to manipulate space-time, or what Freitas called gravitational beings. However, because gravitation is such a weak field, a lot of mass and density must be present to obtain significant effects. What are the densest objects in the universe? Black holes. They are fascinating attractors, not only because of their staggering gravitational field, but also because they are an intelligence's greatest potential. I now invite curious readers to explore this idea via a short adventure into the speculative topic of black hole technology.

### 9.3.2 Energetic

Black holes are the densest objects in the universe. If we want to address the need of continually increasing energy consumption, it would be beneficial to be able to store or extract energy from black holes. Roger Penrose (1969, 270–272) imagined the following extraction mechanism. It consists of injecting matter into a black hole in a carefully chosen way, thereby extracting its rotational energy (see also Misner, Thorne, and Wheeler 1973, 908 for more details). Blandford and Znajek (1977) suggested a similar process with electrically charged and rotating black holes. Other proposals suggest collecting energy from gravitational waves of colliding black holes. Misner imagined this in 1968 as a personal communication to Penrose (1969). Frautschi (1982) also proposed to merge black holes as a way to produce a power source. Louis Crane (2010, 370) has also suggested that small black holes could be used as an energy source, since they can convert matter into energy via the Hawking radiation with great efficiency.

13This is an untranslatable wordgame... let's try anyway. Literally it means “Black holes are troubling”; but in French, “troublant” (troubling) has the same pronunciation as “trou blanc”, which means “white hole”.

### 9.3.3 Societal

The Hawking radiation, Crane argues, could make them the perfect waste-disposal device. Chaisson (1988, 197–198) also envisioned that a black hole would be an ideal solution for a civilization like ours, short on energy and long on garbage. Crane and Westmoreland (2009) conducted an extensive study on the possibility of black hole starships. Paul Davies (2010, 142) also speculated that a black hole could be used to launch a spacecraft at a relativistic speed, by smartly using its gravitational field. Furthermore, general relativity leads to the fascinating topic of time travel via worm holes, theoretical cousins of black holes. Although their existence is extremely implausible, they could in theory provide shortcuts for traveling in space-time (for popular accounts see Thorne 1994; Randall 2005).

### 9.3.4 Scientific

*Of all the entities I have encountered in my life in physics, none approaches the black hole in fascination. And none, I think, is a more important constituent of this universe we call home.*

John A. Wheeler in (Taylor and Wheeler 2000, F–31)

Let us assume that terrestrial and ETIs are curious and continue to develop science. Black holes, especially their interiors, currently challenge our knowledge of the three fundamental physical theories: quantum mechanics, general relativity and thermodynamics. For scientific purposes, there might be an incentive to artificially produce black holes to better understand them. Although it remains an engineering challenge, Westmoreland (2010) showed how artificial optical black holes could be created out of electromagnetic radiation. Even though improbable sources of danger, some concerns have been raised regarding the accidental production of micro black holes in particle accelerators (Giddings and Thomas 2002). Still, we might want to produce them *intentionally* in the future.

A more concrete scientific application of black hole technology is to use them as telescopes or communication devices. How is it possible? An established consequence of general relativity theory is that light is bended by massive objects. This is known as *gravitational lensing*. For a few decades, researchers have proposed to use the Sun as a gravitational lens (see e.g. Von Eshleman 1979; Drake 1988). At 22.45AU and 29.59AU we have a focus for gravitational waves and neutrinos. Starting from 550AU, electromagnetic waves converge. Those focus regions offer one of the greatest opportunity for astronomy and astrophysics, offering gains from 2 to 9 orders of magnitude compared to Earth-based telescopes. Over the years, Claudio Maccone (2009) has detailed with great technical precision such a scientific mission, called FOCAL. It is also worth noting that such gravitational lensing could be used not only for observation, but also for communication. If we want to continue and improve our quest for understanding the cosmos, this mission is a great opportunity to complete our fuzzy astronomy with a focused one. In other words, the time may be ripe to put on our cosmic glasses and to use cosmic loudspeakers.

But other ETIs may already have binoculars. Indeed, it is easy to extrapolate the maximal capacity of gravitational lensing using instead of the Sun a much more massive object, i.e. a neutron star or a black hole. This would probably constitute the most powerful possible telescope. This possibility was envisioned –yet not

developed– by Von Eshleman in (1991). Recently, Claudio Maccone (2012) studied the gravitational lensing potential of supermassive black holes, and showed that even intergalactic communication would be feasible thanks to them. He writes (Maccone 2012, 119–120) that

> this line of thought clearly shows that the central massive black hole of every galaxy is by far the most important "resource" of that galaxy for SETI purposes. In fact, it is like the "central radio station" of that galaxy that every civilization living in that galaxy would like to control in order to keep in touch with other aliens living in nearby galaxies.

Since objects observed by gravitational lensing must be aligned, we can imagine an additional dilating and contracting focal sphere or artificial swarm around a black hole, thereby observing the universe in all directions and depths. Maybe such focal spheres are already in operation. The gains offered by such devices are largely unknown, but this is an exciting topic for an open minded researcher or a PhD student in general relativity.

### 9.3.5 Computational

What is the maximal information that can be processed by an advanced ETI? Elaborating on the work of Bremermann (1982), Robert A. Freitas Jr. (1984) introduced the *sentience quotient*, which is a quantitative "scale of cosmic sentience universally applicable to any intelligent entity in the cosmos". It is defined as $I/M$, the ratio of the information processing rate $I$ to the entity's mass $M$.

At its limits, we have the maximal computational density of matter, what Seth Lloyd (2000) more recently called the "ultimate computer". What does such a computer look like? Lloyd argues that it is a black hole. Interestingly, if Moore's law is extrapolated, we attain such a maximal computational power by 2205 (Lloyd 2005, 162).

Kurzweil argued that the "computational power of a computer is a function of its mass and its computational efficiency". He adds (Kurzweil 2006, 362):

> Once we achieve an optimal computational efficiency, the only way to increase the computational power of a computer would be to increase its mass. If we increase the mass enough, its gravitational force becomes strong enough to cause it to collapse into a black hole. So a black hole can be regarded as the ultimate computer.

But black holes can be even more than ultimate computers. At the edge of theoretical computer science, some models of computation outperform Turing's original definition. Such devices are called *hypercomputers* (see e.g. Earman and Norton 1993). They are theoretically possible assuming particular space-time structures or with slowly rotating black holes (see e.g. Etesi and Németi 2002; Andréka, Németi, and Németi 2009). If the construction of such hypercomputers is successful and indeed possible, this would bring qualitatively new ways to understand and model our universe. A breakthrough perhaps comparable to the invention of the computer itself.

### 9.3.6 Philosophical

Intelligence is the capacity to solve problems. It is by focusing on universal and long-term problems that we have the highest chances to understand the purpose of presumed ETIs. I see only two such serious problems. The first is the already mentioned red giant phase of a star, capable of wiping out life in a solar system like ours. This is a fundamental challenge any civilization born on the shore of a Sun-like star will have to face. A promising SETI strategy is thus to search for civilizations refusing this fate, by looking at artificially modified stars. According to Criswell (1985, 83) star lifting can considerably extend a civilization's time with matter to energy conversion, up to 2 millions times the present age of the universe, assuming the civilization stays at ~KI. Yet, even this runs out in the long term because the star will ultimately run out of usable energy.

What happens then? Possibly migration to other solar systems, but that also cannot continue forever, because new star formation comes to an end in the very long term (F. C. Adams and Laughlin 1997). After realizing that the fate of stars is doomed, the longest term and truly universal problem is the continuation of the universe as a whole, to avoid its inevitable global entropy increase and death (see Ćirković 2003 for a review paper on physical eschatology).

The second challenge is thus, "How can we make life, intelligence and evolution survive infinitely?" We already mentioned two proposals which include a role for black holes (see section 8.3.13 The Case for CAS, p197 and Chapter 8). Freeman Dyson proposed in his landmark (1979) paper that a civilization could hibernate and exploit the time dilation effects generated by black holes to survive forever.

Another speculative solution is to reproduce the universe (see Chapter 8 and e.g. E. R. Harrison 1995; J. N. Gardner 2003; Baláz 2005; Smart 2009; John Gribbin 2009; Stewart 2010). The authors of such scenarios combine the origin and future of the universe with a role for intelligent life. Based on these, I have developed a scenario which I call *Cosmological Artificial Selection* (see Chapter 8 and Vidal 2008b; 2010a; Vaas 2009; 2012 for critical commentaries; and Vidal 2012a for a response) as it is a philosophical extension of Smolin's (1992; 1997) theory of *Cosmological Natural Selection*. It is also worth noting that the emerging discipline of Artificial Cosmogenesis (see Chapters 6, 7 and Vidal 2008b), analogous to Artificial Life but extended to the cosmos, would benefit the power of putative ultimate computers, to run simulations of whole universes.

Finally, if we assume that our universe is a black hole (e.g. Pathria 1972), the puzzling fine-tuning of universal constants could itself be interpreted as an intelligent signal from previous universe makers (Pagels 1989, 155–156; J. N. Gardner 2003). As we mentioned, if this radical Search for ExtraUniversal Intelligence would succeed, it would lead to a refutation of Universecentrism.

## 9.4 Signs of Starivores?

*The later the stage we look at in a binary's evolution,*
*the more difficult it is to see how a binary could have so evolved.*

(Lipunov 1989, 311)

### 9.4.1 Two Scales Argument

In a SETI mindset, considering seriously that black holes are attractors for intelligence, we can now start to ask the following questions. What are the observable manifestations of a black hole when it's used as an energy source? as waste disposal? as a time-machine? as a starship engine? as an ultimate or hyper computer? as a universe production facility?

The exercise is highly speculative, and raises the *efficiency objection*. We saw that the Barrow scale trend makes civilizations develop with more and more efficiency. This would make small black holes more useful and thus hard or impossible to detect. It would be like trying to detect from Earth the existence of nanotechnology on the Moon. This line of argument is the essence of Smart's (2009; 2012) response to Fermi's paradox. We don't see other ETIs because they are confined inside black holes. Another example is if a civilization develops the capacity to perform thermodynamically reversible computation. It would then generate almost no entropy, and therefore be undetectable. Yet, as Krauss and Starkman (2000) have argued, erasure of unnecessary memories are essential for something like consciousness to continue in the universe, and this operation has an entropic cost (Landauer 1961). So, we can expect that a very efficient civilization would still generate entropy.

Does a civilization has to choose between energy intense technology and energy efficiency, as Donald Tarter (1996, 3) suggested? No! At least we never did. Our technology has always been more efficient, yet it has also always been more energy hungry. The two trends of more energy use and more energy efficiency are definitely not incompatible. The key lies in the availability of energy. If it is poor, efficiency will strongly constrain civilizational development. If energy is largely available, then efficiency matters less and civilizations can also grow on the Kardashev scale.

To summarize, on the Kardashev scale, we saw that a Type KII civilization would be able to use an amount of energy of the order of a star, with an endeavor called star lifting. Considering the magnitude of such an undertaking, it has good chances to be observable. On the Barrow scale, we have argued that density attracts intelligence, up to black hole organization. We call such a civilization type BΩ. It is the culmination of a civilization on that scale.

Now, can we derive a concrete astrobiological search strategy, combining both the Kardashev and the Barrow scale? Could a civilization harness with great efficiency the energy of a star, to run its organization at black hole –or lower– density? Can we imagine one day detecting such a configuration?

We don't need to imagine or to wait because such configurations already exist! Indeed, about 20 systems composed of a black hole accreting gas from a star have been found today (e.g. GRO J1655-40, GRS 1915+105, 1659-487, SS433, etc.). They are part of the family of binary systems, called X-Ray Binaries (XRB) because of their emissions in the X-Ray electromagnetic spectrum. For decades, they have been studied as natural astrophysical systems. However, we will now take an astrobiological stance and see to what extent they can be considered as artificial astrophysical systems.

Importantly, researchers have concluded that a thin accretion disk around a rotating black hole is one of the *most efficient power source in the universe*, a process up to ~50 times more efficient than nuclear fusion occurring in stars (e.g. Thorne 1974; Narayan and Quataert 2005). There is only one more efficient process, which is the reaction between matter and its corresponding anti-matter particle, which is 100% efficient, converting all of the mass into energy. If any civilization is to climb the Kardashev scale, it would certainly at some point want to master those energetic sources.

The trend towards black hole density was aimed at finding *maximally* advanced civilizations. But we can easily downgrade the argument to less energy use and less density. So we can replace the black hole (BH) with a neutron star (NS) or a white dwarf (WD) or even a planet. And this opens the door of the fascinating binary zoo.

### 9.4.2 A Partial Visit to the Binary Zoo

Traditional astrophysics sees white dwarfs, neutron stars or black holes as the stellar graveyard, because in most cases such dense bodies are theorized to be the remains of dead stars. However, these bodies are perplexing through the variety of their behavior. Binary astrophysicists speak of the "binary zoo", because of their staggering variety. As Lipunov (1989, 206) puts it:

> Looking back at the late 1960s, the study of variable stars seemed a wonderland. Articles, books, and catalogues swarmed with the types of variables whose diversity terrified the theorist. There were novae, novae-like stars, recurrent novae, dwarf novae, flare stars, cataclysmic stars, eruptive stars, etc. Some stars were known simply as irregular variables, while some stars of the same type were often called different things. Any attempt at classification seemed hopeless.

If you don't know anything about the subject, an excellent introduction at a popular science level is the book *In the World of Binary Stars* (Lipunov 1989).

The population of binary stars is large. Although it is commonly believed that two third of stars are multiple, it is based on a quick generalization of the study of G stars by Duquennoy and Mayor (1991). Indeed, recent research also counting faint red dwarf stars shows that *two third of stars are actually single* (Lada 2006).

The phenomenological diversity of binaries makes the subject very difficult and technical. Why is it so difficult? After all, shouldn't Newton's equations close the issue and easily deal with two gravitational bodies? The many complications arise because, in some rarer cases, the two bodies interact. They exchange matter-energy, in rough or subtle ways, which changes their evolutionary course. This makes our knowledge of single star physics insufficient. How and why do they exchange matter?

What is the evolutionary outcome of such interactions? These are some of the core questions of binary stars astrophysics.

Let us try to draw a map of the binary zoo with an astrobiological mindset. Some classify binaries by the method by which they are discovered (e.g. visual, spectroscopic, eclipsing, astrometric). This classification is useful for astronomers, but not directly relevant for theoreticians. A better classification is the one made by Kopal (1955) which classifies binaries in three types: *detached binaries, semi-detached binaries* and *contact binaries* (see Figure 17). This depends on how they fill their *Roche Lobe*. What is the Roche lobe? It is the largest volume that a star can occupy without disturbing its neighbor companion (see Figure 18 below). If the two stars do not fill their Roche lobe, they don't interact and are called *detached binaries*. If one star expands or if the binary orbit shrinks, material is lost from the point nearest the companion, the *Lagrangian point* $L_1$. The binary is *semi-detached*. If the star continues to extend its Roche lobe, exchange of matter can also occur through the Lagrangian points $L_2$ or $L_3$. Such stars are called *contact binaries*.

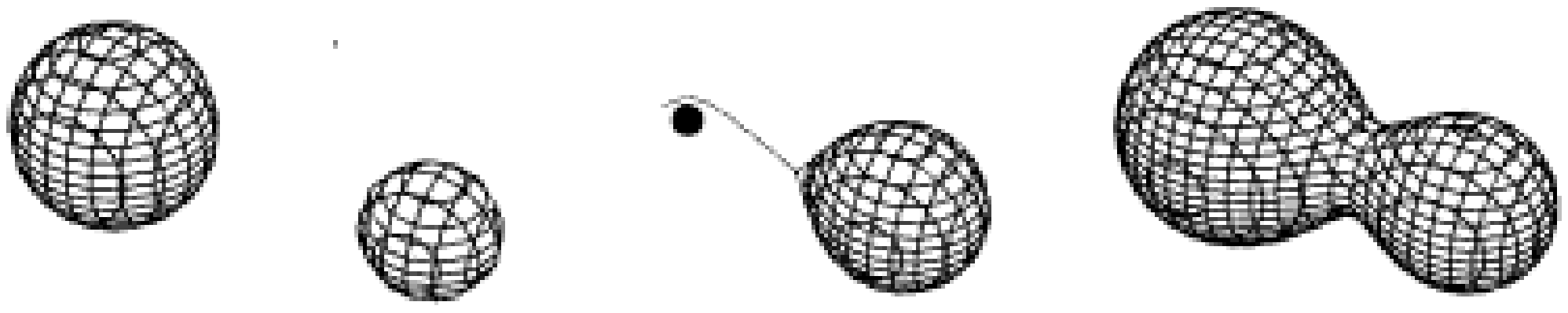

Figure 17 - "*Left*: A detached binary with both stars within their Roche lobes. *Middle*: A semi-detached binary: the secondary fills its Roche lobe emitting a stream of material from $L_1$. If the primary is small enough, the stream will orbit around it. If it were larger, the stream would hit the primary, as occurs in some Algol-type binaries. *Right*: A contact binary, with both stars overfilling their Roche lobes." From (Hellier 2001, 22)

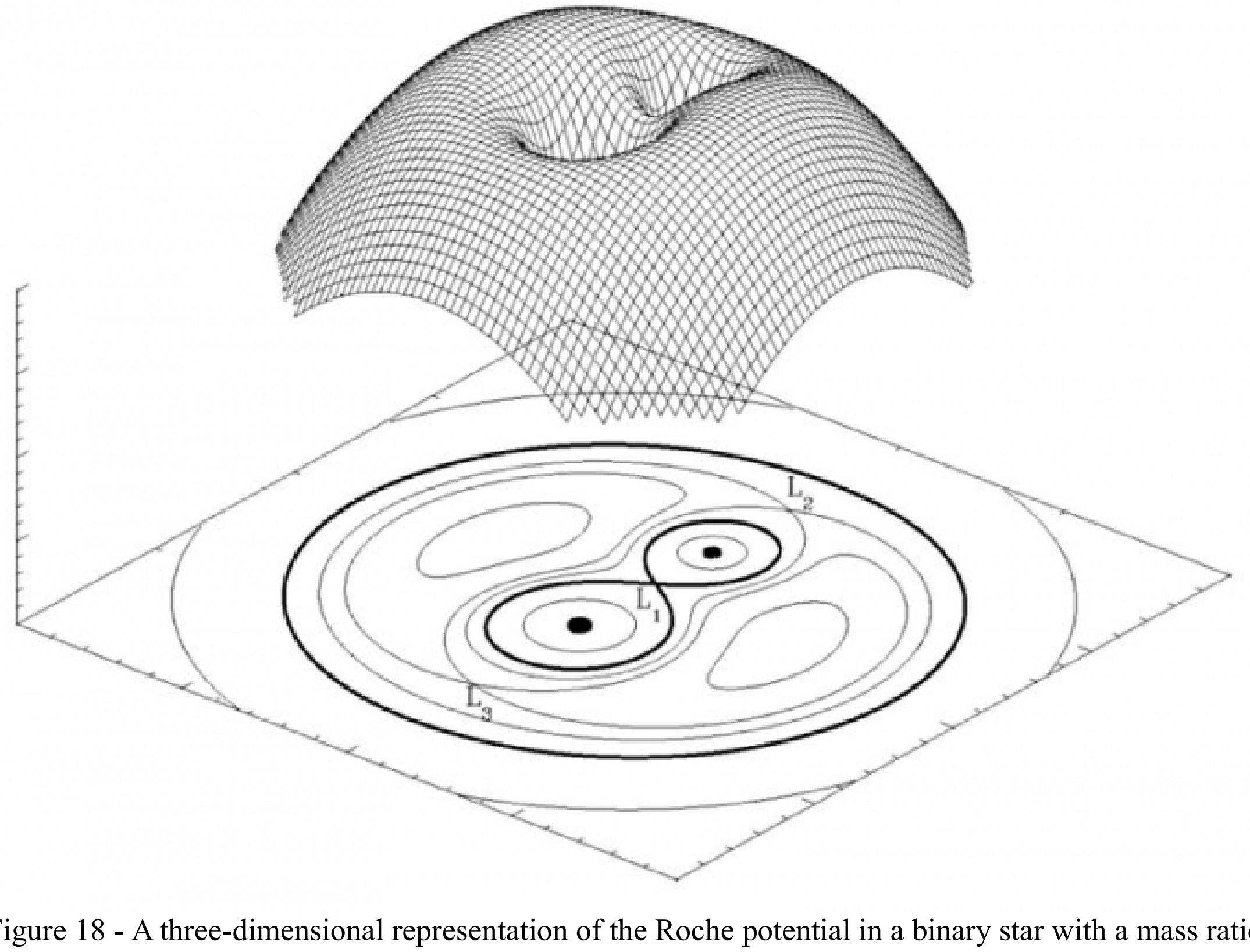

Figure 18 - A three-dimensional representation of the Roche potential in a binary star with a mass ratio of 2, in the co-rotating frame. The droplet-shaped figures in the equipotential plot at the bottom of the figure are called the Roche lobes of each star. $L_1$, $L_2$ and $L_3$ are the points of Lagrange where forces cancel out. Mass can flow through the saddle point $L_1$ from one star to its companion, if the star fills its Roche lobe. Author: Marc van der Sluys, 2006, Wikimedia.

What does this all mean for SETI? Let us remember the thermodynamical criteria. Broadly speaking, *detached binaries* are like two stones, they do not exchange matter and influence each other only through gravitational pull. *Contact binaries* often evolve to a *common envelope* phase, where stars exchange matter unstably and rapidly until the system reaches an equilibrium. This is alike a wild forest fire, which involves a lot of energy and reaches a state of equilibrium when the fire has nothing left to burn. However, *semi-detached binaries* are suspect. The energy flow exists, it is irregular but does not appear out of control. Let us have a closer look at these interacting binaries.

In his (2006) book, Peter Eggleton distinguishes three main ways binaries can interact. They can interact via a *conservative process*, where the overall mass of the binary system is conserved during interactions. The *rapid non-conservative processes* involves mass which is expelled out of the system. The most famous case in this category are type Ia supernovae, triggered when a compact star (*white dwarf* or *WD*) accretes more matter than it can support, and explodes. The third category are the *slow non-conservative processes*, where mass is expelled out of the system, but in a slow way.

Let us consider again our thermodynamical criteria for living systems. Which ones of the three interactions are most alike a living system? Conservative processes are not good candidates because no matter is expelled in a sink out of the system. Rapid non-conservative processes are also not promising because the duration is short and the end point is the total destruction of the system. The third category is more promising because all the conditions of a metabolism are put together. There is a regulation of the energy flow from one star to the compact object, an energy gradient between the two components, and some matter is regularly ejected through cataclysms (novae) or jets. Importantly, unlike supernovae, novae and jets do not destroy the system. So, it makes sense to narrow again our focus to this category.

Eggleton reviews many different processes driving slow non-conservative processes, such as gravitational radiation, tidal friction, wind processes, magnetic braking or stellar dynamos. It is beyond my expertise to explain or explore in details the complicated physics of such systems.

Indeed, such processes become very hard to model. They require to take into account simultaneously the most important physical theories. Thermodynamics for the energetic exchanges, magnetohydrodynamics to describe the flux of gas driven and channelled by magnetic fields from one component to another, relativistic effects when bodies are very dense (white dwarf, neutron stars, black holes) and quantum effects because of their high densities. This is without mentioning the indispensable prerequisite of stellar (single stars) astrophysics. Such extreme regimes are unique opportunities to test physical theories, because we have no way to even approach such conditions by setting up experiments here on Earth. This is why there are so important and interesting to physicists and astrophysicists. In most cases, the gainer of mass (usually the dense body) develops an *accretion disc,* where matter is stored and rotates before interacting with the dense body. The physics of accretion discs is very challenging, notably because it involves phenomena of turbulence and viscosity, still poorly understood.

Accretion is an ubiquitous astrophysical process in galaxy and planetary formation, so we may object that all binaries may simply always be natural. Let me however introduce an analogy. Fission can be found in natural forms, as well as fusion, which is one of the core energetic processes in stellar evolution. Yet, humans try to copy them, and would greatly benefit to –always– control them. So it is not because a process is known to be natural that its actual use is not driven by an intelligence. In fact, the situation may even be more subtle. The formation of XRBs might be natural, but controlled or taken over by ETIs, like a river flowing down a mountain is a natural gravitational energy source humans can harness with hydroelectric power stations.

We now have enough concepts to define more precisely a putative ETI in a binary system. It is *an extraterrestrial civilization using stellar energy (Type KII), in the configuration of a slow non-conservative transient accreting binary (thermodynamic criterion), with the dense primary (Barrow scale) being either a planet, a white dwarf, a neutron star or a black hole*. I call such hypothetical civilizations *starivores*, defined simply as:

> **starivore**: a civilization that actively feeds on stars.

This is a convenient shorthand for the more accurate definition above. Note that if such binary systems are starivores, then we should find the primitive versions of them, extracting energy from a star with a planet, which is *not* dense compared to WDs, NSs or BHs. This would happen at a low accretion rate, and planetary accretion is one of the concrete prediction from the starivore hypothesis (and indeed planet-star interactions have recently been discovered, see e.g. Lecavelier des Etangs et al. 2012).

Up to now, we have only raised suspicion on some peculiar binary systems, out of thermodynamical equilibrium. But again, this is only a necessary condition for extraterrestrial life, not sufficient. We will now apply the criteria we developed to further probe the starivore hypothesis.

### 9.4.3 General Arguments

Putative starivores comply with the *strangeness* criterion, because they are difficult to model and predict, and display an impressive variety of behavior. This invites us to study them with an astrobiological stance.

The heuristic of *non-exclusiveness* is clearly triggered because the literature on binaries shows a wide variety of them. The variety of white dwarfs in so-called "cataclysmic variables" is particularly puzzling (see the reference book, Warner 1995; and Hellier 2001 for a more accessible account). The accretion process itself can be quite varied. There are of course accretion discs, but Warner (1995, 334) also reports cataclysmic variables accreting magnetically not on 1 or 2 magnetic poles, but on 3 or 4!

The *equilibrium heuristic* seems also to be respected because there are already plenty of binaries (and putative starivores) out there.

The *inverse distance-development principle* leads to a concrete testable prediction. We should see less and less putative starivores as we look in further and further away galaxies. We can even refine the prediction and state that higher accretion rate, i.e. energy use, or higher development on the Kardashev scale should be less and less prominent as we look further away. The same holds with density or the Barrow scale. We should see less and less semi-detached binaries in transient accretion as we look in further and further away galaxies. For example, black holes or neutron stars in transient accretion should be even rarer than cataclysmic variables. Of course, such "artificial" evolution reasoning must be confronted with "natural" evolution models, to see which of the two explains and predicts more.

### 9.4.4 Thermodynamical Arguments

*Biological complexification leads to* high energy, far-from-equilibrium *systems, rather than the lower energy, equilibrium systems that are the target of non-biological complexification, so in that fundamental sense the two are quite distinct.*

(Pross 2005, 153–154)

The non-equilibrium thermodynamical argument for the existence of starivores can be summarized as follows:

1. Living systems are far-from-equilibrium metabolic systems
2. We can look in the universe for such systems
3. A subclass of binary systems are the most obvious such systems

This straightforward argument is purely thermodynamical and totally independent from the two-scales civilizational development extrapolation and reasoning (section 9.4.1 Two Scales Argument, p232). Of course, it only points to a necessary and not a sufficient condition for the existence of starivores, but it is remarkable that it leads to the same result as the two scales argument.

Furthermore, as Pross writes in the quote above, the evolutionary trend of complexification is towards high-energy living systems. Could life's complexification indeed lead in the long-term to remarkably high-energy systems such as cataclysmic variables or microquasars?

Of course the weakness of the argument is in step 1. Not only living systems are out of equilibrium. Also dissipative self-organized systems can achieve this state. This is why we need to inquire more about binaries, notably to establish if there is an *energy flow control* which allows the regulation of metabolic processes.

Imagine that you go hiking. You stumble upon a waterfall, which is quiet. As you peacefully look at it, suddenly an enormous strain of water flows, and the next second, it becomes quiet again. And this would go again and again at regular yet not quite predictable intervals. Would you bet the phenomenon is natural or artificial? This is today the main challenge of proving or disproving the starivore hypothesis. Is the accretion behavior natural, like a sudden ice melting strengthening the waterfall? Or is the accretion due to a purposeful behavior, like the sudden water release from a dam to regulate a hydroelectric power station?

When I browse through the binary literature in astrophysics, I find it striking that the question is not *whether* there is accretion control, but *how* it operates. As an example, it is worth quoting Hellier (2001, 111–112) on magnetically-controlled accretion of the system AR Uma:

As the stream approaches the white dwarf, the increasing magnetic pressure of the converging field lines first squeezes the stream, causing it to break up into dense 'blobs' of material. The field cannot easily penetrate such blobs because of screening, so they continue ballistically for a while. As the magnetic pressure climbs the blobs are forced to change direction; collisions in the stream form shocks, energy is dissipated and radiated away, and a pool of material can collect in a 'stagnation' or 'threading' region. This region extends over a range of radii, owing to the range of blob densities. Material from the pool –a mixture of blobs and a fine 'mist' of material stripped from the surface of the blobs– then diverts along field lines and flows onto the white dwarf.

It is also worth mentioning that the accretion is not rough in the sense that it reaches a small region of the white dwarf, 1/100$^{th}$ of its radii (Hellier 2001, 111). Figure 19 shows different accretion types in accreting white dwarfs. They are called *cataclysmic variables*, and if a strong or less strong magnetic field is present, *polars* or *intermediate polars.*

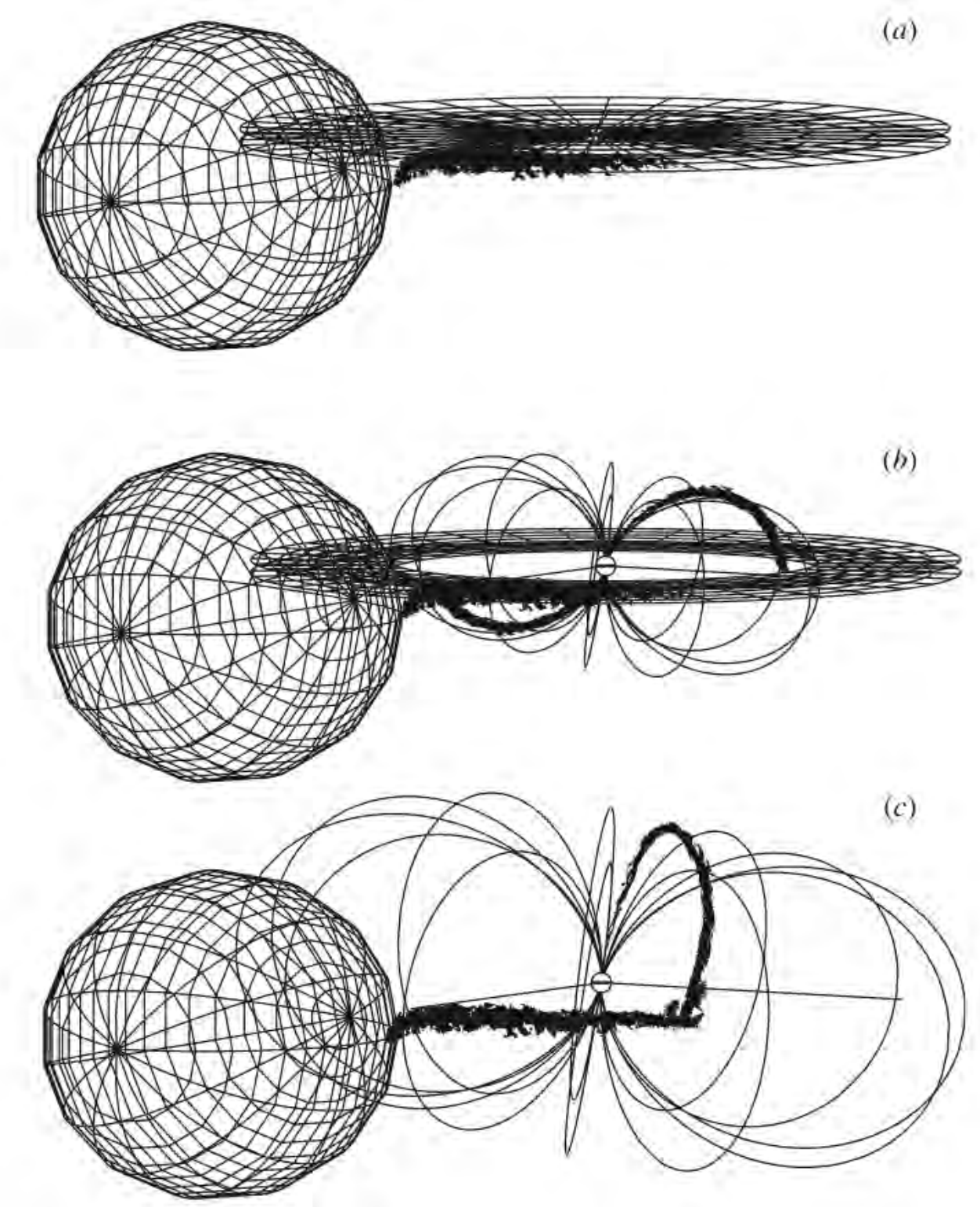

Figure 19 Schematic illustrating the different types of accretion flow in cataclysmic variables. ( a) Non-magnetic systems; (b) intermediate polars and (c) polars. From (Cropper et al. 2002, 2915).

To sum up, a robust necessary condition for life is a metabolism, that is, the utilization of a flow of energy to draw order from chaos and build internal complexity, while dissipating entropy. This situation appears to be fulfilled in some binary systems like cataclysmic variables, X-Ray pulsars or microquasars. They all display an energy flow coming from their companion star, and dissipate entropy in the form of regular cataclysms (in cataclysmic variables) or jets (X-ray pulsars and black holes). Such binaries therefore have a kind of *metabolism*, a fundamental property of living systems. Admittedly, we have not established whether or not such system build internal complexity. Up to now the arguments presented are only suggestive and prove nothing definitely, so let us now see if some binaries can do better and fulfill more living systems criteria.

### 9.4.5 Living Systems Arguments

Let us take again the first 10 critical subsystems that all living system have, and see if putative starivores display them (see table 15 below).

| | |
|---|---|
| MATTER + ENERGY + INFORMATION | |
| *1. Reproducer* | The hypothesized cosmological artificial selection scenario, where *black holes* play a key role in universe reproduction (see Chapter 8). |
| *2. Boundary* | *White dwarfs* have atmospheres (hydrogen and helium layers), which regulates the energetic outflow of the star.<br>*Neutron stars* have outer and inner crusts.<br>*Black holes* in rotation have an ergosphere and an event horizon, which delimit boundaries for radiation or light to escape or not. |
| MATTER + ENERGY | |
| 3. Ingestor | Binaries display many different types of accretion methods. We saw that magnetic white dwarfs make the accretion follow fields lines. Other accretion types include Roche-lobe overflow, tidal friction, gravitational radiation, magnetic activity driven by rapid rotation, stellar winds, magnetic braking, accretion disc and accretion curtain. |
| 4. Distributor | This component is unclear, unless a mechanism to distribute the accreted energy is found in WD, NS and BH. |
| 5. Converter | Conversion of energy extracted from the secondary, for changing the orbit, changing the rotation; increasing the magnetic field; regulating the accretion flow or maintaining an hypothetical internal organization. In white dwarfs, note that the material extruded (nova ejecta) has a different chemical composition from the accreted material. |
| 6. Producer | Subject to different interpretations. In white dwarf, the system of accretion and recurrent dwarf novae outbursts.<br>In neutron stars and black holes, the system of accretion and periodic jets. |
| 7. Matter-energy storage | Matter-energy storage in binaries is mainly in the accretion disc. The disc can also act as an energy buffer. However, energy can also be stored in the rotation of the dense component, or in its sheer mass. |
| 8. Extruder | In *white dwarfs*, recurrent novae, or classical novae.<br>In *neutron stars* and *black holes*, the relativistic jets. Their composition remains a matter of debate. |
| 9. Motor | We should expect small motor control, such as orbital period, rotation speed, inclination (e.g. in polars to move the magnetic fields lines and thus control the accretion rate).<br>Some binaries also move at high speed through the galaxy (e.g. the *neutron star* IGR J1104-6103 or the *black hole* XTE J1118+480). |
| 10. Supporter | Unclear. But accretion control may require adjustments of many parameters in the binary system to be effective. |

Table 15: Tentative living systems interpretation of some binary systems, from a high energy astrobiological perspective. Ten critical living subsystems are suggested to apply to interacting binaries composed of a primary white dwarf, neutron star or black hole .

I will now discuss and comment on this tentative living systems interpretation, starting with white dwarfs then neutron stars and black holes. I will italicize critical subsystems when I mention them.

One could think that the accretion rate of white dwarfs is simply determined by the orbital period and the masses of the two stars. This is wrong. As Hellier (2001, 171) writes, “systems that are very similar in these respects can have $\dot{M}$ [accretion rate] values differing by factors of 100-1000”. This discrepancy is huge and accretion rate is really a difficult parameter to explain.

But let us hypothesize that starivores use strong magnetic fields as their *ingestor*. Then there is at least a way to control accretion. It is “simply” to tilt the inclination of the white dwarf, as shown in Figure 20.

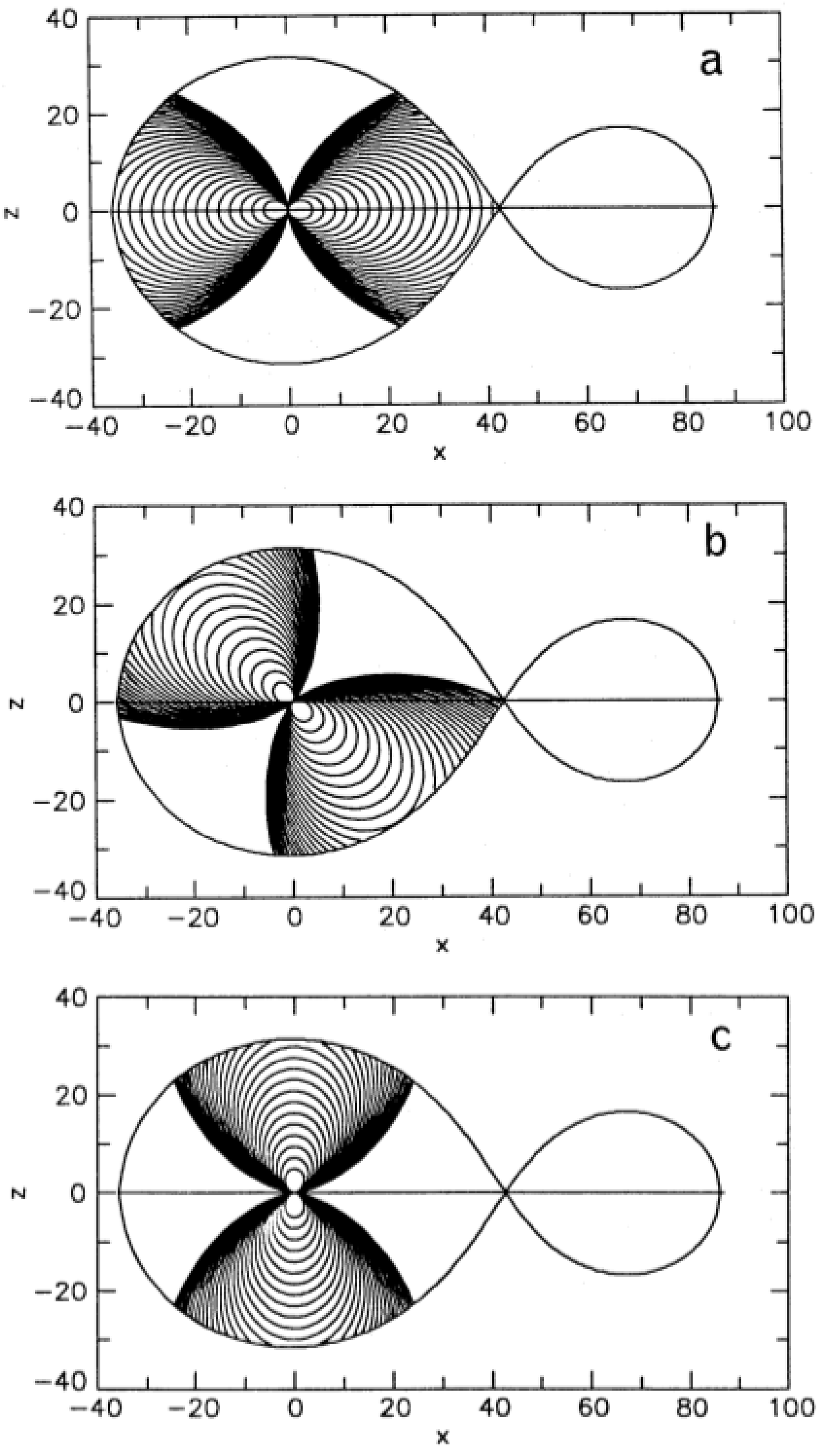

Figure 20 - Field lines of the white dwarf which are contained within the Roche lobe for different values of dipole inclination: (a) 0° (b) 45° (c) 90°. From (Ferrario, Tuohy, and Wickramasinghe 1989). Under the starivore interpretation, case (a) provides a low accretion rate, (b) a high accretion rate and (c) no accretion.

Another important strategy to understand living systems is to look at their waste products expelled through the *extruder*. If we apply this to white dwarfs, we are invited to look at the novae ejectas, which are expelled during novae or recurrent novae. We follow here a review by Prialnik (2001). At first we could think that it is just the accreted matter which is ejected. However, analysis of the composition of nova ejecta shows that it is not possible, since they display unusual heavy-elements abundance. And such heavy elements are not present in the accreted star. Another possibility is that the heavy elements are produced during the nova. This also is unlikely because temperature is not high enough to produce heavier elements than helium. There remain two possibilities. Either the accreted matter is somehow mixed with white dwarf material, or the accreted material is used as fuel to perform work and produces waste as heavy elements.

But how could work be performed at such extreme conditions of temperatures or magnetic fields? At least strong magnetic fields do not seem to prevent organization. In fact, they even open up new way of organizing matter. Indeed, Lange and collaborators (2012) recently identified a third mechanism for chemical bonding in strong magnetic fields. This is a remarkable and fundamental result because chemistry classically distinguishes only two kinds of strong bonds between atoms in molecules: the *covalent bond* and the *ionic bond*. This new *paramagnetic bond,* the authors remark, plays a role in the magnetized atmospheres of white dwarfs. Such a result shows that new and complex chemistry is available under strong magnetic fields.

Let us now turn to the no less fascinating neutron stars. The density of a neutron star (about $2.8 \times 10^{4}$ $g.cm^{-3}$) is equivalent to the one you would get if you would put a Boeing 747 not in a shoebox, but in a grain of sand. As with white dwarfs, the accretion flow is a critical parameter. As Shapiro and Teukolsky (1983, 450) write:

> One crucial, but complicating, feature of neutron star accretion is the presence of a strong magnetic field extending from the stellar surface outward into the stellar magnetosphere. This field can control the manner in which gas flows onto the stellar surface, the torques exerted on the spinning star, the pulse shape and spectrum of the emitted radiation, and so on.

It is possible to interpret the different manners in which gas flows from a living systems perspective. Figure Error: Reference source not found shows three such different states of a neutron star, corresponding to the *extruder*, *motor* and *ingestor* subsystems.

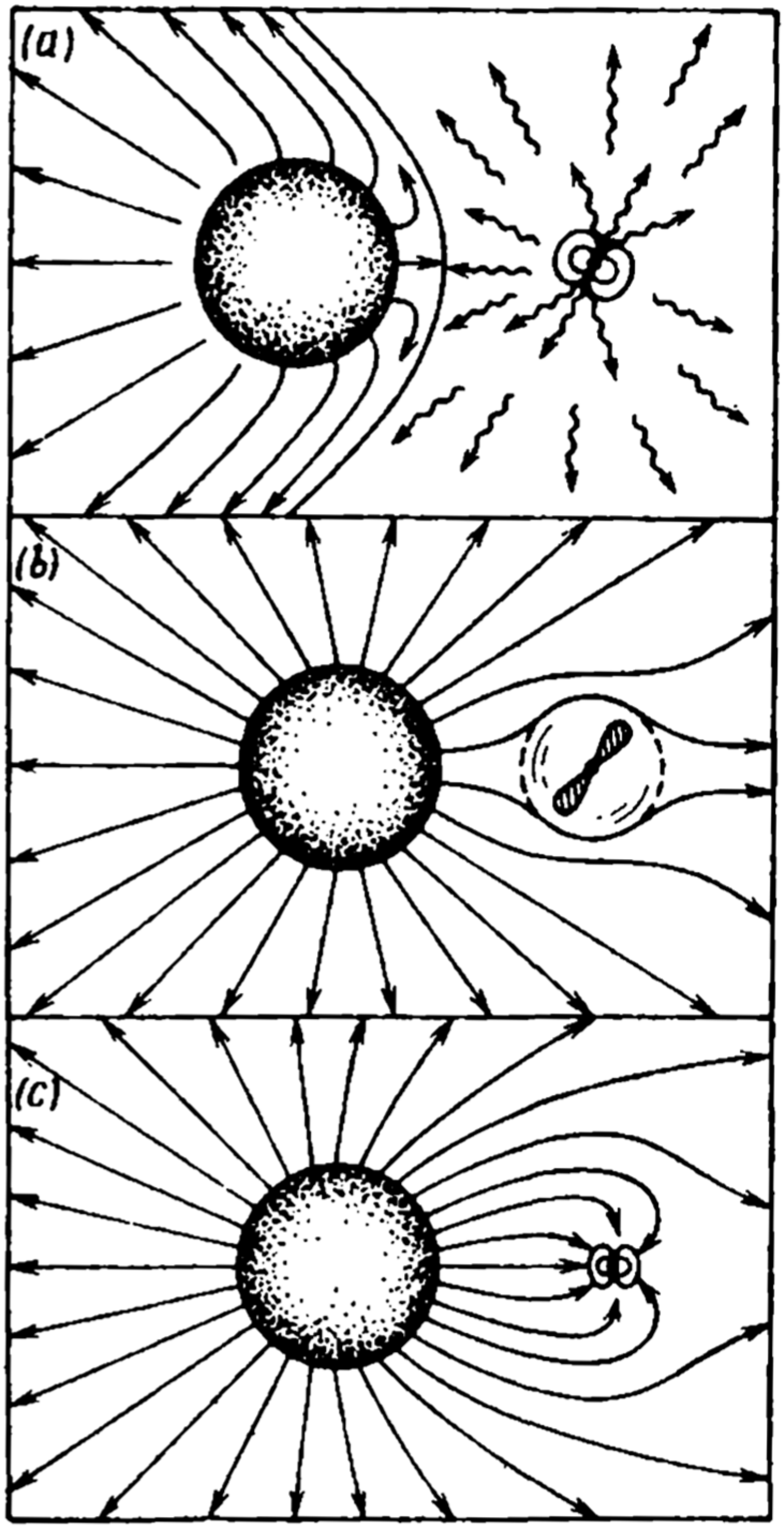

Figure 21 “Three states of a neutron star in a binary system: (a) an ejecting pulsar; (b) a “propeller”; (c) an accreting neutron star.” From (Lipunov 1989, 173). With a living system perspective, state (a) corresponds to the *extruder* function; (b) to the *motor* function; (c) to the *ingestor* function.

The behavior of neutron stars can be varied and challenging to explain. In particular, they display *quasi periodic oscillations* (QPOs). Those QPOs are traditionally attributed to instabilities in the accretion disc. However, other models propose that a special mode of nuclear burning takes place on the surface of neutron stars, which are responsible for the QPOs (see e.g. Revnivtsev et al. 2001; Heger, Cumming, and Woosley 2007). Interestingly, in some systems the bursts are not frequent enough to burn all the accreted fuel. A possible –yet of course very speculative– explanation is that this free energy is being used for the internal organization of the neutron star.

Let us now turn to black holes again. Black holes in binaries which accrete matter and emit jets are called *microquasars.* Their names comes from their behavior, which resembles the behavior of supermassive black holes found at the center of galaxies, called *quasars* and which also emit jets. Again, the accretion pattern is varying, a property challenging to explain. As Meszaros (2010, 101) puts it, the "most puzzling property found in micro-quasars and in a subset of X-ray binaries is that the accretion is variable". This can be illustrated by a sample of X-ray lightcurves of GRS 1915+105 in Figure 22.

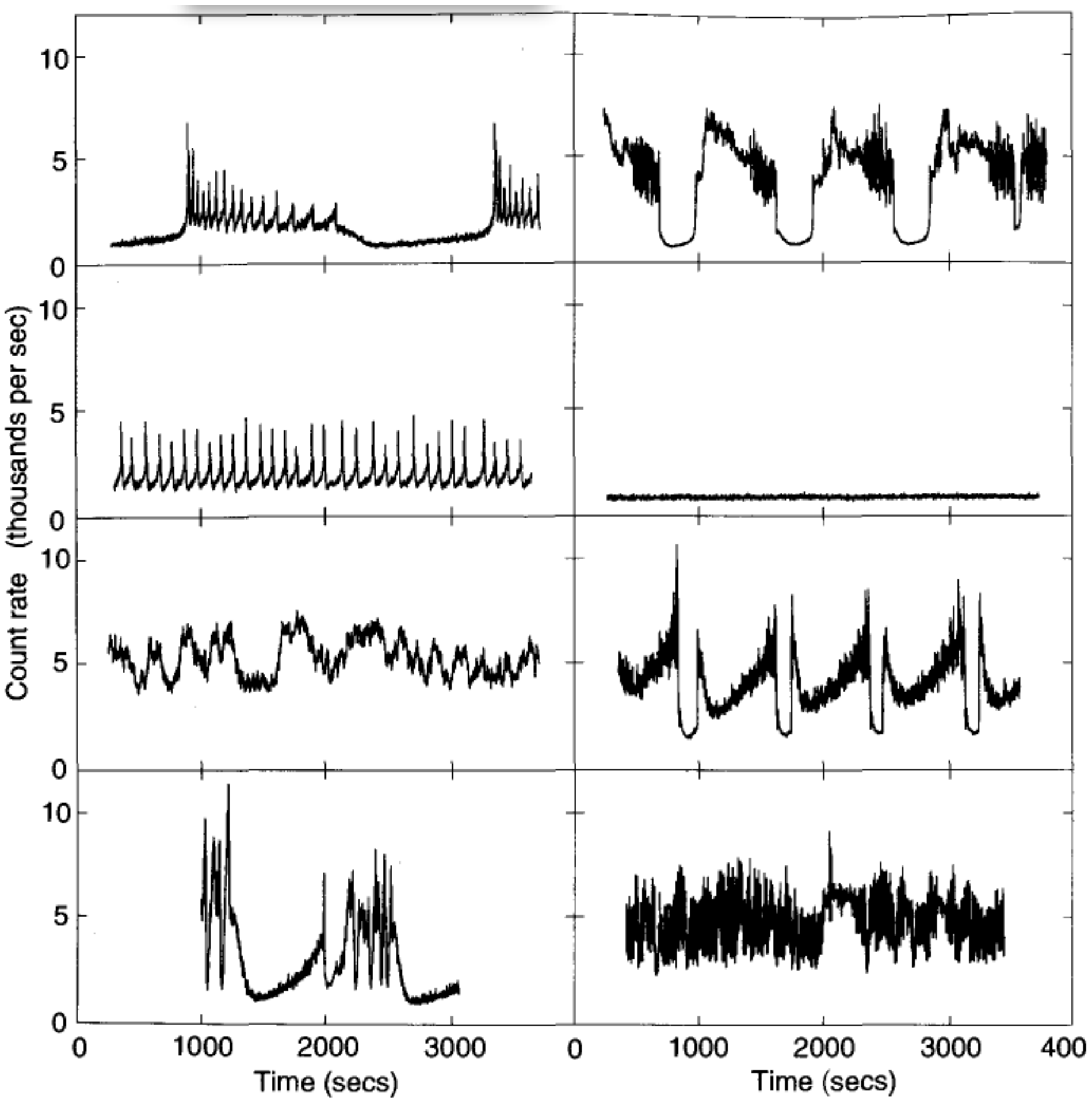

Figure 22 - The varieties of X-ray lightcurves coming from black hole binary GRS 1915+105. Picture from (Hellier 2001, 186), who writes: "A sample of the X-ray lightcurves of GRS 1915+105, obtained with the RXTE satellite in the 2-30 keV band. All the plots are on the same scale, and illustrate the astonishing range of behaviour this one star exhibits. (Based on work by Michael Muno, Edward Morgan & Ronald Remillard (1999))".

The *ingestor* function of black holes is arguably more efficient than the one of white dwarfs. Indeed, black holes accrete more carefully energy than white dwarfs. Black holes in binaries display soft X-ray, periodically. This is why such sources are called Soft X-ray Transient (SXT). As Hellier (2001, 184) writes:

> In dwarf novae the outburst ends when a cooling wave moves through the disc in 1-2 days (the 'thermal timescale' on which disc material can heat or cool the next annulus). In SXTs the irradiation prevents the cooling wave from moving inwards, and so the disc is maintained in a high state. It is gradually depleted of material on a longer timescale of ~ 1 month (the 'viscous timescale' on which material flows through the disc). Thus SXT discs are nearly completely accreted by each outburst (whereas only ~ 10% of a dwarf-nova disc is accreted during an outburst), and so decades can elapse before the disc is sufficiently replenished for the next outburst (compared to the ~ monthly recurrence of dwarf novae).

Regarding its hypothesized *reproducer* function, although very speculative, it is possible to make a quantitative estimate of the time needed to reproduce a universe. There are few universal laws in biology, but the replication time ($\tau$) in function of the replicator mass (M) is one of them. It actually also applies to manmade self-replicating machines. Freitas and Merkle (2004, 175) did gather data about biological and manmade self-replicating machines in 126 species and 9 manmade self-replicating machines. They found a trend across 20 orders of magnitudes, in which replication time appears to follow a 1/4-power law function of replicator mass (see Figure 23). The trend line is more precisely:

$$\tau = 1.78 \times 10^7 M^{1/4}.$$

In their systematic study of self-replicating machines, they also considered the extreme case of the maximum possible self-replication time: the duration of the universe's existence. It exists since $1.37 \times 10^{10}$ years ($4.30 \times 10^{17}$ sec), so the power law above gives $M \sim 3.4 \times 10^{41}$ kg or about one third of the mass of the milky way.

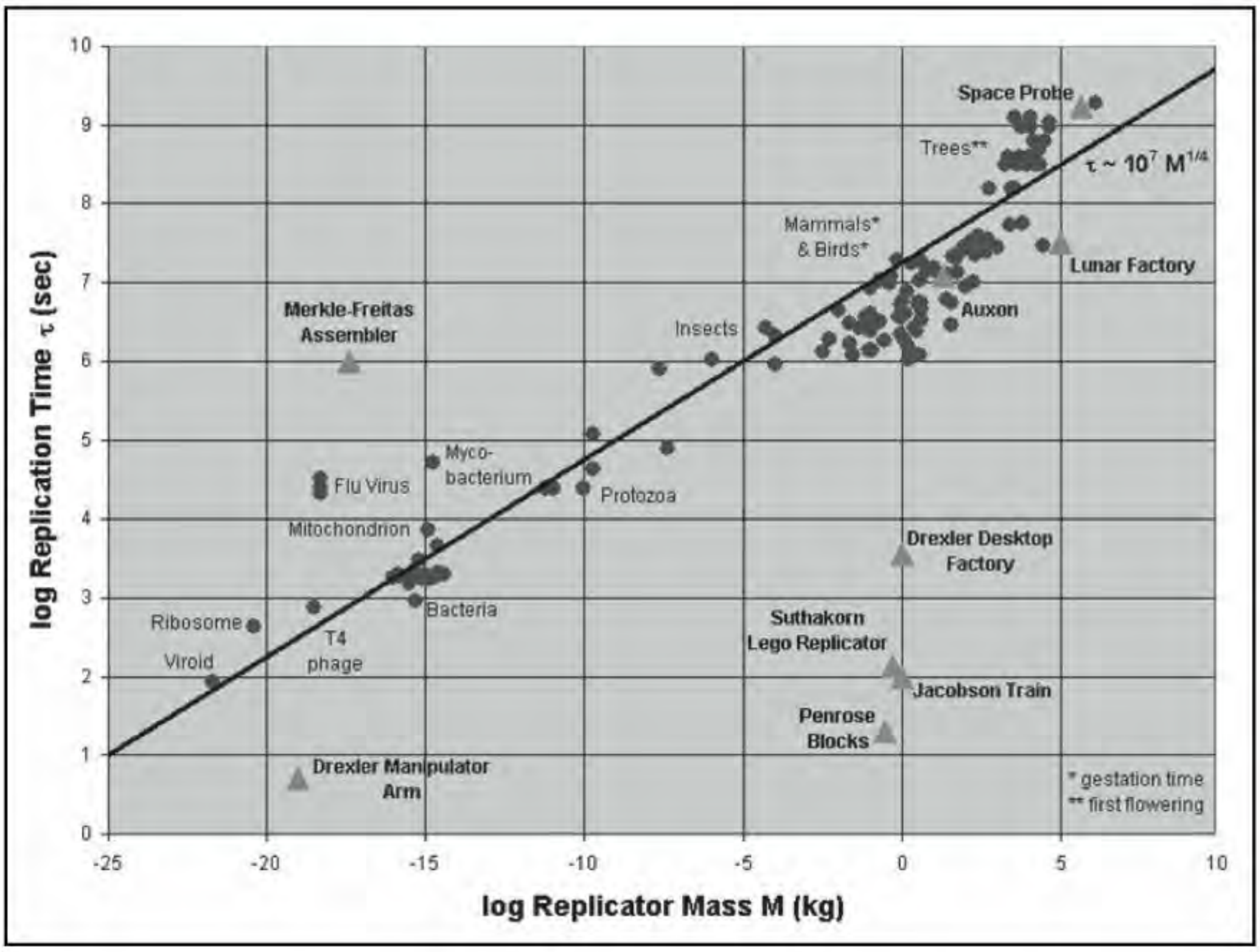

Figure 23 The 1/4-power law for replication time as a function of replicator mass, for biological replicators (circles= biological replicators, triangles = manmade replicators). From (Freitas Jr and Merkle 2004, 175).

But we can also apply the law the other way around and ask the following question. Assuming black holes indeed fulfill a *reproducer* function, how long would it take to make a universe? It depends on the mass of the black hole, stellar or galactic. Let us consider these two cases. If we take a stellar mass black hole of 4 solar masses (7.95 x $10^{33}$ kg) it takes $\tau$ = 30 million years. A black hole of 10 solar masses takes just a bit more time since $\tau \sim$ 37,7 million years. This might seem a lot, but it is not in comparison to the 13.7 billion years of cosmic evolution.

Now let us consider a supermassive black hole, such as the one at the center of our galaxy, Sagittarius A*. Its mass is more than 6 million solar masses (8.57 x $10^{36}$ kg) (Gillessen et al. 2009). The time for a hypothetical replication is then much longer, about 966 million years ($\tau \sim 9{,}66 \times 10^8$ years).

Why is it interesting to speculate about supermassive black holes? The Chandra X-Ray spatial telescope provided data showing that Low-Mass XRB (LMXB) are overabundant within 1 parsec of the galactic center (M. P. Muno et al. 2005). Could these be civilizations migrating toward the supermassive black hole?

It seems worth exploring. Indeed, the galactic center displays an asymmetry when observed at high energies, in the gamma-ray line of 511-keV. This energy level is the signature of electron-positron annihilation, which is, let us remember, a way to have 100% efficiency in matter to energy conversion. A very promising explanation to explain this asymmetry is the similar asymmetry exhibited by X-ray binaries with strong emission, so called hard "low mass x-ray binaries" (LMXBs), as displayed in Figure 24 (Weidenspointner et al. 2008). Importantly, the authors notice that "High-mass X-ray binaries do not, by contrast, show any significant imbalance", which is consistent with the starivore hypothesis, since contrary to low mass x-ray binaries (LMXBs) most high mass x-ray binaries (HMXRBs) do not display accretion control, and therefore are not good ETI candidates.

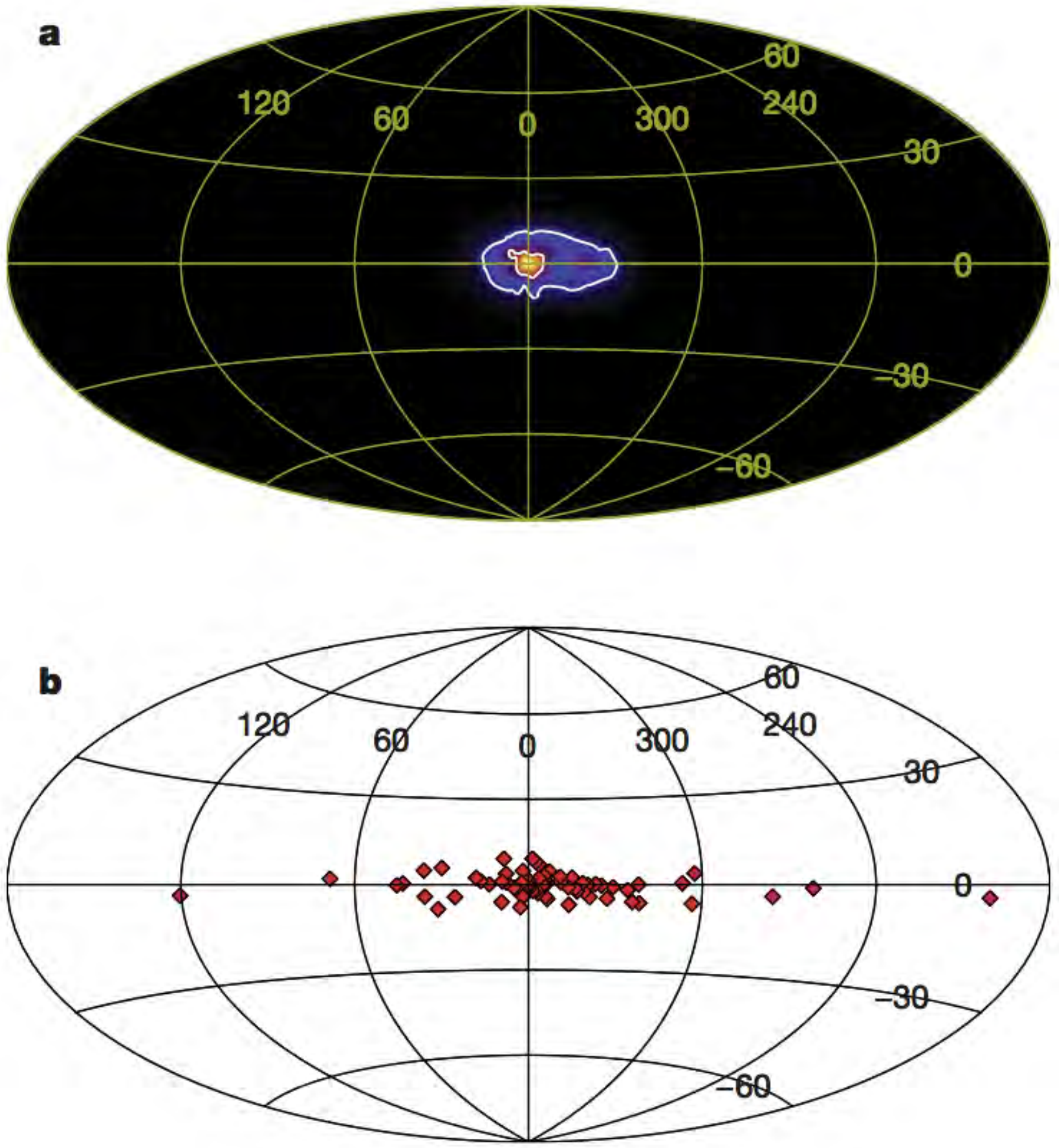

Figure 24 A sky map in the 511-keV electron–positron annihilation line (a), and the sky distribution of hard LMXBs (b). From (Weidenspointner et al. 2008)

Could it be that supermassive black holes are a huge energetic source needed to secure universe making? This brings us back to the highly speculative idea of *Cosmological Sexual Selection* (CSS) mentioned in Chapter 8. We might imagine competition in Cosmological Artificial Selection, in the sense that starivores compete for energetic resources to produce universes. The analog of sexual selection would be if starivores would engage in something like *male combat*, to pass on their (universal) genome through the "female" supermassive black hole. The properties of the supermassive black hole could be interpreted as an entity of the opposite sex constraining universe reproduction. If you find this too misogynist, you might prefer the *female choice* of sexual selection, where the supermassive black hole is the feminine matrix which chooses the male who will have the chance to reproduce his universal genome. But this interpretation leads to a difficulty, namely that it further requires that supermassive black holes have some mechanisms to select.

Sexual selection is a strong mechanism to improve biological offspring and analogously, CSS would also improve cosmological offspring. John M. Smart (2009) was the first to speculate about this rivalry in using supermassive black holes. A following prediction is that reproductively mature civilizations migrate towards the supermassive black hole. So, we could expect LMXBs to migrate towards the galactic center in ways that are different from a “natural” scenario, like the migration by dynamical friction (see e.g. Miralda−Escude and Gould 2000). Note that a competition is not strictly necessary, and we could imagine as well ETIs collaborations, for example in merging or comparing their universe simulations before replication. Again, even if these LMXBs are worth keeping an eye on, for now CSS is fitter for science-fiction than science.

It is possible to critique this living systems analysis because the different subsystems we mentioned mostly concern different binary systems. And displaying only *a few* critical living subsystems does not confirm at all that the system as a whole is living. This is a valid objection. However, we have a problem of observational time-scale. Hypothetical starivores are probably very long-lived compared to a human life. So taking apart one system and trying to guess its evolution is nearly impossible, for high energy astrophysicists as well as for high energy astrobiologists. But observing different systems can give us some clue to reconstruct evolutionary trajectories.

Imagine you are a butterfly and your life lasts only one day. Your mission is to study the human species. If you decide to follow one person during this day, you will learn some details about his life. But if in this one day, you fly over a city, observe hundreds of people of all ages, from birth to death, you will get a clearer picture of the human species. Humans are like butterflies compared to putative starivores. If we want to understand them, our best strategy is to observe many of them and try to construct a global picture.

Now even if we would find a binary system displaying all the 10 matter-energy subsystems, would it constitute a proof that it is a living ETI? It would still be debatable. This brings us to an other objection, namely that we did not take into account the 9 remaining *informational* critical subsystems. The reason is that they are all unknown, and maybe unknowable without actually establishing contact. We could speculate that the *input transducer* would be implemented via gravitational lensing with the dense object, but this is nearly impossible to check.

Now, what about the *output transducer*, which is the focus of orthodox SETI? Could we search for information transmission from hypothetical starivores? Certainly, and this brings us to *pulsars*, and X-Ray pulsars in particular. Let us have a closer look at them.

### 9.4.6 Are Pulsars Artificial Output Transducers?

All our arguments so far do converge towards pulsars as the best ETI candidates. Let us see why. First, many pulsars are found in binary systems, so all our arguments so far apply to them. Amongst binary pulsars are the most puzzling ones, *millisecond pulsars* and *X-Ray pulsars* (see e.g. Ghosh 2007). They are thought to be very dense neutron stars, so they comply with the Barrow scale civilizational developmental trend towards small scales and high density. Note that many pulsars are not in binaries, so the artificiality of *isolated* pulsars is less in line with our argumentation.

Second, we proposed living systems arguments regarding matter-energy processes in binaries. However even if we grant the arguments, they may still be insufficient to prove the existence of ETI. Information processing is key to all life and complexity, and if we can't find any proof of it, it will remain debatable if we have found ETI. In that sense, we are back to orthodox SETI, which focuses on information. More specifically, we conjecture that some pulsars may fulfill one critical living subsystem function, the *output transducer* which transmits information in the environment.

Since their discovery, pulsars have often flirted with the possibility of being artificial (see McNamara 2008 for a popular science book about the history of pulsars). It is well known that Jocelyn Bell Burnell and Antony Hewish observed the first pulsar on November 28, 1967. They were astonished to observe such regular pulses and suspected it might be of artificial origin. They were so puzzled that they nicknamed the source LGM for “Little Green Man” (Burnell 2004).

Despite the strangeness of the phenomenon, astrophysicists' job is to produce natural models, not artificial ones. So they quickly settled on a natural model called the *Lighthouse model* (Gold 1968). The model shows that the pulsating source is a neutron star. It further proposes that pulses come from the two poles of the rotating neutron star, which beam magnetodipole radiation. These beams sweep across the sky generating one observed pulse per rotation period.

However, the pulsar emission mechanism remains largely unsettled. As Shapiro (1983, 289) writes:

> The actual mechanism by which pulsars convert the rotational energy of the neutron star into the observed *pulses* is poorly understood. Many theoretical models have been proposed, but no single one is compelling [footnote: See Manchester and Taylor (1977), Ruderman (1980), Sieber and Wielebinski (1981), or Michel (1982) for a review and critique of some of them.] This is so despite the seemingly universal characteristics of the radio emission from different pulsars; a single basic model probably applies to all pulsars.
>
> On the other hand, the energy observed in pulses is only a small fraction [footnote: The fraction is $\leq 10^{-9}$ for the Crab, and $\leq 10^{-2}$ for some old pulsars] of the total rotational energy dissipated, so that ignorance of the actual pulsed emission process *may* be decoupled from the gross energetics of radiating neutron stars.

Accordingly, Shapiro wrote this in 1983, but recent accounts still leave many open problems in the study of pulsars. For example, Beskin (2010, 89) writes:

> The general view of the radio pulsar activity seems to have been established over many years. On the other hand, some fundamental problems are still to be solved. It is, first of all, the problem of the physical nature of the coherent radio emission of pulsars. In particular, as in the 1970s, there is no common view of the problem of the coherent radio emission mechanism of a maser or an antenna type. Moreover, there is no common view of the pulsar magnetosphere structure. The point is that the initial hypothesis for the magnetodipole energy loss mechanism is, undoubtedly, unrealistic. Therefore, the problem of the slowing-down mechanism can be solved only if the magnetosphere structure of neutron stars is established. However, a consistent theory of radio pulsar magnetospheres has not yet been developed. Thus, the structure of longitudinal currents circulating in the magnetosphere has not been specified and, hence, the problems of neutron star braking, particle acceleration, and energy transport beyond the light cylinder have not been solved either. The theory of the inner structure of neutron stars is also far from completion.

Leaders in SETI have also repeatedly attracted attention to the strangeness of pulsars and the worth to study them with an astrobiological stance. For example Carl Sagan (in Dyson et al. 1973, 228) wrote:

> The pulsar story clearly shows that phenomena which at first closely resemble expected manifestation of ETI may nevertheless turn out to be natural objects – although of a very bizarre sort. But even here there are interesting unexamined possibilities. Has anyone examined systematically the sequencing of pulsar amplitude and polarization nulls? One would need only a very small movable shield above a pulsar surface to modulate emission to Earth. This seems much easier than generating an entire pulsar for communications. For signaling at night it is easier to wave a blanket in front of an existing fire than to start and douse a set of fires in a pattern which communicates a desired message.

To my knowledge nobody has taken Sagan's invitation of a systematic examination... yet. Regarding the issue of how to power the broadcast of interstellar messages, Paul Davies (2010, 105) made a similar remark when he wrote:

> A more technologically savvy civilization might try using the pulsar emission itself to convey the message, by modulating the natural pulses in some way. That would neatly solve the power problem – pulsars are so powerful they can be detected across the entire galaxy with a modest radio telescope. The signal would then show up as a pattern in the frequency, intensity or polarization of the radio pulses.

It is outside our scope to treat the difficult question of what would constitute an artificial informational signal. We discussed principally *matter-energy criteria*, but *informational criteria* should also be further discussed and developed. However, in orthodox SETI something more than a regular pulse is needed to suspect an ETI presence. Let us mention two features. First, in their landmark article, Cocconi and Morrison (1959) argued that the radio emission line around 1.420 Ghz is a unique objective standard of frequency. Indeed, in this part of the radio spectrum there is little noise from natural celestial sources, so an artificial transmission at this frequency would easily be distinguished. Second, in their textbook *Life in the Universe*, Bennett and Shostak (2011, 412) write:

> Most natural radio emissions have fairly broad bandwidths, so a signal confined to one narrow spot on the radio dial would immediately offer a hint that it might be artificial (made by intelligent beings). If the signal was also flashing on and off or switching between two nearby frequencies, we would suspect that it was a coded message. We would undoubtedly record the pattern and try to analyze what was being "said."

It is remarkable that some pulsars fit very well those last two features. For example, the highly magnetized object (XTE J1810−197 or PSR J1809−1943, also called a *magnetar*) emits at 1.4 Ghz. Research on pulsars has also immensely progressed since 1967, and many pulsars are indeed flashing on and off for long periods (*pulse nulling*) or switch between frequencies (*mode changing*). They also display other properties such as *polarization*, *subpulse drifting*, *giant pulses* or *pulse microstructure* (see e.g. Manchester 2009 for a review). Our time resolution may still be largely insufficient to decode meaningfully a hypothetical message. Indeed, the minimum time resolution is Planck's time, which is 5.4 x $10^{-44}$ sec and modern pulsar survey reach 6.4 x $10^{-5}$ sec (Keith et al. 2010). Of course it is not an excuse to postpone the analysis of data we currently have. We can hypothesize that as our time resolution will increase, we will continue to discover in pulses more and more complexity, microstructures and subpulses. Interestingly, it is by developing more sensitive detectors, and thus technology on a smaller scale that we will be able to approach the Planck time resolution. So, the full breadth of a hypothetical message in pulsars may be accessible only to a civilization very developed on the Barrow scale.

Note that these pulsar behaviors are hard to explain within the Lighthouse model. For example, how can it explain pulse nulling? Does it mean that the neutron star suddenly stops rotating, and then starts again? Or "just" that its pole emissions stop from time to time. If so, why? how?

LaViolette (2006) did speculate about pulsars as message broadcasters. The book has some value in its critique of standard models of pulsars and indeed raises suspicion about their artificiality. LaViolette speculates that pulsars beam preferentially to the Earth to indicate the coming of (or a previous) galactic superwave resulting from the activity of the supermassive black hole at the center of the galaxy. However, in addition to anthropomorphism implied by preferential beaming towards the Earth, the model lacks scientific testability. In spite of a claimed message about a coming galactic superwave catastrophe, no prediction is made to inform us when this will happen. No possible refutation is proposed either, and the pulses themselves are also not analyzed for putative messages. Furthermore, it is a pity that analyses in the book may be quickly discredited by other extraordinary claims such as an incredible measurement of 64 times faster than light speed, with as evidence to support the claim... an ordinary personal communication (LaViolette 2006, 48)!

Anyhow, natural or artificial, pulsars are arguably very helpful for galactic navigation. Indeed, the Pioneer plaques placed on board of the 1972 Pioneer 10 and 1973 Pioneer 11 spacecrafts were designed to be understandable if ever intercepted by extraterrestrial life, which is why Carl and Linda Sagan with Drake (1972) used pulsars and the center of the galaxy to localize the position of the Sun (see Figure 25).

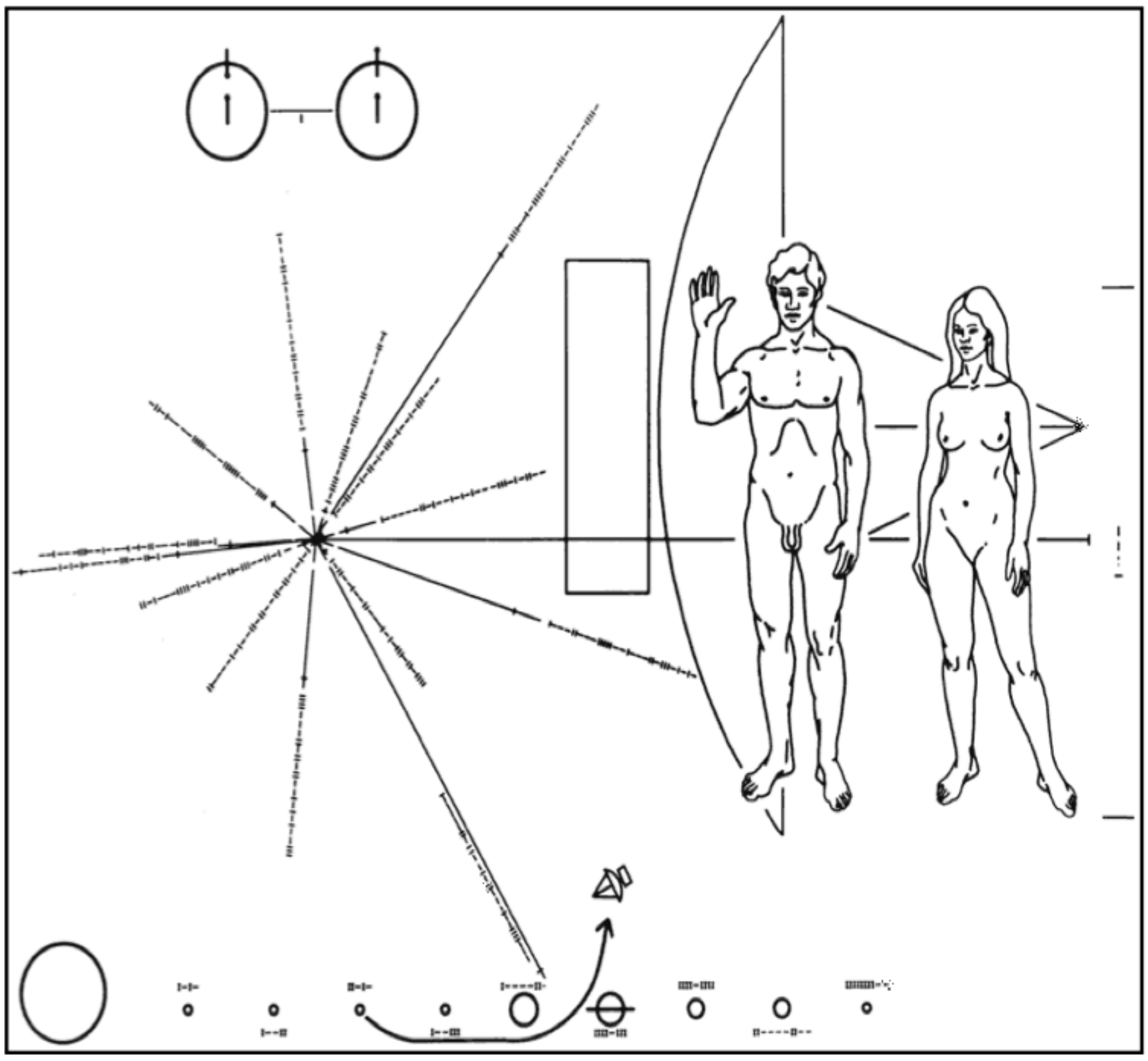

Figure 25 - The Pioneer 10 plate. On the left, the position of the Sun is shown relative to 14 pulsars and the center of the galaxy.

But pulsars can still be much more useful. Emadzadeh and Speyer (2011) recently showed that absolute and relative navigation is possible thanks to X-ray pulsars. It is indeed not at all a trivial matter when you navigate at a significant portion of the speed of light to keep track of time, because of the time-dilation and contraction effects predicted by relativity theory. Why X-ray pulsars specifically? The authors elaborate (Emadzadeh and Speyer 2011, 10):

> The main advantage of spacecraft navigation using X-ray sources is that small sized detectors can be employed (P. S. Ray, Wood, and Phlips 2006). This provides savings in power and mass for spacecraft operations. Another advantage of using X-ray sources is that they are widely distributed. The geometric dispersion of pulsars in the sky is important to enhance accuracy of three-dimensional position estimation since the observability of the source is an important issue. An important complication that must be addressed in utilizing an X-ray source in a navigation system is the timing glitches in its rotation rates. Of X-ray pulsars, ones that are bright and have extremely stable and predictable rotation rates are suitable candidates for the purpose of navigation. These sources are usually older pulsars that have rotation periods on the order of several milliseconds.

It is worth noting that *small* x-ray detectors are an important feature for civilizations climbing the Barrow scale. They would not want to send big spacecrafts and carry huge detectors but rather send small scale spacecrafts with small detectors.

In recent years, pulsars have attracted attention in SETI. For example, in an unpublished preprint, Gregory, James and Dominic Benford (2008) suggested that the strange radio emission from GCRT J17445-3009 could be an artificial beacon (see also LaViolette 2006, 91–92). Edmondson and Stevens (2003) also proposed a research program to find habitable stars based on pulsar alignments.

To conclude, we are not sure how to file pulsars, as natural or artificial. The file of X-ray pulsars are better classified as X-files. We need to figure out with an astrobiological stance if the pulses contain information and structure. After all, could it be that Jocelyn Bell's feminine intuition that pulsars are artificial was correct?

I used to think that SETI was for failed science fiction authors. But watching Carl Sagan in the TV series Cosmos changed my mind. I was impressed by his passionate search coupled with scientific and philosophical rigor.

In 2010, Martin Dominik invited me to speak at a Royal Society meeting entitled "Towards a scientific and societal agenda on extra-terrestrial life", held 6-7 October 2010[14]. The meeting was quite exceptional, with a rare intelligence density combined with a marvelous surrounding. Although I believed, like most scientists, that it is unlikely that we are alone in the universe, I had not yet given a serious thought on *how* to find extraterrestrial life. The organizers invited me to take part on a discussion panel regarding the question: "What are the implications of SETI for the future of humanity?". I thought that to answer this question we not only needed to actually find extraterrestrials, but also our senior in the grand scheme of cosmic evolution. Indeed, it is only in this case that we will have insights on our possible futures. The extraterrestrial perspective also encourages the development of a cosmological ethics (see Chapter 10).

The scenario of CAS and discussions with John Smart lead me to seriously consider the importance of black holes for the future of intelligence. But, as John noticed, if ETIs live off black holes, we would not stand a chance to detect them because we can't see any leakage from them. Still, in a scientific spirit I later went to look more closely at where the few confirmed stellar black holes were observed. I then discovered with fascination microquasars and gradually the variety of accreting binary systems. They immediately struck me as displaying all features of non-equilibrium systems. Non-equilibrium systems, whether simple dissipative structures, living systems or societies, share common properties. The energy flow, a maintenance of an internal organization and an exportation of entropy. All these features appeared to be present in some binary systems. Of course, proving the existence of an internal organization in a white dwarf, neutron star or black hole remains a core open question.

14 http://royalsociety.org/events/2010/extra-terrestrial-life-satellite/

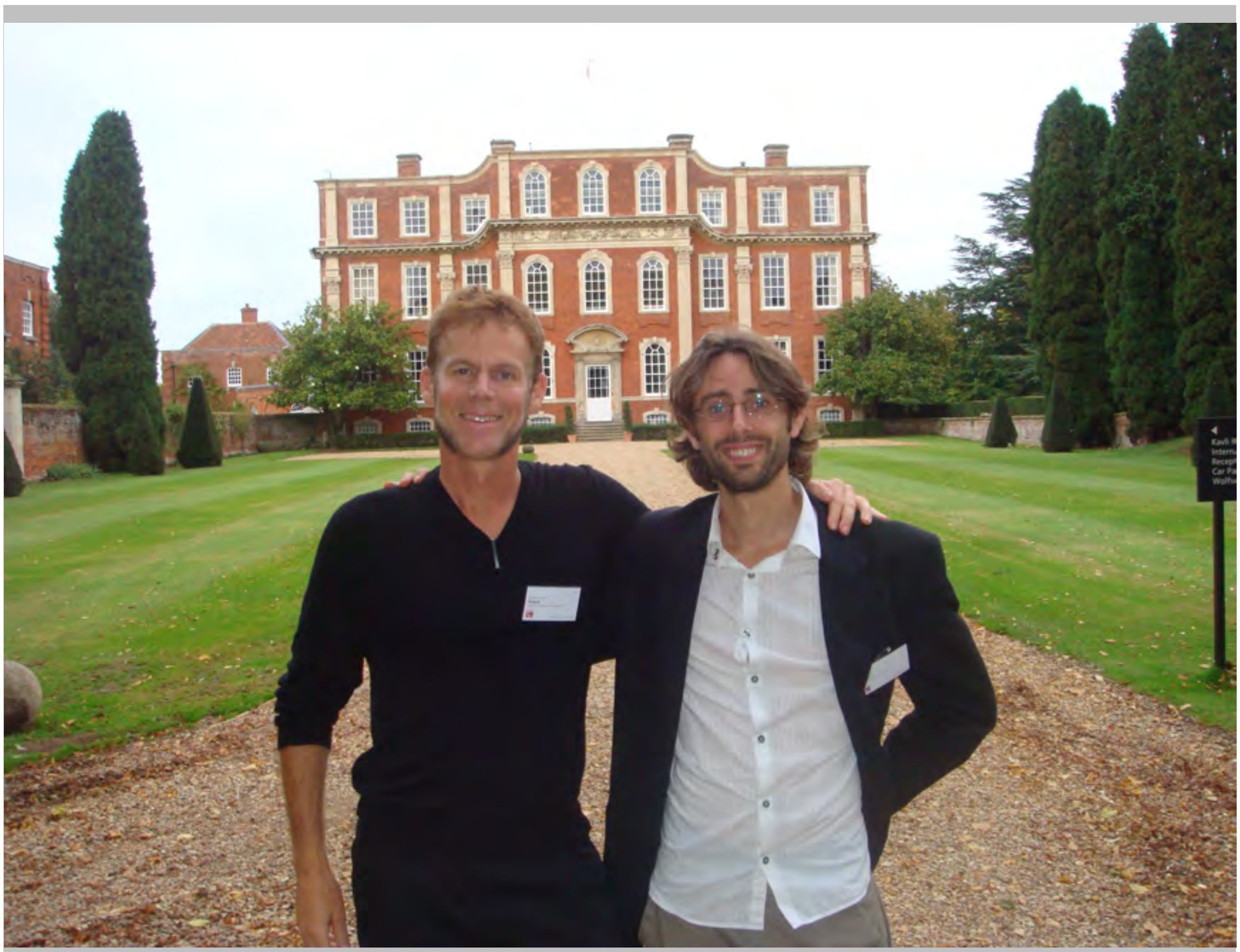

Figure 26 John Smart and Clément Vidal at the Royal Society at Chicheley Hall, home of the Kavli Royal Society International Centre, Buckinghamshire, October 2010.

Nevertheless, as I took my own ideas more and more seriously, I suffered from a "cosmic depression". Indeed, I faced the possibility that we, life on Earth, may be late, or even too late to make any meaningful contribution to the universe. What if humanity was born too late? What if life which has a future in the cosmos must be born around the habitability zone of a binary system (planets in binaries are possible, see e.g. Turnbull and Tarter 2003; Haghighipour 2010)? What if promising life should start in the primordial soup of a neutron star in a binary? Even worse, maybe that life mostly starts in binary systems, which allows high levels of development, thanks to the possibility of harnessing binary accretion. What if life on Earth would be the exception rather than the rule? I mean, not the exception in its wonderfulness, but in its deformity? What if life on Earth was a galactic defect? Life on Earth would be like a cosmic trisomy, and successful life would strive in binary systems. I hope this questioning was a temporary crisis towards a new stage of psychological development, to hopefully take a new cosmic perspective, as analyzed by Kohlberg and Power (1981, 234).

In a more optimistic view, could being late be an advantage? Could we be “spoiled children of the cosmos”, just having to follow the path of our elderly cosmic cousins? Is it really terrific if a child realizes that adults are smarter and stronger than him, that they know more and do strange things which disgusts him –for now? No, we may act as children eager to learn from our cosmic cousins to see what is possible and desirable to do in the long-term future.

All these are wild speculations, but plausible. All in all, I am now thinking that the best thing to do now is to try to know rather than to die stupid. That is why we should keep searching. The famous X-Files series and the 2008 subsequent movie have a subtitle: “I want to believe”. This expression refers to the difficulty to conciliate faith with science. But I and astrobiologists don't want to believe anything. By contrast, the scientific astrobiological mindset is best summarized with “I want to know”. With this cosmic depression, I learned one thing; the question *are we alone?* has its psychological shadow: *are we ready?*

Story 10: A cosmic depression

## 9.5 Objections

> *Hypotheses about extraterrestrial intelligence are hypotheses, not facts. A fine line separates the rational process of extrapolating our knowledge of life on Earth to life elsewhere and the irrational process of projecting fantasies, wishes, or fears onto unknown entities whose very existence is in doubt. We try to do the former without lapsing into the latter, but given our present state of knowledge, the question is not whether but how often we slip across the line.*

(A. A. Harrison 1997, 313)

You certainly have some or many objections against the arguments above. As Harrison analyzes, it is likely that some, many or even all extrapolations I propose are wrong. But at least the existence of binaries is a fact, so the starivore hypothesis is scientifically testable. I will now formulate and anticipate objections, and suggest replies. If you have more objections or suggestions, please let me know!

### 9.5.1 Is it New?

No, it's not! Actually, forming a binary system is an obvious way to always have energy at hand. In his influential science fiction novel *The Star Maker*, Olaf Stapledon (1953, 128) saw the formation of a binary as a solution for long interstellar travel. Let us quote him at large:

> Actual interstellar voyaging was first effected by detaching a planet from its natural orbit by a series of well-timed and well-placed rocket impulsions, and thus projecting it into outer space at a speed far greater than the normal planetary and stellar speeds.
>
> Something more than this was necessary, since life on a sunless planet would have been impossible. For short interstellar voyages the difficulty was sometimes overcome by the generation of sub-atomic energy from the planet's own substance; but for longer voyages, lasting for many thousands of years, the only method was to form a small artificial sun, and project it into space as a blazing satellite of the living world.
>
> For this purpose an uninhabited planet would be brought into proximity with the home planet to form a binary system. A mechanism would then be contrived for the controlled disintegration of the atoms of the lifeless planet, to provide a constant source of light and heat. The two bodies, revolving round one another, would be launched among the stars.

If the starivore hypothesis is correct, we could say on the one hand that Stapledon had anticipated the form extraterrestrials would take, but on the other hand that he did not push the idea far enough in asserting that *known* binaries may be extraterrestrials. It would be an interesting case where science fiction does not go far enough, where reality is stranger than fiction. The plausibility of using a star as an engine for interstellar travel has also been researched scientifically (see e.g. the pioneering work of Shkadov 1987). Interestingly, as we mentioned earlier, we do actually observe binaries which move.

Martin Beech (2008, 157) also hinted at the possibility that a Low Mass X-Ray Binary (LMXRB) could be formed in the context of Sun rejuvenation engineering, to avoid its red giant phase.

### 9.5.2 Dyson Spheres versus Starivores

In a stimulating exchange about putative starivores with astrophysicist and science fiction author David Brin, he objected that accretion is not using the whole radiation of a star. Therefore, starivores would not be as advanced as a civilization using a Dyson sphere. I'll keep using "Dyson sphere", although Dyson had in mind more a "swarm" than a sphere (see Dyson 1996, 27). Starivores would be wasteful because they would still allow an enormous amount of energy to radiate through space.

My first reply is that a Dyson sphere is a passive technology. It is important to distinguish *passive* and *active* energy extraction. For example, in order to grow, plants passively receive the energy of the Sun. They are passive and subject to environmental vagaries. A fundamental innovation of the animal kingdom is that animals are able to seek their own energy source. They actively search, hunt or compete for food. This makes them much more adaptive. Although it is often overlooked, evolutionary progress strongly depends on innovations in energy extraction mechanisms; see for example the (2005) book by Peter A. Corning and the energy innovations which occurred through cosmic evolution, in table 6, p133.

My point here is that a Dyson sphere is like a plant or a solar panel. There is no new energy extraction mechanism involved. Whereas starivores hypothetically use an extremely efficient way to extract energy from a star: accretion. This is a radically new way to use energy, indeed comparable to the difference between plants and animals.

Now, does a Dyson sphere really harnesses more energy than accretion? Although it remains to be checked with calculations, the amount of energy extracted through actively accreting mass-energy from a star could represent even more energy than what is radiated from the whole star. Furthermore, starivores might be able to harvest not only the outer layer of a star, but its inner fusion core, which contains about 70% of a star's mass-energy. If such a thing is possible, it might be that accretion brings much more energy than a Dyson sphere. Yet it is an open question whether accretion can happen on deeper layers of a star. Since these quantitative questions remain open, for the sake of the argument, let us grant that a Dyson sphere still harnesses more energy than an accretion-powered starivore.

The race to more energy is not the whole story. In general it is most parsimonious to harvest just the energy you need for your goals. Again, to take a human example, if you eat too much or too little you die. Another example. I guess galactic-wide civilizations are unlikely because the finiteness of the speed of light would make such a civilization communicate, improve, evolve, cooperate and progress at a very slow pace. The same applies to Dyson sphere versus accretion. The objection of a starivore using less energy than a Dyson sphere takes into account only the Kardashev scale, not the Barrow scale. A Dyson sphere of 1 AU radius is totally anti-Barrow scale because it is a huge structure. The energy needs to be stored and transferred on very large distances before performing useful work. A starivore civilization would accrete energy very near its boundary, ready to be used, processed or stored.

Additionally, starivores may be able to control the accretion rate, thereby controlling the energy use depending on what their needs. The accretion disk or the rotation of the dense body (WD, NS, BH) could be used as a buffer to store additional energy. No additional resources are needed in transferring energy through a megastructure, the energy is directly available. Although not inconceivable, it is not clear whether or how a Dyson sphere could control the energy intake efficiently.

Furthermore, there is the issue of entropy production and the *extruder* function. Dyson indeed hypothesized that infrared radiation would leak out of a Dyson sphere and that this would be an observable feature. But this is not an efficient way to get rid of waste products, at least not as efficient as the putative waste products of novae or jets regularly being expelled by some binaries, sometimes at velocities near the speed of light.

Finally, there is a closing objection. Dyson spheres are purely hypothetical constructs, *with no observational counterpart found*, despite some searches (see e.g. Jugaku, Noguchi, and Nishimura 1995; Carrigan 2009). By contrast, astrophysicists and amateur astronomers observe binaries daily.

### 9.5.3 Are all Binaries Extraterrestrial Life?

Certainly not. The binary configuration obviously occurs naturally. Only a small subset of binaries are 'candidates'. You may want to take a second partial visit to the binary zoo (section 9.4.2 A Partial Visit to the Binary Zoo, p233).

### 9.5.4 “Postbiology?”, Are You Serious?

Many people find it hard to believe that completely postbiological entities could really exist. In other words, they think life must be based on carbon and water. Functional and systems theoretic perspectives on the world make us see the material support as one parameter like an other in a system. Most of us have no idea of the material support and organization underlying computers we use. We only want that they perform the functions we want (see also section 9.1.3 The Case for Postbiology, p215 above). What matters is the organization, structure and function of a system. Matter doesn't matter as much.

### 9.5.5 Is High Temperature Livable?

Would not anything burn at stellar temperatures? Can we seriously hold that starivores would operate at hundreds of millions of degrees?

We tend to think that life has strict boundaries, and that high temperature or high magnetic field would destroy any kind of living organization. But these are only prejudices. The discovery of extremophiles living with no light or under high temperatures greatly surprised us. We tend to dismiss quickly as impossible or dangerous an area we do not know. The first map drawers drew dragons behind what they had explored. We have not explored the possibility of life or complex organization at very high temperatures or under very strong magnetic fields. Yet regarding the latter we saw that under strong magnetic fields, a new stable chemical bond emerges beyond the only two chemical bonds known (Lange et al. 2012).

For the sake of the argument, let us grant that life under high temperature is indeed impossible. We could still see the binary as a power station, and suppose that intelligent beings would stay at cold on a planet orbiting the accreting binary. However, I think it is unlikely, because of additional complications and costs of energy transfer associated with such a configuration.

### 9.5.6 Will We Become Starivores?

Probably. One plausible speculative scenario is the following. We continue to climb the Kardashev scale and use more and more energy. In parallel, our technologies function on smaller and smaller scales, and we climb the Barrow scale. Those small and dense technologies demand more and more energy. Once we cover the Earth with solar panels –having understood that all other energy sources are not sustainable on the long term– we still need more energy. We then follow Stapledon's scenario of carefully changing Earth's orbit to bring it closer to the Sun. Then our solar panels receive much more energy. At this stage, our descendants have evolved to a postbiological substrate and are able to live under high temperatures. Still, the passive energy received (e.g. by stellar wind) is not enough to meet our ever growing energetic need driven by computational or technological progress. Stellar engineers set up the first active accretion from the Sun. We have become starivores.

The Sun's energy is used and transformed for building ever smaller and denser organizations. This goes through hundreds, thousands or millions of years. Eventually, the Sun runs out of energy. Only then it makes sense to migrate to the nearest –preferably single– star, and to continue. The density of the evolved Earth is now comparable to the one of a white dwarf –which indeed are Earth-sized– , and the new binary we form resembles a cataclysmic variable.

### 9.5.7 Starivores' Missing Mass

There is a lot of mass involved in a binary. Where does this mass come from? The scenario above provides a possible answer. Remember that the maximum age of extraterrestrials is 2 billions or more years older than us. They might thus have had plenty of time for complete depletion of stellar energy of one or several stars (if we include migration). Accordingly, more precise estimates need to be calculated, taking into account accretion rates to assess if this scenario is plausible.

However, our closest relative would be very low mass binaries. So we could focus on searching the less advanced starivores. Not surprisingly, thanks to the variety of binaries we can indeed find very low mass ones, such as LB 3459 (AA Dor) or DN WZ Sge. Warner (1995, 452) writes about the latter:

> Another possible origin of CV [cataclysmic variables] precursors is through engulfment of a planet or very low mass star by expansion of a red giant. It is possible to find circumstances in which the planet will accrete sufficient gas to produce a low mass dwarf in a short period orbit, perhaps resembling the DN WZ Sge (Livio 1982; Soker, Livio, and Harpaz 1984).

Migration, and some rare events of a single dense body traveling to a new star should be observable. However, we should keep in mind that those events would be relatively rare.

These speculations lead to predictions that we can test with our current telescopes. We can test new scenarios for binary formation and binary dislocation. Putative starivores would migrate when the companion star runs out of fuel, and the dense body would not be ejected in a random direction, but would most likely go towards the nearest star. So we should monitor binaries whose companion has almost no energy, to see what comes next. We can also monitor single dense bodies with high velocities, to see if they get attached delicately to another star and start accreting gas.

An other reply is that starivores are instances of life which has started on a planet orbiting a binary system, and then gradually started to harness the energy of naturally occurring accretion.

#### 9.5.8 The Explosive Objection

I already introduced Eric Chaisson's universal complexity metric (section 7.2 Increase of Computing Resources, p158). Could we apply it to binaries, to see how well they score? It is a research program well worth pursuing on its own. It would be helpful to classify the putative starivore family. However, we can already ask what is the theoretical maximum energy rate density that a binary could achieve. In astrophysics, we can make such a crude estimate using the *Eddington limit* for luminosity (Frank, King, and Raine 2002, 3) :

$$\sim 1.3 \times 10^{38}(M/M_{\odot}) \text{ erg. s}^{-1}$$

We reach a theoretical maximum of free energy rate density (noted $\Phi_M$):

$$\Phi_M \sim 6.54 \text{ x } 10^4 \text{ erg.s}^{-1}.\text{g}^{-1}$$

Now, how do actual white dwarfs, neutron stars and black holes score? Surprisingly, their luminosity can break this limit! They are amongst the few systems which display *super-Eddington* luminosity. The question of how a binary can break the Eddington limit is difficult and technical.

Nevertheless, those values of energy rate densities are *extremely high for astrophysical systems*. Other astrophysical systems such as the Sun has a value ~2 and planets have ~$10^2$. Higher values are otherwise known only for complex system such as a human body ($\Phi_M \sim 2 \text{ x } 10^4$ erg.$s^{-1}$.$g^{-1}$; (Chaisson 2001, 138)).

Intrigued about these high values of $\Phi_M$, I wrote to Eric Chaisson because I very much value this energetic metric. I claimed that we may be dealing with an anomaly in cosmic evolution. Although he admitted the high $\Phi_M$ values of binaries, he was rather skeptical about the ETI interpretation. His counterargument is that such systems display unstable states, indicative of explosiveness, destructiveness and not constructive complexity. Indeed, supernovae, which are definitely destructive processes also do display high $\Phi_M$ values (>>$10^6$; (Chaisson 2001, 157)).

However, classical or recurrent novae, or jets emitted by neutron stars or black holes are not supernovae, and the binaries systems are not at all destroyed and generally not even disturbed by such events. The case of recurrent novae is particularly clear: even though the system undergoes impressive cataclysms, it does so recurrently, on a regular basis.

In sum, we should remember that what we observe are gross systemic features, and novae or jets are the products of a hypothesized *extruder* subsystem functioning. We admittedly do not observe the hypothesized constructive information processing happening at very small scales. Maybe we could compare the explosiveness we observe from binaries to the sudden release of water from a hydroelectric power station. If you are a fish watching this, you would certainly think that is a cataclysmic event. But if you are a human, you know that it contributes to power the subtle electron exchanges which make our computers process, store and communicate information on a global scale.

### 9.5.9 The Candle Objection

An other objection is that displaying a kind of metabolism is certainly not enough to say that a system is living or intelligent. For example, a candle has a kind of metabolism, ingesting wax and extruding smoke. In the process, chemical energy is converted to heat. This is a valid objection, and this is why thermodynamical criteria and considerations are insufficient. If we take into account living systems criteria, fire as found in nature has no *boundary*, so it lacks a fundamental sub-system of Miller's living systems.

What is more, fire can't control its energy flow. It has no way to seek energy if it runs out of fuel. Maybe I am unlucky, but I've never seen candles switching themselves on and off at regular intervals; neither unlighted candles trying to scratch a match. In sum, it is true that we can attribute basic metabolism to a fire or a candle, but putative starivores display also non-trivial energy flow control and living systems features.

### 9.5.10 The Fridge Objection

A fridge is a very simple system, which can be argued to have some life-qualities. As Freitas (1979, chap. 6.2.3) writes:

> The refrigerator in my house technically should be considered a “live” system in the very broadest sense, as it is a well-defined intermediate system which uses an energy flow to decrease entropy within (the icebox gets colder, and well-ordered ice crystals collect on the freezer walls) at the expense of increasing the entropy in the external environment (the kitchen air gets warmer). Yet its organizational structure is minimal.

Is a binary in accretion like a fridge, with no organizational structure inside? Or is the dense body's organization increasing? If there is no organization, binaries might be like a cosmic fridge, exporting entropy and using energy, but doing no interesting work in between. So it remains a crucial open question whether there is information processing or not. This is of course related to the informational subsystems in Miller's living system. We saw that the way to answer this issue is to study whether or not there is information in binary pulsars' signals (see section 9.4.6 Are Pulsars Artificial Output Transducers?, p248).

## 9.6 Conclusion

> *It would be a tragedy of literally cosmic proportions if we succeeded in annihilating the one truly intelligent species in the entire universe.*
>
> (Davies 2010, 206)

In a 1972 NASA symposium on “Life beyond Earth and the Mind of Man”, Philip Morrison argued that the discovery of extraterrestrials will be a slow discovery like agriculture, not like America (see Dick 1996, 506). History of science supports this view. For example, Percival Lowell elaborated a theory that canals on Mars were artificial. In 1895, the issue of natural or artificial canals was clearly formulated. However, as Dick (1996, 78) reports:

> Thrust into the open, an issue full of public interest, a golden opportunity now presented itself for science either to confirm an earthshaking theory or to crush it quickly under the glare of objective argument. To the dismay of the public and the embarrassment of astronomers, science was unable to do either for almost two (some would say seven) decades.

Let us hope that we will do better and take less time to assess the starivore hypothesis. In the meantime, from an ethical perspective it is not bad to assume that we are alone. Indeed, as Davies ponders, it invites us to take the greatest care of an incredible cosmic phenomenon: life on Earth. But of course we should also prepare ourselves that recognizing our cosmic cousins will change our science, philosophy and religions forever.

Concluding his *Critique of Practical Reason,* Kant (2009) famously wrote that two “things fill the mind with ever new and increasing admiration and awe, the oftener and the more steadily we reflect on them: the starry heavens above and the moral law within.” When I look at the starry heavens above, I see plenty of possibilities through available energy. We still need to learn how to use it, and, more importantly, know what to do with it. Can we combine Kant's two objects of reverence, and develop a starry moral law within? Let us try in our next Chapter 10, with some reflections about a *cosmological ethics*.

# Open Questions – A High Energy Astrobiology Agenda: The Starivore Hypothesis

Current research in astrobiology focuses on searching less advanced extraterrestrial life, such as bacteria or traces of a biosphere in an exoplanet. It makes sense, because these are features of life on Earth that we know and that we know how to recognize. A naive symmetrical argument would require to divide resources in two, one half for searching less advanced extraterrestrials, the other half for more advanced. But even that is not enough since we saw that putative extraterrestrials are on average two billion years older than us. So it would actually make sense to spend *much more resources* in the search for advanced extraterrestrials.

The works of Dyson (1960) and Kardashev (1964) advanced the idea that advanced extraterrestrials use much more energy than us. This is the assumption of *high energy astrobiology*. The starivore hypothesis invites us to look back at high energy astrophysics with a fresh astrobiological perspective. We already encountered many open questions to test the hypothesis. Let us now summarize and specify them to propose a high energy astrobiological agenda.

Many ideas in this chapter are *necessarily* highly speculative, for how else could we aim at searching putative ETIs arguably billion years more advanced? Yet, the hypothesis that some binary systems in accretion are ETIs is testable. We have plenty of data about binaries, and we can gather more. This is to be contrasted with other proposals such as Dyson spheres or Bracewell probes which have so far no observational counterpart.

What is the ideal profile of the high energy astrobiologist? She is not subject to artificiality- or naturality-of-the-gaps. Instead, she takes the more careful astrobiological stance. She understands high energy astrophysics models and theories. However, she spends as much time with artificial and natural models to tackle yet poorly understood high-energy phenomena in the cosmos. Furthermore, she has knowledge and interest in systems theory, living systems theory and complexity sciences. Other research interest can include biology and ecology, especially general biological laws, and the field of *energetics*. Experts in energetics would indeed be able to take a fresh look at energetic exchanges in binaries, from a more biologically-inspired perspective.

## General Agenda

Let us summarize the specific predictions regarding putative starivores:

- the finding of a black hole less than 3 solar masses. Indeed, in astrophysics, stellar black holes are the result of gravitational collapse of a massive star. If the remnant exceed the Tolman–Oppenheimer–Volkoff limit (3-4 solar masses), it will implode in a black hole. Finding a black hole less than 3 solar masses would thus mean that the black hole formation took another road, possibly an artificial one.
- low-rate planetary accretion.
- an instance of migration, where the direction of dense bodies (WD/NS/BH). navigating in the galaxy is not random, but targeted to the nearest star.

An ongoing project is to continue to develop criteria for artificiality. Our list has no exhaustive pretension, and much more work is required to answer this question. We saw that informational criteria are demanded if we want to test whether or not there is a message in pulsars. Even with an excellent criteria list, this will not be enough to convincingly prove the existence of extraterrestrials. What we need are alternative models leading to different predictions. But what motivation would we have to build alternative models other than the natural models we already have?

The strongest motivation comes from showing the limitations, contradictions, unsolved problems or shortcomings of astrophysical models of binaries. Then to propose alternative high energy astrobiological models, which not only solve those problems, but which also lead to new different predictions. Astrophysical or astrobiological models leading to the most successful predictions will progressively be favored. Although I am not expert in binary astrophysics, let me mention a few more examples of open issues in the field, and then propose more precise research proposals to test the starivore hypothesis with an astrobiological stance.

We saw that nova ejecta display heavy elements abundances. It remains a difficult scientific challenge to explain the enrichment mechanism (see Prialnik 2001).

In binaries with accretion discs, the origin of disc viscosity remains very uncertain. As Hellier (2001, 59) writes:

> Although viscosity is essential to the operation of an accretion disc, the physical origin of the viscosity has been uncertain, defying theoretical investigation for many years. We know that molecules are sticky, attempting to form chemical bonds with their neighbours (this accounts for the viscosity of everyday materials such as treacle); however, discs are so diffuse that molecular viscosity is too feeble by a factor of a billion to explain their behaviour.

Recurrent novae usually happen after a brightening due to an excess in accretion, and their luminosity fades *after* the outburst. But as in the biological world, the binary world is full of exceptions. In his impressive review of recurrent novae, Schaeffer (2010) attracted attention to a fading which happened *before* the outburst:

> Why did T CrB suffer a distinct, significant, and unique fading in the year *before* its 1946 eruption? And why would this fading behavior be different in the B and V bands? The fading is by around one magnitude below the usual level of the system, with this going to two magnitudes below the usual level in the B-band at a time 29 days before the eruption. My first thought is that the accretion turned off (for unknown reason) hence making the system lose the light from the accretion disk, but maybe the depth of the drop will require the red giant companion to be dimmed somehow. And what is the physical connection between this fading and the subsequent nova eruption? That is, how can the turning off of accretion *anticipate* or *trigger* the nova event?

Let us go back to the possibility of extragalactic SETI. If there is a developmental pattern in civilizations then we should see less and less binary systems in accretion as we look at stars in younger and younger galaxies, deeper into space. Is it the case? This is an application of the inverse distance-development principle.

For example, what about extragalactic pulsars? The starivore hypothesis predicts that they should be very rare. Note that they are detectable, because many pulsars display giant pulses (see McLaughlin and Cordes 2003). Is it the case? The search for extragalactic pulsars is still in its infancy, yet out of 1500 radio pulsars known, only about 25 are extragalactic, all located in the nearby dwarf galaxies of the Magellanic Clouds (McLaughlin and Cordes 2003, 983).

If density is an indicator of development, a prediction is that the proportion of accreting WD/NS/BH decreases as we look in more distant galaxies. Is it the case?

Are we ready to contact starivores? If we take the starivore hypothesis seriously, Communication with Extraterrestrial Intelligence (CETI) has more chances to succeed by targeting binary systems. Of course, huge ethical issues need to be discussed before sending a signal to putatively so advanced civilizations. Humanity should assess the risk benefit tradeoff. The maximum risk is they coming and destroying the Earth or the Sun. The maximum benefit is they fully collaborating with us, e.g. learning us about cosmic mysteries, and boosting our evolutionary development incredibly thanks to their technology, knowledge and wisdom.

## Research Proposals

Let us now turn to more concrete research proposals. Some of my readers may think that it is easy and fun to speculate, but that we did more science fiction than science. They are partly right, and this is why it is essential to conduct new research with actual data on binaries to assess their naturality or artificiality.

Below are seeds of four scientific research proposals which, if successful, will corroborate the existence of starivores. Only the last one however promises to give indisputable proof of extraterrestrial intelligence.

### Gamma-ray bursts and binaries

Are starivores protected from gamma-ray bursts? Such events are of a rare violence and a galactic gamma-ray burst could wipe out eukaryotes in a range of 14 kpc from the explosion (see Scalo and Wheeler 2002; and also Ćirković, Vukotić, and Dragićević 2009). Long-lived civilizations would certainly anticipate such rare but probable catastrophes. We should be able to show either that binaries get strongly disturbed, dislocated or destroyed by gamma ray bursts (which would tend to falsify the starivore hypothesis), or that they are very robust to such disturbances (which would tend to corroborate the starivore hypothesis).

### Keiber's laws and semi-detached binaries

When we analyzed binaries with living system theory, or when we applied a scaling law of reproduction time to black holes, we applied biological concepts and theories to astrophysical systems. This is indeed part of a general biological (or evolutionary developmental) view of the universe. It remains a challenge for future generations to assess if this view is correct. As Dick (1996, 1–2) writes,

> The whole thrust of physical science since the seventeenth-century scientific revolution has been to demonstrate the role of physical law in the universe, a mission admirably carried out by Kepler, Galileo, Newton, and their successors. The question at stake in the extraterrestrial-life debate is whether an analogous "biological law" reigns throughout the universe

What other robust biological laws could be applied to binaries? Kleiber's law (Kleiber 1932) is the observation that in living organisms the metabolic rate scales to the ¾ power of its mass. It is remarkable, because it holds over 16 orders of magnitude – although the scaling exponent changes slightly (see DeLong et al. 2010 for a recent review). This validity across so many scales suggests that it could hold even for macroscopic living systems, such as cities (Isalgue, Coch, and Serra 2007) or putative starivores. Figure 27 below illustrates this law.

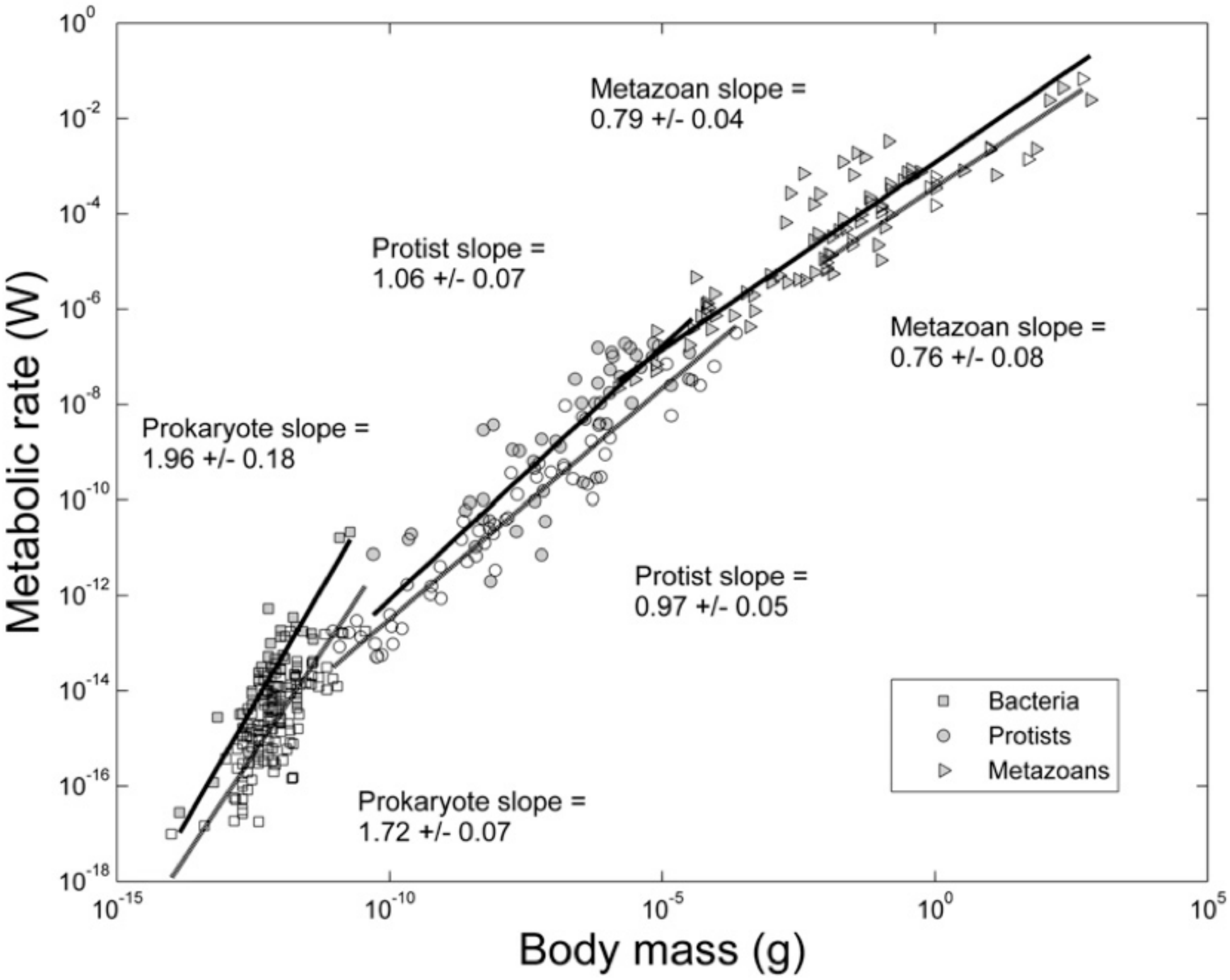

Figure 27 - "Relationship between whole organism metabolic rate and body mass for heterotrophic prokaryotes, protists, and metazoans plotted on logarithmic axes." (DeLong et al. 2010, 12942)

Now, does this law apply to transient accreting binaries? The hypothesis is that, if binaries are starivores, they should fit with this law. If not, it is less likely. How can we test this? It is easy. We need to gather the relevant data for binaries. We can simply interpret the accretion rate as a metabolic rate (both are energy flow metrics); and keep on the x-axis the mass of the primary. Kepler's law famously applies to planets, but does Kleiber's law apply to binaries?

Note that this research program could also be coupled with a systematic calculation of the free energy rate density complexity metric proposed by Chaisson. The only additional parameter to add is the age of the system in consideration.

**Scale Relativity and Binaries**

Scale relativity (see e.g. L. Nottale 2011) generates probability distributions for the formation of gravitational structures. It gives probabilities to have single, double, triple or n-body systems. Preliminary results explain why pairs of galaxies are so common (L. Nottale 2011, 654–658).

This project consists in applying scale relativity to the formation of binary systems. If binaries are starivores, the prediction of scale relativity should fail. Indeed, we should find more binary systems than what would be formed by natural gravitational formation. Or there should be proportionally less pairs of galaxies than binary systems. Note that the picture could be more complicated if putative starivores migrate and leave single depleted stars.

Furthermore, applying the inverse distance-development principle, further and further away galaxies should fit more and more the predictions of scale relativity. Of course, this project represents a lot of work, but it is a global approach which, even if it ends up dismissing the starivore hypothesis, would teach us a lot about star formation.

If it succeeds, we would have an estimate of the number of intelligent civilizations in the galaxy, simply by subtracting the observed number of binaries with the predicted number.

**Pulsars decoding**

Finally and most importantly, a convincing proof of ETI should include information processing. This is why I insisted that the assessment of whether there are messages in pulsars should be a priority (see section 9.4.6 Are Pulsars Artificial Output Transducers?, p248).

### The High Energy Astrobiology Prize

After examination by a scientific jury, the Evo Devo Universe (EDU) community will be happy to deliver a 500€ “High Energy Astrobiology Prize” for the first peer-reviewed paper on either of these projects (Gamma-ray bursts and binaries; Kleiber's law; scale relativity and binaries; assessing or decoding the informational complexity of pulsars' pulses). Of course the prize will be delivered for both a positive or a negative result. New research proposals to test all the ways there could be intelligence in interacting binaries are also most welcome. If you would like to contribute but do not have the scientific expertise to do so, we welcome donations to make the high energy astrobiology prize even more attractive.

I invite you to http://www.highenergyastrobiology.com/theprize.htm for more details.

# *CHAPTER 10 - Cosmological Ethics and Immortality*

**Abstract**: Most ethical principles, religious or not, are based on wisdom acquired through a few millennia. This may seem a long time but once we take a cosmological perspective, even millennia are insignificant. The field of evolutionary ethics makes a big leap by embracing evolutionary time scales (millions of years). Can we continue to extend our ethical reflections, principles and theories up to the 13.7 billion years of cosmic evolution? What is the ultimate good in the universe? Evolutionary ethics concludes that survival is the most important value. But survival of what? and for how long? How can we aim for infinite survival, that is, for immortality? We first outline thermodynamical values, which are truly universal because they depend only on the concept of energy. Then we criticize the naturalistic fallacy and, inspired by Aristotle's theory of moral virtues, we outline evolutionary trade-offs (*egoism-altruism*, *stability-adaptability*, *specialist-generalist*, *exploration-exploitation*, *competition-cooperation* and *r-K selection*) sketching a theory of evolutionary virtues. However, evolutionary values are insufficient for ethical purposes, since they give insights in how to adapt to any circumstance, for any purpose. To remedy this limitation, we outline developmental values for individuals (e.g. cognitive and moral development); developmental values for societies (e.g. rationality increase, violence decrease). Thermodynamical, evolutionary and developmental values promise to be robust ethical principles, because proven through the wisdom of billion years of cosmic evolution. As an application, we examine the age-old will to immortality and propose a voyage to five kinds of immortalities: *spiritual*, *individual*, *creative*, *evolutionary* and *cosmological*. We conclude that *the ultimate good is the infinite continuation of the evolutionary process*.

*To romance of the far future, then, is to attempt to see the human race in its cosmic setting, and to mould our hearts to entertain new values.*

Preface of Olaf Stapledon's (1931) *Last and First Men.*

Until now our intellectual journey has been mainly *descriptive* and *critical*, exploring hypotheses for tackling issues such as the beginning of the universe and scenarios for its long-term future. In order to aim as much as possible for a comprehensive worldview, we need to address the issue of values, or the *normative* dimension of philosophy I identified in Chapter 1. This is what we will explore now. As usual, our scope in space-time is maximal, that is, cosmological. We are thus interested about values general enough to hold for cosmological scales. Philosopher Archie J. Bahm (1980, 4) dubbed the science inquiring into the ultimate values of life as a whole “religiology”. It concerns the ultimate value of life, how to find it, preserve it and enjoy it. The neologism makes sense because answering such questions is traditionally a religious endeavor. But it needs not to be.

Values and ethics can be built without relying on God or revelations (see e.g. Nielsen 1973; Jantsch 1980, 264). Actually, human ethical principles, religious or not, are based on the historical wisdom of a few millennia, which is totally insignificant from a cosmological perspective. We need to increase the scope of our thinking towards the cosmological scale (see table 16).

| Scale | Duration |
|---|---|
| Human | 100 years |
| Historical | 10 000 years |
| Anthropological (Evolutionary) | 10 million years |
| Geological | 5 billion years |
| Cosmological | 13.7 billion years |

Table 16 Time Scales in Human Thought. Adapted from (Dick 2009a, 464).
Since it is difficult for humans to appreciate orders of magnitude greater than historical, see also the remarkable "Chronozoom" project for a dynamical view on time scales: http://www.chronozoomproject.org/

Evolutionary ethics (see e.g. Matt Ridley 1997; E. O. Wilson 1998) encompasses millions of years, a remarkable thousandfold advance compared to history-based ethics. A major insight of evolutionary ethics is that our existing beliefs, motivations and values have no special or universal validity. They have been selected from past evolutionary needs (Stewart 2000, 16).

Yet, evolutionary ethics is insufficient if we are concerned about cosmological timescales. It means that the origin of human nature is not our central concern here, as it is in evolutionary ethics (see e.g. the excellent books of Wright 1994; Waal 1996; Matt Ridley 1997). From a cosmological perspective, human nature is just but one step in cosmic evolution. I am interested in an ethical system which can also withstand evolution *before* and *beyond* humanity. I thus focus in this chapter on the next thousandfold advance from evolutionary to cosmological ethics. As Stapledon wrote above when prefacing his science fiction novel, such thinking opens the possibility to decipher new values. Romance is an approach, but not the only one. Here I will stay with the rational tradition of science and philosophy.

As I already wrote earlier (Story 7: Global warming and universal cooling. This story is fictional., p154), when I tell friends or colleagues that I am busy with cosmological issues, they often do not understand why or how I could seriously care about such a far future. My best reply is to draw an analogy. We are definitely growing in our global awareness and are becoming aware and worried about global economical, ecological or climatological problems concerning the Earth as a whole.

We also do care for phenomena beyond planet Earth. As you read these lines, scientists are monitoring the nearest star, our Sun. Indeed, tracking solar weather is critical to take care of our electrical and communication networks. The famous solar storm of 1859 was so powerful that it created aurorae all around the world, so bright they even woke up gold miners during the night. The consequences of such a solar storm today would be billions of dollars of damage to satellites, power grids and radio communications (see e.g. Odenwald and Green 2008). So increasing our cosmological awareness is not just an intellectual curiosity and a solar storm is just one example of global catastrophic risks (see contributions in Nick Bostrom and Ćirković 2008 for more examples).

The more we progress with our understanding of the Universe, the more global our awareness develops. Peter Russell (1982, 83) remarked that life forms build more and more extended models of their environment. He wrote, an "important characteristic attributed to conscious beings is the ability to form internal models of the world they experience; the greater the consciousness the more complex the models.” Actually this activity of model building is necessary to regulate efficiently the environment (Conant and Ashby 1970). Why would modelling stop at planet Earth? From a cosmological perspective, it is a tiny slice of space. What is then the limit of the process of model building? We saw (Chapter 6.3; Chapter 7) that it is the modelling of our and other possible universes.

In this cosmological context, what are the ultimate values of intelligence? What is the ultimate good for intelligent life in the universe? Evolutionary ethics concludes that it is *survival*. But this almost trivial statement raises two questions. First, *survival of what?* Myself? My family? My culture? My planet? The solar system? Our galaxy? The universe? Second, *survival for how long?* For human, historical, evolutionary, geological or cosmological time scales? Ethical theories aim at producing good results (Bahm 1953, 310). But good for whom? For oneself, for others, for God, for the best people, for most people, for higher principles such as justice or the process of evolution? Could we and should we integrate goodness for these different options?

I argue in this chapter that *the ultimate good is the infinite continuation of the evolutionary process*. Note that the “ultimate good” is an essential component of ethical theories and is also called *summum bonum*, supreme good, supreme goal, supreme value, highest good (Adler 1985, 478a). My thesis implies that I answer “survival of what?” with “the evolutionary process”; and “survival for how long?” with no less than “infinitely”. This is in line with the *Principia Cybernetica Project*, when Heylighen, Joslyn and Turchin (1993) wrote:

> Thus we can from there derive the ultimate good as the continuation of the process of evolution itself, in the negative sense of avoiding evolutionary “dead ends” and general extinction, in the positive sense of constantly increasing our fitness, and thus our intelligence, degree of organization and general mastery over the universe.

It is important to distinguish two different kinds of ethics, *descriptive* and *normative*. Descriptive ethics is a second-order knowledge inquiry consisting in recording empirically what people *do seek*; while normative (or prescriptive) ethics is a first-order knowledge defining what people *should seek*.

Generally, evolutionary ethics focuses on *descriptive ethics*, explaining why we hold such or such values, or how such or such moral behavior could have evolved. Can evolution also justify *normative ethics*? That is, can evolutionary insights tell us that such or such value *should* be followed? Such an endeavor of normative ethics is of course much more delicate and controversial than descriptive ethics. The spectrum of eugenism and its horrible political associations looms. Eugenism is indeed an evolutionary normative ethics, because it attempts to justify its value system based on genetics. Yet, avoiding normative ethics is no option, because it is the only practically helpful kind of ethics, translatable into action. Descriptive ethics, however insightful intellectually doesn't help much to give values and act in the world.

My aim in this Chapter is to take inspirations from some major scientific theories to propose values or guidelines. I am not pretending to deduce absolute moral theorems based on scientific models. Since scientific models evolve and progress and the price to pay is that there is no certainty in morality, as there are no certainties that our scientific models are "true". Yet, it is possible to be relativist without falling into relativism (see 2.1.4 Relativity, not Relativism, p33).

The organization of this chapter is simple. First, we seek universal ethical principles in universal laws of thermodynamics. We thus discuss fundamental insights of thermodynamical values, or *thermoethics*. Then we inquire about evolutionary theory to extract evolutionary values. We debunk the fallacious naturalistic fallacy, discuss fitness as an evolutionary value and show what evolutionary trade-offs such as egoism-altruism, stability-adaptability, specialist-generalist, exploration-exploitation, competition-cooperation or r-K selection tell us about ethics. Because of the large generality and practical applicability of evolutionary values, we conclude that they are more providing *means* rather than *ends*, goals or values. We need further insight about a general evolutionary *development*, *progress* and *direction*. Although the direction of evolution is a controversial topic, the case for it is strong. And this leads to the much more useful *developmental values*, which we discuss for humans, societies and the universe.

As an illustration of the thermodynamical, evolutionary and developmental ethical framework, I apply it to a central longing of humanity: the will to *immortality*. I invite you to a voyage to five kinds of immortalities from spiritual, individual, creative, evolutionary up to cosmological immortality.

Although our analyses toward cosmological ethics and immortality are only precursory, I hope they will lay the ground to further work and eventually offer a meaning of life in harmony with cosmic evolution.

## 10.1 Thermodynamical Values

*Prigogine "wanted the entropy ethics to be taught to children all over the planet"*

(Hammond 2005, v, preface to the Eulogy Edition)

As Dick E. Hammond argues in his (2005) book, entropy ethics brings important insights to children all over the *planet*, but also all over the *universe*.
Is there a way to think about universal values from an energetic perspective? If organisms develop new ways to extract energy, they will have resources to increase their complexity and thus become fitter. The engineering of new ways to extract energy constitutes a fundamental step in major evolutionary transitions (see e.g. Aunger 2007a). In the modern evolutionary picture, Peter Corning (2005) also insists on the importance of *thermoeconomics* in the darwinian synthesis. But what does it means for ethics? How should we best extract, use, store and distribute energy? To answer these questions, we need an ethics based on energetic exchanges, i.e. based on thermodynamics.

What is the underlying principle of entropy ethics? It is "the need to learn to utilize energy efficiently to bring about order in the environment" (Hammond 2005, 67). Importantly, we need to distinguish two entropies, *informational* and *thermodynamical*. Informational entropy has to do with statistics and order in any system. Thermodynamical entropy has to do with heat and energy. The expression

"entropy ethics" can thus be confusing because entropy can be informational or thermodynamical. The two corresponding ethics are *infoethics* (Floridi 2008) and *thermoethics* (Freitas Jr 1979, chap. 25.1.3). Let us outline their core principles.

### 10.1.1 Informational Ethics

Floridi (2008, 58–59) did articulate foundations of infoethics, by proposing four principles:

> 0. entropy ought not to be caused in the infosphere (null law);
> 1. entropy ought to be prevented in the infosphere;
> 2. entropy ought to be removed from the infosphere;
> 3. the flourishing of informational entities as well as of the whole infosphere ought to be promoted by preserving, cultivating and enriching their properties.

We can illustrate the (0)-principle with email spam, because it creates disorder or entropy in our email boxes. Antivirus softwares are faithful to principle (1) because they prevent information destruction on our hard drives from malicious softwares. Examples of applying principle (2) include removing and correcting false statements in Wikipedia; or cleaning, merging databases, which are good actions bringing more order, less redundancy and more integration in the infosphere. Principle (3) is less objective, but can be related to valuing creativity at large.

We will not further elaborate on infoethics, because the concept of information is far from being clear, its definition is highly debated, and information is largely dependent on our goals. There is no such thing as objective information. From a cybernetic perspective, information is information only if it reduces uncertainty and helps thereby to achieve goals. An information about a semi-detached binary system might be very exciting to me, while it would have absolutely no interest to you – unless you are also intrigued about ideas exposed in Chapter 9.

### 10.1.2 Thermodynamical Ethics

Although Freitas coined the term "thermoethics", its origin can be traced back to Wilhelm Ostwald, nobel prize winner in chemistry in 1909. Ostwald (1912; translated by Bayliss 1915, 28) advanced the *thermodynamic imperative*:

> Waste not free energy; treasure it and make the best use of it.

We can derive a lot of value from this seemingly trivial principle. On the negative side, it means that we should avoid randomizing, actions which lead to confusion, conflict or chaos. For example, killing is clearly bad because it is the destruction of a trillion-cells organization. Burning libraries is also a very destructive and bad act, because it destroys knowledge accumulated by a civilization. The thermodynamic imperative is implicitly in application in most businesses. If a firm can do the same amount of work with less energy, it will outcompete others. The same holds in biological evolution. The capacity to extract efficiently energy from the environment is absolutely crucial to stay fit.

Hammond also contrasts making random noise with the orderly coordination of an orchestra. The one produces stressful activation patterns on human brains, while the other can provoke aesthetic experiences and highly coherent brainwaves. Of course, this does not prevent modern artists to change and challenge the rules and will create pseudo-random works in order to trigger negative emotions and feelings.

Why is thermoethics so fundamental? Because it does not make any assumption about the substrate of living systems to which it applies. It it thus valid for humans and animals, but also for posthumans, postbiological or even extraterrestrials. Freitas (1979, chap. 25.1.3) proposed a similar *principal thermoethic*:

> all living beings should always act so as to minimize the total entropy of the universe, or so as to maximize the total negentropy.

Freitas explains that in other words, “living beings should always act to further the mission of life in the cosmos, which is to reduce the universe to order by building the maximum complexity into the mass-energy available.” He summarizes three thermoethical duties, to *avoid harming*, to *preserve* and to *create*, which indeed are very similar to the infoethics principles of Floridi. A corollary of the thermoethical principal is what Freitas calls the *Corollary of Negentropic Equality*:

> All entities of equal negentropy have equal rights and responsibilities; the more negentropic an entity, the greater are its rights and the deeper are its responsibilities. (See Cocca 1962; Fasan 1968; 1970; Haley 1963; 1956; Nicolson 1978; Rhyne 1958.)

Let us take an example of humanity's failing to be thermoethical, which was pointed out very clearly by David Criswell (1985, 63–64):

> The industrially advanced nations are accused and self-accused of wasting the 20 KW (or 0.00002 GW)/person they consume to support their materially extravagant life-styles. However, we also live in an era of enormous waste of solar energy. Not only are we failing to use the 81,000,000 GW/person (averaged over everyone on Earth) of power the sun is currently sending irretrievably to cold deep space but we are also vigorously wasting the extremely inefficiently obtained power that has fallen on Earth over geologic time to produce our fossil fuels and power our inefficient biosphere.
>
> We currently burn several billion tons of coal, oil and wood (fossil solar energy) a year worldwide (equivalent to 1.78 tons of fusion mass burning) to meet our meager energy needs, while we permit the sun to completely fusion burn nearly 136,500,000,000,000 tons of matter each year. We do have present-day limits to our technology, our will, and our immediate needs to tackle this stellar task. However, we can frame the basic challenges and begin imagining the general approaches to the conservation and upgrading of our stellar resources in the Solar System.

What Criswell lays down is a *12 orders of magnitude difference* between personal energy consumption ($2 \times 10^{-5}$ GW/person) and the wasted solar energy ($8.1 \times 10^{7}$ GW/person)! Humanity (and beyond) has a long way to go to cover this gap. Along these lines, it is thermoethical to go to the stars and use their energy to do intelligent work, instead of letting them dissipate their energy in space. But first things first, we need to make the most of Sun, the nearest star.

Could extraterrestrials be more thermoethical than us? Seeing how much we waste solar energy, putative starivores would not be proud of us. But different putative starivores would also be more or less thermoethical. For example, we saw that putative starivores with an accreting black holes are very thermoethical because they deplete the whole material of the accretion disc, while similar systems with white dwarfs accrete only 10% of the disc (see section 9.4.5 Living Systems Arguments, p239).

One could object that, albeit its universality, the thermoethics picture may be too simple. Indeed, moments of chaos, crises, conflict and destruction can lead to new developments. For example, in developmental psychology, states of disorder are often the way towards higher stages (see e.g. Scharmer 2007). In human societies, a revolution creates a lot of disorder, but can make a bloodbath ultimately useful, establishing a new fairer societal organization. Natural catastrophes also shake ecosystems and can give the way to new evolutionary branches to flourish –such as mammals some 65 millions years ago.

Even in such examples, one could reply that thermoethics still holds because those crises in the the long term bring a better organization or energy budgeting. This may be true, but it remains difficult *while in a crisis* to know whether one goes towards destruction or construction of something better. Will the new system really become more thermoethical than the one currently being destroyed? The prospect of regression is not excluded. Thermoethics is universal and therefore sometimes difficult to translate into more precise and concrete ethical principles. Let us now explore insights from evolutionary theory and developmental trends to enrich our cosmological ethics.

## 10.2 Evolutionary Values

*If, even in the long run, ethical behavior were self-defeating, eventually we would not call it ethical, but foolish.*

(Sagan 1997, 218)

### 10.2.1 The Fallacious Naturalistic Fallacy

It is hard to enter evolutionary ethics without tackling the critique of the *naturalistic fallacy.* It states that we cannot derive an "ought" from an "is". In other words, that normative and descriptive dimensions of philosophizing are unbridgeable (see also the is-ought test, section 2.3.2 Testing the Dimensions, p46). Although the spotting of the fallacy is often attributed to Moore (1903), he actually did not claim that we can not derive an "ought" from an "is". He was concerned by the distinctiveness of the content of ethical judgments (see the introduction of Thomas Baldwin in Moore 1993, xviii). So it is better to trace it back to the source, Hume's (1739, bk. III, Part I.1) classical work, *A Treatise of Human Nature*. Although Hume's position is indeed that we can not derive moral judgments from reason alone, he is chiefly criticizing that moral philosophers slip from "is" to "ought" *without justification*.

Let us give two examples of deriving moral judgments from natural facts. First, Herbert Spencer (1851, 324) who, inspired by the functioning of the natural world, harshly criticized social thinkers who advocate a protective law for the poors:

> Blind to the fact, that under the natural order of things society is constantly excreting its unhealthy, imbecile, slow, vacillating, faithless members, these unthinking, though well-meaning, men advocate an interference which not only stops the purifying process, but even increases the vitiation—absolutely encourages the multiplication of the reckless and incompetent by offering them an unfailing provision, and *dis*courages the multiplication of the competent and provident by heightening the prospective difficulty of maintaining a family. And thus, in their eagerness to prevent the really salutary sufferings that surround us, these sigh-wise and groan-foolish people bequeath to posterity a continually increasing curse.

This is an example of drifting from what the process of natural selection "is", to its application in society and how it "ought" to be. To say the least, this "bad naturalistic fallacy" would seem politically incorrect even to the most radical ideologies. Maybe not all, since Adolf Hitler (cited in G. L. Weinberg 2003, 21) would agree:

> The abandonment of sick, frail, deformed children—in other words, their destruction—demonstrated greater human dignity and was in reality a thousand times more humane than the pathetic insanity of our time, which attempts to preserve the lives of the sickest subjects—at any price—while taking the lives of a hundred thousand healthy children through a decrease in the birth rate or through abortifacient agents, subsequently breeding a race of degenerates burdened with illness.

Let us now turn to a second example, which is a "good naturalistic fallacy". It starts from the "is" proposition that "humans need biodiversity" and derives the "ought" that "we should promote biodiversity". Who would challenge this reasoning?

My point with these two examples is only formal. Why accept one fallacy and not the other? *When should we imitate nature, and when should we not?* The real issue is *how* to commit the naturalistic fallacy, not whether we can avoid it. We are going to commit it anyway. This was the point of the *is-ought test* I proposed in Part I.

For sure it is difficult, if not impossible, to deduce with scientific methods moral theorems. But we can and must use insights from science to build ethical theories. For, what are the alternative sources of inspiration? Common sense? Introspection? Art? Existing laws? Supernatural revelation? Axiologies based on these other options are of course possible, but here we hold tight to objective criteria as much as we can. Subjective, intersubjective (social) or supernatural insights are thus not our focus because they hold no promise for a naturalistic cosmological extension.

Let us think for a moment how we can extract the virtues and vices of evolutionary processes. Aristotle has developed a classical theory of ethics in *Nichomachean Ethics*, which is a theory of moral virtues. Virtues fall between two extremes vices of deficiency and excess. These trade-offs lead to a virtuous mean, which can be summarized with table 17 below:

| *Vice of Deficiency* | *Virtuous Mean* | *Vice of Excess* |
|---|---|---|
| *Cowardice* | *Courage* | *Rashness* |
| *Insensibility* | *Temperance* | *Intemperance* |
| *Illiberality* | *Liberality* | *Prodigality* |
| *Pettiness* | *Munificence* | *Vulgarity* |
| *Humble-mindedness* | *High-mindedness* | *Vaingloriness* |
| *Want of Ambition* | *Right Ambition* | *Over-ambition* |
| *Spiritlessness* | *Good Temper* | *Irascibility* |
| *Surliness* | *Friendly Civility* | *Obsequiousness* |
| *Ironical Depreciation* | *Sincerity* | *Boastfulness* |
| *Boorishness* | *Wittiness* | *Buffoonery* |
| *Shamelessness* | *Modesty* | *Bashfulness* |
| *Callousness* | *Just Resentment* | *Spitefulness* |

Table 17 - According to Aristotle, moral virtues fall at the mean between two accompanying vices.

Aristotle did focus on human virtues, but could we apply a similar reasoning to the virtues of different evolutionary mechanisms? We will attempt to reason in these lines and study the importance of six fundamental evolutionary trade-offs. But first a word about fitness.

Fitness is tautologically a good value, because it implies survival. As Sagan writes above, a self-defeating behavior can only be foolish, never ethical. The fitness of a system is most generally its probability of continuity, i.e. the number of agents in the future divided by the number of agents in the present. However it is important to make explicit the level of selection on which fitness applies. Do we try to promote fitness of cells, organs, individuals, groups? Generally, there are many ways and strategies to be fit which involve evolutionary trade-offs that we will now outline from an ethical perspective.

### 10.2.2 Egoism – Altruism Trade-off

That organisms strive to survive is a truism. But some organisms help others to survive and this is a puzzle for evolutionary biologists. Indeed, the competition induced by natural selection selects the fittest individuals, regardless of what happens to others. But some behaviors increase the fitness of others and decreases the fitness of the actor. It is a challenge for evolutionary theorists to put forth mechanisms explaining how such a behavior can evolve (see e.g. Sober and Wilson 1998).

We are not going to tackle this question surrounded by controversies. We can mention that multiple levels selection theory is a promising way to resolve it (see e.g. D. S. Wilson and Sober 1994). The main idea is to look at the actions of natural selection not only at the genetical level, but at other smaller or bigger *units of selection* such as molecules, cells, individuals, groups or species.

Let us simply point out here the trade-off between egoism and altruism. If an individual is too egoistic, she will have few chances to be integrated into a larger group. If he is too altruistic while giving and getting nothing in return, his own benefit and survival might suffer. It seems therefore that a mean has to be found. There are many possible middle grounds. The first is *kin selection* where altruistic behavior is only displayed towards one's group, and not other groups. For example, ants do help each others, but won't help worms.

If we aim at a universal ethics, extrapolating our increasing awareness and compassion, we should arrive at a point where we consider elementary particles and star dust as part of our cosmological group. So it is an interesting case where we aim at enlarging our circle of compassion to become altruistic at higher levels. But the hope that such a disinterested moral sentiment would emerge spontaneously is naive. A more pragmatic form of altruism is *reciprocal altruism*, or the more general *tit-for-tat rule*. Its essence is to avoid blind altruism. An individual will first be altruistic (cooperate), then subsequently replicate an opponent's previous action. It is intuitively fair and indeed an excellent evolutionary strategy. We will come back to it when discussing the competition – cooperation trade-off.

### 10.2.3 Stability – Adaptability Trade-off

Evolution keeps stable what works and adapts what doesn't. Let us say a few words about values associated with *stability* and *adaptability*.

A system's stability amounts to its internal or intrinsic fitness. For example, an organism has an intrinsic stability and capacity for (re)production. The related human values are equilibrium, robustness, strength, durability, autonomy or health. Such an increase in internal fitness requires the system to increase bonds and linkages between its different parts. As Heylighen (1997b; 1999) argues, this is accompanied by the increase of *structural complexity*.

However, building and reinforcing what is already here is a conservative strategy. If others are progressing and getting better at extracting resources in new creative ways, too much of stability will prove unfit.

*Adaptability* values relate to the capacity to cope with specific environmental perturbations or make use of changing external resources. In humans, *wealth* is typically an access to external resources. But adaptability values or the general adaptation to the environment also include variation, innovation, exploration, experimentation, diversity, growth and (re)production. A question naturally arises: for how long are we trying to adapt? Again, when our self and awareness grow, we are more and more inclined to increase the adaptability of larger systems (family, company, nation, humanity, the ecosystem, the Earth or the Universe).

This emerges not only out of altruistic motives, but also out of a concern for longer term adaptability, since when our awareness of interconnectedness increases, our care for external fitness increases accordingly. We strive to fit with a greater whole and to make it fitter.

As a system becomes more and more adaptive, it increases the variety of environmental perturbations that it can deal with, and this irreversibly increases its *functional complexity.* Heylighen (1999) elaborates:

All other things being equal, a system that can survive situations A, B and C, is *absolutely* fitter than a system that can only survive A and B. Such an increase in absolute fitness is necessarily accompanied by an increase in functional complexity. Thus, evolution will tend to irreversibly produce increases of functional complexity.

How do we balance stability and adaptability? In evolution, the combination of variation and selection brings such a balance. It is also very similar to the trial and error method in learning, which is the most basic learning mechanism. Let us now see two kinds of well adapted organisms: the specialist and the generalist.

### 10.2.4 Specialist – Generalist Trade-off

There are two ways to adapt and become fit. Either an organism becomes a *specialist* and is uniquely adapted to its niche or environment. For example, morgan's sphinx moth uses his long tongue specifically adapted to pollinate and feed at comet orchid found only in Madagascar (see Figure 28 below). Darwin and Wallace predicted the existence of such an insect with a long tongue, when they discovered the peculiar orchid. The moth was later discovered. The moth is fit because it is the only organism adapted to this flower.

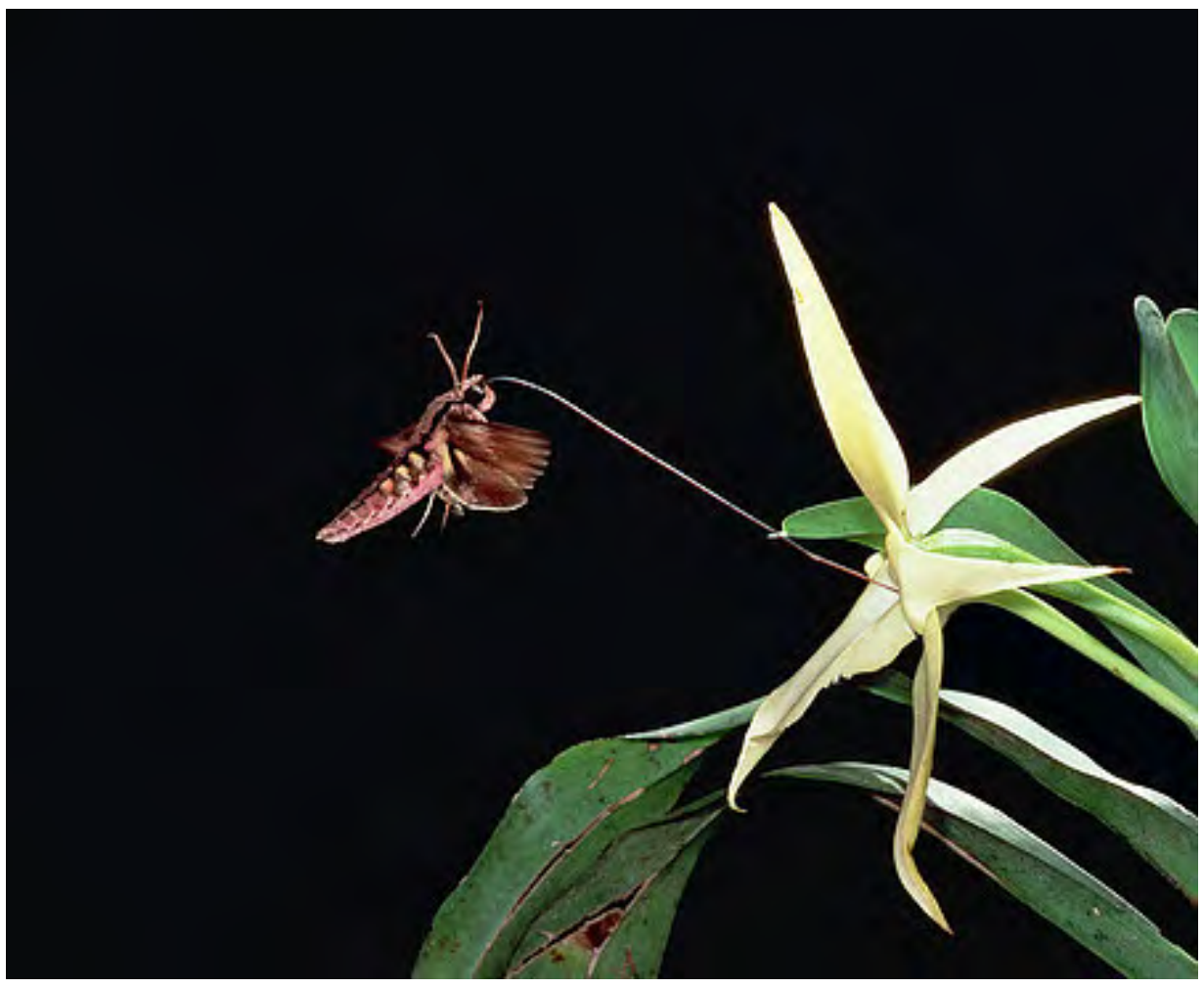

Figure 28 Morgan's Sphinx moth is highly specialized to feed at Comet orchid. Photo © Mitsuhiko Imamori

The *generalist* by contrast is fit because it can adapt to very different niches. For example in human societies, doctors, engineers or scientists can be generalists or specialists. Both generalists and specialists make their evolutionary way differently.

### 10.2.5 Exploration – Exploitation Trade-off

A very important trade-off in evolutionary theory is between exploration and exploitation. Exploration consists in *seeking* affordances or opportunities. Explorators are proactive and do not imitate what others do. By contrast, exploiters *use* what is already here, what has already occurred or been learnt. The attitude behind exploitation is to use known affordances such as resources, in order to maximize utility.

The key question underlying this trade-off is, how to divide the energy expenditure between exploration and exploitation? This can be illustrated in management (see e.g. March 1991) but also in social insects. For example in ants, the evaporation rate of pheromones is a critical parameter which determines the fitness of the ant colony. If the evaporation rate is high, exploration is favored because they will not follow existing pheromones tracks for a long time and will be forced to explore new paths. If the evaporation rate is low, exploitation is favored because ants will tend to follow the same tracks.

### 10.2.6 Competition – Cooperation Trade-off

Competition is a, if not *the* fundamental driving force in evolution. Plants often compete to grow higher to receive more sunlight; biologists even suspect that plants can also poison the roots of their neighbors (see e.g. Keddy 2001, 6). Living organisms constantly compete for territory, resources or mates. Higher animals also compete for prestige, recognition, awards, or social status. Competition has both pros and cons. The pros are that only the most adapted, efficient or clever organisms survive. They are fitter than the outcompeted ones. If resources are limited, those organisms able to make the most of available resources will survive and thus be able to reproduce more successfully and become fitter than others.

The disadvantage of competition is that the more energy invested to survive the competition, the less remains for growth, reproduction or higher purposes. The mentality behind competition is a primitive one that game theorists call “win-lose”. In competition, I win, you lose; or if you prefer, you win, I lose.

In organizations, the trade-off between competition and cooperation is also well studied. Khanna, Gulati and Nohria (1998) have argued that as a company focuses on egoisitic (private) benefits at the expense of altruistic (common) benefits, it tends to be more competitive. The interesting part is of course to find a balance between the two. And we thus see that this trade-off is also related to the egoism-altruism trade-off.

In the outdated evolutionary picture, evolution is a merciless process where all what matters is this competition for survival. But this view is very limited because those who win the competition in the long term are those who cooperate (see e.g. Axelrod 1984; Wright 2000; Stewart 2000).

Cooperation is more commonly referred to as *mutuality* in biology. Ecologist Robert Ulanowicz (2009, 75) turned competition and mutuality upside down, arguing that mutuality is essential and competition is derivative of it. Indeed, mutuality is “the platform from which competition can launch: without mutuality at some lower level, competition at higher levels simply cannot occur.” For example, the “reason one rodent is able to strive against its competitors is that any individual animal is a walking 'orgy of mutual benefaction' (May 1981, 95) within itself.”

So, what are the advantages of cooperation? First, it allows greater cooperative systems to emerge. Second, the underlying value system is “win-win” where both parties – or more – profit. The win-win game can be between two individuals (e.g. we both win) or between two levels (e.g. my individual benefit and our societal benefit). Such mutuality or cooperation is more and more possible and suitable in our information society, because the informational resources are *non-rival*. If I forward a paper to a colleague, I don't lose the paper. The costs in matter-energy are negligible compared to an economy based on matter-energy, where if we want to share a piece of bread, we can't multiply it –without help. Matter-energy exchanges are thus *rival*. Still, information exchanges do have an energetic cost, becoming more and more apparent as we exponentially increase our information processing.

What are the cons of cooperation? There are at least two. First, the danger of cooperation is that it can freeze competition, and thus novelty and adaptability. For example, phone companies cooperate when they agree to keep high prizes on the market. The cooperation is between companies, not between companies and end-users. This limited cooperation at one hierarchical level is detrimental to another level, the end-user. This is why societies have competition laws (or antitrust law) to make sure that market competition endures. So it is not enough to promote cooperation, but also to specify who cooperation should concern, or to balance it with competition.

Second, what if the other party wants to take profit of my cooperative attitude? What should I do? Carl Sagan (1997, chap. 16) did a great job also in ethics when he clearly summarized popular rules to live by, valued world-wide (see table 18 summarizing what follows). Let us summarize these rules, in order to better understand the need for cooperative ones.

| **Rule** | **Description** |
|---|---|
| The Golden rule | *Do unto others as you would have them do unto you* |
| The Silver rule | *Do not do unto others what you would not have them do unto you* |
| The Platinum rule | *Do unto others as they would like done unto them* |
| The Brazen rule | *Do unto others as they do unto you* |
| The Zinc rule | *Do unto others as they would like done unto them to reward them; or as they would not like done unto them to punish them* |
| The Iron rule | *Do unto others as you like, before they do it unto you* |
| The Tin rule | *Do the Golden rule to superiors, do as you like to others (Iron rule)* |
| The Kin rule | *Do the Golden rule to close relatives, do as you like to others (Iron rule)* |
| The Reciprocal rule | *Do the Golden rule to others, then do unto them as they do unto you (Brazen rule), but stop cooperation if they do not reciprocate* |
| The Tit-for-tat rule | *Do the Golden rule to others, then do unto them as they do unto you (Brazen rule) and try again cooperation (forgive)* |

Table 18 - Rules to live by.
Inspired by and adapted from (Sagan 1997, 228). Note that we could group similar rules: Golden, Silver and Platinum; Brazen and Zinc; Tin and Kin; Reciprocal and Tit-for-tat.

The famous *Golden rule* of behavior is attributed to Jesus and in the Bible it says: *Do unto others as you would have them do unto you* (Matthew 7:12). The problem is that it does not take into account human differences. For example, should the masochist inflict pain to his neighbor? Doing so, he would faithfully apply the Golden rule.

The negative formulation of the rule is called the *Silver rule* and is a restrictive rule, because it tells you what *not* to do, but not what to do. Its formulation by Conficius (*Analects* 15:23) says: *Do not do unto others what you would not have them do unto you*. This version is also known worldwide, from the writings of Hillel the Elder, to Gandhi and Martin Luther King Jr. But the same critique of human differences applies. By inflicting pleasure to a masochist, I do her no good. Those two rules would thus benefit to take into account the needs of the neighbor before their practical application. Otherwise they lack empathy.

We thus need to *Do unto others as they would like done unto them* (Alessandra 1996). In other words, to treat others in the way they like to be treated, and not just to apply our values blindly unto others. Now, thanks to this *Platinum rule*, the masochists can be happy.

Enough for lofty and indulgent rules. What do we do when someone is repeatedly violent? What if what she really wants is to hurt us? Should we just be empathic and self-sacrifice? Conficius replies that we should “repay evil with justice, and kindness with kindness” (*Analects* 14:36). Sagan calls it the Brazen rule: *Do unto others as they do unto you*. In the Bible, it is the *lex talionis* “an eye for an eye, and a tooth for a tooth” combined with the proverb “one good turn deserves another”. The Brazen rule contrasts with the Golden, Silver and Platinum rules because it is also retaliating. When the interactions are positive, such as lovers offering each others nicer and nicer presents, very well! But the Brazen rule also fuels the negative fire of endless vengeance. The Brazen rule crudely lacks *forgiveness*.

The Brazen rule can be made more generous or tougher. It just needs to include empathy or the insight of the Platinum rule. It then becomes what I call the *Zinc rule*: *Do unto others as they would like done unto them to reward them; or as they would not like done unto them to punish them.*

Sagan also describes three other widely applied rules of behavior. The first is the *Iron rule* which is at the opposite philosophy of the Golden rule. It says: *Do unto others as you like, before they do it unto you.* It is a rule of the powerful, also known as “he who has the gold makes the rules”.

The other rule widely found in nature illustrates a submission and dominance game. It says *Do the Golden rule to superiors, the Iron rule to inferiors*. Sagan calls this bullying rule the *Tin rule*, because it is an alloy between the Golden and the Iron rule.

In nature we also find the *Kin rule*, which says: *Do the Golden rule to close relatives, do as you like to others*. It is a Golden rule selectively applied for one's kin, and the Iron rule for the rest. It is thus a nepotism rule.

Now, let us summarize the pros and cons of the different rules. The Golden, Silver and Platinum rules are too complacent, unable to punish cruelty and exploitation. The Brazen and Zinc rules are too unforgiving, while the Iron rule is clearly not applicable by everyone and will fail when encountering a superior other. The Tin rule is smarter in this respect, but runs the risks of retaliation from the inferiors, if for whatever reason they become superiors. The kin rule promotes the survival and benefit of the group, but is by definition not scalable. So, can we combine the different rules to still better ones?

There are three types of games we can play, *win-lose*, *lose-lose* or *win-win*. Sagan notes that most competitive games we play, such as Monopoly, boxing, football, hockey, basketball, baseball, tennis, chess or partisan politics are win-lose, because only one party can win while the other loses. In such games, there is room only for the Iron and the Tin rules. Lose-lose games such as nuclear war or economic depression are even worse because all parties lose. The Iron and Tin rules focus only on playing such win-lose or lose-lose games. But we should emphasize that the most important and fruitful games of life are win-win, essential to make love, friendship, business, parenthood, music, art and science flourish.

Games theory studies these three kinds of games. It is not only a formal and theoretical science helping strategic decision making, it also has an experimental aspect. Indeed, sociologist Robert Axelrod (1984) did organize a computer tournament to find out which rules where winning in a repeated prisoner dilemma game (if you are not familiar with this game, see e.g. the summary of Sagan 1997, 225–229). It is clear that players using the Golden rule will tend to be exploited by the Iron rule players. Such players are both playing win-lose games, and thus will have a tendency to self-annihilate. The remaining ones are the win-win players.

A promising rule to play the win-win game is widely found in nature as *reciprocal altruism.* We can formulate the rule as the *Reciprocal rule: Do the Golden rule to others, then do unto them as they do unto you (Brazen rule), but stop cooperation if they do not reciprocate.* It is a very smart rule because you don't let people take advantage of you, yet you seek cooperation with everyone and not only with your kin. However, the cooperation will tend to be restricted to agents which reciprocate in the first place. What if the other player is not inclined to reciprocate right from the start? A broader more forgiving rule is needed.

This leads us to the Tit-for-tat rule, which states: *Do the Golden rule to others, then do unto them as they do unto you (Brazen rule) and try again cooperation (forgive)*. This rule has proven very robust and won the game-theoretical tournament. As Axelrod (1984, 54) summarizes it:

> What accounts for Tit-for-tat's robust success is its combination of being nice, retaliatory, forgiving and clear. Its niceness prevents it of getting into unnecessary trouble. Its retaliation discourages the other side from persisting whenever defection is tried. Its forgiveness helps restore mutual cooperation. And its clarity makes it intelligible to the other player, thereby eliciting long-term cooperation.

It is important to emphasize that this conclusion was reached from cold-blooded rational, strategic and opportunistic players: competing game theorists. There were no ethical presuppositions or inclinations to think that Tit-for-tat would be winning. Tit-for-tat wins because it works, not because its similarity with almost religious values.

There are some variations on Tit-for-tat, and finer analysis shows that it can be improved. One idea is first to exploit full cooperators (a selective Iron rule), and then switch to tit-for-tat (see Matt Ridley 1997 for a popular account of these variants). However, in the real world if you display a behavior of profiteer, you will raise suspicion regarding your ability to cooperate. In a more sophisticated and realistic game theory tournament, a variation on tit-for-tat also won. It is the Generous-tit-for-tat, which, one third of the time, overlooks a defection (see Nowak, May, and Sigmund 1995).

The robustness of the tit-for-tat rule is a successful example of what evolutionary ethics can teach us. Of course, unlike religions, scientists are careful not to make this conclusion an ethical dogma. Indeed, we should be reminded that they have proven the worth of tit-for-tat in simplified contexts. Yet, it certainly contains much more wisdom than the more naïve Golden or Silver rules.

To sum up, we have learned that blind cooperation or blind competition do not make their evolutionary way, but will self-annihilate sooner or later. A more careful trade-off between cooperation and competition is needed, of which tit-for-tat is a prime example.

### 10.2.7 r – K Selection Trade-off

At bottom, fitness is about survival and reproduction. Since no organism is immortal, albeit all its efforts to stay alive and develop, if it wants to continue it needs to reproduce. So it needs to invest resources into reproduction. A fundamental trade-off emerges as organisms need to split their resources between survival (or development) and reproduction. In practice, should organisms aim for more *quantitative* or *qualitative* survival and reproduction? The r-K selection theory (see e.g. Pianka 1970) concerns this trade off. The "r" and "K" stand for the parameters of the ecological model, where "r" refers to the maximum growth rate of the population, and "K" to the carrying capacity of the environment. Organisms in unstable environments will tend to be r-selected and invest in quick reproduction and produce many offspring, while in stable environments, surviving organisms will be K-selected and will rather invest in prolonged development, long life and produce few offspring.

Typical examples of r-selected organisms are bacteria, insects, frogs or mice. They reproduce quickly because it is crucial to survive in their unstable environment. Examples of K-selected organisms are trees, elephants or humans.

Since the last few decades, the r-K selection theory is less in fashion. Indeed not all organisms can be classified in the r-K spectrum. For example, trees display long life and compete for light, making them K-strategists, while they spread thousands of seeds, a characteristic of r-strategists. Also, r-K selection is not the whole of the story and other factors such as environmental variability and predation should be taken into account. But similar trade-offs such as between maintenance and reproduction are fundamental for modeling dynamics in evolutionary biology (see e.g. the disposable soma theory of Kirkwood 1977; 1999).

What does r-K selection implies for ethics? It is difficult to tell because it is a biological theory, and our human world is now dominated by culture. However, some authors have attempted to model *cultural* r-k selection (see e.g. Heylighen and Bernheim 2004; Fog 1997; 1999). As Fog argues, cultural r-selection happens when a group is dominated by external conflicts and wars; while cultural k-selection happens when a group is peaceful and spends more resources on satisfying the individuals than on strengthening the group. Fog further develops applications to religion, politics, music, art, architecture, clothing fashion, sexual behavior, sport and play.

### 10.2.8 Discussion

If evolution is a trial-and-error process capable of solving problems of survival and adaptation, then the values behind the different trade-offs are foremost *practical.*

Being aware of such important trade-offs can allow us to play with their spectrum in order to solve any kind of problem. Evolutionary values thus give insights in how to adapt to *any circumstances*, for *any purpose*. So it is easy to use and abuse those values to reach *any* objective. As with any powerful tool, there are “good” or “bad” applications of evolutionary principles. Furthermore, even balancing properly a particular trade-off is not enough. We need to make explicit the level of selection we want to promote. Am I trying to increase the fitness of my body or my society? We certainly need to trade-off also the different levels of selection. For example, I agree to pay taxes if the government pays to build roads.

By contrast with Aristotle's theory of virtue, the attitude of finding a mean between each trade-off is not necessarily the smartest one. Indeed, there are some environments or situations where it may be better or necessary to be totally altruistic, maximally explorative, etc. We could thus propose a rule of *metadaptation*, adapting each evolutionary trade-off depending on the environmental situation. This remains of course very pragmatic.

One critique of evolutionary ethics is that it may provide necessary knowledge to build an ethics, but this is insufficient (Gewirth 1993). We are evolved beings who can make free choices, often constrained by our evolutionary past, but not always. What we choose is not always in line with basic evolutionary principles. We have the power and freedom to choose to go against evolution. For example, in developed countries, under certain conditions, one can choose euthanasia, abortion or to be homosexual.

Interestingly, these three choices are morally wrong in conservative societies. Indeed, they go most clearly against the most primitive evolutionary values of survival and reproduction.

Only in a more global, societal or cultural context can they make sense. What is the meaning of struggling to keep alive a 95-year old man in the coma, costing several thousands euros per day of palliative care? What sense does it make to have a child resulting from a rape when one is a 14 years girl still going to school? What is the big deal of being homosexual and having no children, knowing that we are more than 7 billion humans on Earth? Shouldn't on the contrary societies further develop the rights to euthanasia, abortion or homosexuality?

The reasoning is tempting, but the problem is of course to set boundaries. Euthanasia or abortion are not decisions to be taken lightly and indeed countries which permit it still strongly regulate those practices. They may be argued to be ethical if they contribute to evolution on a greater scale, contributing to social progress. But what do we mean with the notion of *progress*?

In fact, as Nitecki (1993, 15) argues, we can't answer ethical questions without a notion of progress:

> We believe in correctness of ethical prescriptions only when we see that there will be, or is, progress in human behavior. Thus evolutionary ethics has an unspoken, and hidden, claim to a meaning or a direction to biological process. Since, according to evolutionary ethicists, natural selection is responsible for morality, the more moral will be selected, and hence there is progress in morality. If evolutionary biology rejects progress then it must also reject evolutionary ethics. Just as to remain alive requires regeneration, or repair, so evolutionary ethics requires progress. Progress is a process that produces improvement. It is a process that must continue and cannot end. By definition, the process that ends, dies. Thus progress reflects the drama of the

> evolutionary intellectual, who must either accept evolutionary ethics *and* progress or must reject both –to many of us a difficult choice.

Evolutionary values are thus more the basis of a praxeology, a theory of action, rather than an axiology, a theory of values. We need clearer goals, values, purposes, directions or an idea of progress if we aim for a normative ethics. Indeed where do we want to go with these evolutionary strategies and management of trade-offs? Or maybe evolution already shows some general trends, that we could benefit from following? The tit-for-tat rule showed that cooperation pays in the long term, which is a remarkable and promising result.

To go further, we need broader insights into evolutionary theory, namely into the direction of evolution. Evolution and development go hand in hand, but developmental values are more promising than evolutionary ones because development implies a direction. Let us have a short examination of developmental values in humans, societies and a possible extrapolation to the universe.

## 10.3 Developmental Values

*Cooperators will inherit the earth, and eventually the universe.*

(Stewart 2000, 9)

### 10.3.1 Cybernetic Values

As organisms grow in complexity, they more and more become goal-directed and therefore can be modelled as *cybernetic* systems. They are able to choose which goal to pursue, to compare alternative courses of actions, to remember patterns which do or do not work, that is, to learn. So we can legitimately extend evolutionary values to cybernetic values. Such values are essential for a goal-directed system to survive. For example, homeostatic animals are able to *control* and maintain their temperature. But such control mechanisms are more effective if the system has reserves or buffers, such as a fair amount of fat … or provision in the refrigerator.

An effective cybernetic system also values power of *observation* (*input sensitivity* of its receptors) in order to better anticipate the environment. A blind animal in the wild will have poor chances to survive. Once it makes accurate observations, it should also be able to store past experiences and process present inputs through an intelligent mechanism (e.g. neural network) thereby displaying *knowledge* and *intelligence*.

The system should also be able to act with power, and sufficient energy through its effectors. An animal seeing a predator coming, but mute and memberless cripple (cruelly lacking effectors) will be useless and hopeless because unable to raise the alarm, fight or flight.

As intelligence and knowledge rise, the amount of possible actions and goals growths. It becomes more and more difficult to choose what to do. In our information overloaded society the problem is stringent and we need to have insight into our own goals or preferences to make our way. Two core cybernetic principles are needed to steer a complex system. First, *feedforward* which consists in a proactive anticipation of a future course of action. Unfortunately in a complex environment, such anticipation is not always possible, and systems are easier to regulate using *feedback* mechanisms.

Positive feedback helps to choose goals by avoiding a locked-in situation. For example, in the classical Buridan's ass dilemma, an ass is equally hungry and thirsty, has water and hay at his disposal, but is unable to choose which to drink or eat first, and tragically dies from hunger and thirst. Philosophers like to discuss this thought experiment but from a cognitive point of view, it is unlikely to happen. Indeed, the brain functions with positive feedback mechanisms, and if the ass looks one millisecond more at the water, it will trigger neural activation which will make the ass decide to drink first, and momentarily inhibit his hunger.

Cybernetic values provide a great addition to evolutionary values because they promote values which improve not only survival and reproduction but also control, observation, knowledge, intelligence and effective action. The richness and effectiveness of cybernetic values are impressive. A cybernetic philosophy of time could be:

> Learn from the past,
> Be in the present,
> Predict the future.

It means to acknowledge lessons from the past (feedback); to deal with the present in the present (effective action); and to anticipate to our best the future (feedforward and model making).

But let us keep a macroscopic perspective. Is there a general direction or a developmental pattern in evolution that could help to build a normative ethics? This is what we now examine.

### 10.3.2 Progress in Evolutionary Progress

The notion of evolutionary progress was probably an essential component of the Nazi policy. As Weikart (2009, 2) writes:

> Evolutionary ethics underlay or influenced almost every major feature of Nazi policy: eugenics (i.e., measures to improve human heredity, including compulsory sterilization), euthanasia, racism, population expansion, offensive warfare, and racial extermination. The drive to foster evolutionary progress—and to avoid biological degeneration— was fundamental to Hitler’s ideology and policies.

Weikart (2009, 16) adds that Hitler “thought that by killing certain people he could improve the moral stature of humanity. Thus he committed some of the worst atrocities in world history in the name of morality.” Weikart argues that Hitler's political program was actually based on a consistent and stable evolutionary ethics throughout his career. However, Weikart's work is highly controversial since he explicitly avoided political, social, psychological, and economic factors that may have played key roles in the post-Darwinian development of Nazi eugenics and racism (see e.g. Gooday, Kenneth G. Wilson, and Barsky 2008; Richards 2012; 2013 for critiques). What is more, Weikart's study is academically not neutral, as he is associated with the “Discovery Institute”, which promotes intelligent design and creationism. Furthermore, Hitler as a consistent evolutionary ethicist is a doubtful claim, since Hitler positions himself explicitly as a creationist, when he writes (Hitler 1939, 310) : “it was by the Will of God that men were made of a certain bodily shape, were given their natures and their faculties.”

Nevertheless, the notion of progress may have played a role in the nazi policy, and remains essential to define, debate and criticize in any evolutionary ethics.

Of course, Hitler and his contemporaries had a limited and wrong picture of evolution. Hitler's evolutionary picture is outdated. Today, all biologists would see it naive to strive towards the biological uniformity of the Aryan “race” instead of valuing diversity at the human species level; naïve to believe that moral characters are biologically inherited or naïve to think that the best way to improve humanity is by changing its genetic population. On the last point, evolutionary theorists agree that genetic change is slow compared to cultural change. Therefore, it makes much more sense to attempt to change culture by fostering creativity, innovation and education rather than changing genomes.

But there is another assumption in Hitler's view, which is that competition is *the* driving force of evolution, and that it will help to give it a hand. Since nature is cruel and involves a merciless struggle for survival, should not a political program based on evolutionary ethics also promote such values and mechanisms? Of course we saw that this view is quite limited and the modern evolutionary picture is different.

So, what is progress? The idea of evolutionary progress is very controversial. We find both ardent proponents that evolution is a random process with no preferred direction (see e.g. Williams 1966; Hull 1988; Gould 1996); as well as arguments that there is evolutionary progress (see e.g. Dawkins 2006; 2003, chap. 5.4; Stewart 1997; 2000; Wright 2000; Corning 2003; McShea and Brandon 2010).

There is a very intuitive argument about evolutionary progress through time. A single cell is less complex than a multicellular fish, itself less complex than a human being composed of trillions of cells, capable to adapt to any climate and to surf the web he is weaving with millions of social partners. This is a straightforward argument that more and more complex organisms *did evolve* through time.

Depending on the definition of progress, we might conclude that the same empirical evidence involves progress or not. As Steven Pinker (1995, 332–333) illustrates, a wise elephant would reason that progress defined as nose elongation is rare in the animal kingdom. According to this definition, elephants are arguably the most advanced species on Earth.

We shall rather use a broad yet non trivial definition of progress proposed by Richard Dawkins (2003, 208):

> a tendency for lineages to improve cumulatively their adaptive fit to their particular way of life, by increasing the numbers of features which combine together in adaptive complexes.

This definition is not anthropocentric in the sense that it does not focus on any particularly human train such as brain size. Following this definition, there is a basic evolutionary progress thanks to a *ratchet effect*, where natural selection keeps small gains. Let us mention a few other progressive directions of evolution.

In the modern evolutionary picture, evolution is based not only on competition, but also on cooperation. Cooperation is to be understood as competition through cooperation, in the spirit of tit-for-tat, not a kind of gullible golden rule. The point is that what cooperates eventually outcompetes others, so evolution can't go in any direction. It will go in the direction of greater cooperation... or stop (Stewart 2000)!

A cybernetic analysis of evolution let us see clearly that it involves higher and higher hierarchical control levels, or metasystem transitions (Turchin 1977). Evolutionary progress leads to an increase in complexity of higher and higher control systems working together.

We also mentioned the irreversible *functional increase of complexity* (section 10.2.3 Stability – Adaptability Trade-off, p276). It means that evolving systems will be fitter if they are able to perform more useful functions. There is a cybernetic reason why this provides an evolutionary advantage, which can be explained by Ashby's law of requisite variety (Ashby 1956). Indeed, possessing a wide variety of functions allows to deal with more external perturbations, and thus makes the system fitter. Importantly, there is no anthropocentric agenda behind such arguments. As Heylighen (1999) explains,

> This preferred direction must not be mistaken for a preordained course that evolution has to follow. Though systems can be absolutely ordered by their functional complexity, the resulting relation is not a linear order but a *partial order*: in general, it is not possible to determine which of two arbitrarily chosen systems is most functionally complex. For example, there is no absolute way in which one can decide whether a system that can survive situations A, B and C is more or less complex or fit than a system that can survive C, D and E. Yet, one can state that both systems are absolutely less fit than a system that can survive all A, B, C, D and E. Mathematically, such a partial order can be defined by the inclusion relation operating on the set of all sets of situations or perturbations that the system can survive. This also implies that there are many, mutually incomparable ways in which a system can increase its absolute fitness. For example, the first mentioned system might add either D or E to the set of situations it can cope with. The number of possibilities is infinite. This leaves evolution wholly unpredictable and open-ended.

Another general evolutionary trend can be studied through thermodynamics. Stanley Salthe (1993), Schneider and Dorian Sagan (2005), all argued for a preferred thermodynamical direction to the universe. Eric Chaisson (2001) showed empirically that physical, biological and technological systems increase their capacity to sustain higher and higher flows of energy per mass. Importantly, Schneider and Kay (1994) reviewed arguments that more developed ecosystems degrade more energy. They suggest this can be extended to evolution at large, where species and ecosystems with the highest global dissipation rate are selected.

Granting those arguments, is there a preferential direction through which evolution goes? Let us now examine the sister concept of evolution, *development*. We will explore its application for ethics, starting with humans, society and speculating to a possible extension at a universal scale.

### 10.3.3 Developmental Values for Humans

Let us now turn to humans. Is there a developmental path for humans, as there is for the embryo? The embryo indeed develops from a single fertilized cell to a trillion-cells organized human being. But its development doesn't stop at birth. Our children go to school, and adults value continuous education. Does our development really stops after school or university? Our world is changing at an accelerating pace, and in more and more jobs it is necessary to always learn new knowledge and know-how. What are the ultimate moral and cognitive developments that intelligent beings could achieve in this universe? What are the highest stages of knowledge, intelligence and morality?

To approach these ambitious questions we need to extrapolate some insights of *developmental psychology*. Robert Kegan (1982) did synthesize the common ground of several developmental theories, by noticing how the notions of subject and object evolve at each developmental stage (see table 19, p290 below). His core insight was to notice that *the subject of one stage becomes the object in the next stage*. This process happens recursively. For example in stage 0, the subject in a baby between 6 months and 2 years *is* his reflexes –such as sensing or moving. Later at stage 1, he *has* reflexes as an object of control, and his subject is something new, impulses and perceptions. At stage 2, impulses and perceptions become the object of the individual in stage 3, and so on. Such a recursive growth is not without resemblance with the general evolutionary cybernetic view of Valentin Turchin (1977), and it is an promising research program to study developmental psychology with the thread of “self metasystem transitions”.

The recursive aspect of this theory makes its extrapolation to unknown future stages of development possible. I have attempted such extrapolations to stages 7, 8 and 9 which I call *evolutionary*, *cosmological* and *infinite*. Obviously those stages are not in any way supported by empirical psychological research, like the others are. But they open the door for humanity and its successor to develop.

Cognitively and logically, the evolutionary stage 7 involves awareness of the importance and power of evolutionary mechanisms. The highest logic is not just a static classical logic, with the scientific hypothetico-deductive attitude (as it was in Piaget 1954). The logic includes this attitude, but adds more dynamical reasonings, using principles from complexity theory and evolution. Otto Laske (2006; 2008) did argue that adults can be helped to develop such higher levels of cognitive capabilities. In particular he summarized 28 thought forms regarding process, context, relationship and transformation of systems. Morally, Kohlberg's (1981) classical work focuses on moral principles to make a good society, but the evolutionary stage goes further. This stage further involves a detachment from the human species, which is indeed but one species in the living world. There is an awareness that species have a limited duration. This does not exclude that humans could participate in higher cooperative wholes in the future, like bacteria cooperated with cells to form mitochondria or with our guts to help us digest, but there is no reason why humans would stay the most advanced. Psychologically, in this stage, the individual is ready to change, to transcend its self in order to also embrace larger evolutionary systems.

What could possibly come after the evolutionary stage? I propose the *cosmological* stage 8, where the subject becomes something greater than dynamical evolutionary systems on Earth. He identifies with the universe, that is, the greater context in which the evolutionary dynamics takes place. If the subject is the universe, the object on which he acts are evolving and developing systems of all kinds. Morally, truly universal issues are of central concern, such as the red giant phase of the Sun, the sustainability of new star generation in the galaxy and ultimately a lurking cosmic doom. It is hard to imagine a stage beyond the cosmological, but there is one, that I propose to call the *infinite* stage.

At stage 9, individuals are concerned with infinity. Maybe the mental attitude at this stage can be approached by James Carse's (1987) unique book *Finite and Infinite Games: A Vision of Life as Play and Possibility*. But we should add the lessons of the previous stage, and integrate insights from a cosmic culture. So the individual realizes that our universe is something temporary, subject to death. To go beyond, one must identify with something greater than the universe. Let us illustrate the psychology of this stage with the metaphysics behind Cosmological Artificial Selection (CAS, see Chapter 8). At this stage, the ultimate good is *the infinite continuation of the evolutionary process*. It is the thesis of this chapter, and now we understand that what matters most are *processes* which sustain life at large and our universe, rather that the universe conceptualized as a *static object*. At stage 8, the subject *was* the universe but in stage 9, he *has* the universe. The subject is aware and worried about the importance of the infinite continuation of cosmic reproduction. The universe becomes the object of the subject. He is no more attached to our particular universe. In CAS, what matters is the *recursively fertile infinite production of intelligent universes*. In other words, to make sure that universes continue to evolve and reproduce with intelligence, even beyond our own universe. Thus artificial cosmogenesis becomes a primary way to cognize, not only to decipher to what extent our universe is robust or fine-tuned, but to produce an artificial blueprint of a fertile offspring universe.

This stage maybe the penultimate one, because the next one would simply close the developmental logic, the subject and the object becoming the same. Why should universe reproduction be infinite? The issue here is to achieve a cosmological immortality of the universal reproduction process (see also section 10.4.5 Cosmological Immortality, p304). So reproduction is not just about producing *one* new universe but to generalize to a *fertile* reproduction mechanism, that is, to *n* generations as *n* tends towards infinity. This particular universe reproduction is just one step in the infinite cycle of evolving universes, like having one child is one small step in the chain of humanity. A central motto of infinite psychological beings is “L'infini et l'autre continue”[15].

Again, it is totally impossible to worry or to seriously consider stages 7, 8 and 9 without a strong evolutionary and cosmic culture. It was also totally impossible to worry about climate change or global economy a few thousand years ago. So, it means that those stages can only develop if our awareness extends beyond human societies to include the whole of evolution on Earth (stage 7), to cosmic evolution (stage 8) and to the infinite continuation of cosmic evolutionary processes (stage 9).

15 Again, an untranslatable french word game that I attempt to translate. “L'infini” means “the infinite”, but sounds like “L'un fini”, which literally means “The one finishes”; while “et l'autre continue” means “and the other continues”. So, we could translate (loosing the word game with infinity): “While one finishes, the other continues”.

| STAGES -> | 0 – Incorporative (0,5 – 2 years) | 1 – Impulsive (5 - 7 years) | 2 – Imperial (12 – 16 years) | 3 – Interpersonal (no age) | 4 – Institutional (no age) |
|---|---|---|---|---|---|
| CULTURE | Mothering | Parenting | Role-recognizing | Mutuality | Self-authorship |
| SUBJECT | Reflexes (sensing, moving) | Impulses, perceptions | Needs, interests, wishes | The interpersonal mutuality | Authorship, identity, psychic administration, ideology |
| OBJECT | None | Reflexes (sensing, moving) | Impulses, perceptions | Needs, interests, wishes | The interpersonal mutuality |
| PIAGET (logic) | None | Preoperational | Concrete operational | Early formal operations | Formal operations (dichotomizing) |
| KOHLBERG (moral) | None | Punishment – Reward | Instrumental hedonism (reciprocal fairness) | Interpersonal expectations and concordance | Societal perspective, reflective relativism or class-biased universalism |
| | | | | | |
| **STAGES** | **5 – 6 Interindividual (no age)** | **7 – Evolutionary (?) (no age)** | **8 – Cosmological (?) (no age)** | **9 – Infinite (?) (no age)** | |
| CULTURE | Intimacy | Evolutionary | Cosmic | Infinite | |
| SUBJECT | Interindividuality, interpenetrability of self systems | Interpenetrability of all systems, evolutionary developmental dynamics | The universe | Recursively fertile infinite production of intelligent universes | |
| OBJECT | Authorship, identity, psychic administration, ideology | Interindividuality, interpenetrability of self systems | Interpenetrability of all systems, evolutionary developmental dynamics | The universe | |
| PIAGET (logic) | Formal operation (5- dialectical; 6- Synthetic) | Evolutionary Developmental logic and dynamic | Evolutionary Developmental logic and dynamic applied to the universe | Infinite and recursive | |
| KOHLBERG (moral) | 5- Prior to society, principled higher law (universal and critical); 6- Loyalty to being | Loyalty to evolution and cooperation | Care for the universe and worry about cosmic doom | Loyalty to infinity | |

Table 19 Stages of cognitive and moral development. Inspired and adapted from (Kegan 1982, 86–87). Kegan stops at stage 5-6, while Kohlberg wrote just one paper about the possibility of a 7$^{th}$ stage, which he calls “cosmic” but that we call here evolutionary (Kohlberg and Power 1981). The content of stages 7, 8 and 9 are my proposed speculative extrapolations.

### 10.3.4 Developmental Values for Societies

Jean Gebser's (1986) remarkable book *The Ever-Present Origin* analyzed and documented developmental path for human societies. He analyzes five kinds of societal structures, from archaic, magic, mythical, mental to integral. Each time, the conceptions of space-time, the notions of signs, essence, properties, potentiality, emphasis, consciousness, forms of manifestation, the agency of energy, the organ emphasis, the forms of realization and thought, of expression, assertion or articulation, the relationships, the localization of the soul, the forms of bond or the general motto change.

But there are also much more empirical arguments that societies follow developmental trends. The large scale study of the world values survey shows that secular-rational values instead of traditional values, as well as self-expression values instead of survival values increase through time (R. Inglehart and Welzel 2005). The study shows that countries tend to develop towards to upper right region of the space, where a snapshot in the period 1999-2004 is represented in Figure 29 below.

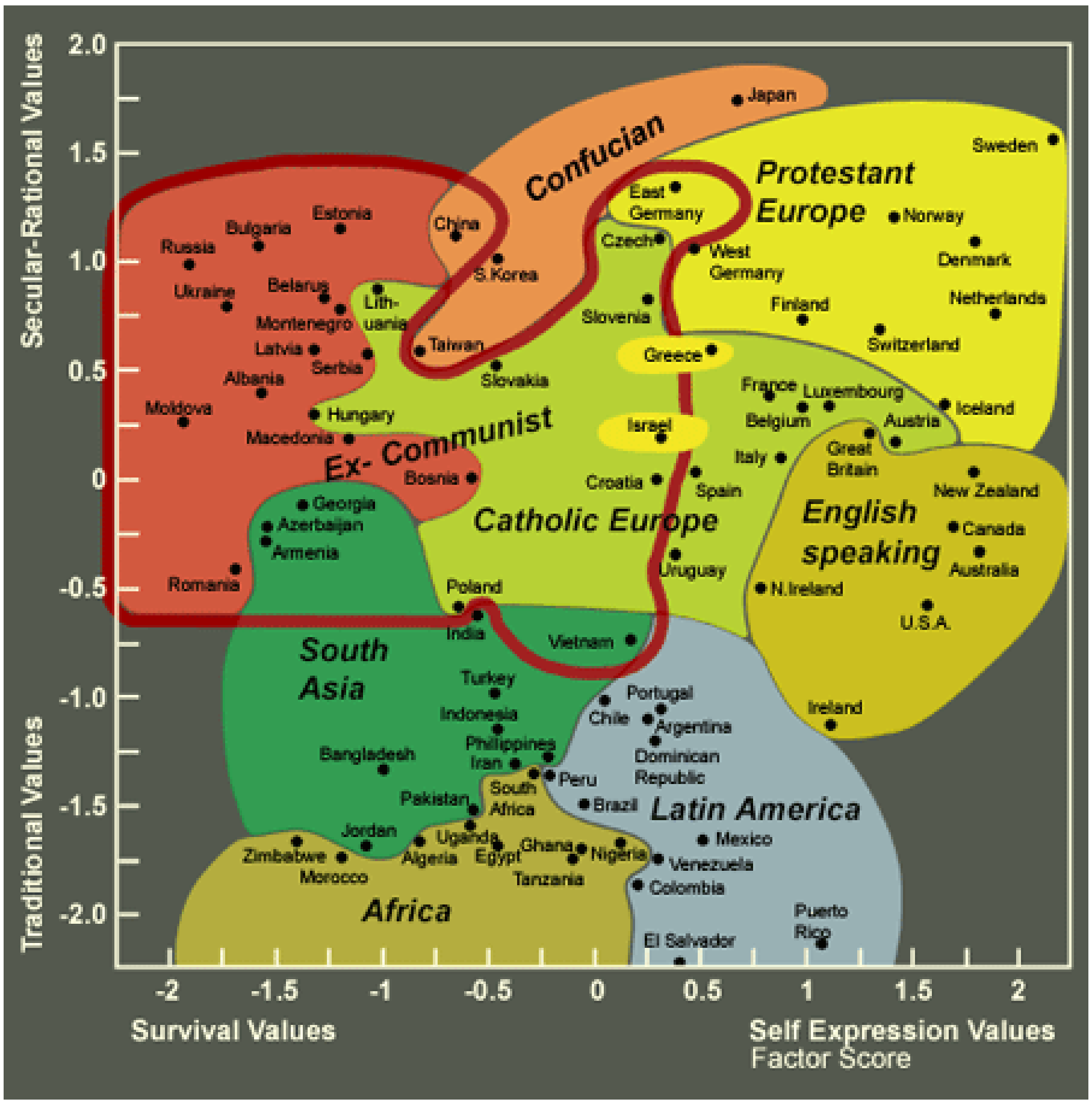

Figure 30:
Figure 29 The World Value Survey Cultural Map 1999-2004. From (R. Inglehart and Welzel 2005, 63)

Steven Pinker in his recent and extensive (2011) book argued at length that there is a decrease in violence in societies, a developmental trend valid on time scales from millennia to years. Paul Hawken (2008) also showed the rise and growing impact of nonprofit groups defending ecological sustainability and social justice. Another factor which increases is the quality of life –or happiness (see e.g. Heylighen and Bernheim 2000; Ronald Inglehart et al. 2008).

So, where is the development of our technological society headed? We already mentioned that a case can be made for the metaphor of a global brain as a next evolutionary transition, fostering cooperation at a planetary scale (see e.g. P. Russell 1982; Mayer-Kress and Barczys 1995; Heylighen 2005). This would also eventually make a new level of intelligence emerge from humans and computers interacting on the internet.

What are the next evolutionary transitions after the global brain? The starivore hypothesis (Chapter 9) provides a long-term vision, from switching all our energy source to solar, to migrating the Earth nearer to the Sun to get more of its energy, up to black hole accretion and universe making.

### 10.3.5 Universal Thermodynamical Evolutionary Developmental Values?

Can we extrapolate psychological and societal development up to universal development? John Smart (2009) already saw the emergence of chemical elements as a progressive differentiation or development through cosmic evolution. We already mentioned the open question of whether there is what Teilhard de Chardin's (1959, 78) called a “cosmic embryogenesis” (see the Open Questions of Chapter 7, p163). Is there a development at play in the universe, like the development of an embryo? We might accept a developmental path for chemical elements, for the child going towards adulthood, but are we ready to accept it for our psychology, societies or our technologies? The debate is likely to remain lively. These are open question I invite you to explore within the “Evo Devo Universe” research community I co-founded with John Smart.

To sum up, evolutionary values are extremely valuable because pragmatically useful. However, we saw that they are rather short term, contextual, adaptive and all purpose. For these reasons they are more practical values, and good basis to explain the source of our behavior, or *descriptive ethics*.

Developmental values on the contrary focus on long term growth. They provide direction for humans to develop towards higher stages of cognitive, social, and emotional functioning; direction for societies to develop towards less violence, greater cooperation and integration, towards a peaceful global society. Such developmental values are thus more promising to develop a *normative ethics*. Without direction, it is hard if not impossible to propose a normative ethics.

To sum up, ethical principles go toward the survival of more and more cooperative, hierarchically complex systems, which use more and more efficiently free energy. But for how long? Isn't the energy of the universe ultimately limited and bond to dissipate towards a heat death? Can we still hope in this cosmological context for a kind of *immortality*? This is what we explore now.

## 10.4 Voyage to Five Immortalities

*The mortality of man defines by contrast the immortality which some men hope for, some men fear, some men scoff at, but no man ever fails sooner or later to consider.*

(Adler 1990, 606)

*All living things seek to perpetuate themselves into the future, but humans seek to perpetuate themselves forever. This seeking –this will to immortality– is the foundation of human achievement; it is the wellspring of religion, the muse of philosophy, the architect of our cities and the impulse behind the arts. It is embedded in our very nature, and its result is what we know as civilization.*

(Cave 2012, 2)

Why do we die? Do we have to die? Does everything has to die? Should we accept death as a blunt fact? This would still leave open the question of how or why we should learn to accept death. Stephen Cave (2012) articulated the *mortality paradox* which goes as follows. *Objectively*, we see death everywhere and know for sure that we will die; but *subjectively*, we cannot conceive death because it is beyond any experience we have. So we tend to accept the former and reject the latter, which is an inconsistency in our worldview, where the *it-I* test (see section 2.3.3 Testing the Big Three, p49) fails. In other words, the subjective world is in contradiction with the objective world.

As usual in our approach, we like to apply concepts to themselves. The death of death leads to the fascinating idea of *immortality*. However the situation is logically peculiar here, because the death of death also annihilates the very concept it was applied to. Immortality means that there is no death anymore, whereas doing philosophy of philosophy or history of history does not destroy philosophy or history.

The desire for immortality contrasts with the fear of death. Implicitly or explicitly this ultimate fear of death triggers a conscious or unconscious desire for immortality. However, outside religions, topics of death and immortality are often taboo. Immortality is seldom found separate from belief in a supernatural order. Can we develop an idea of immortality without supernatural order? This is the axis we focus on now.

Awareness of mortality is not obvious. As Jorge Luis Borges put it in his 1949 novel *The Immortal*, "with the exception of mankind, all creatures are immortal, for they know nothing of death" (Borges 2004, 13). Only few animals seem to understand death, and even children lack the idea of death. Their discovery of mortality leads to great anxiety (see e.g. Yalom 1980). This means that the idea of immortality requires awareness of mortality! This is not so trivial as it seems and suggests that at different cognitive and moral stages of human development, we have different ideas of death and immortality.

For example, what about you? Which death do you care about? Human, creative, evolutionary or cosmological death? To clarify what I mean by these different deaths, let me tell you a fictional story. Imagine that you are sitting in a bus, and the person next to you starts to have difficulty breathing, makes an epilepsy crisis, and is imminently going to die. To your astonishment, nobody in the bus cares about that person. Nobody tries to do something, ask if there is a doctor in the bus, call an ambulance, try to stop the bus or to do any other emergency procedure. This horrible situation is barely imaginable, because humans are social creatures able of *empathy*. We care about human death and won't let a fellow human die.

Now imagine that you watch the news, and you learn that the government of your country has decided to abolish human rights. What would you do? Would you protest, mobilize, sign petitions, go demonstrate in the street? It would make sense because human rights are values inherited through a lot of social fight. Many people would do it, but you'll certainly find less people reacting than in the first example. Indeed, letting an idea die has not the same psychological impact as letting a human being die. Probably even less people will be sensitive to the destruction of a work of art through an act of vandalism. This is because the empathy we have with fellow humans is an instinct, while if you leave in a small village in China you might never have heard about human rights or strange western forms of art. Empathy for ideas requires education and culture.

Let us continue. Do you care about a form of evolutionary death, where biological diversity dramatically drops, in the end disturbing or destroying the planetary ecosystem? Maybe if you have seen the destruction of the amazonian forest, seen some worrying television documentary or read some book about such phenomena. But my guess is that even less people do care, or at least used to care a few decades ago. The good news is that today, our awareness more and more rapidly reaches these kind of global issues.

Now, who cares about cosmological death? How often do you worry about the predictable death of our Sun which will already make oceans boil in one billion years? And why stay solar-system-centric? What about other stars? And the health of the galaxy? And the ultimate death of our Universe? Almost nobody cares anymore, only a tiny percentage of researchers which I can count on my fingers.

Of course, to be motivated to fight against each kind of death requires each time a different idea of immortality. We will use and extrapolate developmental psychology and suggest that there are *stages of immortality*. This idea would be interesting to test empirically, but here we stick to setting the theoretical framework. We hypothesize that the definition of the self is critical in how we apprehend death and immortality. Do I identify with the boundary delineated by my skin, my family, my community, my nation, mankind, the Earth, the solar system, the galaxy or even the universe?

The idea of the self rarely extends beyond a nation. Yet we are living an exceptional age in which the self of humanity is emerging. But the way to a cosmic culture where the self is identified with the universe is still in its infancy. Yet it is the logical development of our psychology, and sooner or later, it will emerge.

Let us now start our voyage to five kinds of immortalities, in two parts. The first has to do with *personal immortalities* where the person longs for an afterlife or hopes to live forever. I call them respectively *spiritual* and *individual* immortality.

The second part deals with *transpersonal immortalities*, where the person longs for an immortality beyond the self. We will see three such immortalities that I call *creative*, *evolutionary* and *cosmological*. Through this voyage we hope to convey cosmological immortality as an ultimate value to interpret human's ultimate fear, death.

### 10.4.1 Spiritual Immortality

Spiritual immortality is the belief in a magical or supernatural realm where the soul goes after death. The idea of an afterlife is present in the major world religions like Christianity, Islam or Hinduism. Of course, the definition above assumes old-fashioned religious interpretations such as the separability of body and soul. We can distinguish a literal belief in the existence of an afterlife from a need to symbolize immortality as an answer to death (Lifton and Olson 2004). Spiritual immortality thus assumes the immortality of the soul.

This belief is quite widespread since, as Dawkins (2008, 356) reports, 95 per cent of the population of the United States believe they will survive their own death. This is not so surprising because the idea of immortality is a point-cognitive attractor for the future. It is appealing for our mind. In his recent book about immortality, Stephen Cave (2012) argued that there are only four narratives around immortality that culture have been transmitting. He calls this first belief the *Soul narrative* of immortality.

Of course, we find also the cyclic cognitive attractor, where death is part of a reincarnation cycle. For example, in the Buddhist tradition, this involves the belief that you have already lived, that you will die but be reborn, over and over again. This happens according to the quality of a person's actions. Cave calls this cyclic perspective on death the *Resurrection narrative*.

Now, what does the rational materialistic mind think about these ideas? In 1925, Bertrand Russell (2004, 7) famously wrote: “I believe that when I die I shall rot, and nothing of my ego will survive”. The critique of spiritual immortality argues that after the body dies, there is nothing. Being dead will be no different from being unborn and anyway, there is no proof of an afterlife. We need to find and define new kinds of immortalities. Can we use and extrapolate our knowledge of science and technology to achieve a materialistic and naturalistic immortality?

### 10.4.2 Individual Immortality

*I don't want to achieve immortality through my work,*
*I want to achieve it through not dying.*

Woody Allen.

Woody Allen made his will for individual immortality crystal clear. Individual (or personal) immortality is the continuation of life in a biological or digital substrate. Cave calls this third option the immortality narrative of *staying alive*. He argues that “almost all cultures contain legends of sages, golden-age heroes or remote peasants who discovered the secret to defeating aging and death" (Cave 2012, 4). Our civilization is no exception. As Cave (2012, 283) puts it:

> The difference between those who swallow 250 dietary supplements per day and the rest of us is not that they will live forever and we will not. No: we will all die, even the transhumanists. The difference is that they tell themselves a story about achieving "longevity escape velocity," which helps them to alleviate their existential angst. They are therefore following a long tradition of elixir seekers, resurrectionists, reincarnationists and others who have attempted to deny the fact of death.

What is the motivation of individual immortality? It is simply the refusal of one's personal death. Death is absurd. Indeed, as John E. Stewart (2000, 316) writes, from "the point of view of an isolated individual, the inevitability of death makes anything he does during his life irrelevant and meaningless". An evolutionary informed argument even goes as follows. Biological death was acceptable when evolution was primarily genetic, so that information could be preserved in the genes of offspring. But in our culture-driven society, most of the information we gather during our lifetime is cultural and gets lost at the time of death. And this is pure waste.

There are two ways to solve this problem. Either to drastically extend our life span (individual biological immortality) or to transfer our "self" in a digital substrate (individual cybernetic immortality).

Committed researchers in life extension are aware that true immortality (infinite lifespan) is not realistic. This is why they often use the more careful expression of "indefinite lifespan". The objective here is to stop all aging processes and thereby live for extremely long timespans (see e.g. Kyriazis 2005). One strategy De Grey (2007, 43) proposes is to develop rejuvenation therapies acting on the seven major classes of cellular and molecular damage, in order to stop them.

Other approaches include cryonics or plastination (see e.g. Olson 1988), whose philosophy is to conserve now, and pray that future science will be able to resurrect later.

We can make an even stronger theoretical case against death. Indeed, if we remember the very general definition of life as a thermodynamically open system, there is no necessity for a living system to die, as long as it can process an energy flow and extrude waste products. There are even examples of plants or animals that some scientists speculate to be immortal. For example, the *bristlecone pine* "Methuselah" had his 4723$^{rd}$ birthday already in 1957 (Stern, Bidlack, and Jansky 2010, 87). We could indeed speculate that such a tree, already living for millennia, could continue indefinitely. The speculation of an immortal animal is no less fascinating. The jellyfish *turritopsis nutricula* can reverse its life cycle and may have an indefinite lifespan. These medusae can perform an ontogeny reversal where they "rearrange their tissues and go back to the polyp stage if subjected to sub-lethal stress and also at the end of their lifespan, after spawning" (Grzimek 2004, 1:139) .

To sum up, as long as we can find free energy in the universe, we can hope to live forever. A cosmological perspective brings two critiques however. First, even assuming this tree and jellyfish *will live forever in the future*, they did not *live forever in the past*, since their atoms were cooked up during the big bang era and in stars. The second critique is that there are limits of available free energy if we aim to live truly forever. Modern physical eschatology teaches us that (see Part III).

Another approach to individually live forever is to bet on the digital substrate. This is called cybernetic immortality (Turchin 1990), mind uploading, digital immortality, virtual immortality or immortality in silico. Imagine you can upload all the information which constitutes your self in a computer. You could then continue to live in a digital world, for as long as your computer hardware can run without failure. But mixed approaches are also conceivable. For example, if future biotechnologies can grow a clone of yourself, then you could re-upload the content of your brain or what ever constitute your "self" in this new you. So it seems storing your self in a digital substrate is a more versatile strategy than just focusing on maintaining your biological substrate. The ethical side of such practices remains of course to be debated.

There are many critiques to these modern yearnings for individual immortality. Let us mention a few. First, the body is embodied (see e.g. Clark 1998). Cybernetic immortality assumes that preserving brain patterns is the only challenge. Yet, it is not clear if memories, thoughts, emotions or consciousness can be reconstructed without a body embedded in an environment. Additionally, if you resurrect a neural pattern in 200 years, all the context will be lost, and the resurrected brain has great chances to be totally maladapted. A possible way out is to run your preserved self in a simulation of your time. But this means you need to also preserve your environment, and not only your brain. And what's the point of being resurrected in a fake world?

A second critique is that the body cognizes. For example, we know that the heart is actually full of neurons, which influence intuitive decision making (see e.g. McCraty, Atkinson, and Bradley 2004). So to be on the safe side, individual immortalists would be better off to preserve the whole body and not only the brain. A computational analogy teaches us that to resurrect a 20 years old microprocessor or harddrive (rough analogs of the brain), you need a compatible motherboard, cables, other chips and software. Otherwise, it's useless.

A third critique of immortality is its futility. In a private communication, my colleague Carlos Gershenson explained that the old alchemic dream of getting gold from lead is now possible thanks to particle accelerators, but actually more expensive than mining gold. The situation with personal immortality will most likely be similar. Once we will be able to achieve it, it will cost so much that we will finally realize its futility.

Does death contribute to the meaning of life? A common critique is that immortality undermines motivation. Why do things now if you have eternity in front of you? There would be no urgency.

Is individual immortality a societal cancer? The analog of an immortal entity in biology is a cancer cell which has gone wild because it will not die anymore. The daily 50 to 70 billions cells dying in our body (Karam and Hsieh 2009, 27) is essential for our bodies to function. If we see society as a superorganism, we should definitely stop individuals wanting to become immortal! Are scientists and futurists passionate about individual immortality working hard towards growing a societal cancer? Maybe.

But there is a brighter option. Imagine you have achieved individual immortality. Your ethics broadens radically, because the time scope of your life has changed radically. It then really makes sense for you to care about far future issues, such as climate change, the red giant phase of our Sun or the heath death of the universe. You know that sooner or later, those issues will affect you.

So, individual immortality may actually be a necessary step for ethics to extend on very long time scales. People wouldn't be able to say anymore “climate change? I don't care because I know it will not affect me”. The same holds for cosmological issues. But could we care for greater things than one's self without living forever? Of course we can, and this is the basis of wisdom and heroism, higher ethical functioning where the individual is able to self-sacrifice for a greater whole (see e.g. Murphy and Ellis 1996). This sacrifice is in fact natural if you have transcended yourself, if you identify with something greater than the boundary of your skin. And this leads us to transpersonal immortalities. But win-win cooperation can also be explored, and sacrifice seen as the last option.

Transpersonal immortalities focus on the future of life beyond the self. They can be summarized with the motto: “Life after death exists. It is the one of others.” Life continues to exist not only through the existence of other human beings, but also other systems or cultural items like a nation, a work of art or a scientific achievement. Stephen Cave (2012) calls this fourth way of dealing with death the *legacy narratives*. Like other immortality narratives, they are present in human societies at least since Antiquity. Those worried about transpersonal immortality see beyond their selves, and this arguably corresponds to higher stages of development and needs. Indeed, in Maslow's (1954) hierarchy of needs, self-actualization is in fact not the latest need. In his later work, Maslow identified a new need beyond self-actualization, namely, self-transcendence (see Koltko-Rivera 2006 for a detailed account of Maslow’s conception of self-transcendence). It means that the self, having climbed the hierarchy of needs through satisfying physiological, safety, love, esteem needs, will not stay self-actualizing and satisfying his individual needs. The self will strive to extend its own boundaries, to identify with greater things. Koltko-Rivera summarizes that the individual now seeks “to further a cause beyond the self and to experience a communion beyond the boundaries of the self through peak experience”. A cause beyond the self “may involve service to others, devotion to an ideal (e.g., truth, art) or a cause (e.g., social justice, environmentalism, the pursuit of science, a religious faith), and/or a desire to be united with what is perceived as transcendent or divine”, while the communion beyond the self may “involve mystical experiences and certain experiences with nature, aesthetic experiences, sexual experiences, and/or other transpersonal experiences, in which the person experiences a sense of identity that transcends or extends beyond the personal self.”

Philosopher Bertrand Russell did actually not stop at the materialistic critique of spiritual immortality quoted above. In his essay *How to Grow Old* (1956, 52–53), he embraced the idea of a transpersonal immortality:

> [T]he fear of death is somewhat abject and ignoble. The best way to overcome it –so at least it seems to me– is to make your interests gradually wider and more impersonal, until bit by bit the walls of the ego recede, and your life becomes increasingly merged in the universal life. An individual human existence should be like a river –small at first, narrowly contained within its banks, and rushing passionately past boulders and over waterfalls. Gradually the river grows wider, the banks recede, the waters flow more quietly, and in the end, without any visible break, they become merged in the sea, and painlessly lose their individual being. The man who, in old age, can see his life in this way, will not suffer from the fear of death, since the things he cares for will continue. And if, with the decay of vitality, weariness increases, the thought of rest will be not unwelcome.

Psychologist Roy Baumeister, specialized in the notion of the self, wrote that the "most effective solution to this threat [of death] is to place one's life in some context that will outlast the self. If one's efforts are devoted to goals and values that project many generations into the future, then death does not undermine them" (Baumeister 1991, 292).

Notwithstanding Woody Allen's wish, in 50 years from now, I am more interested in watching again and again his great movies rather than shaking his shaky hand. Despite his personal preference for individual immortality, his creative work has an enormous value and is a stunning creative achievement. Can we shift from an idea of immortality reacting to the fear of individual death to an immortality expressing the love of life as a whole? This is what we will explore now through three kinds of transpersonal immortalities: *creative*, *evolutionary* and *cosmological*.

### 10.4.3 Creative Immortality

*unless there is an accepted structure into which each new finding can be fitted, the "immortality" of scientists' ideas will vanish.*

(J. G. Miller 1978, 5)

Remarkably, one of the deep motivations of James Grier Miller to develop living systems theory is to provide a conceptual framework to unify and maintain the legacy of scientific ideas. Miller is thus concerned with the noble mission of securing the creative legacy of the scientific enterprise.

Stephen Cave calls this creative immortality the path of *cultural legacy*. Of course, a cultural legacy can also be a work of art, a graffiti, an invention or a contribution to societal progress. Aristotle's, Shakespeare's or Darwin's achievements are all known despite their individual bodies not existing anymore. They are remembered because through their work they have offered a legacy for humanity, a creative immortality. As Baumeister (1991, 292) argued, to feel secure in facing death, "one must draw the value of one's actions from a religious, political, artistic, scientific, or other cause that transcends decades and even centuries". The motto of creative immortalists can be summarized by: "the more you give, the more you stay" (René Berger, cited in de Rosnay 2012, 221).

Another legacy is the sociological one. One may be motivated to pass on a family name; a material legacy to one's family. However, such a legacy is rather limited, similarly to kin selection in biology. The benefits of the legacy is restricted to a small group, not the whole of humanity or the whole of the tree of life.

However, the timescales of creative immortality are uncertain. Even Aristotle, Shakespeare or Darwin did make their legacy through "only" centuries or millennia. What about millions or billions of years? They might well be totally forgotten. Can we hope and aim for a kind of immortality which endures through thousands, millions or even billions of years?

### 10.4.4 Evolutionary Immortality

Evolutionary immortality is the striving to continue or maintain life by replication, or any other process in harmony with evolutionary principles. In the case of our particular biology, Stephen Cave calls it the *biological legacy* narrative, to be contrasted with the *cultural legacy* narrative (creative immortality).

There is immediately an objection here, namely that animals and adults at all stages of development do have children. So there is nothing particularly advanced or wise about evolutionary immortality. The most primitive sexual drive that we –or some?– share with animals leads us to achieve immortality through offspring. Even in some human tribes, the connection between sexuality and offspring is unknown. The situation is similar to the golden rule in developmental psychology. Kohlberg (1981) found consistently that it was held at all stages of moral development. What changes is thus the *justification* of the golden rule.

Similarly, what matters is the justification and *motivation* behind procreation. There is a continuum between unconscious procreation up to the most carefully planned in vitro fecundation for a homosexual couple. Is the motivation only the result of an instinctual drive? Of a social pressure? Of the fruit of love? Of a will to contribute to the tree of life? Here we consider only this last option, in line with a supposed evolutionary stage 7 of psychological development that we outlined earlier (section 10.3.3 Developmental Values for Humans, p287).

The evolutionary thinker will critique individual immortality. Arguments could go as follows. Most importantly, individual immortality freezes variation. Indeed, as we saw with evolutionary values (section 10.2.3 Stability – Adaptability Trade-off, p276), the stability-adaptability trade-off is essential for systems to dynamically adapt. Individual immortality means to get rid of adaptability, which can only lead to long-term extinction of the human species. Indeed, changes, exploration, variations are required to provide adaptation and innovation. Individual immortality is thus unfit biologically. Kirkwood (1999) did a thought experiment about a supposed immortal species and showed that it quickly becomes biologically unfit. Individual immortality is against evolutionary thinking. As John E. Stewart (2000, 320) writes,

> Individuals with restricted adaptability and evolvability would stall evolution if they live forever. For this reason, organisms who are not self-evolving but whose overriding objective is to contribute to the successful evolution of life would not attempt to achieve immortality.

Some studies even show that aging may have been selected for its own sake because genetic systems which limit their lifespan have been conserved over evolutionary time-scales (Mitteldorf 2004). Which means that organism death (e.g. with telomeric aging) might be as important as cellular death (apoptosis), but at the species level. Not without irony, it is in fact good news for those seeking individual immortality, since the possibility of a programmatic mechanism triggering death rises hopes to simply shut it down (see e.g. Bredesen 2004).

For someone seeking individual immortality, evolutionary immortality is not immortality anymore. Indeed, it involves the death of individuals, so how could this be immortality? The answer of evolutionary immortalists is twofolds. First, they have transcended their individual self, so they can accept to individually die. Second, evolution needs genetic and cultural change, that is death of old systems, in order to leave room for fitter new systems.

Evolutionarily, there is another major critique of individual immortality. Even if we assume that an immortal human being can evolve, learn new things, it is generally not conceived that she could go through major evolutionary transitions like the transition from non-living molecules to first cells, from cells to multicellular organisms or from plants to animals. An immortal human is similar to a plant which would have decided to hamper the emergence of animals and therefore to stall evolution. So, individual immortalists may be in a process of hampering the development of higher levels of evolution, and thus going against developmental values we outlined (section 10.3 Developmental Values, p284).

If we see life in a broader picture as the tree of life, the mechanism of reproduction ensures that life always continues to explore novel solutions. So death is beneficial relative to an individual (e.g. with apoptosis) or a species (with individual deaths). Maybe death could be beneficial on a yet larger scale of a whole (post)biosphere? For example on a cosmological scale, might a putative starivore civilization sacrifice itself to leave room for another more advanced civilization? That would probably mean billion of years of evolutionary trial-and-error wasted. But this would be no big news that nature can be wasteful.

Evolutionary immortalists are skeptic not only about individual immortality, but also about the immortality of the human species. Mark Lupisella (2009) coined the term "speciesism", which he defined as "a kind of blind, unethical delusion engendered by biologically driven affinities for one's own likeness". Evolutionary immortalists know too well that 99.9 % of species which have existed are now disappeared (see e.g. Prothero 2003, 83). There is no reason why humans should be an exception. Put an other way, there is 0.1% chance that humanity as a species will continue in the far future. In mammals, species average 1 million years of duration (May 1994, 15) and homo sapiens appeared only about 200 000 years ago. So, if paleobiology is of any help, we might still have a few hundred thousands years to live as a species. Which is very few from geological or cosmological perspectives. Note also that a species endures longer if the environment does not change, and might endure for shorter periods of time if the environment changes a lot. Currently, it would be hard to argue that life on Earth is not changing!

These arguments do not necessarily suggest that everything we do is meaningless and useless. It is not a reason to make the end of humanity a self-fulfilling prophecy and to accelerate our self-annihilation. It simply means again that we are not the center of cosmic evolution. Evolutionary processes often build scaffolding, which are useful for next generations to build on. To illustrate this, let us introduce an analogy which describes the evolutionary process, that ecologist Robert Ulanowicz (2009, 100–101) describes as his only "Eureka" experience in his life. Here we apply it to the particular branch of humanity. Consider a grapevine as in Figure 31 below. Suppose humanity is the central stem and is producing branches.

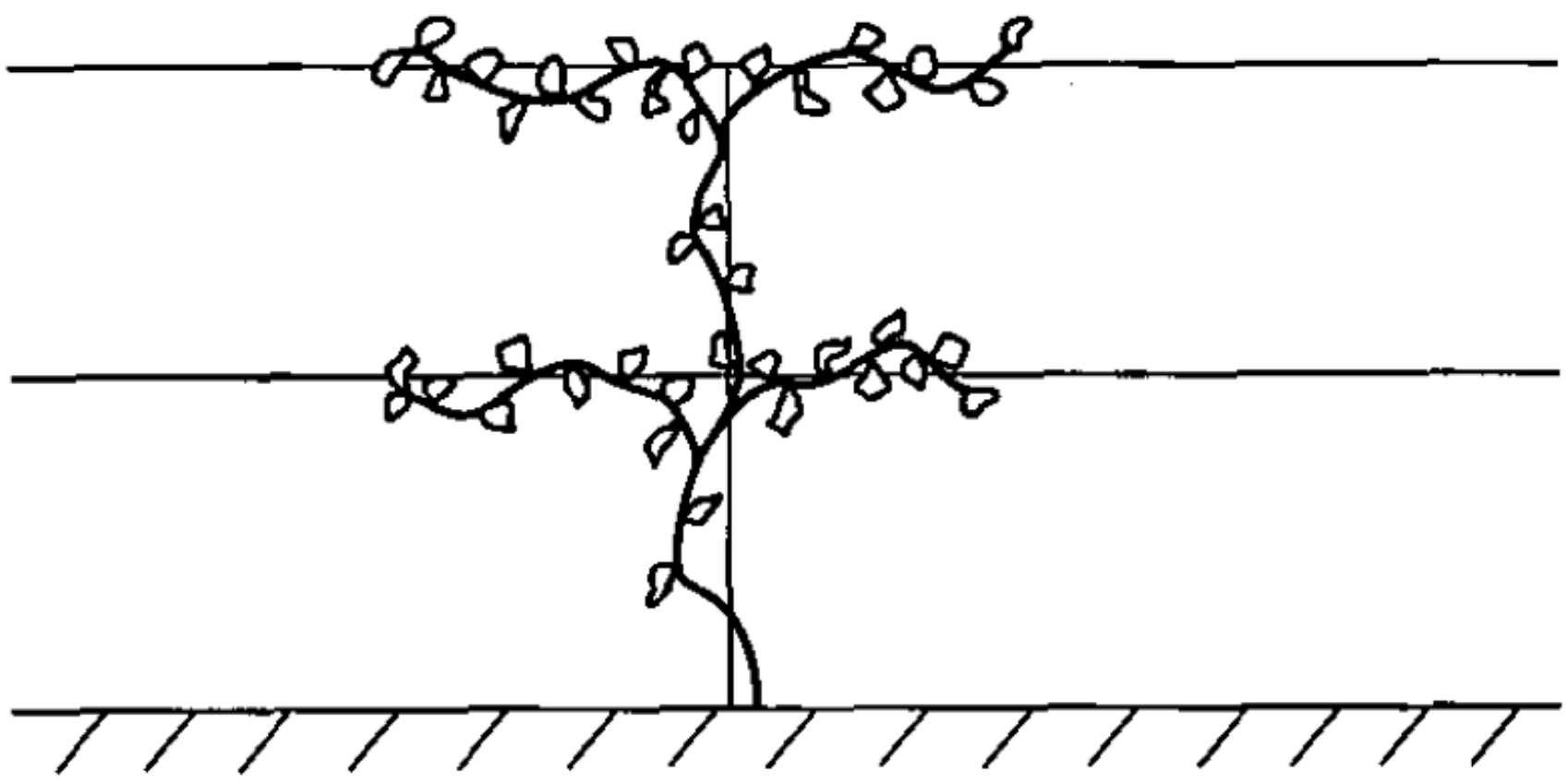

Figure 31 Young muscadine grapevine with central stem and branche. (From Ulanowicz 2009)

Those branches could be seen as our cultural and technological extensions, as the additional branches and roots of Figure 32.

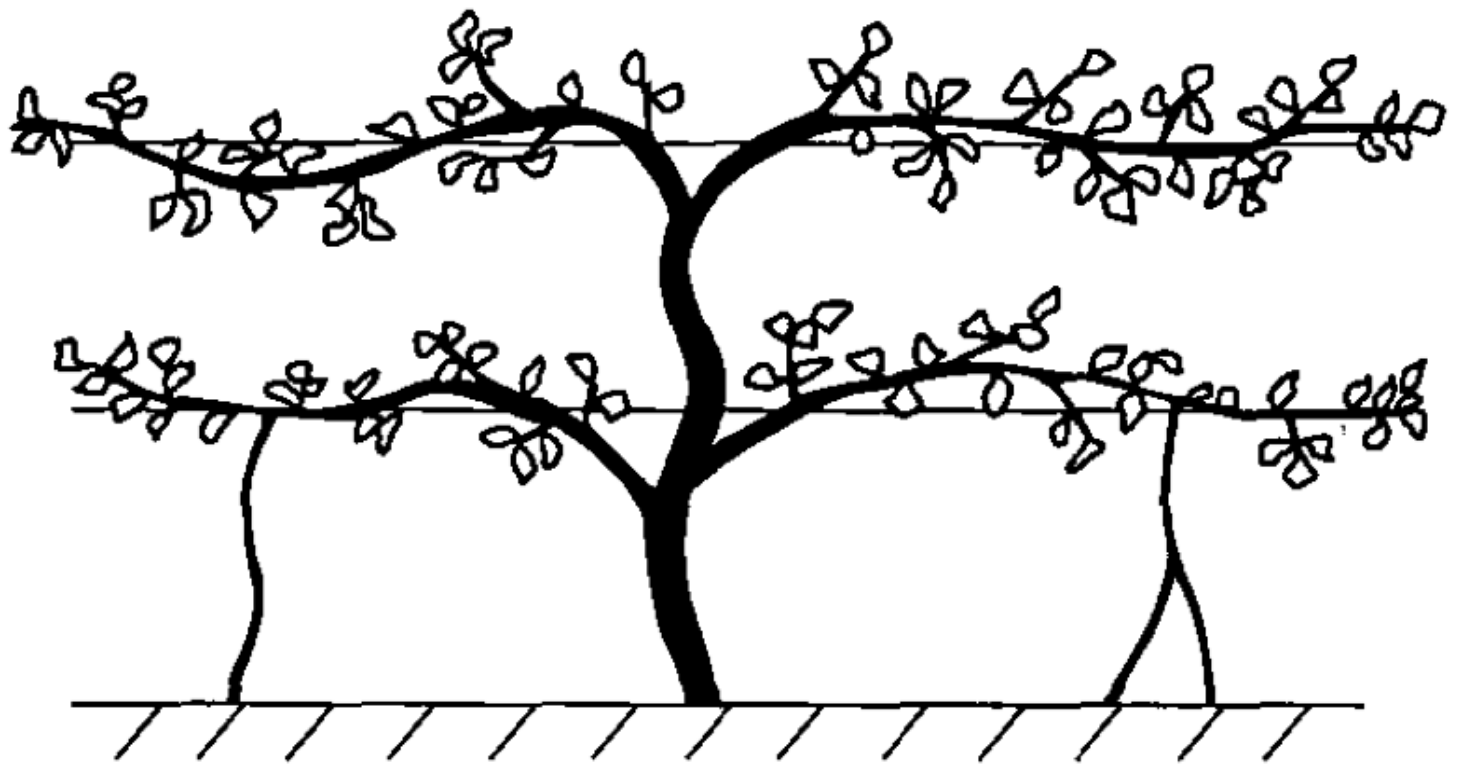

Figure 32 Grapevine several years later, having developed adventitious roots to the sides of the main trunk. (From Ulanowicz 2009)

Finally, at a third stage illustrated in Figure 33, the initial root –humanity– disappears. However, the grapevine has considerably grown since Figure 31, and could not have grown without the initial root.

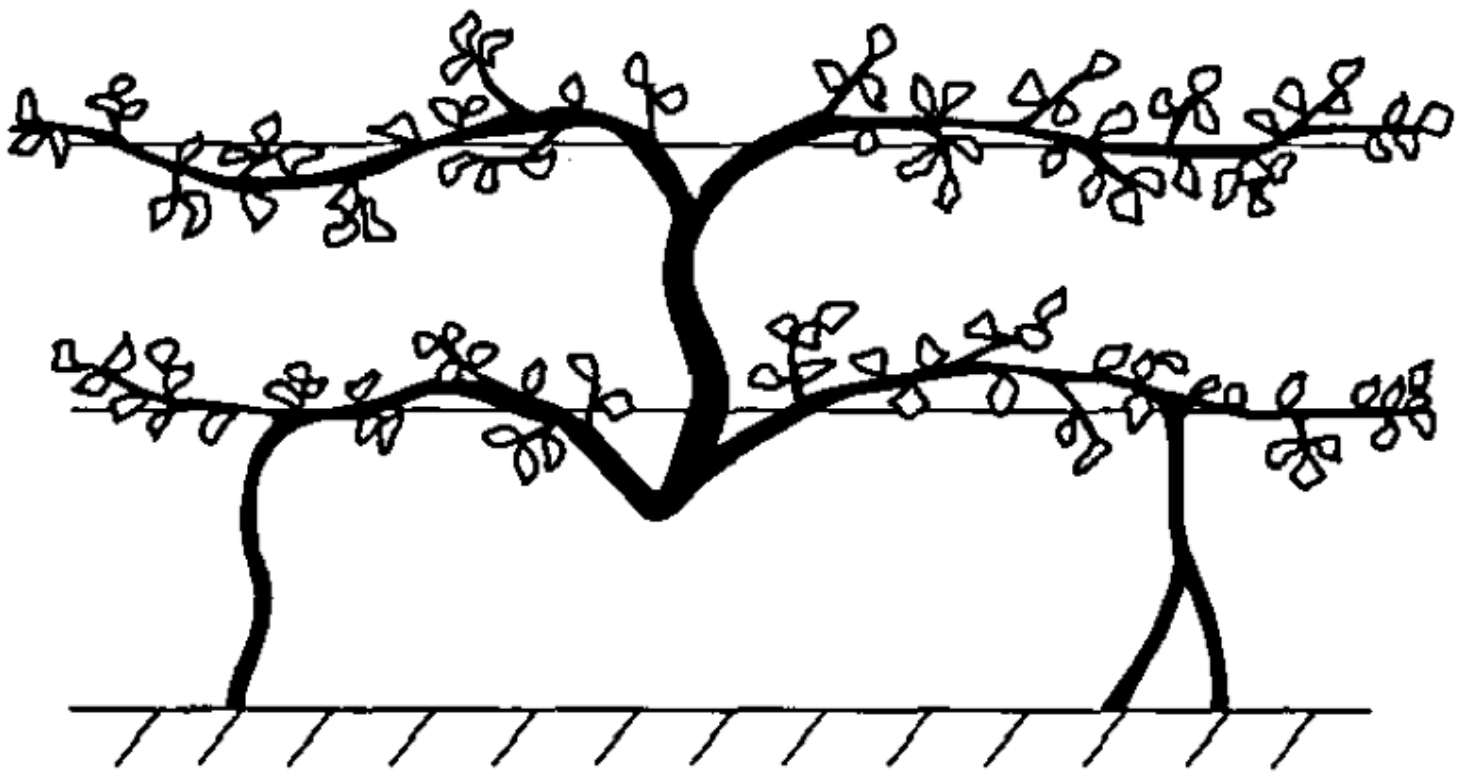

Figure 33 Same grapevine two decades later. Original trunk has rotted away, but vines are sustained by adventitious root system. (From Ulanowicz 2009)

To sum up this critique, those aware and endorsing evolutionary immortality will find the striving of individual immortality selfish, meaningless and potentially harmful to the future of evolution. It is our evolutionary responsibility to be ready and willing to *individually* die. "The secrets of evolution are death and time" famously wrote and said Carl Sagan (1985, 20). He further explained: "the deaths of enormous numbers of lifeforms that were imperfectly adapted to the environment; and time for a long succession of small mutations that were *by accident* adaptive; time for the slow accumulation of patterns of favorable mutations." If my "I", my self identity is transcended towards becoming greater than the biological self, it is no problem to die. Indeed, it is only a small part of me which dies, namely, its biological body part. But others continue to live; the influence I have had on Earth or my cultural or biological legacy will continue. And the Sun will continue to shine... if only for a few remaining billion years. Alan Segal (2004, 78) showed that traditionally, "wisdom and mortality are unconditionally wed." This also holds for evolutionary immortalists.

There are two broader evolutionary legacies, the legacy of genes and of the legacy of the global brain. Richard Dawkins (2006, 34) identified the genes as being the closest entity we could call immortal:

> The genes are the immortals, or rather, they are defined as genetic entities that come close to deserving the title. We, the individual survival machines in the world, can expect to live a few more decades. But the genes in the world have an expectation of life that must be measured not in decades but in thousands and millions of years.

The life expectancy at the level of genes is indeed impressive, but could we find something close to immortal at a macroevolutionary scale?

Very importantly, Stephen Cave showed that immortality narratives *are not mutually incompatible*. He illustrates this with the Egyptian civilization, which used all four immortality narratives (soul, resurrection, staying alive and legacy). So we could see a version of the global brain legacy as a combination of individual, creative and evolutionary immortality. As Francis Heylighen elaborated (cited from the Global Brain Mailing list, 6/7/2006):

> personal knowledge and experience would ultimately be integrated into a collective consciousness, or what I usually call a "global brain", which is itself immortal (or at least does not have an a priori limited life span). This combines the advantages of continuity (none of the good ideas are lost) and innovation (personal experiences may be combined with other personal experiences in order to produce something that is more than the sum of the parts).

So, a wise evolutionary thinker will embrace the whole tree of life. But is it enough, even considering a promising global brain legacy? No, it is not.

The scalability issue arises. Why not extend the tree of life to its deepest roots in cosmic evolution? Even the grandeur of a global brain legacy is not enough. It will at some point run out of energy because the Sun has a limited fuel reserve. We can not ignore the cosmological context. So we must expand our will to immortality from the planetary level to the cosmological. Can we aim for an even more enduring form of immortality, taking into account the predictable end of the Sun and other cosmological constraints?

Why stop our identification with the 4.5 billion years of evolution on Earth since we know that the cosmological context to give birth to humanity took 13.7 billion years? We can be proud to be 13.7 billions years old if we identify with the atoms which compose us. And who knows what those atoms will become in the far future? There are good chances that they will be recycled by our descendants for purposes we can barely imagine.

### 10.4.5 Cosmological Immortality

*A human being is part of a whole, called by us the "Universe" a part limited in time and space. He experiences himself, his thoughts and feelings, as something separated from the rest—a kind of optical delusion of his consciousness. This delusion is a kind of prison for us, restricting us to our personal desires and to affection for a few persons nearest us. Our task must be to free ourselves from this prison by widening our circles of compassion to embrace all living creatures and the whole of nature in its beauty.*

Albert Einstein (quoted in 1972 by the New York Times)

Ultimately everything depends on the fate of the cosmos. As Einstein writes above, we can aim to widen our circles of compassion to embrace the whole universe. This cosmic connection with everything in the universe is an awe inspiring feeling that deep reflection about death brings. As Sogyal Rinpoche (1994, 191) writes in *The Tibetan Book of Living and Dying*,

> when we finally know we are dying, and all other sentient beings are dying with us, we start to have a burning, almost heartbreaking sense of the fragility and preciousness of each moment and each being, and from this can grow a deep, clear, limitless compassion for all beings.

This feeling is indeed very similar to what Einstein eludes to in the above quote. Such connections of our *subjective* experience with the universe are very inspiring and fundamental for cosmological immortalists. In particular, cosmological issues are the most important ones to focus on if one wishes literally to achieve immortality. But those almost religious feelings do not solve the *objective* problem of a lurking cosmic doom.

Once we extend our awareness to the universe, thermoethical issues arise, namely the limited amount of free energy available in the Sun, stars, galaxies and ultimately in the universe. Additionally, feeling compassion with the universe implies knowledge of threatening consequences of cosmological models, such as big crunch, big rip or heat death. To make the cosmos hospitable for life forever is no trivial matter at all.

However, a critique may go as follows. Cosmic doom scenarios may reveal a failure of our imagination, because of the intrinsic limitation of deterministic models to end in stable attractors of disorder or equilibrium. We just fail to model innovation and creativity. This may be valid, but then it is up to us to imagine more creative scenarios and models for the future of life in the universe.

Let us see a few options to continue the evolutionary process infinitely. First, let us muse philosophically about an argument we call *metaphysical immortality.* The reasoning goes as follows. Let us remember the metaphysical challenge for ultimate explanations (section 4.1.2 Metaphysical, p76). When we ask "why not nothing?" we implicitly admit that "there is something". The postulate "nothing comes from nothing" is hard to challenge, at least if we define "nothing" in a philosophical way and not as a quantum vacuum full of potential. It means that "everything always endures, in one form or another". For immortality, it means that something, some of our material-energetic "us" may have had an infinite history, and may have an infinite future! So we may all be truly immortal in this metaphysical sense, both in the past and in the future. But if you are murdered tomorrow, even though the atoms composing your body will be recycled, I am not sure if this metaphysical immortality will really comfort you. So, this metaphysical immortality does not dispense more psychological and philosophical considerations.

We can now turn to the central issue of cosmological immortality. More accurately, the question is, how to *perpetuate* life in the universe infinitely?

A straightforward answer is to actively promote space colonization. To export life or humanity to outer space is a way to achieve immortality (see e.g. Baird 1989, 48). As with any future prospect for humanity, we can speculate that advanced extraterrestrials would have had similar ideas, and do actually colonize the galaxy with life (Crick and Orgel 1973). Intriguingly, this idea of directed panspermia leads to the proposition that bacteria or viruses on Earth might actually have an extraterrestrial origin (Fred Hoyle and Wickramasinghe 1990). However, even space colonization doesn't work in the very long term since it does not deal with the thermodynamical constraints.

So, how, if only possible, is the universe going to survive as a whole? Globally, there are two ways to perpetuate life, either by *survival*, or by *reproduction*. Let us apply those two options to the universe.

We use again the biological theory of r-K selection and see how it could apply to the universe. An r-reproduction strategy would involve fast reproduction and short survival. It doesn't seem to apply well to our universe, which is globally quite stable and slowly changing. As far as we know, there are not billions of baby universes which are produced every second. The K-strategy involves a slow reproduction and a long survival. It seems more appropriate to apply it to our universe. The extreme case is to have no reproduction at all. This is the essence of Freeman Dyson's (1979) proposal to make life –defined as information processing– survive forever.

The steady-state theory also provided a never aging cosmos. Indeed, even if galaxies get born, age and die, new galaxies do appear. In the steady-state universe, there is no beginning, no end and the universe as a whole never gets old. Of course, there is a catch, which is the issue of *how* new galaxies do appear. Paul Davies (1994, 152) remarks that such a process would require to add $10^{50}$ tons of matter to the universe every few billion years!

The reversible computation scenario also offers –in principle– a way to compute forever, because no new energy would be needed thanks to the use of reversible computational gates. But as we saw (section 9.4.1 Two Scales Argument, p232), a civilization choosing this path would not be allowed to forget (erase states) because this operation has an energetic cost.

Tipler's (1997) omega point scenario, although it only works with a closed universe, which is not supported by current cosmology, points towards a very slow reproduction mechanism, where the contracting energy towards the big crunch is re-used for producing a next generation universe. Interestingly, Tipler's omega point theory is actually a mix of individual and cosmological immortality. Indeed, near the omega point, intelligence is so powerful that it has the computational capacity to resurrect all of us –cybernetically. A very unlikely outcome for sure, since a civilization developing cognitively and morally up to a cosmological stage would certainly not care resurrecting all the possible life histories of a particular species, and even less of a particular human being which has lived some billion years ago.

Another reproductive scenario is cosmological artificial selection, a speculative philosophical scenario integrating the origin and future of the universe with a role for intelligent life (see Chapter 8). Cosmological immortality via cosmological artificial selection is analogous to biological immortality through a chain of reproducing universes instead of a chain of living entities.

Interestingly, Stephen Cave (2012, 249) speculated about cosmological artificial selection as a way to realize a cosmic legacy narrative. He apparently had the insight independently of the literature we reviewed (section 8.3.1 History, p176):

> Perhaps one day we—or some far more evolved successor—will be able to seed new universes that are fit for life. Indeed, perhaps we are already in one, seeded by some earlier civilization.

Besides the r-K trade-off, let us now be inspired by an evolutionary theory of aging in biology, the *disposable soma theory* (see Kirkwood 1977; 1999). It simply says that it is more efficient to invest energy in reproduction than in indefinite upkeep of the organism. Indeed, once the best chances for reproduction are used, thus ensuring the survival of the almost immortal genes (germline), the mortal body (soma) can be disposed of. Futurist John M. Smart (2009, 224–226) speculated that this theory may also apply to the universe. The soma is analogous to the constituents of the universe, with its mortal galaxies, stars and planets while the germline is analogous to free parameters which determine immortal physical laws.

If we take seriously this fundamental trade-off in energy expenditure between soma and germline, the ultimate death of the “universe's soma” may indeed be inevitable. But its germline, it's most delicate physical structure may be saved if they are replicable. If CAS holds, it means that intelligent civilizations will spend much energy into preparing and making universes, and few energy in maintaining our disposable universe. If –and this is a big if– the starivore hypothesis (see Chapter 9) is correct, there is already huge amount of energy invested in securing the germline of our universe. A famous maxim by La Rochefoucauld's (1868) says that “Neither the sun nor death can be looked at steadily.” Cosmological immortalists, and maybe starivores, are proud to defy it. They first look steadily at the death of our universe, and then look as steadily at our Sun and other stars only to see them as energy bounties necessary to the task of universe making.

Yet, cosmological immortalists would not be totally content with the making of *one* new universe. They would worry: would the newly formed universe develop life, intelligence, up to the level where universe making becomes again possible? Saving our particular universe by making another universe is not enough. It would just shift the problem of immortality to the next universe. Making a sterile new universe would be as depressing as having one only sterile child. This realization motivates the next cognitive stage 9, concerned about issues regarding infinite evolution.

There is a difference between "indefinitely" and "infinitely". While it seems more careful to say that the evolutionary process would continue indefinitely, a highly advanced intelligence would not be satisfied with the uncertainty implied by the word "indefinitely". It would attempt to find a way to achieve true immortality, through a *provably* infinite process. Let us speculate on how a stage 9 being could approach this issue and accept to die in peace.

There is first a statistical way. Caring universe makers will do extensive simulations to prove statistically that universes will recursively reproduce. Securing this property would be a major challenge of "hard artificial cosmogenesis" (see also 8.3.9 Objection – Are Simulation and Realization Possible? , p190).

The second way is the most rigorous way because it involves the mathematical logic technique of *model checking*. Indeed, statistics based on computer simulations have always a small probability to fail. A good old solid logical proof is much more desirable. Caring universe makers will do model checking of the $n+1$ universe to prove that it will recursively reproduce. However, considering the difficulty and computationally intensive process of model checking even for very simple systems, it seems quite unrealistic that it is achievable, even for extremely advanced civilizations.

The third way is if hypercomputation is unlocked. First it would allow to compute much more, and thus might provide the missing tool to validate the model checking approach. In all likelihood, the proposition "universe $n+1$ is fertile" will be Gödel-undecidable. So universe makers would never be able to be sure that the universe they make would have this desired property of fertility. Hypercomputation may solve this issue, because it could in principle decide issues which are not Turing-computable. This would be a proof of cosmological immortality. The availability or not of these speculative options will depend on the level of universe making which is available (see section 8.3.3 Six Levels of Universe Making, p183). Of course the thermodynamical issue should not be underestimated: where does the energy come from to make universe $n+1$ from universe $n$ (see 8.3.7 Objection – The Thermodynamical Issue, p189)?

In the conclusion of his book, Stephen Cave argued that a combination of virtues which lead to accept death and a hope for a legacy narrative constitute a wise way to deal with death. Cosmological immortalists would largely agree. A multi-scale wisdom towards mortality can even be summarized with the following insight. Microcosmically, we owe our life to the death of cells; macroscopically, we owe our life to the death of stars. Indeed, we saw that cellular death is essential for our body to regenerate itself and it is standard astrophysics that the carbon atoms which constitute our bodies where cooked up in second-generation stars which exploded in supernovae.

But how can we imagine to seriously care for such an issue as cosmological immortality? We can summarize five steps towards it. The first is to realize that your individual death is normal and inevitable in the long term. The second is to develop psychologically, and fulfill all your needs to grow the hierarchy of needs up to the need of self-transcendence. You then surpass your self to become compassionate and identify with the process of cosmic evolution. Even if you accept individual death, you still refuse death as a whole, namely the idea that nothing would continue to evolve after the predictable death of your body, society, species, Sun, galaxy and universe. You then set the immortality of the evolutionary process as a goal. I let the final words of this section to Freeman Dyson (1988, 121) who wrote:

We know very little yet about the potentialities and the destiny of life in the universe. In speculating about these matters we follow a great tradition. We are in the same company with Bernal and Newton, Tsiolkovsky and Thomas Wright. Letting our imagination wander among the stars, we too may hear whispers of immortality.

## Open Questions

Let us mention a few lines of open questions.

- Developmental psychology could be enriched applying the insights of Turchin's metasystem transition to individuals, in order to construct a theory of “self metasystem transitions” (see Stewart 2001 for a starting point).
- The evolutionary ethics based on virtuous trade-offs can be further developed, detailed, compared to Aristotle's virtue ethics. The six trade-off dimensions might be reduced to few more fundamental ones.
- Are there stages of immortality, as there are stages of morality (Kohlberg 1981) or stages of faith (Fowler 1981)? There is already quite some work on the conceptions of death through the child's development. We hypothesize that different hopes and narratives of immortality will be hold at different stages of psychological development. Such a research is a difficult endeavor because scientists must also be clear on their own status in resolving their death anxiety. Otherwise, they risk to bias experiments and their interpretation. In any case, they will not be inclined to do the research, because researchers, like every human being, do not like to face death. As Charles Wahl (1958) wrote, the study of the fear of death is “conspicuous by its absence.”
- Immense work remains to be done to build our cosmic culture. More work in this area is needed if we want to foster personal development to very high levels and develop sustainable societies on the very long term.

# Conclusion

*Science cannot solve the ultimate mystery of nature.*
*And that is because, in the last analysis,*
*we ourselves are part of nature and therefore*
*part of the mystery that we are trying to solve.*

Max Planck (1932)

It's time to take a big picture perspective on our intellectual adventure.

In Part I, I built scaffolding for answering our childishly simple big questions. Adults usually fail to answer such deeply philosophical questions. And for a very good reason, answering the big philosophical questions demands both philosophical care and scientific expertise.

The elucidation of the nature of philosophy and its method is a delicate and arduous endeavor because through its history philosophy has entertained more or less strong ties with art, religion and science. As a result of this mixing with other disciplines and because the scientific enterprise becomes increasingly interdisciplinary, I focused on developing the concept of *worldview*, which holds the promise to meaningfully integrate human knowledge.

We saw in Chapter 1 that we all have and need a worldview, even if it is implicit. To make it explicit requires introspection and philosophizing, but the gain in perspective is priceless. Doing so, we look at how we look at the world. I argued in favor of the tradition of philosophical systems which strive to build *coherent* and *comprehensive* worldviews.

Philosophy is a rich domain constituted by six dimensions, each concerned with different kinds of questions. The *descriptive*, *normative* and *practical* dimensions try to answer what is, what is good and bad, and how to act in the world. The practice of the *dialectical* dimension consists in stating and reconstructing issues and a variety of positions towards them. It is also essential to avoid forming premature doctrines and sinking into dogmatism. The *critical* dimension of philosophy is valued both by continental and analytical philosophy. It is an intellectual acid that can attack any proposition. Finally, the *synthetical* dimension is the climax of philosophizing, the one aiming at coherent and comprehensive worldviews, but also its most demanding dimension. To be successfully handled, it requires mastering and juggling all of the other five dimensions.

Furthermore, given that we all develop different worldviews depending on our culture, education or psychology, in Chapter 2 I tackled the question of how to compare worldviews. Recognizing six philosophical dimensions and making the worldview agenda explicit were the first steps. I then developed in detail nine criteria to compare worldviews, classified in three broad categories: objective criteria (*objective consistency*, *scientificity*, *scope*), subjective criteria (*subjective consistency*, *personal utility*, *emotionality*), and intersubjective criteria (*intersubjective consistency*, *collective utility*, *narrativity*). From the criteria and the agenda, I derived worldview assessment tests (the *is-ought*, *ought-act*, and *is-act* first-order tests; the *critical* and *dialectical* second-order tests; the *mixed-questions* and *first-second-order* third-order tests; and the *we-I*, *we-it*, and *it-I* tests).

I see the six dimensions, the worldview agenda, the nine criteria and the seven tests as a metaphilosophical apparatus to understand, improve, compare, and constructively criticize different religious, scientific or philosophical worldviews. I outlined in Chapter 3 the major strengths and weaknesses of those three kinds of worldviews, and argued that the way to synthesis is through building comprehensive theological or philosophical worldviews.

As William James noted, an important outcome of philosophical activity is to give birth to new scientific disciplines. The almost Oedipal stage of a scientist saying that philosophy is dead, that he doesn't need it, is actually a sure sign that the scientific field can sustain itself. Philosophy has done its job, and the umbilical cord can indeed be cut. The second stage of scientific maturity is the realization that killing philosophy, like killing one's parent, was finally not such a good idea. At least if we want to remain creative, to question the foundations of science and tackle the deepest mysteries of nature.

With hindsight, I proposed giving birth to two new research fields. The first regards the exploration of possible universes, up to now chiefly a metaphysical recreation. I defined the field of *artificial cosmogenesis* (Chapter 6.3 and 7) in order to scientifically study possible universes. The second regards the search for advanced extraterrestrials. Here too, the field is usually very speculative and actually most often explored not by philosophers or scientists, but by science fiction authors. With the starivore hypothesis (Chapter 9) and the more general field of *high energy astrobiology*, I showed that existing knowledge in astrophysics demands a reassessment from an astrobiological viewpoint.

In Part II, I focused on the origin of the universe. I first outlined seven fundamental challenges underlying the quest for ultimate explanations: *epistemological*, *metaphysical*, *thermodynamical*, *causal*, *infinity*, *free parameters* and *fine-tuning*. I only conducted a detailed study on the last two, the free parameters in Chapter 5 and the fine-tuning conjecture in Chapter 6. Of course, the awareness of the five other challenges was in the background and helpful in clarifying analyses and conclusions throughout this thesis.

In Chapter 4, I applied the concept of the origin to itself, and asked what are the origins of the origin? In other words, what do we cognitively expect to be a satisfying answer to the ultimate origin of the universe? I argued that our explanations fall into two kinds of cognitive attractors: the point-explanation and the cycle-explanation. An analogy from dynamical systems theory clearly shows that these correspond to the simplest attractors: the fixed point (0-dimensional) and the limit cycle (1-dimensional). We have no reason to exclude $n$-dimensional attractors or strange attractors (noninteger-dimensional) whose nature are fractal. Admittedly, it becomes challenging for our brains to think about the origin in such terms.

The major thesis of Chapter 5 was that free parameters in particle physics models will be reduced to free parameters in a cosmological model. It is a fundamental issue in physics and cosmology to reduce or explain those remaining free parameters. I analyzed the issue with physical, mathematical, computational and biological backgrounds. But are those free parameters further fine-tuned for life or complexity?

To answer this much debated fine-tuning issue, I started with a review of probabilistic, logical and physical fallacies which surround it. Then I distinguished the issue with seven other closely related issues: *free parameters*, *parameter sensitivity*, *metaphysical issues*, *anthropic principles*, *observational selection effects*, *teleology* and *God's existence*.

I introduced the *Cosmic Evolution Equation* as a central conceptual tool to study how *robust* our universe is when it comes to the emergence of complexity and to what extent it is *fine-tuned* compared to other possible universes. The fine-tuning issue can then be formulated as: "are fecund universes rare or common in the space of possible universes?". The straightforward way to answer such a difficult question is to explore the space of possible universes, through *artificial cosmogenesis*. Albeit extremely ambitious and computationally intensive, in Chapter 7 I gave more arguments to show why artificial cosmogenesis is a natural outcome of future scientific activity. Since comparing our universe to other possible ones has just gotten under way, I concluded that the fine-tuning of our universe can at most be a conjecture.

Studying the fine-tuning conjecture is one thing, explaining it is another. I showed the shortcomings of eight classical explanations: *skepticism*, *necessity*, *fecundity*, *god-of-the-gaps*, *chance-of-the-gaps*, *weak-anthropic-principle-of-the-gaps*, *multiverse* and *design*. This left two additional explanations inspired by evolutionary theory: *Cosmological Natural Selection* (CNS) and *Cosmological Artificial Selection* (CAS) that I discussed and critically analyzed in detail in Chapter 8.

I started by reviewing the history of cosmological natural selection and formulated objections to it. To remedy these objections, I introduced cosmological artificial selection and reviewed its generally unknown history. This evolutionary scenario is my core thesis, a philosophical speculation aimed at explaining the fine-tuning issue and the meaning of life and intelligence in the far future universe.

Of course such an ambitious thesis ignites many objections, and I formulated and addressed eight of them. I then summarized four different roads leading to CAS and further substantiated this scenario by showing that because it is so broad, it has limited alternatives. Furthermore, I showed that the nine other alternatives encounter problems and difficulties.

Once the development of intelligence is seen as a central feature of our universe, we can address the issue Max Planck evokes in the quote above. We need to take into consideration ourselves and intelligence in general if we want to solve the "ultimate mystery of nature". However, according to CAS, the greatest outcome of the scientific enterprise is not an almost spiritual quest to find the key to the mysteries of nature, but actually a more practical activity: to make universes in order to avoid a cosmic doom.

CAS finally invites us to focus more on the future of the universe rather than on its past. Indeed, we will probably never know what happened at the big bang or "before" it. But thinking about the far future leads to increasing understanding of the beginning. This led us to the architect point of view: *the more we are in a position to make a new universe, the more we will understand our own universe.*

We saw that assessing the robustness of the emergence of life and intelligence in the universe is a central issue. I showed how it can be tackled with extensive runs of computer simulations. However there is a shortcut, which is to search for extraterrestrials, or the natural "re-runs" of the tape of life.

In Chapter 9, I thus focused on the search for *advanced* extraterrestrials. Why advanced? Because it would be much more informative, insightful and disruptive to find extraterrestrials two billion years older, rather than finding an extraterrestrial bacterium. On the historical side, I clarified the fundamental importance of astrobiology, whose outcome will lead to major scientific worldview changes. On the methodological side, I first debunked many implicit and limiting assumptions in past and current search and then summarized and compiled criteria for artificiality.

When I started to think about extraterrestrials, it was as an intellectual challenge, trying to explore how CAS could help. This started without scientific pretension but to my own surprise, such extreme speculations quickly turned into a scientific hypothesis, the possible existence of *starivores*, civilizations feeding actively on stars. The hypothesis is now ready for rigorous empirical and scientific assessment. You certainly remember the open questions section at the end of Chapter 9, the concrete scientific research proposals I proposed and the associated High Energy Astrobiology prize. I am looking forward to congratulating the winner of the prize.

Even proving that the hypothesis is wrong, which of course is also entitling to the prize, speculating about advanced extraterrestrials gives a unique cosmological perspective on humanity and its role in the universe. In the meantime, what about us and our values?

In the last Chapter 10, I outlined foundations for a cosmological ethics. I focused on very general ethical principles, as much as possible applicable to all living things. I thus critically outlined thermodynamical, evolutionary, cybernetic and developmental values. As an application of the cosmological ethical framework, I discussed in details five conceptions of immortality, from the personal *spiritual* and *individual* immortalities, to the transpersonal *creative*, *evolutionary* and *cosmological* immortalities. I argued that the ultimate good is the *infinite continuation of the evolutionary process*.

Our time is unique. Humans are connecting via and with more and more networked and pervasive computers, creating a new level of planetary intelligence best conceptualized as a global brain. We are also on the brink of confirming the existence of extraterrestrial life, via astrobiology or high energy astrobiology, which would refute *biocentrism* or *intellicentrism*.

This event will change our worldviews forever, and thanks to the cosmological perspective developed in Part III, we are ready. But the scenario of cosmological artificial selection also prepares us to be ready for the eventual refutation of *universecentrism*, the belief that our universe is somehow central and unique. This would defeat the very last bastion of anthropocentrism.

Is it a tragedy that we will probably die before witnessing such major and magnificent evolutionary or worldview transitions? No, because if we become wise enough to endorse a cosmological ethics and grow towards a will to cosmological immortality, we are also ready to –individually– die.

All in all, what is the meaning of *your* life in a cosmological perspective? Of course, it was not the purpose of this thesis to tell you what the direction of your personal life should take. Becoming a doctor, a dancer or a high energy astrobiologist must be your own choice. Yet, you might want to better harmonize your life with the whole of cosmic evolution. This practical shift remains to be worked out, but at least we have set some theoretical foundations. So, what is the meaning of life in a cosmological perspective? It is to replicate at the grandest scale, through an intimate connection of intelligence with the universe.

Instead of seeing the cosmos as hostile to life and intelligence, I would like to end with a short poem, *Cosmosis*, conveying a vision of cosmic evolution as a love story; a love story between the cosmos and its precious intelligence:

To begin, love is ego.
Then opening to alter ego,
now growing to Earth.

Yet, only from the osmosis
between wisdom and the cosmos
ends love in an infinite cosmosis.

# Appendix I – A Cosmic Evolutionary Worldview: Short Answers to the Big Questions

## *Introduction*[16]

Across centuries, humanity has been wondering about its existence and its place in the universe. Humans employed insights from myths, religions, art, philosophy and science to make sense of the world around them.

However, in the current era of accelerating scientific, cultural and social developments, all the old certainties are put into question. The confusion and fragmentation associated with this often lead to pessimism and uncertainty, and the need for psychological guidance in the form of a clear and reliable system of thought.

This is why it is important to search for a *coherent* and *comprehensive* worldview, by answering today the big questions of this quest for understanding. Answering them explicitly is an enterprise which is traditionally philosophy's task. This took the form of comprehensive and coherent systematic philosophical treatises. The great philosophical systems are of this sort. Regrettably, this trend seems to have fallen out of fashion, since most of today's philosophy is busy with second-order problems (Adler 1965).

In contrast to most of contemporary philosophy's practices, below are tentative and provisional responses to first-order philosophical questions. The answers to these questions together determine a worldview, i.e. a comprehensive philosophical system, a coherent vision of the whole. A worldview gives meaning to our life, and helps us to understand the world around us.

Each worldview question would need at least a book to be properly answered. More than that, the most appropriate way to answer them is with a systematic philosophical system (e.g. Bunge 1974; Rescher 1992). I do not have that objective here. Instead, I provide below very short responses as *positions*, *not arguments.* I give some main references to the works which influenced me, where the curious reader will be able to find many detailed arguments. It is worth reminding the many advantages of explicitly stating one's philosophical position.

First, these short responses will obviously let the reader quickly grasp my position. The position is stated transparently, straightforwardly, with a few technical concepts involved.

Secondly, the task of answering those questions is a daring effort. I balance this *great ambition* with *great caution* in answers I provide. They are non-dogmatic, provisional, revisable and sometimes falsifiable. The responses proposed here are mixed philosophical and scientific conjectures to make sense of the world. Accordingly, some of them are speculative. They are of course not definitive. In such a short format, I also do not make justice of the pros and cons of alternative positions (the dialectical dimension of philosophy). It doesn't mean that I'm not aware of them. Still, if you think I've missed something important, or a position clearly better than the ones presented here, please contact me. As every good philosopher and scientist, I very much value and warmly welcome criticism and further reflection you might have reading this text.

---

16 Eventual updates to this “philosophical identity card” can be found at: http://www.evodevouniverse.com/wiki/A_Cosmic_Evolutionary_Worldview:_Short_Responses_to_the_Big_Questions

Thirdly, this transparency in responding to basic questions allows *efficient debate and communication*. Many debates and disagreements get lost in details, without touching the heart of issues at stake. This practice of answering first-order questions can save an enormous amount of time of confusing debates, because enduring disagreements always end up in disagreements about such fundamental questions. I invite you to do the same exercise before reading what follows, and simply answer the worldview questions for yourself. Feel free to use the following space to outline your main worldview answers. Good luck!

## Make Your Worldview Explicit!

(g) Where to start from?

(a) What is? *Ontology* (model of being);

(b) Where does it all come from? *Explanation* (model of the past);

(c) Where are we going? *Prediction* (model of the future);

(d) What is good and what is evil? *Axiology* (theory of values);

(e) How should we act? *Praxeology* (theory of actions).

(f) What is true and what is false? *Epistemology* (theory of knowledge)

# My Worldview Made Explicit

## *(g) Where to start from?*

Before proposing responses to those big worldview questions, here are some preliminary considerations, laying bare how I start this enterprise. The (meta)philosophical framework and method are mainly inspired by the works of Adler (1965; 1993), Rescher (1985; 2001; 2006) and Bahm (1979).

If I had to choose a philosophical stream, I would say I am mostly influenced by *systems philosophy* (esp. Laszlo 1972b; Heylighen 2000b; and 2010b on which this text is based). To summarize it in one sentence, its "data come from the empirical sciences; its problems from the history of philosophy; and its concepts from modern systems research" (Laszlo 1972a, 12). We may add to *systems theory* an interdisciplinary *problem solving* approach and *evolutionary-developmental theory*, applied on many scales (Vidal 2008a).

**(i) The worldview agenda**

I start with the philosophical agenda described in Chapter 1.

**(ii) The metaphilosophical criteria**

Once the questions are asked, we obviously need to answer them and use evaluation standards to assess their strengths and weaknesses. I developed in Chapter 2, nine criteria and a battery of tests to compare and assess different worldviews. Which criteria do I value most?

The aim in Chapter 2 was descriptive. Now, how do I use the criteria prescriptively to answer the agenda of the worldview questions? Here, I use in priority *objective criteria* (*objective consistency*, *scientificity* and *scope*) to construct a coherent and comprehensive cosmological worldview. In this cosmic evolutionary worldview the *scope in level depth* is maximally wide in time and space, concerning the whole universe. As objective criteria are maximally satisfied, I turn to *subjective* and *intersubjective* criteria to make the worldview successfully applicable in the conduct of a good life and in the organization of a good society. The pursuit of a good life and a good society is then harmonized with cosmic evolution.

## *(a) What is?*

As a preliminary remark, I am generally skeptic with reductionistic ontological statements. Reality is complex, evolving and multi-layered, and different ontologies are more or less appropriate to analyze and solve different problems. Dooyeweerd's (1953) fifteen aspects, although static and not dynamic, offer an example of a non-reductionistic ontology.

My ontological commitment goes towards systems theory, which aims to offer a universal language for sciences (e.g. von Bertalanffy 1968; Boulding 1956). It is also very fruitful for philosophizing (e.g. Laszlo 1972a). It is best combined with evolutionary reasonings, which gives rise to an evolutionary-systemic approach (Heylighen 2000b).

I choose an ontology of actions and agents, i.e. elementary processes and relations, not independent, static pieces of matter (in the spirit of Whitehead (1930), Lazslo (1972a), Jantzch (1980), etc). Out of their interactions, organization emerges. Through evolutionary processes, these systems become more complex and adaptive, they start to exhibit cognition or intelligence, i.e. the ability to make informed choices.

### *(b) Where does it all come from?*

Modern science explains –at least in parts– the harmony within nature, connecting physical, chemical, biological and technological evolution (e.g. Chaisson 2001; De Duve 1995). Regarding the origin of the universe, although Big Bang models are a success of modern cosmology, the initial conditions remain mysteriously fine-tuned (e.g. Leslie 1989; Leslie 1998; Rees 1999; Davies 2008). In Chapter 6, I concluded that fine-tuning is hard to prove, and that at most it is a conjecture. Whatever possible explanation we favor, we need to cope with difficult metaphysical choices (Vidal 2012a). The scenario of Cosmological Artificial Selection (CAS) developed in Chapter 8 connects the origin and future of the universe with a role for intelligent life (Vidal 2008b; 2010a; 2012a).

### *(c) Where are we going?*

Modern science has shown that there are two trends at play in cosmic evolution. First, a tendency to produce *more order*, with the emergence of more and more complex systems, from galaxies, stars, planets, to plants, humans and our technological society (Chaisson 2001; Kurzweil 2006; Morowitz 2002; Livio 2000). Secondly, the second law of thermodynamics applied to the universe as a whole implies that in the far-future the universe will irreversibly go toward a state of *maximum disorder*, or heat death (e.g. F. C. Adams and Laughlin 1997). The outcome of those two opposite trends remains unsettled.

The discovery of the heat death generated a widely spread pessimistic worldview which sees the existence of humanity as purposeless and accidental in the universe (B. Russell 1923; S. Weinberg 1993b). With Darwin (1887b, 70), I estimate that "it is an intolerable thought that he [man] and all other sentient beings are doomed to complete annihilation after such long-continued slow progress".

Hopefully, the first trend will prove to be more promising. The process of on-going complexification and adaptation can reasonably be extrapolated towards the future. This allows us to predict that in middle course, conflict and friction within human society will diminish, cooperation will expand to the planetary level, individual and collective intelligence will spectacularly augment.

Generally, more advanced biological organisms build more and more sophisticated representations of their surroundings (P. Russell 1995). The research field of *artificial cosmogenesis* pushes this trend to its limit, to the point where intelligent life constructs a model of the whole universe. This modeling capacity can be used to understand not only our own universe, but also other possible universes (see section 6.3 The Cosmic Evolution Equation, p122). The radical proposal of Cosmological Artificial Selection (CAS) developed in Chapter 8 is that in order to avoid the effect of the second law of thermodynamics, those toy-universes could become a blueprint for a new universe (Vidal 2008b; Vidal 2010a; Vaas 2012; Vidal 2012a).

This scenario is a mixed scientific and philosophical speculation. It is philosophical because it involves a role for intelligent life and so the success of CAS depends on our conscious choices for the future of cosmic evolution. It thus requires an axiological dimension, proper to philosophy and that we explored in Chapter 10.

CAS has also scientific aspects since its general perspective leads to far-reaching consequences and reasonings to search and maybe find advanced extraterrestrials. In order to confirm or infirm the existence starivores I introduced in

Chapter 9, we must scientifically approach the question within the field of *high energy astrobiology.*

## *(d) What is good and what is evil?*

The inner drive or implicit value governing all life is fitness, i.e. survival, growth, development and reproduction. From a human perspective, this fundamental value includes a sustainable quality-of-life, well-being or happiness. Evolutionary, developmental, thermodynamical, psychological, and cybernetic theories allow us to derive a number of more concrete objectives from this overarching value, i.e. properties that are necessary for long-term well-being. These include openness, diversity, intelligence, knowledge, cooperation, freedom, personal control, health, and a coherent and comprehensive worldview.

In the longer term, fitness implies increasing adaptiveness and evolvability beyond human society as we know it. Actions that promote these values with the less friction as possible are intrinsically good, actions that suppress them are bad.

As our psychology grows in higher stages of development, we make sure our values do not conflict with higher evolutionary systems. Not only do I try to improve my happiness, but my happiness becomes more and more tightly linked with my family, my country, society, humanity, the planet, and the cosmos. Ultimately I should act being aware and compassionate with such a hierarchy, combining the values of my own life with the sustainability of larger and larger evolutionary systems.

At heart, humans have a will to immortality (e.g. Turchin 1990; Lifton and Olson 2004). In my worldview, it takes the form of an endless, infinite cosmic evolution. Indeed, the metaphysical and speculative part of Cosmological Artificial Selection translates this will for immortality in an infinite process of evolution sustained by intelligence making offspring universes (Vidal 2008b; 2010a; 2012; esp. Vidal 2012a).

## *(e) How should we act?*

To maximally achieve these values in real life, we will need to overcome a variety of problems and obstacles. Cognitive sciences, cybernetics, and complex systems science suggest various tools and strategies to tackle complex problems, and to stimulate self-organization so as to be as efficient as possible. These methods include feedback control, anticipation, hierarchical decomposition, heuristic search, stigmergic coordination, extended mind and memetic engineering.

At the level of society, these methods define a strategy for effective governance, for the maximization of collective intelligence, and the minimization of friction and conflicts.

There is a trend in cosmic evolution to do ever more with less energy, space and time (Smart 2009). Using less energy and resources to achieve more is also at the heart of productivity principles. On a personal productivity side, The Getting Things Done method combines high productivity with low-stress (Allen 2001; Heylighen and Vidal 2008).

## *(f) What is true and what is false?*

Let us note that this is a second-order question which concerns knowledge about knowledge. Also, the domains of epistemology and ontology are closely related. We can divide this question in the following two questions (Heylighen 2000b, 15):

- What is knowledge? This question defines the domain of epistemology.

Science can be seen as a natural outcome of the more general evolutionary pressure to get more and more accurate knowledge (D. T. Campbell 1974). Knowledge is the existence in a system of a model, which allows that system to make predictions, that is, to anticipate processes in its environment. Thus, the system gets control over its environment. Such a model is a construction, not an objective reflection of outside reality (Turchin 1993; Heylighen 1997a).

- What is truth?

There are no absolute truths. The truth of a theory is merely its power to produce predictions that are confirmed by observations (Turchin 1993). The scientific enterprise is one of conjectures and refutations (Popper 1962) and there is a natural selection of ideas, theories, which give more power, i.e. prediction and control (D. T. Campbell 1974).

Ultimately, what is the meaning of the phenomenon of science in this pragmatic, constructive and evolutionary epistemology? It is not to seek an ideal "truth", but a pragmatic goal of acquiring knowledge. In the scenario of CAS, it is to build a model of our and other possible universes that could become, with some variation, a blueprint for a future universe, thereby escaping a predicable cosmic doom.

# Appendix II – Argumentative Maps

This appendix presents the logical structure of the main arguments presented in this thesis. The core problems of our concern are mapped in the first map and the proposed solution in the second map. To facilitate readability, I present two versions of each map, a collapsed and a expanded version.

This approach provides an *externalization of reasoning* so that arguments can be clearly visualized. This brings many benefits, such as:

- Allowing the reader to quickly and clearly grasp the logic of the argumentation.
- Presenting an alternative structure of the content of the thesis. The table of content and the abstract tend to present a rhetorical and less logical structure.
- Allowing the possibility of a constructive discussion of assumptions and deductions. For example, a critique can say "the core problem is not P but Q"; or "I disagree that hypothesis X leads to Y, you need implicit hypothesis Z, ..." or "hypothesis W is wrong because"; or "there is another solution to your problem, which is..." etc.

It should be clear however that reading those maps can not replace the reading of the thesis. Only the core reasoning is mapped, often in a simplified way. It also represents what I consider the core issues and proposed solutions, and has no ambition of comprehensiveness. Many more arguments are developed and discussed in the text. You as a reader can distill many other insights from the text. I am not claiming that these trees should be considered as my dogmatic position. I am in principle open to other interpretation and emphasizes. Those having worked with argumentative maps know too well that drawing them is first of all a basis for continuous improvement.

To draw those maps I used some of the insights of Eliyahu Goldratt's Theory of Constraints (TOC) and its "Thinking Process" (see Goldratt and Cox 1984; Goldratt Institute 2001; Scheinkopf 1999). The TOC is a well proven management technique widely used in finance, distribution, project management, people management, strategy, sales and marketing. I see it and use it as part of a generic problem solving toolbox, where causes and effects are mapped in a transparent way. In this TOC framework, three fundamental questions are employed to tackle a problem:

(1) What to change?

A core problem is identified, leading to *undesirable effects*, and mapped in a "Current Reality Tree" (CRT).

(2) To what to change?

A solution is proposed and mapped in a "Future Reality Tree" (FRT), which leads to *desirable effects*.

(3) How to cause the change?

A plan is developed to change from CRT to FRT. This third step involves drawing a transition tree. Such trees are important for practical problems, but in the more theoretical context of this thesis, I did not map it.

To tackle the problem in practice, six important questions should be addressed, constituting the "six layers of resistance to change". These questions can be used to trigger discussions (Goldratt Institute 2001, 6):

(1) Has the right problem been identified?
(2) Is this solution leading us in the right direction?
(3) Will the solution really solve the problems?
(4) What could go wrong with the solution? Are there any negative side-effects?
(5) Is this solution implementable?
(6) Are we all really up to this?

The following pages map the core problems (CRT) and my core arguments and thesis (FRT). Note that in the collapsed CRT, I put chronological links between the different worldview questions so that the diagram follows the linear order of this work. You can also consult the maps in their original expandable-collapsable-zoomable format thanks to the free Flying Logic reader software and the maps[17].

17 http://flyinglogic.com/download/ for the software as well as http://student.vub.ac.be/~clvidal/writings/Vidal-CRT-PhD.logic and http://student.vub.ac.be/~clvidal/writings/Vidal-FRT-PhD.logic for the CRT and FRT maps.

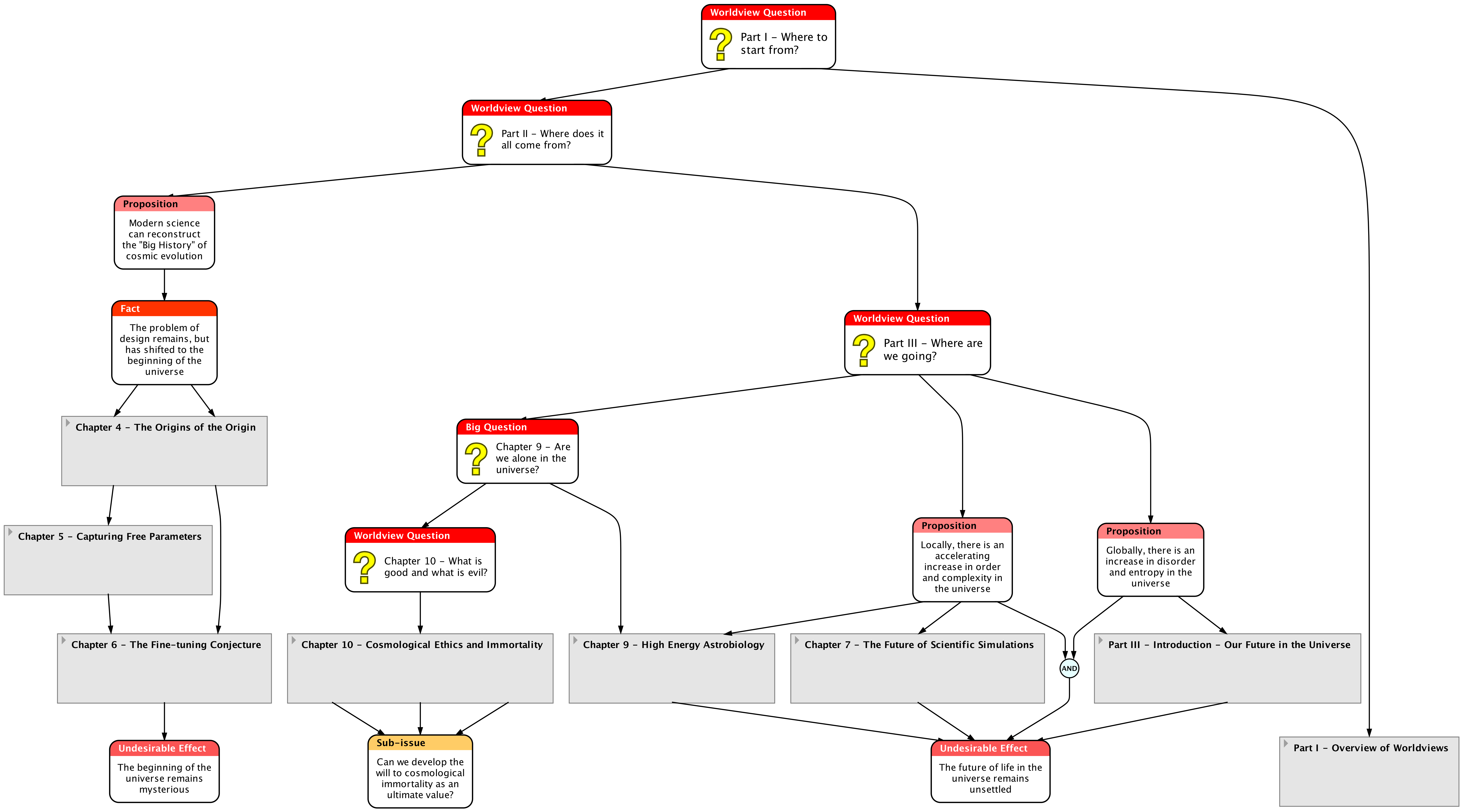

Worldview Question
Part I – Where to start from?
Worldview Question
Part II – Where does it all come from?
Proposition
Modern science can reconstruct the "Big History" of cosmic evolution
Fact
The problem of design remains, but has shifted to the beginning of the universe
Chapter 4 – The Origins of the Origin
Chapter 5 – Capturing Free Parameters
Chapter 6 – The Fine-tuning Conjecture
Undesirable Effect
The beginning of the universe remains mysterious
Worldview Question
Part III – Where are we going?
Big Question
Chapter 9 – Are we alone in the universe?
Worldview Question
Chapter 10 – What is good and what is evil?
Chapter 10 – Cosmological Ethics and Immortality
Sub-issue
Can we develop the will to cosmological immortality as an ultimate value?
Chapter 9 – High Energy Astrobiology
Proposition
Locally, there is an accelerating increase in order and complexity in the universe
Proposition
Globally, there is an increase in disorder and entropy in the universe
Chapter 7 – The Future of Scientific Simulations
AND
Part III – Introduction – Our Future in the Universe
Undesirable Effect
The future of life in the universe remains unsettled
Part I – Overview of Worldviews

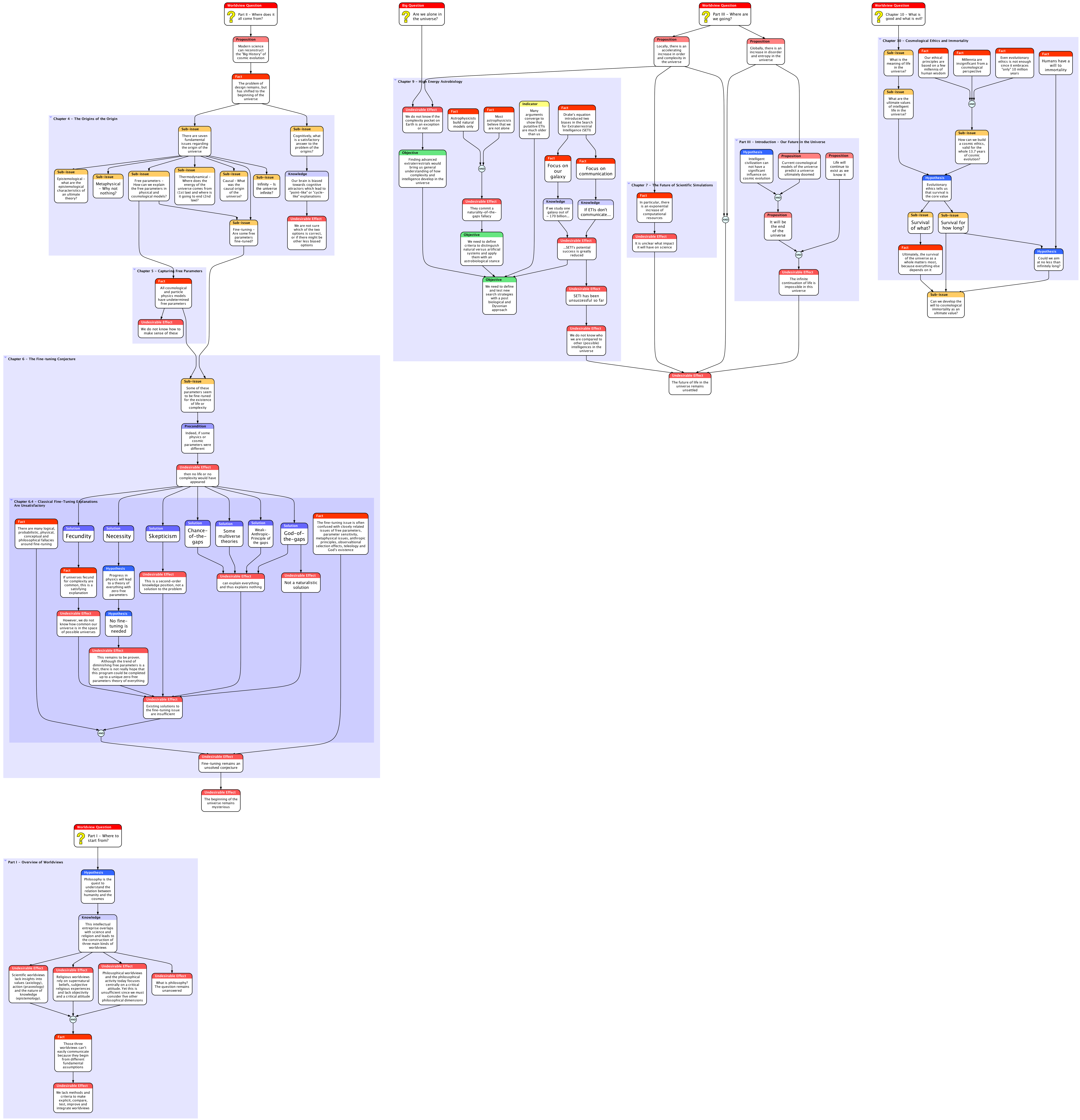
Worldview Question
Part II – Where does it all come from?
Proposition
Modern science can reconstruct the "Big History" of cosmic evolution
Fact
The problem of design remains, but has shifted to the beginning of the universe
Chapter 4 – The Origins of the Origin
Sub-issue
There are seven fundamental issues regarding the origin of the universe
Sub-issue
Cognitively, what is a satisfactory answer to the problem of the origins?
Sub-issue
Epistemological – what are the epistemological characteristics of an ultimate theory?
Sub-issue
Metaphysical – Why not nothing?
Sub-issue
Free parameters – How can we explain the free parameters in the physical and cosmological models?
Sub-issue
Thermodynamical – Where does the energy of the universe come from (1st law) and where is it going to end (2nd law)?
Sub-issue
Causal – What was the causal origin of the universe?
Sub-issue
Infinity – Is the universe infinite?
Knowledge
Our brain is biased towards cognitive attractors which lead to "point-like" or "cycle-like" explanations
Sub-issue
Fine-tuning – Are some free parameters fine-tuned?
Undesirable Effect
We are not sure which of the two options is correct, or if there might be other less biased options
Chapter 5 – Capturing Free Parameters
Fact
All cosmological and particle physics models have undetermined free parameters
Undesirable Effect
We do not know how to make sense of these
Chapter 6 – The Fine-tuning Conjecture
Sub-issue
Some of these parameters seem to be fine-tuned for the existence of life or complexity
Precondition
Indeed, if some physics or cosmic parameters were different
Undesirable Effect
then no life or no complexity would have appeared
Chapter 6.4 – Classical Fine-Tuning Explanations Are Unsatisfactory
Fact
There are many logical, probabilistic, physical, conceptual and philosophical fallacies around fine-tuning
Solution
Fecundity
Solution
Necessity
Solution
Skepticism
Solution
Chance-of-the-gaps
Solution
Some multiverse theories
Solution
Weak-Anthropic-Principle of the gaps
Solution
God-of-the-gaps
Fact
The fine-tuning issue is often confused with closely related issues of free parameters, parameter sensitivity, metaphysical issues, anthropic principles, observaltional selection effects, teleology and God's existence
Fact
If universes fecund for complexity are common, this is a satisfying explanation
Hypothesis
Progress in physics will lead to a theory of everything with zero free parameters
Undesirable Effect
This is a second-order knowledge position, not a solution to the problem
Undesirable Effect
can explain everything and thus explains nothing
Undesirable Effect
Not a naturalistic solution
Undesirable Effect
However, we do not know how common our universe is in the space of possible universes
Hypothesis
No fine-tuning is needed
Undesirable Effect
This remains to be proven. Although the trend of diminishing free parameters is a fact, there is not really hope that this program could be completed up to a unique zero free parameters theory of everything
Undesirable Effect
Existing solutions to the fine-tuning issue are insufficient
AND
Undesirable Effect
Fine-tuning remains an unsolved conjecture
Undesirable Effect
The beginning of the universe remains mysterious
Worldview Question
Part I – Where to start from?
Part I – Overview of Worldviews
Hypothesis
Philosophy is the quest to understand the relation between humanity and the cosmos
Knowledge
This intellectual entreprise overlaps with science and religion and leads to the construction of three main kinds of worldviews
Undesirable Effect
Scientific worldviews lack insights into values (axiology), action (praxeology) and the nature of knowledge (epistemology).
Undesirable Effect
Religious worldviews rely on supernatural beliefs, subjective religious experiences and lack objectivity and a critical attitude
Undesirable Effect
Philosophical worldviews and the philosophical activity today focuses centrally on a critical attitude. Yet this is unsufficient since we must consider five other philosophical dimensions
Undesirable Effect
What is philosophy? The question remains unanswered
AND
Fact
Those three worldviews can't easily communicate because they begin from different fundamental assumptions
Undesirable Effect
We lack methods and criteria to make explicit, compare, test, improve and integrate worldviews
Big Question
Are we alone in the universe?
Chapter 9 – High Energy Astrobiology
Undesirable Effect
We do not know if the complexity pocket on Earth is an exception or not
Fact
Astrophysicists build natural models only
Fact
Most astrophysicists believe that we are not alone
Indicator
Many arguments converge to show that putatitve ETIs are much older than us
Fact
Drake's equation introduced two biases in the Search for Extraterrestrial Intelligence (SETI)
Objective
Finding advanced extraterrestrials would bring us general understanding of how complexity and intelligence develop in the universe
AND
Fact
Focus on our galaxy
Fact
Focus on communication
Undesirable Effect
They commit a naturality-of-the-gaps fallacy
Knowledge
If we study one galaxy out of ~ 170 billion...
Knowledge
If ETIs don't communicate...
Objective
We need to define criteria to distinguish natural versus artificial systems and apply them with an astrobiological stance
Undesirable Effect
...SETI's potential success is greatly reduced
Objective
We need to define and test new search strategies with a post biological and Dysonian approach
Undesirable Effect
SETI has been unsuccessful so far
Undesirable Effect
We do not know who we are compared to other (possible) intelligences in the universe
Worldview Question
Part III – Where are we going?
Proposition
Locally, there is an accelerating increase in order and complexity in the universe
Proposition
Globally, there is an increase in disorder and entropy in the universe
Chapter 7 – The Future of Scientific Simulations
Fact
In particular, there is an exponential increase of computational resources
Undesirable Effect
It is unclear what impact it will have on science
AND
Part III – Introduction – Our Future in the Universe
Hypothesis
Intelligent civilization can not have a significant influence on cosmic evolution
Proposition
Current cosmological models of the universe predict a universe ultimately doomed
Proposition
Life will continue to exist as we know it
AND
Proposition
It will be the end of the universe
AND
Undesirable Effect
The infinite continuation of life is impossible in this universe
Undesirable Effect
The future of life in the universe remains unsettled
Worldview Question
Chapter 10 – What is good and what is evil?
Chapter 10 – Cosmological Ethics and Immortality
Sub-issue
What is the meaning of life in the universe?
Fact
Our ethical principles are based on a few millennia of human wisdom
Fact
Millennia are insignificant from a cosmological perspective
Fact
Even evolutionary ethics is not enough since it embraces "only" 10 million years
Fact
Humans have a will to immortality
Sub-issue
What are the ultimate values of intelligent life in the universe?
AND
Sub-issue
How can we build a cosmic ethics, valid for the whole 13.7 years of cosmic evolution?
Hypothesis
Evolutionary ethics tells us that survival is the core value
Sub-issue
Survival of what?
Sub-issue
Survival for how long?
Fact
Ultimately, the survival of the universe as a whole matters most, because everything else depends on it
Hypothesis
Could we aim at no less than infinitely long?
Sub-issue
Can we develop the will to cosmological immortality as an ultimate value?

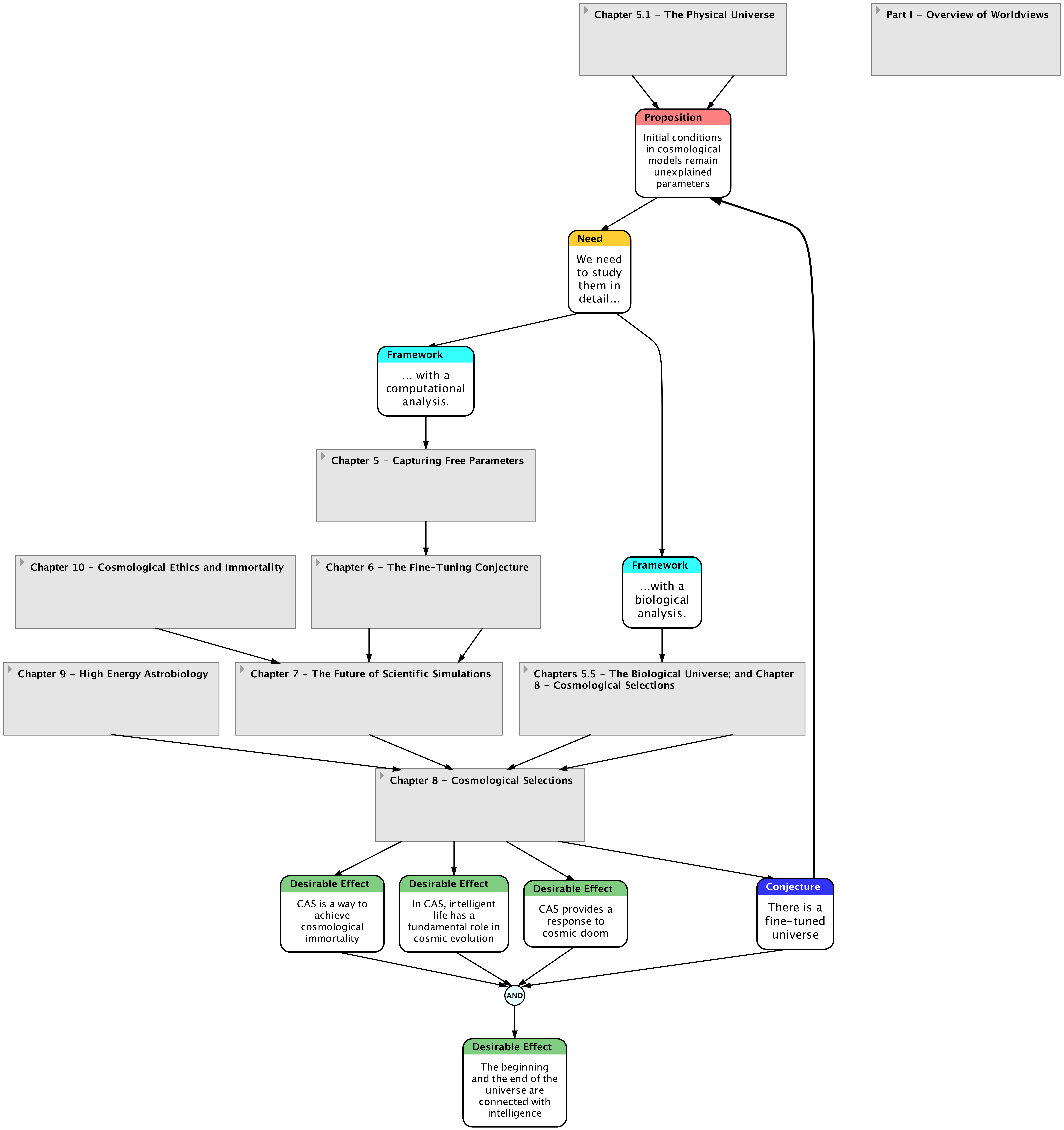
Chapter 5.1 – The Physical Universe
Part I – Overview of Worldviews
Proposition
Initial conditions in cosmological models remain unexplained parameters
Need
We need to study them in detail...
Framework
... with a computational analysis.
Chapter 5 – Capturing Free Parameters
Chapter 10 – Cosmological Ethics and Immortality
Chapter 6 – The Fine–Tuning Conjecture
Framework
...with a biological analysis.
Chapter 9 – High Energy Astrobiology
Chapter 7 – The Future of Scientific Simulations
Chapters 5.5 – The Biological Universe; and Chapter 8 – Cosmological Selections
Chapter 8 – Cosmological Selections
Desirable Effect
CAS is a way to achieve cosmological immortality
Desirable Effect
In CAS, intelligent life has a fundamental role in cosmic evolution
Desirable Effect
CAS provides a response to cosmic doom
Conjecture
There is a fine–tuned universe
AND
Desirable Effect
The beginning and the end of the universe are connected with intelligence

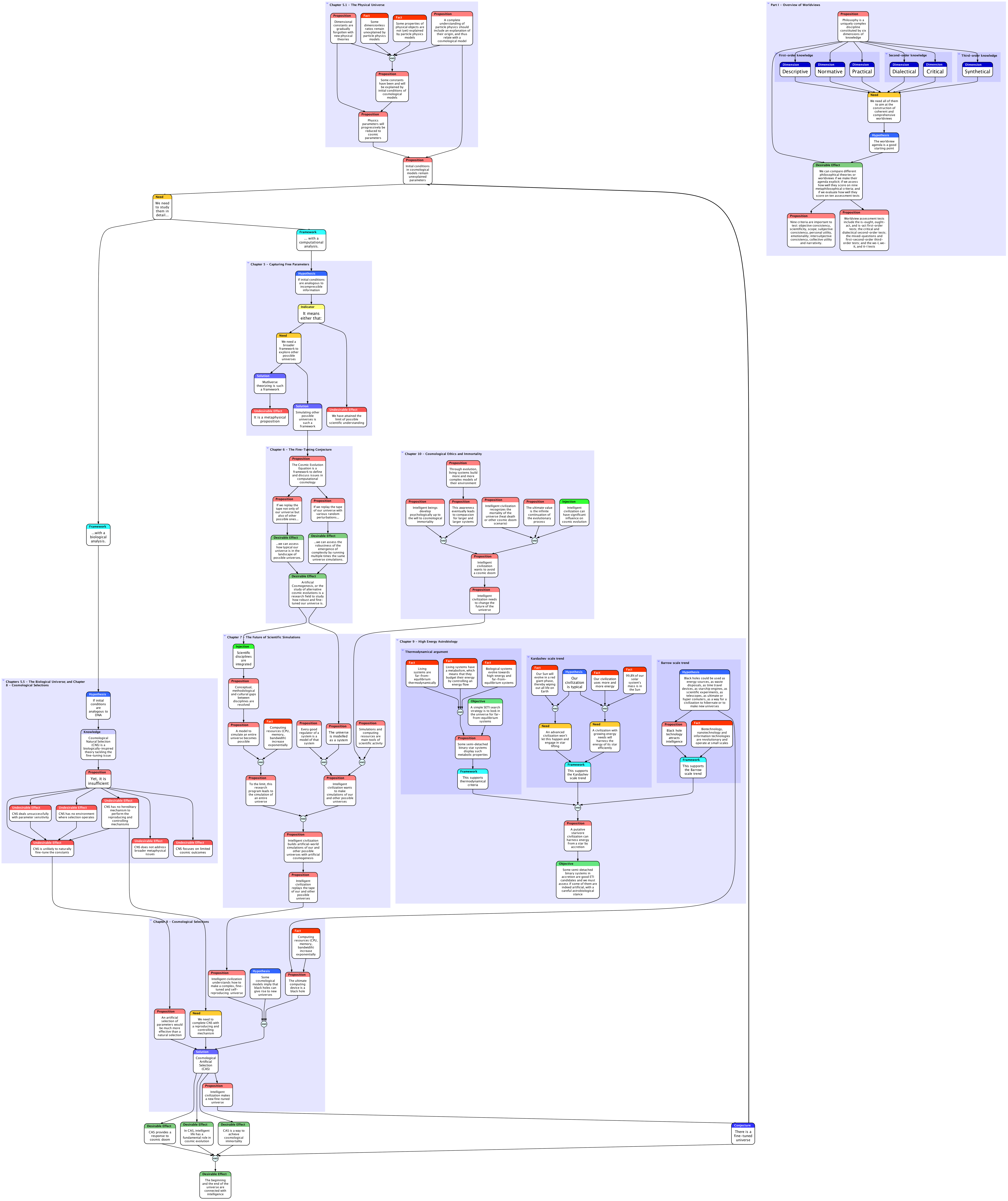

Part I – Overview of Worldviews
Proposition
Philosophy is a uniquely complex discipline constituted by six dimensions of knowledge
First-order knowledge
Second-order knowledge
Third-order knowledge
Dimension
Descriptive
Dimension
Normative
Dimension
Practical
Dimension
Dialectical
Dimension
Critical
Dimension
Synthetical
Need
We need all of them to aim at the construction of coherent and comprehensive worldviews
Hypothesis
The worldview agenda is a good starting point
Desirable Effect
We can compare different philosophical theories or worldviews if we make their agenda explicit; if we assess how well they score on nine metaphilosophical criteria; and if we evaluate how well they score on ten assessment tests
Proposition
Nine criteria are important to test: objective consistency, scientificity, scope; subjective consistency, personal utility, emotionality; intersubjective consistency, collective utility and narrativity.
Proposition
Worldview assessment tests include the is–ought, ought–act, and is–act first-order tests; the critical and dialectical second-order tests; the mixed-questions and first–second–order third-order tests; and the we–I, we–it, and it–I tests
Chapter 5.1 – The Physical Universe
Proposition
Dimensional constants are gradually forgotten with new physical theories
Fact
Some dimensionless ratios remain unexplained by particle physics models
Fact
Some properties of physical objects are not (yet) explained by particle physics models
Proposition
A complete understanding of particle physics should include an explanation of their origin, and thus relate with a cosmological model
AND
Proposition
Some constants have been and will be explained by initial conditions of cosmological models
Proposition
Physics parameters will progressively be reduced to cosmic parameters
Proposition
Initial conditions in cosmological models remain unexplained parameters
Need
We need to study them in detail...
Framework
... with a computational analysis.
Chapter 5 – Capturing Free Parameters
Hypothesis
If initial conditions are analogous to incompressible information
Indicator
It means either that:
Need
We need a broader framework to explore other possible universes
Solution
Mutliverse theorizing is such a framework
Undesirable Effect
It is a metaphysical proposition
Solution
Simulating other possible universes is such a framework
Undesirable Effect
We have attained the limit of possible scientific understanding
Chapter 6 – The Fine-Tuning Conjecture
Proposition
The Cosmic Evolution Equation is a framework to define and discuss issues in computational cosmology
Proposition
If we replay the tape not only of our universe but also of other possible ones...
Proposition
If we replay the tape of our universe with various random perturbations...
Desirable Effect
...we can assess how typical our universe is in the landscape of possible universes.
Desirable Effect
...we can assess the robustness of the emergence of complexity by running multiple times the same universe simulations.
Desirable Effect
Artificial Cosmogenesis, or the study of alternative cosmic evolutions is a research field to study how robust and fine-tuned our universe is.
Chapter 10 – Cosmological Ethics and Immortality
Proposition
Through evolution, living systems build more and more complex models of their environment
Proposition
Intelligent beings develop psychologically up to the will to cosmological immortality
Proposition
This awareness eventually leads to compassion for larger and larger systems
Proposition
Intelligent civilization recognizes the mortality of the universe (heat death or other cosmic doom scenario)
Proposition
The ultimate value is the infinite continuation of the evolutionary process
Injection
Intelligent civilization can have significant influence on cosmic evolution
AND
AND
Proposition
Intelligent civilization wants to avoid a cosmic doom
Proposition
Intelligent civilization needs to change the future of the universe
Framework
...with a biological analysis.
Chapter 7 – The Future of Scientific Simulations
Injection
Scientific disciplines are integrated
Proposition
Conceptual, methodological and cultural gaps between disciplines are resolved
Proposition
A model to simulate an entire universe becomes possible
Fact
Computing resources (CPU, memory, bandwidth) increase exponentially
Proposition
Every good regulator of a system is a model of that system
Proposition
The universe is modelled as a system
Proposition
Simulations and computing resources are main tools of scientific activity
AND
AND
AND
Proposition
To the limit, this research program leads to the simulation of an entire universe
Proposition
Intelligent civilization wants to make simulations of our and other possible universes
AND
Proposition
Intelligent civilization builds artificial-world simulations of our and other possible universes with artificial cosmogenesis
Proposition
Intelligent civilization replays the tape of our and other possible universes
Chapter 9 – High Energy Astrobiology
Thermodynamical argument
Fact
Living systems are far-from-equilibrium thermodynamically
Fact
Living systems have a metabolism, which means that they budget their energy by controlling an energy flow
Fact
Biological systems evolve towards high-energy and far-from-equilibrium systems
AND
Objective
A simple SETI search strategy is to look in the universe for far-from-equilibrium systems
Proposition
Some semi-detached binary star systems display such metabolic properties
Framework
This supports thermodynamical criteria
Kardashev scale trend
Fact
Our Sun will evolve in a red giant phase, thereby wiping out all life on Earth
Hypothesis
Our civilization is typical
Fact
Our civilization uses more and more energy
Fact
99,8% of our solar system's mass is in the Sun
AND
AND
Need
An advanced civilization won't let this happen and engage in star lifting
Need
A civilization with growing energy needs will harness the energy of its star efficiently
Framework
This supports the Kardashev scale trend
Barrow scale trend
Hypothesis
Black holes could be used as energy sources, as waste disposals, as time travel devices, as starship engines, as scientific experiments, as telescopes, as ultimate or hyper comuters, as a way for a civilization to hibernate or to make new universes
Proposition
Black hole technology attracts intelligence
Fact
Biotechnology, nanotechnology and information technologies are revolutionary and operate at small scales
Framework
This supports the Barrow scale trend
AND
Proposition
A putative starivore civilization can harness energy from a star by accretion
Objective
Some semi-detached binary systems in accretion are good ETI candidates and we must assess if some of them are indeed artificial, with a careful astrobiological stance
Chapters 5.5 – The Biological Universe; and Chapter 8 – Cosmological Selections
Hypothesis
If initial conditions are analogous to DNA
Knowledge
Cosmological Natural Selection (CNS) is a biologically-inspired theory tackling the fine-tuning issue
Proposition
Yet, it is insufficient
Undesirable Effect
CNS deals unsuccessfully with parameter sensitivity
Undesirable Effect
CNS has no environment where selection operates
Undesirable Effect
CNS has no hereditary mechanism to perform the reproducing and controlling mechanisms
Undesirable Effect
CNS is unlikely to naturally fine-tune the constants
Undesirable Effect
CNS does not address broader metaphysical issues
Undesirable Effect
CNS focuses on limited cosmic outcomes
Chapter 8 – Cosmological Selections
Fact
Computing resources (CPU, memory, bandwidth) increase exponentially
Proposition
Intelligent civilization understands how to make a complex, fine-tuned and self-reproducing universe
Hypothesis
Some cosmological models imply that black holes can give rise to new universes
Proposition
The ultimate computing device is a black hole
Proposition
An artificial selection of parameters would be much more effective than a natural selection
Need
We need to complete CNS with a reproducing and controlling mechanism
AND
Solution
Cosmological Artificial Selection (CAS)
Proposition
Intelligent civilization makes a new fine-tuned universe
Desirable Effect
CAS provides a response to cosmic doom
Desirable Effect
In CAS, intelligent life has a fundamental role in cosmic evolution
Desirable Effect
CAS is a way to achieve cosmological immortality
Conjecture
There is a fine-tuned universe
AND
Desirable Effect
The beginning and the end of the universe are connected with intelligence

# Appendix III – Revision History

This section will track the main changes of this document, as is practiced in software development. I will be glad to acknowledge you here if you help to correct or improve this essay in one way or an other.

| Revision | Date | Description |
|---|---|---|
| 1.0 | 12/2012 | First public version. |
| 1.0.1 | 05/01/2013 | Abstract improved; thanks to Martin Monperrus! ArXiv v1 |
| 1.0.2 | 10/01/2013 | The distinction in section 6.1 between multiverse fine-tuning argument and theological fine-tuning argument was made, thanks to Jonathan Colvin. |
| 1.0.3 | 16/01/2013 | Typos fixed. |
| 1.1 | 04/02/2013 | Abstracts at the beginning of each chapter added, more typos fixed, definition of starivores made more accurate with the word "actively". Reference to (R. L. Kuhn 2007) added to the metaphysical challenge. Link to original argumentative maps added. |
| 1.1.1 | 06/02/2013 | A short paragraph on directed panspermia as a way to achieve cosmological immortality was added. Figure numbering problem of version 1.1 fixed. |
| 1.1.2 | 12/04/2013 | Typos corrected. The gamma ray resiliency research proposal was corrected (at the end of Chapter 9). Gamma rays are actually unaffected by magnetic fields, so other resiliency mechanisms have yet to be found, different from a strong magnetic field. Thanks to Bernard Goossens! |
| 2.0 | 06/06/2013 | Version after official PhD defense at the Vrije Universiteit Brussel. New cover. Reflections on different kinds of speculations were developed in the preface. An open question about the way to philosophical worldviews was added. My conclusion on the fine-tuning conjecture was expanded to contrast it with the positions of Stenger and Barnes. The non-absolutist aspect of cosmological ethics was emphasized more in Chapter 10. Criticisms to Weikart were added. More typos were fixed, and other improvements were made. ArXiv v2. |

Table 20 - Revision history